\documentclass[11pt,a4paper]{article}
\usepackage{isabelle,isabellesym}

\usepackage{mdwlist}

\usepackage{focus}
\usepackage{pdfsetup}

\isabellestyle{it}

\begin{document}

 \newcommand{\isah}{Isabelle/HOL\xspace}
\newcommand{\tlangle}{\text{\textlbrackdbl}}
\newcommand{\trangle}{\text{\textrbrackdbl}}
\newcommand{\ntlangle}{{\normalfont\text{\textlbrackdbl}}}
\newcommand{\ntrangle}{{\normalfont\text{\textrbrackdbl}}}

\newcommand{\fconsts}{\ (\nconstant c_1,...,c_n \in Config)}
\newcommand{\fconst}{\ (\nconstant c \in Config)}

\newcommand{\mybox}{~\\
\vspace{-15mm}\\ 
\begin{flushright} {\ensuremath \Box} \end{flushright}}

\renewcommand{\labelitemii}{$\diamond$} 
\renewcommand{\labelitemiii}{$\ast$}
\renewcommand{\labelitemiv}{$\cdot$}
 
\newcommand{\ist}[1]{{\isastyle #1}}
\newcommand{\tbf}[1]{\textbf{#1}}
\newcommand{\tit}[1]{\textit{#1}}
\newcommand{\msfn}[1]{\textsf{\small#1}}
\newcommand{\msf}[1]{\textsf{\footnotesize#1}}
\newcommand{\mbsf}[1]{\textsf{\textbf{\footnotesize#1}}}
\newcommand{\pfeil}{\ding{229}}
\newcommand{\erweit}{\ding{239}}
\newcommand{\dto}{\Rightarrow}
\newcommand{\lto}{\longrightarrow}
\newcommand{\andl}{\wedge}
\newcommand{\orl}{\vee}
\newcommand{\isabh}{Isabelle/HOL\xspace}
\newcommand{\timesync}{time-synchronous\xspace}
\newcommand{\istream}{\msf{inStream}}
\newcommand{\ostream}{\msf{outStream}}
\newcommand{\lstream}{\msf{locStream}}
\newcommand{\isheaf}{\msf{inSheaf}}
\newcommand{\osheaf}{\msf{outSheaf}}
\newcommand{\sinout}[1]{\msf{#1\_InOut}}
\newcommand{\fr}{FlexRay\ }
 
\newcommand{\autofocusString}{\textsc{AutoFocus}}
\newcommand{\autofocus}{\autofocusString\xspace}
\newcommand{\af}{\autofocus}
\newcommand{\autofocusIII}{\autofocus~3\xspace}
\newcommand{\afIII}{\autofocusIII}
\newcommand{\aft}{\autofocusIII}
 
\newcommand{\ntimedeq}[2]{\ensuremath{#1 \dashrightarrow #2}}
\newcommand{\ntimedeqd}[3]{\ensuremath{#1 \stackrel{#3}{\dashrightarrow} #2}} 
\newcommand{\nmod}[2]{\ensuremath{\mathsf{mod}(#1,#2)}}

\newcommand{\isabsem}[1]{\textlbrackdbl\xspace$#1$\xspace\textrbrackdbl$_{Isab}$\xspace}
\newcommand{\isabsemm}[1]{\ensuremath{[\![#1]\!]_{Isab}}}

\newcommand{\ttl}[1]{\textsf{ttl}(#1)} 
\newcommand{\ftfin}[1]{\textsf{fti}\ensuremath{^\textsf{fin}}(#1)}
\newcommand{\ftinf}[1]{\textsf{fti}\ensuremath{^\textsf{inf}}(#1)}
\newcommand{\indffti}[1]{\textsf{ind}_\textsf{fti}^\textsf{fin}(#1)}
\newcommand{\indifti}[1]{\textsf{ind}_\textsf{fti}^\textsf{inf}(#1)}
\newcommand{\ndrop}[2]{\ensuremath{{{#1}\uparrow_{#2}}}}
\newcommand{\infdisjS}[1]{\textsf{disj}\ensuremath{^\textsf{inf}_\textsf{S}(#1)}}
\newcommand{\infdisj}[1]{\textsf{disj}\ensuremath{^\textsf{inf}(#1)}}

\title{Stream processing components:\\ Isabelle/HOL formalisation and case studies}
\author{Maria Spichkova}
\maketitle

\begin{abstract}
This set of theories presents an Isabelle/HOL+Isar formalisation  of stream processing components introduces in  
  \Focus, 
a framework for formal specification and development of interactive systems.
This is an extended and updated version of the formalisation, which was 
elaborated within the methodology ``\Focus on Isabelle''. 
In addition, we also applied the formalisation on three case studies 
that cover different application areas: 
process control (Steam Boiler System),
data transmission (FlexRay communication protocol),  
memory and processing components (Automotive-Gateway System). 
\end{abstract}
\tableofcontents

\newpage
\section{Introduction}
The set of theories presented in this paper is an extended and updated Isabelle/HOL+Isar \cite{npw,IsabelleManual} formalisation of stream processing components 
elaborated within the methodology ``\Focus on Isabelle'' \cite{spichkova}. 
This paper is organised as follows: in the first section we give a general introduction to 
the \Focus stream processing components \cite{focus} and briefly describe three case studies 
 to show how the formalisation can be used for specification and verification of system properties applying the idea of refinement-based verification~\cite{ArchReqDecRef}.
After that we present the Isabelle/HOL representation of these concepts and a number of auxiliary theories on lists and natural numbers useful for the proofs in the case studies. 
The last three sections introduce the case studies, where system properties are verified formally using the Isabelle theorem prover. 

This approach can be used as a basis for the abstract modelling level within the development of cyber-physical systems, suggested in our other work \cite{Spichkova_Campetelli2012,issec_cps2013}.

\subsection{Stream processing components}

The central concept in \Focus is a \emph{stream} representing a 
communication history of  a \emph{directed channel} between components. 
A system in \Focus is specified by its components that are 
connected by channels, 
and are described in terms of its input/output behavior.   
The channels in this specification framework are \emph{asynchronous communication links} 
without delays. They are \emph{directed} and generally assumed to be \emph{reliable},
 and \emph{order preserving}. Via these channels components
 exchange information in terms of \emph{messages} of specified types. 
 For any set of messages $M$,  
$M^\infty$ and $M^*$ denote  the sets of all infinite and all finite untimed
streams respectively:
\[ 
\begin{array}{lclcl} 
M^\infty \stackrel{\mathrm{def}}{=} \mathbb{N}_{+} \to M
&  &
M^* \stackrel{\mathrm{def}}{=} {\cup}_{n \in \mathbb{N}}([1..n]\to M)
\end{array}\]
A \emph{timed stream}, as suggested in  our previous work~\cite{spichkova},  
is represented by a sequence of \emph{time intervals} counted from 0, each of them is a finite sequence of messages that are listed in their order of
transmission:  %
\[ \begin{array}{lclcl}
M^{\underline{\infty}} \stackrel{\mathrm{def}}{=} 
\mathbb{N}_+ \to M^* 
&&
M^{\underline{*}} \stackrel{\mathrm{def}}{=} 
\cup_{n \in \mathbb{N}}([1..n]\to M^* )
\end{array}
\]
A specification can be elementary or composite -- composite specifications are
built hierarchically from the elementary ones. 
Any specification characterises the relation between the
\emph{communication histories} for the external \emph{input} and \emph{output channels}: 
the formal meaning of a specification is exactly the \emph{input/output relation}. 
This is specified by the lists of input and output channel identifiers, $I$ and $O$, while
the syntactic interface of the specification $S$ is denoted by $\nint{I_S}{O_S}$. 

To specify the behaviour of a real-time system we  use 
\emph{infinite timed streams} to represent the input and the output streams. 
The type of \emph{finite timed streams} will be used 
only if some argumentation about a timed stream that was truncated 
at some point of time is needed. 
The type of \emph{finite untimed streams} will be used to argue about a sequence of messages 
that are transmitted during a time interval.
The type of \emph{infinite untimed streams} will be used in the case of timed specifications 
only to represent local variables of \Focus specification.
Our definition in Isabelle/HOL of corresponding types is given below: 
\begin{itemize*}
\item 
Finite timed streams of type  \ist{$'$a} are represented by the type \ist{$'$a fstream},  
which is an abbreviation for the type \ist{'a list list}.
\item 
Finite untimed streams of type  \ist{$'$a} are represented by the list type:~
\ist{$'$a~list}.
\item 
Infinite timed streams of type  \ist{$'$a} are represented by the type \ist{$'$a istream},  
which represents the functional type \ist{nat $\dto$ $'$a list}.
\item 
Infinite untimed streams of type  \ist{$'$a} are represented by 
the functional type \ist{nat $\dto$ $'$a}.
\end{itemize*}
All the operators defined in the presented theories are based on the standard Isabelle/HOL library.

\subsection{Case Study 1: Steam Boiler System}

A steam boiler control system  can be represent as a distributed system 
consisting of a number of communicating components and must fulfil real time requirements. 
This case study shows how we can deal with local variables (system's states) and 
in which way we can represent mutually recursive functions to avoid problems in proofs.  
The main idea of the steam boiler specification was taken from \cite{focus}: 
The steam boiler has a water tank, which contains a number of gallons of water, and
a pump, which adds $10$ gallons of water per time unit to its water tank, 
if the pump is on. At most $10$ gallons of water are consumed per time unit by
the steam production, if the pump is off.
The steam boiler has a sensor that measures the water level. 

We specified the following components: \emph{ControlSystem} (general requirements specification), 
\emph{ControlSystemArch} (system architecture),  
\emph{SteamBoiler}, \emph{Converter}, and \emph{Controller}. 
We present here the following \isah theories for this system:
\begin{itemize*} 
	\item \ist{SteamBoiler.thy} --  specifications of the system components, 
	\item \ist{SteamBoiler\_proof} --  proof of refinement relation between the requirements and the architecture specifications.
\end{itemize*}
The specification \emph{ControlSystem} describes  
the requirements for the steam boiler system: 
in each time interval the system outputs it current water level in gallons
 and this level should always be between $200$ and $800$ gallons 
 (the system works in the  time-synchronous manner). 
 
The specification \emph{ControlSystemArch} describes a general  
architecture of the steam boiler system. 
The system consists of three components: a steam boiler,
a converter, and a controller. \\

{\footnotesize
\begin{spec}{\spc{ControlSystemArch}}{gb}
\centering
\includegraphics[width=7cm]{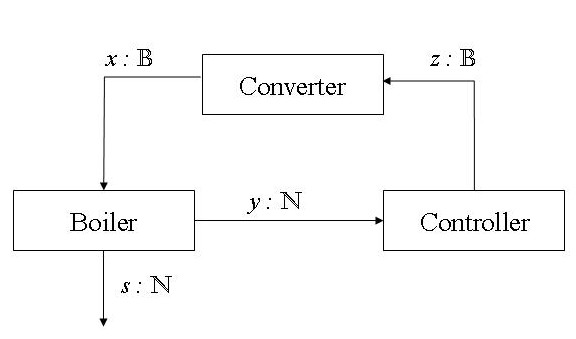}
\end{spec}
}

The  \emph{SteamBoiler}  component works in time-synchronous manner: 
the current water level is controlled every time interval.
The boiler has two output channels with equal streams ($y = s$) and it
 fixes the initial water level to be $500$ gallons.
For every point of time the following must be true: if the pump is off, 
the boiler consumes at most $10$ gallons of water, 
otherwise (the pump is on) at most $10$ gallons of water will be added to its water tank.

The  \emph{Converter}   component converts the asynchronous output produced by the controller
to time-synchronous input for the steam boiler. 
Initially the pump is off, 
and at every later point of time (from receiving the first instruction 
from the controller) the output will be the last input from the controller. 
The   \emph{Controller}  component, 
contrary to the steam boiler component, 
 behaves in a purely asynchronous manner to keep
the number of control signals small, it means it might not be
desirable to switch the pump on and off more often than necessary.
The controller is responsible for
switching the steam boiler pump on and off. 
If the pump is off: 
if the current water level is above $300$ gallons 
the pump stays off, otherwise the pump is started 
and will run until the water level reaches $700$ gallons. 
If the pump is on: if the current water level is below $700$ gallons 
the pump stays on, otherwise the pump is turned off 
and will be off until the water level reaches $300$ gallons.

To show that the specified system fulfills the requirements we need to show that
the specification \emph{ControlSystemArch} is a refinement of the specification \emph{ControlSystem}.
It follows from the definition of behavioral refinement (cf. \cite{broy_refinement2}) that in order to verify that 
$
ControlSystem~ \leadsto ~ ControlSystemArch
$
it is enough to prove that
$$
\tlangle ControlSystemArch\trangle~ \Rightarrow ~\tlangle ControlSystem\trangle
$$
Therefore, we have to  prove a \emph{lemma}  
that says the specification \emph{ControlSystemArch} is a refinement of the specification \emph{ControlSystem}:\\

\begin{isabellebody}%
\isacommand{lemma}\isamarkupfalse%
\ L{\isadigit{0}}{\isacharunderscore}ControlSystem{\isacharcolon} 
{\isasymlbrakk}\ ControlSystemArch\ s{\isasymrbrakk}\ {\isasymLongrightarrow}\ ControlSystem\ s\\
\end{isabellebody}%


\subsection{Case Study 2: FlexRay Communication Protocol}

In this section we present a case study on FlexRay, 
communication protocol for safety-critical real-time applications. 
This protocol has been developed by the \fr Consortium \cite{FlexRayConsortium} 
for embedded systems in vehicles, and its advantages  
are deterministic real-time message transmission, fault tolerance, 
integrated functionality for clock synchronisation and higher bandwidth.

\fr contains a set of complex algorithms to provide the communication
services. From the view of the software layers above \fr only a few 
of these properties become visible. The most important ones are static 
cyclic communication schedules and system-wide synchronous clocks. 
These provide a suitable platform for distributed control algorithms 
as used e.g.\ in drive-by-wire applications. The formalization described 
here is based on the ``Protocol Specification 2.0''\cite{FlexRayProt}.
 
The static message transmission model of \fr is based
on \textit{rounds}. \fr rounds consist of a constant number of
time slices of the same length, so called \emph{slots}.
A node can broadcast its messages to other nodes at
statically defined slots. At most one node can
do it during any slot.

For the formalisation of \fr in \Focus we would like to refer to  \cite{efts_book,spichkova2006flexray,kuhnel2006upcoming}. 
To reduce the complexity of the system several aspects of \fr have been abstracted in this formalisation:
\begin{itemize}
\item[(1)] There is no clock synchronization or start-up phase since clocks are assumed to be synchronous. 
This corresponds very well with the \emph{time-synchronous} notion of \Focus.
\item[(2)] The model does not contain bus guardians that protect channels on the physical layer from interference caused by communication that is not aligned with FlexRay schedules.
\item[(3)] Only the static segment of the communication cycle has been included not the dynamic, 
as we are mainly interested in time-triggered systems.
\item[(4)] The time-basis for the system is one slot i.e.\ one slot \fr corresponds to one tick in in the formalisation.
\item[(5)] The system contains only one \fr channel. Adding a second channel 
would mean simply doubling the \fr component with a different configuration 
and adding extra channels for the access to the \emph{CNI\_Buffer} component.
\end{itemize}
The system architecture consists of the following components, 
which describe the \fr components accordingly to the \fr standard~\cite{FlexRayProt}: 
\begin{itemize*}
	\item \emph{FlexRay} (general requirements specification), 
	\item \emph{FlexRayArch} (system architecture),   
	\item \emph{FlexRayArchitecture} 
(guarantee part of the system architecture), 
  \item \emph{Cable}, 
  \item \emph{Controller}, 
  \item \emph{Scheduler}, and 
  \item \emph{BusInterface}. 
  \end{itemize*}
We present the following \isah theories in this case study:
\begin{itemize*}
  \item \ist{FR\_types.thy} -- datatype definitions, 
	\item \ist{FR.thy} --  specifications of the system components and auxiliary functions and predicates, 
	\item \ist{FR\_proof} --   proof of refinement relation between the requirements and the architecture specifications.
\end{itemize*}
The type \emph{Frame} that describes a \fr frame 
 consists of a slot identifier of type \Nat~and the payload.
The type of payload is defined as a finite list of type \emph{Message}.
  The type \emph{Config} represents the bus configuration and contains the 
scheduling table \emph{schedule} of a node and the length 
of the communication round \emph{cycleLength}. 
A scheduling table of a node consists of a number of slots 
in which this node should be sending a frame with the corresponding identifier
(identifier that is equal to the slot).
\[
\begin{array}{lcl}
	\ntype Message & = & msg~(message\_id: \Nat, ftcdata : Data) \\
	\ntype Frame  & = & frm~(slot : \Nat, data : Data) \\
	\ntype Config & = & conf~(schedule: \nfst{\Nat}, cycleLength: \Nat )
\end{array}
\]
We do not specify the type \emph{Data} here to have a polymorphic specification 
of \fr (this type can be underspecified later to any datatype), therefore, in \isah it will be also defined as a polymorphic type \ist{$'$a}. 
The types \ist{$'$a~nFrame}, \ist{nNat} and \ist{nConfig} 
are used to represent sheaves of channels of types 
\emph{Frame}, \Nat~and \emph{Config} respectively.
In the specification group will be used 
channels \emph{recv} and \emph{activations}, as well as sheaves of channels 
(\emph{return$_1$, \dots,return$_n$}), ($c_1, \dots, c_n$), 
(\emph{store}$_1$, \dots, \emph{store}$_n$), (\emph{get}$_1$, \dots, \emph{get}$_n$), and 
(\emph{send}$_1$, \dots, \emph{send}$_n$).
We also need to declare some constant, \ist{sN}, for the number of specification replication and the corresponding number of channels in sheaves, as well as to define the list of sheaf upper 
bounds, \ist{sheafNumbers}.

The architecture of the \fr communication protocol  is specified
as the \Focus specification \emph{FlexRayArch}. 
Its assumption-part consists of three constraints:
(i) all bus configurations have disjoint scheduling tables, 
(ii) all bus configurations have the equal length of the communication round,
(iii) each \fr controller can receive tab most one data frame each time interval from the environment' of the \fr system.
The guarantee-part of \emph{FlexRayArch} is represented by the specification 
\emph{FlexRayArchitecture} (see below).  
  
{\footnotesize
\begin{spec}{\spc{FlexRayArch} \fconsts}{td}
\InOut{return_1 ,..., return_n : Frame}{store_1 ,..., store_n : Frame ;	get_1 ,..., get_n : \Nat}
\begin{array}{ll}
\uasm 
 &\forall i \in [1..n]: \msg{1}{return_i}\\
 &	DisjointSchedules(c_1 ,\dots, c_n)\\
 & IdenticCycleLength(c_1 ,\dots, c_n) \\
\ugar
 & (store_1 ,\dots, store_n, get_1 ,\dots, get_n) := \\
 & ~~~~~~FlexRayArchitecture(c_1 ,\dots, c_n)(return_1 ,dots, return_n)
\end{array}
\end{spec}
}

{\footnotesize
\begin{spec}{\spc{FlexRayArchitecture}\fconsts}{gb}
\centering
\includegraphics[width=8cm]{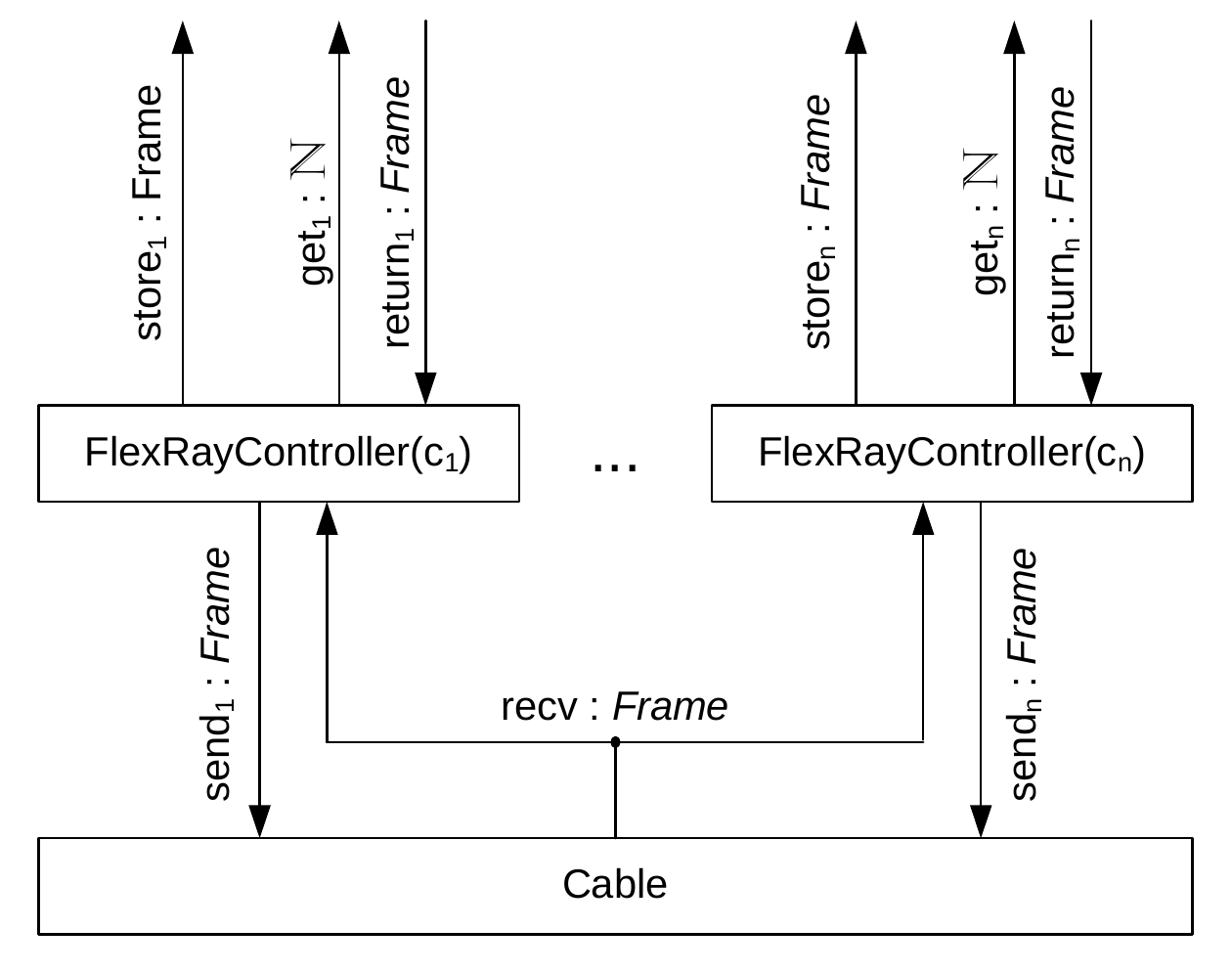}
\end{spec}
}

\noindent
The component \emph{Cable} simulate the broadcast properties of the 
physical network cable -- 
every received \fr frame is resent to all connected nodes. 
Thus, if one \emph{FlexRayController} send some frame, 
this frame will be resent to all nodes 
(to all \emph{FlexRayController}s of the system).
The assumption is that all input streams of the component \emph{Cable} 
 are disjoint -- this holds by the properties of the 
\emph{FlexRayController} components and the overall system assumption that 
the scheduling tables of all nodes are disjoint.  
The guarantee is specified by the predicate \emph{Broadcast}. 

The \Focus specification \emph{FlexRayController} represent the controller component 
for a single node of the system. It consists of the components \emph{Scheduler} 
and \emph{BusInterface}. 
The \emph{Scheduler} signals the \emph{BusInterface}, that is responsible for 
the interaction with other nodes of the system (i.e. for the real send and receive of frames),  
on which time which \fr frames must be send from the node. 
The \emph{Scheduler} describes the communication scheduler. 
It sends at every time $t$ interval, which is equal modulo the length 
of the communication cycle to some \fr frame identifier (that corresponds to the number of the slot in the communication round) from the scheduler table, 
this frame identifier.  

The  specification \emph{FlexRay}  represents requirements on the protocol: 
If the scheduling tables are correct in terms of the predicates 
\emph{Disjoint\-Schedules} (all bus configurations have disjoint scheduling tables) 
and \emph{Identic\-Cycle\-Length} (all bus configurations have the equal length of the communication round), and 
also the \fr component receives in every time interval at most one message from each node 
(via channels $return_i$, $1 \le i \le n$), then
\begin{itemize}
	\item 
	the frame transmission by \fr must be correct in terms of the predicate 
	\emph{FrameTransmission}: if the time $t$ is equal modulo the length of the cycle (\fr communication round) 
to the element of the scheduler table of the node $k$, then this and only this node 
can send a data atn the $t$th time interval;
	\item 
	\fr component sends in every time interval at most one message to each node 
 via channels $get_i$ and $store_i$, $1 \le i \le n$).
\end{itemize}
To show that the specified system fulfill the requirements we need to show that
the specification \emph{FlexRayArch} is a refinement of the specification \emph{FlexRay}. 
It follows from the definition of behavioral refinement that in order to verify that 
$
FlexRay~ \leadsto ~ FlexRayArch
$
it is enough to prove that
$$
\tlangle FlexRayArch\trangle~ \Rightarrow ~\tlangle FlexRay\trangle
$$
Therefore, we have to define and to prove a lemma, that says the specification 
\emph{FlexRayArch} is a refinement of the specification \emph{FlexRay}:
\\

\begin{isabellebody}%
\isacommand{lemma}\isamarkupfalse%
\ main{\isacharunderscore}fr{\isacharunderscore}refinement{\isacharcolon}\isanewline
FlexRayArch\ n\ nReturn\ nC\ nStore\ nGet\ 
{\isasymLongrightarrow}\ FlexRay\ n\ nReturn\ nC\ nStore\ nGet\isanewline
\end{isabellebody}%


\subsection{Case Study 3: Automotive-Gateway}

This section introduces the case study on telematics 
(electronic data transmission) gateway that was done for the 
Verisoft project\cite{verisoft}. 
If the gateway receives from a ECall application of a vehicle
a signal about crash 
(more precise, the command to initiate the call to the Emergency Service Center, ESC), 
and after the establishing the connection it receives 
 the command to send the crash data, received from sensors.
These data are restored 
 in the internal buffer of the gateway  and should  
 be resent to the ESC and 
 the voice communication will be established, assuming that
	there is no connection fails.  
The system description consists of the following specifications: 
\begin{itemize*}
	\item \emph{GatewaySystem} (gateway system architecture), 
	\item \emph{GatewaySystemReq} (gateway system requirements), 
	\item \emph{ServiceCenter} (Emergency Service Center),
	\item \emph{Gateway} (gateway architecture), 
	\item \emph{GatewayReq} (gateway requirements),
	\item \emph{Sample} (the main  component describing its logic), 
	\item \emph{Delay} (the  component   modelling the communication delay), and
	\item \emph{Loss} (the  component modelling the communication loss).  
\end{itemize*}
We present the following \isah theories in this case study:
\begin{itemize*}
  \item \ist{Gateway\_types.thy} --  datatype definitions, 
	\item \ist{Gateway.thy} --  specifications of the system components, 
	\item \ist{Gateway\_proof} --  proofs of refinement relations between the requirements and the architecture specifications (for the components \emph{Gateway} and \emph{GatewaySystem}).
\end{itemize*}
The datatype \emph{ECall\_Info} represents a tuple, 
consisting of the data that the Emergency Service Center needs -- here we specify these data to contain the vehicle coordinates and the collision speed, they can also extend by some other information.  
The datatype \emph{GatewayStatus} represents the status (internal state) of the gateway.
\[
\begin{array}{lcl}
  \ntype Coordinates &=& \Nat \times \Nat
  \\
  \ntype CollisionSpeed &=& \Nat
  \\
	\ntype ECall\_Info &=& ecall(coord \in Coordinates, speed \in CollisionSpeed)
	\\
	\ntype GatewayStatus &=& 
	\{~ init\_state,~ call,~ connection\_ok,\\
	&& ~ ~sending\_data,~ voice\_com ~\}
\end{array}
\]
To specify the automotive gateway we will use a number of datatypes 
consisting of one or two elements: 
$\{init, send\}$, $\{stop\_vc\}$, $\{vc\_com\}$ and $\{sc\_ack\}$. 
We name these types \emph{reqType}, \emph{stopType}, \emph{vcType} and \emph{aType} correspondingly.

The \Focus specification of the general gateway system architecture is presented below:

\begin{spec}{\spc{GatewaySystem}(\nconst d \in \Nat)}{gb}
\centering \includegraphics[scale=0.7]{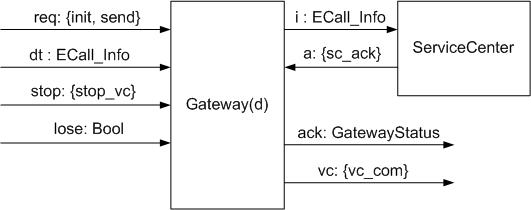}
\end{spec} 

\noindent
The stream \emph{loss} is specified to be a time-synchronous one (exactly one message each time interval). 
It represents the connection status: 
the message \ntrue~at the time interval $t$ corresponds to the connection 
failure at this time interval, the message \nfalse~at the time interval $t$ means that 
at this time interval no data loss on the gateway connection.

The specification \emph{GatewaySystemReq} specifies the requirements for the component \emph{GatewaySystem}:
Assuming that the input streams \emph{req} and \emph{stop} can contain at every time interval at most one message, and assuming that the stream \emph{lose} contains at every time interval exactly one message. 
If 
\begin{itemize*}
	\item 
	at any time interval $t$ the gateway system is in the initial state, 
	\item
	at time interval $t+1$ the signal about crash comes at first time 
	(more precise, the command to initiate the call to the ESC,
	\item 
	after $3+m$ time intervals the command to send the crash data comes at first time,
	\item 
	the gateway system has received until the time interval $t+2$ the crash data,  
	\item 
	there is no connection fails from the time interval $t$ until the time interval $t + 4 + k + 2d$,
	\end{itemize*}
then at time interval $t + 4 + k + 2d$ the voice communication is established.

The component \emph{ServiceCenter} represents the interface behaviour of the ESC 
(wrt. connection to the gateway): 
if at time $t$ a message about a vehicle crash comes, it acknowledges this event by sending 
the at time $t+1$ message \emph{sc\_ack} that represents the 
attempt to establish the voice communication with the driver or a passenger of the vehicle. 
if there is no connection failure, after $d$ time intervals the voice communication will be started. 

We specify the gateway requirements (\emph{GatewayReq}) as follows: 
	\begin{enumerate}
	\item 
	If at time $t$ the gateway is in the initial state $init\_state$, and it gets 
	the command to establish the connection with the central station, and also there is no 
	environment connection problems during the next 2 time intervals, 
	it establishes the connection at the time interval $t+2$. 
	\item 
	If at time $t$  the gateway has establish the connection, 
	and it gets the command to send the ECall data to the central station, and also there is no 
	environment connection problems during the next $d+1$ time intervals, 
	then it sends the last corresponding data. 
        The central station becomes these date at the time $t+d$.
	\item 
	If the gateway becomes the acknowledgment from the central station 
	that it has receives the sent ECall data, and also there is no 
	environment connection problems, then the voice communication is started.
  \end{enumerate}
 The specification of the gateway architecture, \emph{Gateway}, is parameterised one: 
the parameter $d \in \Nat$ denotes the communication delay 
between the central station and a vehicle. 
This component consists of three subcomponents: \emph{Sample}, \emph{Delay}, 
and \emph{Loss}: 
 
\begin{spec}{\spc{Gateway}(\nconst d \in \Nat)}{td}
\centering \includegraphics[scale=0.63]{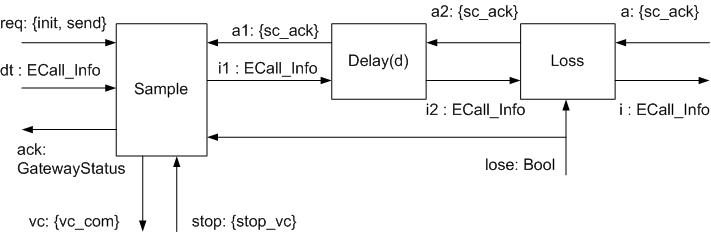}
\end{spec} 
  
\noindent
The component \emph{Delay} models the communication delay. 
Its specification is parameterised one: 
it inherits the parameter of the component \emph{Gateway}. 
This component simply delays all input messages on $d$ time intervals. 
During the first $d$ time intervals no output message will be produced. 

The component \emph{Loss} models the communication loss between 
the central station and the vehicle gateway: 
if during time interval $t$ from the component \emph{Loss} no message about a 
lost connection comes, the messages come during time interval 
$t$ via the input channels $a$ and $i2$ will be forwarded without any delay via channels 
$a2$ and $i$ respectively. 
Otherwise all messages come during time interval 
$t$ will be lost.  
  
The component \emph{Sample} represents the logic of the gateway component. 
If it receives from a ECall application of a vehicle 
 the command to initiate the call to the ESC it tries 
 to establish the connection.
If the connection is established, and the component \emph{Sample} receives 
from a ECall application of a vehicle
 the command to send the crash data, which were already received and stored 
 in the internal buffer of the gateway, 
 these data will be resent to the ESC. 
 After that this component waits to the acknowledgment from the ESC. 
 If the acknowledgment is received, the voice communication will be established, assuming that
	there is no connection fails.

For the component \emph{Sample} we have the assumption, that the 
streams $req$, $a1$, and $stop$ can contain at every time interval at most one message, 
and also that the stream $loss$ must contain at every time interval exactly one message. 
This component uses local variables \emph{st} and \emph{buffer} (more precisely, 
a local variable \emph{buffer} and a state variable \emph{st}). 
The guarantee part of  the component \emph{Sample} can be specified as a timed state transition diagram  (TSTS, cf. also \cite{spichkova2012towards}) 
and an expression which says how the local variable 
\emph{buffer} is computed, or using the corresponding table representation, which is semantically equivalent to the TSTD.

 \begin{figure}[h]
	\begin{center}
		\includegraphics[scale=0.7]{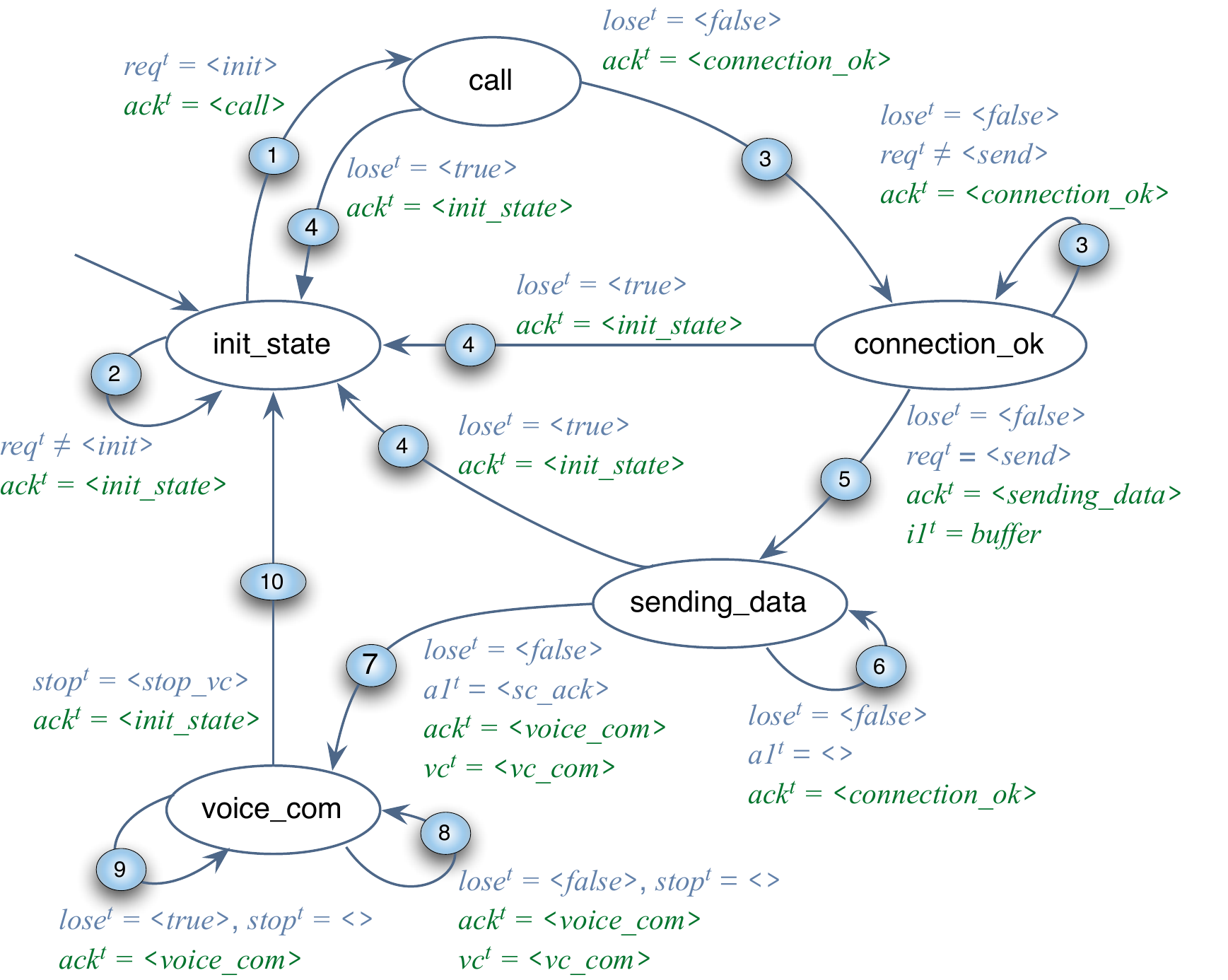}
	\end{center}
	\caption{Timed state transition diagram for the component Sample}
	\label{fig:gateway_std}
\end{figure}

  To show that the specified gateway architecture fulfils the requirements we need to show that
the specification \emph{Gateway} is a refinement of the specification \emph{GatewayReq}. 
Therefore, we need to define and to prove the following lemma:
\\

\begin{isabellebody}%
\isacommand{lemma}\isamarkupfalse%
\ Gateway{\isacharunderscore}L{\isadigit{0}}{\isacharcolon}\isanewline
\ Gateway\ req\ dt\ a\ stop\ lose\ d\ ack\ i\ vc\isanewline
\ {\isasymLongrightarrow} 
\ GatewayReq\ req\ dt\ a\ stop\ lose\ d\ ack\ i\ vc
\end{isabellebody}%

~\\
To show that the specified gateway architecture fulfills the requirements we need to show that
the specification \emph{GatewaySystem} is a refinement of the specification \emph{GatewaySystemReq}. 
Therefore, we need to define and to prove the following lemma:
\\

\begin{isabellebody}%
\isacommand{lemma}\isamarkupfalse%
\ GatewaySystem{\isacharunderscore}L{\isadigit{0}}{\isacharcolon}\isanewline
\ GatewaySystem\ req\ dt\ stop\ lose\ d\ ack\ vc\isanewline
\ {\isasymLongrightarrow} 
\ GatewaySystemReq\ req\ dt\ stop\ lose\ d\ ack\ vc\isanewline
\end{isabellebody}%

\parindent 0pt\parskip 0.5ex

\newpage
\begin{isabellebody}%
\def\isabellecontext{ArithExtras}%
\isamarkupheader{Theory ArithExtras.thy%
}
\isamarkuptrue%
\isadelimtheory
\endisadelimtheory
\isatagtheory
\isacommand{theory}\isamarkupfalse%
\ ArithExtras\ \isanewline
\isakeyword{imports}\ Main\ \isanewline
\isakeyword{begin}%
\endisatagtheory
{\isafoldtheory}%
\isadelimtheory
\endisadelimtheory
\ \isanewline
\isanewline
\isacommand{datatype}\isamarkupfalse%
\ natInf\ {\isacharequal}\ Fin\ nat\ \isanewline
\ \ \ \ \ \ \ \ \ \ \ \ \ \ \ \ {\isacharbar}\ Infty\ \ \ \ \ \ \ \ \ \ \ \ \ \ \ \ \ \ \ \ \ {\isacharparenleft}{\isachardoublequoteopen}{\isasyminfinity}{\isachardoublequoteclose}{\isacharparenright}\isanewline
\isacommand{primrec}\isamarkupfalse%
\isanewline
nat{\isadigit{2}}inat\ {\isacharcolon}{\isacharcolon}\ {\isachardoublequoteopen}nat\ list\ {\isasymRightarrow}\ natInf\ list{\isachardoublequoteclose}\isanewline
\isakeyword{where}\isanewline
\ \ {\isachardoublequoteopen}nat{\isadigit{2}}inat\ {\isacharbrackleft}{\isacharbrackright}\ {\isacharequal}\ {\isacharbrackleft}{\isacharbrackright}{\isachardoublequoteclose}\ {\isacharbar}\isanewline
\ \ {\isachardoublequoteopen}nat{\isadigit{2}}inat\ {\isacharparenleft}x{\isacharhash}xs{\isacharparenright}\ {\isacharequal}\ {\isacharparenleft}Fin\ x{\isacharparenright}\ {\isacharhash}\ {\isacharparenleft}nat{\isadigit{2}}inat\ xs{\isacharparenright}{\isachardoublequoteclose}\isanewline
\isadelimtheory
\isanewline
\endisadelimtheory
\isatagtheory
\isacommand{end}\isamarkupfalse%
\endisatagtheory
{\isafoldtheory}%
\isadelimtheory
\endisadelimtheory
\end{isabellebody}%

%
\begin{isabellebody}%
\def\isabellecontext{ListExtras}%
\isamarkupheader{Auxiliary Theory ListExtras.thy%
}
\isamarkuptrue%
\isadelimtheory
\endisadelimtheory
\isatagtheory
\isacommand{theory}\isamarkupfalse%
\ ListExtras\ \isanewline
\isakeyword{imports}\ Main\isanewline
\isakeyword{begin}%
\endisatagtheory
{\isafoldtheory}%
\isadelimtheory
\endisadelimtheory
\isanewline
\isanewline
\isacommand{definition}\isamarkupfalse%
\isanewline
\ \ disjoint\ {\isacharcolon}{\isacharcolon}\ {\isachardoublequoteopen}{\isacharprime}a\ list\ {\isasymRightarrow}\ {\isacharprime}a\ list\ {\isasymRightarrow}\ bool{\isachardoublequoteclose}\isanewline
\isakeyword{where}\isanewline
\ {\isachardoublequoteopen}disjoint\ x\ y\ {\isasymequiv}\ \ {\isacharparenleft}set\ x{\isacharparenright}\ {\isasyminter}\ {\isacharparenleft}set\ y{\isacharparenright}\ {\isacharequal}\ {\isacharbraceleft}{\isacharbraceright}{\isachardoublequoteclose}\isanewline
\isanewline
\isacommand{primrec}\isamarkupfalse%
\isanewline
\ \ mem\ {\isacharcolon}{\isacharcolon}\ \ {\isachardoublequoteopen}{\isacharprime}a\ {\isasymRightarrow}\ {\isacharprime}a\ list\ {\isasymRightarrow}\ bool{\isachardoublequoteclose}\ {\isacharparenleft}\isakeyword{infixr}\ {\isachardoublequoteopen}mem{\isachardoublequoteclose}\ {\isadigit{6}}{\isadigit{5}}{\isacharparenright}\isanewline
\isakeyword{where}\isanewline
\ \ {\isachardoublequoteopen}x\ mem\ {\isacharbrackleft}{\isacharbrackright}\ {\isacharequal}\ False{\isachardoublequoteclose}\ {\isacharbar}\isanewline
\ \ {\isachardoublequoteopen}x\ mem\ {\isacharparenleft}y\ {\isacharhash}\ l{\isacharparenright}\ {\isacharequal}\ {\isacharparenleft}{\isacharparenleft}x\ {\isacharequal}\ y{\isacharparenright}\ {\isasymor}\ {\isacharparenleft}x\ mem\ l{\isacharparenright}{\isacharparenright}{\isachardoublequoteclose}\isanewline
\isanewline
\isacommand{definition}\isamarkupfalse%
\isanewline
\ \ memS\ {\isacharcolon}{\isacharcolon}\ \ {\isachardoublequoteopen}{\isacharprime}a\ {\isasymRightarrow}\ {\isacharprime}a\ list\ {\isasymRightarrow}\ bool{\isachardoublequoteclose}\isanewline
\isakeyword{where}\isanewline
\ {\isachardoublequoteopen}memS\ x\ l\ \ {\isasymequiv}\ \ x\ {\isasymin}\ {\isacharparenleft}set\ l{\isacharparenright}{\isachardoublequoteclose}\isanewline
\isanewline
\isanewline
\isacommand{lemma}\isamarkupfalse%
\ mem{\isacharunderscore}memS{\isacharunderscore}eq{\isacharcolon}\ \ {\isachardoublequoteopen}x\ mem\ l\ {\isasymequiv}\ memS\ x\ l{\isachardoublequoteclose}\isanewline
\isadelimproof
\endisadelimproof
\isatagproof
\isacommand{proof}\isamarkupfalse%
\ {\isacharparenleft}induct\ l{\isacharparenright}\isanewline
\ \ \isacommand{case}\isamarkupfalse%
\ Nil\isanewline
\ \ \isacommand{from}\isamarkupfalse%
\ this\ \isacommand{show}\isamarkupfalse%
\ {\isacharquery}case\isanewline
\ \ \isacommand{by}\isamarkupfalse%
\ {\isacharparenleft}simp\ add{\isacharcolon}\ memS{\isacharunderscore}def{\isacharparenright}\isanewline
\isacommand{next}\isamarkupfalse%
\isanewline
\ \ \ \ \isacommand{fix}\isamarkupfalse%
\ a\ la\ \isacommand{case}\isamarkupfalse%
\ {\isacharparenleft}Cons\ a\ la{\isacharparenright}\isanewline
\ \ \ \ \isacommand{from}\isamarkupfalse%
\ Cons\ \isacommand{show}\isamarkupfalse%
\ {\isacharquery}case\isanewline
\ \ \ \ \isacommand{by}\isamarkupfalse%
\ {\isacharparenleft}simp\ add{\isacharcolon}\ memS{\isacharunderscore}def{\isacharparenright}\isanewline
\ \isacommand{qed}\isamarkupfalse%
\endisatagproof
{\isafoldproof}%
\isadelimproof
\isanewline
\endisadelimproof
\isanewline
\isanewline
\isacommand{lemma}\isamarkupfalse%
\ mem{\isacharunderscore}set{\isacharunderscore}{\isadigit{1}}{\isacharcolon}\isanewline
\ \ \isakeyword{assumes}\ h{\isadigit{1}}{\isacharcolon}{\isachardoublequoteopen}a\ mem\ l{\isachardoublequoteclose}\isanewline
\ \ \isakeyword{shows}\ {\isachardoublequoteopen}a\ {\isasymin}\ set\ l{\isachardoublequoteclose}\isanewline
\isadelimproof
\endisadelimproof
\isatagproof
\isacommand{using}\isamarkupfalse%
\ assms\isanewline
\isacommand{proof}\isamarkupfalse%
\ {\isacharparenleft}induct\ l{\isacharparenright}\isanewline
\ \ \isacommand{case}\isamarkupfalse%
\ Nil\isanewline
\ \ \isacommand{from}\isamarkupfalse%
\ this\ \isacommand{show}\isamarkupfalse%
\ {\isacharquery}case\isanewline
\ \ \isacommand{by}\isamarkupfalse%
\ auto\isanewline
\isacommand{next}\isamarkupfalse%
\isanewline
\ \ \isacommand{fix}\isamarkupfalse%
\ a\ la\ \isacommand{case}\isamarkupfalse%
\ {\isacharparenleft}Cons\ a\ la{\isacharparenright}\isanewline
\ \ \isacommand{from}\isamarkupfalse%
\ Cons\ \isacommand{show}\isamarkupfalse%
\ {\isacharquery}case\isanewline
\ \ \isacommand{by}\isamarkupfalse%
\ auto\isanewline
\isacommand{qed}\isamarkupfalse%
\endisatagproof
{\isafoldproof}%
\isadelimproof
\isanewline
\endisadelimproof
\isanewline
\isacommand{lemma}\isamarkupfalse%
\ mem{\isacharunderscore}set{\isacharunderscore}{\isadigit{2}}{\isacharcolon}\isanewline
\ \ \isakeyword{assumes}\ h{\isadigit{1}}{\isacharcolon}{\isachardoublequoteopen}a\ {\isasymin}\ set\ l{\isachardoublequoteclose}\isanewline
\ \ \isakeyword{shows}\ {\isachardoublequoteopen}a\ mem\ l{\isachardoublequoteclose}\isanewline
\isadelimproof
\endisadelimproof
\isatagproof
\isacommand{using}\isamarkupfalse%
\ assms\isanewline
\isacommand{proof}\isamarkupfalse%
\ {\isacharparenleft}induct\ l{\isacharparenright}\isanewline
\ \ \isacommand{case}\isamarkupfalse%
\ Nil\isanewline
\ \ \isacommand{from}\isamarkupfalse%
\ this\ \isacommand{show}\isamarkupfalse%
\ {\isacharquery}case\isanewline
\ \ \isacommand{by}\isamarkupfalse%
\ auto\isanewline
\isacommand{next}\isamarkupfalse%
\isanewline
\ \ \isacommand{fix}\isamarkupfalse%
\ a\ la\ \isacommand{case}\isamarkupfalse%
\ {\isacharparenleft}Cons\ a\ la{\isacharparenright}\isanewline
\ \ \isacommand{from}\isamarkupfalse%
\ Cons\ \isacommand{show}\isamarkupfalse%
\ {\isacharquery}case\isanewline
\ \ \isacommand{by}\isamarkupfalse%
\ auto\isanewline
\isacommand{qed}\isamarkupfalse%
\endisatagproof
{\isafoldproof}%
\isadelimproof
\isanewline
\endisadelimproof
\isanewline
\isacommand{lemma}\isamarkupfalse%
\ set{\isacharunderscore}inter{\isacharunderscore}mem{\isacharcolon}\ \isanewline
\ \ \isakeyword{assumes}\ h{\isadigit{1}}{\isacharcolon}{\isachardoublequoteopen}x\ mem\ l{\isadigit{1}}{\isachardoublequoteclose}\isanewline
\ \ \ \ \ \ \isakeyword{and}\ h{\isadigit{2}}{\isacharcolon}{\isachardoublequoteopen}x\ mem\ l{\isadigit{2}}{\isachardoublequoteclose}\isanewline
\ \ \isakeyword{shows}\ {\isachardoublequoteopen}set\ l{\isadigit{1}}\ {\isasyminter}\ set\ l{\isadigit{2}}\ {\isasymnoteq}\ {\isacharbraceleft}{\isacharbraceright}{\isachardoublequoteclose}\isanewline
\isadelimproof
\endisadelimproof
\isatagproof
\isacommand{using}\isamarkupfalse%
\ assms\isanewline
\isacommand{proof}\isamarkupfalse%
\ {\isacharparenleft}induct\ l{\isadigit{1}}{\isacharparenright}\isanewline
\ \ \isacommand{case}\isamarkupfalse%
\ Nil\isanewline
\ \ \isacommand{from}\isamarkupfalse%
\ this\ \isacommand{show}\isamarkupfalse%
\ {\isacharquery}case\isanewline
\ \ \isacommand{by}\isamarkupfalse%
\ auto\isanewline
\isacommand{next}\isamarkupfalse%
\isanewline
\ \ \isacommand{fix}\isamarkupfalse%
\ a\ la\ \isacommand{case}\isamarkupfalse%
\ {\isacharparenleft}Cons\ a\ la{\isacharparenright}\isanewline
\ \ \isacommand{from}\isamarkupfalse%
\ Cons\ \isacommand{show}\isamarkupfalse%
\ {\isacharquery}case\isanewline
\ \ \isacommand{by}\isamarkupfalse%
\ {\isacharparenleft}auto{\isacharcomma}\ simp\ add{\isacharcolon}\ mem{\isacharunderscore}set{\isacharunderscore}{\isadigit{1}}{\isacharparenright}\isanewline
\ \isacommand{qed}\isamarkupfalse%
\endisatagproof
{\isafoldproof}%
\isadelimproof
\isanewline
\endisadelimproof
\isanewline
\isacommand{lemma}\isamarkupfalse%
\ mem{\isacharunderscore}notdisjoint{\isacharcolon}\ \isanewline
\ \ \isakeyword{assumes}\ h{\isadigit{1}}{\isacharcolon}{\isachardoublequoteopen}x\ mem\ l{\isadigit{1}}{\isachardoublequoteclose}\isanewline
\ \ \ \ \ \ \isakeyword{and}\ h{\isadigit{2}}{\isacharcolon}{\isachardoublequoteopen}x\ mem\ l{\isadigit{2}}{\isachardoublequoteclose}\isanewline
\ \ \isakeyword{shows}\ {\isachardoublequoteopen}{\isasymnot}\ disjoint\ l{\isadigit{1}}\ l{\isadigit{2}}{\isachardoublequoteclose}\isanewline
\isadelimproof
\endisadelimproof
\isatagproof
\isacommand{proof}\isamarkupfalse%
\ \isanewline
\ \ \isacommand{assume}\isamarkupfalse%
\ sg{\isadigit{0}}{\isacharcolon}{\isachardoublequoteopen}disjoint\ l{\isadigit{1}}\ l{\isadigit{2}}{\isachardoublequoteclose}\isanewline
\ \ \isacommand{from}\isamarkupfalse%
\ h{\isadigit{1}}\ \isakeyword{and}\ h{\isadigit{2}}\ \isacommand{have}\isamarkupfalse%
\ \ sg{\isadigit{1}}{\isacharcolon}{\isachardoublequoteopen}set\ l{\isadigit{1}}\ {\isasyminter}\ set\ l{\isadigit{2}}\ {\isasymnoteq}\ {\isacharbraceleft}{\isacharbraceright}{\isachardoublequoteclose}\isanewline
\ \ \ \ \isacommand{by}\isamarkupfalse%
\ {\isacharparenleft}simp\ add{\isacharcolon}\ set{\isacharunderscore}inter{\isacharunderscore}mem{\isacharparenright}\isanewline
\ \ \isacommand{from}\isamarkupfalse%
\ h{\isadigit{1}}\ \isakeyword{and}\ h{\isadigit{2}}\ \isakeyword{and}\ sg{\isadigit{1}}\ \isakeyword{and}\ sg{\isadigit{0}}\ \isacommand{show}\isamarkupfalse%
\ {\isachardoublequoteopen}False{\isachardoublequoteclose}\isanewline
\ \ \ \isacommand{by}\isamarkupfalse%
\ {\isacharparenleft}simp\ add{\isacharcolon}\ disjoint{\isacharunderscore}def{\isacharparenright}\isanewline
\isacommand{qed}\isamarkupfalse%
\endisatagproof
{\isafoldproof}%
\isadelimproof
\isanewline
\endisadelimproof
\isanewline 
\isacommand{lemma}\isamarkupfalse%
\ mem{\isacharunderscore}notdisjoint{\isadigit{2}}{\isacharcolon}\isanewline
\ \ \isakeyword{assumes}\ h{\isadigit{1}}{\isacharcolon}{\isachardoublequoteopen}disjoint\ {\isacharparenleft}schedule\ A{\isacharparenright}\ {\isacharparenleft}schedule\ B{\isacharparenright}{\isachardoublequoteclose}\isanewline
\ \ \ \ \ \ \isakeyword{and}\ h{\isadigit{2}}{\isacharcolon}{\isachardoublequoteopen}x\ mem\ schedule\ A{\isachardoublequoteclose}\isanewline
\ \ \isakeyword{shows}\ {\isachardoublequoteopen}{\isasymnot}\ x\ mem\ schedule\ B{\isachardoublequoteclose}\isanewline
\isadelimproof
\endisadelimproof
\isatagproof
\isacommand{proof}\isamarkupfalse%
\ {\isacharminus}\ \isanewline
\ \ \isacommand{{\isacharbraceleft}}\isamarkupfalse%
\ \isacommand{assume}\isamarkupfalse%
\ a{\isadigit{1}}{\isacharcolon}{\isachardoublequoteopen}\ x\ mem\ schedule\ B{\isachardoublequoteclose}\isanewline
\ \ \ \ \isacommand{from}\isamarkupfalse%
\ h{\isadigit{2}}\ \isakeyword{and}\ a{\isadigit{1}}\ \isacommand{have}\isamarkupfalse%
\ sg{\isadigit{1}}{\isacharcolon}{\isachardoublequoteopen}{\isasymnot}\ \ disjoint\ {\isacharparenleft}schedule\ A{\isacharparenright}\ {\isacharparenleft}schedule\ B{\isacharparenright}{\isachardoublequoteclose}\ \isanewline
\ \ \ \ \ \ \isacommand{by}\isamarkupfalse%
\ {\isacharparenleft}simp\ add{\isacharcolon}\ mem{\isacharunderscore}notdisjoint{\isacharparenright}\isanewline
\ \ \ \ \isacommand{from}\isamarkupfalse%
\ h{\isadigit{1}}\ \isakeyword{and}\ sg{\isadigit{1}}\ \isacommand{have}\isamarkupfalse%
\ {\isachardoublequoteopen}False{\isachardoublequoteclose}\ \isacommand{by}\isamarkupfalse%
\ simp\isanewline
\ \ \ \isacommand{{\isacharbraceright}}\isamarkupfalse%
\ \isacommand{from}\isamarkupfalse%
\ this\ \isacommand{have}\isamarkupfalse%
\ sg{\isadigit{2}}{\isacharcolon}{\isachardoublequoteopen}{\isasymnot}\ x\ mem\ schedule\ B{\isachardoublequoteclose}\ \isacommand{by}\isamarkupfalse%
\ blast\isanewline
\ \ \isacommand{from}\isamarkupfalse%
\ this\ \isacommand{show}\isamarkupfalse%
\ {\isacharquery}thesis\ \isacommand{by}\isamarkupfalse%
\ simp\isanewline
\isacommand{qed}\isamarkupfalse%
\endisatagproof
{\isafoldproof}%
\isadelimproof
\isanewline
\endisadelimproof
\isanewline 
\isacommand{lemma}\isamarkupfalse%
\ Add{\isacharunderscore}Less{\isacharcolon}\ \isanewline
\ \ \isakeyword{assumes}\ {\isachardoublequoteopen}{\isadigit{0}}\ {\isacharless}\ b{\isachardoublequoteclose}\isanewline
\ \ \isakeyword{shows}\ {\isachardoublequoteopen}{\isacharparenleft}Suc\ a\ {\isacharminus}\ b\ {\isacharless}\ Suc\ a{\isacharparenright}\ {\isacharequal}\ True{\isachardoublequoteclose}\isanewline
\isadelimproof
\endisadelimproof
\isatagproof
\isacommand{using}\isamarkupfalse%
\ assms\ \isacommand{by}\isamarkupfalse%
\ arith%
\endisatagproof
{\isafoldproof}%
\isadelimproof
\isanewline
\endisadelimproof
\isanewline 
\isacommand{lemma}\isamarkupfalse%
\ list{\isacharunderscore}length{\isacharunderscore}hint{\isadigit{1}}{\isacharcolon}\ \isanewline
\ \ \isakeyword{assumes}\ {\isachardoublequoteopen}l\ {\isachartilde}{\isacharequal}\ {\isacharbrackleft}{\isacharbrackright}{\isachardoublequoteclose}\isanewline
\ \ \isakeyword{shows}\ {\isachardoublequoteopen}{\isadigit{0}}\ {\isacharless}\ length\ l{\isachardoublequoteclose}\ \isanewline
\isadelimproof
\endisadelimproof
\isatagproof
\isacommand{using}\isamarkupfalse%
\ assms\ \isacommand{by}\isamarkupfalse%
\ simp%
\endisatagproof
{\isafoldproof}%
\isadelimproof
\isanewline
\endisadelimproof 
\isacommand{lemma}\isamarkupfalse%
\ list{\isacharunderscore}length{\isacharunderscore}hint{\isadigit{1}}a{\isacharcolon}\ \isanewline
\ \ \isakeyword{assumes}\ {\isachardoublequoteopen}l\ {\isachartilde}{\isacharequal}\ {\isacharbrackleft}{\isacharbrackright}{\isachardoublequoteclose}\isanewline
\ \ \isakeyword{shows}\ {\isachardoublequoteopen}{\isadigit{0}}\ {\isacharless}\ length\ l{\isachardoublequoteclose}\ \isanewline
\isadelimproof
\endisadelimproof
\isatagproof
\isacommand{using}\isamarkupfalse%
\ assms\ \isacommand{by}\isamarkupfalse%
\ simp%
\endisatagproof
{\isafoldproof}%
\isadelimproof
\isanewline
\endisadelimproof
\isanewline 
\isacommand{lemma}\isamarkupfalse%
\ list{\isacharunderscore}length{\isacharunderscore}hint{\isadigit{2}}{\isacharcolon}\ \isanewline
\ \ \isakeyword{assumes}\ h{\isadigit{1}}{\isacharcolon}{\isachardoublequoteopen}length\ x\ \ {\isacharequal}\ Suc\ {\isadigit{0}}{\isachardoublequoteclose}\isanewline
\ \ \isakeyword{shows}\ {\isachardoublequoteopen}{\isacharbrackleft}hd\ x{\isacharbrackright}\ {\isacharequal}\ x{\isachardoublequoteclose}\isanewline
\isadelimproof
\endisadelimproof
\isatagproof
\isacommand{using}\isamarkupfalse%
\ assms\ \isanewline
\isacommand{proof}\isamarkupfalse%
\ {\isacharparenleft}induct\ x{\isacharparenright}\isanewline
\ \ \isacommand{case}\isamarkupfalse%
\ Nil\isanewline
\ \ \isacommand{from}\isamarkupfalse%
\ this\ \isacommand{show}\isamarkupfalse%
\ {\isacharquery}case\isanewline
\ \ \isacommand{by}\isamarkupfalse%
\ auto\isanewline
\isacommand{next}\isamarkupfalse%
\isanewline
\ \ \isacommand{fix}\isamarkupfalse%
\ a\ la\ \isacommand{case}\isamarkupfalse%
\ {\isacharparenleft}Cons\ a\ la{\isacharparenright}\isanewline
\ \ \isacommand{from}\isamarkupfalse%
\ Cons\ \isacommand{show}\isamarkupfalse%
\ {\isacharquery}case\isanewline
\ \ \isacommand{by}\isamarkupfalse%
\ auto\isanewline
\isacommand{qed}\isamarkupfalse%
\endisatagproof
{\isafoldproof}%
\isadelimproof
\isanewline
\endisadelimproof
\isanewline 
\isacommand{lemma}\isamarkupfalse%
\ list{\isacharunderscore}length{\isacharunderscore}hint{\isadigit{2}}a{\isacharcolon}\ \isanewline
\ \ \isakeyword{assumes}\ h{\isadigit{1}}{\isacharcolon}{\isachardoublequoteopen}length\ l\ {\isacharequal}\ Suc\ {\isadigit{0}}{\isachardoublequoteclose}\isanewline
\ \ \isakeyword{shows}\ {\isachardoublequoteopen}tl\ l\ {\isacharequal}\ {\isacharbrackleft}{\isacharbrackright}{\isachardoublequoteclose}\isanewline
\isadelimproof
\endisadelimproof
\isatagproof
\isacommand{using}\isamarkupfalse%
\ assms\isanewline
\isacommand{proof}\isamarkupfalse%
\ {\isacharparenleft}induct\ l{\isacharparenright}\isanewline
\ \ \isacommand{case}\isamarkupfalse%
\ Nil\isanewline
\ \ \isacommand{from}\isamarkupfalse%
\ this\ \isacommand{show}\isamarkupfalse%
\ {\isacharquery}case\isanewline
\ \ \isacommand{by}\isamarkupfalse%
\ auto\isanewline
\isacommand{next}\isamarkupfalse%
\isanewline
\ \ \isacommand{fix}\isamarkupfalse%
\ a\ la\ \isacommand{case}\isamarkupfalse%
\ {\isacharparenleft}Cons\ a\ la{\isacharparenright}\isanewline
\ \ \isacommand{from}\isamarkupfalse%
\ Cons\ \isacommand{show}\isamarkupfalse%
\ {\isacharquery}case\isanewline
\ \ \isacommand{by}\isamarkupfalse%
\ auto\isanewline
\isacommand{qed}\isamarkupfalse%
\endisatagproof
{\isafoldproof}%
\isadelimproof
\isanewline
\endisadelimproof
\isanewline 
\isacommand{lemma}\isamarkupfalse%
\ list{\isacharunderscore}length{\isacharunderscore}hint{\isadigit{3}}{\isacharcolon}\ \isanewline
\ \ \isakeyword{assumes}\ {\isachardoublequoteopen}length\ l\ {\isacharequal}\ Suc\ {\isadigit{0}}{\isachardoublequoteclose}\isanewline
\ \ \isakeyword{shows}\ {\isachardoublequoteopen}l\ {\isasymnoteq}\ {\isacharbrackleft}{\isacharbrackright}{\isachardoublequoteclose}\isanewline
\isadelimproof
\endisadelimproof
\isatagproof
\isacommand{using}\isamarkupfalse%
\ assms\isanewline
\isacommand{proof}\isamarkupfalse%
\ {\isacharparenleft}induct\ l{\isacharparenright}\isanewline
\ \ \isacommand{case}\isamarkupfalse%
\ Nil\isanewline
\ \ \isacommand{from}\isamarkupfalse%
\ this\ \isacommand{show}\isamarkupfalse%
\ {\isacharquery}case\isanewline
\ \ \isacommand{by}\isamarkupfalse%
\ auto\isanewline
\isacommand{next}\isamarkupfalse%
\isanewline
\ \ \isacommand{fix}\isamarkupfalse%
\ a\ la\ \isacommand{case}\isamarkupfalse%
\ {\isacharparenleft}Cons\ a\ la{\isacharparenright}\isanewline
\ \ \isacommand{from}\isamarkupfalse%
\ Cons\ \isacommand{show}\isamarkupfalse%
\ {\isacharquery}case\isanewline
\ \ \isacommand{by}\isamarkupfalse%
\ auto\isanewline
\isacommand{qed}\isamarkupfalse%
\endisatagproof
{\isafoldproof}%
\isadelimproof
\isanewline
\endisadelimproof
\isanewline 
\isacommand{lemma}\isamarkupfalse%
\ list{\isacharunderscore}length{\isacharunderscore}hint{\isadigit{4}}{\isacharcolon}\ \isanewline
\ \ \isakeyword{assumes}\ h{\isadigit{1}}{\isacharcolon}{\isachardoublequoteopen}length\ x\ {\isasymle}\ Suc\ {\isadigit{0}}{\isachardoublequoteclose}\isanewline
\ \ \ \ \ \ \isakeyword{and}\ h{\isadigit{2}}{\isacharcolon}{\isachardoublequoteopen}x\ {\isasymnoteq}\ {\isacharbrackleft}{\isacharbrackright}{\isachardoublequoteclose}\isanewline
\ \ \ \isakeyword{shows}\ {\isachardoublequoteopen}length\ x\ {\isacharequal}\ Suc\ {\isadigit{0}}{\isachardoublequoteclose}\isanewline
\isadelimproof
\endisadelimproof
\isatagproof
\isacommand{using}\isamarkupfalse%
\ assms\isanewline
\isacommand{proof}\isamarkupfalse%
\ {\isacharparenleft}induct\ x{\isacharparenright}\isanewline
\ \ \isacommand{case}\isamarkupfalse%
\ Nil\isanewline
\ \ \isacommand{from}\isamarkupfalse%
\ this\ \isacommand{show}\isamarkupfalse%
\ {\isacharquery}case\isanewline
\ \ \isacommand{by}\isamarkupfalse%
\ auto\isanewline
\isacommand{next}\isamarkupfalse%
\isanewline
\ \ \isacommand{fix}\isamarkupfalse%
\ a\ la\ \isacommand{case}\isamarkupfalse%
\ {\isacharparenleft}Cons\ a\ la{\isacharparenright}\isanewline
\ \ \isacommand{from}\isamarkupfalse%
\ Cons\ \isacommand{show}\isamarkupfalse%
\ {\isacharquery}case\isanewline
\ \ \isacommand{by}\isamarkupfalse%
\ auto\isanewline
\isacommand{qed}\isamarkupfalse%
\endisatagproof
{\isafoldproof}%
\isadelimproof
\isanewline
\endisadelimproof
\isanewline 
\isacommand{lemma}\isamarkupfalse%
\ length{\isacharunderscore}nonempty{\isacharcolon}\ \isanewline
\ \ \isakeyword{assumes}\ h{\isadigit{1}}{\isacharcolon}{\isachardoublequoteopen}x\ {\isasymnoteq}\ {\isacharbrackleft}{\isacharbrackright}{\isachardoublequoteclose}\ \isanewline
\ \ \isakeyword{shows}\ {\isachardoublequoteopen}Suc\ {\isadigit{0}}\ {\isasymle}\ length\ x{\isachardoublequoteclose}\isanewline
\isadelimproof
\endisadelimproof
\isatagproof
\isacommand{using}\isamarkupfalse%
\ assms\isanewline
\isacommand{proof}\isamarkupfalse%
\ {\isacharparenleft}induct\ x{\isacharparenright}\isanewline
\ \ \isacommand{case}\isamarkupfalse%
\ Nil\isanewline
\ \ \isacommand{from}\isamarkupfalse%
\ this\ \isacommand{show}\isamarkupfalse%
\ {\isacharquery}case\isanewline
\ \ \isacommand{by}\isamarkupfalse%
\ auto\isanewline
\isacommand{next}\isamarkupfalse%
\isanewline
\ \ \isacommand{fix}\isamarkupfalse%
\ a\ la\ \isacommand{case}\isamarkupfalse%
\ {\isacharparenleft}Cons\ a\ la{\isacharparenright}\isanewline
\ \ \isacommand{from}\isamarkupfalse%
\ Cons\ \isacommand{show}\isamarkupfalse%
\ {\isacharquery}case\isanewline
\ \ \isacommand{by}\isamarkupfalse%
\ auto\isanewline
\isacommand{qed}\isamarkupfalse%
\endisatagproof
{\isafoldproof}%
\isadelimproof
\ \isanewline
\endisadelimproof
\isanewline
\isacommand{lemma}\isamarkupfalse%
\ last{\isacharunderscore}nth{\isacharunderscore}length{\isacharcolon}\ \isanewline
\ \ \isakeyword{assumes}\ {\isachardoublequoteopen}x\ {\isasymnoteq}\ {\isacharbrackleft}{\isacharbrackright}{\isachardoublequoteclose}\isanewline
\ \ \isakeyword{shows}\ {\isachardoublequoteopen}x\ {\isacharbang}\ {\isacharparenleft}{\isacharparenleft}length\ x{\isacharparenright}\ {\isacharminus}\ Suc\ {\isadigit{0}}{\isacharparenright}\ {\isacharequal}\ last\ x{\isachardoublequoteclose}\isanewline
\isadelimproof
\endisadelimproof
\isatagproof
\isacommand{using}\isamarkupfalse%
\ assms\isanewline
\isacommand{proof}\isamarkupfalse%
\ {\isacharparenleft}induct\ x{\isacharparenright}\isanewline
\ \ \isacommand{case}\isamarkupfalse%
\ Nil\isanewline
\ \ \isacommand{from}\isamarkupfalse%
\ this\ \isacommand{show}\isamarkupfalse%
\ {\isacharquery}case\isanewline
\ \ \isacommand{by}\isamarkupfalse%
\ auto\isanewline
\isacommand{next}\isamarkupfalse%
\isanewline
\ \ \isacommand{fix}\isamarkupfalse%
\ a\ la\ \isacommand{case}\isamarkupfalse%
\ {\isacharparenleft}Cons\ a\ la{\isacharparenright}\isanewline
\ \ \isacommand{from}\isamarkupfalse%
\ Cons\ \isacommand{show}\isamarkupfalse%
\ {\isacharquery}case\isanewline
\ \ \isacommand{by}\isamarkupfalse%
\ auto\isanewline
\isacommand{qed}\isamarkupfalse%
\endisatagproof
{\isafoldproof}%
\isadelimproof
\ \isanewline
\endisadelimproof
\isanewline 
\isacommand{lemma}\isamarkupfalse%
\ list{\isacharunderscore}nth{\isacharunderscore}append{\isadigit{0}}{\isacharcolon}\isanewline
\ \ \isakeyword{assumes}\ h{\isadigit{1}}{\isacharcolon}{\isachardoublequoteopen}i\ {\isacharless}\ length\ x{\isachardoublequoteclose}\isanewline
\ \ \isakeyword{shows}\ {\isachardoublequoteopen}x\ {\isacharbang}\ i\ {\isacharequal}\ {\isacharparenleft}x\ {\isacharat}\ z{\isacharparenright}\ {\isacharbang}\ i{\isachardoublequoteclose}\isanewline
\isadelimproof
\endisadelimproof
\isatagproof
\isacommand{proof}\isamarkupfalse%
\ {\isacharparenleft}cases\ i{\isacharparenright}\isanewline
\ \ \isacommand{assume}\isamarkupfalse%
\ {\isachardoublequoteopen}i{\isacharequal}{\isadigit{0}}{\isachardoublequoteclose}\ \isanewline
\ \ \isacommand{with}\isamarkupfalse%
\ h{\isadigit{1}}\ \ \isacommand{show}\isamarkupfalse%
\ {\isacharquery}thesis\ \isacommand{by}\isamarkupfalse%
\ {\isacharparenleft}simp\ add{\isacharcolon}\ nth{\isacharunderscore}append{\isacharparenright}\isanewline
\isacommand{next}\isamarkupfalse%
\isanewline
\ \ \isacommand{fix}\isamarkupfalse%
\ ii\ \isacommand{assume}\isamarkupfalse%
\ {\isachardoublequoteopen}i\ {\isacharequal}\ Suc\ ii{\isachardoublequoteclose}\isanewline
\ \ \isacommand{with}\isamarkupfalse%
\ h{\isadigit{1}}\ \ \isacommand{show}\isamarkupfalse%
\ {\isacharquery}thesis\ \ \isacommand{by}\isamarkupfalse%
\ {\isacharparenleft}simp\ add{\isacharcolon}\ nth{\isacharunderscore}append{\isacharparenright}\isanewline
\isacommand{qed}\isamarkupfalse%
\endisatagproof
{\isafoldproof}%
\isadelimproof
\isanewline
\endisadelimproof
\isanewline 
\isacommand{lemma}\isamarkupfalse%
\ list{\isacharunderscore}nth{\isacharunderscore}append{\isadigit{1}}{\isacharcolon}\isanewline
\ \ \isakeyword{assumes}\ h{\isadigit{1}}{\isacharcolon}{\isachardoublequoteopen}i\ {\isacharless}\ length\ x{\isachardoublequoteclose}\isanewline
\ \ \isakeyword{shows}\ {\isachardoublequoteopen}{\isacharparenleft}b\ {\isacharhash}\ x{\isacharparenright}\ {\isacharbang}\ i\ {\isacharequal}\ {\isacharparenleft}b\ {\isacharhash}\ x\ {\isacharat}\ y{\isacharparenright}\ {\isacharbang}\ i{\isachardoublequoteclose}\isanewline
\isadelimproof
\endisadelimproof
\isatagproof
\isacommand{proof}\isamarkupfalse%
\ {\isacharminus}\isanewline
\ \ \isacommand{from}\isamarkupfalse%
\ h{\isadigit{1}}\ \isacommand{have}\isamarkupfalse%
\ sg{\isadigit{1}}{\isacharcolon}{\isachardoublequoteopen}i\ {\isacharless}\ length\ {\isacharparenleft}b\ {\isacharhash}\ x{\isacharparenright}{\isachardoublequoteclose}\ \isacommand{by}\isamarkupfalse%
\ auto\isanewline
\ \ \isacommand{from}\isamarkupfalse%
\ this\ \isacommand{have}\isamarkupfalse%
\ sg{\isadigit{2}}{\isacharcolon}\ {\isachardoublequoteopen}{\isacharparenleft}b\ {\isacharhash}\ x{\isacharparenright}\ {\isacharbang}\ i\ {\isacharequal}\ {\isacharparenleft}{\isacharparenleft}b\ {\isacharhash}\ x{\isacharparenright}\ {\isacharat}\ y{\isacharparenright}\ {\isacharbang}\ i{\isachardoublequoteclose}\ \isanewline
\ \ \ \ \isacommand{by}\isamarkupfalse%
\ {\isacharparenleft}rule\ list{\isacharunderscore}nth{\isacharunderscore}append{\isadigit{0}}{\isacharparenright}\isanewline
\ \ \isacommand{from}\isamarkupfalse%
\ this\ \isacommand{show}\isamarkupfalse%
\ {\isacharquery}thesis\ \isacommand{by}\isamarkupfalse%
\ simp\isanewline
\isacommand{qed}\isamarkupfalse%
\endisatagproof
{\isafoldproof}%
\isadelimproof
\isanewline
\endisadelimproof
\ \ \isanewline 
\isacommand{lemma}\isamarkupfalse%
\ list{\isacharunderscore}nth{\isacharunderscore}append{\isadigit{2}}{\isacharcolon}\isanewline
\ \ \isakeyword{assumes}\ h{\isadigit{1}}{\isacharcolon}{\isachardoublequoteopen}i\ {\isacharless}\ Suc\ {\isacharparenleft}length\ x{\isacharparenright}{\isachardoublequoteclose}\isanewline
\ \ \isakeyword{shows}\ {\isachardoublequoteopen}{\isacharparenleft}b\ {\isacharhash}\ x{\isacharparenright}\ {\isacharbang}\ i\ {\isacharequal}\ {\isacharparenleft}b\ {\isacharhash}\ x\ {\isacharat}\ a\ {\isacharhash}\ y{\isacharparenright}\ {\isacharbang}\ i{\isachardoublequoteclose}\isanewline
\isadelimproof
\endisadelimproof
\isatagproof
\isacommand{proof}\isamarkupfalse%
\ {\isacharminus}\ \isanewline
\ \ \isacommand{from}\isamarkupfalse%
\ h{\isadigit{1}}\ \isacommand{have}\isamarkupfalse%
\ sg{\isadigit{1}}{\isacharcolon}{\isachardoublequoteopen}i\ {\isacharless}\ length\ {\isacharparenleft}b\ {\isacharhash}\ x{\isacharparenright}{\isachardoublequoteclose}\ \isacommand{by}\isamarkupfalse%
\ auto\isanewline
\ \ \isacommand{from}\isamarkupfalse%
\ this\ \isacommand{have}\isamarkupfalse%
\ sg{\isadigit{2}}{\isacharcolon}\ {\isachardoublequoteopen}{\isacharparenleft}b\ {\isacharhash}\ x{\isacharparenright}\ {\isacharbang}\ i\ {\isacharequal}\ {\isacharparenleft}{\isacharparenleft}b\ {\isacharhash}\ x{\isacharparenright}\ {\isacharat}\ {\isacharparenleft}a\ {\isacharhash}\ y{\isacharparenright}{\isacharparenright}\ {\isacharbang}\ i{\isachardoublequoteclose}\isanewline
\ \ \ \ \isacommand{by}\isamarkupfalse%
\ {\isacharparenleft}rule\ list{\isacharunderscore}nth{\isacharunderscore}append{\isadigit{0}}{\isacharparenright}\isanewline
\ \ \isacommand{from}\isamarkupfalse%
\ this\ \isacommand{show}\isamarkupfalse%
\ {\isacharquery}thesis\ \isacommand{by}\isamarkupfalse%
\ simp\isanewline
\isacommand{qed}\isamarkupfalse%
\endisatagproof
{\isafoldproof}%
\isadelimproof
\isanewline
\endisadelimproof
\isanewline 
\isacommand{lemma}\isamarkupfalse%
\ list{\isacharunderscore}nth{\isacharunderscore}append{\isadigit{3}}{\isacharcolon}\isanewline
\ \ \isakeyword{assumes}\ h{\isadigit{1}}{\isacharcolon}{\isachardoublequoteopen}{\isasymnot}\ i\ {\isacharless}\ Suc\ {\isacharparenleft}length\ x{\isacharparenright}{\isachardoublequoteclose}\isanewline
\ \ \ \ \ \ \isakeyword{and}\ h{\isadigit{2}}{\isacharcolon}{\isachardoublequoteopen}i\ {\isacharminus}\ Suc\ {\isacharparenleft}length\ x{\isacharparenright}\ {\isacharless}\ Suc\ {\isacharparenleft}length\ y{\isacharparenright}{\isachardoublequoteclose}\isanewline
\ \ \isakeyword{shows}\ {\isachardoublequoteopen}{\isacharparenleft}a\ {\isacharhash}\ y{\isacharparenright}\ {\isacharbang}\ {\isacharparenleft}i\ {\isacharminus}\ Suc\ {\isacharparenleft}length\ x{\isacharparenright}{\isacharparenright}\ {\isacharequal}\ {\isacharparenleft}b\ {\isacharhash}\ x\ {\isacharat}\ a\ {\isacharhash}\ y{\isacharparenright}\ {\isacharbang}\ i{\isachardoublequoteclose}\isanewline
\isadelimproof
\endisadelimproof
\isatagproof
\isacommand{proof}\isamarkupfalse%
\ {\isacharparenleft}cases\ i{\isacharparenright}\isanewline
\ \ \isacommand{assume}\isamarkupfalse%
\ {\isachardoublequoteopen}i{\isacharequal}{\isadigit{0}}{\isachardoublequoteclose}\ \isanewline
\ \ \isacommand{with}\isamarkupfalse%
\ h{\isadigit{1}}\ \ \isacommand{show}\isamarkupfalse%
\ {\isacharquery}thesis\ \isacommand{by}\isamarkupfalse%
\ {\isacharparenleft}simp\ add{\isacharcolon}\ nth{\isacharunderscore}append{\isacharparenright}\isanewline
\isacommand{next}\isamarkupfalse%
\isanewline
\ \ \isacommand{fix}\isamarkupfalse%
\ ii\ \isacommand{assume}\isamarkupfalse%
\ {\isachardoublequoteopen}i\ {\isacharequal}\ Suc\ ii{\isachardoublequoteclose}\isanewline
\ \ \isacommand{with}\isamarkupfalse%
\ h{\isadigit{1}}\ \ \isacommand{show}\isamarkupfalse%
\ {\isacharquery}thesis\ \ \isacommand{by}\isamarkupfalse%
\ {\isacharparenleft}simp\ add{\isacharcolon}\ nth{\isacharunderscore}append{\isacharparenright}\isanewline
\isacommand{qed}\isamarkupfalse%
\endisatagproof
{\isafoldproof}%
\isadelimproof
\isanewline
\endisadelimproof
\isanewline 
\isacommand{lemma}\isamarkupfalse%
\ list{\isacharunderscore}nth{\isacharunderscore}append{\isadigit{4}}{\isacharcolon}\isanewline
\ \ \isakeyword{assumes}\ h{\isadigit{1}}{\isacharcolon}{\isachardoublequoteopen}i\ {\isacharless}\ Suc\ {\isacharparenleft}length\ x\ {\isacharplus}\ length\ y{\isacharparenright}{\isachardoublequoteclose}\isanewline
\ \ \ \ \ \ \isakeyword{and}\ h{\isadigit{2}}{\isacharcolon}{\isachardoublequoteopen}{\isasymnot}\ i\ {\isacharminus}\ Suc\ {\isacharparenleft}length\ x{\isacharparenright}\ {\isacharless}\ Suc\ {\isacharparenleft}length\ y{\isacharparenright}{\isachardoublequoteclose}\ \isanewline
\ \ \isakeyword{shows}\ {\isachardoublequoteopen}False{\isachardoublequoteclose}\isanewline
\isadelimproof
\endisadelimproof
\isatagproof
\isacommand{using}\isamarkupfalse%
\ assms\ \ \isacommand{by}\isamarkupfalse%
\ arith%
\endisatagproof
{\isafoldproof}%
\isadelimproof
\isanewline
\endisadelimproof
\isanewline 
\isacommand{lemma}\isamarkupfalse%
\ list{\isacharunderscore}nth{\isacharunderscore}append{\isadigit{5}}{\isacharcolon}\isanewline
\ \ \isakeyword{assumes}\ h{\isadigit{1}}{\isacharcolon}{\isachardoublequoteopen}i\ {\isacharminus}\ length\ x\ {\isacharless}\ Suc\ {\isacharparenleft}length\ y{\isacharparenright}{\isachardoublequoteclose}\ \isanewline
\ \ \ \ \ \ \isakeyword{and}\ h{\isadigit{2}}{\isacharcolon}{\isachardoublequoteopen}{\isasymnot}\ i\ {\isacharminus}\ Suc\ {\isacharparenleft}length\ x{\isacharparenright}\ {\isacharless}\ Suc\ {\isacharparenleft}length\ y{\isacharparenright}{\isachardoublequoteclose}\isanewline
\ \ \isakeyword{shows}\ {\isachardoublequoteopen}{\isasymnot}\ \ i\ {\isacharless}\ Suc\ {\isacharparenleft}length\ x\ {\isacharplus}\ length\ y{\isacharparenright}{\isachardoublequoteclose}\isanewline
\isadelimproof
\endisadelimproof
\isatagproof
\isacommand{using}\isamarkupfalse%
\ assms\ \ \isacommand{by}\isamarkupfalse%
\ arith%
\endisatagproof
{\isafoldproof}%
\isadelimproof
\isanewline
\endisadelimproof
\isanewline 
\isacommand{lemma}\isamarkupfalse%
\ list{\isacharunderscore}nth{\isacharunderscore}append{\isadigit{6}}{\isacharcolon}\isanewline
\ \ \isakeyword{assumes}\ h{\isadigit{1}}{\isacharcolon}{\isachardoublequoteopen}{\isasymnot}\ i\ {\isacharminus}\ length\ x\ {\isacharless}\ Suc\ {\isacharparenleft}length\ y{\isacharparenright}{\isachardoublequoteclose}\isanewline
\ \ \ \ \ \ \isakeyword{and}\ h{\isadigit{2}}{\isacharcolon}{\isachardoublequoteopen}{\isasymnot}\ i\ {\isacharminus}\ Suc\ {\isacharparenleft}length\ x{\isacharparenright}\ {\isacharless}\ Suc\ {\isacharparenleft}length\ y{\isacharparenright}{\isachardoublequoteclose}\isanewline
\ \ \isakeyword{shows}\ {\isachardoublequoteopen}{\isasymnot}\ i\ {\isacharless}\ Suc\ {\isacharparenleft}length\ x\ {\isacharplus}\ length\ y{\isacharparenright}{\isachardoublequoteclose}\isanewline
\isadelimproof
\endisadelimproof
\isatagproof
\isacommand{using}\isamarkupfalse%
\ assms\ \isacommand{by}\isamarkupfalse%
\ arith%
\endisatagproof
{\isafoldproof}%
\isadelimproof
\isanewline
\endisadelimproof
\isanewline 
\isacommand{lemma}\isamarkupfalse%
\ list{\isacharunderscore}nth{\isacharunderscore}append{\isadigit{6}}a{\isacharcolon}\isanewline
\ \ \isakeyword{assumes}\ h{\isadigit{1}}{\isacharcolon}{\isachardoublequoteopen}i\ {\isacharless}\ Suc\ {\isacharparenleft}length\ x\ {\isacharplus}\ length\ y{\isacharparenright}{\isachardoublequoteclose}\isanewline
\ \ \ \ \ \ \isakeyword{and}\ h{\isadigit{2}}{\isacharcolon}{\isachardoublequoteopen}{\isasymnot}\ i\ {\isacharminus}\ length\ x\ {\isacharless}\ Suc\ {\isacharparenleft}length\ y{\isacharparenright}{\isachardoublequoteclose}\isanewline
\ \ \isakeyword{shows}\ {\isachardoublequoteopen}False{\isachardoublequoteclose}\isanewline
\isadelimproof
\endisadelimproof
\isatagproof
\isacommand{using}\isamarkupfalse%
\ assms\ \isacommand{by}\isamarkupfalse%
\ arith%
\endisatagproof
{\isafoldproof}%
\isadelimproof
\ \isanewline
\endisadelimproof
\isanewline 
\isacommand{lemma}\isamarkupfalse%
\ list{\isacharunderscore}nth{\isacharunderscore}append{\isadigit{7}}{\isacharcolon}\isanewline
\ \ \isakeyword{assumes}\ h{\isadigit{1}}{\isacharcolon}{\isachardoublequoteopen}i\ {\isacharminus}\ length\ x\ {\isacharless}\ Suc\ {\isacharparenleft}length\ y{\isacharparenright}{\isachardoublequoteclose}\isanewline
\ \ \ \ \ \ \isakeyword{and}\ h{\isadigit{2}}{\isacharcolon}{\isachardoublequoteopen}i\ {\isacharminus}\ Suc\ {\isacharparenleft}length\ x{\isacharparenright}\ {\isacharless}\ Suc\ {\isacharparenleft}length\ y{\isacharparenright}{\isachardoublequoteclose}\isanewline
\ \ \isakeyword{shows}\ {\isachardoublequoteopen}i\ {\isacharless}\ Suc\ {\isacharparenleft}Suc\ {\isacharparenleft}length\ x\ {\isacharplus}\ length\ y{\isacharparenright}{\isacharparenright}{\isachardoublequoteclose}\isanewline
\isadelimproof
\endisadelimproof
\isatagproof
\isacommand{using}\isamarkupfalse%
\ assms\ \ \isacommand{by}\isamarkupfalse%
\ arith%
\endisatagproof
{\isafoldproof}%
\isadelimproof
\isanewline
\endisadelimproof
\isanewline
\isacommand{lemma}\isamarkupfalse%
\ list{\isacharunderscore}nth{\isacharunderscore}append{\isadigit{8}}{\isacharcolon}\isanewline
\ \ \isakeyword{assumes}\ h{\isadigit{1}}{\isacharcolon}{\isachardoublequoteopen}{\isasymnot}\ i\ {\isacharless}\ Suc\ {\isacharparenleft}length\ x\ {\isacharplus}\ length\ y{\isacharparenright}{\isachardoublequoteclose}\isanewline
\ \ \ \ \ \ \isakeyword{and}\ h{\isadigit{2}}{\isacharcolon}{\isachardoublequoteopen}i\ {\isacharless}\ Suc\ {\isacharparenleft}Suc\ {\isacharparenleft}length\ x\ {\isacharplus}\ length\ y{\isacharparenright}{\isacharparenright}{\isachardoublequoteclose}\isanewline
\ \ \isakeyword{shows}\ {\isachardoublequoteopen}i\ {\isacharequal}\ Suc\ {\isacharparenleft}length\ x\ {\isacharplus}\ length\ y{\isacharparenright}{\isachardoublequoteclose}\isanewline
\isadelimproof
\endisadelimproof
\isatagproof
\isacommand{using}\isamarkupfalse%
\ assms\ \ \isacommand{by}\isamarkupfalse%
\ arith%
\endisatagproof
{\isafoldproof}%
\isadelimproof
\isanewline
\endisadelimproof
\isanewline 
\isacommand{lemma}\isamarkupfalse%
\ list{\isacharunderscore}nth{\isacharunderscore}append{\isadigit{9}}{\isacharcolon}\isanewline
\ \ \isakeyword{assumes}\ h{\isadigit{1}}{\isacharcolon}{\isachardoublequoteopen}i\ {\isacharminus}\ Suc\ {\isacharparenleft}length\ x{\isacharparenright}\ {\isacharless}\ Suc\ {\isacharparenleft}length\ y{\isacharparenright}{\isachardoublequoteclose}\isanewline
\ \ \isakeyword{shows}\ {\isachardoublequoteopen}i\ {\isacharless}\ Suc\ {\isacharparenleft}Suc\ {\isacharparenleft}length\ x\ {\isacharplus}\ length\ y{\isacharparenright}{\isacharparenright}{\isachardoublequoteclose}\isanewline
\isadelimproof
\endisadelimproof
\isatagproof
\isacommand{using}\isamarkupfalse%
\ assms\ \isacommand{by}\isamarkupfalse%
\ arith%
\endisatagproof
{\isafoldproof}%
\isadelimproof
\isanewline
\endisadelimproof
\ \ \isanewline 
\isacommand{lemma}\isamarkupfalse%
\ list{\isacharunderscore}nth{\isacharunderscore}append{\isadigit{1}}{\isadigit{0}}{\isacharcolon}\isanewline
\ \ \isakeyword{assumes}\ h{\isadigit{1}}{\isacharcolon}{\isachardoublequoteopen}{\isasymnot}\ i\ {\isacharless}\ Suc\ {\isacharparenleft}length\ x{\isacharparenright}{\isachardoublequoteclose}\isanewline
\ \ \ \ \ \ \isakeyword{and}\ h{\isadigit{2}}{\isacharcolon}{\isachardoublequoteopen}{\isasymnot}\ i\ {\isacharminus}\ Suc\ {\isacharparenleft}length\ x{\isacharparenright}\ {\isacharless}\ Suc\ {\isacharparenleft}length\ y{\isacharparenright}{\isachardoublequoteclose}\isanewline
\ \ \isakeyword{shows}\ {\isachardoublequoteopen}{\isasymnot}\ i\ {\isacharless}\ Suc\ {\isacharparenleft}Suc\ {\isacharparenleft}length\ x\ {\isacharplus}\ length\ y{\isacharparenright}{\isacharparenright}{\isachardoublequoteclose}\isanewline
\isadelimproof
\endisadelimproof
\isatagproof
\isacommand{using}\isamarkupfalse%
\ assms\ \isacommand{by}\isamarkupfalse%
\ arith%
\endisatagproof
{\isafoldproof}%
\isadelimproof
\isanewline
\endisadelimproof
\isadelimtheory
\isanewline
\endisadelimtheory
\isatagtheory
\isacommand{end}\isamarkupfalse%
\endisatagtheory
{\isafoldtheory}%
\isadelimtheory
\endisadelimtheory
\end{isabellebody}%

%
\begin{isabellebody}%
\def\isabellecontext{arith{\isacharunderscore}hints}%
\isamarkupheader{Auxiliary arithmetic lemmas%
}
\isamarkuptrue%
\isadelimtheory
\endisadelimtheory
\isatagtheory
\isacommand{theory}\isamarkupfalse%
\ arith{\isacharunderscore}hints\isanewline
\isakeyword{imports}\ Main\isanewline
\isakeyword{begin}%
\endisatagtheory
{\isafoldtheory}%
\isadelimtheory
\endisadelimtheory
\isanewline
\isanewline 
\isacommand{lemma}\isamarkupfalse%
\ arith{\isacharunderscore}mod{\isacharunderscore}neq{\isacharcolon}\isanewline
\ \ \isakeyword{assumes}\ h{\isadigit{1}}{\isacharcolon}{\isachardoublequoteopen}a\ mod\ n\ {\isasymnoteq}\ b\ mod\ n{\isachardoublequoteclose}\isanewline
\ \ \isakeyword{shows}\ {\isachardoublequoteopen}a\ {\isasymnoteq}\ b{\isachardoublequoteclose}\isanewline
\isadelimproof
\endisadelimproof
\isatagproof
\isacommand{using}\isamarkupfalse%
\ assms\ \isacommand{by}\isamarkupfalse%
\ auto%
\endisatagproof
{\isafoldproof}%
\isadelimproof
\ \isanewline
\endisadelimproof
\isanewline 
\isacommand{lemma}\isamarkupfalse%
\ arith{\isacharunderscore}mod{\isacharunderscore}nzero{\isacharcolon}\ \isanewline
\ \ \isakeyword{fixes}\ i{\isacharcolon}{\isacharcolon}nat\isanewline
\ \ \isakeyword{assumes}\ h{\isadigit{1}}{\isacharcolon}\ {\isachardoublequoteopen}i\ {\isacharless}\ n{\isachardoublequoteclose}\ \isanewline
\ \ \ \ \ \ \isakeyword{and}\ h{\isadigit{2}}{\isacharcolon}{\isachardoublequoteopen}{\isadigit{0}}\ {\isacharless}\ i{\isachardoublequoteclose}\isanewline
\ \ \isakeyword{shows}\ {\isachardoublequoteopen}{\isadigit{0}}\ {\isacharless}\ {\isacharparenleft}n\ {\isacharasterisk}\ t\ {\isacharplus}\ i{\isacharparenright}\ mod\ n{\isachardoublequoteclose}\isanewline
\isadelimproof
\endisadelimproof
\isatagproof
\isacommand{proof}\isamarkupfalse%
\ {\isacharminus}\isanewline
\ \ \isacommand{from}\isamarkupfalse%
\ h{\isadigit{1}}\ \isakeyword{and}\ h{\isadigit{2}}\ \isacommand{have}\isamarkupfalse%
\ \ sg{\isadigit{1}}{\isacharcolon}{\isachardoublequoteopen}{\isacharparenleft}i\ {\isacharplus}\ n\ {\isacharasterisk}\ t{\isacharparenright}\ mod\ n\ {\isacharequal}\ i{\isachardoublequoteclose}\isanewline
\ \ \ \ \isacommand{by}\isamarkupfalse%
\ {\isacharparenleft}simp\ add{\isacharcolon}\ mod{\isacharunderscore}mult{\isacharunderscore}self{\isadigit{2}}{\isacharparenright}\isanewline
\ \ \isacommand{also}\isamarkupfalse%
\ \isacommand{have}\isamarkupfalse%
\ sg{\isadigit{2}}{\isacharcolon}{\isachardoublequoteopen}n\ {\isacharasterisk}\ t\ {\isacharplus}\ i\ {\isacharequal}\ i\ {\isacharplus}\ n\ {\isacharasterisk}\ t{\isachardoublequoteclose}\ \ \isacommand{by}\isamarkupfalse%
\ simp\isanewline
\ \ \isacommand{from}\isamarkupfalse%
\ this\ \isakeyword{and}\ h{\isadigit{1}}\ \isakeyword{and}\ h{\isadigit{2}}\ \isacommand{show}\isamarkupfalse%
\ {\isacharquery}thesis\isanewline
\ \ \ \ \isacommand{by}\isamarkupfalse%
\ {\isacharparenleft}simp\ {\isacharparenleft}no{\isacharunderscore}asm{\isacharunderscore}simp{\isacharparenright}{\isacharparenright}\isanewline
\isacommand{qed}\isamarkupfalse%
\endisatagproof
{\isafoldproof}%
\isadelimproof
\isanewline
\endisadelimproof
\isanewline 
\isacommand{lemma}\isamarkupfalse%
\ arith{\isacharunderscore}mult{\isacharunderscore}neq{\isacharunderscore}nzero{\isadigit{1}}{\isacharcolon}\isanewline
\ \ \isakeyword{fixes}\ i{\isacharcolon}{\isacharcolon}nat\isanewline
\ \ \isakeyword{assumes}\ h{\isadigit{1}}{\isacharcolon}{\isachardoublequoteopen}i\ {\isacharless}\ n{\isachardoublequoteclose}\isanewline
\ \ \ \ \ \ \isakeyword{and}\ h{\isadigit{2}}{\isacharcolon}{\isachardoublequoteopen}{\isadigit{0}}\ {\isacharless}\ i{\isachardoublequoteclose}\isanewline
\ \ \isakeyword{shows}\ {\isachardoublequoteopen}i\ {\isacharplus}\ n\ {\isacharasterisk}\ t\ {\isasymnoteq}\ n\ {\isacharasterisk}\ q{\isachardoublequoteclose}\isanewline
\isadelimproof
\endisadelimproof
\isatagproof
\isacommand{proof}\isamarkupfalse%
\ {\isacharminus}\isanewline
\ \ \isacommand{from}\isamarkupfalse%
\ h{\isadigit{1}}\ \isakeyword{and}\ h{\isadigit{2}}\ \isacommand{have}\isamarkupfalse%
\ sg{\isadigit{1}}{\isacharcolon}{\isachardoublequoteopen}{\isacharparenleft}i\ {\isacharplus}\ n\ {\isacharasterisk}\ t{\isacharparenright}\ mod\ n\ {\isacharequal}\ i{\isachardoublequoteclose}\isanewline
\ \ \ \ \isacommand{by}\isamarkupfalse%
\ {\isacharparenleft}simp\ add{\isacharcolon}\ mod{\isacharunderscore}mult{\isacharunderscore}self{\isadigit{2}}{\isacharparenright}\isanewline
\ \ \isacommand{also}\isamarkupfalse%
\ \isacommand{have}\isamarkupfalse%
\ sg{\isadigit{2}}{\isacharcolon}{\isachardoublequoteopen}{\isacharparenleft}n\ {\isacharasterisk}\ q{\isacharparenright}\ mod\ n\ {\isacharequal}\ {\isadigit{0}}{\isachardoublequoteclose}\ \ \isacommand{by}\isamarkupfalse%
\ simp\isanewline
\ \ \isacommand{from}\isamarkupfalse%
\ this\ \isakeyword{and}\ h{\isadigit{1}}\ \isakeyword{and}\ h{\isadigit{2}}\ \isacommand{have}\isamarkupfalse%
\ {\isachardoublequoteopen}{\isacharparenleft}i\ {\isacharplus}\ n\ {\isacharasterisk}\ t{\isacharparenright}\ mod\ n\ {\isasymnoteq}\ {\isacharparenleft}n\ {\isacharasterisk}\ q{\isacharparenright}\ mod\ n{\isachardoublequoteclose}\isanewline
\ \ \ \ \isacommand{by}\isamarkupfalse%
\ simp\isanewline
\ \ \isacommand{from}\isamarkupfalse%
\ this\ \isacommand{show}\isamarkupfalse%
\ {\isacharquery}thesis\ \ \isacommand{by}\isamarkupfalse%
\ {\isacharparenleft}rule\ arith{\isacharunderscore}mod{\isacharunderscore}neq{\isacharparenright}\isanewline
\isacommand{qed}\isamarkupfalse%
\endisatagproof
{\isafoldproof}%
\isadelimproof
\isanewline
\endisadelimproof
\isanewline 
\isacommand{lemma}\isamarkupfalse%
\ arith{\isacharunderscore}mult{\isacharunderscore}neq{\isacharunderscore}nzero{\isadigit{2}}{\isacharcolon}\isanewline
\ \ \isakeyword{fixes}\ i{\isacharcolon}{\isacharcolon}nat\isanewline
\ \ \isakeyword{assumes}\ h{\isadigit{1}}{\isacharcolon}{\isachardoublequoteopen}i\ {\isacharless}\ n{\isachardoublequoteclose}\isanewline
\ \ \ \ \ \ \isakeyword{and}\ h{\isadigit{2}}{\isacharcolon}{\isachardoublequoteopen}{\isadigit{0}}\ {\isacharless}\ i{\isachardoublequoteclose}\isanewline
\ \ \isakeyword{shows}\ {\isachardoublequoteopen}n\ {\isacharasterisk}\ t\ {\isacharplus}\ i\ {\isasymnoteq}\ n\ {\isacharasterisk}\ q{\isachardoublequoteclose}\isanewline
\isadelimproof
\endisadelimproof
\isatagproof
\isacommand{proof}\isamarkupfalse%
\ {\isacharminus}\ \isanewline
\ \ \isacommand{from}\isamarkupfalse%
\ h{\isadigit{1}}\ \isakeyword{and}\ h{\isadigit{2}}\ \isacommand{have}\isamarkupfalse%
\ {\isachardoublequoteopen}i\ {\isacharplus}\ n\ {\isacharasterisk}\ t\ {\isasymnoteq}\ n\ {\isacharasterisk}\ q{\isachardoublequoteclose}\ \isanewline
\ \ \ \ \isacommand{by}\isamarkupfalse%
\ {\isacharparenleft}rule\ arith{\isacharunderscore}mult{\isacharunderscore}neq{\isacharunderscore}nzero{\isadigit{1}}{\isacharparenright}\isanewline
\ \ \isacommand{from}\isamarkupfalse%
\ this\ \isacommand{show}\isamarkupfalse%
\ {\isacharquery}thesis\ \isacommand{by}\isamarkupfalse%
\ simp\isanewline
\isacommand{qed}\isamarkupfalse%
\endisatagproof
{\isafoldproof}%
\isadelimproof
\isanewline
\endisadelimproof
\isanewline 
\isacommand{lemma}\isamarkupfalse%
\ arith{\isacharunderscore}mult{\isacharunderscore}neq{\isacharunderscore}nzero{\isadigit{3}}{\isacharcolon}\isanewline
\ \ \isakeyword{fixes}\ i{\isacharcolon}{\isacharcolon}nat\isanewline
\ \ \isakeyword{assumes}\ h{\isadigit{1}}{\isacharcolon}{\isachardoublequoteopen}i\ {\isacharless}\ n{\isachardoublequoteclose} 
 \ \isakeyword{and}\ h{\isadigit{2}}{\isacharcolon}{\isachardoublequoteopen}{\isadigit{0}}\ {\isacharless}\ i{\isachardoublequoteclose}\isanewline
\ \ \isakeyword{shows}\ {\isachardoublequoteopen}n\ {\isacharplus}\ n\ {\isacharasterisk}\ t\ {\isacharplus}\ i\ {\isasymnoteq}\ n\ {\isacharasterisk}\ qc{\isachardoublequoteclose}\isanewline
\isadelimproof
\endisadelimproof
\isatagproof
\isacommand{proof}\isamarkupfalse%
\ {\isacharminus}\isanewline
\ \ \ \isacommand{from}\isamarkupfalse%
\ h{\isadigit{1}}\ \isakeyword{and}\ h{\isadigit{2}}\ \isacommand{have}\isamarkupfalse%
\ sg{\isadigit{1}}{\isacharcolon}\ {\isachardoublequoteopen}n\ {\isacharasterisk}{\isacharparenleft}Suc\ t{\isacharparenright}\ {\isacharplus}\ i\ \ {\isasymnoteq}\ n\ {\isacharasterisk}\ qc{\isachardoublequoteclose}\isanewline
\ \ \ \ \ \isacommand{by}\isamarkupfalse%
\ {\isacharparenleft}rule\ arith{\isacharunderscore}mult{\isacharunderscore}neq{\isacharunderscore}nzero{\isadigit{2}}{\isacharparenright}\isanewline
\ \ \ \isacommand{have}\isamarkupfalse%
\ sg{\isadigit{2}}{\isacharcolon}\ {\isachardoublequoteopen}n\ {\isacharplus}\ n\ {\isacharasterisk}\ t\ {\isacharplus}\ i\ {\isacharequal}\ n\ {\isacharasterisk}{\isacharparenleft}Suc\ t{\isacharparenright}\ {\isacharplus}\ i{\isachardoublequoteclose}\ \isacommand{by}\isamarkupfalse%
\ simp\isanewline
\ \ \ \isacommand{from}\isamarkupfalse%
\ sg{\isadigit{1}}\ \isakeyword{and}\ sg{\isadigit{2}}\ \ \isacommand{show}\isamarkupfalse%
\ {\isacharquery}thesis\ \ \isacommand{by}\isamarkupfalse%
\ arith\isanewline
\isacommand{qed}\isamarkupfalse%
\endisatagproof
{\isafoldproof}%
\isadelimproof
\isanewline
\endisadelimproof
\isanewline 
\isacommand{lemma}\isamarkupfalse%
\ arith{\isacharunderscore}modZero{\isadigit{1}}{\isacharcolon}\isanewline
\ \ {\isachardoublequoteopen}{\isacharparenleft}t\ {\isacharplus}\ n\ {\isacharasterisk}\ t{\isacharparenright}\ mod\ Suc\ n\ {\isacharequal}\ {\isadigit{0}}{\isachardoublequoteclose}\isanewline
\isadelimproof
\endisadelimproof
\isatagproof
\isacommand{proof}\isamarkupfalse%
\ {\isacharminus}\ \isanewline
\ \ \isacommand{have}\isamarkupfalse%
\ {\isachardoublequoteopen}{\isacharparenleft}{\isacharparenleft}Suc\ n{\isacharparenright}\ {\isacharasterisk}\ t{\isacharparenright}\ mod\ Suc\ n\ {\isacharequal}\ {\isadigit{0}}{\isachardoublequoteclose}\ \isacommand{by}\isamarkupfalse%
\ {\isacharparenleft}rule\ mod{\isacharunderscore}mult{\isacharunderscore}self{\isadigit{1}}{\isacharunderscore}is{\isacharunderscore}{\isadigit{0}}{\isacharparenright}\isanewline
\ \ \isacommand{from}\isamarkupfalse%
\ this\ \isacommand{show}\isamarkupfalse%
\ {\isacharquery}thesis\ \isacommand{by}\isamarkupfalse%
\ simp\isanewline
\isacommand{qed}\isamarkupfalse%
\endisatagproof
{\isafoldproof}%
\isadelimproof
\isanewline
\endisadelimproof
\isanewline 
\isacommand{lemma}\isamarkupfalse%
\ arith{\isacharunderscore}modZero{\isadigit{2}}{\isacharcolon}\isanewline
\ \ {\isachardoublequoteopen}Suc\ {\isacharparenleft}n\ {\isacharplus}\ {\isacharparenleft}t\ {\isacharplus}\ n\ {\isacharasterisk}\ t{\isacharparenright}{\isacharparenright}\ mod\ Suc\ n\ {\isacharequal}\ {\isadigit{0}}{\isachardoublequoteclose}\isanewline
\isadelimproof
\endisadelimproof
\isatagproof
\isacommand{proof}\isamarkupfalse%
\ {\isacharminus}\isanewline
\ \ \isacommand{have}\isamarkupfalse%
\ {\isachardoublequoteopen}{\isacharparenleft}{\isacharparenleft}Suc\ n{\isacharparenright}\ {\isacharasterisk}\ {\isacharparenleft}Suc\ t{\isacharparenright}{\isacharparenright}\ mod\ Suc\ n\ {\isacharequal}\ {\isadigit{0}}{\isachardoublequoteclose}\ \isacommand{by}\isamarkupfalse%
\ {\isacharparenleft}rule\ mod{\isacharunderscore}mult{\isacharunderscore}self{\isadigit{1}}{\isacharunderscore}is{\isacharunderscore}{\isadigit{0}}{\isacharparenright}\isanewline
\ \ \isacommand{from}\isamarkupfalse%
\ this\ \isacommand{show}\isamarkupfalse%
\ {\isacharquery}thesis\ \isacommand{by}\isamarkupfalse%
\ simp\isanewline
\isacommand{qed}\isamarkupfalse%
\endisatagproof
{\isafoldproof}%
\isadelimproof
\isanewline
\endisadelimproof
\isanewline 
\isacommand{lemma}\isamarkupfalse%
\ arith{\isadigit{1}}{\isacharcolon}\isanewline
\ \ \isakeyword{assumes}\ h{\isadigit{1}}{\isacharcolon}{\isachardoublequoteopen}Suc\ n\ {\isacharasterisk}\ t\ {\isacharequal}\ Suc\ n\ {\isacharasterisk}\ q{\isachardoublequoteclose}\isanewline
\ \ \isakeyword{shows}\ {\isachardoublequoteopen}t\ {\isacharequal}\ q{\isachardoublequoteclose}\isanewline
\isadelimproof
\endisadelimproof
\isatagproof
\isacommand{proof}\isamarkupfalse%
\ {\isacharminus}\isanewline
\ \ \isacommand{have}\isamarkupfalse%
\ {\isachardoublequoteopen}Suc\ n\ {\isacharasterisk}\ t\ {\isacharequal}\ Suc\ n\ {\isacharasterisk}\ q\ {\isacharequal}\ {\isacharparenleft}t\ {\isacharequal}\ q\ {\isacharbar}\ {\isacharparenleft}Suc\ n{\isacharparenright}\ {\isacharequal}\ {\isacharparenleft}{\isadigit{0}}{\isacharcolon}{\isacharcolon}nat{\isacharparenright}{\isacharparenright}{\isachardoublequoteclose}\isanewline
\ \ \ \ \isacommand{by}\isamarkupfalse%
\ {\isacharparenleft}rule\ mult{\isacharunderscore}cancel{\isadigit{1}}{\isacharparenright}\isanewline
\ \ \isacommand{from}\isamarkupfalse%
\ this\ \isakeyword{and}\ h{\isadigit{1}}\ \isacommand{show}\isamarkupfalse%
\ {\isacharquery}thesis\ \isacommand{by}\isamarkupfalse%
\ simp\isanewline
\isacommand{qed}\isamarkupfalse%
\endisatagproof
{\isafoldproof}%
\isadelimproof
\isanewline
\endisadelimproof
\isanewline 
\isacommand{lemma}\isamarkupfalse%
\ arith{\isadigit{2}}{\isacharcolon}\isanewline
\ \ \isakeyword{fixes}\ t\ n\ q\ {\isacharcolon}{\isacharcolon}\ {\isachardoublequoteopen}nat{\isachardoublequoteclose}\isanewline
\ \ \isakeyword{assumes}\ h{\isadigit{1}}{\isacharcolon}{\isachardoublequoteopen}t\ {\isacharplus}\ n\ {\isacharasterisk}\ t\ {\isacharequal}\ q\ {\isacharplus}\ n\ {\isacharasterisk}\ q{\isachardoublequoteclose}\isanewline
\ \ \isakeyword{shows}\ {\isachardoublequoteopen}t\ {\isacharequal}\ q{\isachardoublequoteclose}\isanewline
\isadelimproof
\endisadelimproof
\isatagproof
\isacommand{proof}\isamarkupfalse%
\ {\isacharminus}\isanewline
\ \ \isacommand{have}\isamarkupfalse%
\ sg{\isadigit{1}}{\isacharcolon}{\isachardoublequoteopen}t\ {\isacharplus}\ n\ {\isacharasterisk}\ t\ {\isacharequal}\ {\isacharparenleft}Suc\ n{\isacharparenright}\ {\isacharasterisk}\ t{\isachardoublequoteclose}\ \isacommand{by}\isamarkupfalse%
\ auto\isanewline
\ \ \isacommand{have}\isamarkupfalse%
\ sg{\isadigit{2}}{\isacharcolon}{\isachardoublequoteopen}q\ {\isacharplus}\ n\ {\isacharasterisk}\ q\ {\isacharequal}\ {\isacharparenleft}Suc\ n{\isacharparenright}\ {\isacharasterisk}\ q{\isachardoublequoteclose}\ \isacommand{by}\isamarkupfalse%
\ auto\isanewline
\ \ \isacommand{from}\isamarkupfalse%
\ h{\isadigit{1}}\ \isakeyword{and}\ sg{\isadigit{1}}\ \isakeyword{and}\ sg{\isadigit{2}}\ \isacommand{have}\isamarkupfalse%
\ {\isachardoublequoteopen}Suc\ n\ {\isacharasterisk}\ t\ {\isacharequal}\ Suc\ n\ {\isacharasterisk}\ q{\isachardoublequoteclose}\ \isacommand{by}\isamarkupfalse%
\ arith\isanewline
\ \ \isacommand{from}\isamarkupfalse%
\ this\ \isacommand{show}\isamarkupfalse%
\ {\isacharquery}thesis\ \isacommand{by}\isamarkupfalse%
\ {\isacharparenleft}rule\ arith{\isadigit{1}}{\isacharparenright}\isanewline
\isacommand{qed}\isamarkupfalse%
\endisatagproof
{\isafoldproof}%
\isadelimproof
\isanewline
\endisadelimproof
\isadelimtheory
\isanewline
\endisadelimtheory
\isatagtheory
\isacommand{end}\isamarkupfalse%
\endisatagtheory
{\isafoldtheory}%
\isadelimtheory
\endisadelimtheory
\end{isabellebody}%

%
\begin{isabellebody}%
\def\isabellecontext{stream}%
\isamarkupheader{FOCUS streams: operators and lemmas%
}
\isamarkuptrue%
\isadelimtheory
\endisadelimtheory
\isatagtheory
\isacommand{theory}\isamarkupfalse%
\ stream\isanewline
\ \ \isakeyword{imports}\ ListExtras\ ArithExtras\isanewline
\isakeyword{begin}%
\endisatagtheory
{\isafoldtheory}%
\isadelimtheory
\endisadelimtheory
\isamarkupsubsection{Definition of the FOCUS stream types%
}
\isamarkuptrue%
\isamarkupcmt{Finite timed FOCUS stream%
}
\isanewline
\isacommand{type{\isacharunderscore}synonym}\isamarkupfalse%
\ {\isacharprime}a\ fstream\ {\isacharequal}\ {\isachardoublequoteopen}{\isacharprime}a\ list\ list{\isachardoublequoteclose}\isanewline
\isanewline
\isamarkupcmt{Infinite timed FOCUS stream%
}
\isanewline
\isacommand{type{\isacharunderscore}synonym}\isamarkupfalse%
\ {\isacharprime}a\ istream\ {\isacharequal}\ {\isachardoublequoteopen}nat\ {\isasymRightarrow}\ {\isacharprime}a\ list{\isachardoublequoteclose}\isanewline
\isanewline
\isamarkupcmt{Infinite untimed FOCUS stream%
}
\isanewline
\isacommand{type{\isacharunderscore}synonym}\isamarkupfalse%
\ {\isacharprime}a\ iustream\ {\isacharequal}\ {\isachardoublequoteopen}nat\ {\isasymRightarrow}\ {\isacharprime}a{\isachardoublequoteclose}\isanewline
\isanewline
\isamarkupcmt{FOCUS stream (general)%
}
\isanewline
\isacommand{datatype}\isamarkupfalse%
\ {\isacharprime}a\ stream\ {\isacharequal}\ \isanewline
\ \ \ \ \ \ \ \ \ \ FinT\ {\isachardoublequoteopen}{\isacharprime}a\ fstream{\isachardoublequoteclose}\ %
\isamarkupcmt{finite timed streams%
}
\isanewline
\ \ \ \ \ \ \ \ {\isacharbar}\ FinU\ {\isachardoublequoteopen}{\isacharprime}a\ list{\isachardoublequoteclose}\ %
\isamarkupcmt{finite untimed streams%
}
\isanewline
\ \ \ \ \ \ \ \ {\isacharbar}\ InfT\ {\isachardoublequoteopen}{\isacharprime}a\ istream{\isachardoublequoteclose}\ %
\isamarkupcmt{infinite timed streams%
}
\isanewline
\ \ \ \ \ \ \ \ {\isacharbar}\ InfU\ {\isachardoublequoteopen}{\isacharprime}a\ iustream{\isachardoublequoteclose}\ %
\isamarkupcmt{infinite untimed streams%
}
\isamarkupsubsection{Definitions of operators%
}
\isamarkuptrue%
\isamarkupcmt{domain of an infinite untimed stream%
}
\isanewline
\isacommand{definition}\isamarkupfalse%
\isanewline
\ \ \ infU{\isacharunderscore}dom\ {\isacharcolon}{\isacharcolon}\ {\isachardoublequoteopen}natInf\ set{\isachardoublequoteclose}\isanewline
\ \isakeyword{where}\isanewline
\ \ {\isachardoublequoteopen}infU{\isacharunderscore}dom\ {\isasymequiv}\ {\isacharbraceleft}x{\isachardot}\ {\isasymexists}\ i{\isachardot}\ x\ {\isacharequal}\ {\isacharparenleft}Fin\ i{\isacharparenright}{\isacharbraceright}\ {\isasymunion}\ {\isacharbraceleft}{\isasyminfinity}{\isacharbraceright}{\isachardoublequoteclose}\isanewline
\ \ \isanewline
\isamarkupcmt{domain of a finite untimed stream (using natural numbers enriched by Infinity)%
}
\isanewline
\isacommand{definition}\isamarkupfalse%
\isanewline
\ \ \ finU{\isacharunderscore}dom{\isacharunderscore}natInf\ {\isacharcolon}{\isacharcolon}\ {\isachardoublequoteopen}{\isacharprime}a\ list\ {\isasymRightarrow}\ natInf\ set{\isachardoublequoteclose}\isanewline
\ \ \ \isakeyword{where}\isanewline
\ \ {\isachardoublequoteopen}finU{\isacharunderscore}dom{\isacharunderscore}natInf\ s\ {\isasymequiv}\ {\isacharbraceleft}x{\isachardot}\ {\isasymexists}\ i{\isachardot}\ x\ {\isacharequal}\ {\isacharparenleft}Fin\ i{\isacharparenright}\ {\isasymand}\ i\ {\isacharless}\ {\isacharparenleft}length\ s{\isacharparenright}{\isacharbraceright}{\isachardoublequoteclose}\isanewline
\isanewline
\isamarkupcmt{domain of a finite untimed stream%
}
\isanewline
\isacommand{primrec}\isamarkupfalse%
\isanewline
finU{\isacharunderscore}dom\ {\isacharcolon}{\isacharcolon}\ {\isachardoublequoteopen}{\isacharprime}a\ list\ {\isasymRightarrow}\ nat\ set{\isachardoublequoteclose}\isanewline
\isakeyword{where}\isanewline
\ \ {\isachardoublequoteopen}finU{\isacharunderscore}dom\ {\isacharbrackleft}{\isacharbrackright}\ {\isacharequal}\ {\isacharbraceleft}{\isacharbraceright}{\isachardoublequoteclose}\ {\isacharbar}\isanewline
\ \ {\isachardoublequoteopen}finU{\isacharunderscore}dom\ {\isacharparenleft}x{\isacharhash}xs{\isacharparenright}\ {\isacharequal}\ {\isacharbraceleft}length\ xs{\isacharbraceright}\ {\isasymunion}\ {\isacharparenleft}finU{\isacharunderscore}dom\ xs{\isacharparenright}{\isachardoublequoteclose}\isanewline
\isanewline
\isamarkupcmt{range of a finite timed stream%
}
\isanewline
\isacommand{primrec}\isamarkupfalse%
\isanewline
\ \ finT{\isacharunderscore}range\ {\isacharcolon}{\isacharcolon}\ {\isachardoublequoteopen}{\isacharprime}a\ fstream\ {\isasymRightarrow}\ {\isacharprime}a\ set{\isachardoublequoteclose}\isanewline
\isakeyword{where}\ \ \isanewline
\ \ {\isachardoublequoteopen}finT{\isacharunderscore}range\ {\isacharbrackleft}{\isacharbrackright}\ {\isacharequal}\ {\isacharbraceleft}{\isacharbraceright}{\isachardoublequoteclose}\ {\isacharbar}\isanewline
\ \ {\isachardoublequoteopen}finT{\isacharunderscore}range\ {\isacharparenleft}x{\isacharhash}xs{\isacharparenright}\ {\isacharequal}\ {\isacharparenleft}set\ x{\isacharparenright}\ {\isasymunion}\ finT{\isacharunderscore}range\ xs{\isachardoublequoteclose}\isanewline
\isanewline
\isamarkupcmt{range of a finite untimed stream%
}
\isanewline
\isacommand{definition}\isamarkupfalse%
\isanewline
\ \ \ finU{\isacharunderscore}range\ {\isacharcolon}{\isacharcolon}\ {\isachardoublequoteopen}{\isacharprime}a\ list\ {\isasymRightarrow}\ {\isacharprime}a\ set{\isachardoublequoteclose}\isanewline
\isakeyword{where}\isanewline
\ \ {\isachardoublequoteopen}finU{\isacharunderscore}range\ x\ {\isasymequiv}\ set\ x{\isachardoublequoteclose}\isanewline
\isanewline
\isamarkupcmt{range of an infinite timed stream%
}
\isanewline
\isacommand{definition}\isamarkupfalse%
\isanewline
\ \ \ infT{\isacharunderscore}range\ {\isacharcolon}{\isacharcolon}\ {\isachardoublequoteopen}{\isacharprime}a\ istream\ {\isasymRightarrow}\ {\isacharprime}a\ set{\isachardoublequoteclose}\isanewline
\isakeyword{where}\isanewline
\ \ {\isachardoublequoteopen}infT{\isacharunderscore}range\ s\ {\isasymequiv}\ {\isacharbraceleft}y{\isachardot}\ {\isasymexists}\ i{\isacharcolon}{\isacharcolon}nat{\isachardot}\ y\ mem\ {\isacharparenleft}s\ i{\isacharparenright}{\isacharbraceright}{\isachardoublequoteclose}\isanewline
\isanewline
\isamarkupcmt{range of a finite untimed stream%
}
\isanewline
\isacommand{definition}\isamarkupfalse%
\isanewline
\ \ \ infU{\isacharunderscore}range\ {\isacharcolon}{\isacharcolon}\ {\isachardoublequoteopen}{\isacharparenleft}nat\ {\isasymRightarrow}\ {\isacharprime}a{\isacharparenright}\ {\isasymRightarrow}\ {\isacharprime}a\ set{\isachardoublequoteclose}\isanewline
\isakeyword{where}\isanewline
\ \ {\isachardoublequoteopen}infU{\isacharunderscore}range\ s\ {\isasymequiv}\ {\isacharbraceleft}\ y{\isachardot}\ {\isasymexists}\ i{\isacharcolon}{\isacharcolon}nat{\isachardot}\ y\ {\isacharequal}\ {\isacharparenleft}s\ i{\isacharparenright}\ {\isacharbraceright}{\isachardoublequoteclose}\isanewline
\isanewline
\isamarkupcmt{range of a (general) stream%
}
\isanewline
\isacommand{definition}\isamarkupfalse%
\isanewline
\ \ \ stream{\isacharunderscore}range\ {\isacharcolon}{\isacharcolon}\ {\isachardoublequoteopen}{\isacharprime}a\ stream\ {\isasymRightarrow}\ {\isacharprime}a\ set{\isachardoublequoteclose}\isanewline
\isakeyword{where}\isanewline
\ {\isachardoublequoteopen}stream{\isacharunderscore}range\ s\ {\isasymequiv}\ case\ s\ of\isanewline
\ \ \ \ \ \ \ FinT\ x\ {\isasymRightarrow}\ finT{\isacharunderscore}range\ x\ \isanewline
\ \ \ \ \ {\isacharbar}\ FinU\ x\ {\isasymRightarrow}\ finU{\isacharunderscore}range\ x\ \isanewline
\ \ \ \ \ {\isacharbar}\ InfT\ x\ {\isasymRightarrow}\ infT{\isacharunderscore}range\ x\ \ \isanewline
\ \ \ \ \ {\isacharbar}\ InfU\ x\ {\isasymRightarrow}\ infU{\isacharunderscore}range\ x{\isachardoublequoteclose}\ \isanewline
\isanewline
\isamarkupcmt{finite timed stream that consists of n empty time intervals%
}
\ \isanewline
\isacommand{primrec}\isamarkupfalse%
\isanewline
\ \ \ nticks\ {\isacharcolon}{\isacharcolon}\ {\isachardoublequoteopen}nat\ {\isasymRightarrow}\ {\isacharprime}a\ fstream{\isachardoublequoteclose}\isanewline
\isakeyword{where}\isanewline
\ \ {\isachardoublequoteopen}nticks\ {\isadigit{0}}\ {\isacharequal}\ {\isacharbrackleft}{\isacharbrackright}{\isachardoublequoteclose}\ {\isacharbar}\isanewline
\ \ {\isachardoublequoteopen}nticks\ {\isacharparenleft}Suc\ i{\isacharparenright}\ {\isacharequal}\ {\isacharbrackleft}{\isacharbrackright}\ {\isacharhash}\ {\isacharparenleft}nticks\ i{\isacharparenright}{\isachardoublequoteclose}\isanewline
\isanewline
\isamarkupcmt{removing the first element from an infinite stream%
}
\isanewline
\isamarkupcmt{in the case of an untimed stream: removing the first data element%
}
\ \ \isanewline
\isamarkupcmt{in the case of a timed stream: removing the first time interval%
}
\ \isanewline
\isacommand{definition}\isamarkupfalse%
\isanewline
\ \ \ inf{\isacharunderscore}tl\ {\isacharcolon}{\isacharcolon}\ {\isachardoublequoteopen}{\isacharparenleft}nat\ {\isasymRightarrow}\ {\isacharprime}a{\isacharparenright}\ {\isasymRightarrow}\ {\isacharparenleft}nat\ {\isasymRightarrow}\ {\isacharprime}a{\isacharparenright}{\isachardoublequoteclose}\isanewline
\isakeyword{where}\isanewline
\ \ {\isachardoublequoteopen}inf{\isacharunderscore}tl\ s\ {\isasymequiv}\ {\isacharparenleft}{\isasymlambda}\ i{\isachardot}\ s\ {\isacharparenleft}Suc\ i{\isacharparenright}{\isacharparenright}{\isachardoublequoteclose}\isanewline
\isanewline
\isamarkupcmt{removing i first elements from an infinite stream s%
}
\ \ \isanewline
\isamarkupcmt{in the case of an untimed stream: removing i first data elements%
}
\ \ \isanewline
\isamarkupcmt{in the case of a timed stream: removing i first time intervals%
}
\ \isanewline
\isacommand{definition}\isamarkupfalse%
\isanewline
\ \ \ inf{\isacharunderscore}drop\ {\isacharcolon}{\isacharcolon}\ {\isachardoublequoteopen}nat\ {\isasymRightarrow}\ {\isacharparenleft}nat\ {\isasymRightarrow}\ {\isacharprime}a{\isacharparenright}\ {\isasymRightarrow}\ {\isacharparenleft}nat\ {\isasymRightarrow}\ {\isacharprime}a{\isacharparenright}{\isachardoublequoteclose}\isanewline
\isakeyword{where}\isanewline
\ \ {\isachardoublequoteopen}inf{\isacharunderscore}drop\ i\ s\ {\isasymequiv}\ \ {\isasymlambda}\ j{\isachardot}\ s\ {\isacharparenleft}i{\isacharplus}j{\isacharparenright}{\isachardoublequoteclose}\ \ \isanewline
\isanewline
\isamarkupcmt{finding the first nonempty time interval in a finite timed stream%
}
\isanewline
\isacommand{primrec}\isamarkupfalse%
\isanewline
\ fin{\isacharunderscore}find{\isadigit{1}}nonemp\ {\isacharcolon}{\isacharcolon}\ {\isachardoublequoteopen}{\isacharprime}a\ fstream\ {\isasymRightarrow}\ {\isacharprime}a\ list{\isachardoublequoteclose}\isanewline
\isakeyword{where}\ \isanewline
\ {\isachardoublequoteopen}fin{\isacharunderscore}find{\isadigit{1}}nonemp\ {\isacharbrackleft}{\isacharbrackright}\ {\isacharequal}\ {\isacharbrackleft}{\isacharbrackright}{\isachardoublequoteclose}\ {\isacharbar}\isanewline
\ {\isachardoublequoteopen}fin{\isacharunderscore}find{\isadigit{1}}nonemp\ {\isacharparenleft}x{\isacharhash}xs{\isacharparenright}\ {\isacharequal}\isanewline
\ \ \ \ {\isacharparenleft}\ if\ x\ {\isacharequal}\ {\isacharbrackleft}{\isacharbrackright}\isanewline
\ \ \ \ \ \ then\ fin{\isacharunderscore}find{\isadigit{1}}nonemp\ xs\isanewline
\ \ \ \ \ \ else\ x\ {\isacharparenright}{\isachardoublequoteclose}\isanewline
\isanewline
\isamarkupcmt{finding the first nonempty time interval in an infinite timed stream%
}
\isanewline
\isacommand{definition}\isamarkupfalse%
\ \isanewline
\ \ inf{\isacharunderscore}find{\isadigit{1}}nonemp\ {\isacharcolon}{\isacharcolon}\ {\isachardoublequoteopen}{\isacharprime}a\ istream\ {\isasymRightarrow}\ {\isacharprime}a\ list{\isachardoublequoteclose}\isanewline
\isakeyword{where}\isanewline
\ {\isachardoublequoteopen}inf{\isacharunderscore}find{\isadigit{1}}nonemp\ s\ \isanewline
\ \ {\isasymequiv}\isanewline
\ \ {\isacharparenleft}\ if\ {\isacharparenleft}{\isasymexists}\ i{\isachardot}\ s\ i\ {\isasymnoteq}\ {\isacharbrackleft}{\isacharbrackright}{\isacharparenright}\isanewline
\ \ \ \ then\ s\ {\isacharparenleft}LEAST\ i{\isachardot}\ s\ i\ {\isasymnoteq}\ {\isacharbrackleft}{\isacharbrackright}{\isacharparenright}\isanewline
\ \ \ \ else\ {\isacharbrackleft}{\isacharbrackright}\ {\isacharparenright}{\isachardoublequoteclose}\isanewline
\isanewline
\isamarkupcmt{finding the index of the first nonempty time interval in a finite timed stream%
}
\isanewline
\isacommand{primrec}\isamarkupfalse%
\isanewline
\ fin{\isacharunderscore}find{\isadigit{1}}nonemp{\isacharunderscore}index\ {\isacharcolon}{\isacharcolon}\ {\isachardoublequoteopen}{\isacharprime}a\ fstream\ {\isasymRightarrow}\ nat{\isachardoublequoteclose}\isanewline
\isakeyword{where}\isanewline
\ {\isachardoublequoteopen}fin{\isacharunderscore}find{\isadigit{1}}nonemp{\isacharunderscore}index\ {\isacharbrackleft}{\isacharbrackright}\ {\isacharequal}\ {\isadigit{0}}{\isachardoublequoteclose}\ {\isacharbar}\isanewline
\ {\isachardoublequoteopen}fin{\isacharunderscore}find{\isadigit{1}}nonemp{\isacharunderscore}index\ {\isacharparenleft}x{\isacharhash}xs{\isacharparenright}\ {\isacharequal}\isanewline
\ \ \ \ {\isacharparenleft}\ if\ x\ {\isacharequal}\ {\isacharbrackleft}{\isacharbrackright}\isanewline
\ \ \ \ \ \ then\ Suc\ {\isacharparenleft}fin{\isacharunderscore}find{\isadigit{1}}nonemp{\isacharunderscore}index\ xs{\isacharparenright}\isanewline
\ \ \ \ \ \ else\ {\isadigit{0}}\ {\isacharparenright}{\isachardoublequoteclose}\isanewline
\isanewline
\isamarkupcmt{finding the index of the first nonempty time interval in an infinite timed stream%
}
\isanewline
\isacommand{definition}\isamarkupfalse%
\isanewline
\ \ inf{\isacharunderscore}find{\isadigit{1}}nonemp{\isacharunderscore}index\ {\isacharcolon}{\isacharcolon}\ {\isachardoublequoteopen}{\isacharprime}a\ istream\ {\isasymRightarrow}\ nat{\isachardoublequoteclose}\isanewline
\isakeyword{where}\isanewline
\ {\isachardoublequoteopen}inf{\isacharunderscore}find{\isadigit{1}}nonemp{\isacharunderscore}index\ s\ \isanewline
\ \ {\isasymequiv}\isanewline
\ \ {\isacharparenleft}\ if\ {\isacharparenleft}{\isasymexists}\ i{\isachardot}\ s\ i\ {\isasymnoteq}\ {\isacharbrackleft}{\isacharbrackright}{\isacharparenright}\isanewline
\ \ \ \ then\ {\isacharparenleft}LEAST\ i{\isachardot}\ s\ i\ {\isasymnoteq}\ {\isacharbrackleft}{\isacharbrackright}{\isacharparenright}\isanewline
\ \ \ \ else\ {\isadigit{0}}\ {\isacharparenright}{\isachardoublequoteclose}\isanewline
\isanewline
\isamarkupcmt{length of a finite timed stream: number of data elements in this stream%
}
\ \ \isanewline
\isacommand{primrec}\isamarkupfalse%
\isanewline
\ \ fin{\isacharunderscore}length\ {\isacharcolon}{\isacharcolon}\ {\isachardoublequoteopen}{\isacharprime}a\ fstream\ {\isasymRightarrow}\ nat{\isachardoublequoteclose}\isanewline
\isakeyword{where}\isanewline
\ \ {\isachardoublequoteopen}fin{\isacharunderscore}length\ {\isacharbrackleft}{\isacharbrackright}\ {\isacharequal}\ {\isadigit{0}}{\isachardoublequoteclose}\ {\isacharbar}\isanewline
\ \ {\isachardoublequoteopen}fin{\isacharunderscore}length\ {\isacharparenleft}x{\isacharhash}xs{\isacharparenright}\ {\isacharequal}\ {\isacharparenleft}length\ x{\isacharparenright}\ {\isacharplus}\ {\isacharparenleft}fin{\isacharunderscore}length\ xs{\isacharparenright}{\isachardoublequoteclose}\isanewline
\isanewline
\isamarkupcmt{length of a (general) stream%
}
\isanewline
\isacommand{definition}\isamarkupfalse%
\isanewline
\ \ \ stream{\isacharunderscore}length\ {\isacharcolon}{\isacharcolon}\ {\isachardoublequoteopen}{\isacharprime}a\ stream\ {\isasymRightarrow}\ natInf{\isachardoublequoteclose}\isanewline
\isakeyword{where}\isanewline
\ \ {\isachardoublequoteopen}stream{\isacharunderscore}length\ s\ {\isasymequiv}\ \isanewline
\ \ \ \ \ \ case\ s\ of\ \isanewline
\ \ \ \ \ \ \ \ \ \ \ \ \ \ \ \ {\isacharparenleft}FinT\ x{\isacharparenright}\ {\isasymRightarrow}\ Fin\ {\isacharparenleft}fin{\isacharunderscore}length\ x{\isacharparenright}\ \isanewline
\ \ \ \ \ \ \ \ \ \ \ \ \ \ {\isacharbar}\ {\isacharparenleft}FinU\ x{\isacharparenright}\ {\isasymRightarrow}\ Fin\ {\isacharparenleft}length\ x{\isacharparenright}\ \ \isanewline
\ \ \ \ \ \ \ \ \ \ \ \ \ \ {\isacharbar}\ {\isacharparenleft}InfT\ x{\isacharparenright}\ {\isasymRightarrow}\ {\isasyminfinity}\ \isanewline
\ \ \ \ \ \ \ \ \ \ \ \ \ \ {\isacharbar}\ {\isacharparenleft}InfU\ x{\isacharparenright}\ {\isasymRightarrow}\ {\isasyminfinity}{\isachardoublequoteclose}\isanewline
\isanewline
\isamarkupcmt{removing the first k elements from a finite (nonempty) timed stream%
}
\isanewline
\isacommand{primrec}\isamarkupfalse%
\isanewline
\ \ fin{\isacharunderscore}nth\ {\isacharcolon}{\isacharcolon}\ {\isachardoublequoteopen}{\isacharprime}a\ fstream\ {\isasymRightarrow}\ nat\ {\isasymRightarrow}\ {\isacharprime}a{\isachardoublequoteclose}\isanewline
\isakeyword{where}\ \isanewline
\ \ \ fin{\isacharunderscore}nth{\isacharunderscore}Cons{\isacharcolon}\isanewline
\ \ {\isachardoublequoteopen}fin{\isacharunderscore}nth\ {\isacharparenleft}hds\ {\isacharhash}\ tls{\isacharparenright}\ k\ {\isacharequal}\ \isanewline
\ \ \ \ \ \ {\isacharparenleft}\ if\ hds\ {\isacharequal}\ {\isacharbrackleft}{\isacharbrackright}\isanewline
\ \ \ \ \ \ \ \ then\ fin{\isacharunderscore}nth\ tls\ k\isanewline
\ \ \ \ \ \ \ \ else\ {\isacharparenleft}\ if\ {\isacharparenleft}k\ {\isacharless}\ {\isacharparenleft}length\ hds{\isacharparenright}{\isacharparenright}\isanewline
\ \ \ \ \ \ \ \ \ \ \ \ \ \ \ then\ nth\ hds\ k\isanewline
\ \ \ \ \ \ \ \ \ \ \ \ \ \ \ else\ fin{\isacharunderscore}nth\ tls\ {\isacharparenleft}k\ {\isacharminus}\ length\ hds{\isacharparenright}\ {\isacharparenright}{\isacharparenright}{\isachardoublequoteclose}\isanewline
\isanewline
\isamarkupcmt{removing i first data elements from an infinite timed stream s%
}
\ \ \ \isanewline
\isacommand{primrec}\isamarkupfalse%
\isanewline
\ \ inf{\isacharunderscore}nth\ {\isacharcolon}{\isacharcolon}\ {\isachardoublequoteopen}{\isacharprime}a\ istream\ {\isasymRightarrow}\ nat\ {\isasymRightarrow}\ {\isacharprime}a{\isachardoublequoteclose}\isanewline
\isakeyword{where}\ \isanewline
\ {\isachardoublequoteopen}inf{\isacharunderscore}nth\ s\ {\isadigit{0}}\ {\isacharequal}\ \ hd\ {\isacharparenleft}s\ {\isacharparenleft}LEAST\ i{\isachardot}{\isacharparenleft}s\ i{\isacharparenright}\ {\isasymnoteq}\ {\isacharbrackleft}{\isacharbrackright}{\isacharparenright}{\isacharparenright}{\isachardoublequoteclose}\ \ {\isacharbar}\isanewline
\ {\isachardoublequoteopen}inf{\isacharunderscore}nth\ s\ {\isacharparenleft}Suc\ k{\isacharparenright}\ {\isacharequal}\ \isanewline
\ \ \ \ \ \ {\isacharparenleft}\ if\ {\isacharparenleft}{\isacharparenleft}Suc\ k{\isacharparenright}\ {\isacharless}\ {\isacharparenleft}length\ {\isacharparenleft}s\ {\isadigit{0}}{\isacharparenright}{\isacharparenright}{\isacharparenright}\ \isanewline
\ \ \ \ \ \ \ \ then\ {\isacharparenleft}nth\ {\isacharparenleft}s\ {\isadigit{0}}{\isacharparenright}\ {\isacharparenleft}Suc\ k{\isacharparenright}{\isacharparenright}\isanewline
\ \ \ \ \ \ \ \ else\ {\isacharparenleft}\ if\ {\isacharparenleft}s\ {\isadigit{0}}{\isacharparenright}\ {\isacharequal}\ {\isacharbrackleft}{\isacharbrackright}\isanewline
\ \ \ \ \ \ \ \ \ \ \ \ \ \ \ then\ {\isacharparenleft}inf{\isacharunderscore}nth\ {\isacharparenleft}inf{\isacharunderscore}tl\ {\isacharparenleft}inf{\isacharunderscore}drop\ \isanewline
\ \ \ \ \ \ \ \ \ \ \ \ \ \ \ \ \ \ \ \ \ {\isacharparenleft}LEAST\ i{\isachardot}\ {\isacharparenleft}s\ i{\isacharparenright}\ {\isasymnoteq}\ {\isacharbrackleft}{\isacharbrackright}{\isacharparenright}\ s{\isacharparenright}{\isacharparenright}\ k\ {\isacharparenright}\isanewline
\ \ \ \ \ \ \ \ \ \ \ \ \ \ \ else\ inf{\isacharunderscore}nth\ {\isacharparenleft}inf{\isacharunderscore}tl\ s{\isacharparenright}\ k\ {\isacharparenright}{\isacharparenright}{\isachardoublequoteclose}\isanewline
\isanewline
\isamarkupcmt{removing the first k data elements from a (general) stream%
}
\isanewline
\isacommand{definition}\isamarkupfalse%
\isanewline
\ \ \ \ stream{\isacharunderscore}nth\ {\isacharcolon}{\isacharcolon}\ {\isachardoublequoteopen}{\isacharprime}a\ stream\ {\isasymRightarrow}\ nat\ {\isasymRightarrow}\ {\isacharprime}a{\isachardoublequoteclose}\isanewline
\isakeyword{where}\isanewline
\ \ \ {\isachardoublequoteopen}stream{\isacharunderscore}nth\ s\ k\ {\isasymequiv}\ \isanewline
\ \ \ \ \ \ \ \ \ case\ s\ of\ {\isacharparenleft}FinT\ x{\isacharparenright}\ {\isasymRightarrow}\ fin{\isacharunderscore}nth\ x\ k\isanewline
\ \ \ \ \ \ \ \ \ \ \ \ \ \ \ \ \ {\isacharbar}\ {\isacharparenleft}FinU\ x{\isacharparenright}\ {\isasymRightarrow}\ nth\ x\ k\isanewline
\ \ \ \ \ \ \ \ \ \ \ \ \ \ \ \ \ {\isacharbar}\ {\isacharparenleft}InfT\ x{\isacharparenright}\ {\isasymRightarrow}\ inf{\isacharunderscore}nth\ x\ k\ \isanewline
\ \ \ \ \ \ \ \ \ \ \ \ \ \ \ \ \ {\isacharbar}\ {\isacharparenleft}InfU\ x{\isacharparenright}\ {\isasymRightarrow}\ x\ k{\isachardoublequoteclose}\isanewline
\isanewline
\isamarkupcmt{prefix of an infinite stream%
}
\isanewline
\isacommand{primrec}\isamarkupfalse%
\ \isanewline
\ \ inf{\isacharunderscore}prefix\ {\isacharcolon}{\isacharcolon}\ {\isachardoublequoteopen}{\isacharprime}a\ list\ {\isasymRightarrow}\ {\isacharparenleft}nat\ {\isasymRightarrow}\ {\isacharprime}a{\isacharparenright}\ {\isasymRightarrow}\ nat\ {\isasymRightarrow}\ bool{\isachardoublequoteclose}\isanewline
\isakeyword{where}\ \ \ \isanewline
\ \ {\isachardoublequoteopen}inf{\isacharunderscore}prefix\ {\isacharbrackleft}{\isacharbrackright}\ s\ k\ {\isacharequal}\ True{\isachardoublequoteclose}\ {\isacharbar}\isanewline
\ \ {\isachardoublequoteopen}inf{\isacharunderscore}prefix\ {\isacharparenleft}x{\isacharhash}xs{\isacharparenright}\ s\ k\ {\isacharequal}\ {\isacharparenleft}\ {\isacharparenleft}x\ {\isacharequal}\ {\isacharparenleft}s\ k{\isacharparenright}{\isacharparenright}\ {\isasymand}\ {\isacharparenleft}inf{\isacharunderscore}prefix\ xs\ s\ {\isacharparenleft}Suc\ k{\isacharparenright}{\isacharparenright}\ {\isacharparenright}{\isachardoublequoteclose}\isanewline
\isanewline
\isamarkupcmt{prefix of a finite stream%
}
\isanewline
\isacommand{primrec}\isamarkupfalse%
\ \isanewline
\ \ fin{\isacharunderscore}prefix\ {\isacharcolon}{\isacharcolon}\ {\isachardoublequoteopen}{\isacharprime}a\ list\ {\isasymRightarrow}\ {\isacharprime}a\ list\ {\isasymRightarrow}\ bool{\isachardoublequoteclose}\isanewline
\isakeyword{where}\ \ \ \isanewline
\ \ {\isachardoublequoteopen}fin{\isacharunderscore}prefix\ {\isacharbrackleft}{\isacharbrackright}\ s\ {\isacharequal}\ True{\isachardoublequoteclose}\ {\isacharbar}\isanewline
\ \ {\isachardoublequoteopen}fin{\isacharunderscore}prefix\ {\isacharparenleft}x{\isacharhash}xs{\isacharparenright}\ s\ {\isacharequal}\ \isanewline
\ \ \ \ \ {\isacharparenleft}if\ {\isacharparenleft}s\ {\isacharequal}\ {\isacharbrackleft}{\isacharbrackright}{\isacharparenright}\ \isanewline
\ \ \ \ \ \ then\ False\isanewline
\ \ \ \ \ \ else\ {\isacharparenleft}x\ {\isacharequal}\ {\isacharparenleft}hd\ s{\isacharparenright}{\isacharparenright}\ {\isasymand}\ {\isacharparenleft}fin{\isacharunderscore}prefix\ xs\ s{\isacharparenright}\ {\isacharparenright}{\isachardoublequoteclose}\isanewline
\isanewline
\isamarkupcmt{prefix of a (general) stream%
}
\isanewline
\isacommand{definition}\isamarkupfalse%
\isanewline
\ \ \ stream{\isacharunderscore}prefix\ {\isacharcolon}{\isacharcolon}\ {\isachardoublequoteopen}{\isacharprime}a\ stream\ {\isasymRightarrow}\ {\isacharprime}a\ stream\ {\isasymRightarrow}\ bool{\isachardoublequoteclose}\isanewline
\isakeyword{where}\isanewline
\ \ {\isachardoublequoteopen}stream{\isacharunderscore}prefix\ p\ s\ {\isasymequiv}\isanewline
\ \ \ {\isacharparenleft}case\ p\ of\ \isanewline
\ \ \ \ \ \ \ \ {\isacharparenleft}FinT\ x{\isacharparenright}\ {\isasymRightarrow}\ \isanewline
\ \ \ \ \ \ \ \ {\isacharparenleft}case\ s\ of\ {\isacharparenleft}FinT\ y{\isacharparenright}\ {\isasymRightarrow}\ \ {\isacharparenleft}fin{\isacharunderscore}prefix\ x\ y{\isacharparenright}\isanewline
\ \ \ \ \ \ \ \ \ \ \ \ \ \ \ \ \ {\isacharbar}\ {\isacharparenleft}FinU\ y{\isacharparenright}\ {\isasymRightarrow}\ False\isanewline
\ \ \ \ \ \ \ \ \ \ \ \ \ \ \ \ \ {\isacharbar}\ {\isacharparenleft}InfT\ y{\isacharparenright}\ {\isasymRightarrow}\ inf{\isacharunderscore}prefix\ x\ y\ {\isadigit{0}}\isanewline
\ \ \ \ \ \ \ \ \ \ \ \ \ \ \ \ \ {\isacharbar}\ {\isacharparenleft}InfU\ y{\isacharparenright}\ {\isasymRightarrow}\ False\ {\isacharparenright}\isanewline
\ \ \ \ \ \ {\isacharbar}\ {\isacharparenleft}FinU\ x{\isacharparenright}\ {\isasymRightarrow}\ \isanewline
\ \ \ \ \ \ \ \ {\isacharparenleft}case\ s\ of\ {\isacharparenleft}FinT\ y{\isacharparenright}\ {\isasymRightarrow}\ False\isanewline
\ \ \ \ \ \ \ \ \ \ \ \ \ \ \ \ \ {\isacharbar}\ {\isacharparenleft}FinU\ y{\isacharparenright}\ {\isasymRightarrow}\ \ {\isacharparenleft}fin{\isacharunderscore}prefix\ x\ y{\isacharparenright}\isanewline
\ \ \ \ \ \ \ \ \ \ \ \ \ \ \ \ \ {\isacharbar}\ {\isacharparenleft}InfT\ y{\isacharparenright}\ {\isasymRightarrow}\ False\isanewline
\ \ \ \ \ \ \ \ \ \ \ \ \ \ \ \ \ {\isacharbar}\ {\isacharparenleft}InfU\ y{\isacharparenright}\ {\isasymRightarrow}\ \ inf{\isacharunderscore}prefix\ x\ y\ {\isadigit{0}}\ {\isacharparenright}\ \ \ \ \isanewline
\ \ \ \ \ \ {\isacharbar}\ {\isacharparenleft}InfT\ x{\isacharparenright}\ {\isasymRightarrow}\isanewline
\ \ \ \ \ \ \ \ {\isacharparenleft}case\ s\ of\ {\isacharparenleft}FinT\ y{\isacharparenright}\ {\isasymRightarrow}\ False\isanewline
\ \ \ \ \ \ \ \ \ \ \ \ \ \ \ \ \ {\isacharbar}\ {\isacharparenleft}FinU\ y{\isacharparenright}\ {\isasymRightarrow}\ False\isanewline
\ \ \ \ \ \ \ \ \ \ \ \ \ \ \ \ \ {\isacharbar}\ {\isacharparenleft}InfT\ y{\isacharparenright}\ {\isasymRightarrow}\ {\isacharparenleft}{\isasymforall}\ i{\isachardot}\ x\ i\ {\isacharequal}\ y\ i{\isacharparenright}\isanewline
\ \ \ \ \ \ \ \ \ \ \ \ \ \ \ \ \ {\isacharbar}\ {\isacharparenleft}InfU\ y{\isacharparenright}\ {\isasymRightarrow}\ False\ {\isacharparenright}\ \isanewline
\ \ \ \ \ \ {\isacharbar}\ {\isacharparenleft}InfU\ x{\isacharparenright}\ {\isasymRightarrow}\isanewline
\ \ \ \ \ \ \ \ {\isacharparenleft}case\ s\ of\ {\isacharparenleft}FinT\ y{\isacharparenright}\ {\isasymRightarrow}\ False\isanewline
\ \ \ \ \ \ \ \ \ \ \ \ \ \ \ \ \ {\isacharbar}\ {\isacharparenleft}FinU\ y{\isacharparenright}\ {\isasymRightarrow}\ False\isanewline
\ \ \ \ \ \ \ \ \ \ \ \ \ \ \ \ \ {\isacharbar}\ {\isacharparenleft}InfT\ y{\isacharparenright}\ {\isasymRightarrow}\ False\isanewline
\ \ \ \ \ \ \ \ \ \ \ \ \ \ \ \ \ {\isacharbar}\ {\isacharparenleft}InfU\ y{\isacharparenright}\ {\isasymRightarrow}\ {\isacharparenleft}{\isasymforall}\ i{\isachardot}\ x\ i\ {\isacharequal}\ y\ i{\isacharparenright}\ {\isacharparenright}\ \ {\isacharparenright}{\isachardoublequoteclose}\isanewline
\isanewline
\isamarkupcmt{truncating a finite stream after the n-th element%
}
\isanewline
\isacommand{primrec}\isamarkupfalse%
\ \ \isanewline
fin{\isacharunderscore}truncate\ {\isacharcolon}{\isacharcolon}\ {\isachardoublequoteopen}{\isacharprime}a\ list\ {\isasymRightarrow}\ nat\ {\isasymRightarrow}\ {\isacharprime}a\ list{\isachardoublequoteclose}\isanewline
\isakeyword{where}\ \isanewline
\ \ {\isachardoublequoteopen}fin{\isacharunderscore}truncate\ {\isacharbrackleft}{\isacharbrackright}\ n\ {\isacharequal}\ {\isacharbrackleft}{\isacharbrackright}{\isachardoublequoteclose}\ {\isacharbar}\isanewline
\ \ {\isachardoublequoteopen}fin{\isacharunderscore}truncate\ {\isacharparenleft}x{\isacharhash}xs{\isacharparenright}\ i\ {\isacharequal}\ \isanewline
\ \ \ \ \ \ {\isacharparenleft}case\ i\ of\ {\isadigit{0}}\ {\isasymRightarrow}\ {\isacharbrackleft}{\isacharbrackright}\ \isanewline
\ \ \ \ \ \ \ \ \ {\isacharbar}\ {\isacharparenleft}Suc\ n{\isacharparenright}\ {\isasymRightarrow}\ x\ {\isacharhash}\ {\isacharparenleft}fin{\isacharunderscore}truncate\ xs\ n{\isacharparenright}{\isacharparenright}{\isachardoublequoteclose}\isanewline
\isanewline
\isamarkupcmt{truncating a finite stream after the n-th element%
}
\isanewline
\isamarkupcmt{n is of type of natural numbers enriched by Infinity%
}
\isanewline
\isacommand{definition}\isamarkupfalse%
\isanewline
\ \ fin{\isacharunderscore}truncate{\isacharunderscore}plus\ {\isacharcolon}{\isacharcolon}\ {\isachardoublequoteopen}{\isacharprime}a\ list\ {\isasymRightarrow}\ natInf\ {\isasymRightarrow}\ {\isacharprime}a\ list{\isachardoublequoteclose}\isanewline
\ \isakeyword{where}\isanewline
\ {\isachardoublequoteopen}fin{\isacharunderscore}truncate{\isacharunderscore}plus\ s\ n\ \isanewline
\ \ {\isasymequiv}\ \isanewline
\ \ case\ n\ of\ {\isacharparenleft}Fin\ i{\isacharparenright}\ {\isasymRightarrow}\ fin{\isacharunderscore}truncate\ s\ i\ \ \ \ {\isacharbar}\ \ {\isasyminfinity}\ \ \ \ \ {\isasymRightarrow}\ s\ {\isachardoublequoteclose}\isanewline
\isanewline
\isamarkupcmt{truncating an infinite stream after the n-th element%
}
\isanewline
\isacommand{primrec}\isamarkupfalse%
\isanewline
\ \ inf{\isacharunderscore}truncate\ {\isacharcolon}{\isacharcolon}\ {\isachardoublequoteopen}{\isacharparenleft}nat\ {\isasymRightarrow}\ {\isacharprime}a{\isacharparenright}\ {\isasymRightarrow}\ nat\ {\isasymRightarrow}\ {\isacharprime}a\ list{\isachardoublequoteclose}\isanewline
\isakeyword{where}\ \ \ \isanewline
\ \ \ {\isachardoublequoteopen}inf{\isacharunderscore}truncate\ s\ {\isadigit{0}}\ {\isacharequal}\ {\isacharbrackleft}\ s\ {\isadigit{0}}\ {\isacharbrackright}{\isachardoublequoteclose}\ {\isacharbar}\isanewline
\ \ \ {\isachardoublequoteopen}inf{\isacharunderscore}truncate\ s\ {\isacharparenleft}Suc\ k{\isacharparenright}\ {\isacharequal}\ {\isacharparenleft}inf{\isacharunderscore}truncate\ s\ k{\isacharparenright}\ {\isacharat}\ {\isacharbrackleft}s\ {\isacharparenleft}Suc\ k{\isacharparenright}{\isacharbrackright}{\isachardoublequoteclose}\isanewline
\isanewline
\isamarkupcmt{truncating an infinite stream after the n-th element%
}
\isanewline
\isamarkupcmt{n is of type of natural numbers enriched by Infinity%
}
\isanewline
\isacommand{definition}\isamarkupfalse%
\isanewline
\ \ inf{\isacharunderscore}truncate{\isacharunderscore}plus\ {\isacharcolon}{\isacharcolon}\ {\isachardoublequoteopen}{\isacharprime}a\ istream\ {\isasymRightarrow}\ natInf\ {\isasymRightarrow}\ {\isacharprime}a\ stream{\isachardoublequoteclose}\ \isanewline
\ \isakeyword{where}\isanewline
\ {\isachardoublequoteopen}inf{\isacharunderscore}truncate{\isacharunderscore}plus\ s\ n\ \isanewline
\ \ {\isasymequiv}\ \isanewline
\ \ case\ n\ of\ {\isacharparenleft}Fin\ i{\isacharparenright}\ {\isasymRightarrow}\ FinT\ {\isacharparenleft}inf{\isacharunderscore}truncate\ s\ i{\isacharparenright}\isanewline
\ \ \ \ \ \ \ \ \ \ \ {\isacharbar}\ {\isasyminfinity}\ \ \ \ \ {\isasymRightarrow}\ InfT\ s{\isachardoublequoteclose}\isanewline
\isanewline
\isamarkupcmt{concatanation of a finite and an infinite stream%
}
\isanewline
\isacommand{definition}\isamarkupfalse%
\isanewline
\ \ \ \ fin{\isacharunderscore}inf{\isacharunderscore}append\ {\isacharcolon}{\isacharcolon}\ \isanewline
\ \ \ \ \ \ \ \ {\isachardoublequoteopen}{\isacharprime}a\ list\ {\isasymRightarrow}\ {\isacharparenleft}nat\ {\isasymRightarrow}\ {\isacharprime}a{\isacharparenright}\ {\isasymRightarrow}\ {\isacharparenleft}nat\ {\isasymRightarrow}\ {\isacharprime}a{\isacharparenright}{\isachardoublequoteclose}\isanewline
\isakeyword{where}\isanewline
{\isachardoublequoteopen}fin{\isacharunderscore}inf{\isacharunderscore}append\ us\ s\ {\isasymequiv}\ \ \isanewline
\ {\isacharparenleft}{\isasymlambda}\ i{\isachardot}\ {\isacharparenleft}\ if\ {\isacharparenleft}i\ {\isacharless}\ {\isacharparenleft}length\ us{\isacharparenright}{\isacharparenright}\isanewline
\ \ \ \ \ \ \ \ \ then\ {\isacharparenleft}nth\ us\ i{\isacharparenright}\isanewline
\ \ \ \ \ \ \ \ \ else\ s\ {\isacharparenleft}i\ {\isacharminus}\ {\isacharparenleft}length\ us{\isacharparenright}{\isacharparenright}\ {\isacharparenright}{\isacharparenright}{\isachardoublequoteclose}\ \isanewline
\ \isanewline
\isamarkupcmt{insuring that the infinite timed stream is time-synchronous%
}
\isanewline
\isacommand{definition}\isamarkupfalse%
\isanewline
\ \ \ ts\ {\isacharcolon}{\isacharcolon}\ {\isachardoublequoteopen}{\isacharprime}a\ istream\ {\isasymRightarrow}\ bool{\isachardoublequoteclose}\isanewline
\isakeyword{where}\isanewline
{\isachardoublequoteopen}ts\ s\ {\isasymequiv}\ \ {\isasymforall}\ i{\isachardot}\ {\isacharparenleft}length\ {\isacharparenleft}s\ i{\isacharparenright}\ {\isacharequal}\ {\isadigit{1}}{\isacharparenright}{\isachardoublequoteclose}\isanewline
\isanewline
\isamarkupcmt{insuring that each time interval of an infinite timed stream contains at most n data elements%
}
\isanewline
\isacommand{definition}\isamarkupfalse%
\isanewline
\ \ msg\ {\isacharcolon}{\isacharcolon}\ {\isachardoublequoteopen}nat\ {\isasymRightarrow}\ {\isacharprime}a\ istream\ {\isasymRightarrow}\ bool{\isachardoublequoteclose}\isanewline
\isakeyword{where}\isanewline
\ {\isachardoublequoteopen}msg\ n\ s\ {\isasymequiv}\ \ {\isasymforall}\ t{\isachardot}\ length\ {\isacharparenleft}s\ t{\isacharparenright}\ {\isasymle}\ n{\isachardoublequoteclose}\isanewline
\isanewline
\isamarkupcmt{insuring that each time interval of a finite timed stream contains at most n data elements%
}
\isanewline
\isacommand{primrec}\isamarkupfalse%
\isanewline
\ \ fin{\isacharunderscore}msg\ {\isacharcolon}{\isacharcolon}\ {\isachardoublequoteopen}nat\ {\isasymRightarrow}\ {\isacharprime}a\ list\ list\ {\isasymRightarrow}\ bool{\isachardoublequoteclose}\isanewline
\isakeyword{where}\ \ \ \isanewline
\ {\isachardoublequoteopen}fin{\isacharunderscore}msg\ n\ {\isacharbrackleft}{\isacharbrackright}\ {\isacharequal}\ True{\isachardoublequoteclose}\ {\isacharbar}\isanewline
\ {\isachardoublequoteopen}fin{\isacharunderscore}msg\ n\ {\isacharparenleft}x{\isacharhash}xs{\isacharparenright}\ {\isacharequal}\ {\isacharparenleft}{\isacharparenleft}{\isacharparenleft}length\ x{\isacharparenright}\ {\isasymle}\ n{\isacharparenright}\ {\isasymand}\ {\isacharparenleft}fin{\isacharunderscore}msg\ n\ xs{\isacharparenright}{\isacharparenright}{\isachardoublequoteclose}\ \isanewline
\isanewline
\isamarkupcmt{making a finite timed stream to a finite untimed stream%
}
\isanewline
\isacommand{definition}\isamarkupfalse%
\isanewline
\ \ \ fin{\isacharunderscore}make{\isacharunderscore}untimed\ {\isacharcolon}{\isacharcolon}\ {\isachardoublequoteopen}{\isacharprime}a\ fstream\ \ {\isasymRightarrow}\ {\isacharprime}a\ list{\isachardoublequoteclose}\isanewline
\isakeyword{where}\isanewline
\ \ {\isachardoublequoteopen}fin{\isacharunderscore}make{\isacharunderscore}untimed\ x\ {\isasymequiv}\ \ concat\ x{\isachardoublequoteclose}\isanewline
\isanewline
\isamarkupcmt{making an infinite timed stream to an infinite untimed stream%
}
\isanewline
\isamarkupcmt{(auxiliary function)%
}
\isanewline
\isacommand{primrec}\isamarkupfalse%
\isanewline
\ \ inf{\isacharunderscore}make{\isacharunderscore}untimed{\isadigit{1}}\ {\isacharcolon}{\isacharcolon}\ {\isachardoublequoteopen}{\isacharprime}a\ istream\ {\isasymRightarrow}\ nat\ {\isasymRightarrow}\ {\isacharprime}a\ {\isachardoublequoteclose}\isanewline
\isakeyword{where}\ \isanewline
inf{\isacharunderscore}make{\isacharunderscore}untimed{\isadigit{1}}{\isacharunderscore}{\isadigit{0}}{\isacharcolon}\isanewline
\ \ {\isachardoublequoteopen}inf{\isacharunderscore}make{\isacharunderscore}untimed{\isadigit{1}}\ s\ {\isadigit{0}}\ {\isacharequal}\ \ hd\ {\isacharparenleft}s\ {\isacharparenleft}LEAST\ i{\isachardot}{\isacharparenleft}s\ i{\isacharparenright}\ {\isasymnoteq}\ {\isacharbrackleft}{\isacharbrackright}{\isacharparenright}{\isacharparenright}{\isachardoublequoteclose}\ \ {\isacharbar}\isanewline
inf{\isacharunderscore}make{\isacharunderscore}untimed{\isadigit{1}}{\isacharunderscore}Suc{\isacharcolon}\isanewline
\ \ {\isachardoublequoteopen}inf{\isacharunderscore}make{\isacharunderscore}untimed{\isadigit{1}}\ s\ {\isacharparenleft}Suc\ k{\isacharparenright}\ {\isacharequal}\isanewline
\ \ \ \ {\isacharparenleft}\ if\ {\isacharparenleft}{\isacharparenleft}Suc\ k{\isacharparenright}\ {\isacharless}\ length\ {\isacharparenleft}s\ {\isadigit{0}}{\isacharparenright}{\isacharparenright}\isanewline
\ \ \ \ \ \ then\ nth\ {\isacharparenleft}s\ {\isadigit{0}}{\isacharparenright}\ {\isacharparenleft}Suc\ k{\isacharparenright}\isanewline
\ \ \ \ \ \ else\ {\isacharparenleft}\ if\ {\isacharparenleft}s\ {\isadigit{0}}{\isacharparenright}\ {\isacharequal}\ {\isacharbrackleft}{\isacharbrackright}\isanewline
\ \ \ \ \ \ \ \ \ \ \ \ \ then\ {\isacharparenleft}inf{\isacharunderscore}make{\isacharunderscore}untimed{\isadigit{1}}\ {\isacharparenleft}inf{\isacharunderscore}tl\ {\isacharparenleft}inf{\isacharunderscore}drop\ \isanewline
\ \ \ \ \ \ \ \ \ \ \ \ \ \ \ \ \ \ \ \ \ \ \ \ \ \ {\isacharparenleft}LEAST\ i{\isachardot}\ {\isasymforall}\ j{\isachardot}\ j\ {\isacharless}\ i\ {\isasymlongrightarrow}\ {\isacharparenleft}s\ j{\isacharparenright}\ {\isacharequal}\ {\isacharbrackleft}{\isacharbrackright}{\isacharparenright}\isanewline
\ \ \ \ \ \ \ \ \ \ \ \ \ \ \ \ \ \ \ \ \ \ \ \ \ \ \ s{\isacharparenright}{\isacharparenright}\ k\ {\isacharparenright}\isanewline
\ \ \ \ \ \ \ \ \ \ \ \ \ else\ inf{\isacharunderscore}make{\isacharunderscore}untimed{\isadigit{1}}\ {\isacharparenleft}inf{\isacharunderscore}tl\ s{\isacharparenright}\ k\ {\isacharparenright}{\isacharparenright}{\isachardoublequoteclose}\isanewline
\ \ \ \ \ \ \ \ \ \ \ \ \ \isanewline
\isamarkupcmt{making an infinite timed stream to an infinite untimed stream%
}
\isanewline
\isamarkupcmt{(main function)%
}
\isanewline
\isacommand{definition}\isamarkupfalse%
\isanewline
\ \ \ inf{\isacharunderscore}make{\isacharunderscore}untimed\ {\isacharcolon}{\isacharcolon}\ {\isachardoublequoteopen}{\isacharprime}a\ istream\ {\isasymRightarrow}\ {\isacharparenleft}nat\ {\isasymRightarrow}\ {\isacharprime}a{\isacharparenright}\ {\isachardoublequoteclose}\isanewline
\isakeyword{where}\isanewline
\ \ {\isachardoublequoteopen}inf{\isacharunderscore}make{\isacharunderscore}untimed\ s\isanewline
\ \ \ {\isasymequiv}\ \isanewline
\ \ \ {\isasymlambda}\ i{\isachardot}\ inf{\isacharunderscore}make{\isacharunderscore}untimed{\isadigit{1}}\ s\ i{\isachardoublequoteclose}\isanewline
\isanewline
\isanewline
\isamarkupcmt{making a (general) stream untimed%
}
\isanewline
\isacommand{definition}\isamarkupfalse%
\isanewline
\ \ \ \ make{\isacharunderscore}untimed\ {\isacharcolon}{\isacharcolon}\ {\isachardoublequoteopen}{\isacharprime}a\ stream\ \ {\isasymRightarrow}\ {\isacharprime}a\ stream{\isachardoublequoteclose}\isanewline
\isakeyword{where}\isanewline
\ \ \ {\isachardoublequoteopen}make{\isacharunderscore}untimed\ s\ {\isasymequiv}\ \isanewline
\ \ \ \ \ \ case\ s\ of\ {\isacharparenleft}FinT\ x{\isacharparenright}\ {\isasymRightarrow}\ FinU\ {\isacharparenleft}fin{\isacharunderscore}make{\isacharunderscore}untimed\ x{\isacharparenright}\isanewline
\ \ \ \ \ \ \ \ \ \ \ \ \ \ {\isacharbar}\ {\isacharparenleft}FinU\ x{\isacharparenright}\ {\isasymRightarrow}\ FinU\ x\isanewline
\ \ \ \ \ \ \ \ \ \ \ \ \ \ {\isacharbar}\ {\isacharparenleft}InfT\ x{\isacharparenright}\ {\isasymRightarrow}\ \isanewline
\ \ \ \ \ \ \ \ \ \ \ \ \ \ \ \ {\isacharparenleft}if\ {\isacharparenleft}{\isasymexists}\ i{\isachardot}{\isasymforall}\ j{\isachardot}\ i\ {\isacharless}\ j\ {\isasymlongrightarrow}\ {\isacharparenleft}x\ j{\isacharparenright}\ {\isacharequal}\ {\isacharbrackleft}{\isacharbrackright}{\isacharparenright}\isanewline
\ \ \ \ \ \ \ \ \ \ \ \ \ \ \ \ \ then\ FinU\ {\isacharparenleft}fin{\isacharunderscore}make{\isacharunderscore}untimed\ {\isacharparenleft}inf{\isacharunderscore}truncate\ x\ \isanewline
\ \ \ \ \ \ \ \ \ \ \ \ \ \ \ \ \ \ \ \ \ \ \ \ \ \ \ \ \ {\isacharparenleft}LEAST\ i{\isachardot}{\isasymforall}\ j{\isachardot}\ i\ {\isacharless}\ j\ {\isasymlongrightarrow}\ {\isacharparenleft}x\ j{\isacharparenright}\ {\isacharequal}\ {\isacharbrackleft}{\isacharbrackright}{\isacharparenright}{\isacharparenright}{\isacharparenright}\isanewline
\ \ \ \ \ \ \ \ \ \ \ \ \ \ \ \ \ else\ InfU\ {\isacharparenleft}inf{\isacharunderscore}make{\isacharunderscore}untimed\ x{\isacharparenright}{\isacharparenright}\isanewline
\ \ \ \ \ \ \ \ \ \ \ \ \ \ {\isacharbar}\ {\isacharparenleft}InfU\ x{\isacharparenright}\ {\isasymRightarrow}\ InfU\ x{\isachardoublequoteclose}\isanewline
\isanewline
\isanewline
\isamarkupcmt{finding the index of the time interval that contains the k-th data element%
}
\isanewline
\isamarkupcmt{defined over a finite timed stream%
}
\isanewline
\isacommand{primrec}\isamarkupfalse%
\isanewline
\ \ fin{\isacharunderscore}tm\ {\isacharcolon}{\isacharcolon}\ {\isachardoublequoteopen}{\isacharprime}a\ fstream\ {\isasymRightarrow}\ nat\ {\isasymRightarrow}\ nat{\isachardoublequoteclose}\isanewline
\isakeyword{where}\isanewline
\ \ {\isachardoublequoteopen}fin{\isacharunderscore}tm\ {\isacharbrackleft}{\isacharbrackright}\ k\ {\isacharequal}\ k{\isachardoublequoteclose}\ {\isacharbar}\isanewline
\ \ {\isachardoublequoteopen}fin{\isacharunderscore}tm\ {\isacharparenleft}x{\isacharhash}xs{\isacharparenright}\ k\ {\isacharequal}\ \isanewline
\ \ \ \ {\isacharparenleft}if\ k\ {\isacharequal}\ {\isadigit{0}}\isanewline
\ \ \ \ \ then\ {\isadigit{0}}\isanewline
\ \ \ \ \ else\ {\isacharparenleft}if\ {\isacharparenleft}k\ {\isasymle}\ length\ x{\isacharparenright}\isanewline
\ \ \ \ \ \ \ \ \ \ \ then\ {\isacharparenleft}Suc\ {\isadigit{0}}{\isacharparenright}\isanewline
\ \ \ \ \ \ \ \ \ \ \ else\ Suc{\isacharparenleft}fin{\isacharunderscore}tm\ xs\ {\isacharparenleft}k\ {\isacharminus}\ length\ x{\isacharparenright}{\isacharparenright}{\isacharparenright}{\isacharparenright}{\isachardoublequoteclose}\isanewline
\isanewline
\isamarkupcmt{auxiliary lemma for the definition of the truncate operator%
}
\isanewline
\isacommand{lemma}\isamarkupfalse%
\ inf{\isacharunderscore}tm{\isacharunderscore}hint{\isadigit{1}}{\isacharcolon}\isanewline
\ \ \isakeyword{assumes}\ {\isachardoublequoteopen}i{\isadigit{2}}\ {\isacharequal}\ Suc\ i\ {\isacharminus}\ length\ a{\isachardoublequoteclose}\isanewline
\ \ \ \ \ \ \isakeyword{and}\ {\isachardoublequoteopen}{\isasymnot}\ Suc\ i\ {\isasymle}\ length\ a{\isachardoublequoteclose}\ \isanewline
\ \ \ \ \ \ \isakeyword{and}\ {\isachardoublequoteopen}a\ {\isasymnoteq}\ {\isacharbrackleft}{\isacharbrackright}{\isachardoublequoteclose}\ \isanewline
\ \ \isakeyword{shows}\ {\isachardoublequoteopen}i{\isadigit{2}}\ {\isacharless}\ Suc\ i{\isachardoublequoteclose}\isanewline
\isadelimproof
\endisadelimproof
\isatagproof
\isacommand{using}\isamarkupfalse%
\ assms\isanewline
\isacommand{by}\isamarkupfalse%
\ auto\isanewline
\isanewline
\isanewline
\isamarkupcmt{filtering a finite timed stream%
}
\endisatagproof
{\isafoldproof}%
\isadelimproof
\isanewline
\endisadelimproof
\isacommand{definition}\isamarkupfalse%
\isanewline
\ \ \ finT{\isacharunderscore}filter\ {\isacharcolon}{\isacharcolon}\ {\isachardoublequoteopen}{\isacharprime}a\ set\ {\isacharequal}{\isachargreater}\ {\isacharprime}a\ fstream\ {\isacharequal}{\isachargreater}\ {\isacharprime}a\ fstream{\isachardoublequoteclose}\ \ \ \isanewline
\isakeyword{where}\ \isanewline
\ \ {\isachardoublequoteopen}finT{\isacharunderscore}filter\ m\ s\ {\isasymequiv}\ \ map\ {\isacharparenleft}{\isasymlambda}\ s{\isachardot}\ filter\ {\isacharparenleft}{\isasymlambda}\ y{\isachardot}\ y\ {\isasymin}\ m{\isacharparenright}\ s{\isacharparenright}\ s{\isachardoublequoteclose}\isanewline
\isanewline
\isamarkupcmt{filtering an infinite timed stream%
}
\isanewline
\isacommand{definition}\isamarkupfalse%
\isanewline
\ \ \ infT{\isacharunderscore}filter\ {\isacharcolon}{\isacharcolon}\ {\isachardoublequoteopen}{\isacharprime}a\ set\ {\isacharequal}{\isachargreater}\ {\isacharprime}a\ istream\ {\isacharequal}{\isachargreater}\ {\isacharprime}a\ istream{\isachardoublequoteclose}\ \ \isanewline
\isakeyword{where}\isanewline
\ \ {\isachardoublequoteopen}infT{\isacharunderscore}filter\ m\ s\ {\isasymequiv}\ \ {\isacharparenleft}{\isasymlambda}i{\isachardot}{\isacharparenleft}\ filter\ {\isacharparenleft}{\isasymlambda}\ x{\isachardot}\ x\ {\isasymin}\ m{\isacharparenright}\ {\isacharparenleft}s\ i{\isacharparenright}{\isacharparenright}{\isacharparenright}{\isachardoublequoteclose}\isanewline
\isanewline
\isamarkupcmt{removing duplications from a finite timed stream%
}
\isanewline
\isacommand{definition}\isamarkupfalse%
\isanewline
\ \ \ finT{\isacharunderscore}remdups\ {\isacharcolon}{\isacharcolon}\ {\isachardoublequoteopen}{\isacharprime}a\ fstream\ {\isacharequal}{\isachargreater}\ {\isacharprime}a\ fstream{\isachardoublequoteclose}\isanewline
\isakeyword{where}\ \ \isanewline
\ \ {\isachardoublequoteopen}finT{\isacharunderscore}remdups\ s\ {\isasymequiv}\ \ map\ {\isacharparenleft}{\isasymlambda}\ s{\isachardot}\ remdups\ s{\isacharparenright}\ s{\isachardoublequoteclose}\isanewline
\isanewline
\isamarkupcmt{removing duplications from an infinite timed stream%
}
\isanewline
\isacommand{definition}\isamarkupfalse%
\isanewline
\ \ \ infT{\isacharunderscore}remdups\ {\isacharcolon}{\isacharcolon}\ {\isachardoublequoteopen}{\isacharprime}a\ istream\ {\isacharequal}{\isachargreater}\ {\isacharprime}a\ istream{\isachardoublequoteclose}\ \ \isanewline
\isakeyword{where}\isanewline
\ \ {\isachardoublequoteopen}infT{\isacharunderscore}remdups\ s\ {\isasymequiv}\ \ {\isacharparenleft}{\isasymlambda}i{\isachardot}{\isacharparenleft}\ remdups\ {\isacharparenleft}s\ i{\isacharparenright}{\isacharparenright}{\isacharparenright}{\isachardoublequoteclose}\isanewline
\isanewline
\isamarkupcmt{removing duplications from a time interval of a stream%
}
\isanewline
\isacommand{primrec}\isamarkupfalse%
\isanewline
\ \ \ fst{\isacharunderscore}remdups\ {\isacharcolon}{\isacharcolon}\ {\isachardoublequoteopen}{\isacharprime}a\ list\ {\isasymRightarrow}\ {\isacharprime}a\ list{\isachardoublequoteclose}\isanewline
\isakeyword{where}\isanewline
\ {\isachardoublequoteopen}fst{\isacharunderscore}remdups\ {\isacharbrackleft}{\isacharbrackright}\ {\isacharequal}\ {\isacharbrackleft}{\isacharbrackright}{\isachardoublequoteclose}\ {\isacharbar}\isanewline
\ {\isachardoublequoteopen}fst{\isacharunderscore}remdups\ {\isacharparenleft}x{\isacharhash}xs{\isacharparenright}\ {\isacharequal}\ \isanewline
\ \ \ \ {\isacharparenleft}if\ xs\ {\isacharequal}\ {\isacharbrackleft}{\isacharbrackright}\ \isanewline
\ \ \ \ \ then\ {\isacharbrackleft}x{\isacharbrackright}\isanewline
\ \ \ \ \ else\ {\isacharparenleft}if\ x\ {\isacharequal}\ {\isacharparenleft}hd\ xs{\isacharparenright}\isanewline
\ \ \ \ \ \ \ \ \ \ \ then\ fst{\isacharunderscore}remdups\ xs\isanewline
\ \ \ \ \ \ \ \ \ \ \ else\ {\isacharparenleft}x{\isacharhash}xs{\isacharparenright}{\isacharparenright}{\isacharparenright}{\isachardoublequoteclose}\isanewline
\isanewline
\ \isanewline
\isamarkupcmt{time interval operator%
}
\isanewline
\isacommand{definition}\isamarkupfalse%
\isanewline
\ \ ti\ {\isacharcolon}{\isacharcolon}\ {\isachardoublequoteopen}{\isacharprime}a\ fstream\ {\isasymRightarrow}\ nat\ {\isasymRightarrow}\ {\isacharprime}a\ list{\isachardoublequoteclose}\isanewline
\isakeyword{where}\isanewline
\ {\isachardoublequoteopen}ti\ s\ i\ {\isasymequiv}\ \ \isanewline
\ \ \ \ {\isacharparenleft}if\ s\ {\isacharequal}\ {\isacharbrackleft}{\isacharbrackright}
 \ then\ {\isacharbrackleft}{\isacharbrackright}
 \ else\ {\isacharparenleft}nth\ s\ i{\isacharparenright}{\isacharparenright}{\isachardoublequoteclose}\isanewline
\isanewline
\isamarkupcmt{insuring that a sheaf of channels is correctly defined%
}
\isanewline
\isacommand{definition}\isamarkupfalse%
\isanewline
\ \ \ CorrectSheaf\ {\isacharcolon}{\isacharcolon}\ {\isachardoublequoteopen}nat\ {\isasymRightarrow}\ bool{\isachardoublequoteclose}\isanewline
\isakeyword{where}\isanewline
\ \ {\isachardoublequoteopen}CorrectSheaf\ n\ {\isasymequiv}\ {\isadigit{0}}\ {\isacharless}\ n{\isachardoublequoteclose}\isanewline
\ \isanewline
\isamarkupcmt{insuring that all channels in a sheaf are disjunct%
}
\isanewline
\isamarkupcmt{indices in the sheaf are represented using an extra specified set%
}
\isanewline
\isacommand{definition}\isamarkupfalse%
\isanewline
\ \ \ inf{\isacharunderscore}disjS\ {\isacharcolon}{\isacharcolon}\ {\isachardoublequoteopen}{\isacharprime}b\ set\ {\isasymRightarrow}\ {\isacharparenleft}{\isacharprime}b\ {\isasymRightarrow}\ {\isacharprime}a\ istream{\isacharparenright}\ {\isasymRightarrow}\ bool{\isachardoublequoteclose}\isanewline
\isakeyword{where}\isanewline
\ \ {\isachardoublequoteopen}inf{\isacharunderscore}disjS\ IdSet\ nS\isanewline
\ \ \ {\isasymequiv}\isanewline
\ \ \ {\isasymforall}\ {\isacharparenleft}t{\isacharcolon}{\isacharcolon}nat{\isacharparenright}\ i\ j{\isachardot}\ {\isacharparenleft}i{\isacharcolon}IdSet{\isacharparenright}\ {\isasymand}\ {\isacharparenleft}j{\isacharcolon}IdSet{\isacharparenright}\ {\isasymand}\ \ \isanewline
\ \ \ {\isacharparenleft}{\isacharparenleft}nS\ i{\isacharparenright}\ t{\isacharparenright}\ {\isasymnoteq}\ {\isacharbrackleft}{\isacharbrackright}\ \ {\isasymlongrightarrow}\ {\isacharparenleft}{\isacharparenleft}nS\ j{\isacharparenright}\ t{\isacharparenright}\ {\isacharequal}\ {\isacharbrackleft}{\isacharbrackright}{\isachardoublequoteclose}\ \ \isanewline
\isanewline
\ \ \ \ \isanewline
\isamarkupcmt{insuring that all channels in a sheaf are disjunct%
}
\isanewline
\isamarkupcmt{indices in the sheaf are represented using natural numbers%
}
\isanewline
\isacommand{definition}\isamarkupfalse%
\ \ \ \ \ \ \ \ \ \ \ \ \ \ \isanewline
\ \ \ inf{\isacharunderscore}disj\ {\isacharcolon}{\isacharcolon}\ {\isachardoublequoteopen}nat\ {\isasymRightarrow}\ {\isacharparenleft}nat\ {\isasymRightarrow}\ {\isacharprime}a\ istream{\isacharparenright}\ {\isasymRightarrow}\ bool{\isachardoublequoteclose}\isanewline
\isakeyword{where}\isanewline
\ \ {\isachardoublequoteopen}inf{\isacharunderscore}disj\ n\ nS\isanewline
\ \ \ {\isasymequiv}\isanewline
\ \ \ {\isasymforall}\ {\isacharparenleft}t{\isacharcolon}{\isacharcolon}nat{\isacharparenright}\ {\isacharparenleft}i{\isacharcolon}{\isacharcolon}nat{\isacharparenright}\ {\isacharparenleft}j{\isacharcolon}{\isacharcolon}nat{\isacharparenright}{\isachardot}\ \isanewline
\ \ \ i\ {\isacharless}\ n\ \ {\isasymand}\ \ j\ {\isacharless}\ n\ {\isasymand}\ i\ {\isasymnoteq}\ j\ {\isasymand}\ {\isacharparenleft}{\isacharparenleft}nS\ i{\isacharparenright}\ t{\isacharparenright}\ {\isasymnoteq}\ {\isacharbrackleft}{\isacharbrackright}\ \ {\isasymlongrightarrow}\ \isanewline
\ \ \ {\isacharparenleft}{\isacharparenleft}nS\ j{\isacharparenright}\ t{\isacharparenright}\ {\isacharequal}\ {\isacharbrackleft}{\isacharbrackright}{\isachardoublequoteclose}\ \ \isanewline
\isanewline
\isamarkupcmt{taking the prefix of n data elements from a finite timed stream%
}
\isanewline
\isamarkupcmt{(defined over natural numbers)%
}
\isanewline
\isacommand{fun}\isamarkupfalse%
\ fin{\isacharunderscore}get{\isacharunderscore}prefix\ \ {\isacharcolon}{\isacharcolon}\ {\isachardoublequoteopen}{\isacharparenleft}{\isacharprime}a\ fstream\ {\isasymtimes}\ nat{\isacharparenright}\ {\isasymRightarrow}\ {\isacharprime}a\ fstream{\isachardoublequoteclose}\isanewline
\isakeyword{where}\isanewline
\ \ {\isachardoublequoteopen}fin{\isacharunderscore}get{\isacharunderscore}prefix{\isacharparenleft}{\isacharbrackleft}{\isacharbrackright}{\isacharcomma}\ n{\isacharparenright}\ {\isacharequal}\ {\isacharbrackleft}{\isacharbrackright}{\isachardoublequoteclose}\ {\isacharbar}\isanewline
\ \ {\isachardoublequoteopen}fin{\isacharunderscore}get{\isacharunderscore}prefix{\isacharparenleft}x{\isacharhash}xs{\isacharcomma}\ i{\isacharparenright}\ {\isacharequal}\ \isanewline
\ \ \ \ \ {\isacharparenleft}\ if\ {\isacharparenleft}length\ x{\isacharparenright}\ {\isacharless}\ i\ \isanewline
\ \ \ \ \ \ \ then\ x\ {\isacharhash}\ fin{\isacharunderscore}get{\isacharunderscore}prefix{\isacharparenleft}xs{\isacharcomma}\ {\isacharparenleft}i\ {\isacharminus}\ {\isacharparenleft}length\ x{\isacharparenright}{\isacharparenright}{\isacharparenright}\isanewline
\ \ \ \ \ \ \ else\ {\isacharbrackleft}take\ i\ x{\isacharbrackright}\ {\isacharparenright}\ {\isachardoublequoteclose}\isanewline
\isanewline
\isamarkupcmt{taking the prefix of n data elements from a finite timed stream%
}
\isanewline
\isamarkupcmt{(defined over natural numbers enriched by Infinity)%
}
\isanewline
\isacommand{definition}\isamarkupfalse%
\isanewline
\ \ fin{\isacharunderscore}get{\isacharunderscore}prefix{\isacharunderscore}plus\ {\isacharcolon}{\isacharcolon}\ {\isachardoublequoteopen}{\isacharprime}a\ fstream\ {\isasymRightarrow}\ natInf\ {\isasymRightarrow}\ {\isacharprime}a\ fstream{\isachardoublequoteclose}\isanewline
\isakeyword{where}\isanewline
\ {\isachardoublequoteopen}fin{\isacharunderscore}get{\isacharunderscore}prefix{\isacharunderscore}plus\ s\ n\ \isanewline
\ \ {\isasymequiv}\ \isanewline
\ \ case\ n\ of\ {\isacharparenleft}Fin\ i{\isacharparenright}\ {\isasymRightarrow}\ fin{\isacharunderscore}get{\isacharunderscore}prefix{\isacharparenleft}s{\isacharcomma}\ i{\isacharparenright}\isanewline
\ \ \ \ \ \ \ \ \ \ \ {\isacharbar}\ {\isasyminfinity}\ \ \ \ \ \ {\isasymRightarrow}\ s\ {\isachardoublequoteclose}\isanewline
\isanewline
\isamarkupcmt{auxiliary lemmas%
}
\isanewline
\isacommand{lemma}\isamarkupfalse%
\ length{\isacharunderscore}inf{\isacharunderscore}drop{\isacharunderscore}hint{\isadigit{1}}{\isacharcolon}\ \isanewline
\ \ \isakeyword{assumes}\ {\isachardoublequoteopen}s\ k\ {\isasymnoteq}\ {\isacharbrackleft}{\isacharbrackright}{\isachardoublequoteclose}\isanewline
\ \ \isakeyword{shows}\ {\isachardoublequoteopen}length\ {\isacharparenleft}inf{\isacharunderscore}drop\ k\ s\ {\isadigit{0}}{\isacharparenright}\ {\isasymnoteq}\ {\isadigit{0}}{\isachardoublequoteclose}\isanewline
\isadelimproof
\endisadelimproof
\isatagproof
\isacommand{using}\isamarkupfalse%
\ assms\isanewline
\isacommand{by}\isamarkupfalse%
\ {\isacharparenleft}auto\ simp{\isacharcolon}\ inf{\isacharunderscore}drop{\isacharunderscore}def{\isacharparenright}%
\endisatagproof
{\isafoldproof}%
\isadelimproof
\isanewline
\endisadelimproof
\isanewline
\isanewline
\isacommand{lemma}\isamarkupfalse%
\ length{\isacharunderscore}inf{\isacharunderscore}drop{\isacharunderscore}hint{\isadigit{2}}{\isacharcolon}\isanewline
{\isachardoublequoteopen}{\isacharparenleft}s\ {\isadigit{0}}\ {\isasymnoteq}\ {\isacharbrackleft}{\isacharbrackright}\ {\isasymlongrightarrow}\ length\ {\isacharparenleft}inf{\isacharunderscore}drop\ {\isadigit{0}}\ s\ {\isadigit{0}}{\isacharparenright}\ {\isacharless}\ Suc\ i\ \isanewline
\ \ {\isasymlongrightarrow}\ Suc\ i\ {\isacharminus}\ length\ {\isacharparenleft}inf{\isacharunderscore}drop\ {\isadigit{0}}\ s\ {\isadigit{0}}{\isacharparenright}\ {\isacharless}\ Suc\ i{\isacharparenright}{\isachardoublequoteclose}\isanewline
\isadelimproof
\ \ %
\endisadelimproof
\isatagproof
\isacommand{by}\isamarkupfalse%
\ {\isacharparenleft}simp\ add{\isacharcolon}\ inf{\isacharunderscore}drop{\isacharunderscore}def\ list{\isacharunderscore}length{\isacharunderscore}hint{\isadigit{1}}{\isacharparenright}\isanewline
\isanewline
\isanewline
\isamarkupcmt{taking the prefix of n data elements from an infinite timed stream%
}
\isanewline
\isamarkupcmt{(defined over natural numbers)%
}
\endisatagproof
{\isafoldproof}%
\isadelimproof
\isanewline
\endisadelimproof
\isacommand{fun}\isamarkupfalse%
\ infT{\isacharunderscore}get{\isacharunderscore}prefix\ \ {\isacharcolon}{\isacharcolon}\ {\isachardoublequoteopen}{\isacharparenleft}{\isacharprime}a\ istream\ {\isasymtimes}\ nat{\isacharparenright}\ {\isasymRightarrow}\ {\isacharprime}a\ fstream{\isachardoublequoteclose}\isanewline
\isakeyword{where}\ \ \ \isanewline
\ \ {\isachardoublequoteopen}infT{\isacharunderscore}get{\isacharunderscore}prefix{\isacharparenleft}s{\isacharcomma}\ {\isadigit{0}}{\isacharparenright}\ {\isacharequal}\ {\isacharbrackleft}{\isacharbrackright}{\isachardoublequoteclose}\isanewline
{\isacharbar}\isanewline
\ \ {\isachardoublequoteopen}infT{\isacharunderscore}get{\isacharunderscore}prefix{\isacharparenleft}s{\isacharcomma}\ Suc\ i{\isacharparenright}\ {\isacharequal}\ \isanewline
\ \ \ \ {\isacharparenleft}\ if\ {\isacharparenleft}s\ {\isadigit{0}}{\isacharparenright}\ {\isacharequal}\ {\isacharbrackleft}{\isacharbrackright}\isanewline
\ \ \ \ \ \ then\ {\isacharparenleft}\ if\ {\isacharparenleft}{\isasymforall}\ i{\isachardot}\ s\ i\ {\isacharequal}\ {\isacharbrackleft}{\isacharbrackright}{\isacharparenright}\isanewline
\ \ \ \ \ \ \ \ \ \ \ \ \ then\ {\isacharbrackleft}{\isacharbrackright}\isanewline
\ \ \ \ \ \ \ \ \ \ \ \ \ else\ {\isacharparenleft}let\ \isanewline
\ \ \ \ \ \ \ \ \ \ \ \ \ \ \ \ \ \ \ \ \ k\ {\isacharequal}\ {\isacharparenleft}LEAST\ k{\isachardot}\ s\ k\ {\isasymnoteq}\ {\isacharbrackleft}{\isacharbrackright}\ {\isasymand}\ {\isacharparenleft}{\isasymforall}i{\isachardot}\ i\ {\isacharless}\ k\ {\isasymlongrightarrow}\ s\ i\ {\isacharequal}\ {\isacharbrackleft}{\isacharbrackright}{\isacharparenright}{\isacharparenright}{\isacharsemicolon}\isanewline
\ \ \ \ \ \ \ \ \ \ \ \ \ \ \ \ \ \ \ \ \ s{\isadigit{2}}\ {\isacharequal}\ inf{\isacharunderscore}drop\ {\isacharparenleft}k{\isacharplus}{\isadigit{1}}{\isacharparenright}\ s\isanewline
\ \ \ \ \ \ \ \ \ \ \ \ \ \ \ \ \ \ \ in\ {\isacharparenleft}if\ {\isacharparenleft}length\ {\isacharparenleft}s\ k{\isacharparenright}{\isacharequal}{\isadigit{0}}{\isacharparenright}\ \isanewline
\ \ \ \ \ \ \ \ \ \ \ \ \ \ \ \ \ \ \ \ \ \ \ then\ {\isacharbrackleft}{\isacharbrackright}\ \isanewline
\ \ \ \ \ \ \ \ \ \ \ \ \ \ \ \ \ \ \ \ \ \ \ else\ {\isacharparenleft}if\ {\isacharparenleft}length\ {\isacharparenleft}s\ k{\isacharparenright}\ {\isacharless}\ {\isacharparenleft}Suc\ i{\isacharparenright}{\isacharparenright}\ \isanewline
\ \ \ \ \ \ \ \ \ \ \ \ \ \ \ \ \ \ \ \ \ \ \ \ \ \ \ \ \ then\ s\ k\ {\isacharhash}\ infT{\isacharunderscore}get{\isacharunderscore}prefix\ {\isacharparenleft}s{\isadigit{2}}{\isacharcomma}Suc\ i\ {\isacharminus}\ length\ {\isacharparenleft}s\ k{\isacharparenright}{\isacharparenright}\isanewline
\ \ \ \ \ \ \ \ \ \ \ \ \ \ \ \ \ \ \ \ \ \ \ \ \ \ \ \ \ else\ {\isacharbrackleft}take\ {\isacharparenleft}Suc\ i{\isacharparenright}\ {\isacharparenleft}s\ k{\isacharparenright}{\isacharbrackright}\ {\isacharparenright}{\isacharparenright}{\isacharparenright}\isanewline
\ \ \ \ \ \ \ \ \ \ \ {\isacharparenright}\isanewline
\ \ \ \ \ \ else\ \isanewline
\ \ \ \ \ \ \ \ {\isacharparenleft}if\ {\isacharparenleft}{\isacharparenleft}length\ {\isacharparenleft}s\ {\isadigit{0}}{\isacharparenright}{\isacharparenright}\ {\isacharless}\ {\isacharparenleft}Suc\ i{\isacharparenright}{\isacharparenright}\ \isanewline
\ \ \ \ \ \ \ \ \ then\ {\isacharparenleft}s\ {\isadigit{0}}{\isacharparenright}\ {\isacharhash}\ infT{\isacharunderscore}get{\isacharunderscore}prefix{\isacharparenleft}\ inf{\isacharunderscore}drop\ {\isadigit{1}}\ s{\isacharcomma}\ {\isacharparenleft}Suc\ i{\isacharparenright}\ {\isacharminus}\ {\isacharparenleft}length\ {\isacharparenleft}s\ {\isadigit{0}}{\isacharparenright}{\isacharparenright}{\isacharparenright}\isanewline
\ \ \ \ \ \ \ \ \ else\ {\isacharbrackleft}take\ {\isacharparenleft}Suc\ i{\isacharparenright}\ {\isacharparenleft}s\ {\isadigit{0}}{\isacharparenright}{\isacharbrackright}\ \isanewline
\ \ \ \ \ \ \ \ \ {\isacharparenright}\isanewline
\ \ \ \ \ {\isacharparenright}{\isachardoublequoteclose}\isanewline
\isanewline
\isanewline
\isamarkupcmt{taking the prefix of n data elements from an infinite untimed stream%
}
\isanewline
\isamarkupcmt{(defined over natural numbers)%
}
\isanewline
\isacommand{primrec}\isamarkupfalse%
\isanewline
\ \ infU{\isacharunderscore}get{\isacharunderscore}prefix\ \ {\isacharcolon}{\isacharcolon}\ {\isachardoublequoteopen}{\isacharparenleft}nat\ {\isasymRightarrow}\ {\isacharprime}a{\isacharparenright}\ {\isasymRightarrow}\ nat\ {\isasymRightarrow}\ {\isacharprime}a\ list{\isachardoublequoteclose}\isanewline
\isakeyword{where}\isanewline
\ \ {\isachardoublequoteopen}infU{\isacharunderscore}get{\isacharunderscore}prefix\ s\ {\isadigit{0}}\ {\isacharequal}\ {\isacharbrackleft}{\isacharbrackright}{\isachardoublequoteclose}\ {\isacharbar}\isanewline
\ \ {\isachardoublequoteopen}infU{\isacharunderscore}get{\isacharunderscore}prefix\ s\ {\isacharparenleft}Suc\ i{\isacharparenright}\ \isanewline
\ \ \ \ \ \ \ \ \ {\isacharequal}\ {\isacharparenleft}infU{\isacharunderscore}get{\isacharunderscore}prefix\ s\ i{\isacharparenright}\ {\isacharat}\ {\isacharbrackleft}s\ i{\isacharbrackright}{\isachardoublequoteclose}\isanewline
\isanewline
\isamarkupcmt{taking the prefix of n data elements from an infinite timed stream%
}
\isanewline
\isamarkupcmt{(defined over natural numbers enriched by Infinity)%
}
\isanewline
\isacommand{definition}\isamarkupfalse%
\isanewline
\ \ infT{\isacharunderscore}get{\isacharunderscore}prefix{\isacharunderscore}plus\ {\isacharcolon}{\isacharcolon}\ {\isachardoublequoteopen}{\isacharprime}a\ istream\ {\isasymRightarrow}\ natInf\ {\isasymRightarrow}\ {\isacharprime}a\ stream{\isachardoublequoteclose}\isanewline
\isakeyword{where}\isanewline
{\isachardoublequoteopen}infT{\isacharunderscore}get{\isacharunderscore}prefix{\isacharunderscore}plus\ s\ n\ \isanewline
\ \ {\isasymequiv}\ \isanewline
\ \ case\ n\ of\ {\isacharparenleft}Fin\ i{\isacharparenright}\ {\isasymRightarrow}\ FinT\ {\isacharparenleft}infT{\isacharunderscore}get{\isacharunderscore}prefix{\isacharparenleft}s{\isacharcomma}\ i{\isacharparenright}{\isacharparenright}\ \isanewline
\ \ \ \ \ \ \ \ \ \ \ {\isacharbar}\ {\isasyminfinity}\ \ \ \ \ {\isasymRightarrow}\ InfT\ s{\isachardoublequoteclose}\isanewline
\isanewline
\isamarkupcmt{taking the prefix of n data elements from an infinite untimed stream%
}
\isanewline
\isamarkupcmt{(defined over natural numbers enriched by Infinity)%
}
\isanewline
\isacommand{definition}\isamarkupfalse%
\isanewline
\ \ infU{\isacharunderscore}get{\isacharunderscore}prefix{\isacharunderscore}plus\ {\isacharcolon}{\isacharcolon}\ {\isachardoublequoteopen}{\isacharparenleft}nat\ {\isasymRightarrow}\ {\isacharprime}a{\isacharparenright}\ {\isasymRightarrow}\ natInf\ {\isasymRightarrow}\ {\isacharprime}a\ stream{\isachardoublequoteclose}\isanewline
\isakeyword{where}\isanewline
\ {\isachardoublequoteopen}infU{\isacharunderscore}get{\isacharunderscore}prefix{\isacharunderscore}plus\ s\ n\ \isanewline
\ \ {\isasymequiv}\ \isanewline
\ \ case\ n\ of\ {\isacharparenleft}Fin\ i{\isacharparenright}\ {\isasymRightarrow}\ FinU\ {\isacharparenleft}infU{\isacharunderscore}get{\isacharunderscore}prefix\ s\ i{\isacharparenright}\ \isanewline
\ \ \ \ \ \ \ \ \ \ \ {\isacharbar}\ {\isasyminfinity}\ \ \ \ \ {\isasymRightarrow}\ InfU\ s{\isachardoublequoteclose}\isanewline
\isanewline
\isamarkupcmt{taking the prefix of n data elements from an infinite stream%
}
\isanewline
\isamarkupcmt{(defined over natural numbers enriched by Infinity)%
}
\isanewline
\isacommand{definition}\isamarkupfalse%
\isanewline
\ \ take{\isacharunderscore}plus\ {\isacharcolon}{\isacharcolon}\ {\isachardoublequoteopen}natInf\ {\isasymRightarrow}\ {\isacharprime}a\ list\ {\isasymRightarrow}\ {\isacharprime}a\ list{\isachardoublequoteclose}\isanewline
\isakeyword{where}\isanewline
\ {\isachardoublequoteopen}take{\isacharunderscore}plus\ n\ s\ \isanewline
\ \ {\isasymequiv}\ \isanewline
\ \ case\ n\ of\ {\isacharparenleft}Fin\ i{\isacharparenright}\ {\isasymRightarrow}\ {\isacharparenleft}take\ i\ s{\isacharparenright}\ \isanewline
\ \ \ \ \ \ \ \ \ \ \ {\isacharbar}\ {\isasyminfinity}\ \ \ \ \ \ {\isasymRightarrow}\ s{\isachardoublequoteclose}\isanewline
\isanewline
\isamarkupcmt{taking the prefix of n data elements from a (general) stream%
}
\isanewline
\isamarkupcmt{(defined over natural numbers enriched by Infinity)%
}
\isanewline
\isacommand{definition}\isamarkupfalse%
\ \isanewline
\ \ \ get{\isacharunderscore}prefix\ {\isacharcolon}{\isacharcolon}\ {\isachardoublequoteopen}{\isacharprime}a\ stream\ {\isasymRightarrow}\ natInf\ {\isasymRightarrow}\ {\isacharprime}a\ stream{\isachardoublequoteclose}\isanewline
\isakeyword{where}\isanewline
\ \ \ {\isachardoublequoteopen}get{\isacharunderscore}prefix\ s\ k\ {\isasymequiv}\ \isanewline
\ \ \ \ \ \ \ \ case\ s\ of\ \ {\isacharparenleft}FinT\ x{\isacharparenright}\ {\isasymRightarrow}\ FinT\ {\isacharparenleft}fin{\isacharunderscore}get{\isacharunderscore}prefix{\isacharunderscore}plus\ x\ k{\isacharparenright}\isanewline
\ \ \ \ \ \ \ \ \ \ \ \ \ \ \ \ \ {\isacharbar}\ {\isacharparenleft}FinU\ x{\isacharparenright}\ {\isasymRightarrow}\ FinU\ {\isacharparenleft}take{\isacharunderscore}plus\ k\ x{\isacharparenright}\isanewline
\ \ \ \ \ \ \ \ \ \ \ \ \ \ \ \ \ {\isacharbar}\ {\isacharparenleft}InfT\ x{\isacharparenright}\ {\isasymRightarrow}\ infT{\isacharunderscore}get{\isacharunderscore}prefix{\isacharunderscore}plus\ x\ k\isanewline
\ \ \ \ \ \ \ \ \ \ \ \ \ \ \ \ \ {\isacharbar}\ {\isacharparenleft}InfU\ x{\isacharparenright}\ {\isasymRightarrow}\ infU{\isacharunderscore}get{\isacharunderscore}prefix{\isacharunderscore}plus\ x\ k{\isachardoublequoteclose}\isanewline
\isanewline
\isamarkupcmt{merging time intervals of two finite timed streams%
}
\isanewline
\isacommand{primrec}\isamarkupfalse%
\isanewline
\ \ \ fin{\isacharunderscore}merge{\isacharunderscore}ti\ {\isacharcolon}{\isacharcolon}\ {\isachardoublequoteopen}{\isacharprime}a\ fstream\ {\isasymRightarrow}\ {\isacharprime}a\ fstream\ {\isasymRightarrow}\ {\isacharprime}a\ fstream{\isachardoublequoteclose}\isanewline
\isakeyword{where}\isanewline
\ {\isachardoublequoteopen}fin{\isacharunderscore}merge{\isacharunderscore}ti\ {\isacharbrackleft}{\isacharbrackright}\ y\ {\isacharequal}\ y{\isachardoublequoteclose}\ {\isacharbar}\isanewline
\ {\isachardoublequoteopen}fin{\isacharunderscore}merge{\isacharunderscore}ti\ {\isacharparenleft}x{\isacharhash}xs{\isacharparenright}\ y\ {\isacharequal}\ \isanewline
\ \ \ \ {\isacharparenleft}\ case\ y\ of\ {\isacharbrackleft}{\isacharbrackright}\ {\isasymRightarrow}\ {\isacharparenleft}x{\isacharhash}xs{\isacharparenright}\isanewline
\ \ \ \ \ \ \ \ \ {\isacharbar}\ {\isacharparenleft}z{\isacharhash}zs{\isacharparenright}\ {\isasymRightarrow}\ {\isacharparenleft}x{\isacharat}z{\isacharparenright}\ {\isacharhash}\ {\isacharparenleft}fin{\isacharunderscore}merge{\isacharunderscore}ti\ xs\ zs{\isacharparenright}{\isacharparenright}{\isachardoublequoteclose}\isanewline
\isanewline
\isamarkupcmt{merging time intervals of two infinite timed streams%
}
\isanewline
\isacommand{definition}\isamarkupfalse%
\isanewline
\ inf{\isacharunderscore}merge{\isacharunderscore}ti\ {\isacharcolon}{\isacharcolon}\ {\isachardoublequoteopen}{\isacharprime}a\ istream\ {\isasymRightarrow}\ {\isacharprime}a\ istream\ {\isasymRightarrow}\ {\isacharprime}a\ istream{\isachardoublequoteclose}\isanewline
\isakeyword{where}\isanewline
\ {\isachardoublequoteopen}inf{\isacharunderscore}merge{\isacharunderscore}ti\ x\ y\ \isanewline
\ \ {\isasymequiv}\ \isanewline
\ \ {\isasymlambda}\ i{\isachardot}\ {\isacharparenleft}x\ i{\isacharparenright}{\isacharat}{\isacharparenleft}y\ i{\isacharparenright}{\isachardoublequoteclose}\isanewline
\isanewline
\isamarkupcmt{the last time interval of a finite timed stream%
}
\isanewline
\isacommand{primrec}\isamarkupfalse%
\isanewline
\ \ fin{\isacharunderscore}last{\isacharunderscore}ti\ {\isacharcolon}{\isacharcolon}\ {\isachardoublequoteopen}{\isacharparenleft}{\isacharprime}a\ list{\isacharparenright}\ list\ {\isasymRightarrow}\ nat\ {\isasymRightarrow}\ {\isacharprime}a\ list{\isachardoublequoteclose}\isanewline
\isakeyword{where}\ \ \ \isanewline
\ {\isachardoublequoteopen}fin{\isacharunderscore}last{\isacharunderscore}ti\ s\ {\isadigit{0}}\ {\isacharequal}\ hd\ s{\isachardoublequoteclose}\ {\isacharbar}\isanewline
\ {\isachardoublequoteopen}fin{\isacharunderscore}last{\isacharunderscore}ti\ s\ {\isacharparenleft}Suc\ i{\isacharparenright}\ {\isacharequal}\ \isanewline
\ \ \ \ \ {\isacharparenleft}\ if\ s{\isacharbang}{\isacharparenleft}Suc\ i{\isacharparenright}\ {\isasymnoteq}\ {\isacharbrackleft}{\isacharbrackright}\isanewline
\ \ \ \ \ \ \ then\ s{\isacharbang}{\isacharparenleft}Suc\ i{\isacharparenright}\ \isanewline
\ \ \ \ \ \ \ else\ fin{\isacharunderscore}last{\isacharunderscore}ti\ s\ i{\isacharparenright}{\isachardoublequoteclose}\isanewline
\isanewline
\isamarkupcmt{the last nonempty time interval of a finite timed stream%
}
\isanewline
\isamarkupcmt{(can be applied to the streams which time intervals are empty from some moment)%
}
\isanewline
\isacommand{primrec}\isamarkupfalse%
\isanewline
\ \ inf{\isacharunderscore}last{\isacharunderscore}ti\ {\isacharcolon}{\isacharcolon}\ {\isachardoublequoteopen}{\isacharprime}a\ istream\ {\isasymRightarrow}\ nat\ {\isasymRightarrow}\ {\isacharprime}a\ list{\isachardoublequoteclose}\isanewline
\isakeyword{where}\ \ \isanewline
\ {\isachardoublequoteopen}inf{\isacharunderscore}last{\isacharunderscore}ti\ s\ {\isadigit{0}}\ {\isacharequal}\ s\ {\isadigit{0}}{\isachardoublequoteclose}\ {\isacharbar}\isanewline
\ {\isachardoublequoteopen}inf{\isacharunderscore}last{\isacharunderscore}ti\ s\ {\isacharparenleft}Suc\ i{\isacharparenright}\ {\isacharequal}\ \isanewline
\ \ \ \ \ {\isacharparenleft}\ if\ s\ {\isacharparenleft}Suc\ i{\isacharparenright}\ {\isasymnoteq}\ {\isacharbrackleft}{\isacharbrackright}\isanewline
\ \ \ \ \ \ \ then\ s\ {\isacharparenleft}Suc\ i{\isacharparenright}\ \isanewline
\ \ \ \ \ \ \ else\ inf{\isacharunderscore}last{\isacharunderscore}ti\ s\ i{\isacharparenright}{\isachardoublequoteclose}%
\isamarkupsubsection{Properties of operators%
}
\isamarkuptrue%
\isacommand{lemma}\isamarkupfalse%
\ inf{\isacharunderscore}last{\isacharunderscore}ti{\isacharunderscore}nonempty{\isacharunderscore}k{\isacharcolon}\isanewline
\ \ \isakeyword{assumes}\ {\isachardoublequoteopen}inf{\isacharunderscore}last{\isacharunderscore}ti\ dt\ t\ {\isasymnoteq}\ {\isacharbrackleft}{\isacharbrackright}{\isachardoublequoteclose}\isanewline
\ \ \isakeyword{shows}\ {\isachardoublequoteopen}inf{\isacharunderscore}last{\isacharunderscore}ti\ dt\ {\isacharparenleft}t\ {\isacharplus}\ k{\isacharparenright}\ {\isasymnoteq}\ {\isacharbrackleft}{\isacharbrackright}{\isachardoublequoteclose}\isanewline
\isadelimproof
\endisadelimproof
\isatagproof
\isacommand{using}\isamarkupfalse%
\ assms\ \ \isacommand{by}\isamarkupfalse%
\ {\isacharparenleft}induct\ k{\isacharcomma}\ auto{\isacharparenright}%
\endisatagproof
{\isafoldproof}%
\isadelimproof
\isanewline
\endisadelimproof
\isanewline
\isanewline
\isacommand{lemma}\isamarkupfalse%
\ inf{\isacharunderscore}last{\isacharunderscore}ti{\isacharunderscore}nonempty{\isacharcolon}\isanewline
\ \ \isakeyword{assumes}\ {\isachardoublequoteopen}s\ t\ {\isasymnoteq}\ {\isacharbrackleft}{\isacharbrackright}{\isachardoublequoteclose}\isanewline
\ \ \isakeyword{shows}\ {\isachardoublequoteopen}inf{\isacharunderscore}last{\isacharunderscore}ti\ s\ {\isacharparenleft}t\ {\isacharplus}\ k{\isacharparenright}\ {\isasymnoteq}\ {\isacharbrackleft}{\isacharbrackright}{\isachardoublequoteclose}\isanewline
\isadelimproof
\endisadelimproof
\isatagproof
\isacommand{using}\isamarkupfalse%
\ assms\ \ \isacommand{by}\isamarkupfalse%
\ {\isacharparenleft}induct\ k{\isacharcomma}\ auto{\isacharcomma}\ induct\ t{\isacharcomma}\ auto{\isacharparenright}%
\endisatagproof
{\isafoldproof}%
\isadelimproof
\isanewline
\endisadelimproof
\isanewline
\isacommand{lemma}\isamarkupfalse%
\ arith{\isacharunderscore}sum{\isacharunderscore}t{\isadigit{2}}k{\isacharcolon}\isanewline
{\isachardoublequoteopen}t\ {\isacharplus}\ {\isadigit{2}}\ {\isacharplus}\ k\ {\isacharequal}\ {\isacharparenleft}Suc\ t{\isacharparenright}\ {\isacharplus}\ {\isacharparenleft}Suc\ k{\isacharparenright}{\isachardoublequoteclose}\ \isanewline
\isadelimproof
\endisadelimproof
\isatagproof
\isacommand{by}\isamarkupfalse%
\ arith%
\endisatagproof
{\isafoldproof}%
\isadelimproof
\ \isanewline
\endisadelimproof
\isanewline
\isacommand{lemma}\isamarkupfalse%
\ inf{\isacharunderscore}last{\isacharunderscore}ti{\isacharunderscore}Suc{\isadigit{2}}{\isacharcolon}\isanewline
\ \ \isakeyword{assumes}\ h{\isadigit{1}}{\isacharcolon}{\isachardoublequoteopen}dt\ {\isacharparenleft}Suc\ t{\isacharparenright}\ {\isasymnoteq}\ {\isacharbrackleft}{\isacharbrackright}\ {\isasymor}\ dt\ {\isacharparenleft}Suc\ {\isacharparenleft}Suc\ t{\isacharparenright}{\isacharparenright}\ {\isasymnoteq}\ {\isacharbrackleft}{\isacharbrackright}{\isachardoublequoteclose}\isanewline
\ \ \isakeyword{shows}\ \ \ \ \ \ {\isachardoublequoteopen}inf{\isacharunderscore}last{\isacharunderscore}ti\ dt\ {\isacharparenleft}t\ {\isacharplus}\ {\isadigit{2}}\ {\isacharplus}\ k{\isacharparenright}\ {\isasymnoteq}\ {\isacharbrackleft}{\isacharbrackright}{\isachardoublequoteclose}\isanewline
\isadelimproof
\endisadelimproof
\isatagproof
\isacommand{proof}\isamarkupfalse%
\ {\isacharparenleft}cases\ {\isachardoublequoteopen}dt\ {\isacharparenleft}Suc\ t{\isacharparenright}\ {\isasymnoteq}\ {\isacharbrackleft}{\isacharbrackright}{\isachardoublequoteclose}{\isacharparenright}\isanewline
\ \ \isacommand{assume}\isamarkupfalse%
\ a{\isadigit{1}}{\isacharcolon}{\isachardoublequoteopen}dt\ {\isacharparenleft}Suc\ t{\isacharparenright}\ {\isasymnoteq}\ {\isacharbrackleft}{\isacharbrackright}{\isachardoublequoteclose}\isanewline
\ \ \isacommand{from}\isamarkupfalse%
\ a{\isadigit{1}}\ \isacommand{have}\isamarkupfalse%
\ sg{\isadigit{2}}{\isacharcolon}{\isachardoublequoteopen}inf{\isacharunderscore}last{\isacharunderscore}ti\ dt\ {\isacharparenleft}{\isacharparenleft}Suc\ t{\isacharparenright}\ {\isacharplus}\ {\isacharparenleft}Suc\ k{\isacharparenright}{\isacharparenright}\ {\isasymnoteq}\ {\isacharbrackleft}{\isacharbrackright}{\isachardoublequoteclose}\ \isanewline
\ \ \ \ \isacommand{by}\isamarkupfalse%
\ {\isacharparenleft}rule\ inf{\isacharunderscore}last{\isacharunderscore}ti{\isacharunderscore}nonempty{\isacharparenright}\isanewline
\ \ \isacommand{from}\isamarkupfalse%
\ sg{\isadigit{2}}\ \isacommand{show}\isamarkupfalse%
\ {\isacharquery}thesis\ \isacommand{by}\isamarkupfalse%
\ {\isacharparenleft}simp\ add{\isacharcolon}\ arith{\isacharunderscore}sum{\isacharunderscore}t{\isadigit{2}}k{\isacharparenright}\ \isanewline
\isacommand{next}\isamarkupfalse%
\isanewline
\ \ \isacommand{assume}\isamarkupfalse%
\ a{\isadigit{2}}{\isacharcolon}{\isachardoublequoteopen}{\isasymnot}\ dt\ {\isacharparenleft}Suc\ t{\isacharparenright}\ {\isasymnoteq}\ {\isacharbrackleft}{\isacharbrackright}{\isachardoublequoteclose}\isanewline
\ \ \isacommand{from}\isamarkupfalse%
\ a{\isadigit{2}}\ \isakeyword{and}\ h{\isadigit{1}}\ \isacommand{have}\isamarkupfalse%
\ sg{\isadigit{1}}{\isacharcolon}{\isachardoublequoteopen}dt\ {\isacharparenleft}Suc\ {\isacharparenleft}Suc\ t{\isacharparenright}{\isacharparenright}\ {\isasymnoteq}\ {\isacharbrackleft}{\isacharbrackright}{\isachardoublequoteclose}\ \isacommand{by}\isamarkupfalse%
\ simp\isanewline
\ \ \isacommand{from}\isamarkupfalse%
\ sg{\isadigit{1}}\ \isacommand{have}\isamarkupfalse%
\ sg{\isadigit{2}}{\isacharcolon}{\isachardoublequoteopen}inf{\isacharunderscore}last{\isacharunderscore}ti\ dt\ {\isacharparenleft}Suc\ {\isacharparenleft}Suc\ t{\isacharparenright}\ {\isacharplus}\ k{\isacharparenright}\ {\isasymnoteq}\ {\isacharbrackleft}{\isacharbrackright}{\isachardoublequoteclose}\ \isanewline
\ \ \ \ \isacommand{by}\isamarkupfalse%
\ {\isacharparenleft}rule\ inf{\isacharunderscore}last{\isacharunderscore}ti{\isacharunderscore}nonempty{\isacharparenright}\isanewline
\ \ \isacommand{from}\isamarkupfalse%
\ sg{\isadigit{2}}\ \isacommand{show}\isamarkupfalse%
\ {\isacharquery}thesis\ \isacommand{by}\isamarkupfalse%
\ auto\isanewline
\isacommand{qed}\isamarkupfalse%
\endisatagproof
{\isafoldproof}%
\isadelimproof
\endisadelimproof
\isamarkupsubsubsection{Lemmas for concatenation operator%
}
\isamarkuptrue%
\isacommand{lemma}\isamarkupfalse%
\ fin{\isacharunderscore}length{\isacharunderscore}append{\isacharcolon}\isanewline
{\isachardoublequoteopen}fin{\isacharunderscore}length\ {\isacharparenleft}x{\isacharat}y{\isacharparenright}\ {\isacharequal}\ {\isacharparenleft}fin{\isacharunderscore}length\ x{\isacharparenright}\ {\isacharplus}\ {\isacharparenleft}fin{\isacharunderscore}length\ y{\isacharparenright}{\isachardoublequoteclose}\isanewline
\isadelimproof
\ \ %
\endisadelimproof
\isatagproof
\isacommand{by}\isamarkupfalse%
\ {\isacharparenleft}induct\ x{\isacharcomma}\ auto{\isacharparenright}%
\endisatagproof
{\isafoldproof}%
\isadelimproof
\isanewline
\endisadelimproof
\isanewline
\isacommand{lemma}\isamarkupfalse%
\ fin{\isacharunderscore}append{\isacharunderscore}Nil{\isacharcolon}\isanewline
{\isachardoublequoteopen}fin{\isacharunderscore}inf{\isacharunderscore}append\ {\isacharbrackleft}{\isacharbrackright}\ z\ {\isacharequal}\ z{\isachardoublequoteclose}\isanewline
\isadelimproof
\ \ %
\endisadelimproof
\isatagproof
\isacommand{by}\isamarkupfalse%
\ {\isacharparenleft}simp\ add{\isacharcolon}\ fin{\isacharunderscore}inf{\isacharunderscore}append{\isacharunderscore}def{\isacharparenright}%
\endisatagproof
{\isafoldproof}%
\isadelimproof
\isanewline
\endisadelimproof
\isanewline
\isacommand{lemma}\isamarkupfalse%
\ correct{\isacharunderscore}fin{\isacharunderscore}inf{\isacharunderscore}append{\isadigit{1}}{\isacharcolon}\isanewline
\ \ \isakeyword{assumes}\ {\isachardoublequoteopen}s{\isadigit{1}}\ {\isacharequal}\ fin{\isacharunderscore}inf{\isacharunderscore}append\ {\isacharbrackleft}x{\isacharbrackright}\ s{\isachardoublequoteclose}\isanewline
\ \ \isakeyword{shows}\ {\isachardoublequoteopen}s{\isadigit{1}}\ {\isacharparenleft}Suc\ i{\isacharparenright}\ {\isacharequal}\ s\ i{\isachardoublequoteclose}\isanewline
\isadelimproof
\endisadelimproof
\isatagproof
\isacommand{using}\isamarkupfalse%
\ assms\ \ \isacommand{by}\isamarkupfalse%
\ {\isacharparenleft}simp\ add{\isacharcolon}\ fin{\isacharunderscore}inf{\isacharunderscore}append{\isacharunderscore}def{\isacharparenright}%
\endisatagproof
{\isafoldproof}%
\isadelimproof
\isanewline
\endisadelimproof
\isanewline
\isacommand{lemma}\isamarkupfalse%
\ correct{\isacharunderscore}fin{\isacharunderscore}inf{\isacharunderscore}append{\isadigit{2}}{\isacharcolon}\isanewline
{\isachardoublequoteopen}fin{\isacharunderscore}inf{\isacharunderscore}append\ {\isacharbrackleft}x{\isacharbrackright}\ s\ {\isacharparenleft}Suc\ i{\isacharparenright}\ {\isacharequal}\ s\ i{\isachardoublequoteclose}\isanewline
\isadelimproof
\ \ %
\endisadelimproof
\isatagproof
\isacommand{by}\isamarkupfalse%
\ {\isacharparenleft}simp\ add{\isacharcolon}\ fin{\isacharunderscore}inf{\isacharunderscore}append{\isacharunderscore}def{\isacharparenright}%
\endisatagproof
{\isafoldproof}%
\isadelimproof
\isanewline
\endisadelimproof
\isanewline
\isacommand{lemma}\isamarkupfalse%
\ fin{\isacharunderscore}append{\isacharunderscore}com{\isacharunderscore}Nil{\isadigit{1}}{\isacharcolon}\isanewline
{\isachardoublequoteopen}fin{\isacharunderscore}inf{\isacharunderscore}append\ {\isacharbrackleft}{\isacharbrackright}\ {\isacharparenleft}fin{\isacharunderscore}inf{\isacharunderscore}append\ y\ z{\isacharparenright}\ \isanewline
\ {\isacharequal}\ fin{\isacharunderscore}inf{\isacharunderscore}append\ {\isacharparenleft}{\isacharbrackleft}{\isacharbrackright}\ {\isacharat}\ y{\isacharparenright}\ z{\isachardoublequoteclose}\isanewline
\isadelimproof
\ \ %
\endisadelimproof
\isatagproof
\isacommand{by}\isamarkupfalse%
\ {\isacharparenleft}simp\ add{\isacharcolon}\ fin{\isacharunderscore}append{\isacharunderscore}Nil{\isacharparenright}%
\endisatagproof
{\isafoldproof}%
\isadelimproof
\isanewline
\endisadelimproof
\isanewline
\isacommand{lemma}\isamarkupfalse%
\ fin{\isacharunderscore}append{\isacharunderscore}com{\isacharunderscore}Nil{\isadigit{2}}{\isacharcolon}\isanewline
{\isachardoublequoteopen}fin{\isacharunderscore}inf{\isacharunderscore}append\ x\ {\isacharparenleft}fin{\isacharunderscore}inf{\isacharunderscore}append\ {\isacharbrackleft}{\isacharbrackright}\ z{\isacharparenright}\ {\isacharequal}\ fin{\isacharunderscore}inf{\isacharunderscore}append\ {\isacharparenleft}x\ {\isacharat}\ {\isacharbrackleft}{\isacharbrackright}{\isacharparenright}\ z{\isachardoublequoteclose}\isanewline
\isadelimproof
\ \ %
\endisadelimproof
\isatagproof
\isacommand{by}\isamarkupfalse%
\ {\isacharparenleft}simp\ add{\isacharcolon}\ fin{\isacharunderscore}append{\isacharunderscore}Nil{\isacharparenright}%
\endisatagproof
{\isafoldproof}%
\isadelimproof
\isanewline
\endisadelimproof
\isanewline
\isanewline
\isacommand{lemma}\isamarkupfalse%
\ fin{\isacharunderscore}append{\isacharunderscore}com{\isacharunderscore}i{\isacharcolon}\isanewline
{\isachardoublequoteopen}fin{\isacharunderscore}inf{\isacharunderscore}append\ x\ {\isacharparenleft}fin{\isacharunderscore}inf{\isacharunderscore}append\ y\ z{\isacharparenright}\ i\ {\isacharequal}\ fin{\isacharunderscore}inf{\isacharunderscore}append\ {\isacharparenleft}x\ {\isacharat}\ y{\isacharparenright}\ z\ i\ {\isachardoublequoteclose}\isanewline
\isadelimproof
\endisadelimproof
\isatagproof
\isacommand{proof}\isamarkupfalse%
\ {\isacharparenleft}cases\ x{\isacharparenright}\isanewline
\ \ \isacommand{assume}\isamarkupfalse%
\ Nil{\isacharcolon}{\isachardoublequoteopen}x\ {\isacharequal}\ {\isacharbrackleft}{\isacharbrackright}{\isachardoublequoteclose}\isanewline
\ \ \isacommand{thus}\isamarkupfalse%
\ {\isacharquery}thesis\ \isacommand{by}\isamarkupfalse%
\ {\isacharparenleft}simp\ add{\isacharcolon}\ fin{\isacharunderscore}append{\isacharunderscore}com{\isacharunderscore}Nil{\isadigit{1}}{\isacharparenright}\isanewline
\isacommand{next}\isamarkupfalse%
\isanewline
\ \ \isacommand{fix}\isamarkupfalse%
\ a\ l\ \isacommand{assume}\isamarkupfalse%
\ Cons{\isacharcolon}{\isachardoublequoteopen}x\ {\isacharequal}\ a\ {\isacharhash}\ l{\isachardoublequoteclose}\isanewline
\ \ \isacommand{thus}\isamarkupfalse%
\ {\isacharquery}thesis\isanewline
\ \ \isacommand{proof}\isamarkupfalse%
\ {\isacharparenleft}cases\ y{\isacharparenright}\isanewline
\ \ \ \ \isacommand{assume}\isamarkupfalse%
\ {\isachardoublequoteopen}y\ {\isacharequal}\ {\isacharbrackleft}{\isacharbrackright}{\isachardoublequoteclose}\isanewline
\ \ \ \ \isacommand{thus}\isamarkupfalse%
\ {\isacharquery}thesis\ \isacommand{by}\isamarkupfalse%
\ {\isacharparenleft}simp\ add{\isacharcolon}\ fin{\isacharunderscore}append{\isacharunderscore}com{\isacharunderscore}Nil{\isadigit{2}}{\isacharparenright}\isanewline
\ \ \isacommand{next}\isamarkupfalse%
\isanewline
\ \ \ \ \isacommand{fix}\isamarkupfalse%
\ aa\ la\ \isacommand{assume}\isamarkupfalse%
\ Cons{\isadigit{2}}{\isacharcolon}{\isachardoublequoteopen}y\ {\isacharequal}\ aa\ {\isacharhash}\ la{\isachardoublequoteclose}\isanewline
\ \ \ \ \isacommand{show}\isamarkupfalse%
\ {\isacharquery}thesis\ \isanewline
\ \ \ \ \isacommand{apply}\isamarkupfalse%
\ {\isacharparenleft}simp\ add{\isacharcolon}\ fin{\isacharunderscore}inf{\isacharunderscore}append{\isacharunderscore}def{\isacharcomma}\ auto{\isacharcomma}\ simp\ add{\isacharcolon}\ list{\isacharunderscore}nth{\isacharunderscore}append{\isadigit{0}}{\isacharparenright}\isanewline
\ \ \ \ \isacommand{by}\isamarkupfalse%
\ {\isacharparenleft}simp\ add{\isacharcolon}\ nth{\isacharunderscore}append{\isacharparenright}\isanewline
\ \ \isacommand{qed}\isamarkupfalse%
\isanewline
\isacommand{qed}\isamarkupfalse%
\endisatagproof
{\isafoldproof}%
\isadelimproof
\endisadelimproof
\isamarkupsubsubsection{Lemmas for operators $ts$ and $msg$%
}
\isamarkuptrue%
\isacommand{lemma}\isamarkupfalse%
\ ts{\isacharunderscore}msg{\isadigit{1}}{\isacharcolon}\isanewline
\ \ \isakeyword{assumes}\ {\isachardoublequoteopen}ts\ p{\isachardoublequoteclose}\isanewline
\ \ \isakeyword{shows}\ {\isachardoublequoteopen}msg\ {\isadigit{1}}\ p{\isachardoublequoteclose}\isanewline
\isadelimproof
\endisadelimproof
\isatagproof
\isacommand{using}\isamarkupfalse%
\ assms\ \isacommand{by}\isamarkupfalse%
\ {\isacharparenleft}simp\ add{\isacharcolon}\ ts{\isacharunderscore}def\ msg{\isacharunderscore}def{\isacharparenright}%
\endisatagproof
{\isafoldproof}%
\isadelimproof
\isanewline
\endisadelimproof
\isanewline 
\isacommand{lemma}\isamarkupfalse%
\ ts{\isacharunderscore}inf{\isacharunderscore}tl{\isacharcolon}\isanewline
\ \ \isakeyword{assumes}\ {\isachardoublequoteopen}ts\ x{\isachardoublequoteclose}\isanewline
\ \ \isakeyword{shows}\ {\isachardoublequoteopen}ts\ {\isacharparenleft}inf{\isacharunderscore}tl\ x{\isacharparenright}{\isachardoublequoteclose}\isanewline
\isadelimproof
\endisadelimproof
\isatagproof
\isacommand{using}\isamarkupfalse%
\ assms\ \ \isacommand{by}\isamarkupfalse%
\ {\isacharparenleft}simp\ add{\isacharcolon}\ ts{\isacharunderscore}def\ inf{\isacharunderscore}tl{\isacharunderscore}def{\isacharparenright}%
\endisatagproof
{\isafoldproof}%
\isadelimproof
\isanewline
\endisadelimproof
\isanewline
\isacommand{lemma}\isamarkupfalse%
\ ts{\isacharunderscore}length{\isacharunderscore}hint{\isadigit{1}}{\isacharcolon}\isanewline
\ \isakeyword{assumes}\ h{\isadigit{1}}{\isacharcolon}{\isachardoublequoteopen}ts\ x{\isachardoublequoteclose}\isanewline
\ \ \isakeyword{shows}\ {\isachardoublequoteopen}x\ i\ {\isasymnoteq}\ {\isacharbrackleft}{\isacharbrackright}{\isachardoublequoteclose}\isanewline
\isadelimproof
\endisadelimproof
\isatagproof
\isacommand{proof}\isamarkupfalse%
\ {\isacharminus}\ \isanewline
\ \ \isacommand{from}\isamarkupfalse%
\ h{\isadigit{1}}\ \isacommand{have}\isamarkupfalse%
\ sg{\isadigit{1}}{\isacharcolon}{\isachardoublequoteopen}length\ {\isacharparenleft}x\ i{\isacharparenright}\ {\isacharequal}\ Suc\ {\isadigit{0}}{\isachardoublequoteclose}\ \ \isacommand{by}\isamarkupfalse%
\ {\isacharparenleft}simp\ add{\isacharcolon}\ ts{\isacharunderscore}def{\isacharparenright}\isanewline
\ \ \isacommand{thus}\isamarkupfalse%
\ {\isacharquery}thesis\ \isacommand{by}\isamarkupfalse%
\ auto\isanewline
\isacommand{qed}\isamarkupfalse%
\endisatagproof
{\isafoldproof}%
\isadelimproof
\isanewline
\endisadelimproof
\isanewline
\isacommand{lemma}\isamarkupfalse%
\ ts{\isacharunderscore}length{\isacharunderscore}hint{\isadigit{2}}{\isacharcolon}\isanewline
\ \isakeyword{assumes}\ h{\isadigit{1}}{\isacharcolon}{\isachardoublequoteopen}ts\ x{\isachardoublequoteclose}\isanewline
\ \ \isakeyword{shows}\ {\isachardoublequoteopen}length\ {\isacharparenleft}x\ i{\isacharparenright}\ {\isacharequal}\ Suc\ {\isacharparenleft}{\isadigit{0}}{\isacharcolon}{\isacharcolon}nat{\isacharparenright}{\isachardoublequoteclose}\isanewline
\isadelimproof
\endisadelimproof
\isatagproof
\isacommand{using}\isamarkupfalse%
\ assms\isanewline
\ \ \isacommand{by}\isamarkupfalse%
\ {\isacharparenleft}simp\ add{\isacharcolon}\ ts{\isacharunderscore}def{\isacharparenright}%
\endisatagproof
{\isafoldproof}%
\isadelimproof
\isanewline
\endisadelimproof
\isanewline
\isacommand{lemma}\isamarkupfalse%
\ ts{\isacharunderscore}Least{\isacharunderscore}{\isadigit{0}}{\isacharcolon}\isanewline
\ \ \isakeyword{assumes}\ h{\isadigit{1}}{\isacharcolon}{\isachardoublequoteopen}ts\ x{\isachardoublequoteclose}\isanewline
\ \ \isakeyword{shows}\ {\isachardoublequoteopen}{\isacharparenleft}LEAST\ i{\isachardot}\ {\isacharparenleft}x\ i{\isacharparenright}\ {\isasymnoteq}\ {\isacharbrackleft}{\isacharbrackright}\ {\isacharparenright}\ {\isacharequal}\ {\isacharparenleft}{\isadigit{0}}{\isacharcolon}{\isacharcolon}nat{\isacharparenright}{\isachardoublequoteclose}\isanewline
\isadelimproof
\endisadelimproof
\isatagproof
\isacommand{using}\isamarkupfalse%
\ assms\isanewline
\isacommand{proof}\isamarkupfalse%
\ {\isacharminus}\ \isanewline
\ \ \isacommand{from}\isamarkupfalse%
\ h{\isadigit{1}}\ \isacommand{have}\isamarkupfalse%
\ sg{\isadigit{1}}{\isacharcolon}{\isachardoublequoteopen}x\ {\isadigit{0}}\ {\isasymnoteq}\ {\isacharbrackleft}{\isacharbrackright}{\isachardoublequoteclose}\ \isacommand{by}\isamarkupfalse%
\ {\isacharparenleft}rule\ ts{\isacharunderscore}length{\isacharunderscore}hint{\isadigit{1}}{\isacharparenright}\isanewline
\ \ \isacommand{thus}\isamarkupfalse%
\ {\isacharquery}thesis\isanewline
\ \ \isacommand{apply}\isamarkupfalse%
\ {\isacharparenleft}simp\ add{\isacharcolon}\ Least{\isacharunderscore}def{\isacharparenright}\isanewline
\ \ \isacommand{by}\isamarkupfalse%
\ auto\isanewline
\isacommand{qed}\isamarkupfalse%
\endisatagproof
{\isafoldproof}%
\isadelimproof
\isanewline
\endisadelimproof
\isanewline
\isacommand{lemma}\isamarkupfalse%
\ inf{\isacharunderscore}tl{\isacharunderscore}Suc{\isacharcolon}\isanewline
{\isachardoublequoteopen}inf{\isacharunderscore}tl\ x\ i\ {\isacharequal}\ x\ {\isacharparenleft}Suc\ i{\isacharparenright}{\isachardoublequoteclose}\isanewline
\isadelimproof
\ \ %
\endisadelimproof
\isatagproof
\isacommand{by}\isamarkupfalse%
\ {\isacharparenleft}simp\ add{\isacharcolon}\ inf{\isacharunderscore}tl{\isacharunderscore}def{\isacharparenright}%
\endisatagproof
{\isafoldproof}%
\isadelimproof
\ \isanewline
\endisadelimproof
\isanewline
\isanewline
\isacommand{lemma}\isamarkupfalse%
\ ts{\isacharunderscore}Least{\isacharunderscore}Suc{\isadigit{0}}{\isacharcolon}\isanewline
\ \ \isakeyword{assumes}\ h{\isadigit{1}}{\isacharcolon}{\isachardoublequoteopen}ts\ x{\isachardoublequoteclose}\isanewline
\ \ \isakeyword{shows}\ {\isachardoublequoteopen}{\isacharparenleft}LEAST\ i{\isachardot}\ x\ {\isacharparenleft}Suc\ i{\isacharparenright}\ {\isasymnoteq}\ {\isacharbrackleft}{\isacharbrackright}{\isacharparenright}\ {\isacharequal}\ {\isadigit{0}}{\isachardoublequoteclose}\isanewline
\isadelimproof
\endisadelimproof
\isatagproof
\isacommand{proof}\isamarkupfalse%
\ {\isacharminus}\isanewline
\ \ \isacommand{from}\isamarkupfalse%
\ h{\isadigit{1}}\ \isacommand{have}\isamarkupfalse%
\ sg{\isadigit{1}}{\isacharcolon}{\isachardoublequoteopen}x\ {\isacharparenleft}Suc\ {\isadigit{0}}{\isacharparenright}\ {\isasymnoteq}\ {\isacharbrackleft}{\isacharbrackright}{\isachardoublequoteclose}\ \isacommand{by}\isamarkupfalse%
\ {\isacharparenleft}simp\ add{\isacharcolon}\ ts{\isacharunderscore}length{\isacharunderscore}hint{\isadigit{1}}{\isacharparenright}\isanewline
\ \ \isacommand{thus}\isamarkupfalse%
\ {\isacharquery}thesis\ \isacommand{by}\isamarkupfalse%
\ {\isacharparenleft}simp\ add{\isacharcolon}\ Least{\isacharunderscore}def{\isacharcomma}\ auto{\isacharparenright}\isanewline
\isacommand{qed}\isamarkupfalse%
\endisatagproof
{\isafoldproof}%
\isadelimproof
\isanewline
\endisadelimproof
\isanewline
\isanewline
\isacommand{lemma}\isamarkupfalse%
\ ts{\isacharunderscore}inf{\isacharunderscore}make{\isacharunderscore}untimed{\isacharunderscore}inf{\isacharunderscore}tl{\isacharcolon}\isanewline
\ \ \isakeyword{assumes}\ h{\isadigit{1}}{\isacharcolon}{\isachardoublequoteopen}ts\ x{\isachardoublequoteclose}\isanewline
\ \ \isakeyword{shows}\ {\isachardoublequoteopen}inf{\isacharunderscore}make{\isacharunderscore}untimed\ {\isacharparenleft}inf{\isacharunderscore}tl\ x{\isacharparenright}\ i\ {\isacharequal}\ inf{\isacharunderscore}make{\isacharunderscore}untimed\ x\ {\isacharparenleft}Suc\ i{\isacharparenright}{\isachardoublequoteclose}\isanewline
\isadelimproof
\endisadelimproof
\isatagproof
\isacommand{using}\isamarkupfalse%
\ assms\isanewline
\ \ \isacommand{apply}\isamarkupfalse%
\ {\isacharparenleft}simp\ add{\isacharcolon}\ inf{\isacharunderscore}make{\isacharunderscore}untimed{\isacharunderscore}def{\isacharparenright}\ \isanewline
\ \ \isacommand{proof}\isamarkupfalse%
\ {\isacharparenleft}induct\ i{\isacharparenright}\isanewline
\ \ \ \ \isacommand{case}\isamarkupfalse%
\ {\isadigit{0}}\isanewline
\ \ \ \ \isacommand{from}\isamarkupfalse%
\ h{\isadigit{1}}\ \isacommand{show}\isamarkupfalse%
\ {\isacharquery}case\isanewline
\ \ \ \ \ \ \isacommand{by}\isamarkupfalse%
\ {\isacharparenleft}simp\ add{\isacharcolon}\ ts{\isacharunderscore}length{\isacharunderscore}hint{\isadigit{1}}\ ts{\isacharunderscore}length{\isacharunderscore}hint{\isadigit{2}}{\isacharparenright}\isanewline
\ \ \isacommand{next}\isamarkupfalse%
\isanewline
\ \ \ \ \isacommand{case}\isamarkupfalse%
\ {\isacharparenleft}Suc\ i{\isacharparenright}\isanewline
\ \ \ \ \isacommand{from}\isamarkupfalse%
\ h{\isadigit{1}}\ \isacommand{show}\isamarkupfalse%
\ {\isacharquery}case\isanewline
\ \ \ \ \ \ \isacommand{by}\isamarkupfalse%
\ {\isacharparenleft}simp\ add{\isacharcolon}\ ts{\isacharunderscore}length{\isacharunderscore}hint{\isadigit{1}}\ ts{\isacharunderscore}length{\isacharunderscore}hint{\isadigit{2}}{\isacharparenright}\isanewline
\ \isacommand{qed}\isamarkupfalse%
\endisatagproof
{\isafoldproof}%
\isadelimproof
\isanewline
\endisadelimproof
\isanewline
\isanewline
\isacommand{lemma}\isamarkupfalse%
\ ts{\isacharunderscore}inf{\isacharunderscore}make{\isacharunderscore}untimed{\isadigit{1}}{\isacharunderscore}inf{\isacharunderscore}tl{\isacharcolon}\isanewline
\ \ \isakeyword{assumes}\ h{\isadigit{1}}{\isacharcolon}{\isachardoublequoteopen}ts\ x{\isachardoublequoteclose}\isanewline
\ \ \isakeyword{shows}\ {\isachardoublequoteopen}inf{\isacharunderscore}make{\isacharunderscore}untimed{\isadigit{1}}\ {\isacharparenleft}inf{\isacharunderscore}tl\ x{\isacharparenright}\ i\ {\isacharequal}\ inf{\isacharunderscore}make{\isacharunderscore}untimed{\isadigit{1}}\ x\ {\isacharparenleft}Suc\ i{\isacharparenright}{\isachardoublequoteclose}\isanewline
\isadelimproof
\endisadelimproof
\isatagproof
\isacommand{using}\isamarkupfalse%
\ assms\isanewline
\ \ \isacommand{proof}\isamarkupfalse%
\ {\isacharparenleft}induct\ i{\isacharparenright}\isanewline
\ \ \ \ \isacommand{case}\isamarkupfalse%
\ {\isadigit{0}}\isanewline
\ \ \ \ \isacommand{from}\isamarkupfalse%
\ h{\isadigit{1}}\ \isacommand{show}\isamarkupfalse%
\ {\isacharquery}case\isanewline
\ \ \ \ \ \ \isacommand{by}\isamarkupfalse%
\ {\isacharparenleft}simp\ add{\isacharcolon}\ ts{\isacharunderscore}length{\isacharunderscore}hint{\isadigit{1}}\ ts{\isacharunderscore}length{\isacharunderscore}hint{\isadigit{2}}{\isacharparenright}\isanewline
\ \ \isacommand{next}\isamarkupfalse%
\isanewline
\ \ \ \ \isacommand{case}\isamarkupfalse%
\ {\isacharparenleft}Suc\ i{\isacharparenright}\isanewline
\ \ \ \ \isacommand{from}\isamarkupfalse%
\ h{\isadigit{1}}\ \isacommand{show}\isamarkupfalse%
\ {\isacharquery}case\isanewline
\ \ \ \ \ \ \isacommand{by}\isamarkupfalse%
\ {\isacharparenleft}simp\ add{\isacharcolon}\ ts{\isacharunderscore}length{\isacharunderscore}hint{\isadigit{1}}\ ts{\isacharunderscore}length{\isacharunderscore}hint{\isadigit{2}}{\isacharparenright}\isanewline
\ \isacommand{qed}\isamarkupfalse%
\endisatagproof
{\isafoldproof}%
\isadelimproof
\isanewline
\endisadelimproof
\isanewline
\isanewline
\isacommand{lemma}\isamarkupfalse%
\ msg{\isacharunderscore}nonempty{\isadigit{1}}{\isacharcolon}\isanewline
\ \ \isakeyword{assumes}\ h{\isadigit{1}}{\isacharcolon}{\isachardoublequoteopen}msg\ {\isacharparenleft}Suc\ {\isadigit{0}}{\isacharparenright}\ a{\isachardoublequoteclose}\ \isakeyword{and}\ h{\isadigit{2}}{\isacharcolon}{\isachardoublequoteopen}a\ t\ {\isacharequal}\ aa\ {\isacharhash}\ l{\isachardoublequoteclose}\isanewline
\ \ \isakeyword{shows}\ {\isachardoublequoteopen}l\ {\isacharequal}\ {\isacharbrackleft}{\isacharbrackright}{\isachardoublequoteclose}\isanewline
\isadelimproof
\endisadelimproof
\isatagproof
\isacommand{proof}\isamarkupfalse%
\ {\isacharminus}\ \isanewline
\ \ \isacommand{from}\isamarkupfalse%
\ h{\isadigit{1}}\ \isacommand{have}\isamarkupfalse%
\ sg{\isadigit{1}}{\isacharcolon}{\isachardoublequoteopen}length\ {\isacharparenleft}a\ t{\isacharparenright}\ {\isasymle}\ Suc\ {\isadigit{0}}{\isachardoublequoteclose}\ \isacommand{by}\isamarkupfalse%
\ {\isacharparenleft}simp\ add{\isacharcolon}\ msg{\isacharunderscore}def{\isacharparenright}\isanewline
\ \ \isacommand{from}\isamarkupfalse%
\ h{\isadigit{2}}\ \isakeyword{and}\ sg{\isadigit{1}}\ \isacommand{show}\isamarkupfalse%
\ {\isacharquery}thesis\ \isacommand{by}\isamarkupfalse%
\ auto\isanewline
\isacommand{qed}\isamarkupfalse%
\endisatagproof
{\isafoldproof}%
\isadelimproof
\isanewline
\endisadelimproof
\isanewline
\isanewline
\isacommand{lemma}\isamarkupfalse%
\ msg{\isacharunderscore}nonempty{\isadigit{2}}{\isacharcolon}\isanewline
\ \ \isakeyword{assumes}\ h{\isadigit{1}}{\isacharcolon}{\isachardoublequoteopen}msg\ {\isacharparenleft}Suc\ {\isadigit{0}}{\isacharparenright}\ a{\isachardoublequoteclose}\ \isakeyword{and}\ h{\isadigit{2}}{\isacharcolon}{\isachardoublequoteopen}a\ t\ \ {\isasymnoteq}\ {\isacharbrackleft}{\isacharbrackright}{\isachardoublequoteclose}\isanewline
\ \ \isakeyword{shows}\ {\isachardoublequoteopen}length\ {\isacharparenleft}a\ t{\isacharparenright}\ {\isacharequal}\ {\isacharparenleft}Suc\ {\isadigit{0}}{\isacharparenright}{\isachardoublequoteclose}\isanewline
\isadelimproof
\endisadelimproof
\isatagproof
\isacommand{proof}\isamarkupfalse%
\ {\isacharminus}\ \isanewline
\ \ \isacommand{from}\isamarkupfalse%
\ h{\isadigit{1}}\ \isacommand{have}\isamarkupfalse%
\ sg{\isadigit{1}}{\isacharcolon}{\isachardoublequoteopen}length\ {\isacharparenleft}a\ t{\isacharparenright}\ {\isasymle}\ Suc\ {\isadigit{0}}{\isachardoublequoteclose}\ \isacommand{by}\isamarkupfalse%
\ {\isacharparenleft}simp\ add{\isacharcolon}\ msg{\isacharunderscore}def{\isacharparenright}\isanewline
\ \ \isacommand{from}\isamarkupfalse%
\ h{\isadigit{2}}\ \isacommand{have}\isamarkupfalse%
\ sg{\isadigit{2}}{\isacharcolon}{\isachardoublequoteopen}length\ {\isacharparenleft}a\ t{\isacharparenright}\ {\isasymnoteq}\ {\isadigit{0}}{\isachardoublequoteclose}\ \isacommand{by}\isamarkupfalse%
\ auto\isanewline
\ \ \isacommand{from}\isamarkupfalse%
\ sg{\isadigit{1}}\ \isakeyword{and}\ sg{\isadigit{2}}\ \isacommand{show}\isamarkupfalse%
\ {\isacharquery}thesis\ \isacommand{by}\isamarkupfalse%
\ arith\ \isanewline
\isacommand{qed}\isamarkupfalse%
\endisatagproof
{\isafoldproof}%
\isadelimproof
\endisadelimproof
\isamarkupsubsubsection{Lemmas for $inf\_truncate$%
}
\isamarkuptrue%
\isacommand{lemma}\isamarkupfalse%
\ inf{\isacharunderscore}truncate{\isacharunderscore}nonempty{\isacharcolon}\isanewline
\ \ \isakeyword{assumes}\ h{\isadigit{1}}{\isacharcolon}{\isachardoublequoteopen}z\ i\ {\isasymnoteq}\ {\isacharbrackleft}{\isacharbrackright}{\isachardoublequoteclose}\isanewline
\ \ \isakeyword{shows}\ {\isachardoublequoteopen}inf{\isacharunderscore}truncate\ z\ i\ {\isasymnoteq}\ {\isacharbrackleft}{\isacharbrackright}{\isachardoublequoteclose}\isanewline
\isadelimproof
\endisadelimproof
\isatagproof
\isacommand{proof}\isamarkupfalse%
\ {\isacharparenleft}induct\ i{\isacharparenright}\isanewline
\ \ \ \ \isacommand{case}\isamarkupfalse%
\ {\isadigit{0}}\isanewline
\ \ \ \ \isacommand{from}\isamarkupfalse%
\ h{\isadigit{1}}\ \isacommand{show}\isamarkupfalse%
\ {\isacharquery}case\ \isacommand{by}\isamarkupfalse%
\ auto\isanewline
\ \ \isacommand{next}\isamarkupfalse%
\isanewline
\ \ \ \ \isacommand{case}\isamarkupfalse%
\ {\isacharparenleft}Suc\ i{\isacharparenright}\isanewline
\ \ \ \ \ \isacommand{from}\isamarkupfalse%
\ h{\isadigit{1}}\ \isacommand{show}\isamarkupfalse%
\ {\isacharquery}case\ \isacommand{by}\isamarkupfalse%
\ auto\isanewline
\isacommand{qed}\isamarkupfalse%
\endisatagproof
{\isafoldproof}%
\isadelimproof
\isanewline
\endisadelimproof
\isanewline
\isanewline
\isacommand{lemma}\isamarkupfalse%
\ concat{\isacharunderscore}inf{\isacharunderscore}truncate{\isacharunderscore}nonempty{\isacharcolon}\isanewline
\ \ \isakeyword{assumes}\ h{\isadigit{1}}{\isacharcolon}\ {\isachardoublequoteopen}z\ i\ {\isasymnoteq}\ {\isacharbrackleft}{\isacharbrackright}{\isachardoublequoteclose}\isanewline
\ \ \isakeyword{shows}\ {\isachardoublequoteopen}concat\ {\isacharparenleft}inf{\isacharunderscore}truncate\ z\ i{\isacharparenright}\ {\isasymnoteq}\ {\isacharbrackleft}{\isacharbrackright}{\isachardoublequoteclose}\isanewline
\isadelimproof
\endisadelimproof
\isatagproof
\isacommand{using}\isamarkupfalse%
\ assms\isanewline
\isacommand{proof}\isamarkupfalse%
\ {\isacharparenleft}induct\ i{\isacharparenright}\isanewline
\ \ \ \ \isacommand{case}\isamarkupfalse%
\ {\isadigit{0}}\isanewline
\ \ \ \ \isacommand{thus}\isamarkupfalse%
\ {\isacharquery}case\ \isacommand{by}\isamarkupfalse%
\ auto\isanewline
\ \ \isacommand{next}\isamarkupfalse%
\isanewline
\ \ \ \ \isacommand{case}\isamarkupfalse%
\ {\isacharparenleft}Suc\ i{\isacharparenright}\isanewline
\ \ \ \ \isacommand{thus}\isamarkupfalse%
\ {\isacharquery}case\ \isacommand{by}\isamarkupfalse%
\ auto\isanewline
\isacommand{qed}\isamarkupfalse%
\endisatagproof
{\isafoldproof}%
\isadelimproof
\isanewline
\endisadelimproof
\ \ \isanewline
\isanewline
\isacommand{lemma}\isamarkupfalse%
\ concat{\isacharunderscore}inf{\isacharunderscore}truncate{\isacharunderscore}nonempty{\isacharunderscore}a{\isacharcolon}\isanewline
\ \ \isakeyword{assumes}\ h{\isadigit{1}}{\isacharcolon}{\isachardoublequoteopen}z\ i\ {\isacharequal}\ {\isacharbrackleft}a{\isacharbrackright}{\isachardoublequoteclose}\ \isanewline
\ \ \isakeyword{shows}\ {\isachardoublequoteopen}concat\ {\isacharparenleft}inf{\isacharunderscore}truncate\ z\ i{\isacharparenright}\ {\isasymnoteq}\ {\isacharbrackleft}{\isacharbrackright}{\isachardoublequoteclose}\isanewline
\isadelimproof
\endisadelimproof
\isatagproof
\isacommand{using}\isamarkupfalse%
\ assms\isanewline
\isacommand{proof}\isamarkupfalse%
\ {\isacharparenleft}induct\ i{\isacharparenright}\isanewline
\ \ \ \ \isacommand{case}\isamarkupfalse%
\ {\isadigit{0}}\isanewline
\ \ \ \ \isacommand{thus}\isamarkupfalse%
\ {\isacharquery}case\ \isacommand{by}\isamarkupfalse%
\ auto\isanewline
\ \ \isacommand{next}\isamarkupfalse%
\isanewline
\ \ \ \ \isacommand{case}\isamarkupfalse%
\ {\isacharparenleft}Suc\ i{\isacharparenright}\isanewline
\ \ \ \ \isacommand{thus}\isamarkupfalse%
\ {\isacharquery}case\ \isacommand{by}\isamarkupfalse%
\ auto\isanewline
\isacommand{qed}\isamarkupfalse%
\endisatagproof
{\isafoldproof}%
\isadelimproof
\isanewline
\endisadelimproof
\ \ \isanewline
\isanewline
\isacommand{lemma}\isamarkupfalse%
\ concat{\isacharunderscore}inf{\isacharunderscore}truncate{\isacharunderscore}nonempty{\isacharunderscore}el{\isacharcolon}\isanewline
\ \ \isakeyword{assumes}\ h{\isadigit{1}}{\isacharcolon}{\isachardoublequoteopen}z\ i\ {\isasymnoteq}\ {\isacharbrackleft}{\isacharbrackright}{\isachardoublequoteclose}\ \isanewline
\ \ \isakeyword{shows}\ {\isachardoublequoteopen}concat\ {\isacharparenleft}inf{\isacharunderscore}truncate\ z\ i{\isacharparenright}\ {\isasymnoteq}\ {\isacharbrackleft}{\isacharbrackright}{\isachardoublequoteclose}\isanewline
\isadelimproof
\endisadelimproof
\isatagproof
\isacommand{using}\isamarkupfalse%
\ assms\isanewline
\isacommand{proof}\isamarkupfalse%
\ {\isacharparenleft}induct\ i{\isacharparenright}\isanewline
\ \ \ \ \isacommand{case}\isamarkupfalse%
\ {\isadigit{0}}\isanewline
\ \ \ \ \isacommand{thus}\isamarkupfalse%
\ {\isacharquery}case\ \isacommand{by}\isamarkupfalse%
\ auto\isanewline
\ \ \isacommand{next}\isamarkupfalse%
\isanewline
\ \ \ \ \isacommand{case}\isamarkupfalse%
\ {\isacharparenleft}Suc\ i{\isacharparenright}\isanewline
\ \ \ \ \isacommand{thus}\isamarkupfalse%
\ {\isacharquery}case\ \isacommand{by}\isamarkupfalse%
\ auto\isanewline
\isacommand{qed}\isamarkupfalse%
\endisatagproof
{\isafoldproof}%
\isadelimproof
\isanewline
\endisadelimproof
\isanewline
\isanewline
\isacommand{lemma}\isamarkupfalse%
\ inf{\isacharunderscore}truncate{\isacharunderscore}append{\isacharcolon}\isanewline
{\isachardoublequoteopen}{\isacharparenleft}inf{\isacharunderscore}truncate\ z\ i\ {\isacharat}\ {\isacharbrackleft}z\ {\isacharparenleft}Suc\ i{\isacharparenright}{\isacharbrackright}{\isacharparenright}\ {\isacharequal}\ inf{\isacharunderscore}truncate\ z\ {\isacharparenleft}Suc\ i{\isacharparenright}{\isachardoublequoteclose}\isanewline
\isadelimproof
\endisadelimproof
\isatagproof
\isacommand{proof}\isamarkupfalse%
\ {\isacharparenleft}induct\ i{\isacharparenright}\isanewline
\ \ \ \ \isacommand{case}\isamarkupfalse%
\ {\isadigit{0}}\isanewline
\ \ \ \ \isacommand{thus}\isamarkupfalse%
\ {\isacharquery}case\ \isacommand{by}\isamarkupfalse%
\ auto\isanewline
\ \ \isacommand{next}\isamarkupfalse%
\isanewline
\ \ \ \ \isacommand{case}\isamarkupfalse%
\ {\isacharparenleft}Suc\ i{\isacharparenright}\isanewline
\ \ \ \ \isacommand{thus}\isamarkupfalse%
\ {\isacharquery}case\ \isacommand{by}\isamarkupfalse%
\ auto\isanewline
\isacommand{qed}\isamarkupfalse%
\endisatagproof
{\isafoldproof}%
\isadelimproof
\endisadelimproof
\isamarkupsubsubsection{Lemmas for $fin\_make\_untimed$%
}
\isamarkuptrue%
\isacommand{lemma}\isamarkupfalse%
\ fin{\isacharunderscore}make{\isacharunderscore}untimed{\isacharunderscore}append{\isacharcolon}\isanewline
\ \ \isakeyword{assumes}\ h{\isadigit{1}}{\isacharcolon}{\isachardoublequoteopen}fin{\isacharunderscore}make{\isacharunderscore}untimed\ x\ {\isasymnoteq}\ {\isacharbrackleft}{\isacharbrackright}{\isachardoublequoteclose}\isanewline
\ \ \isakeyword{shows}\ {\isachardoublequoteopen}fin{\isacharunderscore}make{\isacharunderscore}untimed\ {\isacharparenleft}x\ {\isacharat}\ y{\isacharparenright}\ {\isasymnoteq}\ {\isacharbrackleft}{\isacharbrackright}{\isachardoublequoteclose}\isanewline
\isadelimproof
\endisadelimproof
\isatagproof
\isacommand{using}\isamarkupfalse%
\ assms\ \isacommand{by}\isamarkupfalse%
\ {\isacharparenleft}simp\ add{\isacharcolon}\ fin{\isacharunderscore}make{\isacharunderscore}untimed{\isacharunderscore}def{\isacharparenright}%
\endisatagproof
{\isafoldproof}%
\isadelimproof
\isanewline
\endisadelimproof
\isanewline
\isanewline
\isacommand{lemma}\isamarkupfalse%
\ fin{\isacharunderscore}make{\isacharunderscore}untimed{\isacharunderscore}inf{\isacharunderscore}truncate{\isacharunderscore}Nonempty{\isacharcolon}\isanewline
\ \ \isakeyword{assumes}\ h{\isadigit{1}}{\isacharcolon}{\isachardoublequoteopen}z\ k\ {\isasymnoteq}\ {\isacharbrackleft}{\isacharbrackright}{\isachardoublequoteclose}\isanewline
\ \ \ \ \ \ \isakeyword{and}\ h{\isadigit{2}}{\isacharcolon}{\isachardoublequoteopen}k\ {\isasymle}\ i{\isachardoublequoteclose}\isanewline
\ \ \isakeyword{shows}\ {\isachardoublequoteopen}fin{\isacharunderscore}make{\isacharunderscore}untimed\ {\isacharparenleft}inf{\isacharunderscore}truncate\ z\ i{\isacharparenright}\ {\isasymnoteq}\ {\isacharbrackleft}{\isacharbrackright}{\isachardoublequoteclose}\isanewline
\isadelimproof
\endisadelimproof
\isatagproof
\isacommand{using}\isamarkupfalse%
\ assms\isanewline
\ \ \isacommand{apply}\isamarkupfalse%
\ {\isacharparenleft}simp\ add{\isacharcolon}\ fin{\isacharunderscore}make{\isacharunderscore}untimed{\isacharunderscore}def{\isacharparenright}\isanewline
\ \ \isacommand{proof}\isamarkupfalse%
\ {\isacharparenleft}induct\ i{\isacharparenright}\isanewline
\ \ \ \ \isacommand{case}\isamarkupfalse%
\ {\isadigit{0}}\isanewline
\ \ \ \ \isacommand{thus}\isamarkupfalse%
\ {\isacharquery}case\ \ \isacommand{by}\isamarkupfalse%
\ auto\isanewline
\ \ \isacommand{next}\isamarkupfalse%
\isanewline
\ \ \ \ \isacommand{case}\isamarkupfalse%
\ {\isacharparenleft}Suc\ i{\isacharparenright}\isanewline
\ \ \ \ \isacommand{thus}\isamarkupfalse%
\ {\isacharquery}case\ \isanewline
\ \ \ \ \isacommand{proof}\isamarkupfalse%
\ cases\isanewline
\ \ \ \ \ \ \isacommand{assume}\isamarkupfalse%
\ {\isachardoublequoteopen}k\ {\isasymle}\ i{\isachardoublequoteclose}\isanewline
\ \ \ \ \ \ \isacommand{from}\isamarkupfalse%
\ Suc\ \isakeyword{and}\ this\ \isacommand{show}\isamarkupfalse%
\ {\isachardoublequoteopen}{\isasymexists}xs{\isasymin}set\ {\isacharparenleft}inf{\isacharunderscore}truncate\ z\ {\isacharparenleft}Suc\ i{\isacharparenright}{\isacharparenright}{\isachardot}\ xs\ {\isasymnoteq}\ {\isacharbrackleft}{\isacharbrackright}{\isachardoublequoteclose}\isanewline
\ \ \ \ \ \ \ \ \isacommand{by}\isamarkupfalse%
\ auto\isanewline
\ \ \ \ \isacommand{next}\isamarkupfalse%
\isanewline
\ \ \ \ \ \ \isacommand{assume}\isamarkupfalse%
\ {\isachardoublequoteopen}{\isasymnot}\ k\ {\isasymle}\ i{\isachardoublequoteclose}\isanewline
\ \ \ \ \ \ \isacommand{from}\isamarkupfalse%
\ Suc\ \isakeyword{and}\ this\ \isacommand{have}\isamarkupfalse%
\ sg{\isadigit{1}}{\isacharcolon}{\isachardoublequoteopen}k\ {\isacharequal}\ Suc\ i{\isachardoublequoteclose}\ \isacommand{by}\isamarkupfalse%
\ arith\isanewline
\ \ \ \ \ \ \isacommand{from}\isamarkupfalse%
\ Suc\ \isakeyword{and}\ this\ \isacommand{show}\isamarkupfalse%
\ {\isachardoublequoteopen}{\isasymexists}xs{\isasymin}set\ {\isacharparenleft}inf{\isacharunderscore}truncate\ z\ {\isacharparenleft}Suc\ i{\isacharparenright}{\isacharparenright}{\isachardot}\ xs\ {\isasymnoteq}\ {\isacharbrackleft}{\isacharbrackright}{\isachardoublequoteclose}\isanewline
\ \ \ \ \ \ \ \ \isacommand{by}\isamarkupfalse%
\ auto\isanewline
\ \ \ \ \ \isacommand{qed}\isamarkupfalse%
\isanewline
\isacommand{qed}\isamarkupfalse%
\endisatagproof
{\isafoldproof}%
\isadelimproof
\isanewline
\endisadelimproof
\isanewline
\isanewline
\isacommand{lemma}\isamarkupfalse%
\ last{\isacharunderscore}fin{\isacharunderscore}make{\isacharunderscore}untimed{\isacharunderscore}append{\isacharcolon}\isanewline
{\isachardoublequoteopen}last\ {\isacharparenleft}fin{\isacharunderscore}make{\isacharunderscore}untimed\ {\isacharparenleft}z\ {\isacharat}\ {\isacharbrackleft}{\isacharbrackleft}a{\isacharbrackright}{\isacharbrackright}{\isacharparenright}{\isacharparenright}\ {\isacharequal}\ a{\isachardoublequoteclose}\isanewline
\isadelimproof
\ \ %
\endisadelimproof
\isatagproof
\isacommand{by}\isamarkupfalse%
\ {\isacharparenleft}simp\ add{\isacharcolon}\ fin{\isacharunderscore}make{\isacharunderscore}untimed{\isacharunderscore}def{\isacharparenright}%
\endisatagproof
{\isafoldproof}%
\isadelimproof
\isanewline
\endisadelimproof
\isanewline
\isanewline
\isacommand{lemma}\isamarkupfalse%
\ last{\isacharunderscore}fin{\isacharunderscore}make{\isacharunderscore}untimed{\isacharunderscore}inf{\isacharunderscore}truncate{\isacharcolon}\isanewline
\ \ \isakeyword{assumes}\ h{\isadigit{1}}{\isacharcolon}{\isachardoublequoteopen}z\ i\ {\isacharequal}\ {\isacharbrackleft}a{\isacharbrackright}{\isachardoublequoteclose}\isanewline
\ \ \isakeyword{shows}\ {\isachardoublequoteopen}last\ {\isacharparenleft}fin{\isacharunderscore}make{\isacharunderscore}untimed\ {\isacharparenleft}inf{\isacharunderscore}truncate\ z\ i{\isacharparenright}{\isacharparenright}\ {\isacharequal}\ a{\isachardoublequoteclose}\isanewline
\isadelimproof
\endisadelimproof
\isatagproof
\isacommand{using}\isamarkupfalse%
\ assms\isanewline
\isacommand{proof}\isamarkupfalse%
\ {\isacharparenleft}induction\ i{\isacharparenright}\isanewline
\ \ \isacommand{case}\isamarkupfalse%
\ {\isadigit{0}}\isanewline
\ \ \ \ \isacommand{from}\isamarkupfalse%
\ this\ \isacommand{show}\isamarkupfalse%
\ {\isacharquery}case\ \isacommand{by}\isamarkupfalse%
\ {\isacharparenleft}simp\ add{\isacharcolon}\ fin{\isacharunderscore}make{\isacharunderscore}untimed{\isacharunderscore}def{\isacharparenright}\isanewline
\isacommand{next}\isamarkupfalse%
\isanewline
\ \ \ \ \isacommand{case}\isamarkupfalse%
\ {\isacharparenleft}Suc\ i{\isacharparenright}\isanewline
\ \ \ \ \isacommand{thus}\isamarkupfalse%
\ {\isacharquery}case\ \isanewline
\ \ \ \ \isacommand{by}\isamarkupfalse%
\ {\isacharparenleft}simp\ add{\isacharcolon}\ fin{\isacharunderscore}make{\isacharunderscore}untimed{\isacharunderscore}def{\isacharparenright}\isanewline
\isacommand{qed}\isamarkupfalse%
\endisatagproof
{\isafoldproof}%
\isadelimproof
\isanewline
\endisadelimproof
\isanewline
\isanewline
\isacommand{lemma}\isamarkupfalse%
\ fin{\isacharunderscore}make{\isacharunderscore}untimed{\isacharunderscore}append{\isacharunderscore}empty{\isacharcolon}\isanewline
{\isachardoublequoteopen}fin{\isacharunderscore}make{\isacharunderscore}untimed\ {\isacharparenleft}z\ {\isacharat}\ {\isacharbrackleft}{\isacharbrackleft}{\isacharbrackright}{\isacharbrackright}{\isacharparenright}\ {\isacharequal}\ fin{\isacharunderscore}make{\isacharunderscore}untimed\ z{\isachardoublequoteclose}\isanewline
\isadelimproof
\ \ %
\endisadelimproof
\isatagproof
\isacommand{by}\isamarkupfalse%
\ {\isacharparenleft}simp\ add{\isacharcolon}\ fin{\isacharunderscore}make{\isacharunderscore}untimed{\isacharunderscore}def{\isacharparenright}%
\endisatagproof
{\isafoldproof}%
\isadelimproof
\isanewline
\endisadelimproof
\isanewline
\isanewline
\isacommand{lemma}\isamarkupfalse%
\ fin{\isacharunderscore}make{\isacharunderscore}untimed{\isacharunderscore}inf{\isacharunderscore}truncate{\isacharunderscore}append{\isacharunderscore}a{\isacharcolon}\isanewline
{\isachardoublequoteopen}fin{\isacharunderscore}make{\isacharunderscore}untimed\ {\isacharparenleft}inf{\isacharunderscore}truncate\ z\ i\ {\isacharat}\ {\isacharbrackleft}{\isacharbrackleft}a{\isacharbrackright}{\isacharbrackright}{\isacharparenright}\ {\isacharbang}\ \isanewline
\ \ {\isacharparenleft}length\ {\isacharparenleft}fin{\isacharunderscore}make{\isacharunderscore}untimed\ {\isacharparenleft}inf{\isacharunderscore}truncate\ z\ i\ {\isacharat}\ {\isacharbrackleft}{\isacharbrackleft}a{\isacharbrackright}{\isacharbrackright}{\isacharparenright}{\isacharparenright}\ {\isacharminus}\ Suc\ {\isadigit{0}}{\isacharparenright}\ {\isacharequal}\ a{\isachardoublequoteclose}\isanewline
\isadelimproof
\ \ %
\endisadelimproof
\isatagproof
\isacommand{by}\isamarkupfalse%
\ {\isacharparenleft}simp\ add{\isacharcolon}\ fin{\isacharunderscore}make{\isacharunderscore}untimed{\isacharunderscore}def{\isacharparenright}%
\endisatagproof
{\isafoldproof}%
\isadelimproof
\isanewline
\endisadelimproof
\isanewline
\isanewline
\isacommand{lemma}\isamarkupfalse%
\ fin{\isacharunderscore}make{\isacharunderscore}untimed{\isacharunderscore}inf{\isacharunderscore}truncate{\isacharunderscore}Nonempty{\isacharunderscore}all{\isacharcolon}\isanewline
\ \ \isakeyword{assumes}\ h{\isadigit{1}}{\isacharcolon}{\isachardoublequoteopen}z\ k\ {\isasymnoteq}\ {\isacharbrackleft}{\isacharbrackright}{\isachardoublequoteclose}\ \isanewline
\ \ \isakeyword{shows}\ {\isachardoublequoteopen}{\isasymforall}\ i{\isachardot}\ k\ {\isasymle}\ i\ {\isasymlongrightarrow}\ fin{\isacharunderscore}make{\isacharunderscore}untimed\ {\isacharparenleft}inf{\isacharunderscore}truncate\ z\ i{\isacharparenright}\ {\isasymnoteq}\ {\isacharbrackleft}{\isacharbrackright}{\isachardoublequoteclose}\isanewline
\isadelimproof
\endisadelimproof
\isatagproof
\isacommand{using}\isamarkupfalse%
\ assms\ \isacommand{by}\isamarkupfalse%
\ {\isacharparenleft}simp\ add{\isacharcolon}\ \ fin{\isacharunderscore}make{\isacharunderscore}untimed{\isacharunderscore}inf{\isacharunderscore}truncate{\isacharunderscore}Nonempty{\isacharparenright}%
\endisatagproof
{\isafoldproof}%
\isadelimproof
\isanewline
\endisadelimproof
\isanewline
\isanewline
\isacommand{lemma}\isamarkupfalse%
\ fin{\isacharunderscore}make{\isacharunderscore}untimed{\isacharunderscore}inf{\isacharunderscore}truncate{\isacharunderscore}Nonempty{\isacharunderscore}all{\isadigit{0}}{\isacharcolon}\isanewline
\ \ \isakeyword{assumes}\ h{\isadigit{1}}{\isacharcolon}{\isachardoublequoteopen}z\ {\isadigit{0}}\ {\isasymnoteq}\ {\isacharbrackleft}{\isacharbrackright}{\isachardoublequoteclose}\isanewline
\ \ \isakeyword{shows}\ {\isachardoublequoteopen}{\isasymforall}\ i{\isachardot}\ fin{\isacharunderscore}make{\isacharunderscore}untimed\ {\isacharparenleft}inf{\isacharunderscore}truncate\ z\ i{\isacharparenright}\ {\isasymnoteq}\ {\isacharbrackleft}{\isacharbrackright}{\isachardoublequoteclose}\isanewline
\isadelimproof
\endisadelimproof
\isatagproof
\isacommand{using}\isamarkupfalse%
\ assms\ \isacommand{by}\isamarkupfalse%
\ {\isacharparenleft}simp\ add{\isacharcolon}\ fin{\isacharunderscore}make{\isacharunderscore}untimed{\isacharunderscore}inf{\isacharunderscore}truncate{\isacharunderscore}Nonempty{\isacharparenright}%
\endisatagproof
{\isafoldproof}%
\isadelimproof
\isanewline
\endisadelimproof
\isanewline
\isanewline
\isacommand{lemma}\isamarkupfalse%
\ fin{\isacharunderscore}make{\isacharunderscore}untimed{\isacharunderscore}inf{\isacharunderscore}truncate{\isacharunderscore}Nonempty{\isacharunderscore}all{\isadigit{0}}a{\isacharcolon}\isanewline
\ \ \isakeyword{assumes}\ h{\isadigit{1}}{\isacharcolon}{\isachardoublequoteopen}z\ {\isadigit{0}}\ {\isacharequal}\ {\isacharbrackleft}a{\isacharbrackright}{\isachardoublequoteclose}\isanewline
\ \ \isakeyword{shows}\ {\isachardoublequoteopen}{\isasymforall}\ i{\isachardot}\ fin{\isacharunderscore}make{\isacharunderscore}untimed\ {\isacharparenleft}inf{\isacharunderscore}truncate\ z\ i{\isacharparenright}\ {\isasymnoteq}\ {\isacharbrackleft}{\isacharbrackright}{\isachardoublequoteclose}\isanewline
\isadelimproof
\endisadelimproof
\isatagproof
\isacommand{using}\isamarkupfalse%
\ assms\ \isacommand{by}\isamarkupfalse%
\ {\isacharparenleft}simp\ add{\isacharcolon}\ fin{\isacharunderscore}make{\isacharunderscore}untimed{\isacharunderscore}inf{\isacharunderscore}truncate{\isacharunderscore}Nonempty{\isacharunderscore}all{\isadigit{0}}{\isacharparenright}%
\endisatagproof
{\isafoldproof}%
\isadelimproof
\isanewline
\endisadelimproof
\isanewline
\isanewline
\isacommand{lemma}\isamarkupfalse%
\ fin{\isacharunderscore}make{\isacharunderscore}untimed{\isacharunderscore}inf{\isacharunderscore}truncate{\isacharunderscore}Nonempty{\isacharunderscore}all{\isacharunderscore}app{\isacharcolon}\isanewline
\ \ \isakeyword{assumes}\ h{\isadigit{1}}{\isacharcolon}{\isachardoublequoteopen}z\ {\isadigit{0}}\ {\isacharequal}\ {\isacharbrackleft}a{\isacharbrackright}{\isachardoublequoteclose}\ \isanewline
\ \ \isakeyword{shows}\ {\isachardoublequoteopen}{\isasymforall}\ i{\isachardot}\ fin{\isacharunderscore}make{\isacharunderscore}untimed\ {\isacharparenleft}inf{\isacharunderscore}truncate\ z\ i\ {\isacharat}\ {\isacharbrackleft}z\ {\isacharparenleft}Suc\ i{\isacharparenright}{\isacharbrackright}{\isacharparenright}\ {\isasymnoteq}\ {\isacharbrackleft}{\isacharbrackright}{\isachardoublequoteclose}\isanewline
\isadelimproof
\endisadelimproof
\isatagproof
\isacommand{proof}\isamarkupfalse%
\ \isanewline
\ \ \isacommand{fix}\isamarkupfalse%
\ i\isanewline
\ \ \isacommand{from}\isamarkupfalse%
\ h{\isadigit{1}}\ \isacommand{have}\isamarkupfalse%
\ sg{\isadigit{1}}{\isacharcolon}{\isachardoublequoteopen}fin{\isacharunderscore}make{\isacharunderscore}untimed\ {\isacharparenleft}inf{\isacharunderscore}truncate\ z\ i{\isacharparenright}\ {\isasymnoteq}\ {\isacharbrackleft}{\isacharbrackright}{\isachardoublequoteclose}\isanewline
\ \ \ \ \isacommand{by}\isamarkupfalse%
\ {\isacharparenleft}simp\ add{\isacharcolon}\ fin{\isacharunderscore}make{\isacharunderscore}untimed{\isacharunderscore}inf{\isacharunderscore}truncate{\isacharunderscore}Nonempty{\isacharunderscore}all{\isadigit{0}}a{\isacharparenright}\isanewline
\ \ \isacommand{from}\isamarkupfalse%
\ h{\isadigit{1}}\ \isakeyword{and}\ sg{\isadigit{1}}\ \isacommand{show}\isamarkupfalse%
\ {\isachardoublequoteopen}fin{\isacharunderscore}make{\isacharunderscore}untimed\ {\isacharparenleft}inf{\isacharunderscore}truncate\ z\ i\ {\isacharat}\ {\isacharbrackleft}z\ {\isacharparenleft}Suc\ i{\isacharparenright}{\isacharbrackright}{\isacharparenright}\ {\isasymnoteq}\ {\isacharbrackleft}{\isacharbrackright}{\isachardoublequoteclose}\isanewline
\ \ \ \ \isacommand{by}\isamarkupfalse%
\ {\isacharparenleft}simp\ add{\isacharcolon}\ fin{\isacharunderscore}make{\isacharunderscore}untimed{\isacharunderscore}append{\isacharparenright}\isanewline
\isacommand{qed}\isamarkupfalse%
\endisatagproof
{\isafoldproof}%
\isadelimproof
\isanewline
\endisadelimproof
\isanewline
\isanewline
\isacommand{lemma}\isamarkupfalse%
\ fin{\isacharunderscore}make{\isacharunderscore}untimed{\isacharunderscore}nth{\isacharunderscore}length{\isacharcolon}\isanewline
\ \ \isakeyword{assumes}\ h{\isadigit{1}}{\isacharcolon}{\isachardoublequoteopen}z\ i\ {\isacharequal}\ {\isacharbrackleft}a{\isacharbrackright}{\isachardoublequoteclose}\isanewline
\ \ \isakeyword{shows}\ \isanewline
\ \ {\isachardoublequoteopen}fin{\isacharunderscore}make{\isacharunderscore}untimed\ {\isacharparenleft}inf{\isacharunderscore}truncate\ z\ i{\isacharparenright}\ {\isacharbang}\ \isanewline
\ \ \ \ \ {\isacharparenleft}length\ {\isacharparenleft}fin{\isacharunderscore}make{\isacharunderscore}untimed\ {\isacharparenleft}inf{\isacharunderscore}truncate\ z\ i{\isacharparenright}{\isacharparenright}\ {\isacharminus}\ Suc\ {\isadigit{0}}{\isacharparenright}\isanewline
\ \ \ \ {\isacharequal}\ a{\isachardoublequoteclose}\isanewline
\isadelimproof
\endisadelimproof
\isatagproof
\isacommand{proof}\isamarkupfalse%
\ {\isacharminus}\ \isanewline
\isacommand{from}\isamarkupfalse%
\ h{\isadigit{1}}\ \isacommand{have}\isamarkupfalse%
\ sg{\isadigit{1}}{\isacharcolon}{\isachardoublequoteopen}last\ {\isacharparenleft}fin{\isacharunderscore}make{\isacharunderscore}untimed\ {\isacharparenleft}inf{\isacharunderscore}truncate\ z\ i{\isacharparenright}{\isacharparenright}\ {\isacharequal}\ a{\isachardoublequoteclose}\isanewline
\ \ \isacommand{by}\isamarkupfalse%
\ {\isacharparenleft}simp\ add{\isacharcolon}\ last{\isacharunderscore}fin{\isacharunderscore}make{\isacharunderscore}untimed{\isacharunderscore}inf{\isacharunderscore}truncate{\isacharparenright}\isanewline
\isacommand{from}\isamarkupfalse%
\ h{\isadigit{1}}\ \isacommand{have}\isamarkupfalse%
\ sg{\isadigit{2}}{\isacharcolon}{\isachardoublequoteopen}concat\ {\isacharparenleft}inf{\isacharunderscore}truncate\ z\ i{\isacharparenright}\ {\isasymnoteq}\ {\isacharbrackleft}{\isacharbrackright}{\isachardoublequoteclose}\isanewline
\ \ \isacommand{by}\isamarkupfalse%
\ {\isacharparenleft}rule\ concat{\isacharunderscore}inf{\isacharunderscore}truncate{\isacharunderscore}nonempty{\isacharunderscore}a{\isacharparenright}\isanewline
\isacommand{from}\isamarkupfalse%
\ h{\isadigit{1}}\ \isakeyword{and}\ sg{\isadigit{1}}\ \isakeyword{and}\ sg{\isadigit{2}}\ \isacommand{show}\isamarkupfalse%
\ {\isacharquery}thesis\ \isanewline
\ \ \isacommand{by}\isamarkupfalse%
\ {\isacharparenleft}simp\ add{\isacharcolon}\ fin{\isacharunderscore}make{\isacharunderscore}untimed{\isacharunderscore}def\ last{\isacharunderscore}nth{\isacharunderscore}length{\isacharparenright}\isanewline
\isacommand{qed}\isamarkupfalse%
\endisatagproof
{\isafoldproof}%
\isadelimproof
\endisadelimproof
\isamarkupsubsubsection{Lemmas for $inf\_disj$ and $inf\_disjS$%
}
\isamarkuptrue%
\isacommand{lemma}\isamarkupfalse%
\ inf{\isacharunderscore}disj{\isacharunderscore}index{\isacharcolon}\isanewline
\ \ \isakeyword{assumes}\ h{\isadigit{1}}{\isacharcolon}{\isachardoublequoteopen}inf{\isacharunderscore}disj\ n\ nS{\isachardoublequoteclose}\isanewline
\ \ \ \ \ \ \isakeyword{and}\ h{\isadigit{2}}{\isacharcolon}{\isachardoublequoteopen}nS\ k\ t\ {\isasymnoteq}\ {\isacharbrackleft}{\isacharbrackright}{\isachardoublequoteclose}\isanewline
\ \ \ \ \ \ \isakeyword{and}\ h{\isadigit{3}}{\isacharcolon}{\isachardoublequoteopen}k\ {\isacharless}\ n{\isachardoublequoteclose}\isanewline
\ \ \isakeyword{shows}\ {\isachardoublequoteopen}{\isacharparenleft}SOME\ i{\isachardot}\ i\ {\isacharless}\ n\ {\isasymand}\ \ nS\ i\ t\ {\isasymnoteq}\ {\isacharbrackleft}{\isacharbrackright}{\isacharparenright}\ {\isacharequal}\ k{\isachardoublequoteclose}\isanewline
\isadelimproof
\endisadelimproof
\isatagproof
\isacommand{proof}\isamarkupfalse%
\ {\isacharminus}\ \isanewline
\ \ \isacommand{from}\isamarkupfalse%
\ h{\isadigit{1}}\ \isacommand{have}\isamarkupfalse%
\ {\isachardoublequoteopen}{\isasymforall}\ j{\isachardot}\ k\ {\isacharless}\ n\ {\isasymand}\ j\ {\isacharless}\ n\ {\isasymand}\ k\ {\isasymnoteq}\ j\ {\isasymand}\ nS\ k\ t\ {\isasymnoteq}\ {\isacharbrackleft}{\isacharbrackright}\ {\isasymlongrightarrow}\ nS\ j\ t\ {\isacharequal}\ {\isacharbrackleft}{\isacharbrackright}{\isachardoublequoteclose}\isanewline
\ \ \ \ \isacommand{by}\isamarkupfalse%
\ {\isacharparenleft}simp\ add{\isacharcolon}\ inf{\isacharunderscore}disj{\isacharunderscore}def{\isacharcomma}\ auto{\isacharparenright}\isanewline
\ \ \isacommand{from}\isamarkupfalse%
\ this\ \isakeyword{and}\ assms\ \isacommand{show}\isamarkupfalse%
\ {\isacharquery}thesis\ \isacommand{by}\isamarkupfalse%
\ auto\isanewline
\isacommand{qed}\isamarkupfalse%
\endisatagproof
{\isafoldproof}%
\isadelimproof
\isanewline
\endisadelimproof
\ \isanewline
\isanewline
\isacommand{lemma}\isamarkupfalse%
\ inf{\isacharunderscore}disjS{\isacharunderscore}index{\isacharcolon}\isanewline
\ \ \isakeyword{assumes}\ h{\isadigit{1}}{\isacharcolon}{\isachardoublequoteopen}inf{\isacharunderscore}disjS\ IdSet\ nS{\isachardoublequoteclose}\isanewline
\ \ \ \ \ \ \isakeyword{and}\ h{\isadigit{2}}{\isacharcolon}{\isachardoublequoteopen}k{\isacharcolon}IdSet{\isachardoublequoteclose}\isanewline
\ \ \ \ \ \ \isakeyword{and}\ h{\isadigit{3}}{\isacharcolon}{\isachardoublequoteopen}nS\ k\ t\ {\isasymnoteq}\ {\isacharbrackleft}{\isacharbrackright}{\isachardoublequoteclose}\isanewline
\ \ \isakeyword{shows}\ {\isachardoublequoteopen}{\isacharparenleft}SOME\ i{\isachardot}\ {\isacharparenleft}i{\isacharcolon}IdSet{\isacharparenright}\ {\isasymand}\ nSend\ i\ t\ {\isasymnoteq}\ {\isacharbrackleft}{\isacharbrackright}{\isacharparenright}\ {\isacharequal}\ k{\isachardoublequoteclose}\isanewline
\isadelimproof
\endisadelimproof
\isatagproof
\isacommand{proof}\isamarkupfalse%
\ {\isacharminus}\isanewline
\ \ \isacommand{from}\isamarkupfalse%
\ h{\isadigit{1}}\ \isacommand{have}\isamarkupfalse%
\ {\isachardoublequoteopen}{\isasymforall}\ j{\isachardot}\ k\ {\isasymin}\ IdSet\ {\isasymand}\ j\ {\isasymin}\ IdSet\ {\isasymand}\ nS\ k\ t\ {\isasymnoteq}\ {\isacharbrackleft}{\isacharbrackright}\ {\isasymlongrightarrow}\ nS\ j\ t\ {\isacharequal}\ {\isacharbrackleft}{\isacharbrackright}{\isachardoublequoteclose}\isanewline
\ \ \ \ \isacommand{by}\isamarkupfalse%
\ {\isacharparenleft}simp\ add{\isacharcolon}\ inf{\isacharunderscore}disjS{\isacharunderscore}def{\isacharcomma}\ auto{\isacharparenright}\isanewline
\ \ \isacommand{from}\isamarkupfalse%
\ this\ \isakeyword{and}\ assms\ \isacommand{show}\isamarkupfalse%
\ {\isacharquery}thesis\ \isacommand{by}\isamarkupfalse%
\ auto\isanewline
\isacommand{qed}\isamarkupfalse%
\endisatagproof
{\isafoldproof}%
\isadelimproof
\isanewline
\endisadelimproof
\isanewline
\isadelimtheory
\isanewline
\endisadelimtheory
\isatagtheory
\isacommand{end}\isamarkupfalse%
\endisatagtheory
{\isafoldtheory}%
\isadelimtheory
\endisadelimtheory
\end{isabellebody}%

%
\begin{isabellebody}%
\def\isabellecontext{BitBoolTS}%
\isamarkupheader{Properties of time-synchronous streams of types bool and bit%
}
\isamarkuptrue%
\isadelimtheory
\endisadelimtheory
\isatagtheory
\isacommand{theory}\isamarkupfalse%
\ BitBoolTS\ \isanewline
\isakeyword{imports}\ Main\ stream\isanewline
\isakeyword{begin}%
\endisatagtheory
{\isafoldtheory}%
\isadelimtheory
\endisadelimtheory
\isanewline
\isanewline
\isacommand{datatype}\isamarkupfalse%
\ bit\ {\isacharequal}\ Zero\ {\isacharbar}\ One\isanewline
\isanewline
\isacommand{primrec}\isamarkupfalse%
\ \isanewline
\ \ \ negation\ {\isacharcolon}{\isacharcolon}\ {\isachardoublequoteopen}bit\ {\isasymRightarrow}\ bit{\isachardoublequoteclose}\isanewline
\isakeyword{where}\isanewline
\ \ {\isachardoublequoteopen}negation\ Zero\ {\isacharequal}\ One{\isachardoublequoteclose}\ {\isacharbar}\isanewline
\ \ {\isachardoublequoteopen}negation\ One\ {\isacharequal}\ Zero{\isachardoublequoteclose}\isanewline
\isanewline
\isanewline
\isacommand{lemma}\isamarkupfalse%
\ ts{\isacharunderscore}bit{\isacharunderscore}stream{\isacharunderscore}One{\isacharcolon}\isanewline
\ \ \isakeyword{assumes}\ h{\isadigit{1}}{\isacharcolon}{\isachardoublequoteopen}ts\ x{\isachardoublequoteclose}\isanewline
\ \ \ \ \ \ \isakeyword{and}\ h{\isadigit{2}}{\isacharcolon}{\isachardoublequoteopen}x\ i\ {\isasymnoteq}\ {\isacharbrackleft}Zero{\isacharbrackright}{\isachardoublequoteclose}\isanewline
\ \ \isakeyword{shows}\ {\isachardoublequoteopen}x\ i\ {\isacharequal}\ {\isacharbrackleft}One{\isacharbrackright}{\isachardoublequoteclose}\isanewline
\isadelimproof
\endisadelimproof
\isatagproof
\isacommand{proof}\isamarkupfalse%
\ {\isacharminus}\isanewline
\ \ \isacommand{from}\isamarkupfalse%
\ h{\isadigit{1}}\ \isacommand{have}\isamarkupfalse%
\ sg{\isadigit{1}}{\isacharcolon}{\isachardoublequoteopen}length\ {\isacharparenleft}x\ i{\isacharparenright}\ {\isacharequal}\ Suc\ {\isadigit{0}}{\isachardoublequoteclose}\ \isanewline
\ \ \ \ \isacommand{by}\isamarkupfalse%
\ {\isacharparenleft}simp\ add{\isacharcolon}\ ts{\isacharunderscore}def{\isacharparenright}\isanewline
\ \ \isacommand{from}\isamarkupfalse%
\ this\ \isakeyword{and}\ h{\isadigit{2}}\ \isacommand{show}\isamarkupfalse%
\ {\isacharquery}thesis\isanewline
\ \ \isacommand{proof}\isamarkupfalse%
\ {\isacharparenleft}cases\ {\isachardoublequoteopen}x\ i{\isachardoublequoteclose}{\isacharparenright}\isanewline
\ \ \ \ \isacommand{assume}\isamarkupfalse%
\ Nil{\isacharcolon}{\isachardoublequoteopen}x\ i\ {\isacharequal}\ {\isacharbrackleft}{\isacharbrackright}{\isachardoublequoteclose}\ \isanewline
\ \ \ \ \isacommand{from}\isamarkupfalse%
\ this\ \isakeyword{and}\ sg{\isadigit{1}}\ \isacommand{show}\isamarkupfalse%
\ {\isacharquery}thesis\ \isacommand{by}\isamarkupfalse%
\ simp\isanewline
\ \ \isacommand{next}\isamarkupfalse%
\isanewline
\ \ \isacommand{fix}\isamarkupfalse%
\ a\ l\ \isacommand{assume}\isamarkupfalse%
\ Cons{\isacharcolon}{\isachardoublequoteopen}x\ i\ {\isacharequal}\ a\ {\isacharhash}\ l{\isachardoublequoteclose}\isanewline
\ \ \ \ \isacommand{from}\isamarkupfalse%
\ this\ \isakeyword{and}\ sg{\isadigit{1}}\ \isakeyword{and}\ h{\isadigit{2}}\ \isacommand{show}\isamarkupfalse%
\ {\isacharquery}thesis\isanewline
\ \ \ \ \isacommand{proof}\isamarkupfalse%
\ {\isacharparenleft}cases\ {\isachardoublequoteopen}a{\isachardoublequoteclose}{\isacharparenright}\isanewline
\ \ \ \ \ \ \isacommand{assume}\isamarkupfalse%
\ {\isachardoublequoteopen}a\ {\isacharequal}\ Zero{\isachardoublequoteclose}\isanewline
\ \ \ \ \ \ \isacommand{from}\isamarkupfalse%
\ this\ \isakeyword{and}\ sg{\isadigit{1}}\ \isakeyword{and}\ h{\isadigit{2}}\ \isakeyword{and}\ Cons\ \isacommand{show}\isamarkupfalse%
\ {\isacharquery}thesis\ \isacommand{by}\isamarkupfalse%
\ auto\isanewline
\ \ \ \ \isacommand{next}\isamarkupfalse%
\ \isanewline
\ \ \ \ \ \ \isacommand{assume}\isamarkupfalse%
\ {\isachardoublequoteopen}a\ {\isacharequal}\ One{\isachardoublequoteclose}\isanewline
\ \ \ \ \ \ \isacommand{from}\isamarkupfalse%
\ this\ \isakeyword{and}\ sg{\isadigit{1}}\ \isakeyword{and}\ Cons\ \isacommand{show}\isamarkupfalse%
\ {\isacharquery}thesis\ \isacommand{by}\isamarkupfalse%
\ auto\isanewline
\ \ \ \ \isacommand{qed}\isamarkupfalse%
\isanewline
\ \ \isacommand{qed}\isamarkupfalse%
\isanewline
\isacommand{qed}\isamarkupfalse%
\endisatagproof
{\isafoldproof}%
\isadelimproof
\isanewline
\endisadelimproof
\isanewline
\isanewline
\isacommand{lemma}\isamarkupfalse%
\ ts{\isacharunderscore}bit{\isacharunderscore}stream{\isacharunderscore}Zero{\isacharcolon}\isanewline
\ \ \isakeyword{assumes}\ h{\isadigit{1}}{\isacharcolon}{\isachardoublequoteopen}ts\ x{\isachardoublequoteclose}\isanewline
\ \ \ \ \ \ \isakeyword{and}\ h{\isadigit{2}}{\isacharcolon}{\isachardoublequoteopen}x\ i\ {\isasymnoteq}\ {\isacharbrackleft}One{\isacharbrackright}{\isachardoublequoteclose}\isanewline
\ \ \isakeyword{shows}\ {\isachardoublequoteopen}x\ i\ {\isacharequal}\ {\isacharbrackleft}Zero{\isacharbrackright}{\isachardoublequoteclose}\isanewline
\isadelimproof
\endisadelimproof
\isatagproof
\isacommand{proof}\isamarkupfalse%
\ {\isacharminus}\isanewline
\ \ \isacommand{from}\isamarkupfalse%
\ h{\isadigit{1}}\ \isacommand{have}\isamarkupfalse%
\ sg{\isadigit{1}}{\isacharcolon}{\isachardoublequoteopen}length\ {\isacharparenleft}x\ i{\isacharparenright}\ {\isacharequal}\ Suc\ {\isadigit{0}}{\isachardoublequoteclose}\ \isanewline
\ \ \ \ \isacommand{by}\isamarkupfalse%
\ {\isacharparenleft}simp\ add{\isacharcolon}\ ts{\isacharunderscore}def{\isacharparenright}\isanewline
\ \ \isacommand{from}\isamarkupfalse%
\ this\ \isakeyword{and}\ h{\isadigit{2}}\ \isacommand{show}\isamarkupfalse%
\ {\isacharquery}thesis\isanewline
\ \ \isacommand{proof}\isamarkupfalse%
\ {\isacharparenleft}cases\ {\isachardoublequoteopen}x\ i{\isachardoublequoteclose}{\isacharparenright}\isanewline
\ \ \ \ \isacommand{assume}\isamarkupfalse%
\ Nil{\isacharcolon}{\isachardoublequoteopen}x\ i\ {\isacharequal}\ {\isacharbrackleft}{\isacharbrackright}{\isachardoublequoteclose}\ \isanewline
\ \ \ \ \isacommand{from}\isamarkupfalse%
\ this\ \isakeyword{and}\ sg{\isadigit{1}}\ \isacommand{show}\isamarkupfalse%
\ {\isacharquery}thesis\ \isacommand{by}\isamarkupfalse%
\ simp\isanewline
\ \ \isacommand{next}\isamarkupfalse%
\isanewline
\ \ \isacommand{fix}\isamarkupfalse%
\ a\ l\ \isacommand{assume}\isamarkupfalse%
\ Cons{\isacharcolon}{\isachardoublequoteopen}x\ i\ {\isacharequal}\ a\ {\isacharhash}\ l{\isachardoublequoteclose}\isanewline
\ \ \ \ \isacommand{from}\isamarkupfalse%
\ this\ \isakeyword{and}\ sg{\isadigit{1}}\ \isakeyword{and}\ h{\isadigit{2}}\ \isacommand{show}\isamarkupfalse%
\ {\isacharquery}thesis\isanewline
\ \ \ \ \isacommand{proof}\isamarkupfalse%
\ {\isacharparenleft}cases\ {\isachardoublequoteopen}a{\isachardoublequoteclose}{\isacharparenright}\isanewline
\ \ \ \ \ \ \isacommand{assume}\isamarkupfalse%
\ {\isachardoublequoteopen}a\ {\isacharequal}\ Zero{\isachardoublequoteclose}\isanewline
\ \ \ \ \ \ \isacommand{from}\isamarkupfalse%
\ this\ \isakeyword{and}\ sg{\isadigit{1}}\ \isakeyword{and}\ Cons\ \isacommand{show}\isamarkupfalse%
\ {\isacharquery}thesis\ \isacommand{by}\isamarkupfalse%
\ auto\isanewline
\ \ \ \ \isacommand{next}\isamarkupfalse%
\ \isanewline
\ \ \ \ \ \ \isacommand{assume}\isamarkupfalse%
\ {\isachardoublequoteopen}a\ {\isacharequal}\ One{\isachardoublequoteclose}\isanewline
\ \ \ \ \ \ \isacommand{from}\isamarkupfalse%
\ this\ \isakeyword{and}\ sg{\isadigit{1}}\ \isakeyword{and}\ h{\isadigit{2}}\ \isakeyword{and}\ Cons\ \isacommand{show}\isamarkupfalse%
\ {\isacharquery}thesis\ \isacommand{by}\isamarkupfalse%
\ auto\isanewline
\ \ \ \ \isacommand{qed}\isamarkupfalse%
\isanewline
\ \ \isacommand{qed}\isamarkupfalse%
\isanewline
\isacommand{qed}\isamarkupfalse%
\endisatagproof
{\isafoldproof}%
\isadelimproof
\isanewline
\endisadelimproof
\isanewline
\isanewline
\isacommand{lemma}\isamarkupfalse%
\ ts{\isacharunderscore}bool{\isacharunderscore}True{\isacharcolon}\isanewline
\ \ \isakeyword{assumes}\ h{\isadigit{1}}{\isacharcolon}{\isachardoublequoteopen}ts\ x{\isachardoublequoteclose}\isanewline
\ \ \ \ \ \ \isakeyword{and}\ h{\isadigit{2}}{\isacharcolon}{\isachardoublequoteopen}x\ i\ {\isasymnoteq}\ {\isacharbrackleft}False{\isacharbrackright}{\isachardoublequoteclose}\isanewline
\ \ \isakeyword{shows}\ {\isachardoublequoteopen}x\ i\ {\isacharequal}\ {\isacharbrackleft}True{\isacharbrackright}{\isachardoublequoteclose}\isanewline
\isadelimproof
\endisadelimproof
\isatagproof
\isacommand{proof}\isamarkupfalse%
\ {\isacharminus}\isanewline
\ \ \isacommand{from}\isamarkupfalse%
\ h{\isadigit{1}}\ \isacommand{have}\isamarkupfalse%
\ sg{\isadigit{1}}{\isacharcolon}{\isachardoublequoteopen}length\ {\isacharparenleft}x\ i{\isacharparenright}\ {\isacharequal}\ Suc\ {\isadigit{0}}{\isachardoublequoteclose}\ \isanewline
\ \ \ \ \isacommand{by}\isamarkupfalse%
\ {\isacharparenleft}simp\ add{\isacharcolon}\ ts{\isacharunderscore}def{\isacharparenright}\isanewline
\ \ \isacommand{from}\isamarkupfalse%
\ this\ \isakeyword{and}\ h{\isadigit{2}}\ \isacommand{show}\isamarkupfalse%
\ {\isacharquery}thesis\isanewline
\ \ \isacommand{proof}\isamarkupfalse%
\ {\isacharparenleft}cases\ {\isachardoublequoteopen}x\ i{\isachardoublequoteclose}{\isacharparenright}\isanewline
\ \ \ \ \isacommand{assume}\isamarkupfalse%
\ Nil{\isacharcolon}{\isachardoublequoteopen}x\ i\ {\isacharequal}\ {\isacharbrackleft}{\isacharbrackright}{\isachardoublequoteclose}\ \isanewline
\ \ \ \ \isacommand{from}\isamarkupfalse%
\ this\ \isakeyword{and}\ sg{\isadigit{1}}\ \isacommand{show}\isamarkupfalse%
\ {\isacharquery}thesis\ \isacommand{by}\isamarkupfalse%
\ simp\isanewline
\ \ \isacommand{next}\isamarkupfalse%
\isanewline
\ \ \isacommand{fix}\isamarkupfalse%
\ a\ l\ \isacommand{assume}\isamarkupfalse%
\ Cons{\isacharcolon}{\isachardoublequoteopen}x\ i\ {\isacharequal}\ a\ {\isacharhash}\ l{\isachardoublequoteclose}\isanewline
\ \ \ \ \isacommand{from}\isamarkupfalse%
\ this\ \isakeyword{and}\ sg{\isadigit{1}}\ \isacommand{have}\isamarkupfalse%
\ sg{\isadigit{2}}{\isacharcolon}{\isachardoublequoteopen}x\ i\ {\isacharequal}\ {\isacharbrackleft}a{\isacharbrackright}{\isachardoublequoteclose}\ \isacommand{by}\isamarkupfalse%
\ simp\isanewline
\ \ \ \ \isacommand{from}\isamarkupfalse%
\ this\ \isakeyword{and}\ h{\isadigit{2}}\ \isacommand{show}\isamarkupfalse%
\ {\isacharquery}thesis\ \isacommand{by}\isamarkupfalse%
\ auto\isanewline
\ \ \isacommand{qed}\isamarkupfalse%
\isanewline
\isacommand{qed}\isamarkupfalse%
\endisatagproof
{\isafoldproof}%
\isadelimproof
\isanewline
\endisadelimproof
\isanewline
\isanewline
\isacommand{lemma}\isamarkupfalse%
\ ts{\isacharunderscore}bool{\isacharunderscore}False{\isacharcolon}\isanewline
\ \ \isakeyword{assumes}\ h{\isadigit{1}}{\isacharcolon}{\isachardoublequoteopen}ts\ x{\isachardoublequoteclose}\isanewline
\ \ \ \ \ \ \isakeyword{and}\ h{\isadigit{2}}{\isacharcolon}{\isachardoublequoteopen}x\ i\ {\isasymnoteq}\ {\isacharbrackleft}True{\isacharbrackright}{\isachardoublequoteclose}\isanewline
\ \ \isakeyword{shows}\ {\isachardoublequoteopen}x\ i\ {\isacharequal}\ {\isacharbrackleft}False{\isacharbrackright}{\isachardoublequoteclose}\isanewline
\isadelimproof
\endisadelimproof
\isatagproof
\isacommand{proof}\isamarkupfalse%
\ {\isacharminus}\isanewline
\ \ \isacommand{from}\isamarkupfalse%
\ h{\isadigit{1}}\ \isacommand{have}\isamarkupfalse%
\ sg{\isadigit{1}}{\isacharcolon}{\isachardoublequoteopen}length\ {\isacharparenleft}x\ i{\isacharparenright}\ {\isacharequal}\ Suc\ {\isadigit{0}}{\isachardoublequoteclose}\ \isanewline
\ \ \ \ \isacommand{by}\isamarkupfalse%
\ {\isacharparenleft}simp\ add{\isacharcolon}\ ts{\isacharunderscore}def{\isacharparenright}\isanewline
\ \ \isacommand{from}\isamarkupfalse%
\ this\ \isakeyword{and}\ h{\isadigit{2}}\ \isacommand{show}\isamarkupfalse%
\ {\isacharquery}thesis\isanewline
\ \ \isacommand{proof}\isamarkupfalse%
\ {\isacharparenleft}cases\ {\isachardoublequoteopen}x\ i{\isachardoublequoteclose}{\isacharparenright}\isanewline
\ \ \ \ \isacommand{assume}\isamarkupfalse%
\ Nil{\isacharcolon}{\isachardoublequoteopen}x\ i\ {\isacharequal}\ {\isacharbrackleft}{\isacharbrackright}{\isachardoublequoteclose}\ \isanewline
\ \ \ \ \isacommand{from}\isamarkupfalse%
\ this\ \isakeyword{and}\ sg{\isadigit{1}}\ \isacommand{show}\isamarkupfalse%
\ {\isacharquery}thesis\ \isacommand{by}\isamarkupfalse%
\ simp\isanewline
\ \ \isacommand{next}\isamarkupfalse%
\isanewline
\ \ \isacommand{fix}\isamarkupfalse%
\ a\ l\ \isacommand{assume}\isamarkupfalse%
\ Cons{\isacharcolon}{\isachardoublequoteopen}x\ i\ {\isacharequal}\ a\ {\isacharhash}\ l{\isachardoublequoteclose}\isanewline
\ \ \ \ \isacommand{from}\isamarkupfalse%
\ this\ \isakeyword{and}\ sg{\isadigit{1}}\ \isacommand{have}\isamarkupfalse%
\ sg{\isadigit{2}}{\isacharcolon}{\isachardoublequoteopen}x\ i\ {\isacharequal}\ {\isacharbrackleft}a{\isacharbrackright}{\isachardoublequoteclose}\ \isacommand{by}\isamarkupfalse%
\ simp\isanewline
\ \ \ \ \isacommand{from}\isamarkupfalse%
\ this\ \isakeyword{and}\ h{\isadigit{2}}\ \isacommand{show}\isamarkupfalse%
\ {\isacharquery}thesis\ \isacommand{by}\isamarkupfalse%
\ auto\isanewline
\ \ \isacommand{qed}\isamarkupfalse%
\isanewline
\isacommand{qed}\isamarkupfalse%
\endisatagproof
{\isafoldproof}%
\isadelimproof
\isanewline
\endisadelimproof
\isanewline
\isanewline
\isacommand{lemma}\isamarkupfalse%
\ ts{\isacharunderscore}bool{\isacharunderscore}True{\isacharunderscore}False{\isacharcolon}\isanewline
\ \ \isakeyword{fixes}\ x{\isacharcolon}{\isacharcolon}{\isachardoublequoteopen}bool\ istream{\isachardoublequoteclose}\isanewline
\ \ \isakeyword{assumes}\ h{\isadigit{1}}{\isacharcolon}{\isachardoublequoteopen}ts\ x{\isachardoublequoteclose}\ \isanewline
\ \ \isakeyword{shows}\ {\isachardoublequoteopen}x\ i\ {\isacharequal}\ {\isacharbrackleft}True{\isacharbrackright}\ {\isasymor}\ x\ i\ {\isacharequal}\ {\isacharbrackleft}False{\isacharbrackright}{\isachardoublequoteclose}\isanewline
\isadelimproof
\endisadelimproof
\isatagproof
\isacommand{proof}\isamarkupfalse%
\ {\isacharparenleft}cases\ {\isachardoublequoteopen}x\ i\ {\isacharequal}\ {\isacharbrackleft}True{\isacharbrackright}{\isachardoublequoteclose}{\isacharparenright}\isanewline
\ \ \isacommand{assume}\isamarkupfalse%
\ {\isachardoublequoteopen}x\ i\ {\isacharequal}\ {\isacharbrackleft}True{\isacharbrackright}{\isachardoublequoteclose}\isanewline
\ \ \isacommand{from}\isamarkupfalse%
\ this\ \isakeyword{and}\ h{\isadigit{1}}\ \isacommand{show}\isamarkupfalse%
\ {\isacharquery}thesis\ \isacommand{by}\isamarkupfalse%
\ simp\isanewline
\isacommand{next}\isamarkupfalse%
\isanewline
\ \ \isacommand{assume}\isamarkupfalse%
\ {\isachardoublequoteopen}x\ i\ {\isasymnoteq}\ {\isacharbrackleft}True{\isacharbrackright}{\isachardoublequoteclose}\isanewline
\ \ \isacommand{from}\isamarkupfalse%
\ this\ \isakeyword{and}\ h{\isadigit{1}}\ \isacommand{show}\isamarkupfalse%
\ {\isacharquery}thesis\ \isacommand{by}\isamarkupfalse%
\ {\isacharparenleft}simp\ add{\isacharcolon}\ ts{\isacharunderscore}bool{\isacharunderscore}False{\isacharparenright}\isanewline
\isacommand{qed}\isamarkupfalse%
\endisatagproof
{\isafoldproof}%
\isadelimproof
\isanewline
\endisadelimproof
\isadelimtheory
\isanewline
\endisadelimtheory
\isatagtheory
\isacommand{end}\isamarkupfalse%
\endisatagtheory
{\isafoldtheory}%
\isadelimtheory
\endisadelimtheory
\ \end{isabellebody}%

%
\begin{isabellebody}%
\def\isabellecontext{JoinSplitTime}%
\isamarkupheader{Changing time granularity of the streams%
}
\isamarkuptrue%
\isadelimtheory
\endisadelimtheory
\isatagtheory
\isacommand{theory}\isamarkupfalse%
\ JoinSplitTime\isanewline
\isakeyword{imports}\ stream\ arith{\isacharunderscore}hints\isanewline
\isakeyword{begin}%
\endisatagtheory
{\isafoldtheory}%
\isadelimtheory
\endisadelimtheory
\isamarkupsubsection{Join time units%
}
\isamarkuptrue%
\isacommand{primrec}\isamarkupfalse%
\isanewline
\ \ join{\isacharunderscore}ti\ {\isacharcolon}{\isacharcolon}{\isachardoublequoteopen}{\isacharprime}a\ istream\ {\isasymRightarrow}\ nat\ {\isasymRightarrow}\ nat\ {\isasymRightarrow}\ {\isacharprime}a\ list{\isachardoublequoteclose}\isanewline
\isakeyword{where}\isanewline
join{\isacharunderscore}ti{\isacharunderscore}{\isadigit{0}}{\isacharcolon}\isanewline
\ {\isachardoublequoteopen}join{\isacharunderscore}ti\ s\ x\ {\isadigit{0}}\ {\isacharequal}\ s\ x{\isachardoublequoteclose}\ {\isacharbar}\isanewline
join{\isacharunderscore}ti{\isacharunderscore}Suc{\isacharcolon}\isanewline
\ {\isachardoublequoteopen}join{\isacharunderscore}ti\ s\ x\ {\isacharparenleft}Suc\ i{\isacharparenright}\ {\isacharequal}\ {\isacharparenleft}join{\isacharunderscore}ti\ s\ x\ i{\isacharparenright}\ {\isacharat}\ {\isacharparenleft}s\ {\isacharparenleft}x\ {\isacharplus}\ {\isacharparenleft}Suc\ i{\isacharparenright}{\isacharparenright}{\isacharparenright}{\isachardoublequoteclose}\isanewline
\isanewline
\isanewline
\isacommand{primrec}\isamarkupfalse%
\isanewline
\ \ fin{\isacharunderscore}join{\isacharunderscore}ti\ {\isacharcolon}{\isacharcolon}{\isachardoublequoteopen}{\isacharprime}a\ fstream\ {\isasymRightarrow}\ nat\ {\isasymRightarrow}\ nat\ {\isasymRightarrow}\ {\isacharprime}a\ list{\isachardoublequoteclose}\isanewline
\isakeyword{where}\isanewline
fin{\isacharunderscore}join{\isacharunderscore}ti{\isacharunderscore}{\isadigit{0}}{\isacharcolon}\isanewline
\ {\isachardoublequoteopen}fin{\isacharunderscore}join{\isacharunderscore}ti\ s\ x\ {\isadigit{0}}\ {\isacharequal}\ nth\ s\ x{\isachardoublequoteclose}\ {\isacharbar}\isanewline
fin{\isacharunderscore}join{\isacharunderscore}ti{\isacharunderscore}Suc{\isacharcolon}\isanewline
\ {\isachardoublequoteopen}fin{\isacharunderscore}join{\isacharunderscore}ti\ s\ x\ {\isacharparenleft}Suc\ i{\isacharparenright}\ {\isacharequal}\ {\isacharparenleft}fin{\isacharunderscore}join{\isacharunderscore}ti\ s\ x\ i{\isacharparenright}\ {\isacharat}\ {\isacharparenleft}nth\ s\ {\isacharparenleft}x\ {\isacharplus}\ {\isacharparenleft}Suc\ i{\isacharparenright}{\isacharparenright}{\isacharparenright}{\isachardoublequoteclose}\isanewline
\isanewline
\isacommand{definition}\isamarkupfalse%
\ \isanewline
\ \ join{\isacharunderscore}time\ {\isacharcolon}{\isacharcolon}{\isachardoublequoteopen}{\isacharprime}a\ istream\ {\isasymRightarrow}\ nat\ {\isasymRightarrow}\ {\isacharprime}a\ istream{\isachardoublequoteclose}\isanewline
\isakeyword{where}\isanewline
\ {\isachardoublequoteopen}join{\isacharunderscore}time\ s\ n\ t\ {\isasymequiv}\ \isanewline
\ \ {\isacharparenleft}case\ n\ of\ \isanewline
\ \ \ \ \ \ \ \ {\isadigit{0}}\ \ {\isasymRightarrow}\ {\isacharbrackleft}{\isacharbrackright}\isanewline
\ \ {\isacharbar}{\isacharparenleft}Suc\ i{\isacharparenright}\ {\isasymRightarrow}\ \ join{\isacharunderscore}ti\ s\ {\isacharparenleft}n{\isacharasterisk}t{\isacharparenright}\ i{\isacharparenright}{\isachardoublequoteclose}\isanewline
\isanewline
\isanewline
\isacommand{lemma}\isamarkupfalse%
\ join{\isacharunderscore}ti{\isacharunderscore}hint{\isadigit{1}}{\isacharcolon}\isanewline
\ \ \isakeyword{assumes}\ {\isachardoublequoteopen}join{\isacharunderscore}ti\ s\ x\ {\isacharparenleft}Suc\ i{\isacharparenright}\ {\isacharequal}\ {\isacharbrackleft}{\isacharbrackright}{\isachardoublequoteclose}\isanewline
\ \ \isakeyword{shows}\ \ \ {\isachardoublequoteopen}join{\isacharunderscore}ti\ s\ x\ i\ {\isacharequal}\ {\isacharbrackleft}{\isacharbrackright}{\isachardoublequoteclose}\isanewline
\isadelimproof
\endisadelimproof
\isatagproof
\isacommand{using}\isamarkupfalse%
\ assms\ \isacommand{by}\isamarkupfalse%
\ auto%
\endisatagproof
{\isafoldproof}%
\isadelimproof
\isanewline
\endisadelimproof
\isanewline
\isanewline
\isacommand{lemma}\isamarkupfalse%
\ join{\isacharunderscore}ti{\isacharunderscore}hint{\isadigit{2}}{\isacharcolon}\isanewline
\ \ \isakeyword{assumes}\ {\isachardoublequoteopen}join{\isacharunderscore}ti\ s\ x\ {\isacharparenleft}Suc\ i{\isacharparenright}\ {\isacharequal}\ {\isacharbrackleft}{\isacharbrackright}{\isachardoublequoteclose}\isanewline
\ \ \isakeyword{shows}\ \ \ {\isachardoublequoteopen}s\ {\isacharparenleft}x\ {\isacharplus}\ {\isacharparenleft}Suc\ i{\isacharparenright}{\isacharparenright}\ {\isacharequal}\ {\isacharbrackleft}{\isacharbrackright}{\isachardoublequoteclose}\isanewline
\isadelimproof
\endisadelimproof
\isatagproof
\isacommand{using}\isamarkupfalse%
\ assms\ \isacommand{by}\isamarkupfalse%
\ auto%
\endisatagproof
{\isafoldproof}%
\isadelimproof
\isanewline
\endisadelimproof
\isanewline
\isacommand{lemma}\isamarkupfalse%
\ join{\isacharunderscore}ti{\isacharunderscore}hint{\isadigit{3}}{\isacharcolon}\isanewline
\ \ \isakeyword{assumes}\ {\isachardoublequoteopen}join{\isacharunderscore}ti\ s\ x\ {\isacharparenleft}Suc\ i{\isacharparenright}\ {\isacharequal}\ {\isacharbrackleft}{\isacharbrackright}{\isachardoublequoteclose}\isanewline
\ \ \isakeyword{shows}\ \ \ {\isachardoublequoteopen}s\ {\isacharparenleft}x\ {\isacharplus}\ i{\isacharparenright}\ {\isacharequal}\ {\isacharbrackleft}{\isacharbrackright}{\isachardoublequoteclose}\isanewline
\isadelimproof
\endisadelimproof
\isatagproof
\isacommand{using}\isamarkupfalse%
\ assms\ \isacommand{by}\isamarkupfalse%
\ {\isacharparenleft}induct\ i{\isacharcomma}\ auto{\isacharparenright}%
\endisatagproof
{\isafoldproof}%
\isadelimproof
\isanewline
\endisadelimproof
\ \isanewline
\isanewline
\isacommand{lemma}\isamarkupfalse%
\ join{\isacharunderscore}ti{\isacharunderscore}empty{\isacharunderscore}join{\isacharcolon}\isanewline
\ \ \isakeyword{assumes}\ h{\isadigit{1}}{\isacharcolon}{\isachardoublequoteopen}i\ {\isasymle}\ n{\isachardoublequoteclose}\isanewline
\ \ \ \ \ \ \isakeyword{and}\ h{\isadigit{2}}{\isacharcolon}{\isachardoublequoteopen}join{\isacharunderscore}ti\ s\ x\ n\ {\isacharequal}\ {\isacharbrackleft}{\isacharbrackright}{\isachardoublequoteclose}\isanewline
\ \ \isakeyword{shows}\ \ \ \ \ \ {\isachardoublequoteopen}s\ {\isacharparenleft}x{\isacharplus}i{\isacharparenright}\ {\isacharequal}\ {\isacharbrackleft}{\isacharbrackright}{\isachardoublequoteclose}\isanewline
\isadelimproof
\endisadelimproof
\isatagproof
\isacommand{using}\isamarkupfalse%
\ assms\ \isanewline
\isacommand{proof}\isamarkupfalse%
\ {\isacharparenleft}induct\ n{\isacharparenright}\isanewline
\ \ \isacommand{case}\isamarkupfalse%
\ {\isadigit{0}}\ \isanewline
\ \ \isacommand{from}\isamarkupfalse%
\ this\ \isacommand{show}\isamarkupfalse%
\ {\isacharquery}case\ \isacommand{by}\isamarkupfalse%
\ auto\isanewline
\isacommand{next}\isamarkupfalse%
\ \isanewline
\ \ \isacommand{case}\isamarkupfalse%
\ {\isacharparenleft}Suc\ n{\isacharparenright}\isanewline
\ \ \isacommand{from}\isamarkupfalse%
\ this\ \isacommand{show}\isamarkupfalse%
\ {\isacharquery}case\ \isanewline
\ \ \isacommand{proof}\isamarkupfalse%
\ {\isacharparenleft}cases\ {\isachardoublequoteopen}i\ {\isacharequal}\ Suc\ n{\isachardoublequoteclose}{\isacharparenright}\isanewline
\ \ \ \ \isacommand{assume}\isamarkupfalse%
\ a{\isadigit{1}}{\isacharcolon}{\isachardoublequoteopen}i\ {\isacharequal}\ Suc\ n{\isachardoublequoteclose}\ \ \isanewline
\ \ \ \ \isacommand{from}\isamarkupfalse%
\ a{\isadigit{1}}\ \isakeyword{and}\ Suc\ \isacommand{show}\isamarkupfalse%
\ {\isacharquery}thesis\ \isacommand{by}\isamarkupfalse%
\ simp\isanewline
\ \ \isacommand{next}\isamarkupfalse%
\ \isanewline
\ \ \ \ \isacommand{assume}\isamarkupfalse%
\ a{\isadigit{2}}{\isacharcolon}{\isachardoublequoteopen}i\ {\isasymnoteq}\ Suc\ n{\isachardoublequoteclose}\isanewline
\ \ \ \ \isacommand{from}\isamarkupfalse%
\ a{\isadigit{2}}\ \isakeyword{and}\ Suc\ \isacommand{show}\isamarkupfalse%
\ {\isacharquery}thesis\ \isacommand{by}\isamarkupfalse%
\ simp\isanewline
\ \ \isacommand{qed}\isamarkupfalse%
\isanewline
\isacommand{qed}\isamarkupfalse%
\endisatagproof
{\isafoldproof}%
\isadelimproof
\isanewline
\endisadelimproof
\ \ \isanewline
\isacommand{lemma}\isamarkupfalse%
\ join{\isacharunderscore}ti{\isacharunderscore}empty{\isacharunderscore}ti{\isacharcolon}\isanewline
\ \ \isakeyword{assumes}\ {\isachardoublequoteopen}{\isasymforall}\ i\ {\isasymle}\ n{\isachardot}\ s\ {\isacharparenleft}x{\isacharplus}i{\isacharparenright}\ {\isacharequal}\ {\isacharbrackleft}{\isacharbrackright}{\isachardoublequoteclose}\isanewline
\ \ \isakeyword{shows}\ \ \ \ {\isachardoublequoteopen}join{\isacharunderscore}ti\ s\ x\ n\ {\isacharequal}\ {\isacharbrackleft}{\isacharbrackright}{\isachardoublequoteclose}\isanewline
\isadelimproof
\endisadelimproof
\isatagproof
\isacommand{using}\isamarkupfalse%
\ assms\ \isacommand{by}\isamarkupfalse%
\ {\isacharparenleft}induct\ n{\isacharcomma}\ auto{\isacharparenright}%
\endisatagproof
{\isafoldproof}%
\isadelimproof
\isanewline
\endisadelimproof
\ \isanewline
\isanewline
\isacommand{lemma}\isamarkupfalse%
\ join{\isacharunderscore}ti{\isacharunderscore}{\isadigit{1}}nempty{\isacharcolon}\isanewline
\ \ \isakeyword{assumes}\ {\isachardoublequoteopen}{\isasymforall}\ i{\isachardot}\ {\isadigit{0}}\ {\isacharless}\ i\ {\isasymand}\ i\ {\isacharless}\ Suc\ n\ {\isasymlongrightarrow}\ s\ {\isacharparenleft}x{\isacharplus}i{\isacharparenright}\ {\isacharequal}\ {\isacharbrackleft}{\isacharbrackright}{\isachardoublequoteclose}\ \isanewline
\ \ \isakeyword{shows}\ \ \ {\isachardoublequoteopen}join{\isacharunderscore}ti\ s\ x\ n\ {\isacharequal}\ s\ x{\isachardoublequoteclose}\isanewline
\isadelimproof
\endisadelimproof
\isatagproof
\isacommand{using}\isamarkupfalse%
\ assms\ \isacommand{by}\isamarkupfalse%
\ {\isacharparenleft}induct\ n{\isacharcomma}\ auto{\isacharparenright}%
\endisatagproof
{\isafoldproof}%
\isadelimproof
\isanewline
\endisadelimproof
\ \isanewline
\isanewline
\isacommand{lemma}\isamarkupfalse%
\ join{\isacharunderscore}time{\isadigit{1}}t{\isacharcolon}\ {\isachardoublequoteopen}{\isasymforall}\ t{\isachardot}\ join{\isacharunderscore}time\ s\ {\isacharparenleft}{\isadigit{1}}{\isacharcolon}{\isacharcolon}nat{\isacharparenright}\ t\ {\isacharequal}\ s\ t{\isachardoublequoteclose}\isanewline
\isadelimproof
\endisadelimproof
\isatagproof
\isacommand{by}\isamarkupfalse%
\ {\isacharparenleft}simp\ add{\isacharcolon}\ join{\isacharunderscore}time{\isacharunderscore}def{\isacharparenright}%
\endisatagproof
{\isafoldproof}%
\isadelimproof
\isanewline
\endisadelimproof
\isanewline
\ \isanewline
\isacommand{lemma}\isamarkupfalse%
\ join{\isacharunderscore}time{\isadigit{1}}{\isacharcolon}\ {\isachardoublequoteopen}join{\isacharunderscore}time\ s\ {\isadigit{1}}\ {\isacharequal}\ s{\isachardoublequoteclose}\isanewline
\isadelimproof
\endisadelimproof
\isatagproof
\isacommand{by}\isamarkupfalse%
\ {\isacharparenleft}simp\ add{\isacharcolon}\ fun{\isacharunderscore}eq{\isacharunderscore}iff\ join{\isacharunderscore}time{\isacharunderscore}def{\isacharparenright}%
\endisatagproof
{\isafoldproof}%
\isadelimproof
\isanewline
\endisadelimproof
\isanewline
\isanewline
\isacommand{lemma}\isamarkupfalse%
\ join{\isacharunderscore}time{\isacharunderscore}empty{\isadigit{1}}{\isacharcolon}\isanewline
\ \ \isakeyword{assumes}\ h{\isadigit{1}}{\isacharcolon}{\isachardoublequoteopen}i\ {\isacharless}\ n{\isachardoublequoteclose}\isanewline
\ \ \ \ \ \ \isakeyword{and}\ h{\isadigit{2}}{\isacharcolon}{\isachardoublequoteopen}join{\isacharunderscore}time\ s\ n\ t\ {\isacharequal}\ {\isacharbrackleft}{\isacharbrackright}{\isachardoublequoteclose}\isanewline
\ \ \isakeyword{shows}\ \ \ \ \ \ {\isachardoublequoteopen}s\ {\isacharparenleft}n{\isacharasterisk}t\ {\isacharplus}\ i{\isacharparenright}\ {\isacharequal}\ {\isacharbrackleft}{\isacharbrackright}{\isachardoublequoteclose}\isanewline
\isadelimproof
\endisadelimproof
\isatagproof
\isacommand{proof}\isamarkupfalse%
\ {\isacharparenleft}cases\ n{\isacharparenright}\ \isanewline
\ \ \isacommand{assume}\isamarkupfalse%
\ a{\isadigit{1}}{\isacharcolon}{\isachardoublequoteopen}n\ {\isacharequal}\ {\isadigit{0}}{\isachardoublequoteclose}\isanewline
\ \ \isacommand{from}\isamarkupfalse%
\ assms\ \isakeyword{and}\ a{\isadigit{1}}\ \isacommand{show}\isamarkupfalse%
\ {\isacharquery}thesis\ \isacommand{by}\isamarkupfalse%
\ {\isacharparenleft}simp\ add{\isacharcolon}\ join{\isacharunderscore}time{\isacharunderscore}def{\isacharparenright}\isanewline
\isacommand{next}\isamarkupfalse%
\isanewline
\ \ \isacommand{fix}\isamarkupfalse%
\ x\isanewline
\ \ \isacommand{assume}\isamarkupfalse%
\ a{\isadigit{2}}{\isacharcolon}{\isachardoublequoteopen}n\ {\isacharequal}\ Suc\ x{\isachardoublequoteclose}\isanewline
\ \ \isacommand{from}\isamarkupfalse%
\ assms\ \isakeyword{and}\ a{\isadigit{2}}\ \isacommand{have}\isamarkupfalse%
\ sg{\isadigit{1}}{\isacharcolon}{\isachardoublequoteopen}join{\isacharunderscore}ti\ s\ {\isacharparenleft}t\ {\isacharplus}\ x\ {\isacharasterisk}\ t{\isacharparenright}\ x\ {\isacharequal}\ {\isacharbrackleft}{\isacharbrackright}{\isachardoublequoteclose}\ \ \ \ \isanewline
\ \ \ \ \isacommand{by}\isamarkupfalse%
\ {\isacharparenleft}simp\ add{\isacharcolon}\ join{\isacharunderscore}time{\isacharunderscore}def{\isacharparenright}\isanewline
\ \ \isacommand{from}\isamarkupfalse%
\ a{\isadigit{2}}\ \isakeyword{and}\ h{\isadigit{1}}\ \isacommand{have}\isamarkupfalse%
\ sg{\isadigit{2}}{\isacharcolon}{\isachardoublequoteopen}i\ {\isasymle}\ x{\isachardoublequoteclose}\ \isacommand{by}\isamarkupfalse%
\ simp\isanewline
\ \ \isacommand{from}\isamarkupfalse%
\ sg{\isadigit{2}}\ \isakeyword{and}\ sg{\isadigit{1}}\ \isakeyword{and}\ a{\isadigit{2}}\ \isacommand{show}\isamarkupfalse%
\ {\isacharquery}thesis\ \isacommand{by}\isamarkupfalse%
\ {\isacharparenleft}simp\ add{\isacharcolon}\ join{\isacharunderscore}ti{\isacharunderscore}empty{\isacharunderscore}join{\isacharparenright}\isanewline
\isacommand{qed}\isamarkupfalse%
\endisatagproof
{\isafoldproof}%
\isadelimproof
\isanewline
\endisadelimproof
\isanewline
\isanewline
\isacommand{lemma}\isamarkupfalse%
\ fin{\isacharunderscore}join{\isacharunderscore}ti{\isacharunderscore}hint{\isadigit{1}}{\isacharcolon}\isanewline
\ \ \isakeyword{assumes}\ {\isachardoublequoteopen}fin{\isacharunderscore}join{\isacharunderscore}ti\ s\ x\ {\isacharparenleft}Suc\ i{\isacharparenright}\ {\isacharequal}\ {\isacharbrackleft}{\isacharbrackright}{\isachardoublequoteclose}\isanewline
\ \ \isakeyword{shows}\ \ \ {\isachardoublequoteopen}fin{\isacharunderscore}join{\isacharunderscore}ti\ s\ x\ i\ {\isacharequal}\ {\isacharbrackleft}{\isacharbrackright}{\isachardoublequoteclose}\isanewline
\isadelimproof
\endisadelimproof
\isatagproof
\isacommand{using}\isamarkupfalse%
\ assms\ \isacommand{by}\isamarkupfalse%
\ auto%
\endisatagproof
{\isafoldproof}%
\isadelimproof
\isanewline
\endisadelimproof
\isanewline
\isanewline
\isacommand{lemma}\isamarkupfalse%
\ fin{\isacharunderscore}join{\isacharunderscore}ti{\isacharunderscore}hint{\isadigit{2}}{\isacharcolon}\isanewline
\ \ \isakeyword{assumes}\ {\isachardoublequoteopen}fin{\isacharunderscore}join{\isacharunderscore}ti\ s\ x\ {\isacharparenleft}Suc\ i{\isacharparenright}\ {\isacharequal}\ {\isacharbrackleft}{\isacharbrackright}{\isachardoublequoteclose}\isanewline
\ \ \isakeyword{shows}\ \ \ \ {\isachardoublequoteopen}nth\ s\ {\isacharparenleft}x\ {\isacharplus}\ {\isacharparenleft}Suc\ i{\isacharparenright}{\isacharparenright}\ {\isacharequal}\ {\isacharbrackleft}{\isacharbrackright}{\isachardoublequoteclose}\isanewline
\isadelimproof
\endisadelimproof
\isatagproof
\isacommand{using}\isamarkupfalse%
\ assms\ \isacommand{by}\isamarkupfalse%
\ auto%
\endisatagproof
{\isafoldproof}%
\isadelimproof
\isanewline
\endisadelimproof
\ \isanewline
\isanewline
\isacommand{lemma}\isamarkupfalse%
\ fin{\isacharunderscore}join{\isacharunderscore}ti{\isacharunderscore}hint{\isadigit{3}}{\isacharcolon}\isanewline
\ \ \isakeyword{assumes}\ {\isachardoublequoteopen}fin{\isacharunderscore}join{\isacharunderscore}ti\ s\ x\ {\isacharparenleft}Suc\ i{\isacharparenright}\ {\isacharequal}\ {\isacharbrackleft}{\isacharbrackright}{\isachardoublequoteclose}\isanewline
\ \ \isakeyword{shows}\ \ \ {\isachardoublequoteopen}nth\ s\ {\isacharparenleft}x\ {\isacharplus}\ i{\isacharparenright}\ {\isacharequal}\ {\isacharbrackleft}{\isacharbrackright}{\isachardoublequoteclose}\isanewline
\isadelimproof
\endisadelimproof
\isatagproof
\isacommand{using}\isamarkupfalse%
\ assms\ \isacommand{by}\isamarkupfalse%
\ {\isacharparenleft}induct\ i{\isacharcomma}\ auto{\isacharparenright}%
\endisatagproof
{\isafoldproof}%
\isadelimproof
\isanewline
\endisadelimproof
\isanewline
\isanewline
\isacommand{lemma}\isamarkupfalse%
\ fin{\isacharunderscore}join{\isacharunderscore}ti{\isacharunderscore}empty{\isacharunderscore}join{\isacharcolon}\isanewline
\ \ \isakeyword{assumes}\ h{\isadigit{1}}{\isacharcolon}{\isachardoublequoteopen}i\ {\isasymle}\ n{\isachardoublequoteclose}\isanewline
\ \ \ \ \ \ \isakeyword{and}\ h{\isadigit{2}}{\isacharcolon}{\isachardoublequoteopen}fin{\isacharunderscore}join{\isacharunderscore}ti\ s\ x\ n\ {\isacharequal}\ {\isacharbrackleft}{\isacharbrackright}{\isachardoublequoteclose}\isanewline
\ \ \isakeyword{shows}\ \ \ \ \ \ {\isachardoublequoteopen}nth\ s\ {\isacharparenleft}x{\isacharplus}i{\isacharparenright}\ {\isacharequal}\ {\isacharbrackleft}{\isacharbrackright}{\isachardoublequoteclose}\isanewline
\isadelimproof
\endisadelimproof
\isatagproof
\isacommand{using}\isamarkupfalse%
\ assms\isanewline
\isacommand{proof}\isamarkupfalse%
\ {\isacharparenleft}induct\ n{\isacharparenright}\isanewline
\ \ \isacommand{case}\isamarkupfalse%
\ {\isadigit{0}}\isanewline
\ \ \isacommand{from}\isamarkupfalse%
\ this\ \isacommand{show}\isamarkupfalse%
\ {\isacharquery}case\ \isacommand{by}\isamarkupfalse%
\ auto\isanewline
\isacommand{next}\isamarkupfalse%
\isanewline
\ \ \isacommand{case}\isamarkupfalse%
\ {\isacharparenleft}Suc\ n{\isacharparenright}\isanewline
\ \ \isacommand{from}\isamarkupfalse%
\ this\ \isacommand{show}\isamarkupfalse%
\ {\isacharquery}case\isanewline
\ \ \isacommand{proof}\isamarkupfalse%
\ {\isacharparenleft}cases\ {\isachardoublequoteopen}i\ {\isacharequal}\ Suc\ n{\isachardoublequoteclose}{\isacharparenright}\isanewline
\ \ \ \ \isacommand{assume}\isamarkupfalse%
\ a{\isadigit{1}}{\isacharcolon}{\isachardoublequoteopen}i\ {\isacharequal}\ Suc\ n{\isachardoublequoteclose}\isanewline
\ \ \ \ \isacommand{from}\isamarkupfalse%
\ Suc\ \isakeyword{and}\ a{\isadigit{1}}\ \isacommand{show}\isamarkupfalse%
\ {\isacharquery}thesis\ \isacommand{by}\isamarkupfalse%
\ simp\isanewline
\ \ \isacommand{next}\isamarkupfalse%
\isanewline
\ \ \ \ \isacommand{assume}\isamarkupfalse%
\ a{\isadigit{2}}{\isacharcolon}{\isachardoublequoteopen}i\ {\isasymnoteq}\ Suc\ n{\isachardoublequoteclose}\isanewline
\ \ \ \ \isacommand{from}\isamarkupfalse%
\ Suc\ \isakeyword{and}\ a{\isadigit{2}}\ \isacommand{show}\isamarkupfalse%
\ {\isacharquery}thesis\ \isacommand{by}\isamarkupfalse%
\ simp\isanewline
\ \ \isacommand{qed}\isamarkupfalse%
\isanewline
\isacommand{qed}\isamarkupfalse%
\endisatagproof
{\isafoldproof}%
\isadelimproof
\isanewline
\endisadelimproof
\isanewline
\ \ \isanewline
\isacommand{lemma}\isamarkupfalse%
\ fin{\isacharunderscore}join{\isacharunderscore}ti{\isacharunderscore}empty{\isacharunderscore}ti{\isacharcolon}\isanewline
\ \ \isakeyword{assumes}\ {\isachardoublequoteopen}{\isasymforall}\ i\ {\isasymle}\ n{\isachardot}\ nth\ s\ {\isacharparenleft}x{\isacharplus}i{\isacharparenright}\ {\isacharequal}\ {\isacharbrackleft}{\isacharbrackright}{\isachardoublequoteclose}\isanewline
\ \ \isakeyword{shows}\ \ \ {\isachardoublequoteopen}fin{\isacharunderscore}join{\isacharunderscore}ti\ s\ x\ n\ {\isacharequal}\ {\isacharbrackleft}{\isacharbrackright}{\isachardoublequoteclose}\isanewline
\isadelimproof
\endisadelimproof
\isatagproof
\isacommand{using}\isamarkupfalse%
\ assms\ \isacommand{by}\isamarkupfalse%
\ {\isacharparenleft}induct\ n{\isacharcomma}\ auto{\isacharparenright}%
\endisatagproof
{\isafoldproof}%
\isadelimproof
\isanewline
\endisadelimproof
\isanewline
\isanewline
\isacommand{lemma}\isamarkupfalse%
\ fin{\isacharunderscore}join{\isacharunderscore}ti{\isacharunderscore}{\isadigit{1}}nempty{\isacharcolon}\isanewline
\ \ \isakeyword{assumes}\ {\isachardoublequoteopen}{\isasymforall}\ i{\isachardot}\ {\isadigit{0}}\ {\isacharless}\ i\ {\isasymand}\ i\ {\isacharless}\ Suc\ n\ {\isasymlongrightarrow}\ nth\ s\ {\isacharparenleft}x{\isacharplus}i{\isacharparenright}\ {\isacharequal}\ {\isacharbrackleft}{\isacharbrackright}{\isachardoublequoteclose}\ \isanewline
\ \ \isakeyword{shows}\ {\isachardoublequoteopen}fin{\isacharunderscore}join{\isacharunderscore}ti\ s\ x\ n\ {\isacharequal}\ nth\ s\ x{\isachardoublequoteclose}\isanewline
\isadelimproof
\endisadelimproof
\isatagproof
\isacommand{using}\isamarkupfalse%
\ assms\ \ \isacommand{by}\isamarkupfalse%
\ {\isacharparenleft}induct\ n{\isacharcomma}\ auto{\isacharparenright}%
\endisatagproof
{\isafoldproof}%
\isadelimproof
\endisadelimproof
\isanewline
\isamarkupsubsection{Split time units%
}
\isamarkuptrue%
\isacommand{definition}\isamarkupfalse%
\ \isanewline
\ \ split{\isacharunderscore}time\ {\isacharcolon}{\isacharcolon}{\isachardoublequoteopen}{\isacharprime}a\ istream\ {\isasymRightarrow}\ nat\ {\isasymRightarrow}\ {\isacharprime}a\ istream{\isachardoublequoteclose}\isanewline
\isakeyword{where}\isanewline
\ {\isachardoublequoteopen}split{\isacharunderscore}time\ s\ n\ t\ {\isasymequiv}\ \isanewline
\ \ {\isacharparenleft}\ if\ {\isacharparenleft}t\ mod\ n\ {\isacharequal}\ {\isadigit{0}}{\isacharparenright}\ \isanewline
\ \ \ \ then\ s\ {\isacharparenleft}t\ div\ n{\isacharparenright}\isanewline
\ \ \ \ else\ {\isacharbrackleft}{\isacharbrackright}{\isacharparenright}{\isachardoublequoteclose}\isanewline
\isanewline
\isacommand{lemma}\isamarkupfalse%
\ split{\isacharunderscore}time{\isadigit{1}}t{\isacharcolon}\ {\isachardoublequoteopen}{\isasymforall}\ t{\isachardot}\ split{\isacharunderscore}time\ s\ {\isadigit{1}}\ t\ {\isacharequal}\ s\ t{\isachardoublequoteclose}\isanewline
\isadelimproof
\endisadelimproof
\isatagproof
\isacommand{by}\isamarkupfalse%
\ {\isacharparenleft}simp\ add{\isacharcolon}\ split{\isacharunderscore}time{\isacharunderscore}def{\isacharparenright}%
\endisatagproof
{\isafoldproof}%
\isadelimproof
\isanewline
\endisadelimproof
\isanewline
\isacommand{lemma}\isamarkupfalse%
\ split{\isacharunderscore}time{\isadigit{1}}{\isacharcolon}\ {\isachardoublequoteopen}split{\isacharunderscore}time\ s\ {\isadigit{1}}\ {\isacharequal}\ s{\isachardoublequoteclose}\isanewline
\isadelimproof
\endisadelimproof
\isatagproof
\isacommand{by}\isamarkupfalse%
\ {\isacharparenleft}simp\ add{\isacharcolon}\ fun{\isacharunderscore}eq{\isacharunderscore}iff\ split{\isacharunderscore}time{\isacharunderscore}def{\isacharparenright}%
\endisatagproof
{\isafoldproof}%
\isadelimproof
\isanewline
\endisadelimproof
\isanewline
\isacommand{lemma}\isamarkupfalse%
\ split{\isacharunderscore}time{\isacharunderscore}mod{\isacharcolon}\ \isanewline
\ \ \isakeyword{assumes}\ {\isachardoublequoteopen}t\ mod\ n\ {\isasymnoteq}\ {\isadigit{0}}{\isachardoublequoteclose}\isanewline
\ \ \isakeyword{shows}\ \ \ {\isachardoublequoteopen}split{\isacharunderscore}time\ s\ n\ t\ {\isacharequal}\ {\isacharbrackleft}{\isacharbrackright}{\isachardoublequoteclose}\isanewline
\isadelimproof
\endisadelimproof
\isatagproof
\isacommand{using}\isamarkupfalse%
\ assms\ \isacommand{by}\isamarkupfalse%
\ {\isacharparenleft}simp\ add{\isacharcolon}\ split{\isacharunderscore}time{\isacharunderscore}def{\isacharparenright}%
\endisatagproof
{\isafoldproof}%
\isadelimproof
\isanewline
\endisadelimproof
\isanewline
\isacommand{lemma}\isamarkupfalse%
\ split{\isacharunderscore}time{\isacharunderscore}nempty{\isacharcolon}\ \isanewline
\ \ \isakeyword{assumes}\ {\isachardoublequoteopen}{\isadigit{0}}\ {\isacharless}\ n{\isachardoublequoteclose}\isanewline
\ \ \isakeyword{shows}\ \ \ {\isachardoublequoteopen}split{\isacharunderscore}time\ s\ n\ {\isacharparenleft}n\ {\isacharasterisk}\ t{\isacharparenright}\ {\isacharequal}\ s\ t{\isachardoublequoteclose}\isanewline
\isadelimproof
\endisadelimproof
\isatagproof
\isacommand{using}\isamarkupfalse%
\ assms\ \isacommand{by}\isamarkupfalse%
\ {\isacharparenleft}simp\ add{\isacharcolon}\ split{\isacharunderscore}time{\isacharunderscore}def{\isacharparenright}%
\endisatagproof
{\isafoldproof}%
\isadelimproof
\isanewline
\endisadelimproof
\isanewline
\isacommand{lemma}\isamarkupfalse%
\ split{\isacharunderscore}time{\isacharunderscore}nempty{\isacharunderscore}Suc{\isacharcolon}\isanewline
\ \ \isakeyword{assumes}\ {\isachardoublequoteopen}{\isadigit{0}}\ {\isacharless}\ n{\isachardoublequoteclose}\isanewline
\ \ \isakeyword{shows}\ \ \ {\isachardoublequoteopen}split{\isacharunderscore}time\ s\ {\isacharparenleft}Suc\ n{\isacharparenright}\ {\isacharparenleft}{\isacharparenleft}Suc\ n{\isacharparenright}\ {\isacharasterisk}\ t{\isacharparenright}\ {\isacharequal}\ split{\isacharunderscore}time\ s\ n\ {\isacharparenleft}n\ {\isacharasterisk}\ t{\isacharparenright}{\isachardoublequoteclose}\isanewline
\isadelimproof
\endisadelimproof
\isatagproof
\isacommand{proof}\isamarkupfalse%
\ {\isacharminus}\ \isanewline
\ \ \isacommand{have}\isamarkupfalse%
\ sg{\isadigit{0}}{\isacharcolon}{\isachardoublequoteopen}{\isadigit{0}}\ {\isacharless}\ Suc\ n{\isachardoublequoteclose}\ \isacommand{by}\isamarkupfalse%
\ simp\isanewline
\ \ \isacommand{from}\isamarkupfalse%
\ sg{\isadigit{0}}\ \isacommand{have}\isamarkupfalse%
\ sg{\isadigit{1}}{\isacharcolon}{\isachardoublequoteopen}split{\isacharunderscore}time\ s\ {\isacharparenleft}Suc\ n{\isacharparenright}\ {\isacharparenleft}{\isacharparenleft}Suc\ n{\isacharparenright}\ {\isacharasterisk}\ t{\isacharparenright}\ {\isacharequal}\ \ s\ t{\isachardoublequoteclose}\isanewline
\ \ \ \ \isacommand{by}\isamarkupfalse%
\ {\isacharparenleft}rule\ split{\isacharunderscore}time{\isacharunderscore}nempty{\isacharparenright}\isanewline
\ \ \isacommand{from}\isamarkupfalse%
\ assms\ \isacommand{have}\isamarkupfalse%
\ sg{\isadigit{2}}{\isacharcolon}{\isachardoublequoteopen}split{\isacharunderscore}time\ s\ n\ {\isacharparenleft}n\ {\isacharasterisk}\ t{\isacharparenright}\ {\isacharequal}\ s\ t{\isachardoublequoteclose}\ \ \isanewline
\ \ \ \ \isacommand{by}\isamarkupfalse%
\ {\isacharparenleft}rule\ split{\isacharunderscore}time{\isacharunderscore}nempty{\isacharparenright}\isanewline
\ \ \isacommand{from}\isamarkupfalse%
\ sg{\isadigit{1}}\ \isakeyword{and}\ sg{\isadigit{2}}\ \isacommand{show}\isamarkupfalse%
\ {\isacharquery}thesis\ \isacommand{by}\isamarkupfalse%
\ simp\isanewline
\isacommand{qed}\isamarkupfalse%
\endisatagproof
{\isafoldproof}%
\isadelimproof
\isanewline
\endisadelimproof
\isanewline
\isacommand{lemma}\isamarkupfalse%
\ split{\isacharunderscore}time{\isacharunderscore}empty{\isacharcolon}\isanewline
\ \ \isakeyword{assumes}\ h{\isadigit{1}}{\isacharcolon}{\isachardoublequoteopen}i\ {\isacharless}\ n{\isachardoublequoteclose}\ \isakeyword{and}\ h{\isadigit{2}}{\isacharcolon}{\isachardoublequoteopen}{\isadigit{0}}\ {\isacharless}\ i{\isachardoublequoteclose}\isanewline
\ \ \isakeyword{shows}\ {\isachardoublequoteopen}split{\isacharunderscore}time\ s\ n\ {\isacharparenleft}n\ {\isacharasterisk}\ t\ {\isacharplus}\ i{\isacharparenright}\ {\isacharequal}\ {\isacharbrackleft}{\isacharbrackright}{\isachardoublequoteclose}\isanewline
\isadelimproof
\endisadelimproof
\isatagproof
\isacommand{proof}\isamarkupfalse%
\ {\isacharminus}\ \isanewline
\ \ \isacommand{from}\isamarkupfalse%
\ assms\ \isacommand{have}\isamarkupfalse%
\ sg{\isadigit{1}}{\isacharcolon}{\isachardoublequoteopen}{\isadigit{0}}\ {\isacharless}\ {\isacharparenleft}n\ {\isacharasterisk}\ t\ {\isacharplus}\ i{\isacharparenright}\ mod\ n{\isachardoublequoteclose}\ \isacommand{by}\isamarkupfalse%
\ {\isacharparenleft}simp\ add{\isacharcolon}\ arith{\isacharunderscore}mod{\isacharunderscore}nzero{\isacharparenright}\isanewline
\ \ \isacommand{from}\isamarkupfalse%
\ assms\ \isakeyword{and}\ sg{\isadigit{1}}\ \isacommand{show}\isamarkupfalse%
\ {\isacharquery}thesis\ \isacommand{by}\isamarkupfalse%
\ {\isacharparenleft}simp\ add{\isacharcolon}\ split{\isacharunderscore}time{\isacharunderscore}def{\isacharparenright}\isanewline
\isacommand{qed}\isamarkupfalse%
\endisatagproof
{\isafoldproof}%
\isadelimproof
\isanewline
\endisadelimproof
\isanewline
\isacommand{lemma}\isamarkupfalse%
\ split{\isacharunderscore}time{\isacharunderscore}empty{\isacharunderscore}Suc{\isacharcolon}\isanewline
\ \ \isakeyword{assumes}\ h{\isadigit{1}}{\isacharcolon}{\isachardoublequoteopen}i\ {\isacharless}\ n{\isachardoublequoteclose}\ \isakeyword{and}\ h{\isadigit{2}}{\isacharcolon}{\isachardoublequoteopen}{\isadigit{0}}\ {\isacharless}\ i{\isachardoublequoteclose}\isanewline
\ \ \isakeyword{shows}\ {\isachardoublequoteopen}split{\isacharunderscore}time\ s\ {\isacharparenleft}Suc\ n{\isacharparenright}\ {\isacharparenleft}{\isacharparenleft}Suc\ n{\isacharparenright}{\isacharasterisk}\ t\ {\isacharplus}\ i{\isacharparenright}\ \ {\isacharequal}\ split{\isacharunderscore}time\ s\ n\ {\isacharparenleft}n\ {\isacharasterisk}\ t\ {\isacharplus}\ i{\isacharparenright}{\isachardoublequoteclose}\isanewline
\isadelimproof
\endisadelimproof
\isatagproof
\isacommand{proof}\isamarkupfalse%
\ {\isacharminus}\ \isanewline
\ \ \isacommand{from}\isamarkupfalse%
\ h{\isadigit{1}}\ \isacommand{have}\isamarkupfalse%
\ sg{\isadigit{1}}{\isacharcolon}{\isachardoublequoteopen}i\ {\isacharless}\ Suc\ n{\isachardoublequoteclose}\ \isacommand{by}\isamarkupfalse%
\ simp\isanewline
\ \ \isacommand{from}\isamarkupfalse%
\ sg{\isadigit{1}}\ \isakeyword{and}\ h{\isadigit{2}}\ \isacommand{have}\isamarkupfalse%
\ sg{\isadigit{2}}{\isacharcolon}{\isachardoublequoteopen}split{\isacharunderscore}time\ s\ {\isacharparenleft}Suc\ n{\isacharparenright}\ {\isacharparenleft}Suc\ n\ {\isacharasterisk}\ t\ {\isacharplus}\ i{\isacharparenright}\ {\isacharequal}\ {\isacharbrackleft}{\isacharbrackright}{\isachardoublequoteclose}\isanewline
\ \ \ \ \isacommand{by}\isamarkupfalse%
\ {\isacharparenleft}rule\ split{\isacharunderscore}time{\isacharunderscore}empty{\isacharparenright}\isanewline
\ \ \isacommand{from}\isamarkupfalse%
\ assms\ \isacommand{have}\isamarkupfalse%
\ sg{\isadigit{3}}{\isacharcolon}{\isachardoublequoteopen}split{\isacharunderscore}time\ s\ n\ {\isacharparenleft}n\ {\isacharasterisk}\ t\ {\isacharplus}\ i{\isacharparenright}\ {\isacharequal}\ {\isacharbrackleft}{\isacharbrackright}{\isachardoublequoteclose}\isanewline
\ \ \ \ \isacommand{by}\isamarkupfalse%
\ {\isacharparenleft}rule\ split{\isacharunderscore}time{\isacharunderscore}empty{\isacharparenright}\isanewline
\ \ \isacommand{from}\isamarkupfalse%
\ sg{\isadigit{3}}\ \isakeyword{and}\ sg{\isadigit{2}}\ \isacommand{show}\isamarkupfalse%
\ {\isacharquery}thesis\ \isacommand{by}\isamarkupfalse%
\ simp\isanewline
\isacommand{qed}\isamarkupfalse%
\endisatagproof
{\isafoldproof}%
\isadelimproof
\isanewline
\endisadelimproof
\isanewline
\isacommand{lemma}\isamarkupfalse%
\ split{\isacharunderscore}time{\isacharunderscore}hint{\isadigit{1}}{\isacharcolon}\isanewline
\ \ \isakeyword{assumes}\ {\isachardoublequoteopen}n\ {\isacharequal}\ Suc\ m{\isachardoublequoteclose}\isanewline
\ \ \isakeyword{shows}\ \ \ {\isachardoublequoteopen}split{\isacharunderscore}time\ s\ {\isacharparenleft}Suc\ n{\isacharparenright}\ {\isacharparenleft}i\ {\isacharplus}\ n\ {\isacharasterisk}\ i\ {\isacharplus}\ n{\isacharparenright}\ {\isacharequal}\ {\isacharbrackleft}{\isacharbrackright}{\isachardoublequoteclose}\isanewline
\isadelimproof
\endisadelimproof
\isatagproof
\isacommand{proof}\isamarkupfalse%
\ {\isacharminus}\ \isanewline
\ \ \isacommand{have}\isamarkupfalse%
\ sg{\isadigit{1}}{\isacharcolon}{\isachardoublequoteopen}i\ {\isacharplus}\ n\ {\isacharasterisk}\ i\ {\isacharplus}\ n\ {\isacharequal}\ {\isacharparenleft}Suc\ n{\isacharparenright}\ {\isacharasterisk}\ i\ {\isacharplus}\ n{\isachardoublequoteclose}\ \isacommand{by}\isamarkupfalse%
\ simp\isanewline
\ \ \isacommand{have}\isamarkupfalse%
\ sg{\isadigit{2}}{\isacharcolon}{\isachardoublequoteopen}n\ {\isacharless}\ Suc\ n{\isachardoublequoteclose}\ \isacommand{by}\isamarkupfalse%
\ simp\isanewline
\ \ \isacommand{from}\isamarkupfalse%
\ assms\ \isacommand{have}\isamarkupfalse%
\ sg{\isadigit{3}}{\isacharcolon}{\isachardoublequoteopen}{\isadigit{0}}\ {\isacharless}\ n{\isachardoublequoteclose}\ \isacommand{by}\isamarkupfalse%
\ simp\isanewline
\ \ \isacommand{from}\isamarkupfalse%
\ sg{\isadigit{2}}\ \isakeyword{and}\ sg{\isadigit{3}}\ \isacommand{have}\isamarkupfalse%
\ sg{\isadigit{4}}{\isacharcolon}{\isachardoublequoteopen}split{\isacharunderscore}time\ s\ {\isacharparenleft}Suc\ n{\isacharparenright}\ {\isacharparenleft}Suc\ n\ {\isacharasterisk}\ i\ {\isacharplus}\ n{\isacharparenright}\ {\isacharequal}\ {\isacharbrackleft}{\isacharbrackright}{\isachardoublequoteclose}\isanewline
\ \ \ \ \ \isacommand{by}\isamarkupfalse%
\ {\isacharparenleft}rule\ split{\isacharunderscore}time{\isacharunderscore}empty{\isacharparenright}\isanewline
\ \ \isacommand{from}\isamarkupfalse%
\ sg{\isadigit{1}}\ \isakeyword{and}\ sg{\isadigit{4}}\ \ \isacommand{show}\isamarkupfalse%
\ {\isacharquery}thesis\ \isacommand{by}\isamarkupfalse%
\ auto\isanewline
\isacommand{qed}\isamarkupfalse%
\endisatagproof
{\isafoldproof}%
\isadelimproof
\endisadelimproof
\isamarkupsubsection{Duality of the split and the join operators%
}
\isamarkuptrue%
\isacommand{lemma}\isamarkupfalse%
\ join{\isacharunderscore}split{\isacharunderscore}i{\isacharcolon}\isanewline
\ \ \isakeyword{assumes}\ {\isachardoublequoteopen}{\isadigit{0}}\ {\isacharless}\ n{\isachardoublequoteclose}\isanewline
\ \ \isakeyword{shows}\ \ \ {\isachardoublequoteopen}join{\isacharunderscore}time\ {\isacharparenleft}split{\isacharunderscore}time\ s\ n{\isacharparenright}\ n\ i\ {\isacharequal}\ s\ i{\isachardoublequoteclose}\isanewline
\isadelimproof
\endisadelimproof
\isatagproof
\isacommand{proof}\isamarkupfalse%
\ {\isacharparenleft}cases\ n{\isacharparenright}\isanewline
\ \ \isacommand{assume}\isamarkupfalse%
\ a{\isadigit{1}}{\isacharcolon}{\isachardoublequoteopen}n\ {\isacharequal}\ {\isadigit{0}}{\isachardoublequoteclose}\isanewline
\ \ \isacommand{from}\isamarkupfalse%
\ this\ \isakeyword{and}\ assms\ \isacommand{show}\isamarkupfalse%
\ {\isacharquery}thesis\ \isacommand{by}\isamarkupfalse%
\ simp\isanewline
\isacommand{next}\isamarkupfalse%
\isanewline
\ \ \isacommand{fix}\isamarkupfalse%
\ k\isanewline
\ \ \isacommand{assume}\isamarkupfalse%
\ a{\isadigit{2}}{\isacharcolon}{\isachardoublequoteopen}n\ {\isacharequal}\ Suc\ k{\isachardoublequoteclose}\isanewline
\ \ \isacommand{have}\isamarkupfalse%
\ sg{\isadigit{0}}{\isacharcolon}{\isachardoublequoteopen}{\isadigit{0}}\ {\isacharless}\ Suc\ k{\isachardoublequoteclose}\ \isacommand{by}\isamarkupfalse%
\ simp\isanewline
\ \ \isacommand{from}\isamarkupfalse%
\ sg{\isadigit{0}}\ \isacommand{have}\isamarkupfalse%
\ sg{\isadigit{1}}{\isacharcolon}{\isachardoublequoteopen}{\isacharparenleft}split{\isacharunderscore}time\ s\ {\isacharparenleft}Suc\ k{\isacharparenright}{\isacharparenright}\ {\isacharparenleft}Suc\ k\ {\isacharasterisk}\ i{\isacharparenright}\ {\isacharequal}\ s\ i{\isachardoublequoteclose}\isanewline
\ \ \ \ \isacommand{by}\isamarkupfalse%
\ {\isacharparenleft}rule\ split{\isacharunderscore}time{\isacharunderscore}nempty{\isacharparenright}\isanewline
\ \ \isacommand{have}\isamarkupfalse%
\ sg{\isadigit{2}}{\isacharcolon}{\isachardoublequoteopen}i\ {\isacharplus}\ k\ {\isacharasterisk}\ i\ {\isacharequal}\ {\isacharparenleft}Suc\ k{\isacharparenright}\ {\isacharasterisk}\ i{\isachardoublequoteclose}\ \isacommand{by}\isamarkupfalse%
\ simp\isanewline
\ \ \isacommand{have}\isamarkupfalse%
\ sg{\isadigit{3}}{\isacharcolon}{\isachardoublequoteopen}{\isasymforall}\ j{\isachardot}\ {\isadigit{0}}\ {\isacharless}\ j\ {\isasymand}\ j\ {\isacharless}\ Suc\ k\ {\isasymlongrightarrow}\ split{\isacharunderscore}time\ s\ {\isacharparenleft}Suc\ k{\isacharparenright}\ {\isacharparenleft}Suc\ k\ {\isacharasterisk}\ i\ {\isacharplus}\ j{\isacharparenright}\ {\isacharequal}\ {\isacharbrackleft}{\isacharbrackright}{\isachardoublequoteclose}\isanewline
\ \ \ \ \isacommand{by}\isamarkupfalse%
\ {\isacharparenleft}clarify{\isacharcomma}\ rule\ split{\isacharunderscore}time{\isacharunderscore}empty{\isacharcomma}\ auto{\isacharparenright}\isanewline
\ \ \isacommand{from}\isamarkupfalse%
\ sg{\isadigit{3}}\ \isacommand{have}\isamarkupfalse%
\ sg{\isadigit{4}}{\isacharcolon}{\isachardoublequoteopen}join{\isacharunderscore}ti\ {\isacharparenleft}split{\isacharunderscore}time\ s\ {\isacharparenleft}Suc\ k{\isacharparenright}{\isacharparenright}\ {\isacharparenleft}{\isacharparenleft}Suc\ k{\isacharparenright}\ {\isacharasterisk}\ i{\isacharparenright}\ k\ {\isacharequal}\ \isanewline
\ \ \ \ \ \ \ \ \ \ \ \ \ \ \ \ \ \ \ \ \ {\isacharparenleft}split{\isacharunderscore}time\ s\ {\isacharparenleft}Suc\ k{\isacharparenright}{\isacharparenright}\ {\isacharparenleft}Suc\ k\ {\isacharasterisk}\ i{\isacharparenright}{\isachardoublequoteclose}\isanewline
\ \ \ \ \isacommand{by}\isamarkupfalse%
\ {\isacharparenleft}rule\ join{\isacharunderscore}ti{\isacharunderscore}{\isadigit{1}}nempty{\isacharparenright}\isanewline
\ \ \isacommand{from}\isamarkupfalse%
\ a{\isadigit{2}}\ \isakeyword{and}\ sg{\isadigit{4}}\ \isakeyword{and}\ sg{\isadigit{1}}\ \isacommand{show}\isamarkupfalse%
\ {\isacharquery}thesis\ \isacommand{by}\isamarkupfalse%
\ {\isacharparenleft}simp\ add{\isacharcolon}\ join{\isacharunderscore}time{\isacharunderscore}def{\isacharparenright}\isanewline
\isacommand{qed}\isamarkupfalse%
\endisatagproof
{\isafoldproof}%
\isadelimproof
\isanewline
\endisadelimproof
\isanewline
\isacommand{lemma}\isamarkupfalse%
\ join{\isacharunderscore}split{\isacharcolon}\isanewline
\ \ \isakeyword{assumes}\ {\isachardoublequoteopen}{\isadigit{0}}\ {\isacharless}\ n{\isachardoublequoteclose}\isanewline
\ \ \isakeyword{shows}\ {\isachardoublequoteopen}join{\isacharunderscore}time\ {\isacharparenleft}split{\isacharunderscore}time\ s\ n{\isacharparenright}\ n\ {\isacharequal}\ s{\isachardoublequoteclose}\isanewline
\isadelimproof
\endisadelimproof
\isatagproof
\isacommand{using}\isamarkupfalse%
\ assms\ \ \isacommand{by}\isamarkupfalse%
\ {\isacharparenleft}simp\ add{\isacharcolon}\ fun{\isacharunderscore}eq{\isacharunderscore}iff\ join{\isacharunderscore}split{\isacharunderscore}i{\isacharparenright}%
\endisatagproof
{\isafoldproof}%
\isadelimproof
\isanewline
\endisadelimproof
\isadelimtheory
\isanewline
\endisadelimtheory
\isatagtheory
\isacommand{end}\isamarkupfalse%
\endisatagtheory
{\isafoldtheory}%
\isadelimtheory
\endisadelimtheory
\end{isabellebody}%

%
\begin{isabellebody}%
\def\isabellecontext{SteamBoiler}%
\isamarkupheader{Steam Boiler System: Specification%
}
\isamarkuptrue%
\isadelimtheory
\endisadelimtheory
\isatagtheory
\isacommand{theory}\isamarkupfalse%
\ \ SteamBoiler\ \isanewline
\isakeyword{imports}\ stream\ BitBoolTS\isanewline
\isakeyword{begin}%
\endisatagtheory
{\isafoldtheory}%
\isadelimtheory
\endisadelimtheory
\isanewline
\isanewline
\isacommand{definition}\isamarkupfalse%
\isanewline
\ ControlSystem\ {\isacharcolon}{\isacharcolon}\ {\isachardoublequoteopen}nat\ istream\ {\isasymRightarrow}\ bool{\isachardoublequoteclose}\isanewline
\isakeyword{where}\isanewline
\ {\isachardoublequoteopen}ControlSystem\ s\ {\isasymequiv}\ \ \isanewline
\ \ {\isacharparenleft}ts\ s{\isacharparenright}\ {\isasymand}\ \isanewline
\ \ \ {\isacharparenleft}{\isasymforall}\ {\isacharparenleft}j{\isacharcolon}{\isacharcolon}nat{\isacharparenright}{\isachardot}\ {\isacharparenleft}{\isadigit{2}}{\isadigit{0}}{\isadigit{0}}{\isacharcolon}{\isacharcolon}nat{\isacharparenright}\ {\isasymle}\ hd\ {\isacharparenleft}s\ j{\isacharparenright}\ {\isasymand}\ hd\ {\isacharparenleft}s\ j{\isacharparenright}\ {\isasymle}\ {\isacharparenleft}{\isadigit{8}}{\isadigit{0}}{\isadigit{0}}{\isacharcolon}{\isacharcolon}\ nat{\isacharparenright}{\isacharparenright}{\isachardoublequoteclose}\isanewline
\isanewline
\isacommand{definition}\isamarkupfalse%
\isanewline
\ \ SteamBoiler\ {\isacharcolon}{\isacharcolon}\ {\isachardoublequoteopen}bit\ istream\ {\isasymRightarrow}\ nat\ istream\ {\isasymRightarrow}\ nat\ istream\ {\isasymRightarrow}\ bool{\isachardoublequoteclose}\isanewline
\isakeyword{where}\isanewline
\ {\isachardoublequoteopen}SteamBoiler\ x\ s\ y\ {\isasymequiv}\ \ \isanewline
\ \ ts\ x\ \isanewline
\ \ {\isasymlongrightarrow}\ \isanewline
\ \ {\isacharparenleft}{\isacharparenleft}ts\ y{\isacharparenright}\ {\isasymand}\ {\isacharparenleft}ts\ s{\isacharparenright}\ {\isasymand}\ {\isacharparenleft}y\ {\isacharequal}\ s{\isacharparenright}\ {\isasymand}\ \isanewline
\ \ \ {\isacharparenleft}{\isacharparenleft}s\ {\isadigit{0}}{\isacharparenright}\ {\isacharequal}\ {\isacharbrackleft}{\isadigit{5}}{\isadigit{0}}{\isadigit{0}}{\isacharcolon}{\isacharcolon}nat{\isacharbrackright}{\isacharparenright}\ {\isasymand}\ \isanewline
\ \ \ {\isacharparenleft}{\isasymforall}\ {\isacharparenleft}j{\isacharcolon}{\isacharcolon}nat{\isacharparenright}{\isachardot}\ {\isacharparenleft}{\isasymexists}\ {\isacharparenleft}r{\isacharcolon}{\isacharcolon}nat{\isacharparenright}{\isachardot}\ \isanewline
\ \ \ \ \ \ \ \ \ \ \ \ \ \ \ \ {\isacharparenleft}{\isadigit{0}}{\isacharcolon}{\isacharcolon}nat{\isacharparenright}\ {\isacharless}\ r\ {\isasymand}\ r\ {\isasymle}\ {\isacharparenleft}{\isadigit{1}}{\isadigit{0}}{\isacharcolon}{\isacharcolon}nat{\isacharparenright}\ {\isasymand}\ \isanewline
\ \ \ \ \ \ \ \ \ \ \ \ \ \ \ \ hd\ {\isacharparenleft}s\ {\isacharparenleft}Suc\ j{\isacharparenright}{\isacharparenright}\ {\isacharequal}\ \isanewline
\ \ \ \ \ \ \ \ \ \ \ \ \ \ \ \ \ \ \ {\isacharparenleft}if\ hd\ {\isacharparenleft}x\ j{\isacharparenright}\ {\isacharequal}\ Zero\ \isanewline
\ \ \ \ \ \ \ \ \ \ \ \ \ \ \ \ \ \ \ \ then\ {\isacharparenleft}hd\ {\isacharparenleft}s\ j{\isacharparenright}{\isacharparenright}\ {\isacharminus}\ r\ \isanewline
\ \ \ \ \ \ \ \ \ \ \ \ \ \ \ \ \ \ \ \ else\ {\isacharparenleft}hd\ {\isacharparenleft}s\ j{\isacharparenright}{\isacharparenright}\ {\isacharplus}\ r{\isacharparenright}{\isacharparenright}\ {\isacharparenright}{\isacharparenright}{\isachardoublequoteclose}\isanewline
\isanewline
\isacommand{definition}\isamarkupfalse%
\isanewline
\ \ Converter\ {\isacharcolon}{\isacharcolon}\ {\isachardoublequoteopen}bit\ istream\ {\isasymRightarrow}\ bit\ istream\ {\isasymRightarrow}\ bool{\isachardoublequoteclose}\isanewline
\isakeyword{where}\isanewline
\ {\isachardoublequoteopen}Converter\ z\ x\ \isanewline
\ \ {\isasymequiv}\ \isanewline
\ \ {\isacharparenleft}ts\ x{\isacharparenright}\ \isanewline
\ \ {\isasymand}\ \isanewline
\ \ {\isacharparenleft}{\isasymforall}\ {\isacharparenleft}t{\isacharcolon}{\isacharcolon}nat{\isacharparenright}{\isachardot}\ \isanewline
\ \ \ \ hd\ {\isacharparenleft}x\ t{\isacharparenright}\ {\isacharequal}\ \isanewline
\ \ \ \ \ \ \ \ {\isacharparenleft}if\ \ {\isacharparenleft}fin{\isacharunderscore}make{\isacharunderscore}untimed\ {\isacharparenleft}inf{\isacharunderscore}truncate\ z\ t{\isacharparenright}\ {\isacharequal}\ {\isacharbrackleft}{\isacharbrackright}{\isacharparenright}\isanewline
\ \ \ \ \ \ \ \ \ then\ \ \isanewline
\ \ \ \ \ \ \ \ \ \ \ \ \ Zero\isanewline
\ \ \ \ \ \ \ \ \ else\ \isanewline
\ \ \ \ \ \ \ \ \ \ \ \ \ {\isacharparenleft}fin{\isacharunderscore}make{\isacharunderscore}untimed\ {\isacharparenleft}inf{\isacharunderscore}truncate\ z\ t{\isacharparenright}{\isacharparenright}\ {\isacharbang}\ \isanewline
\ \ \ \ \ \ \ \ \ \ \ \ \ \ \ \ \ {\isacharparenleft}{\isacharparenleft}length\ {\isacharparenleft}fin{\isacharunderscore}make{\isacharunderscore}untimed\ {\isacharparenleft}inf{\isacharunderscore}truncate\ z\ t{\isacharparenright}{\isacharparenright}{\isacharparenright}\ {\isacharminus}\ {\isacharparenleft}{\isadigit{1}}{\isacharcolon}{\isacharcolon}nat{\isacharparenright}{\isacharparenright}\ \isanewline
\ \ \ \ \ \ \ {\isacharparenright}{\isacharparenright}{\isachardoublequoteclose}\isanewline
\isanewline
\isacommand{definition}\isamarkupfalse%
\isanewline
\ \ Controller{\isacharunderscore}L\ {\isacharcolon}{\isacharcolon}\ \isanewline
\ \ \ \ {\isachardoublequoteopen}nat\ istream\ {\isasymRightarrow}\ bit\ iustream\ {\isasymRightarrow}\ bit\ iustream\ {\isasymRightarrow}\ bit\ istream\ {\isasymRightarrow}\ bool{\isachardoublequoteclose}\isanewline
\isakeyword{where}\isanewline
\ {\isachardoublequoteopen}Controller{\isacharunderscore}L\ y\ lIn\ lOut\ z\ \isanewline
\ \ {\isasymequiv}\ \isanewline
\ \ {\isacharparenleft}z\ {\isadigit{0}}\ {\isacharequal}\ {\isacharbrackleft}Zero{\isacharbrackright}{\isacharparenright}\ \isanewline
\ \ {\isasymand}\isanewline
\ \ {\isacharparenleft}{\isasymforall}\ {\isacharparenleft}t{\isacharcolon}{\isacharcolon}nat{\isacharparenright}{\isachardot}\ \isanewline
\ \ {\isacharparenleft}\ if\ {\isacharparenleft}lIn\ t{\isacharparenright}\ {\isacharequal}\ Zero\ \isanewline
\ \ \ \ then\ {\isacharparenleft}\ if\ {\isadigit{3}}{\isadigit{0}}{\isadigit{0}}\ {\isacharless}\ hd\ {\isacharparenleft}y\ t{\isacharparenright}\ \ \isanewline
\ \ \ \ \ \ \ \ \ \ \ then\ \ {\isacharparenleft}z\ t{\isacharparenright}\ {\isacharequal}\ {\isacharbrackleft}{\isacharbrackright}\ \ \ \ {\isasymand}\ {\isacharparenleft}lOut\ t{\isacharparenright}\ {\isacharequal}\ Zero\isanewline
\ \ \ \ \ \ \ \ \ \ \ else\ \ {\isacharparenleft}z\ t{\isacharparenright}\ {\isacharequal}\ {\isacharbrackleft}One{\isacharbrackright}\ {\isasymand}\ {\isacharparenleft}lOut\ t{\isacharparenright}\ {\isacharequal}\ One\ \isanewline
\ \ \ \ \ \ \ \ \ \ {\isacharparenright}\isanewline
\ \ \ \ else\ {\isacharparenleft}\ if\ \ hd\ {\isacharparenleft}y\ t{\isacharparenright}\ {\isacharless}\ {\isadigit{7}}{\isadigit{0}}{\isadigit{0}}\ \ \isanewline
\ \ \ \ \ \ \ \ \ \ \ then\ \ {\isacharparenleft}z\ t{\isacharparenright}\ {\isacharequal}\ {\isacharbrackleft}{\isacharbrackright}\ \ \ \ \ {\isasymand}\ {\isacharparenleft}lOut\ t{\isacharparenright}\ {\isacharequal}\ One\ \ \ \isanewline
\ \ \ \ \ \ \ \ \ \ \ else\ \ {\isacharparenleft}z\ t{\isacharparenright}\ {\isacharequal}\ {\isacharbrackleft}Zero{\isacharbrackright}\ {\isasymand}\ {\isacharparenleft}lOut\ t{\isacharparenright}\ {\isacharequal}\ Zero\ {\isacharparenright}\ {\isacharparenright}{\isacharparenright}{\isachardoublequoteclose}\isanewline
\isanewline
\isacommand{definition}\isamarkupfalse%
\ \isanewline
\ \ Controller\ {\isacharcolon}{\isacharcolon}\ {\isachardoublequoteopen}nat\ istream\ {\isasymRightarrow}\ bit\ istream\ {\isasymRightarrow}\ bool{\isachardoublequoteclose}\isanewline
\isakeyword{where}\isanewline
\ {\isachardoublequoteopen}Controller\ y\ z\ \isanewline
\ \ {\isasymequiv}\ \isanewline
\ \ {\isacharparenleft}ts\ y{\isacharparenright}\isanewline
\ \ {\isasymlongrightarrow}\ \isanewline
\ \ {\isacharparenleft}{\isasymexists}\ l{\isachardot}\ Controller{\isacharunderscore}L\ y\ {\isacharparenleft}fin{\isacharunderscore}inf{\isacharunderscore}append\ {\isacharbrackleft}Zero{\isacharbrackright}\ l{\isacharparenright}\ l\ z{\isacharparenright}{\isachardoublequoteclose}\isanewline
\isanewline
\isacommand{definition}\isamarkupfalse%
\isanewline
\ \ ControlSystemArch\ {\isacharcolon}{\isacharcolon}\ {\isachardoublequoteopen}nat\ istream\ {\isasymRightarrow}\ bool{\isachardoublequoteclose}\isanewline
\isakeyword{where}\isanewline
\ {\isachardoublequoteopen}ControlSystemArch\ s\ \isanewline
\ \ {\isasymequiv}\ \isanewline
\ \ {\isasymexists}\ x\ z\ {\isacharcolon}{\isacharcolon}\ bit\ istream{\isachardot}\ {\isasymexists}\ y\ {\isacharcolon}{\isacharcolon}\ nat\ istream{\isachardot}\isanewline
\ \ \ \ {\isacharparenleft}\ SteamBoiler\ x\ s\ y\ {\isasymand}\ Controller\ y\ z\ {\isasymand}\ Converter\ z\ x\ {\isacharparenright}{\isachardoublequoteclose}\isanewline
\isadelimtheory
\isanewline
\endisadelimtheory
\isatagtheory
\isacommand{end}\isamarkupfalse%
\endisatagtheory
{\isafoldtheory}%
\isadelimtheory
\endisadelimtheory
\ \end{isabellebody}%

%
\begin{isabellebody}%
\def\isabellecontext{SteamBoiler{\isacharunderscore}proof}%
\isamarkupheader{Steam Boiler System: Verification%
}
\isamarkuptrue%
\isadelimtheory
\endisadelimtheory
\isatagtheory
\isacommand{theory}\isamarkupfalse%
\ \ SteamBoiler{\isacharunderscore}proof\ \isanewline
\isakeyword{imports}\ SteamBoiler\isanewline
\isakeyword{begin}%
\endisatagtheory
{\isafoldtheory}%
\isadelimtheory
\endisadelimtheory
\isamarkupsubsection{Properties of the Boiler Component%
}
\isamarkuptrue%
\isacommand{lemma}\isamarkupfalse%
\ L{\isadigit{1}}{\isacharunderscore}Boiler{\isacharcolon}\isanewline
\ \ \isakeyword{assumes}\ h{\isadigit{1}}{\isacharcolon}\ {\isachardoublequoteopen}SteamBoiler\ x\ s\ y{\isachardoublequoteclose}\isanewline
\ \ \ \ \ \ \isakeyword{and}\ h{\isadigit{2}}{\isacharcolon}\ {\isachardoublequoteopen}ts\ x{\isachardoublequoteclose}\isanewline
\ \ \isakeyword{shows}\ {\isachardoublequoteopen}ts\ s{\isachardoublequoteclose}\isanewline
\isadelimproof
\endisadelimproof
\isatagproof
\isacommand{using}\isamarkupfalse%
\ assms\ \ \isacommand{by}\isamarkupfalse%
\ {\isacharparenleft}simp\ add{\isacharcolon}\ SteamBoiler{\isacharunderscore}def{\isacharparenright}%
\endisatagproof
{\isafoldproof}%
\isadelimproof
\isanewline
\endisadelimproof
\isanewline
\isacommand{lemma}\isamarkupfalse%
\ L{\isadigit{2}}{\isacharunderscore}Boiler{\isacharcolon}\isanewline
\ \ \isakeyword{assumes}\ h{\isadigit{1}}{\isacharcolon}\ {\isachardoublequoteopen}SteamBoiler\ x\ s\ y{\isachardoublequoteclose}\isanewline
\ \ \ \ \ \ \isakeyword{and}\ h{\isadigit{2}}{\isacharcolon}\ {\isachardoublequoteopen}ts\ x{\isachardoublequoteclose}\isanewline
\ \ \isakeyword{shows}\ {\isachardoublequoteopen}ts\ y{\isachardoublequoteclose}\isanewline
\isadelimproof
\endisadelimproof
\isatagproof
\isacommand{using}\isamarkupfalse%
\ assms\ \ \isacommand{by}\isamarkupfalse%
\ {\isacharparenleft}simp\ add{\isacharcolon}\ SteamBoiler{\isacharunderscore}def{\isacharparenright}%
\endisatagproof
{\isafoldproof}%
\isadelimproof
\ \isanewline
\endisadelimproof
\isanewline
\isanewline
\isacommand{lemma}\isamarkupfalse%
\ L{\isadigit{3}}{\isacharunderscore}Boiler{\isacharcolon}\isanewline
\ \ \isakeyword{assumes}\ h{\isadigit{1}}{\isacharcolon}{\isachardoublequoteopen}SteamBoiler\ x\ s\ y{\isachardoublequoteclose}\isanewline
\ \ \ \ \ \ \isakeyword{and}\ h{\isadigit{2}}{\isacharcolon}{\isachardoublequoteopen}ts\ x{\isachardoublequoteclose}\ \isanewline
\ \ \isakeyword{shows}\ {\isachardoublequoteopen}{\isadigit{2}}{\isadigit{0}}{\isadigit{0}}\ {\isasymle}\ hd\ {\isacharparenleft}s\ {\isadigit{0}}{\isacharparenright}{\isachardoublequoteclose}\isanewline
\isadelimproof
\endisadelimproof
\isatagproof
\isacommand{using}\isamarkupfalse%
\ assms\ \isacommand{by}\isamarkupfalse%
\ {\isacharparenleft}simp\ add{\isacharcolon}\ SteamBoiler{\isacharunderscore}def{\isacharparenright}%
\endisatagproof
{\isafoldproof}%
\isadelimproof
\isanewline
\endisadelimproof
\isanewline
\isanewline
\isacommand{lemma}\isamarkupfalse%
\ L{\isadigit{4}}{\isacharunderscore}Boiler{\isacharcolon}\isanewline
\ \ \isakeyword{assumes}\ h{\isadigit{1}}{\isacharcolon}{\isachardoublequoteopen}SteamBoiler\ x\ s\ y{\isachardoublequoteclose}\isanewline
\ \ \ \ \ \ \isakeyword{and}\ h{\isadigit{2}}{\isacharcolon}{\isachardoublequoteopen}ts\ x{\isachardoublequoteclose}\ \isanewline
\ \ \isakeyword{shows}\ {\isachardoublequoteopen}hd\ {\isacharparenleft}s\ {\isadigit{0}}{\isacharparenright}\ {\isasymle}\ {\isadigit{8}}{\isadigit{0}}{\isadigit{0}}{\isachardoublequoteclose}\isanewline
\isadelimproof
\endisadelimproof
\isatagproof
\isacommand{using}\isamarkupfalse%
\ assms\ \isacommand{by}\isamarkupfalse%
\ {\isacharparenleft}simp\ add{\isacharcolon}\ SteamBoiler{\isacharunderscore}def{\isacharparenright}%
\endisatagproof
{\isafoldproof}%
\isadelimproof
\isanewline
\endisadelimproof
\isanewline
\isacommand{lemma}\isamarkupfalse%
\ L{\isadigit{5}}{\isacharunderscore}Boiler{\isacharcolon}\isanewline
\ \ \isakeyword{assumes}\ h{\isadigit{1}}{\isacharcolon}{\isachardoublequoteopen}SteamBoiler\ x\ s\ y{\isachardoublequoteclose}\isanewline
\ \ \ \ \ \ \isakeyword{and}\ h{\isadigit{2}}{\isacharcolon}{\isachardoublequoteopen}ts\ x{\isachardoublequoteclose}\ \isanewline
\ \ \ \ \ \ \isakeyword{and}\ h{\isadigit{3}}{\isacharcolon}{\isachardoublequoteopen}hd\ {\isacharparenleft}x\ j{\isacharparenright}\ {\isacharequal}\ Zero{\isachardoublequoteclose}\isanewline
\ \ \isakeyword{shows}\ {\isachardoublequoteopen}{\isacharparenleft}hd\ {\isacharparenleft}s\ j{\isacharparenright}{\isacharparenright}\ {\isasymle}\ hd\ {\isacharparenleft}s\ {\isacharparenleft}Suc\ j{\isacharparenright}{\isacharparenright}\ {\isacharplus}\ {\isacharparenleft}{\isadigit{1}}{\isadigit{0}}{\isacharcolon}{\isacharcolon}nat{\isacharparenright}{\isachardoublequoteclose}\isanewline
\isadelimproof
\endisadelimproof
\isatagproof
\isacommand{proof}\isamarkupfalse%
\ {\isacharminus}\ \ \isanewline
\ \ \ \isacommand{from}\isamarkupfalse%
\ h{\isadigit{1}}\ \isakeyword{and}\ h{\isadigit{2}}\ \isacommand{obtain}\isamarkupfalse%
\ r\ \isakeyword{where}\isanewline
\ \ \ \ \ a{\isadigit{1}}{\isacharcolon}{\isachardoublequoteopen}r\ {\isasymle}\ {\isadigit{1}}{\isadigit{0}}{\isachardoublequoteclose}\ \ \isakeyword{and}\isanewline
\ \ \ \ \ a{\isadigit{2}}{\isacharcolon}{\isachardoublequoteopen}hd\ {\isacharparenleft}s\ {\isacharparenleft}Suc\ j{\isacharparenright}{\isacharparenright}\ {\isacharequal}\ {\isacharparenleft}if\ hd\ {\isacharparenleft}x\ j{\isacharparenright}\ {\isacharequal}\ Zero\ then\ hd\ {\isacharparenleft}s\ j{\isacharparenright}\ {\isacharminus}\ r\ else\ hd\ {\isacharparenleft}s\ j{\isacharparenright}\ {\isacharplus}\ r{\isacharparenright}{\isachardoublequoteclose}\ \isanewline
\ \ \ \ \ \isacommand{by}\isamarkupfalse%
\ {\isacharparenleft}simp\ add{\isacharcolon}\ SteamBoiler{\isacharunderscore}def{\isacharcomma}\ auto{\isacharparenright}\isanewline
\ \ \ \isacommand{from}\isamarkupfalse%
\ a{\isadigit{2}}\ \isakeyword{and}\ h{\isadigit{3}}\ \isacommand{have}\isamarkupfalse%
\ sg{\isadigit{1}}{\isacharcolon}{\isachardoublequoteopen}hd\ {\isacharparenleft}s\ {\isacharparenleft}Suc\ j{\isacharparenright}{\isacharparenright}\ {\isacharequal}\ hd\ {\isacharparenleft}s\ j{\isacharparenright}\ {\isacharminus}\ r{\isachardoublequoteclose}\ \isacommand{by}\isamarkupfalse%
\ simp\isanewline
\ \ \ \isacommand{from}\isamarkupfalse%
\ sg{\isadigit{1}}\ \isakeyword{and}\ a{\isadigit{1}}\ \isacommand{show}\isamarkupfalse%
\ {\isacharquery}thesis\ \isacommand{by}\isamarkupfalse%
\ auto\isanewline
\isacommand{qed}\isamarkupfalse%
\endisatagproof
{\isafoldproof}%
\isadelimproof
\isanewline
\endisadelimproof
\isanewline
\isanewline
\isacommand{lemma}\isamarkupfalse%
\ L{\isadigit{6}}{\isacharunderscore}Boiler{\isacharcolon}\isanewline
\ \ \isakeyword{assumes}\ h{\isadigit{1}}{\isacharcolon}{\isachardoublequoteopen}SteamBoiler\ x\ s\ y{\isachardoublequoteclose}\isanewline
\ \ \ \ \ \ \isakeyword{and}\ h{\isadigit{2}}{\isacharcolon}{\isachardoublequoteopen}ts\ x{\isachardoublequoteclose}\ \isanewline
\ \ \ \ \ \ \isakeyword{and}\ h{\isadigit{3}}{\isacharcolon}{\isachardoublequoteopen}hd\ {\isacharparenleft}x\ j{\isacharparenright}\ {\isacharequal}\ Zero{\isachardoublequoteclose}\isanewline
\ \ \isakeyword{shows}\ {\isachardoublequoteopen}{\isacharparenleft}hd\ {\isacharparenleft}s\ j{\isacharparenright}{\isacharparenright}\ {\isacharminus}\ {\isacharparenleft}{\isadigit{1}}{\isadigit{0}}{\isacharcolon}{\isacharcolon}nat{\isacharparenright}\ {\isasymle}\ hd\ {\isacharparenleft}s\ {\isacharparenleft}Suc\ j{\isacharparenright}{\isacharparenright}{\isachardoublequoteclose}\ \isanewline
\isadelimproof
\endisadelimproof
\isatagproof
\isacommand{proof}\isamarkupfalse%
\ {\isacharminus}\ \ \isanewline
\ \ \ \isacommand{from}\isamarkupfalse%
\ h{\isadigit{1}}\ \isakeyword{and}\ h{\isadigit{2}}\ \isacommand{obtain}\isamarkupfalse%
\ r\ \isakeyword{where}\isanewline
\ \ \ \ \ a{\isadigit{1}}{\isacharcolon}{\isachardoublequoteopen}r\ {\isasymle}\ {\isadigit{1}}{\isadigit{0}}{\isachardoublequoteclose}\ \ \isakeyword{and}\isanewline
\ \ \ \ \ a{\isadigit{2}}{\isacharcolon}{\isachardoublequoteopen}hd\ {\isacharparenleft}s\ {\isacharparenleft}Suc\ j{\isacharparenright}{\isacharparenright}\ {\isacharequal}\ {\isacharparenleft}if\ hd\ {\isacharparenleft}x\ j{\isacharparenright}\ {\isacharequal}\ Zero\ then\ hd\ {\isacharparenleft}s\ j{\isacharparenright}\ {\isacharminus}\ r\ else\ hd\ {\isacharparenleft}s\ j{\isacharparenright}\ {\isacharplus}\ r{\isacharparenright}{\isachardoublequoteclose}\ \isanewline
\ \ \ \ \ \isacommand{by}\isamarkupfalse%
\ {\isacharparenleft}simp\ add{\isacharcolon}\ SteamBoiler{\isacharunderscore}def{\isacharcomma}\ auto{\isacharparenright}\isanewline
\ \ \ \isacommand{from}\isamarkupfalse%
\ a{\isadigit{2}}\ \isakeyword{and}\ h{\isadigit{3}}\ \isacommand{have}\isamarkupfalse%
\ sg{\isadigit{1}}{\isacharcolon}{\isachardoublequoteopen}hd\ {\isacharparenleft}s\ {\isacharparenleft}Suc\ j{\isacharparenright}{\isacharparenright}\ {\isacharequal}\ hd\ {\isacharparenleft}s\ j{\isacharparenright}\ {\isacharminus}\ r{\isachardoublequoteclose}\ \isacommand{by}\isamarkupfalse%
\ simp\isanewline
\ \ \ \isacommand{from}\isamarkupfalse%
\ sg{\isadigit{1}}\ \isakeyword{and}\ a{\isadigit{1}}\ \isacommand{show}\isamarkupfalse%
\ {\isacharquery}thesis\ \isacommand{by}\isamarkupfalse%
\ auto\isanewline
\isacommand{qed}\isamarkupfalse%
\endisatagproof
{\isafoldproof}%
\isadelimproof
\isanewline
\endisadelimproof
\isanewline
\isanewline
\isanewline
\isacommand{lemma}\isamarkupfalse%
\ L{\isadigit{7}}{\isacharunderscore}Boiler{\isacharcolon}\isanewline
\ \ \isakeyword{assumes}\ h{\isadigit{1}}{\isacharcolon}{\isachardoublequoteopen}SteamBoiler\ x\ s\ y{\isachardoublequoteclose}\isanewline
\ \ \ \ \ \ \isakeyword{and}\ h{\isadigit{2}}{\isacharcolon}{\isachardoublequoteopen}ts\ x{\isachardoublequoteclose}\ \isanewline
\ \ \ \ \ \ \isakeyword{and}\ h{\isadigit{3}}{\isacharcolon}{\isachardoublequoteopen}hd\ {\isacharparenleft}x\ j{\isacharparenright}\ {\isasymnoteq}\ Zero{\isachardoublequoteclose}\isanewline
\ \ \isakeyword{shows}\ {\isachardoublequoteopen}{\isacharparenleft}hd\ {\isacharparenleft}s\ j{\isacharparenright}{\isacharparenright}\ {\isasymge}\ hd\ {\isacharparenleft}s\ {\isacharparenleft}Suc\ j{\isacharparenright}{\isacharparenright}\ {\isacharminus}\ {\isacharparenleft}{\isadigit{1}}{\isadigit{0}}{\isacharcolon}{\isacharcolon}nat{\isacharparenright}{\isachardoublequoteclose}\ \isanewline
\isadelimproof
\endisadelimproof
\isatagproof
\isacommand{using}\isamarkupfalse%
\ assms\isanewline
\isacommand{proof}\isamarkupfalse%
\ {\isacharminus}\ \ \isanewline
\ \ \ \isacommand{from}\isamarkupfalse%
\ h{\isadigit{1}}\ \isakeyword{and}\ h{\isadigit{2}}\ \isacommand{obtain}\isamarkupfalse%
\ r\ \isakeyword{where}\isanewline
\ \ \ \ \ a{\isadigit{1}}{\isacharcolon}{\isachardoublequoteopen}r\ {\isasymle}\ {\isadigit{1}}{\isadigit{0}}{\isachardoublequoteclose}\ \ \isakeyword{and}\isanewline
\ \ \ \ \ a{\isadigit{2}}{\isacharcolon}{\isachardoublequoteopen}hd\ {\isacharparenleft}s\ {\isacharparenleft}Suc\ j{\isacharparenright}{\isacharparenright}\ {\isacharequal}\ {\isacharparenleft}if\ hd\ {\isacharparenleft}x\ j{\isacharparenright}\ {\isacharequal}\ Zero\ then\ hd\ {\isacharparenleft}s\ j{\isacharparenright}\ {\isacharminus}\ r\ else\ hd\ {\isacharparenleft}s\ j{\isacharparenright}\ {\isacharplus}\ r{\isacharparenright}{\isachardoublequoteclose}\ \isanewline
\ \ \ \ \ \isacommand{by}\isamarkupfalse%
\ {\isacharparenleft}simp\ add{\isacharcolon}\ SteamBoiler{\isacharunderscore}def{\isacharcomma}\ auto{\isacharparenright}\isanewline
\ \ \ \isacommand{from}\isamarkupfalse%
\ a{\isadigit{2}}\ \isakeyword{and}\ h{\isadigit{3}}\ \isacommand{have}\isamarkupfalse%
\ sg{\isadigit{1}}{\isacharcolon}{\isachardoublequoteopen}hd\ {\isacharparenleft}s\ {\isacharparenleft}Suc\ j{\isacharparenright}{\isacharparenright}\ {\isacharequal}\ hd\ {\isacharparenleft}s\ j{\isacharparenright}\ {\isacharplus}\ r{\isachardoublequoteclose}\ \isacommand{by}\isamarkupfalse%
\ simp\isanewline
\ \ \ \isacommand{from}\isamarkupfalse%
\ sg{\isadigit{1}}\ \isakeyword{and}\ a{\isadigit{1}}\ \isacommand{show}\isamarkupfalse%
\ {\isacharquery}thesis\ \isacommand{by}\isamarkupfalse%
\ auto\isanewline
\isacommand{qed}\isamarkupfalse%
\endisatagproof
{\isafoldproof}%
\isadelimproof
\isanewline
\endisadelimproof
\isanewline
\isanewline
\isacommand{lemma}\isamarkupfalse%
\ L{\isadigit{8}}{\isacharunderscore}Boiler{\isacharcolon}\isanewline
\ \ \isakeyword{assumes}\ h{\isadigit{1}}{\isacharcolon}{\isachardoublequoteopen}SteamBoiler\ x\ s\ y{\isachardoublequoteclose}\isanewline
\ \ \ \ \ \ \isakeyword{and}\ h{\isadigit{2}}{\isacharcolon}{\isachardoublequoteopen}ts\ x{\isachardoublequoteclose}\ \isanewline
\ \ \ \ \ \ \isakeyword{and}\ h{\isadigit{3}}{\isacharcolon}{\isachardoublequoteopen}hd\ {\isacharparenleft}x\ j{\isacharparenright}\ {\isasymnoteq}\ Zero{\isachardoublequoteclose}\isanewline
\ \ \isakeyword{shows}\ {\isachardoublequoteopen}{\isacharparenleft}hd\ {\isacharparenleft}s\ j{\isacharparenright}{\isacharparenright}\ {\isacharplus}\ {\isacharparenleft}{\isadigit{1}}{\isadigit{0}}{\isacharcolon}{\isacharcolon}nat{\isacharparenright}\ {\isasymge}\ hd\ {\isacharparenleft}s\ {\isacharparenleft}Suc\ j{\isacharparenright}{\isacharparenright}{\isachardoublequoteclose}\ \isanewline
\isadelimproof
\endisadelimproof
\isatagproof
\isacommand{proof}\isamarkupfalse%
\ {\isacharminus}\ \ \isanewline
\ \ \ \isacommand{from}\isamarkupfalse%
\ h{\isadigit{1}}\ \isakeyword{and}\ h{\isadigit{2}}\ \isacommand{obtain}\isamarkupfalse%
\ r\ \isakeyword{where}\isanewline
\ \ \ \ \ a{\isadigit{1}}{\isacharcolon}{\isachardoublequoteopen}r\ {\isasymle}\ {\isadigit{1}}{\isadigit{0}}{\isachardoublequoteclose}\ \ \isakeyword{and}\isanewline
\ \ \ \ \ a{\isadigit{2}}{\isacharcolon}{\isachardoublequoteopen}hd\ {\isacharparenleft}s\ {\isacharparenleft}Suc\ j{\isacharparenright}{\isacharparenright}\ {\isacharequal}\ {\isacharparenleft}if\ hd\ {\isacharparenleft}x\ j{\isacharparenright}\ {\isacharequal}\ Zero\ then\ hd\ {\isacharparenleft}s\ j{\isacharparenright}\ {\isacharminus}\ r\ else\ hd\ {\isacharparenleft}s\ j{\isacharparenright}\ {\isacharplus}\ r{\isacharparenright}{\isachardoublequoteclose}\ \isanewline
\ \ \ \ \ \isacommand{by}\isamarkupfalse%
\ {\isacharparenleft}simp\ add{\isacharcolon}\ SteamBoiler{\isacharunderscore}def{\isacharcomma}\ auto{\isacharparenright}\isanewline
\ \ \ \isacommand{from}\isamarkupfalse%
\ a{\isadigit{2}}\ \isakeyword{and}\ h{\isadigit{3}}\ \isacommand{have}\isamarkupfalse%
\ sg{\isadigit{1}}{\isacharcolon}{\isachardoublequoteopen}hd\ {\isacharparenleft}s\ {\isacharparenleft}Suc\ j{\isacharparenright}{\isacharparenright}\ {\isacharequal}\ hd\ {\isacharparenleft}s\ j{\isacharparenright}\ {\isacharplus}\ r{\isachardoublequoteclose}\ \isacommand{by}\isamarkupfalse%
\ simp\isanewline
\ \ \ \isacommand{from}\isamarkupfalse%
\ sg{\isadigit{1}}\ \isakeyword{and}\ a{\isadigit{1}}\ \isacommand{show}\isamarkupfalse%
\ {\isacharquery}thesis\ \isacommand{by}\isamarkupfalse%
\ auto\isanewline
\isacommand{qed}\isamarkupfalse%
\endisatagproof
{\isafoldproof}%
\isadelimproof
\endisadelimproof
\isamarkupsubsection{Properties of the Controller Component%
}
\isamarkuptrue%
\isacommand{lemma}\isamarkupfalse%
\ L{\isadigit{1}}{\isacharunderscore}Controller{\isacharcolon}\isanewline
\ \ \isakeyword{assumes}\ h{\isadigit{1}}{\isacharcolon}{\isachardoublequoteopen}Controller{\isacharunderscore}L\ s\ {\isacharparenleft}fin{\isacharunderscore}inf{\isacharunderscore}append\ {\isacharbrackleft}Zero{\isacharbrackright}\ l{\isacharparenright}\ l\ z{\isachardoublequoteclose}\isanewline
\ \ \isakeyword{shows}\ \ \ \ \ \ {\isachardoublequoteopen}fin{\isacharunderscore}make{\isacharunderscore}untimed\ {\isacharparenleft}inf{\isacharunderscore}truncate\ z\ i{\isacharparenright}\ {\isasymnoteq}\ {\isacharbrackleft}{\isacharbrackright}{\isachardoublequoteclose}\isanewline
\isadelimproof
\endisadelimproof
\isatagproof
\isacommand{proof}\isamarkupfalse%
\ {\isacharminus}\isanewline
\ \ \isacommand{from}\isamarkupfalse%
\ h{\isadigit{1}}\ \isacommand{have}\isamarkupfalse%
\ {\isachardoublequoteopen}{\isasymforall}\ i{\isachardot}\ {\isadigit{0}}\ {\isasymle}\ i\ {\isasymlongrightarrow}\ fin{\isacharunderscore}make{\isacharunderscore}untimed\ {\isacharparenleft}inf{\isacharunderscore}truncate\ z\ i{\isacharparenright}\ {\isasymnoteq}\ {\isacharbrackleft}{\isacharbrackright}{\isachardoublequoteclose}\isanewline
\ \ \ \ \isacommand{by}\isamarkupfalse%
\ {\isacharparenleft}simp\ add{\isacharcolon}\ Controller{\isacharunderscore}L{\isacharunderscore}def\ fin{\isacharunderscore}make{\isacharunderscore}untimed{\isacharunderscore}inf{\isacharunderscore}truncate{\isacharunderscore}Nonempty{\isacharunderscore}all{\isadigit{0}}a{\isacharparenright}\isanewline
\ \ \isacommand{from}\isamarkupfalse%
\ this\ \isacommand{show}\isamarkupfalse%
\ {\isacharquery}thesis\ \isacommand{by}\isamarkupfalse%
\ simp\isanewline
\isacommand{qed}\isamarkupfalse%
\endisatagproof
{\isafoldproof}%
\isadelimproof
\ \isanewline
\endisadelimproof
\isanewline
\isanewline
\isacommand{lemma}\isamarkupfalse%
\ L{\isadigit{2}}{\isacharunderscore}Controller{\isacharunderscore}Zero{\isacharcolon}\isanewline
\ \ \isakeyword{assumes}\ h{\isadigit{1}}{\isacharcolon}{\isachardoublequoteopen}Controller{\isacharunderscore}L\ y\ {\isacharparenleft}fin{\isacharunderscore}inf{\isacharunderscore}append\ {\isacharbrackleft}Zero{\isacharbrackright}\ l{\isacharparenright}\ l\ z{\isachardoublequoteclose}\isanewline
\ \ \ \ \ \ \isakeyword{and}\ h{\isadigit{2}}{\isacharcolon}{\isachardoublequoteopen}l\ t\ {\isacharequal}\ Zero{\isachardoublequoteclose}\isanewline
\ \ \ \ \ \ \isakeyword{and}\ h{\isadigit{3}}{\isacharcolon}{\isachardoublequoteopen}{\isadigit{3}}{\isadigit{0}}{\isadigit{0}}\ {\isacharless}\ hd\ {\isacharparenleft}y\ {\isacharparenleft}Suc\ t{\isacharparenright}{\isacharparenright}{\isachardoublequoteclose}\isanewline
\ \ \isakeyword{shows}\ \ \ \ \ \ {\isachardoublequoteopen}z\ {\isacharparenleft}Suc\ t{\isacharparenright}\ {\isacharequal}\ {\isacharbrackleft}{\isacharbrackright}{\isachardoublequoteclose}\isanewline
\isadelimproof
\endisadelimproof
\isatagproof
\isacommand{proof}\isamarkupfalse%
\ {\isacharminus}\isanewline
\ \ \isacommand{from}\isamarkupfalse%
\ h{\isadigit{2}}\ \isacommand{have}\isamarkupfalse%
\ sg{\isadigit{1}}{\isacharcolon}{\isachardoublequoteopen}fin{\isacharunderscore}inf{\isacharunderscore}append\ {\isacharbrackleft}Zero{\isacharbrackright}\ l\ {\isacharparenleft}Suc\ t{\isacharparenright}\ {\isacharequal}\ Zero{\isachardoublequoteclose}\isanewline
\ \ \ \ \isacommand{by}\isamarkupfalse%
\ {\isacharparenleft}simp\ add{\isacharcolon}\ correct{\isacharunderscore}fin{\isacharunderscore}inf{\isacharunderscore}append{\isadigit{1}}{\isacharparenright}\isanewline
\ \ \isacommand{from}\isamarkupfalse%
\ h{\isadigit{1}}\ \isakeyword{and}\ sg{\isadigit{1}}\ \isakeyword{and}\ h{\isadigit{3}}\ \isacommand{show}\isamarkupfalse%
\ {\isacharquery}thesis\ \isacommand{by}\isamarkupfalse%
\ {\isacharparenleft}simp\ add{\isacharcolon}\ Controller{\isacharunderscore}L{\isacharunderscore}def{\isacharparenright}\isanewline
\isacommand{qed}\isamarkupfalse%
\endisatagproof
{\isafoldproof}%
\isadelimproof
\isanewline
\endisadelimproof
\isanewline
\isanewline
\isacommand{lemma}\isamarkupfalse%
\ L{\isadigit{2}}{\isacharunderscore}Controller{\isacharunderscore}One{\isacharcolon}\isanewline
\ \ \isakeyword{assumes}\ h{\isadigit{1}}{\isacharcolon}{\isachardoublequoteopen}Controller{\isacharunderscore}L\ y\ {\isacharparenleft}fin{\isacharunderscore}inf{\isacharunderscore}append\ {\isacharbrackleft}Zero{\isacharbrackright}\ l{\isacharparenright}\ l\ z{\isachardoublequoteclose}\isanewline
\ \ \ \ \ \ \isakeyword{and}\ h{\isadigit{2}}{\isacharcolon}{\isachardoublequoteopen}l\ t\ {\isacharequal}\ One{\isachardoublequoteclose}\isanewline
\ \ \ \ \ \ \isakeyword{and}\ h{\isadigit{3}}{\isacharcolon}{\isachardoublequoteopen}hd\ {\isacharparenleft}y\ {\isacharparenleft}Suc\ t{\isacharparenright}{\isacharparenright}\ {\isacharless}\ {\isadigit{7}}{\isadigit{0}}{\isadigit{0}}{\isachardoublequoteclose}\isanewline
\ \ \isakeyword{shows}\ \ \ \ \ \ {\isachardoublequoteopen}z\ {\isacharparenleft}Suc\ t{\isacharparenright}\ {\isacharequal}\ {\isacharbrackleft}{\isacharbrackright}{\isachardoublequoteclose}\isanewline
\isadelimproof
\endisadelimproof
\isatagproof
\isacommand{proof}\isamarkupfalse%
\ {\isacharminus}\isanewline
\ \ \isacommand{from}\isamarkupfalse%
\ h{\isadigit{2}}\ \isacommand{have}\isamarkupfalse%
\ sg{\isadigit{1}}{\isacharcolon}{\isachardoublequoteopen}fin{\isacharunderscore}inf{\isacharunderscore}append\ {\isacharbrackleft}Zero{\isacharbrackright}\ l\ {\isacharparenleft}Suc\ t{\isacharparenright}\ {\isasymnoteq}\ Zero{\isachardoublequoteclose}\isanewline
\ \ \ \ \isacommand{by}\isamarkupfalse%
\ {\isacharparenleft}simp\ add{\isacharcolon}\ correct{\isacharunderscore}fin{\isacharunderscore}inf{\isacharunderscore}append{\isadigit{1}}{\isacharparenright}\isanewline
\ \ \isacommand{from}\isamarkupfalse%
\ h{\isadigit{1}}\ \isakeyword{and}\ sg{\isadigit{1}}\ \isakeyword{and}\ h{\isadigit{3}}\ \isacommand{show}\isamarkupfalse%
\ {\isacharquery}thesis\ \isacommand{by}\isamarkupfalse%
\ {\isacharparenleft}simp\ add{\isacharcolon}\ Controller{\isacharunderscore}L{\isacharunderscore}def{\isacharparenright}\isanewline
\isacommand{qed}\isamarkupfalse%
\endisatagproof
{\isafoldproof}%
\isadelimproof
\isanewline
\endisadelimproof
\ \isanewline
\isanewline
\isanewline
\isacommand{lemma}\isamarkupfalse%
\ L{\isadigit{3}}{\isacharunderscore}Controller{\isacharunderscore}Zero{\isacharcolon}\isanewline
\ \ \isakeyword{assumes}\ h{\isadigit{1}}{\isacharcolon}{\isachardoublequoteopen}Controller{\isacharunderscore}L\ y\ {\isacharparenleft}fin{\isacharunderscore}inf{\isacharunderscore}append\ {\isacharbrackleft}Zero{\isacharbrackright}\ l{\isacharparenright}\ l\ z{\isachardoublequoteclose}\isanewline
\ \ \ \ \ \ \isakeyword{and}\ h{\isadigit{2}}{\isacharcolon}{\isachardoublequoteopen}l\ t\ {\isacharequal}\ Zero{\isachardoublequoteclose}\isanewline
\ \ \ \ \ \ \isakeyword{and}\ h{\isadigit{3}}{\isacharcolon}{\isachardoublequoteopen}{\isasymnot}\ {\isadigit{3}}{\isadigit{0}}{\isadigit{0}}\ {\isacharless}\ hd\ {\isacharparenleft}y\ {\isacharparenleft}Suc\ t{\isacharparenright}{\isacharparenright}{\isachardoublequoteclose}\isanewline
\ \ \isakeyword{shows}\ \ \ \ \ \ {\isachardoublequoteopen}z\ {\isacharparenleft}Suc\ t{\isacharparenright}\ {\isacharequal}\ {\isacharbrackleft}One{\isacharbrackright}{\isachardoublequoteclose}\isanewline
\isadelimproof
\endisadelimproof
\isatagproof
\isacommand{proof}\isamarkupfalse%
\ {\isacharminus}\isanewline
\ \ \isacommand{from}\isamarkupfalse%
\ h{\isadigit{2}}\ \isacommand{have}\isamarkupfalse%
\ sg{\isadigit{1}}{\isacharcolon}{\isachardoublequoteopen}fin{\isacharunderscore}inf{\isacharunderscore}append\ {\isacharbrackleft}Zero{\isacharbrackright}\ l\ {\isacharparenleft}Suc\ t{\isacharparenright}\ {\isacharequal}\ Zero{\isachardoublequoteclose}\isanewline
\ \ \ \ \isacommand{by}\isamarkupfalse%
\ {\isacharparenleft}simp\ add{\isacharcolon}\ correct{\isacharunderscore}fin{\isacharunderscore}inf{\isacharunderscore}append{\isadigit{1}}{\isacharparenright}\isanewline
\ \ \isacommand{from}\isamarkupfalse%
\ h{\isadigit{1}}\ \isakeyword{and}\ sg{\isadigit{1}}\ \isakeyword{and}\ h{\isadigit{3}}\ \isacommand{show}\isamarkupfalse%
\ {\isacharquery}thesis\ \isacommand{by}\isamarkupfalse%
\ {\isacharparenleft}simp\ add{\isacharcolon}\ Controller{\isacharunderscore}L{\isacharunderscore}def{\isacharparenright}\isanewline
\isacommand{qed}\isamarkupfalse%
\endisatagproof
{\isafoldproof}%
\isadelimproof
\isanewline
\endisadelimproof
\isanewline
\isacommand{lemma}\isamarkupfalse%
\ L{\isadigit{3}}{\isacharunderscore}Controller{\isacharunderscore}One{\isacharcolon}\isanewline
\ \ \isakeyword{assumes}\ h{\isadigit{1}}{\isacharcolon}{\isachardoublequoteopen}Controller{\isacharunderscore}L\ y\ {\isacharparenleft}fin{\isacharunderscore}inf{\isacharunderscore}append\ {\isacharbrackleft}Zero{\isacharbrackright}\ l{\isacharparenright}\ l\ z{\isachardoublequoteclose}\isanewline
\ \ \ \ \ \ \isakeyword{and}\ h{\isadigit{2}}{\isacharcolon}{\isachardoublequoteopen}l\ t\ {\isacharequal}\ One{\isachardoublequoteclose}\isanewline
\ \ \ \ \ \ \isakeyword{and}\ h{\isadigit{3}}{\isacharcolon}{\isachardoublequoteopen}{\isasymnot}\ hd\ {\isacharparenleft}y\ {\isacharparenleft}Suc\ t{\isacharparenright}{\isacharparenright}\ {\isacharless}\ {\isadigit{7}}{\isadigit{0}}{\isadigit{0}}{\isachardoublequoteclose}\isanewline
\ \ \isakeyword{shows}\ \ \ \ \ \ {\isachardoublequoteopen}z\ {\isacharparenleft}Suc\ t{\isacharparenright}\ {\isacharequal}\ {\isacharbrackleft}Zero{\isacharbrackright}{\isachardoublequoteclose}\isanewline
\isadelimproof
\endisadelimproof
\isatagproof
\isacommand{proof}\isamarkupfalse%
\ {\isacharminus}\isanewline
\ \ \isacommand{from}\isamarkupfalse%
\ h{\isadigit{2}}\ \isacommand{have}\isamarkupfalse%
\ sg{\isadigit{1}}{\isacharcolon}{\isachardoublequoteopen}fin{\isacharunderscore}inf{\isacharunderscore}append\ {\isacharbrackleft}Zero{\isacharbrackright}\ l\ {\isacharparenleft}Suc\ t{\isacharparenright}\ {\isasymnoteq}\ Zero{\isachardoublequoteclose}\isanewline
\ \ \ \ \isacommand{by}\isamarkupfalse%
\ {\isacharparenleft}simp\ add{\isacharcolon}\ correct{\isacharunderscore}fin{\isacharunderscore}inf{\isacharunderscore}append{\isadigit{1}}{\isacharparenright}\isanewline
\ \ \isacommand{from}\isamarkupfalse%
\ h{\isadigit{1}}\ \isakeyword{and}\ sg{\isadigit{1}}\ \isakeyword{and}\ h{\isadigit{3}}\ \isacommand{show}\isamarkupfalse%
\ {\isacharquery}thesis\ \isacommand{by}\isamarkupfalse%
\ {\isacharparenleft}simp\ add{\isacharcolon}\ Controller{\isacharunderscore}L{\isacharunderscore}def{\isacharparenright}\isanewline
\isacommand{qed}\isamarkupfalse%
\endisatagproof
{\isafoldproof}%
\isadelimproof
\ \isanewline
\endisadelimproof
\isanewline
\isanewline
\isacommand{lemma}\isamarkupfalse%
\ L{\isadigit{4}}{\isacharunderscore}Controller{\isacharunderscore}Zero{\isacharcolon}\isanewline
\ \ \isakeyword{assumes}\ h{\isadigit{1}}{\isacharcolon}{\isachardoublequoteopen}Controller{\isacharunderscore}L\ y\ {\isacharparenleft}fin{\isacharunderscore}inf{\isacharunderscore}append\ {\isacharbrackleft}Zero{\isacharbrackright}\ l{\isacharparenright}\ l\ z{\isachardoublequoteclose}\isanewline
\ \ \ \ \ \ \isakeyword{and}\ h{\isadigit{2}}{\isacharcolon}{\isachardoublequoteopen}l\ {\isacharparenleft}Suc\ t{\isacharparenright}\ {\isacharequal}\ Zero{\isachardoublequoteclose}\ \isanewline
\ \ \isakeyword{shows}\ \ \ \ \ \ {\isachardoublequoteopen}{\isacharparenleft}z\ {\isacharparenleft}Suc\ t{\isacharparenright}\ {\isacharequal}\ {\isacharbrackleft}{\isacharbrackright}\ {\isasymand}\ l\ t\ {\isacharequal}\ Zero{\isacharparenright}\ {\isasymor}\ {\isacharparenleft}z\ {\isacharparenleft}Suc\ t{\isacharparenright}\ {\isacharequal}\ {\isacharbrackleft}Zero{\isacharbrackright}\ {\isasymand}\ l\ t\ {\isacharequal}\ One{\isacharparenright}{\isachardoublequoteclose}\isanewline
\isadelimproof
\endisadelimproof
\isatagproof
\isacommand{proof}\isamarkupfalse%
\ {\isacharparenleft}cases\ {\isachardoublequoteopen}l\ t{\isachardoublequoteclose}{\isacharparenright}\isanewline
\ \ \isacommand{assume}\isamarkupfalse%
\ a{\isadigit{1}}{\isacharcolon}{\isachardoublequoteopen}l\ t\ {\isacharequal}\ Zero{\isachardoublequoteclose}\isanewline
\ \ \isacommand{from}\isamarkupfalse%
\ this\ \isakeyword{and}\ h{\isadigit{1}}\ \isakeyword{and}\ h{\isadigit{2}}\ \isacommand{show}\isamarkupfalse%
\ {\isacharquery}thesis\isanewline
\ \ \isacommand{proof}\isamarkupfalse%
\ {\isacharminus}\isanewline
\ \ \ \ \isacommand{from}\isamarkupfalse%
\ a{\isadigit{1}}\ \isacommand{have}\isamarkupfalse%
\ sg{\isadigit{1}}{\isacharcolon}{\isachardoublequoteopen}fin{\isacharunderscore}inf{\isacharunderscore}append\ {\isacharbrackleft}Zero{\isacharbrackright}\ l\ {\isacharparenleft}Suc\ t{\isacharparenright}\ {\isacharequal}\ Zero{\isachardoublequoteclose}\isanewline
\ \ \ \ \ \ \isacommand{by}\isamarkupfalse%
\ {\isacharparenleft}simp\ add{\isacharcolon}\ correct{\isacharunderscore}fin{\isacharunderscore}inf{\isacharunderscore}append{\isadigit{1}}{\isacharparenright}\isanewline
\ \ \ \ \isacommand{from}\isamarkupfalse%
\ h{\isadigit{1}}\ \isakeyword{and}\ sg{\isadigit{1}}\ \isacommand{have}\isamarkupfalse%
\ sg{\isadigit{2}}{\isacharcolon}\isanewline
\ \ \ \ \ \ {\isachardoublequoteopen}if\ {\isadigit{3}}{\isadigit{0}}{\isadigit{0}}\ {\isacharless}\ hd\ {\isacharparenleft}y\ {\isacharparenleft}Suc\ t{\isacharparenright}{\isacharparenright}\ \isanewline
\ \ \ \ \ \ \ then\ z\ {\isacharparenleft}Suc\ t{\isacharparenright}\ {\isacharequal}\ {\isacharbrackleft}{\isacharbrackright}\ {\isasymand}\ l\ {\isacharparenleft}Suc\ t{\isacharparenright}\ {\isacharequal}\ Zero\ \isanewline
\ \ \ \ \ \ \ else\ z\ {\isacharparenleft}Suc\ t{\isacharparenright}\ {\isacharequal}\ {\isacharbrackleft}One{\isacharbrackright}\ {\isasymand}\ l\ {\isacharparenleft}Suc\ t{\isacharparenright}\ {\isacharequal}\ One{\isachardoublequoteclose}\isanewline
\ \ \ \ \ \ \ \isacommand{by}\isamarkupfalse%
\ {\isacharparenleft}simp\ add{\isacharcolon}\ Controller{\isacharunderscore}L{\isacharunderscore}def{\isacharparenright}\isanewline
\ \ \ \ \isacommand{show}\isamarkupfalse%
\ {\isacharquery}thesis\isanewline
\ \ \ \ \isacommand{proof}\isamarkupfalse%
\ {\isacharparenleft}cases\ {\isachardoublequoteopen}{\isadigit{3}}{\isadigit{0}}{\isadigit{0}}\ {\isacharless}\ hd\ {\isacharparenleft}y\ {\isacharparenleft}Suc\ t{\isacharparenright}{\isacharparenright}{\isachardoublequoteclose}{\isacharparenright}\isanewline
\ \ \ \ \ \ \isacommand{assume}\isamarkupfalse%
\ a{\isadigit{1}}{\isadigit{1}}{\isacharcolon}{\isachardoublequoteopen}{\isadigit{3}}{\isadigit{0}}{\isadigit{0}}\ {\isacharless}\ hd\ {\isacharparenleft}y\ {\isacharparenleft}Suc\ t{\isacharparenright}{\isacharparenright}{\isachardoublequoteclose}\isanewline
\ \ \ \ \ \ \isacommand{from}\isamarkupfalse%
\ a{\isadigit{1}}{\isadigit{1}}\ \isakeyword{and}\ sg{\isadigit{2}}\ \isacommand{have}\isamarkupfalse%
\ sg{\isadigit{3}}{\isacharcolon}{\isachardoublequoteopen}z\ {\isacharparenleft}Suc\ t{\isacharparenright}\ {\isacharequal}\ {\isacharbrackleft}{\isacharbrackright}\ {\isasymand}\ l\ {\isacharparenleft}Suc\ t{\isacharparenright}\ {\isacharequal}\ Zero{\isachardoublequoteclose}\ \isacommand{by}\isamarkupfalse%
\ simp\isanewline
\ \ \ \ \ \ \isacommand{from}\isamarkupfalse%
\ this\ \isakeyword{and}\ a{\isadigit{1}}\ \isacommand{show}\isamarkupfalse%
\ {\isacharquery}thesis\ \isacommand{by}\isamarkupfalse%
\ simp\isanewline
\ \ \ \ \isacommand{next}\isamarkupfalse%
\isanewline
\ \ \ \ \ \ \isacommand{assume}\isamarkupfalse%
\ a{\isadigit{1}}{\isadigit{2}}{\isacharcolon}{\isachardoublequoteopen}{\isasymnot}\ {\isadigit{3}}{\isadigit{0}}{\isadigit{0}}\ {\isacharless}\ hd\ {\isacharparenleft}y\ {\isacharparenleft}Suc\ t{\isacharparenright}{\isacharparenright}{\isachardoublequoteclose}\isanewline
\ \ \ \ \ \ \isacommand{from}\isamarkupfalse%
\ a{\isadigit{1}}{\isadigit{2}}\ \isakeyword{and}\ sg{\isadigit{2}}\ \isacommand{have}\isamarkupfalse%
\ sg{\isadigit{4}}{\isacharcolon}{\isachardoublequoteopen}z\ {\isacharparenleft}Suc\ t{\isacharparenright}\ {\isacharequal}\ {\isacharbrackleft}One{\isacharbrackright}\ {\isasymand}\ l\ {\isacharparenleft}Suc\ t{\isacharparenright}\ {\isacharequal}\ One{\isachardoublequoteclose}\ \isacommand{by}\isamarkupfalse%
\ simp\isanewline
\ \ \ \ \ \ \isacommand{from}\isamarkupfalse%
\ this\ \isakeyword{and}\ h{\isadigit{2}}\ \isacommand{show}\isamarkupfalse%
\ {\isacharquery}thesis\ \isacommand{by}\isamarkupfalse%
\ simp\isanewline
\ \ \ \ \isacommand{qed}\isamarkupfalse%
\isanewline
\ \ \isacommand{qed}\isamarkupfalse%
\isanewline
\isacommand{next}\isamarkupfalse%
\isanewline
\ \ \isacommand{assume}\isamarkupfalse%
\ a{\isadigit{2}}{\isacharcolon}{\isachardoublequoteopen}l\ t\ {\isacharequal}\ One{\isachardoublequoteclose}\isanewline
\ \ \isacommand{from}\isamarkupfalse%
\ this\ \isakeyword{and}\ h{\isadigit{1}}\ \isakeyword{and}\ h{\isadigit{2}}\ \isacommand{show}\isamarkupfalse%
\ {\isacharquery}thesis\isanewline
\ \ \isacommand{proof}\isamarkupfalse%
\ {\isacharminus}\isanewline
\ \ \ \ \isacommand{from}\isamarkupfalse%
\ a{\isadigit{2}}\ \isacommand{have}\isamarkupfalse%
\ sg{\isadigit{5}}{\isacharcolon}{\isachardoublequoteopen}fin{\isacharunderscore}inf{\isacharunderscore}append\ {\isacharbrackleft}Zero{\isacharbrackright}\ l\ {\isacharparenleft}Suc\ t{\isacharparenright}\ {\isasymnoteq}\ Zero{\isachardoublequoteclose}\isanewline
\ \ \ \ \ \ \isacommand{by}\isamarkupfalse%
\ {\isacharparenleft}simp\ add{\isacharcolon}\ correct{\isacharunderscore}fin{\isacharunderscore}inf{\isacharunderscore}append{\isadigit{1}}{\isacharparenright}\isanewline
\ \ \ \ \isacommand{from}\isamarkupfalse%
\ h{\isadigit{1}}\ \isakeyword{and}\ sg{\isadigit{5}}\ \isacommand{have}\isamarkupfalse%
\ sg{\isadigit{6}}{\isacharcolon}\isanewline
\ \ \ \ \ \ {\isachardoublequoteopen}if\ hd\ {\isacharparenleft}y\ {\isacharparenleft}Suc\ t{\isacharparenright}{\isacharparenright}\ {\isacharless}\ {\isadigit{7}}{\isadigit{0}}{\isadigit{0}}\ \isanewline
\ \ \ \ \ \ \ then\ z\ {\isacharparenleft}Suc\ t{\isacharparenright}\ {\isacharequal}\ {\isacharbrackleft}{\isacharbrackright}\ {\isasymand}\ l\ {\isacharparenleft}Suc\ t{\isacharparenright}\ {\isacharequal}\ One\ \isanewline
\ \ \ \ \ \ \ else\ z\ {\isacharparenleft}Suc\ t{\isacharparenright}\ {\isacharequal}\ {\isacharbrackleft}Zero{\isacharbrackright}\ {\isasymand}\ l\ {\isacharparenleft}Suc\ t{\isacharparenright}\ {\isacharequal}\ Zero{\isachardoublequoteclose}\isanewline
\ \ \ \ \ \ \ \isacommand{by}\isamarkupfalse%
\ {\isacharparenleft}simp\ add{\isacharcolon}\ Controller{\isacharunderscore}L{\isacharunderscore}def{\isacharparenright}\isanewline
\ \ \ \ \isacommand{show}\isamarkupfalse%
\ {\isacharquery}thesis\isanewline
\ \ \ \ \isacommand{proof}\isamarkupfalse%
\ {\isacharparenleft}cases\ {\isachardoublequoteopen}hd\ {\isacharparenleft}y\ {\isacharparenleft}Suc\ t{\isacharparenright}{\isacharparenright}\ {\isacharless}\ {\isadigit{7}}{\isadigit{0}}{\isadigit{0}}{\isachardoublequoteclose}{\isacharparenright}\isanewline
\ \ \ \ \ \ \isacommand{assume}\isamarkupfalse%
\ a{\isadigit{2}}{\isadigit{1}}{\isacharcolon}{\isachardoublequoteopen}hd\ {\isacharparenleft}y\ {\isacharparenleft}Suc\ t{\isacharparenright}{\isacharparenright}\ {\isacharless}\ {\isadigit{7}}{\isadigit{0}}{\isadigit{0}}{\isachardoublequoteclose}\isanewline
\ \ \ \ \ \ \isacommand{from}\isamarkupfalse%
\ a{\isadigit{2}}{\isadigit{1}}\ \isakeyword{and}\ sg{\isadigit{6}}\ \isacommand{have}\isamarkupfalse%
\ sg{\isadigit{7}}{\isacharcolon}{\isachardoublequoteopen}z\ {\isacharparenleft}Suc\ t{\isacharparenright}\ {\isacharequal}\ {\isacharbrackleft}{\isacharbrackright}\ {\isasymand}\ l\ {\isacharparenleft}Suc\ t{\isacharparenright}\ {\isacharequal}\ One{\isachardoublequoteclose}\ \isacommand{by}\isamarkupfalse%
\ simp\isanewline
\ \ \ \ \ \ \isacommand{from}\isamarkupfalse%
\ this\ \isakeyword{and}\ h{\isadigit{2}}\ \isacommand{show}\isamarkupfalse%
\ {\isacharquery}thesis\ \isacommand{by}\isamarkupfalse%
\ simp\isanewline
\ \ \ \ \isacommand{next}\isamarkupfalse%
\ \isanewline
\ \ \ \ \ \ \isacommand{assume}\isamarkupfalse%
\ a{\isadigit{2}}{\isadigit{2}}{\isacharcolon}{\isachardoublequoteopen}{\isasymnot}\ hd\ {\isacharparenleft}y\ {\isacharparenleft}Suc\ t{\isacharparenright}{\isacharparenright}\ {\isacharless}\ {\isadigit{7}}{\isadigit{0}}{\isadigit{0}}{\isachardoublequoteclose}\isanewline
\ \ \ \ \ \ \isacommand{from}\isamarkupfalse%
\ a{\isadigit{2}}{\isadigit{2}}\ \isakeyword{and}\ sg{\isadigit{6}}\ \isacommand{have}\isamarkupfalse%
\ sg{\isadigit{8}}{\isacharcolon}{\isachardoublequoteopen}z\ {\isacharparenleft}Suc\ t{\isacharparenright}\ {\isacharequal}\ {\isacharbrackleft}Zero{\isacharbrackright}\ {\isasymand}\ l\ {\isacharparenleft}Suc\ t{\isacharparenright}\ {\isacharequal}\ Zero{\isachardoublequoteclose}\ \isacommand{by}\isamarkupfalse%
\ simp\isanewline
\ \ \ \ \ \ \isacommand{from}\isamarkupfalse%
\ this\ \isakeyword{and}\ a{\isadigit{2}}\ \isacommand{show}\isamarkupfalse%
\ {\isacharquery}thesis\ \isacommand{by}\isamarkupfalse%
\ simp\isanewline
\ \ \ \ \isacommand{qed}\isamarkupfalse%
\isanewline
\ \ \isacommand{qed}\isamarkupfalse%
\isanewline
\isacommand{qed}\isamarkupfalse%
\endisatagproof
{\isafoldproof}%
\isadelimproof
\isanewline
\endisadelimproof
\isanewline
\isanewline
\isacommand{lemma}\isamarkupfalse%
\ L{\isadigit{4}}{\isacharunderscore}Controller{\isacharunderscore}One{\isacharcolon}\isanewline
\ \ \isakeyword{assumes}\ h{\isadigit{1}}{\isacharcolon}{\isachardoublequoteopen}Controller{\isacharunderscore}L\ y\ {\isacharparenleft}fin{\isacharunderscore}inf{\isacharunderscore}append\ {\isacharbrackleft}Zero{\isacharbrackright}\ l{\isacharparenright}\ l\ z{\isachardoublequoteclose}\isanewline
\ \ \ \ \ \ \isakeyword{and}\ h{\isadigit{2}}{\isacharcolon}{\isachardoublequoteopen}l\ {\isacharparenleft}Suc\ t{\isacharparenright}\ {\isacharequal}\ One{\isachardoublequoteclose}\ \isanewline
\ \ \isakeyword{shows}\ \ \ \ \ \ {\isachardoublequoteopen}{\isacharparenleft}z\ {\isacharparenleft}Suc\ t{\isacharparenright}\ {\isacharequal}\ {\isacharbrackleft}{\isacharbrackright}\ {\isasymand}\ l\ t\ {\isacharequal}\ One{\isacharparenright}\ {\isasymor}\ {\isacharparenleft}z\ {\isacharparenleft}Suc\ t{\isacharparenright}\ {\isacharequal}\ {\isacharbrackleft}One{\isacharbrackright}\ {\isasymand}\ l\ t\ {\isacharequal}\ Zero{\isacharparenright}{\isachardoublequoteclose}\isanewline
\isadelimproof
\endisadelimproof
\isatagproof
\isacommand{proof}\isamarkupfalse%
\ {\isacharparenleft}cases\ {\isachardoublequoteopen}l\ t{\isachardoublequoteclose}{\isacharparenright}\isanewline
\ \ \isacommand{assume}\isamarkupfalse%
\ a{\isadigit{1}}{\isacharcolon}{\isachardoublequoteopen}l\ t\ {\isacharequal}\ Zero{\isachardoublequoteclose}\isanewline
\ \ \isacommand{from}\isamarkupfalse%
\ this\ \isakeyword{and}\ h{\isadigit{1}}\ \isakeyword{and}\ h{\isadigit{2}}\ \isacommand{show}\isamarkupfalse%
\ {\isacharquery}thesis\isanewline
\ \ \isacommand{proof}\isamarkupfalse%
\ {\isacharminus}\isanewline
\ \ \ \ \isacommand{from}\isamarkupfalse%
\ a{\isadigit{1}}\ \isacommand{have}\isamarkupfalse%
\ sg{\isadigit{1}}{\isacharcolon}{\isachardoublequoteopen}fin{\isacharunderscore}inf{\isacharunderscore}append\ {\isacharbrackleft}Zero{\isacharbrackright}\ l\ {\isacharparenleft}Suc\ t{\isacharparenright}\ {\isacharequal}\ Zero{\isachardoublequoteclose}\isanewline
\ \ \ \ \ \ \isacommand{by}\isamarkupfalse%
\ {\isacharparenleft}simp\ add{\isacharcolon}\ correct{\isacharunderscore}fin{\isacharunderscore}inf{\isacharunderscore}append{\isadigit{1}}{\isacharparenright}\isanewline
\ \ \ \ \isacommand{from}\isamarkupfalse%
\ h{\isadigit{1}}\ \isakeyword{and}\ sg{\isadigit{1}}\ \isacommand{have}\isamarkupfalse%
\ sg{\isadigit{2}}{\isacharcolon}\isanewline
\ \ \ \ \ \ {\isachardoublequoteopen}if\ {\isadigit{3}}{\isadigit{0}}{\isadigit{0}}\ {\isacharless}\ hd\ {\isacharparenleft}y\ {\isacharparenleft}Suc\ t{\isacharparenright}{\isacharparenright}\ \isanewline
\ \ \ \ \ \ \ then\ z\ {\isacharparenleft}Suc\ t{\isacharparenright}\ {\isacharequal}\ {\isacharbrackleft}{\isacharbrackright}\ {\isasymand}\ l\ {\isacharparenleft}Suc\ t{\isacharparenright}\ {\isacharequal}\ Zero\ \isanewline
\ \ \ \ \ \ \ else\ z\ {\isacharparenleft}Suc\ t{\isacharparenright}\ {\isacharequal}\ {\isacharbrackleft}One{\isacharbrackright}\ {\isasymand}\ l\ {\isacharparenleft}Suc\ t{\isacharparenright}\ {\isacharequal}\ One{\isachardoublequoteclose}\isanewline
\ \ \ \ \ \ \ \isacommand{by}\isamarkupfalse%
\ {\isacharparenleft}simp\ add{\isacharcolon}\ Controller{\isacharunderscore}L{\isacharunderscore}def{\isacharparenright}\isanewline
\ \ \ \ \isacommand{show}\isamarkupfalse%
\ {\isacharquery}thesis\isanewline
\ \ \ \ \isacommand{proof}\isamarkupfalse%
\ {\isacharparenleft}cases\ {\isachardoublequoteopen}{\isadigit{3}}{\isadigit{0}}{\isadigit{0}}\ {\isacharless}\ hd\ {\isacharparenleft}y\ {\isacharparenleft}Suc\ t{\isacharparenright}{\isacharparenright}{\isachardoublequoteclose}{\isacharparenright}\isanewline
\ \ \ \ \ \ \isacommand{assume}\isamarkupfalse%
\ a{\isadigit{1}}{\isadigit{1}}{\isacharcolon}{\isachardoublequoteopen}{\isadigit{3}}{\isadigit{0}}{\isadigit{0}}\ {\isacharless}\ hd\ {\isacharparenleft}y\ {\isacharparenleft}Suc\ t{\isacharparenright}{\isacharparenright}{\isachardoublequoteclose}\isanewline
\ \ \ \ \ \ \isacommand{from}\isamarkupfalse%
\ a{\isadigit{1}}{\isadigit{1}}\ \isakeyword{and}\ sg{\isadigit{2}}\ \isacommand{have}\isamarkupfalse%
\ sg{\isadigit{3}}{\isacharcolon}{\isachardoublequoteopen}z\ {\isacharparenleft}Suc\ t{\isacharparenright}\ {\isacharequal}\ {\isacharbrackleft}{\isacharbrackright}\ {\isasymand}\ l\ {\isacharparenleft}Suc\ t{\isacharparenright}\ {\isacharequal}\ Zero{\isachardoublequoteclose}\ \isacommand{by}\isamarkupfalse%
\ simp\isanewline
\ \ \ \ \ \ \isacommand{from}\isamarkupfalse%
\ this\ \isakeyword{and}\ h{\isadigit{2}}\ \isacommand{show}\isamarkupfalse%
\ {\isacharquery}thesis\ \isacommand{by}\isamarkupfalse%
\ simp\isanewline
\ \ \ \ \isacommand{next}\isamarkupfalse%
\isanewline
\ \ \ \ \ \ \isacommand{assume}\isamarkupfalse%
\ a{\isadigit{1}}{\isadigit{2}}{\isacharcolon}{\isachardoublequoteopen}{\isasymnot}\ {\isadigit{3}}{\isadigit{0}}{\isadigit{0}}\ {\isacharless}\ hd\ {\isacharparenleft}y\ {\isacharparenleft}Suc\ t{\isacharparenright}{\isacharparenright}{\isachardoublequoteclose}\isanewline
\ \ \ \ \ \ \isacommand{from}\isamarkupfalse%
\ a{\isadigit{1}}{\isadigit{2}}\ \isakeyword{and}\ sg{\isadigit{2}}\ \isacommand{have}\isamarkupfalse%
\ sg{\isadigit{4}}{\isacharcolon}{\isachardoublequoteopen}z\ {\isacharparenleft}Suc\ t{\isacharparenright}\ {\isacharequal}\ {\isacharbrackleft}One{\isacharbrackright}\ {\isasymand}\ l\ {\isacharparenleft}Suc\ t{\isacharparenright}\ {\isacharequal}\ One{\isachardoublequoteclose}\ \isacommand{by}\isamarkupfalse%
\ simp\isanewline
\ \ \ \ \ \ \isacommand{from}\isamarkupfalse%
\ this\ \isakeyword{and}\ a{\isadigit{1}}\ \isacommand{show}\isamarkupfalse%
\ {\isacharquery}thesis\ \isacommand{by}\isamarkupfalse%
\ simp\isanewline
\ \ \ \ \isacommand{qed}\isamarkupfalse%
\isanewline
\ \ \isacommand{qed}\isamarkupfalse%
\isanewline
\isacommand{next}\isamarkupfalse%
\isanewline
\ \ \isacommand{assume}\isamarkupfalse%
\ a{\isadigit{2}}{\isacharcolon}{\isachardoublequoteopen}l\ t\ {\isacharequal}\ One{\isachardoublequoteclose}\isanewline
\ \ \isacommand{from}\isamarkupfalse%
\ this\ \isakeyword{and}\ h{\isadigit{1}}\ \isakeyword{and}\ h{\isadigit{2}}\ \isacommand{show}\isamarkupfalse%
\ {\isacharquery}thesis\isanewline
\ \ \isacommand{proof}\isamarkupfalse%
\ {\isacharminus}\isanewline
\ \ \ \ \isacommand{from}\isamarkupfalse%
\ a{\isadigit{2}}\ \isacommand{have}\isamarkupfalse%
\ sg{\isadigit{5}}{\isacharcolon}{\isachardoublequoteopen}fin{\isacharunderscore}inf{\isacharunderscore}append\ {\isacharbrackleft}Zero{\isacharbrackright}\ l\ {\isacharparenleft}Suc\ t{\isacharparenright}\ {\isasymnoteq}\ Zero{\isachardoublequoteclose}\isanewline
\ \ \ \ \ \ \isacommand{by}\isamarkupfalse%
\ {\isacharparenleft}simp\ add{\isacharcolon}\ correct{\isacharunderscore}fin{\isacharunderscore}inf{\isacharunderscore}append{\isadigit{1}}{\isacharparenright}\isanewline
\ \ \ \ \isacommand{from}\isamarkupfalse%
\ h{\isadigit{1}}\ \isakeyword{and}\ sg{\isadigit{5}}\ \isacommand{have}\isamarkupfalse%
\ sg{\isadigit{6}}{\isacharcolon}\isanewline
\ \ \ \ \ \ {\isachardoublequoteopen}if\ hd\ {\isacharparenleft}y\ {\isacharparenleft}Suc\ t{\isacharparenright}{\isacharparenright}\ {\isacharless}\ {\isadigit{7}}{\isadigit{0}}{\isadigit{0}}\ \isanewline
\ \ \ \ \ \ \ then\ z\ {\isacharparenleft}Suc\ t{\isacharparenright}\ {\isacharequal}\ {\isacharbrackleft}{\isacharbrackright}\ {\isasymand}\ l\ {\isacharparenleft}Suc\ t{\isacharparenright}\ {\isacharequal}\ One\ \isanewline
\ \ \ \ \ \ \ else\ z\ {\isacharparenleft}Suc\ t{\isacharparenright}\ {\isacharequal}\ {\isacharbrackleft}Zero{\isacharbrackright}\ {\isasymand}\ l\ {\isacharparenleft}Suc\ t{\isacharparenright}\ {\isacharequal}\ Zero{\isachardoublequoteclose}\isanewline
\ \ \ \ \ \ \ \isacommand{by}\isamarkupfalse%
\ {\isacharparenleft}simp\ add{\isacharcolon}\ Controller{\isacharunderscore}L{\isacharunderscore}def{\isacharparenright}\isanewline
\ \ \ \ \isacommand{show}\isamarkupfalse%
\ {\isacharquery}thesis\isanewline
\ \ \ \ \isacommand{proof}\isamarkupfalse%
\ {\isacharparenleft}cases\ {\isachardoublequoteopen}hd\ {\isacharparenleft}y\ {\isacharparenleft}Suc\ t{\isacharparenright}{\isacharparenright}\ {\isacharless}\ {\isadigit{7}}{\isadigit{0}}{\isadigit{0}}{\isachardoublequoteclose}{\isacharparenright}\isanewline
\ \ \ \ \ \ \isacommand{assume}\isamarkupfalse%
\ a{\isadigit{2}}{\isadigit{1}}{\isacharcolon}{\isachardoublequoteopen}hd\ {\isacharparenleft}y\ {\isacharparenleft}Suc\ t{\isacharparenright}{\isacharparenright}\ {\isacharless}\ {\isadigit{7}}{\isadigit{0}}{\isadigit{0}}{\isachardoublequoteclose}\isanewline
\ \ \ \ \ \ \isacommand{from}\isamarkupfalse%
\ a{\isadigit{2}}{\isadigit{1}}\ \isakeyword{and}\ sg{\isadigit{6}}\ \isacommand{have}\isamarkupfalse%
\ sg{\isadigit{7}}{\isacharcolon}{\isachardoublequoteopen}z\ {\isacharparenleft}Suc\ t{\isacharparenright}\ {\isacharequal}\ {\isacharbrackleft}{\isacharbrackright}\ {\isasymand}\ l\ {\isacharparenleft}Suc\ t{\isacharparenright}\ {\isacharequal}\ One{\isachardoublequoteclose}\ \isacommand{by}\isamarkupfalse%
\ simp\isanewline
\ \ \ \ \ \ \isacommand{from}\isamarkupfalse%
\ this\ \isakeyword{and}\ a{\isadigit{2}}\ \isacommand{show}\isamarkupfalse%
\ {\isacharquery}thesis\ \isacommand{by}\isamarkupfalse%
\ simp\isanewline
\ \ \ \ \isacommand{next}\isamarkupfalse%
\ \isanewline
\ \ \ \ \ \ \isacommand{assume}\isamarkupfalse%
\ a{\isadigit{2}}{\isadigit{2}}{\isacharcolon}{\isachardoublequoteopen}{\isasymnot}\ hd\ {\isacharparenleft}y\ {\isacharparenleft}Suc\ t{\isacharparenright}{\isacharparenright}\ {\isacharless}\ {\isadigit{7}}{\isadigit{0}}{\isadigit{0}}{\isachardoublequoteclose}\isanewline
\ \ \ \ \ \ \isacommand{from}\isamarkupfalse%
\ a{\isadigit{2}}{\isadigit{2}}\ \isakeyword{and}\ sg{\isadigit{6}}\ \isacommand{have}\isamarkupfalse%
\ sg{\isadigit{8}}{\isacharcolon}{\isachardoublequoteopen}z\ {\isacharparenleft}Suc\ t{\isacharparenright}\ {\isacharequal}\ {\isacharbrackleft}Zero{\isacharbrackright}\ {\isasymand}\ l\ {\isacharparenleft}Suc\ t{\isacharparenright}\ {\isacharequal}\ Zero{\isachardoublequoteclose}\ \isacommand{by}\isamarkupfalse%
\ simp\isanewline
\ \ \ \ \ \ \isacommand{from}\isamarkupfalse%
\ this\ \isakeyword{and}\ h{\isadigit{2}}\ \isacommand{show}\isamarkupfalse%
\ {\isacharquery}thesis\ \isacommand{by}\isamarkupfalse%
\ simp\isanewline
\ \ \ \ \isacommand{qed}\isamarkupfalse%
\isanewline
\ \ \isacommand{qed}\isamarkupfalse%
\isanewline
\isacommand{qed}\isamarkupfalse%
\endisatagproof
{\isafoldproof}%
\isadelimproof
\isanewline
\endisadelimproof
\isanewline
\isacommand{lemma}\isamarkupfalse%
\ L{\isadigit{5}}{\isacharunderscore}Controller{\isacharunderscore}Zero{\isacharcolon}\isanewline
\ \ \isakeyword{assumes}\ h{\isadigit{1}}{\isacharcolon}{\isachardoublequoteopen}Controller{\isacharunderscore}L\ y\ lIn\ lOut\ z{\isachardoublequoteclose}\isanewline
\ \ \ \ \ \ \isakeyword{and}\ h{\isadigit{2}}{\isacharcolon}{\isachardoublequoteopen}lOut\ t\ {\isacharequal}\ Zero{\isachardoublequoteclose}\isanewline
\ \ \ \ \ \ \isakeyword{and}\ h{\isadigit{3}}{\isacharcolon}{\isachardoublequoteopen}z\ t\ {\isacharequal}\ {\isacharbrackleft}{\isacharbrackright}{\isachardoublequoteclose}\isanewline
\ \ \isakeyword{shows}\ {\isachardoublequoteopen}lIn\ t\ {\isacharequal}\ Zero{\isachardoublequoteclose}\ \isanewline
\isadelimproof
\endisadelimproof
\isatagproof
\isacommand{proof}\isamarkupfalse%
\ {\isacharparenleft}cases\ {\isachardoublequoteopen}lIn\ t{\isachardoublequoteclose}{\isacharparenright}\isanewline
\ \ \isacommand{assume}\isamarkupfalse%
\ a{\isadigit{1}}{\isacharcolon}{\isachardoublequoteopen}lIn\ t\ {\isacharequal}\ Zero{\isachardoublequoteclose}\isanewline
\ \ \isacommand{from}\isamarkupfalse%
\ this\ \isacommand{show}\isamarkupfalse%
\ {\isacharquery}thesis\ \isacommand{by}\isamarkupfalse%
\ simp\ \isanewline
\isacommand{next}\isamarkupfalse%
\isanewline
\ \ \isacommand{assume}\isamarkupfalse%
\ a{\isadigit{2}}{\isacharcolon}{\isachardoublequoteopen}lIn\ t\ {\isacharequal}\ One{\isachardoublequoteclose}\isanewline
\ \ \isacommand{from}\isamarkupfalse%
\ a{\isadigit{2}}\ \isakeyword{and}\ h{\isadigit{1}}\ \isacommand{have}\isamarkupfalse%
\ sg{\isadigit{1}}{\isacharcolon}\isanewline
\ \ \ \ {\isachardoublequoteopen}if\ hd\ {\isacharparenleft}y\ t{\isacharparenright}\ {\isacharless}\ {\isadigit{7}}{\isadigit{0}}{\isadigit{0}}\ \isanewline
\ \ \ \ \ then\ z\ t\ {\isacharequal}\ {\isacharbrackleft}{\isacharbrackright}\ {\isasymand}\ lOut\ t\ {\isacharequal}\ One\ \isanewline
\ \ \ \ \ else\ z\ t\ {\isacharequal}\ {\isacharbrackleft}Zero{\isacharbrackright}\ {\isasymand}\ lOut\ t\ {\isacharequal}\ Zero{\isachardoublequoteclose}\isanewline
\ \ \ \ \ \isacommand{by}\isamarkupfalse%
\ {\isacharparenleft}simp\ add{\isacharcolon}\ Controller{\isacharunderscore}L{\isacharunderscore}def{\isacharparenright}\isanewline
\ \ \isacommand{show}\isamarkupfalse%
\ {\isacharquery}thesis\isanewline
\ \ \isacommand{proof}\isamarkupfalse%
\ {\isacharparenleft}cases\ {\isachardoublequoteopen}hd\ {\isacharparenleft}y\ t{\isacharparenright}\ {\isacharless}\ {\isadigit{7}}{\isadigit{0}}{\isadigit{0}}{\isachardoublequoteclose}{\isacharparenright}\isanewline
\ \ \ \ \isacommand{assume}\isamarkupfalse%
\ a{\isadigit{3}}{\isacharcolon}{\isachardoublequoteopen}hd\ {\isacharparenleft}y\ t{\isacharparenright}\ {\isacharless}\ {\isadigit{7}}{\isadigit{0}}{\isadigit{0}}{\isachardoublequoteclose}\isanewline
\ \ \ \ \isacommand{from}\isamarkupfalse%
\ a{\isadigit{3}}\ \isakeyword{and}\ sg{\isadigit{1}}\ \isacommand{have}\isamarkupfalse%
\ sg{\isadigit{2}}{\isacharcolon}{\isachardoublequoteopen}z\ t\ {\isacharequal}\ {\isacharbrackleft}{\isacharbrackright}\ {\isasymand}\ lOut\ t\ {\isacharequal}\ One{\isachardoublequoteclose}\ \isacommand{by}\isamarkupfalse%
\ simp\isanewline
\ \ \ \ \isacommand{from}\isamarkupfalse%
\ this\ \isakeyword{and}\ h{\isadigit{2}}\ \isacommand{show}\isamarkupfalse%
\ {\isacharquery}thesis\ \isacommand{by}\isamarkupfalse%
\ simp\isanewline
\ \ \isacommand{next}\isamarkupfalse%
\isanewline
\ \ \ \ \isacommand{assume}\isamarkupfalse%
\ a{\isadigit{4}}{\isacharcolon}{\isachardoublequoteopen}{\isasymnot}\ hd\ {\isacharparenleft}y\ t{\isacharparenright}\ {\isacharless}\ {\isadigit{7}}{\isadigit{0}}{\isadigit{0}}{\isachardoublequoteclose}\isanewline
\ \ \ \ \isacommand{from}\isamarkupfalse%
\ a{\isadigit{4}}\ \isakeyword{and}\ sg{\isadigit{1}}\ \isacommand{have}\isamarkupfalse%
\ sg{\isadigit{3}}{\isacharcolon}{\isachardoublequoteopen}z\ t\ {\isacharequal}\ {\isacharbrackleft}Zero{\isacharbrackright}\ {\isasymand}\ lOut\ t\ {\isacharequal}\ Zero{\isachardoublequoteclose}\ \isacommand{by}\isamarkupfalse%
\ simp\isanewline
\ \ \ \ \isacommand{from}\isamarkupfalse%
\ this\ \isakeyword{and}\ h{\isadigit{3}}\ \isacommand{show}\isamarkupfalse%
\ {\isacharquery}thesis\ \isacommand{by}\isamarkupfalse%
\ simp\isanewline
\ \ \isacommand{qed}\isamarkupfalse%
\isanewline
\isacommand{qed}\isamarkupfalse%
\endisatagproof
{\isafoldproof}%
\isadelimproof
\isanewline
\endisadelimproof
\isanewline
\isanewline
\isacommand{lemma}\isamarkupfalse%
\ L{\isadigit{5}}{\isacharunderscore}Controller{\isacharunderscore}One{\isacharcolon}\isanewline
\ \ \isakeyword{assumes}\ h{\isadigit{1}}{\isacharcolon}{\isachardoublequoteopen}Controller{\isacharunderscore}L\ y\ lIn\ lOut\ z{\isachardoublequoteclose}\isanewline
\ \ \ \ \ \ \isakeyword{and}\ h{\isadigit{2}}{\isacharcolon}{\isachardoublequoteopen}lOut\ t\ {\isacharequal}\ One{\isachardoublequoteclose}\isanewline
\ \ \ \ \ \ \isakeyword{and}\ h{\isadigit{3}}{\isacharcolon}{\isachardoublequoteopen}z\ t\ {\isacharequal}\ {\isacharbrackleft}{\isacharbrackright}{\isachardoublequoteclose}\isanewline
\ \ \isakeyword{shows}\ {\isachardoublequoteopen}lIn\ t\ {\isacharequal}\ One{\isachardoublequoteclose}\ \isanewline
\isadelimproof
\endisadelimproof
\isatagproof
\isacommand{proof}\isamarkupfalse%
\ {\isacharparenleft}cases\ {\isachardoublequoteopen}lIn\ t{\isachardoublequoteclose}{\isacharparenright}\isanewline
\ \ \isacommand{assume}\isamarkupfalse%
\ a{\isadigit{1}}{\isacharcolon}{\isachardoublequoteopen}lIn\ t\ {\isacharequal}\ Zero{\isachardoublequoteclose}\isanewline
\ \ \isacommand{from}\isamarkupfalse%
\ a{\isadigit{1}}\ \isakeyword{and}\ h{\isadigit{1}}\ \isacommand{have}\isamarkupfalse%
\ sg{\isadigit{1}}{\isacharcolon}\isanewline
\ \ \ \ {\isachardoublequoteopen}if\ {\isadigit{3}}{\isadigit{0}}{\isadigit{0}}\ {\isacharless}\ hd\ {\isacharparenleft}y\ t{\isacharparenright}\ \isanewline
\ \ \ \ \ then\ z\ t\ {\isacharequal}\ {\isacharbrackleft}{\isacharbrackright}\ {\isasymand}\ lOut\ t\ {\isacharequal}\ Zero\ \isanewline
\ \ \ \ \ else\ z\ t\ {\isacharequal}\ {\isacharbrackleft}One{\isacharbrackright}\ {\isasymand}\ lOut\ t\ {\isacharequal}\ One{\isachardoublequoteclose}\isanewline
\ \ \ \ \ \isacommand{by}\isamarkupfalse%
\ {\isacharparenleft}simp\ add{\isacharcolon}\ Controller{\isacharunderscore}L{\isacharunderscore}def{\isacharparenright}\isanewline
\ \ \isacommand{show}\isamarkupfalse%
\ {\isacharquery}thesis\isanewline
\ \ \isacommand{proof}\isamarkupfalse%
\ {\isacharparenleft}cases\ {\isachardoublequoteopen}{\isadigit{3}}{\isadigit{0}}{\isadigit{0}}\ {\isacharless}\ hd\ {\isacharparenleft}y\ t{\isacharparenright}{\isachardoublequoteclose}{\isacharparenright}\isanewline
\ \ \ \ \isacommand{assume}\isamarkupfalse%
\ a{\isadigit{3}}{\isacharcolon}{\isachardoublequoteopen}{\isadigit{3}}{\isadigit{0}}{\isadigit{0}}\ {\isacharless}\ hd\ {\isacharparenleft}y\ t{\isacharparenright}{\isachardoublequoteclose}\isanewline
\ \ \ \ \isacommand{from}\isamarkupfalse%
\ a{\isadigit{3}}\ \isakeyword{and}\ sg{\isadigit{1}}\ \isacommand{have}\isamarkupfalse%
\ sg{\isadigit{2}}{\isacharcolon}{\isachardoublequoteopen}z\ t\ {\isacharequal}\ {\isacharbrackleft}{\isacharbrackright}\ {\isasymand}\ lOut\ t\ {\isacharequal}\ Zero{\isachardoublequoteclose}\ \isacommand{by}\isamarkupfalse%
\ simp\isanewline
\ \ \ \ \isacommand{from}\isamarkupfalse%
\ this\ \isakeyword{and}\ h{\isadigit{2}}\ \isacommand{show}\isamarkupfalse%
\ {\isacharquery}thesis\ \isacommand{by}\isamarkupfalse%
\ simp\isanewline
\ \ \isacommand{next}\isamarkupfalse%
\isanewline
\ \ \ \ \isacommand{assume}\isamarkupfalse%
\ a{\isadigit{4}}{\isacharcolon}{\isachardoublequoteopen}{\isasymnot}\ {\isadigit{3}}{\isadigit{0}}{\isadigit{0}}\ {\isacharless}\ hd\ {\isacharparenleft}y\ t{\isacharparenright}{\isachardoublequoteclose}\isanewline
\ \ \ \ \isacommand{from}\isamarkupfalse%
\ a{\isadigit{4}}\ \isakeyword{and}\ sg{\isadigit{1}}\ \isacommand{have}\isamarkupfalse%
\ sg{\isadigit{3}}{\isacharcolon}{\isachardoublequoteopen}z\ t\ {\isacharequal}\ {\isacharbrackleft}One{\isacharbrackright}\ {\isasymand}\ lOut\ t\ {\isacharequal}\ One{\isachardoublequoteclose}\ \isacommand{by}\isamarkupfalse%
\ simp\isanewline
\ \ \ \ \isacommand{from}\isamarkupfalse%
\ this\ \isakeyword{and}\ h{\isadigit{3}}\ \isacommand{show}\isamarkupfalse%
\ {\isacharquery}thesis\ \isacommand{by}\isamarkupfalse%
\ simp\isanewline
\ \ \isacommand{qed}\isamarkupfalse%
\ \isanewline
\isacommand{next}\isamarkupfalse%
\isanewline
\ \ \isacommand{assume}\isamarkupfalse%
\ a{\isadigit{2}}{\isacharcolon}{\isachardoublequoteopen}lIn\ t\ {\isacharequal}\ One{\isachardoublequoteclose}\isanewline
\ \ \isacommand{from}\isamarkupfalse%
\ this\ \isacommand{show}\isamarkupfalse%
\ {\isacharquery}thesis\ \isacommand{by}\isamarkupfalse%
\ simp\ \isanewline
\isacommand{qed}\isamarkupfalse%
\endisatagproof
{\isafoldproof}%
\isadelimproof
\isanewline
\endisadelimproof
\isanewline
\isanewline
\isacommand{lemma}\isamarkupfalse%
\ L{\isadigit{5}}{\isacharunderscore}Controller{\isacharcolon}\isanewline
\ \ \isakeyword{assumes}\ h{\isadigit{1}}{\isacharcolon}{\isachardoublequoteopen}Controller{\isacharunderscore}L\ y\ lIn\ lOut\ z{\isachardoublequoteclose}\isanewline
\ \ \ \ \ \ \isakeyword{and}\ h{\isadigit{2}}{\isacharcolon}{\isachardoublequoteopen}lOut\ t\ {\isacharequal}\ a{\isachardoublequoteclose}\isanewline
\ \ \ \ \ \ \isakeyword{and}\ h{\isadigit{3}}{\isacharcolon}{\isachardoublequoteopen}z\ t\ {\isacharequal}\ {\isacharbrackleft}{\isacharbrackright}{\isachardoublequoteclose}\isanewline
\ \ \isakeyword{shows}\ {\isachardoublequoteopen}lIn\ t\ {\isacharequal}\ a{\isachardoublequoteclose}\ \isanewline
\isadelimproof
\endisadelimproof
\isatagproof
\isacommand{proof}\isamarkupfalse%
\ {\isacharparenleft}cases\ {\isachardoublequoteopen}a{\isachardoublequoteclose}{\isacharparenright}\isanewline
\ \ \isacommand{assume}\isamarkupfalse%
\ {\isachardoublequoteopen}a\ {\isacharequal}\ Zero{\isachardoublequoteclose}\isanewline
\ \ \isacommand{from}\isamarkupfalse%
\ this\ \isakeyword{and}\ h{\isadigit{1}}\ \isakeyword{and}\ h{\isadigit{2}}\ \isakeyword{and}\ h{\isadigit{3}}\ \isacommand{show}\isamarkupfalse%
\ {\isacharquery}thesis\isanewline
\ \ \ \ \isacommand{by}\isamarkupfalse%
\ {\isacharparenleft}simp\ add{\isacharcolon}\ L{\isadigit{5}}{\isacharunderscore}Controller{\isacharunderscore}Zero{\isacharparenright}\ \isanewline
\isacommand{next}\isamarkupfalse%
\isanewline
\ \ \isacommand{assume}\isamarkupfalse%
\ {\isachardoublequoteopen}a\ {\isacharequal}\ One{\isachardoublequoteclose}\ \isanewline
\ \ \isacommand{from}\isamarkupfalse%
\ this\ \isakeyword{and}\ h{\isadigit{1}}\ \isakeyword{and}\ h{\isadigit{2}}\ \isakeyword{and}\ h{\isadigit{3}}\ \isacommand{show}\isamarkupfalse%
\ {\isacharquery}thesis\isanewline
\ \ \ \ \isacommand{by}\isamarkupfalse%
\ {\isacharparenleft}simp\ add{\isacharcolon}\ L{\isadigit{5}}{\isacharunderscore}Controller{\isacharunderscore}One{\isacharparenright}\ \ \isanewline
\isacommand{qed}\isamarkupfalse%
\endisatagproof
{\isafoldproof}%
\isadelimproof
\isanewline
\endisadelimproof
\isanewline
\isanewline
\isacommand{lemma}\isamarkupfalse%
\ L{\isadigit{6}}{\isacharunderscore}Controller{\isacharunderscore}Zero{\isacharcolon}\isanewline
\ \ \isakeyword{assumes}\ h{\isadigit{1}}{\isacharcolon}{\isachardoublequoteopen}Controller{\isacharunderscore}L\ y\ {\isacharparenleft}fin{\isacharunderscore}inf{\isacharunderscore}append\ {\isacharbrackleft}Zero{\isacharbrackright}\ l{\isacharparenright}\ l\ z{\isachardoublequoteclose}\isanewline
\ \ \ \ \ \ \isakeyword{and}\ h{\isadigit{2}}{\isacharcolon}{\isachardoublequoteopen}l\ {\isacharparenleft}Suc\ t{\isacharparenright}\ {\isacharequal}\ Zero{\isachardoublequoteclose}\isanewline
\ \ \ \ \ \ \isakeyword{and}\ h{\isadigit{3}}{\isacharcolon}{\isachardoublequoteopen}z\ {\isacharparenleft}Suc\ t{\isacharparenright}\ {\isacharequal}\ {\isacharbrackleft}{\isacharbrackright}{\isachardoublequoteclose}\isanewline
\ \ \isakeyword{shows}\ {\isachardoublequoteopen}l\ t\ {\isacharequal}\ Zero{\isachardoublequoteclose}\isanewline
\isadelimproof
\endisadelimproof
\isatagproof
\isacommand{proof}\isamarkupfalse%
\ {\isacharminus}\isanewline
\ \ \isacommand{from}\isamarkupfalse%
\ h{\isadigit{1}}\ \isakeyword{and}\ h{\isadigit{2}}\ \isakeyword{and}\ h{\isadigit{3}}\ \isacommand{have}\isamarkupfalse%
\ {\isachardoublequoteopen}{\isacharparenleft}fin{\isacharunderscore}inf{\isacharunderscore}append\ {\isacharbrackleft}Zero{\isacharbrackright}\ l{\isacharparenright}\ {\isacharparenleft}Suc\ t{\isacharparenright}\ {\isacharequal}\ Zero{\isachardoublequoteclose}\isanewline
\ \ \ \ \isacommand{by}\isamarkupfalse%
\ {\isacharparenleft}simp\ add{\isacharcolon}\ \ L{\isadigit{5}}{\isacharunderscore}Controller{\isacharunderscore}Zero{\isacharparenright}\isanewline
\ \ \isacommand{from}\isamarkupfalse%
\ this\ \isacommand{show}\isamarkupfalse%
\ {\isacharquery}thesis\ \isanewline
\ \ \ \ \isacommand{by}\isamarkupfalse%
\ {\isacharparenleft}simp\ add{\isacharcolon}\ correct{\isacharunderscore}fin{\isacharunderscore}inf{\isacharunderscore}append{\isadigit{2}}{\isacharparenright}\isanewline
\isacommand{qed}\isamarkupfalse%
\endisatagproof
{\isafoldproof}%
\isadelimproof
\isanewline
\endisadelimproof
\isanewline
\isanewline
\isacommand{lemma}\isamarkupfalse%
\ L{\isadigit{6}}{\isacharunderscore}Controller{\isacharunderscore}One{\isacharcolon}\isanewline
\ \ \isakeyword{assumes}\ h{\isadigit{1}}{\isacharcolon}{\isachardoublequoteopen}Controller{\isacharunderscore}L\ y\ {\isacharparenleft}fin{\isacharunderscore}inf{\isacharunderscore}append\ {\isacharbrackleft}Zero{\isacharbrackright}\ l{\isacharparenright}\ l\ z{\isachardoublequoteclose}\isanewline
\ \ \ \ \ \ \isakeyword{and}\ h{\isadigit{2}}{\isacharcolon}{\isachardoublequoteopen}l\ {\isacharparenleft}Suc\ t{\isacharparenright}\ {\isacharequal}\ One{\isachardoublequoteclose}\isanewline
\ \ \ \ \ \ \isakeyword{and}\ h{\isadigit{3}}{\isacharcolon}{\isachardoublequoteopen}z\ {\isacharparenleft}Suc\ t{\isacharparenright}\ {\isacharequal}\ {\isacharbrackleft}{\isacharbrackright}{\isachardoublequoteclose}\isanewline
\ \ \isakeyword{shows}\ {\isachardoublequoteopen}l\ t\ {\isacharequal}\ One{\isachardoublequoteclose}\isanewline
\isadelimproof
\endisadelimproof
\isatagproof
\isacommand{proof}\isamarkupfalse%
\ {\isacharminus}\isanewline
\ \ \isacommand{from}\isamarkupfalse%
\ h{\isadigit{1}}\ \isakeyword{and}\ h{\isadigit{2}}\ \isakeyword{and}\ h{\isadigit{3}}\ \isacommand{have}\isamarkupfalse%
\ {\isachardoublequoteopen}{\isacharparenleft}fin{\isacharunderscore}inf{\isacharunderscore}append\ {\isacharbrackleft}Zero{\isacharbrackright}\ l{\isacharparenright}\ {\isacharparenleft}Suc\ t{\isacharparenright}\ {\isacharequal}\ One{\isachardoublequoteclose}\isanewline
\ \ \ \ \isacommand{by}\isamarkupfalse%
\ {\isacharparenleft}simp\ add{\isacharcolon}\ \ L{\isadigit{5}}{\isacharunderscore}Controller{\isacharunderscore}One{\isacharparenright}\isanewline
\ \ \isacommand{from}\isamarkupfalse%
\ this\ \isacommand{show}\isamarkupfalse%
\ {\isacharquery}thesis\ \isanewline
\ \ \ \ \isacommand{by}\isamarkupfalse%
\ {\isacharparenleft}simp\ add{\isacharcolon}\ correct{\isacharunderscore}fin{\isacharunderscore}inf{\isacharunderscore}append{\isadigit{2}}{\isacharparenright}\isanewline
\isacommand{qed}\isamarkupfalse%
\endisatagproof
{\isafoldproof}%
\isadelimproof
\isanewline
\endisadelimproof
\isanewline
\isanewline
\isacommand{lemma}\isamarkupfalse%
\ L{\isadigit{6}}{\isacharunderscore}Controller{\isacharcolon}\isanewline
\ \ \isakeyword{assumes}\ h{\isadigit{1}}{\isacharcolon}{\isachardoublequoteopen}Controller{\isacharunderscore}L\ y\ {\isacharparenleft}fin{\isacharunderscore}inf{\isacharunderscore}append\ {\isacharbrackleft}Zero{\isacharbrackright}\ l{\isacharparenright}\ l\ z{\isachardoublequoteclose}\isanewline
\ \ \ \ \ \ \isakeyword{and}\ h{\isadigit{2}}{\isacharcolon}{\isachardoublequoteopen}l\ {\isacharparenleft}Suc\ t{\isacharparenright}\ {\isacharequal}\ a{\isachardoublequoteclose}\isanewline
\ \ \ \ \ \ \isakeyword{and}\ h{\isadigit{3}}{\isacharcolon}{\isachardoublequoteopen}z\ {\isacharparenleft}Suc\ t{\isacharparenright}\ {\isacharequal}\ {\isacharbrackleft}{\isacharbrackright}{\isachardoublequoteclose}\isanewline
\ \ \isakeyword{shows}\ {\isachardoublequoteopen}l\ t\ {\isacharequal}\ a{\isachardoublequoteclose}\isanewline
\isadelimproof
\endisadelimproof
\isatagproof
\isacommand{proof}\isamarkupfalse%
\ {\isacharparenleft}cases\ {\isachardoublequoteopen}a{\isachardoublequoteclose}{\isacharparenright}\isanewline
\ \ \isacommand{assume}\isamarkupfalse%
\ {\isachardoublequoteopen}a\ {\isacharequal}\ Zero{\isachardoublequoteclose}\isanewline
\ \ \isacommand{from}\isamarkupfalse%
\ this\ \isakeyword{and}\ h{\isadigit{1}}\ \isakeyword{and}\ h{\isadigit{2}}\ \isakeyword{and}\ h{\isadigit{3}}\ \isacommand{show}\isamarkupfalse%
\ {\isacharquery}thesis\isanewline
\ \ \ \ \isacommand{by}\isamarkupfalse%
\ {\isacharparenleft}simp\ add{\isacharcolon}\ \ L{\isadigit{6}}{\isacharunderscore}Controller{\isacharunderscore}Zero{\isacharparenright}\ \isanewline
\isacommand{next}\isamarkupfalse%
\isanewline
\ \ \isacommand{assume}\isamarkupfalse%
\ {\isachardoublequoteopen}a\ {\isacharequal}\ One{\isachardoublequoteclose}\isanewline
\ \ \isacommand{from}\isamarkupfalse%
\ this\ \isakeyword{and}\ h{\isadigit{1}}\ \isakeyword{and}\ h{\isadigit{2}}\ \isakeyword{and}\ h{\isadigit{3}}\ \isacommand{show}\isamarkupfalse%
\ {\isacharquery}thesis\isanewline
\ \ \ \ \isacommand{by}\isamarkupfalse%
\ {\isacharparenleft}simp\ add{\isacharcolon}\ \ L{\isadigit{6}}{\isacharunderscore}Controller{\isacharunderscore}One{\isacharparenright}\ \isanewline
\isacommand{qed}\isamarkupfalse%
\endisatagproof
{\isafoldproof}%
\isadelimproof
\isanewline
\endisadelimproof
\isanewline
\isanewline
\isacommand{lemma}\isamarkupfalse%
\ L{\isadigit{7}}{\isacharunderscore}Controller{\isacharunderscore}Zero{\isacharcolon}\isanewline
\ \ \isakeyword{assumes}\ h{\isadigit{1}}{\isacharcolon}{\isachardoublequoteopen}Controller{\isacharunderscore}L\ y\ {\isacharparenleft}fin{\isacharunderscore}inf{\isacharunderscore}append\ {\isacharbrackleft}Zero{\isacharbrackright}\ l{\isacharparenright}\ l\ z{\isachardoublequoteclose}\isanewline
\ \ \ \ \ \ \isakeyword{and}\ h{\isadigit{2}}{\isacharcolon}{\isachardoublequoteopen}l\ t\ {\isacharequal}\ Zero{\isachardoublequoteclose}\isanewline
\ \ \isakeyword{shows}\ \ \ \ \ \ {\isachardoublequoteopen}last\ {\isacharparenleft}fin{\isacharunderscore}make{\isacharunderscore}untimed\ {\isacharparenleft}inf{\isacharunderscore}truncate\ z\ t{\isacharparenright}{\isacharparenright}\ {\isacharequal}\ Zero{\isachardoublequoteclose}\isanewline
\isadelimproof
\endisadelimproof
\isatagproof
\isacommand{using}\isamarkupfalse%
\ assms\isanewline
\isacommand{proof}\isamarkupfalse%
\ {\isacharparenleft}induct\ t{\isacharparenright}\isanewline
\ \ \isacommand{case}\isamarkupfalse%
\ {\isadigit{0}}\ \ \isanewline
\ \ \isacommand{from}\isamarkupfalse%
\ h{\isadigit{1}}\ \isacommand{have}\isamarkupfalse%
\ sg{\isadigit{1}}{\isacharcolon}{\isachardoublequoteopen}z\ {\isadigit{0}}\ {\isacharequal}\ {\isacharbrackleft}Zero{\isacharbrackright}{\isachardoublequoteclose}\ \isacommand{by}\isamarkupfalse%
\ {\isacharparenleft}simp\ add{\isacharcolon}\ Controller{\isacharunderscore}L{\isacharunderscore}def{\isacharparenright}\ \isanewline
\ \ \isacommand{from}\isamarkupfalse%
\ this\ \isacommand{show}\isamarkupfalse%
\ {\isacharquery}case\ \isacommand{by}\isamarkupfalse%
\ {\isacharparenleft}simp\ add{\isacharcolon}\ fin{\isacharunderscore}make{\isacharunderscore}untimed{\isacharunderscore}def{\isacharparenright}\isanewline
\isacommand{next}\isamarkupfalse%
\isanewline
\ \ \ \isacommand{fix}\isamarkupfalse%
\ t\isanewline
\ \ \ \isacommand{case}\isamarkupfalse%
\ {\isacharparenleft}Suc\ t{\isacharparenright}\isanewline
\ \ \ \isacommand{from}\isamarkupfalse%
\ this\ \isacommand{show}\isamarkupfalse%
\ {\isacharquery}case\isanewline
\ \ \ \isacommand{proof}\isamarkupfalse%
\ {\isacharparenleft}cases\ {\isachardoublequoteopen}l\ t{\isachardoublequoteclose}{\isacharparenright}\isanewline
\ \ \ \ \ \isacommand{assume}\isamarkupfalse%
\ a{\isadigit{1}}{\isacharcolon}{\isachardoublequoteopen}l\ t\ {\isacharequal}\ Zero{\isachardoublequoteclose}\isanewline
\ \ \ \ \ \isacommand{from}\isamarkupfalse%
\ Suc\ \isacommand{have}\isamarkupfalse%
\isanewline
\ \ \ \ \ \ \ sg{\isadigit{1}}{\isacharcolon}{\isachardoublequoteopen}{\isacharparenleft}z\ {\isacharparenleft}Suc\ t{\isacharparenright}\ {\isacharequal}\ {\isacharbrackleft}{\isacharbrackright}\ {\isasymand}\ l\ t\ {\isacharequal}\ Zero{\isacharparenright}\ {\isasymor}\ {\isacharparenleft}z\ {\isacharparenleft}Suc\ t{\isacharparenright}\ {\isacharequal}\ {\isacharbrackleft}Zero{\isacharbrackright}\ {\isasymand}\ l\ t\ {\isacharequal}\ One{\isacharparenright}{\isachardoublequoteclose}\isanewline
\ \ \ \ \ \ \ \isacommand{by}\isamarkupfalse%
\ {\isacharparenleft}simp\ add{\isacharcolon}\ L{\isadigit{4}}{\isacharunderscore}Controller{\isacharunderscore}Zero{\isacharparenright}\isanewline
\ \ \ \ \ \isacommand{from}\isamarkupfalse%
\ this\ \isakeyword{and}\ a{\isadigit{1}}\ \isacommand{have}\isamarkupfalse%
\ sg{\isadigit{2}}{\isacharcolon}{\isachardoublequoteopen}z\ {\isacharparenleft}Suc\ t{\isacharparenright}\ {\isacharequal}\ {\isacharbrackleft}{\isacharbrackright}{\isachardoublequoteclose}\isanewline
\ \ \ \ \ \ \ \isacommand{by}\isamarkupfalse%
\ simp\ \isanewline
\ \ \ \ \ \isacommand{from}\isamarkupfalse%
\ Suc\ \isakeyword{and}\ sg{\isadigit{2}}\ \isakeyword{and}\ a{\isadigit{1}}\ \isacommand{show}\isamarkupfalse%
\ {\isacharquery}case\isanewline
\ \ \ \ \ \ \ \isacommand{by}\isamarkupfalse%
\ {\isacharparenleft}simp\ add{\isacharcolon}\ fin{\isacharunderscore}make{\isacharunderscore}untimed{\isacharunderscore}append{\isacharunderscore}empty{\isacharparenright}\isanewline
\ \ \ \isacommand{next}\isamarkupfalse%
\ \isanewline
\ \ \ \ \ \isacommand{assume}\isamarkupfalse%
\ a{\isadigit{1}}{\isacharcolon}{\isachardoublequoteopen}l\ t\ {\isacharequal}\ One{\isachardoublequoteclose}\isanewline
\ \ \ \ \ \isacommand{from}\isamarkupfalse%
\ Suc\ \isacommand{have}\isamarkupfalse%
\isanewline
\ \ \ \ \ \ \ sg{\isadigit{1}}{\isacharcolon}{\isachardoublequoteopen}{\isacharparenleft}z\ {\isacharparenleft}Suc\ t{\isacharparenright}\ {\isacharequal}\ {\isacharbrackleft}{\isacharbrackright}\ {\isasymand}\ l\ t\ {\isacharequal}\ Zero{\isacharparenright}\ {\isasymor}\ {\isacharparenleft}z\ {\isacharparenleft}Suc\ t{\isacharparenright}\ {\isacharequal}\ {\isacharbrackleft}Zero{\isacharbrackright}\ {\isasymand}\ l\ t\ {\isacharequal}\ One{\isacharparenright}{\isachardoublequoteclose}\isanewline
\ \ \ \ \ \ \ \isacommand{by}\isamarkupfalse%
\ {\isacharparenleft}simp\ add{\isacharcolon}\ L{\isadigit{4}}{\isacharunderscore}Controller{\isacharunderscore}Zero{\isacharparenright}\isanewline
\ \ \ \ \ \isacommand{from}\isamarkupfalse%
\ this\ \isakeyword{and}\ a{\isadigit{1}}\ \isacommand{have}\isamarkupfalse%
\ sg{\isadigit{2}}{\isacharcolon}{\isachardoublequoteopen}z\ {\isacharparenleft}Suc\ t{\isacharparenright}\ {\isacharequal}\ {\isacharbrackleft}Zero{\isacharbrackright}{\isachardoublequoteclose}\ 
\ \isacommand{by}\isamarkupfalse%
\ simp\ \isanewline
\ \ \ \ \ \isacommand{from}\isamarkupfalse%
\ a{\isadigit{1}}\ \isakeyword{and}\ Suc\ \isakeyword{and}\ sg{\isadigit{2}}\ \isacommand{show}\isamarkupfalse%
\ {\isacharquery}case\isanewline
\ \ \ \ \ \ \ \isacommand{by}\isamarkupfalse%
\ {\isacharparenleft}simp\ add{\isacharcolon}\ fin{\isacharunderscore}make{\isacharunderscore}untimed{\isacharunderscore}def{\isacharparenright}\ \isanewline
\ \ \ \isacommand{qed}\isamarkupfalse%
\isanewline
\isacommand{qed}\isamarkupfalse%
\endisatagproof
{\isafoldproof}%
\isadelimproof
\isanewline
\endisadelimproof
\isanewline
\isanewline
\isacommand{lemma}\isamarkupfalse%
\ L{\isadigit{7}}{\isacharunderscore}Controller{\isacharunderscore}One{\isacharunderscore}l{\isadigit{0}}{\isacharcolon}\isanewline
\ \ \isakeyword{assumes}\ h{\isadigit{1}}{\isacharcolon}{\isachardoublequoteopen}Controller{\isacharunderscore}L\ y\ {\isacharparenleft}fin{\isacharunderscore}inf{\isacharunderscore}append\ {\isacharbrackleft}Zero{\isacharbrackright}\ l{\isacharparenright}\ l\ z{\isachardoublequoteclose}\ \isanewline
\ \ \ \ \ \ \isakeyword{and}\ h{\isadigit{2}}{\isacharcolon}{\isachardoublequoteopen}y\ {\isadigit{0}}\ {\isacharequal}\ {\isacharbrackleft}{\isadigit{5}}{\isadigit{0}}{\isadigit{0}}{\isacharcolon}{\isacharcolon}nat{\isacharbrackright}{\isachardoublequoteclose}\isanewline
\ \ \isakeyword{shows}\ \ \ \ {\isachardoublequoteopen}l\ {\isadigit{0}}\ {\isacharequal}\ Zero{\isachardoublequoteclose}\isanewline
\isadelimproof
\endisadelimproof
\isatagproof
\isacommand{proof}\isamarkupfalse%
\ {\isacharparenleft}rule\ ccontr{\isacharparenright}\isanewline
\ \ \isacommand{assume}\isamarkupfalse%
\ a{\isadigit{1}}{\isacharcolon}{\isachardoublequoteopen}\ {\isasymnot}\ l\ {\isadigit{0}}\ {\isacharequal}\ Zero{\isachardoublequoteclose}\ \isanewline
\ \ \isacommand{from}\isamarkupfalse%
\ assms\ \isacommand{have}\isamarkupfalse%
\ sg{\isadigit{1}}{\isacharcolon}{\isachardoublequoteopen}z\ {\isadigit{0}}\ {\isacharequal}\ {\isacharbrackleft}Zero{\isacharbrackright}{\isachardoublequoteclose}\ \isacommand{by}\isamarkupfalse%
\ {\isacharparenleft}simp\ add{\isacharcolon}\ Controller{\isacharunderscore}L{\isacharunderscore}def{\isacharparenright}\ \isanewline
\ \ \isacommand{have}\isamarkupfalse%
\ sg{\isadigit{2}}{\isacharcolon}{\isachardoublequoteopen}fin{\isacharunderscore}inf{\isacharunderscore}append\ {\isacharbrackleft}Zero{\isacharbrackright}\ l\ {\isacharparenleft}{\isadigit{0}}{\isacharcolon}{\isacharcolon}nat{\isacharparenright}\ {\isacharequal}\ Zero{\isachardoublequoteclose}\ \isacommand{by}\isamarkupfalse%
\ \ {\isacharparenleft}simp\ add{\isacharcolon}\ fin{\isacharunderscore}inf{\isacharunderscore}append{\isacharunderscore}def{\isacharparenright}\isanewline
\ \ \isacommand{from}\isamarkupfalse%
\ assms\ \isakeyword{and}\ a{\isadigit{1}}\ \isakeyword{and}\ sg{\isadigit{1}}\ \isakeyword{and}\ sg{\isadigit{2}}\ \isacommand{show}\isamarkupfalse%
\ {\isachardoublequoteopen}False{\isachardoublequoteclose}\isanewline
\ \ \ \isacommand{by}\isamarkupfalse%
\ {\isacharparenleft}simp\ add{\isacharcolon}\ Controller{\isacharunderscore}L{\isacharunderscore}def{\isacharparenright}\ \isanewline
\isacommand{qed}\isamarkupfalse%
\endisatagproof
{\isafoldproof}%
\isadelimproof
\isanewline
\endisadelimproof
\isanewline
\isanewline
\isacommand{lemma}\isamarkupfalse%
\ L{\isadigit{7}}{\isacharunderscore}Controller{\isacharunderscore}One{\isacharcolon}\isanewline
\ \ \isakeyword{assumes}\ h{\isadigit{1}}{\isacharcolon}{\isachardoublequoteopen}Controller{\isacharunderscore}L\ y\ {\isacharparenleft}fin{\isacharunderscore}inf{\isacharunderscore}append\ {\isacharbrackleft}Zero{\isacharbrackright}\ l{\isacharparenright}\ l\ z{\isachardoublequoteclose}\isanewline
\ \ \ \ \ \ \isakeyword{and}\ h{\isadigit{2}}{\isacharcolon}{\isachardoublequoteopen}l\ t\ {\isacharequal}\ One{\isachardoublequoteclose}\isanewline
\ \ \ \ \ \ \isakeyword{and}\ h{\isadigit{3}}{\isacharcolon}{\isachardoublequoteopen}y\ {\isadigit{0}}\ {\isacharequal}\ {\isacharbrackleft}{\isadigit{5}}{\isadigit{0}}{\isadigit{0}}{\isacharcolon}{\isacharcolon}nat{\isacharbrackright}{\isachardoublequoteclose}\isanewline
\ \ \isakeyword{shows}\ {\isachardoublequoteopen}last\ {\isacharparenleft}fin{\isacharunderscore}make{\isacharunderscore}untimed\ {\isacharparenleft}inf{\isacharunderscore}truncate\ z\ t{\isacharparenright}{\isacharparenright}\ {\isacharequal}\ One{\isachardoublequoteclose}\isanewline
\isadelimproof
\endisadelimproof
\isatagproof
\isacommand{using}\isamarkupfalse%
\ assms\isanewline
\isacommand{proof}\isamarkupfalse%
\ {\isacharparenleft}induct\ t{\isacharparenright}\isanewline
\ \ \isacommand{case}\isamarkupfalse%
\ {\isadigit{0}}\isanewline
\ \ \isacommand{from}\isamarkupfalse%
\ h{\isadigit{1}}\ \ \isakeyword{and}\ h{\isadigit{3}}\ \isacommand{have}\isamarkupfalse%
\ sg{\isadigit{0}}{\isacharcolon}{\isachardoublequoteopen}l\ {\isadigit{0}}\ {\isacharequal}\ Zero{\isachardoublequoteclose}\ \isacommand{by}\isamarkupfalse%
\ {\isacharparenleft}simp\ add{\isacharcolon}\ L{\isadigit{7}}{\isacharunderscore}Controller{\isacharunderscore}One{\isacharunderscore}l{\isadigit{0}}{\isacharparenright}\ \isanewline
\ \ \isacommand{from}\isamarkupfalse%
\ this\ \isakeyword{and}\ {\isadigit{0}}\ \isacommand{show}\isamarkupfalse%
\ {\isacharquery}case\ \isacommand{by}\isamarkupfalse%
\ simp\isanewline
\isacommand{next}\isamarkupfalse%
\isanewline
\ \ \ \isacommand{fix}\isamarkupfalse%
\ t\isanewline
\ \ \ \isacommand{case}\isamarkupfalse%
\ {\isacharparenleft}Suc\ t{\isacharparenright}\isanewline
\ \ \ \isacommand{from}\isamarkupfalse%
\ this\ \isacommand{show}\isamarkupfalse%
\ {\isacharquery}case\isanewline
\ \ \ \isacommand{proof}\isamarkupfalse%
\ {\isacharparenleft}cases\ {\isachardoublequoteopen}l\ t{\isachardoublequoteclose}{\isacharparenright}\isanewline
\ \ \ \ \ \isacommand{assume}\isamarkupfalse%
\ a{\isadigit{1}}{\isacharcolon}{\isachardoublequoteopen}l\ t\ {\isacharequal}\ Zero{\isachardoublequoteclose}\isanewline
\ \ \ \ \ \isacommand{from}\isamarkupfalse%
\ Suc\ \isacommand{have}\isamarkupfalse%
\isanewline
\ \ \ \ \ \ \ sg{\isadigit{1}}{\isacharcolon}{\isachardoublequoteopen}{\isacharparenleft}z\ {\isacharparenleft}Suc\ t{\isacharparenright}\ {\isacharequal}\ {\isacharbrackleft}{\isacharbrackright}\ {\isasymand}\ l\ t\ {\isacharequal}\ One{\isacharparenright}\ {\isasymor}\ {\isacharparenleft}z\ {\isacharparenleft}Suc\ t{\isacharparenright}\ {\isacharequal}\ {\isacharbrackleft}One{\isacharbrackright}\ {\isasymand}\ l\ t\ {\isacharequal}\ Zero{\isacharparenright}{\isachardoublequoteclose}\ \isanewline
\ \ \ \ \ \ \ \isacommand{by}\isamarkupfalse%
\ {\isacharparenleft}simp\ add{\isacharcolon}\ L{\isadigit{4}}{\isacharunderscore}Controller{\isacharunderscore}One{\isacharparenright}\isanewline
\ \ \ \ \ \isacommand{from}\isamarkupfalse%
\ this\ \isakeyword{and}\ a{\isadigit{1}}\ \isacommand{have}\isamarkupfalse%
\ sg{\isadigit{2}}{\isacharcolon}{\isachardoublequoteopen}z\ {\isacharparenleft}Suc\ t{\isacharparenright}\ {\isacharequal}\ {\isacharbrackleft}One{\isacharbrackright}{\isachardoublequoteclose}\isanewline
\ \ \ \ \ \ \ \isacommand{by}\isamarkupfalse%
\ simp\ \ \isanewline
\ \ \ \ \ \isacommand{from}\isamarkupfalse%
\ Suc\ \isakeyword{and}\ sg{\isadigit{2}}\ \isakeyword{and}\ a{\isadigit{1}}\ \isacommand{show}\isamarkupfalse%
\ {\isacharquery}case\isanewline
\ \ \ \ \ \ \ \isacommand{by}\isamarkupfalse%
\ {\isacharparenleft}simp\ add{\isacharcolon}\ fin{\isacharunderscore}make{\isacharunderscore}untimed{\isacharunderscore}def{\isacharparenright}\ \isanewline
\ \ \ \isacommand{next}\isamarkupfalse%
\ \isanewline
\ \ \ \ \ \isacommand{assume}\isamarkupfalse%
\ a{\isadigit{1}}{\isacharcolon}{\isachardoublequoteopen}l\ t\ {\isacharequal}\ One{\isachardoublequoteclose}\isanewline
\ \ \ \ \ \isacommand{from}\isamarkupfalse%
\ Suc\ \isacommand{have}\isamarkupfalse%
\isanewline
\ \ \ \ \ \ \ sg{\isadigit{1}}{\isacharcolon}{\isachardoublequoteopen}{\isacharparenleft}z\ {\isacharparenleft}Suc\ t{\isacharparenright}\ {\isacharequal}\ {\isacharbrackleft}{\isacharbrackright}\ {\isasymand}\ l\ t\ {\isacharequal}\ One{\isacharparenright}\ {\isasymor}\ {\isacharparenleft}z\ {\isacharparenleft}Suc\ t{\isacharparenright}\ {\isacharequal}\ {\isacharbrackleft}One{\isacharbrackright}\ {\isasymand}\ l\ t\ {\isacharequal}\ Zero{\isacharparenright}{\isachardoublequoteclose}\ \isanewline
\ \ \ \ \ \ \ \isacommand{by}\isamarkupfalse%
\ {\isacharparenleft}simp\ add{\isacharcolon}\ L{\isadigit{4}}{\isacharunderscore}Controller{\isacharunderscore}One{\isacharparenright}\isanewline
\ \ \ \ \ \isacommand{from}\isamarkupfalse%
\ this\ \isakeyword{and}\ a{\isadigit{1}}\ \isacommand{have}\isamarkupfalse%
\ sg{\isadigit{2}}{\isacharcolon}{\isachardoublequoteopen}z\ {\isacharparenleft}Suc\ t{\isacharparenright}\ {\isacharequal}\ {\isacharbrackleft}{\isacharbrackright}{\isachardoublequoteclose}\isanewline
\ \ \ \ \ \ \ \isacommand{by}\isamarkupfalse%
\ simp\ \isanewline
\ \ \ \ \ \isacommand{from}\isamarkupfalse%
\ a{\isadigit{1}}\ \isakeyword{and}\ Suc\ \isakeyword{and}\ sg{\isadigit{2}}\ \isacommand{show}\isamarkupfalse%
\ {\isacharquery}case\isanewline
\ \ \ \ \ \ \ \isacommand{by}\isamarkupfalse%
\ {\isacharparenleft}simp\ add{\isacharcolon}\ fin{\isacharunderscore}make{\isacharunderscore}untimed{\isacharunderscore}def{\isacharparenright}\ \isanewline
\ \ \ \isacommand{qed}\isamarkupfalse%
\isanewline
\isacommand{qed}\isamarkupfalse%
\endisatagproof
{\isafoldproof}%
\isadelimproof
\isanewline
\endisadelimproof
\isanewline
\isanewline
\isacommand{lemma}\isamarkupfalse%
\ L{\isadigit{7}}{\isacharunderscore}Controller{\isacharcolon}\isanewline
\ \ \isakeyword{assumes}\ h{\isadigit{1}}{\isacharcolon}{\isachardoublequoteopen}Controller{\isacharunderscore}L\ y\ {\isacharparenleft}fin{\isacharunderscore}inf{\isacharunderscore}append\ {\isacharbrackleft}Zero{\isacharbrackright}\ l{\isacharparenright}\ l\ z{\isachardoublequoteclose}\isanewline
\ \ \ \ \ \ \isakeyword{and}\ h{\isadigit{2}}{\isacharcolon}{\isachardoublequoteopen}y\ {\isadigit{0}}\ {\isacharequal}\ {\isacharbrackleft}{\isadigit{5}}{\isadigit{0}}{\isadigit{0}}{\isacharcolon}{\isacharcolon}nat{\isacharbrackright}{\isachardoublequoteclose}\isanewline
\ \ \isakeyword{shows}\ \ \ \ \ \ {\isachardoublequoteopen}last\ {\isacharparenleft}fin{\isacharunderscore}make{\isacharunderscore}untimed\ {\isacharparenleft}inf{\isacharunderscore}truncate\ z\ t{\isacharparenright}{\isacharparenright}\ {\isacharequal}\ \ l\ t{\isachardoublequoteclose}\isanewline
\isadelimproof
\endisadelimproof
\isatagproof
\isacommand{proof}\isamarkupfalse%
\ {\isacharparenleft}cases\ {\isachardoublequoteopen}l\ t{\isachardoublequoteclose}{\isacharparenright}\isanewline
\ \ \isacommand{assume}\isamarkupfalse%
\ {\isachardoublequoteopen}\ l\ t\ {\isacharequal}\ Zero{\isachardoublequoteclose}\isanewline
\ \ \isacommand{from}\isamarkupfalse%
\ this\ \isakeyword{and}\ h{\isadigit{1}}\ \isacommand{show}\isamarkupfalse%
\ {\isacharquery}thesis\ \isanewline
\ \ \ \ \ \isacommand{by}\isamarkupfalse%
\ {\isacharparenleft}simp\ add{\isacharcolon}\ L{\isadigit{7}}{\isacharunderscore}Controller{\isacharunderscore}Zero{\isacharparenright}\isanewline
\isacommand{next}\isamarkupfalse%
\isanewline
\ \ \isacommand{assume}\isamarkupfalse%
\ {\isachardoublequoteopen}\ l\ t\ {\isacharequal}\ One{\isachardoublequoteclose}\isanewline
\ \ \isacommand{from}\isamarkupfalse%
\ this\ \isakeyword{and}\ h{\isadigit{1}}\ \isakeyword{and}\ h{\isadigit{2}}\ \isacommand{show}\isamarkupfalse%
\ {\isacharquery}thesis\ \isanewline
\ \ \ \ \ \isacommand{by}\isamarkupfalse%
\ {\isacharparenleft}simp\ add{\isacharcolon}\ L{\isadigit{7}}{\isacharunderscore}Controller{\isacharunderscore}One{\isacharparenright}\isanewline
\isacommand{qed}\isamarkupfalse%
\endisatagproof
{\isafoldproof}%
\isadelimproof
\isanewline
\endisadelimproof
\isanewline
\isanewline
\isacommand{lemma}\isamarkupfalse%
\ L{\isadigit{8}}{\isacharunderscore}Controller{\isacharcolon}\isanewline
\ \ \isakeyword{assumes}\ h{\isadigit{1}}{\isacharcolon}{\isachardoublequoteopen}Controller{\isacharunderscore}L\ y\ {\isacharparenleft}fin{\isacharunderscore}inf{\isacharunderscore}append\ {\isacharbrackleft}Zero{\isacharbrackright}\ l{\isacharparenright}\ l\ z{\isachardoublequoteclose}\isanewline
\ \ \isakeyword{shows}\ \ \ \ \ \ {\isachardoublequoteopen}z\ t\ {\isacharequal}\ {\isacharbrackleft}{\isacharbrackright}\ {\isasymor}\ z\ t\ {\isacharequal}\ {\isacharbrackleft}Zero{\isacharbrackright}\ {\isasymor}\ z\ t\ {\isacharequal}\ {\isacharbrackleft}One{\isacharbrackright}{\isachardoublequoteclose}\isanewline
\isadelimproof
\endisadelimproof
\isatagproof
\isacommand{proof}\isamarkupfalse%
\ {\isacharparenleft}cases\ {\isachardoublequoteopen}fin{\isacharunderscore}inf{\isacharunderscore}append\ {\isacharbrackleft}Zero{\isacharbrackright}\ l\ t\ {\isacharequal}\ Zero{\isachardoublequoteclose}{\isacharparenright}\isanewline
\ \ \isacommand{assume}\isamarkupfalse%
\ a{\isadigit{1}}{\isacharcolon}{\isachardoublequoteopen}fin{\isacharunderscore}inf{\isacharunderscore}append\ {\isacharbrackleft}Zero{\isacharbrackright}\ l\ t\ {\isacharequal}\ Zero{\isachardoublequoteclose}\isanewline
\ \ \isacommand{from}\isamarkupfalse%
\ a{\isadigit{1}}\ \isakeyword{and}\ h{\isadigit{1}}\ \isacommand{have}\isamarkupfalse%
\ sg{\isadigit{1}}{\isacharcolon}\isanewline
\ \ \ {\isachardoublequoteopen}if\ {\isadigit{3}}{\isadigit{0}}{\isadigit{0}}\ {\isacharless}\ hd\ {\isacharparenleft}y\ t{\isacharparenright}\ \isanewline
\ \ \ \ then\ z\ t\ {\isacharequal}\ {\isacharbrackleft}{\isacharbrackright}\ {\isasymand}\ l\ t\ {\isacharequal}\ Zero\ \isanewline
\ \ \ \ else\ z\ t\ {\isacharequal}\ {\isacharbrackleft}One{\isacharbrackright}\ {\isasymand}\ l\ t\ {\isacharequal}\ One{\isachardoublequoteclose}\isanewline
\ \ \ \ \isacommand{by}\isamarkupfalse%
\ {\isacharparenleft}simp\ add{\isacharcolon}\ Controller{\isacharunderscore}L{\isacharunderscore}def{\isacharparenright}\isanewline
\ \ \isacommand{show}\isamarkupfalse%
\ {\isacharquery}thesis\isanewline
\ \ \isacommand{proof}\isamarkupfalse%
\ {\isacharparenleft}cases\ {\isachardoublequoteopen}{\isadigit{3}}{\isadigit{0}}{\isadigit{0}}\ {\isacharless}\ hd\ {\isacharparenleft}y\ t{\isacharparenright}{\isachardoublequoteclose}{\isacharparenright}\isanewline
\ \ \ \ \isacommand{assume}\isamarkupfalse%
\ a{\isadigit{1}}{\isadigit{1}}{\isacharcolon}{\isachardoublequoteopen}{\isadigit{3}}{\isadigit{0}}{\isadigit{0}}\ {\isacharless}\ hd\ {\isacharparenleft}y\ t{\isacharparenright}{\isachardoublequoteclose}\isanewline
\ \ \ \ \isacommand{from}\isamarkupfalse%
\ a{\isadigit{1}}{\isadigit{1}}\ \isakeyword{and}\ sg{\isadigit{1}}\ \isacommand{show}\isamarkupfalse%
\ {\isacharquery}thesis\ \isacommand{by}\isamarkupfalse%
\ simp\isanewline
\ \ \isacommand{next}\isamarkupfalse%
\isanewline
\ \ \ \ \isacommand{assume}\isamarkupfalse%
\ a{\isadigit{1}}{\isadigit{2}}{\isacharcolon}{\isachardoublequoteopen}{\isasymnot}\ {\isadigit{3}}{\isadigit{0}}{\isadigit{0}}\ {\isacharless}\ hd\ {\isacharparenleft}y\ t{\isacharparenright}{\isachardoublequoteclose}\isanewline
\ \ \ \ \isacommand{from}\isamarkupfalse%
\ a{\isadigit{1}}{\isadigit{2}}\ \isakeyword{and}\ sg{\isadigit{1}}\ \isacommand{show}\isamarkupfalse%
\ {\isacharquery}thesis\ \isacommand{by}\isamarkupfalse%
\ simp\isanewline
\ \ \isacommand{qed}\isamarkupfalse%
\isanewline
\isacommand{next}\isamarkupfalse%
\isanewline
\ \ \isacommand{assume}\isamarkupfalse%
\ a{\isadigit{2}}{\isacharcolon}{\isachardoublequoteopen}fin{\isacharunderscore}inf{\isacharunderscore}append\ {\isacharbrackleft}Zero{\isacharbrackright}\ l\ t\ {\isasymnoteq}\ Zero{\isachardoublequoteclose}\isanewline
\ \ \isacommand{from}\isamarkupfalse%
\ a{\isadigit{2}}\ \isakeyword{and}\ h{\isadigit{1}}\ \isacommand{have}\isamarkupfalse%
\ sg{\isadigit{2}}{\isacharcolon}\isanewline
\ \ \ {\isachardoublequoteopen}if\ hd\ {\isacharparenleft}y\ t{\isacharparenright}\ {\isacharless}\ {\isadigit{7}}{\isadigit{0}}{\isadigit{0}}\ \isanewline
\ \ \ \ then\ z\ t\ {\isacharequal}\ {\isacharbrackleft}{\isacharbrackright}\ {\isasymand}\ l\ t\ {\isacharequal}\ One\ \isanewline
\ \ \ \ else\ z\ t\ {\isacharequal}\ {\isacharbrackleft}Zero{\isacharbrackright}\ {\isasymand}\ l\ t\ {\isacharequal}\ Zero{\isachardoublequoteclose}\isanewline
\ \ \ \ \isacommand{by}\isamarkupfalse%
\ {\isacharparenleft}simp\ add{\isacharcolon}\ Controller{\isacharunderscore}L{\isacharunderscore}def{\isacharparenright}\isanewline
\ \ \isacommand{show}\isamarkupfalse%
\ {\isacharquery}thesis\isanewline
\ \ \isacommand{proof}\isamarkupfalse%
\ {\isacharparenleft}cases\ {\isachardoublequoteopen}hd\ {\isacharparenleft}y\ t{\isacharparenright}\ {\isacharless}\ {\isadigit{7}}{\isadigit{0}}{\isadigit{0}}{\isachardoublequoteclose}{\isacharparenright}\isanewline
\ \ \ \ \isacommand{assume}\isamarkupfalse%
\ a{\isadigit{2}}{\isadigit{1}}{\isacharcolon}{\isachardoublequoteopen}hd\ {\isacharparenleft}y\ t{\isacharparenright}\ {\isacharless}\ {\isadigit{7}}{\isadigit{0}}{\isadigit{0}}{\isachardoublequoteclose}\isanewline
\ \ \ \ \isacommand{from}\isamarkupfalse%
\ a{\isadigit{2}}{\isadigit{1}}\ \isakeyword{and}\ sg{\isadigit{2}}\ \isacommand{show}\isamarkupfalse%
\ {\isacharquery}thesis\ \isacommand{by}\isamarkupfalse%
\ simp\isanewline
\ \ \isacommand{next}\isamarkupfalse%
\isanewline
\ \ \ \ \isacommand{assume}\isamarkupfalse%
\ a{\isadigit{2}}{\isadigit{2}}{\isacharcolon}{\isachardoublequoteopen}{\isasymnot}\ hd\ {\isacharparenleft}y\ t{\isacharparenright}\ {\isacharless}\ {\isadigit{7}}{\isadigit{0}}{\isadigit{0}}{\isachardoublequoteclose}\isanewline
\ \ \ \ \isacommand{from}\isamarkupfalse%
\ a{\isadigit{2}}{\isadigit{2}}\ \isakeyword{and}\ sg{\isadigit{2}}\ \isacommand{show}\isamarkupfalse%
\ {\isacharquery}thesis\ \isacommand{by}\isamarkupfalse%
\ simp\isanewline
\ \ \isacommand{qed}\isamarkupfalse%
\isanewline
\isacommand{qed}\isamarkupfalse%
\endisatagproof
{\isafoldproof}%
\isadelimproof
\isanewline
\endisadelimproof
\isanewline
\isanewline
\isanewline
\isanewline
\isacommand{lemma}\isamarkupfalse%
\ L{\isadigit{9}}{\isacharunderscore}Controller{\isacharcolon}\isanewline
\ \ \isakeyword{assumes}\ h{\isadigit{1}}{\isacharcolon}{\isachardoublequoteopen}Controller{\isacharunderscore}L\ s\ {\isacharparenleft}fin{\isacharunderscore}inf{\isacharunderscore}append\ {\isacharbrackleft}Zero{\isacharbrackright}\ l{\isacharparenright}\ l\ z{\isachardoublequoteclose}\isanewline
\ \ \ \ \ \ \isakeyword{and}\ h{\isadigit{2}}{\isacharcolon}{\isachardoublequoteopen}fin{\isacharunderscore}make{\isacharunderscore}untimed\ {\isacharparenleft}inf{\isacharunderscore}truncate\ z\ i{\isacharparenright}\ {\isacharbang}\isanewline
\ \ \ \ \ \ \ \ \ \ \ \ \ \ {\isacharparenleft}length\ {\isacharparenleft}fin{\isacharunderscore}make{\isacharunderscore}untimed\ {\isacharparenleft}inf{\isacharunderscore}truncate\ z\ i{\isacharparenright}{\isacharparenright}\ {\isacharminus}\ Suc\ {\isadigit{0}}{\isacharparenright}\ {\isacharequal}\ \ Zero{\isachardoublequoteclose}\isanewline
\ \ \ \ \ \ \isakeyword{and}\ h{\isadigit{3}}{\isacharcolon}{\isachardoublequoteopen}last\ {\isacharparenleft}fin{\isacharunderscore}make{\isacharunderscore}untimed\ {\isacharparenleft}inf{\isacharunderscore}truncate\ z\ i{\isacharparenright}{\isacharparenright}\ {\isacharequal}\ l\ i{\isachardoublequoteclose}\isanewline
\ \ \ \ \ \ \isakeyword{and}\ h{\isadigit{4}}{\isacharcolon}{\isachardoublequoteopen}{\isadigit{2}}{\isadigit{0}}{\isadigit{0}}\ {\isasymle}\ hd\ {\isacharparenleft}s\ i{\isacharparenright}{\isachardoublequoteclose}\isanewline
\ \ \ \ \ \ \isakeyword{and}\ h{\isadigit{5}}{\isacharcolon}{\isachardoublequoteopen}hd\ {\isacharparenleft}s\ {\isacharparenleft}Suc\ i{\isacharparenright}{\isacharparenright}\ {\isacharequal}\ hd\ {\isacharparenleft}s\ i{\isacharparenright}\ {\isacharminus}\ r{\isachardoublequoteclose}\isanewline
\ \ \ \ \ \ \isakeyword{and}\ h{\isadigit{6}}{\isacharcolon}{\isachardoublequoteopen}fin{\isacharunderscore}make{\isacharunderscore}untimed\ {\isacharparenleft}inf{\isacharunderscore}truncate\ z\ i{\isacharparenright}\ {\isasymnoteq}\ {\isacharbrackleft}{\isacharbrackright}{\isachardoublequoteclose}\isanewline
\ \ \ \ \ \ \isakeyword{and}\ h{\isadigit{7}}{\isacharcolon}{\isachardoublequoteopen}{\isadigit{0}}\ {\isacharless}\ r{\isachardoublequoteclose}\isanewline
\ \ \ \ \ \ \isakeyword{and}\ h{\isadigit{8}}{\isacharcolon}{\isachardoublequoteopen}r\ {\isasymle}\ {\isadigit{1}}{\isadigit{0}}{\isachardoublequoteclose}\isanewline
\ \ \isakeyword{shows}\ {\isachardoublequoteopen}{\isadigit{2}}{\isadigit{0}}{\isadigit{0}}\ {\isasymle}\ hd\ {\isacharparenleft}s\ {\isacharparenleft}Suc\ i{\isacharparenright}{\isacharparenright}{\isachardoublequoteclose}\isanewline
\isadelimproof
\endisadelimproof
\isatagproof
\isacommand{proof}\isamarkupfalse%
\ {\isacharminus}\isanewline
\ \ \isacommand{from}\isamarkupfalse%
\ h{\isadigit{6}}\ \isakeyword{and}\ h{\isadigit{2}}\ \isakeyword{and}\ h{\isadigit{3}}\ \isacommand{have}\isamarkupfalse%
\ sg{\isadigit{0}}{\isacharcolon}{\isachardoublequoteopen}l\ i\ {\isacharequal}\ Zero{\isachardoublequoteclose}\isanewline
\ \ \ \ \isacommand{by}\isamarkupfalse%
\ {\isacharparenleft}simp\ add{\isacharcolon}\ last{\isacharunderscore}nth{\isacharunderscore}length{\isacharparenright}\isanewline
\ \ \isacommand{show}\isamarkupfalse%
\ {\isacharquery}thesis\isanewline
\ \ \isacommand{proof}\isamarkupfalse%
\ {\isacharparenleft}cases\ {\isachardoublequoteopen}fin{\isacharunderscore}inf{\isacharunderscore}append\ {\isacharbrackleft}Zero{\isacharbrackright}\ l\ i\ {\isacharequal}\ Zero{\isachardoublequoteclose}{\isacharparenright}\isanewline
\ \ \ \ \isacommand{assume}\isamarkupfalse%
\ a{\isadigit{1}}{\isacharcolon}{\isachardoublequoteopen}fin{\isacharunderscore}inf{\isacharunderscore}append\ {\isacharbrackleft}Zero{\isacharbrackright}\ l\ i\ {\isacharequal}\ Zero{\isachardoublequoteclose}\isanewline
\ \ \ \ \isacommand{from}\isamarkupfalse%
\ a{\isadigit{1}}\ \isakeyword{and}\ h{\isadigit{1}}\ \isacommand{have}\isamarkupfalse%
\ sg{\isadigit{1}}{\isacharcolon}\isanewline
\ \ \ \ \ \ {\isachardoublequoteopen}if\ {\isadigit{3}}{\isadigit{0}}{\isadigit{0}}\ {\isacharless}\ hd\ {\isacharparenleft}s\ i{\isacharparenright}\ \isanewline
\ \ \ \ \ \ \ then\ z\ i\ {\isacharequal}\ {\isacharbrackleft}{\isacharbrackright}\ {\isasymand}\ l\ i\ {\isacharequal}\ Zero\ \isanewline
\ \ \ \ \ \ \ else\ z\ i\ {\isacharequal}\ {\isacharbrackleft}One{\isacharbrackright}\ {\isasymand}\ l\ i\ {\isacharequal}\ One{\isachardoublequoteclose}\isanewline
\ \ \ \ \ \ \ \isacommand{by}\isamarkupfalse%
\ {\isacharparenleft}simp\ add{\isacharcolon}\ Controller{\isacharunderscore}L{\isacharunderscore}def{\isacharparenright}\isanewline
\ \ \ \ \isacommand{show}\isamarkupfalse%
\ {\isacharquery}thesis\isanewline
\ \ \ \ \isacommand{proof}\isamarkupfalse%
\ {\isacharparenleft}cases\ {\isachardoublequoteopen}{\isadigit{3}}{\isadigit{0}}{\isadigit{0}}\ {\isacharless}\ hd\ {\isacharparenleft}s\ i{\isacharparenright}{\isachardoublequoteclose}{\isacharparenright}\isanewline
\ \ \ \ \ \ \isacommand{assume}\isamarkupfalse%
\ a{\isadigit{1}}{\isadigit{1}}{\isacharcolon}{\isachardoublequoteopen}{\isadigit{3}}{\isadigit{0}}{\isadigit{0}}\ {\isacharless}\ hd\ {\isacharparenleft}s\ i{\isacharparenright}{\isachardoublequoteclose}\isanewline
\ \ \ \ \ \ \isacommand{from}\isamarkupfalse%
\ a{\isadigit{1}}{\isadigit{1}}\ \isakeyword{and}\ h{\isadigit{5}}\ \isakeyword{and}\ h{\isadigit{8}}\ \isacommand{show}\isamarkupfalse%
\ {\isacharquery}thesis\ \isacommand{by}\isamarkupfalse%
\ simp\isanewline
\ \ \ \ \isacommand{next}\isamarkupfalse%
\isanewline
\ \ \ \ \ \ \isacommand{assume}\isamarkupfalse%
\ a{\isadigit{1}}{\isadigit{2}}{\isacharcolon}{\isachardoublequoteopen}{\isasymnot}\ {\isadigit{3}}{\isadigit{0}}{\isadigit{0}}\ {\isacharless}\ hd\ {\isacharparenleft}s\ i{\isacharparenright}{\isachardoublequoteclose}\ \isanewline
\ \ \ \ \ \ \isacommand{from}\isamarkupfalse%
\ a{\isadigit{1}}{\isadigit{2}}\ \isakeyword{and}\ sg{\isadigit{1}}\ \isakeyword{and}\ sg{\isadigit{0}}\ \isacommand{show}\isamarkupfalse%
\ {\isacharquery}thesis\ \isacommand{by}\isamarkupfalse%
\ simp\ \isanewline
\ \ \ \ \isacommand{qed}\isamarkupfalse%
\isanewline
\ \ \isacommand{next}\isamarkupfalse%
\isanewline
\ \ \ \ \isacommand{assume}\isamarkupfalse%
\ a{\isadigit{2}}{\isacharcolon}{\isachardoublequoteopen}fin{\isacharunderscore}inf{\isacharunderscore}append\ {\isacharbrackleft}Zero{\isacharbrackright}\ l\ i\ {\isasymnoteq}\ Zero{\isachardoublequoteclose}\isanewline
\ \ \ \ \isacommand{from}\isamarkupfalse%
\ a{\isadigit{2}}\ \isakeyword{and}\ h{\isadigit{1}}\ \isacommand{have}\isamarkupfalse%
\ sg{\isadigit{2}}{\isacharcolon}\isanewline
\ \ \ \ \ \ {\isachardoublequoteopen}if\ hd\ {\isacharparenleft}s\ i{\isacharparenright}\ {\isacharless}\ {\isadigit{7}}{\isadigit{0}}{\isadigit{0}}\ \isanewline
\ \ \ \ \ \ \ then\ z\ i\ {\isacharequal}\ {\isacharbrackleft}{\isacharbrackright}\ {\isasymand}\ l\ i\ {\isacharequal}\ One\ \isanewline
\ \ \ \ \ \ \ else\ z\ i\ {\isacharequal}\ {\isacharbrackleft}Zero{\isacharbrackright}\ {\isasymand}\ l\ i\ {\isacharequal}\ Zero{\isachardoublequoteclose}\isanewline
\ \ \ \ \ \ \ \isacommand{by}\isamarkupfalse%
\ {\isacharparenleft}simp\ add{\isacharcolon}\ Controller{\isacharunderscore}L{\isacharunderscore}def{\isacharparenright}\isanewline
\ \ \ \ \isacommand{show}\isamarkupfalse%
\ {\isacharquery}thesis\isanewline
\ \ \ \ \isacommand{proof}\isamarkupfalse%
\ {\isacharparenleft}cases\ {\isachardoublequoteopen}hd\ {\isacharparenleft}s\ i{\isacharparenright}\ {\isacharless}\ {\isadigit{7}}{\isadigit{0}}{\isadigit{0}}{\isachardoublequoteclose}{\isacharparenright}\isanewline
\ \ \ \ \ \ \isacommand{assume}\isamarkupfalse%
\ a{\isadigit{2}}{\isadigit{1}}{\isacharcolon}{\isachardoublequoteopen}hd\ {\isacharparenleft}s\ i{\isacharparenright}\ {\isacharless}\ {\isadigit{7}}{\isadigit{0}}{\isadigit{0}}{\isachardoublequoteclose}\isanewline
\ \ \ \ \ \ \isacommand{from}\isamarkupfalse%
\ this\ \isakeyword{and}\ sg{\isadigit{2}}\ \isakeyword{and}\ sg{\isadigit{0}}\ \isacommand{show}\isamarkupfalse%
\ {\isacharquery}thesis\ \isacommand{by}\isamarkupfalse%
\ simp\isanewline
\ \ \ \ \isacommand{next}\isamarkupfalse%
\isanewline
\ \ \ \ \ \ \isacommand{assume}\isamarkupfalse%
\ a{\isadigit{2}}{\isadigit{2}}{\isacharcolon}{\isachardoublequoteopen}{\isasymnot}\ hd\ {\isacharparenleft}s\ i{\isacharparenright}\ {\isacharless}\ {\isadigit{7}}{\isadigit{0}}{\isadigit{0}}{\isachardoublequoteclose}\isanewline
\ \ \ \ \ \ \isacommand{from}\isamarkupfalse%
\ this\ \isakeyword{and}\ h{\isadigit{5}}\ \isakeyword{and}\ h{\isadigit{8}}\ \ \isacommand{show}\isamarkupfalse%
\ {\isacharquery}thesis\ \isacommand{by}\isamarkupfalse%
\ simp\isanewline
\ \ \ \ \isacommand{qed}\isamarkupfalse%
\isanewline
\ \ \isacommand{qed}\isamarkupfalse%
\isanewline
\isacommand{qed}\isamarkupfalse%
\endisatagproof
{\isafoldproof}%
\isadelimproof
\isanewline
\endisadelimproof
\isanewline
\isanewline
\isacommand{lemma}\isamarkupfalse%
\ L{\isadigit{1}}{\isadigit{0}}{\isacharunderscore}Controller{\isacharcolon}\isanewline
\ \ \isakeyword{assumes}\ h{\isadigit{1}}{\isacharcolon}{\isachardoublequoteopen}Controller{\isacharunderscore}L\ s\ {\isacharparenleft}fin{\isacharunderscore}inf{\isacharunderscore}append\ {\isacharbrackleft}Zero{\isacharbrackright}\ l{\isacharparenright}\ l\ z{\isachardoublequoteclose}\isanewline
\ \ \ \ \ \ \isakeyword{and}\ h{\isadigit{2}}{\isacharcolon}{\isachardoublequoteopen}fin{\isacharunderscore}make{\isacharunderscore}untimed\ {\isacharparenleft}inf{\isacharunderscore}truncate\ z\ i{\isacharparenright}\ {\isacharbang}\isanewline
\ \ \ \ \ \ \ \ \ \ \ \ \ \ {\isacharparenleft}length\ {\isacharparenleft}fin{\isacharunderscore}make{\isacharunderscore}untimed\ {\isacharparenleft}inf{\isacharunderscore}truncate\ z\ i{\isacharparenright}{\isacharparenright}\ {\isacharminus}\ Suc\ {\isadigit{0}}{\isacharparenright}\ {\isasymnoteq}\ \ Zero{\isachardoublequoteclose}\isanewline
\ \ \ \ \ \ \isakeyword{and}\ h{\isadigit{3}}{\isacharcolon}{\isachardoublequoteopen}last\ {\isacharparenleft}fin{\isacharunderscore}make{\isacharunderscore}untimed\ {\isacharparenleft}inf{\isacharunderscore}truncate\ z\ i{\isacharparenright}{\isacharparenright}\ {\isacharequal}\ l\ i{\isachardoublequoteclose}\isanewline
\ \ \ \ \ \ \isakeyword{and}\ h{\isadigit{4}}{\isacharcolon}{\isachardoublequoteopen}hd\ {\isacharparenleft}s\ i{\isacharparenright}\ {\isasymle}\ {\isadigit{8}}{\isadigit{0}}{\isadigit{0}}{\isachardoublequoteclose}\isanewline
\ \ \ \ \ \ \isakeyword{and}\ h{\isadigit{5}}{\isacharcolon}{\isachardoublequoteopen}hd\ {\isacharparenleft}s\ {\isacharparenleft}Suc\ i{\isacharparenright}{\isacharparenright}\ {\isacharequal}\ hd\ {\isacharparenleft}s\ i{\isacharparenright}\ {\isacharplus}\ r{\isachardoublequoteclose}\isanewline
\ \ \ \ \ \ \isakeyword{and}\ h{\isadigit{6}}{\isacharcolon}{\isachardoublequoteopen}fin{\isacharunderscore}make{\isacharunderscore}untimed\ {\isacharparenleft}inf{\isacharunderscore}truncate\ z\ i{\isacharparenright}\ {\isasymnoteq}\ {\isacharbrackleft}{\isacharbrackright}{\isachardoublequoteclose}\isanewline
\ \ \ \ \ \ \isakeyword{and}\ h{\isadigit{7}}{\isacharcolon}{\isachardoublequoteopen}{\isadigit{0}}\ {\isacharless}\ r{\isachardoublequoteclose}\isanewline
\ \ \ \ \ \ \isakeyword{and}\ h{\isadigit{8}}{\isacharcolon}{\isachardoublequoteopen}r\ {\isasymle}\ {\isadigit{1}}{\isadigit{0}}{\isachardoublequoteclose}\isanewline
\ \ \isakeyword{shows}\ {\isachardoublequoteopen}hd\ {\isacharparenleft}s\ {\isacharparenleft}Suc\ i{\isacharparenright}{\isacharparenright}\ {\isasymle}\ {\isadigit{8}}{\isadigit{0}}{\isadigit{0}}{\isachardoublequoteclose}\isanewline
\isadelimproof
\endisadelimproof
\isatagproof
\isacommand{proof}\isamarkupfalse%
\ {\isacharminus}\isanewline
\ \ \isacommand{from}\isamarkupfalse%
\ h{\isadigit{6}}\ \isakeyword{and}\ h{\isadigit{2}}\ \isakeyword{and}\ h{\isadigit{3}}\ \isacommand{have}\isamarkupfalse%
\ sg{\isadigit{0}}{\isacharcolon}{\isachardoublequoteopen}l\ i\ {\isasymnoteq}\ \ Zero{\isachardoublequoteclose}\isanewline
\ \ \ \ \isacommand{by}\isamarkupfalse%
\ {\isacharparenleft}simp\ add{\isacharcolon}\ last{\isacharunderscore}nth{\isacharunderscore}length{\isacharparenright}\isanewline
\ \ \isacommand{show}\isamarkupfalse%
\ {\isacharquery}thesis\isanewline
\ \ \isacommand{proof}\isamarkupfalse%
\ {\isacharparenleft}cases\ {\isachardoublequoteopen}fin{\isacharunderscore}inf{\isacharunderscore}append\ {\isacharbrackleft}Zero{\isacharbrackright}\ l\ i\ {\isacharequal}\ Zero{\isachardoublequoteclose}{\isacharparenright}\isanewline
\ \ \ \ \isacommand{assume}\isamarkupfalse%
\ a{\isadigit{1}}{\isacharcolon}{\isachardoublequoteopen}fin{\isacharunderscore}inf{\isacharunderscore}append\ {\isacharbrackleft}Zero{\isacharbrackright}\ l\ i\ {\isacharequal}\ Zero{\isachardoublequoteclose}\isanewline
\ \ \ \ \isacommand{from}\isamarkupfalse%
\ a{\isadigit{1}}\ \isakeyword{and}\ h{\isadigit{1}}\ \isacommand{have}\isamarkupfalse%
\ sg{\isadigit{1}}{\isacharcolon}\isanewline
\ \ \ \ \ \ {\isachardoublequoteopen}if\ {\isadigit{3}}{\isadigit{0}}{\isadigit{0}}\ {\isacharless}\ hd\ {\isacharparenleft}s\ i{\isacharparenright}\ \isanewline
\ \ \ \ \ \ \ then\ z\ i\ {\isacharequal}\ {\isacharbrackleft}{\isacharbrackright}\ {\isasymand}\ l\ i\ {\isacharequal}\ Zero\ \isanewline
\ \ \ \ \ \ \ else\ z\ i\ {\isacharequal}\ {\isacharbrackleft}One{\isacharbrackright}\ {\isasymand}\ l\ i\ {\isacharequal}\ One{\isachardoublequoteclose}\isanewline
\ \ \ \ \ \ \ \isacommand{by}\isamarkupfalse%
\ {\isacharparenleft}simp\ add{\isacharcolon}\ Controller{\isacharunderscore}L{\isacharunderscore}def{\isacharparenright}\isanewline
\ \ \ \ \isacommand{show}\isamarkupfalse%
\ {\isacharquery}thesis\isanewline
\ \ \ \ \isacommand{proof}\isamarkupfalse%
\ {\isacharparenleft}cases\ {\isachardoublequoteopen}{\isadigit{3}}{\isadigit{0}}{\isadigit{0}}\ {\isacharless}\ hd\ {\isacharparenleft}s\ i{\isacharparenright}{\isachardoublequoteclose}{\isacharparenright}\isanewline
\ \ \ \ \ \ \isacommand{assume}\isamarkupfalse%
\ a{\isadigit{1}}{\isadigit{1}}{\isacharcolon}{\isachardoublequoteopen}{\isadigit{3}}{\isadigit{0}}{\isadigit{0}}\ {\isacharless}\ hd\ {\isacharparenleft}s\ i{\isacharparenright}{\isachardoublequoteclose}\isanewline
\ \ \ \ \ \ \isacommand{from}\isamarkupfalse%
\ a{\isadigit{1}}{\isadigit{1}}\ \isakeyword{and}\ sg{\isadigit{1}}\ \isakeyword{and}\ sg{\isadigit{0}}\ \isacommand{show}\isamarkupfalse%
\ {\isacharquery}thesis\ \isacommand{by}\isamarkupfalse%
\ simp\isanewline
\ \ \ \ \isacommand{next}\isamarkupfalse%
\isanewline
\ \ \ \ \ \ \isacommand{assume}\isamarkupfalse%
\ a{\isadigit{1}}{\isadigit{2}}{\isacharcolon}{\isachardoublequoteopen}{\isasymnot}\ {\isadigit{3}}{\isadigit{0}}{\isadigit{0}}\ {\isacharless}\ hd\ {\isacharparenleft}s\ i{\isacharparenright}{\isachardoublequoteclose}\ \isanewline
\ \ \ \ \ \ \isacommand{from}\isamarkupfalse%
\ h{\isadigit{5}}\ \isakeyword{and}\ a{\isadigit{1}}{\isadigit{2}}\ \isakeyword{and}\ h{\isadigit{8}}\ \isacommand{show}\isamarkupfalse%
\ {\isacharquery}thesis\ \isacommand{by}\isamarkupfalse%
\ simp\ \isanewline
\ \ \ \ \isacommand{qed}\isamarkupfalse%
\isanewline
\ \ \isacommand{next}\isamarkupfalse%
\isanewline
\ \ \ \ \isacommand{assume}\isamarkupfalse%
\ a{\isadigit{2}}{\isacharcolon}{\isachardoublequoteopen}fin{\isacharunderscore}inf{\isacharunderscore}append\ {\isacharbrackleft}Zero{\isacharbrackright}\ l\ i\ {\isasymnoteq}\ Zero{\isachardoublequoteclose}\isanewline
\ \ \ \ \isacommand{from}\isamarkupfalse%
\ a{\isadigit{2}}\ \isakeyword{and}\ h{\isadigit{1}}\ \isacommand{have}\isamarkupfalse%
\ sg{\isadigit{2}}{\isacharcolon}\isanewline
\ \ \ \ \ \ {\isachardoublequoteopen}if\ hd\ {\isacharparenleft}s\ i{\isacharparenright}\ {\isacharless}\ {\isadigit{7}}{\isadigit{0}}{\isadigit{0}}\ \isanewline
\ \ \ \ \ \ \ then\ z\ i\ {\isacharequal}\ {\isacharbrackleft}{\isacharbrackright}\ {\isasymand}\ l\ i\ {\isacharequal}\ One\ \isanewline
\ \ \ \ \ \ \ else\ z\ i\ {\isacharequal}\ {\isacharbrackleft}Zero{\isacharbrackright}\ {\isasymand}\ l\ i\ {\isacharequal}\ Zero{\isachardoublequoteclose}\isanewline
\ \ \ \ \ \ \ \isacommand{by}\isamarkupfalse%
\ {\isacharparenleft}simp\ add{\isacharcolon}\ Controller{\isacharunderscore}L{\isacharunderscore}def{\isacharparenright}\isanewline
\ \ \ \ \isacommand{show}\isamarkupfalse%
\ {\isacharquery}thesis\isanewline
\ \ \ \ \isacommand{proof}\isamarkupfalse%
\ {\isacharparenleft}cases\ {\isachardoublequoteopen}hd\ {\isacharparenleft}s\ i{\isacharparenright}\ {\isacharless}\ {\isadigit{7}}{\isadigit{0}}{\isadigit{0}}{\isachardoublequoteclose}{\isacharparenright}\isanewline
\ \ \ \ \ \ \isacommand{assume}\isamarkupfalse%
\ a{\isadigit{2}}{\isadigit{1}}{\isacharcolon}{\isachardoublequoteopen}hd\ {\isacharparenleft}s\ i{\isacharparenright}\ {\isacharless}\ {\isadigit{7}}{\isadigit{0}}{\isadigit{0}}{\isachardoublequoteclose}\isanewline
\ \ \ \ \ \ \isacommand{from}\isamarkupfalse%
\ this\ \isakeyword{and}\ h{\isadigit{5}}\ \isakeyword{and}\ h{\isadigit{8}}\ \isacommand{show}\isamarkupfalse%
\ {\isacharquery}thesis\ \isacommand{by}\isamarkupfalse%
\ simp\isanewline
\ \ \ \ \isacommand{next}\isamarkupfalse%
\isanewline
\ \ \ \ \ \ \isacommand{assume}\isamarkupfalse%
\ a{\isadigit{2}}{\isadigit{2}}{\isacharcolon}{\isachardoublequoteopen}{\isasymnot}\ hd\ {\isacharparenleft}s\ i{\isacharparenright}\ {\isacharless}\ {\isadigit{7}}{\isadigit{0}}{\isadigit{0}}{\isachardoublequoteclose}\isanewline
\ \ \ \ \ \ \isacommand{from}\isamarkupfalse%
\ this\ \isakeyword{and}\ sg{\isadigit{2}}\ \isakeyword{and}\ sg{\isadigit{0}}\ \isacommand{show}\isamarkupfalse%
\ {\isacharquery}thesis\ \isacommand{by}\isamarkupfalse%
\ simp\isanewline
\ \ \ \ \isacommand{qed}\isamarkupfalse%
\isanewline
\ \ \isacommand{qed}\isamarkupfalse%
\isanewline
\isacommand{qed}\isamarkupfalse%
\endisatagproof
{\isafoldproof}%
\isadelimproof
\endisadelimproof
\isamarkupsubsection{Properties of the Converter Component%
}
\isamarkuptrue%
\isacommand{lemma}\isamarkupfalse%
\ L{\isadigit{1}}{\isacharunderscore}Converter{\isacharcolon}\isanewline
\ \ \isakeyword{assumes}\ h{\isadigit{1}}{\isacharcolon}{\isachardoublequoteopen}Converter\ z\ x{\isachardoublequoteclose}\isanewline
\ \ \ \ \ \ \isakeyword{and}\ h{\isadigit{2}}{\isacharcolon}{\isachardoublequoteopen}fin{\isacharunderscore}make{\isacharunderscore}untimed\ {\isacharparenleft}inf{\isacharunderscore}truncate\ z\ t{\isacharparenright}\ {\isasymnoteq}\ {\isacharbrackleft}{\isacharbrackright}{\isachardoublequoteclose}\isanewline
\ \ \isakeyword{shows}\ \ \ \ \ \ {\isachardoublequoteopen}hd\ {\isacharparenleft}x\ t{\isacharparenright}\ {\isacharequal}\ {\isacharparenleft}fin{\isacharunderscore}make{\isacharunderscore}untimed\ {\isacharparenleft}inf{\isacharunderscore}truncate\ z\ t{\isacharparenright}{\isacharparenright}\ {\isacharbang}\ \isanewline
\ \ \ \ \ \ \ \ \ \ \ \ \ \ \ \ \ {\isacharparenleft}{\isacharparenleft}length\ {\isacharparenleft}fin{\isacharunderscore}make{\isacharunderscore}untimed\ {\isacharparenleft}inf{\isacharunderscore}truncate\ z\ t{\isacharparenright}{\isacharparenright}{\isacharparenright}\ {\isacharminus}\ {\isacharparenleft}{\isadigit{1}}{\isacharcolon}{\isacharcolon}nat{\isacharparenright}{\isacharparenright}{\isachardoublequoteclose}\isanewline
\isadelimproof
\endisadelimproof
\isatagproof
\isacommand{using}\isamarkupfalse%
\ assms\isanewline
\ \ \isacommand{by}\isamarkupfalse%
\ {\isacharparenleft}simp\ add{\isacharcolon}\ Converter{\isacharunderscore}def{\isacharparenright}%
\endisatagproof
{\isafoldproof}%
\isadelimproof
\isanewline
\endisadelimproof
\isanewline
\isanewline
\isacommand{lemma}\isamarkupfalse%
\ L{\isadigit{1}}a{\isacharunderscore}Converter{\isacharcolon}\isanewline
\ \ \isakeyword{assumes}\ h{\isadigit{1}}{\isacharcolon}{\isachardoublequoteopen}Converter\ z\ x{\isachardoublequoteclose}\isanewline
\ \ \ \ \ \ \isakeyword{and}\ h{\isadigit{2}}{\isacharcolon}{\isachardoublequoteopen}fin{\isacharunderscore}make{\isacharunderscore}untimed\ {\isacharparenleft}inf{\isacharunderscore}truncate\ z\ t{\isacharparenright}\ {\isasymnoteq}\ {\isacharbrackleft}{\isacharbrackright}{\isachardoublequoteclose}\isanewline
\ \ \ \ \ \ \isakeyword{and}\ h{\isadigit{3}}{\isacharcolon}{\isachardoublequoteopen}hd\ {\isacharparenleft}x\ t{\isacharparenright}\ {\isacharequal}\ Zero{\isachardoublequoteclose}\isanewline
\ \ \isakeyword{shows}\ \ \ \ \ \ {\isachardoublequoteopen}{\isacharparenleft}fin{\isacharunderscore}make{\isacharunderscore}untimed\ {\isacharparenleft}inf{\isacharunderscore}truncate\ z\ t{\isacharparenright}{\isacharparenright}\ {\isacharbang}\ \isanewline
\ \ \ \ \ \ \ \ \ \ \ \ \ \ \ \ \ {\isacharparenleft}{\isacharparenleft}length\ {\isacharparenleft}fin{\isacharunderscore}make{\isacharunderscore}untimed\ {\isacharparenleft}inf{\isacharunderscore}truncate\ z\ t{\isacharparenright}{\isacharparenright}{\isacharparenright}\ {\isacharminus}\ {\isacharparenleft}{\isadigit{1}}{\isacharcolon}{\isacharcolon}nat{\isacharparenright}{\isacharparenright}\ \isanewline
\ \ \ \ \ \ \ \ \ \ \ \ \ \ {\isacharequal}\ Zero{\isachardoublequoteclose}\isanewline
\isadelimproof
\endisadelimproof
\isatagproof
\isacommand{using}\isamarkupfalse%
\ assms\isanewline
\ \ \isacommand{by}\isamarkupfalse%
\ {\isacharparenleft}simp\ add{\isacharcolon}\ L{\isadigit{1}}{\isacharunderscore}Converter{\isacharparenright}%
\endisatagproof
{\isafoldproof}%
\isadelimproof
\endisadelimproof
\isamarkupsubsection{Properties of the System%
}
\isamarkuptrue%
\isacommand{lemma}\isamarkupfalse%
\ L{\isadigit{1}}{\isacharunderscore}ControlSystem{\isacharcolon}\isanewline
\ \ \isakeyword{assumes}\ h{\isadigit{1}}{\isacharcolon}{\isachardoublequoteopen}ControlSystemArch\ s{\isachardoublequoteclose}\isanewline
\ \ \isakeyword{shows}\ {\isachardoublequoteopen}ts\ s{\isachardoublequoteclose}\isanewline
\isadelimproof
\endisadelimproof
\isatagproof
\isacommand{proof}\isamarkupfalse%
\ {\isacharminus}\ \isanewline
\ \ \isacommand{from}\isamarkupfalse%
\ h{\isadigit{1}}\ \isacommand{obtain}\isamarkupfalse%
\ x\ z\ y\ \isanewline
\ \ \ \ \isakeyword{where}\ a{\isadigit{1}}{\isacharcolon}{\isachardoublequoteopen}Converter\ z\ x{\isachardoublequoteclose}\ \isakeyword{and}\ a{\isadigit{2}}{\isacharcolon}\ {\isachardoublequoteopen}SteamBoiler\ x\ s\ y{\isachardoublequoteclose}\isanewline
\ \ \ \ \isacommand{by}\isamarkupfalse%
\ {\isacharparenleft}simp\ only{\isacharcolon}\ ControlSystemArch{\isacharunderscore}def{\isacharcomma}\ auto{\isacharparenright}\ \ \ \ \isanewline
\ \ \isacommand{from}\isamarkupfalse%
\ this\ \isacommand{have}\isamarkupfalse%
\ sg{\isadigit{1}}{\isacharcolon}{\isachardoublequoteopen}ts\ x{\isachardoublequoteclose}\isanewline
\ \ \ \ \isacommand{by}\isamarkupfalse%
\ {\isacharparenleft}simp\ add{\isacharcolon}\ Converter{\isacharunderscore}def{\isacharparenright}\isanewline
\ \ \isacommand{from}\isamarkupfalse%
\ a{\isadigit{2}}\ \isakeyword{and}\ sg{\isadigit{1}}\ \isacommand{show}\isamarkupfalse%
\ {\isacharquery}thesis\ \isacommand{by}\isamarkupfalse%
\ {\isacharparenleft}rule\ L{\isadigit{1}}{\isacharunderscore}Boiler{\isacharparenright}\ \isanewline
\isacommand{qed}\isamarkupfalse%
\endisatagproof
{\isafoldproof}%
\isadelimproof
\isanewline
\endisadelimproof
\ \isanewline 
\isacommand{lemma}\isamarkupfalse%
\ L{\isadigit{2}}{\isacharunderscore}ControlSystem{\isacharcolon}\isanewline
\ \ \isakeyword{assumes}\ h{\isadigit{1}}{\isacharcolon}{\isachardoublequoteopen}ControlSystemArch\ s{\isachardoublequoteclose}\isanewline
\ \ \isakeyword{shows}\ {\isachardoublequoteopen}{\isacharparenleft}{\isadigit{2}}{\isadigit{0}}{\isadigit{0}}{\isacharcolon}{\isacharcolon}nat{\isacharparenright}\ {\isasymle}\ hd\ {\isacharparenleft}s\ i{\isacharparenright}{\isachardoublequoteclose}\isanewline
\isadelimproof
\endisadelimproof
\isatagproof
\isacommand{proof}\isamarkupfalse%
\ {\isacharminus}\ \isanewline
\ \ \isacommand{from}\isamarkupfalse%
\ h{\isadigit{1}}\ \isacommand{obtain}\isamarkupfalse%
\ x\ z\ y\ \isanewline
\ \ \ \ \isakeyword{where}\ a{\isadigit{1}}{\isacharcolon}{\isachardoublequoteopen}Converter\ z\ x{\isachardoublequoteclose}\ \isakeyword{and}\ a{\isadigit{2}}{\isacharcolon}\ {\isachardoublequoteopen}SteamBoiler\ x\ s\ y{\isachardoublequoteclose}\ \isakeyword{and}\ a{\isadigit{3}}{\isacharcolon}{\isachardoublequoteopen}Controller\ y\ z{\isachardoublequoteclose}\isanewline
\ \ \ \ \isacommand{by}\isamarkupfalse%
\ {\isacharparenleft}simp\ only{\isacharcolon}\ ControlSystemArch{\isacharunderscore}def{\isacharcomma}\ auto{\isacharparenright}\ \isanewline
\ \ \isacommand{from}\isamarkupfalse%
\ this\ \isacommand{have}\isamarkupfalse%
\ sg{\isadigit{1}}{\isacharcolon}{\isachardoublequoteopen}ts\ x{\isachardoublequoteclose}\ \ \isacommand{by}\isamarkupfalse%
\ {\isacharparenleft}simp\ add{\isacharcolon}\ Converter{\isacharunderscore}def{\isacharparenright}\isanewline
\ \ \isacommand{from}\isamarkupfalse%
\ sg{\isadigit{1}}\ \isakeyword{and}\ a{\isadigit{2}}\ \isacommand{have}\isamarkupfalse%
\ sg{\isadigit{2}}{\isacharcolon}{\isachardoublequoteopen}ts\ y{\isachardoublequoteclose}\ \ \isacommand{by}\isamarkupfalse%
\ {\isacharparenleft}simp\ add{\isacharcolon}\ L{\isadigit{2}}{\isacharunderscore}Boiler{\isacharparenright}\isanewline
\ \ \isacommand{from}\isamarkupfalse%
\ sg{\isadigit{1}}\ \isakeyword{and}\ a{\isadigit{2}}\ \isacommand{have}\isamarkupfalse%
\ sg{\isadigit{3}}{\isacharcolon}{\isachardoublequoteopen}y\ {\isacharequal}\ s{\isachardoublequoteclose}\ \isacommand{by}\isamarkupfalse%
\ {\isacharparenleft}simp\ add{\isacharcolon}\ SteamBoiler{\isacharunderscore}def{\isacharparenright}\isanewline
\ \ \isacommand{from}\isamarkupfalse%
\ a{\isadigit{1}}\ \isakeyword{and}\ a{\isadigit{2}}\ \isakeyword{and}\ a{\isadigit{3}}\ \isakeyword{and}\ sg{\isadigit{1}}\ \isakeyword{and}\ sg{\isadigit{2}}\ \isakeyword{and}\ sg{\isadigit{3}}\ \isacommand{show}\isamarkupfalse%
\ {\isachardoublequoteopen}{\isadigit{2}}{\isadigit{0}}{\isadigit{0}}\ {\isasymle}\ hd\ {\isacharparenleft}s\ i{\isacharparenright}{\isachardoublequoteclose}\isanewline
\ \ \isacommand{proof}\isamarkupfalse%
\ {\isacharparenleft}induction\ i{\isacharparenright}\isanewline
\ \ \ \ \isacommand{case}\isamarkupfalse%
\ {\isadigit{0}}\isanewline
\ \ \ \ \isacommand{from}\isamarkupfalse%
\ this\ \isacommand{show}\isamarkupfalse%
\ {\isacharquery}case\ \ \isacommand{by}\isamarkupfalse%
\ {\isacharparenleft}simp\ add{\isacharcolon}\ L{\isadigit{3}}{\isacharunderscore}Boiler{\isacharparenright}\isanewline
\ \ \isacommand{next}\isamarkupfalse%
\isanewline
\ \ \ \ \isacommand{fix}\isamarkupfalse%
\ i\isanewline
\ \ \ \ \isacommand{case}\isamarkupfalse%
\ {\isacharparenleft}Suc\ i{\isacharparenright}\isanewline
\ \ \ \ \isacommand{from}\isamarkupfalse%
\ this\ \isacommand{obtain}\isamarkupfalse%
\ l\isanewline
\ \ \ \ \ \ \isakeyword{where}\ a{\isadigit{4}}{\isacharcolon}\ {\isachardoublequoteopen}Controller{\isacharunderscore}L\ s\ {\isacharparenleft}fin{\isacharunderscore}inf{\isacharunderscore}append\ {\isacharbrackleft}Zero{\isacharbrackright}\ l{\isacharparenright}\ l\ z{\isachardoublequoteclose}\isanewline
\ \ \ \ \ \ \isacommand{by}\isamarkupfalse%
\ {\isacharparenleft}simp\ add{\isacharcolon}\ Controller{\isacharunderscore}def{\isacharcomma}\ atomize{\isacharcomma}\ auto{\isacharparenright}\isanewline
\ \ \ \ \isacommand{from}\isamarkupfalse%
\ Suc\ \isakeyword{and}\ a{\isadigit{4}}\ \isacommand{have}\isamarkupfalse%
\ sg{\isadigit{4}}{\isacharcolon}{\isachardoublequoteopen}fin{\isacharunderscore}make{\isacharunderscore}untimed\ {\isacharparenleft}inf{\isacharunderscore}truncate\ z\ i{\isacharparenright}\ {\isasymnoteq}\ {\isacharbrackleft}{\isacharbrackright}{\isachardoublequoteclose}\isanewline
\ \ \ \ \ \ \isacommand{by}\isamarkupfalse%
\ {\isacharparenleft}simp\ add{\isacharcolon}\ \ L{\isadigit{1}}{\isacharunderscore}Controller{\isacharparenright}\isanewline
\ \ \ \ \isacommand{from}\isamarkupfalse%
\ a{\isadigit{2}}\ \isakeyword{and}\ sg{\isadigit{1}}\ \isacommand{have}\isamarkupfalse%
\ y{\isadigit{0}}asm{\isacharcolon}{\isachardoublequoteopen}y\ {\isadigit{0}}\ {\isacharequal}\ {\isacharbrackleft}{\isadigit{5}}{\isadigit{0}}{\isadigit{0}}{\isacharcolon}{\isacharcolon}nat{\isacharbrackright}{\isachardoublequoteclose}\ \ \isacommand{by}\isamarkupfalse%
\ {\isacharparenleft}simp\ add{\isacharcolon}\ SteamBoiler{\isacharunderscore}def{\isacharparenright}\isanewline
\ \ \ \ \isacommand{from}\isamarkupfalse%
\ Suc\ \isakeyword{and}\ a{\isadigit{4}}\ \isakeyword{and}\ sg{\isadigit{4}}\ \isakeyword{and}\ y{\isadigit{0}}asm\ \isacommand{have}\isamarkupfalse%
\ sg{\isadigit{5}}{\isacharcolon}\ {\isachardoublequoteopen}last\ {\isacharparenleft}fin{\isacharunderscore}make{\isacharunderscore}untimed\ {\isacharparenleft}inf{\isacharunderscore}truncate\ z\ i{\isacharparenright}{\isacharparenright}\ {\isacharequal}\ \ l\ i{\isachardoublequoteclose}\isanewline
\ \ \ \ \ \ \isacommand{by}\isamarkupfalse%
\ {\isacharparenleft}simp\ add{\isacharcolon}\ L{\isadigit{7}}{\isacharunderscore}Controller{\isacharparenright}\isanewline
\ \ \ \ \isacommand{from}\isamarkupfalse%
\ a{\isadigit{2}}\ \isakeyword{and}\ sg{\isadigit{1}}\ \isacommand{obtain}\isamarkupfalse%
\ r\ \isakeyword{where}\isanewline
 \ \ \ \ \ \ \ aa{\isadigit{0}}{\isacharcolon}{\isachardoublequoteopen}{\isadigit{0}}\ {\isacharless}\ r{\isachardoublequoteclose}\ \isakeyword{and}\isanewline
\ \ \ \ \ \ \ aa{\isadigit{1}}{\isacharcolon}{\isachardoublequoteopen}r\ {\isasymle}\ {\isadigit{1}}{\isadigit{0}}{\isachardoublequoteclose}\ \isakeyword{and}\ \isanewline
\ \ \ \ \ \ \ aa{\isadigit{2}}{\isacharcolon}{\isachardoublequoteopen}hd\ {\isacharparenleft}s\ {\isacharparenleft}Suc\ i{\isacharparenright}{\isacharparenright}\ {\isacharequal}\ {\isacharparenleft}if\ hd\ {\isacharparenleft}x\ i{\isacharparenright}\ {\isacharequal}\ Zero\ then\ hd\ {\isacharparenleft}s\ i{\isacharparenright}\ {\isacharminus}\ r\ else\ hd\ {\isacharparenleft}s\ i{\isacharparenright}\ {\isacharplus}\ r{\isacharparenright}{\isachardoublequoteclose}\isanewline
\ \ \ \ \ \ \ \ \ \isacommand{by}\isamarkupfalse%
\ {\isacharparenleft}simp\ add{\isacharcolon}\ SteamBoiler{\isacharunderscore}def{\isacharcomma}\ auto{\isacharparenright}\isanewline
\ \ \ \ \isacommand{from}\isamarkupfalse%
\ Suc\ \isakeyword{and}\ a{\isadigit{4}}\ \isakeyword{and}\ sg{\isadigit{4}}\ \isakeyword{and}\ sg{\isadigit{5}}\ \isacommand{show}\isamarkupfalse%
\ {\isacharquery}case\isanewline
\ \ \ \ \isacommand{proof}\isamarkupfalse%
\ {\isacharparenleft}cases\ {\isachardoublequoteopen}hd\ {\isacharparenleft}x\ i{\isacharparenright}\ {\isacharequal}\ Zero{\isachardoublequoteclose}{\isacharparenright}\isanewline
\ \ \ \ \ \ \ \isacommand{assume}\isamarkupfalse%
\ aaZero{\isacharcolon}{\isachardoublequoteopen}hd\ {\isacharparenleft}x\ i{\isacharparenright}\ {\isacharequal}\ Zero{\isachardoublequoteclose}\isanewline
\ \ \ \ \ \ \ \isacommand{from}\isamarkupfalse%
\ a{\isadigit{1}}\ \isakeyword{and}\ sg{\isadigit{4}}\ \isakeyword{and}\ this\ \isacommand{have}\isamarkupfalse%
\isanewline
\ \ \ \ \ \ \ \ \ sg{\isadigit{7}}{\isacharcolon}{\isachardoublequoteopen}{\isacharparenleft}fin{\isacharunderscore}make{\isacharunderscore}untimed\ {\isacharparenleft}inf{\isacharunderscore}truncate\ z\ i{\isacharparenright}{\isacharparenright}\ {\isacharbang}\ \isanewline
\ \ \ \ \ \ \ \ \ \ \ \ \ {\isacharparenleft}{\isacharparenleft}length\ {\isacharparenleft}fin{\isacharunderscore}make{\isacharunderscore}untimed\ {\isacharparenleft}inf{\isacharunderscore}truncate\ z\ i{\isacharparenright}{\isacharparenright}{\isacharparenright}\ {\isacharminus}\ Suc\ {\isadigit{0}}{\isacharparenright}\ \ {\isacharequal}\ Zero{\isachardoublequoteclose}\isanewline
\ \ \ \ \ \ \ \ \ \isacommand{by}\isamarkupfalse%
\ {\isacharparenleft}simp\ add{\isacharcolon}\ L{\isadigit{1}}{\isacharunderscore}Converter{\isacharparenright}\isanewline
\ \ \ \ \ \ \ \isacommand{from}\isamarkupfalse%
\ aa{\isadigit{2}}\ \isakeyword{and}\ aaZero\ \isacommand{have}\isamarkupfalse%
\ sg{\isadigit{8}}{\isacharcolon}{\isachardoublequoteopen}hd\ {\isacharparenleft}s\ {\isacharparenleft}Suc\ i{\isacharparenright}{\isacharparenright}\ {\isacharequal}\ hd\ {\isacharparenleft}s\ i{\isacharparenright}\ {\isacharminus}\ r{\isachardoublequoteclose}\ \isacommand{by}\isamarkupfalse%
\ simp\isanewline
\ \ \ \ \ \ \ \isacommand{from}\isamarkupfalse%
\ Suc\ \isacommand{have}\isamarkupfalse%
\ sgSuc{\isacharcolon}{\isachardoublequoteopen}{\isadigit{2}}{\isadigit{0}}{\isadigit{0}}\ {\isasymle}\ hd\ {\isacharparenleft}s\ i{\isacharparenright}{\isachardoublequoteclose}\ \isacommand{by}\isamarkupfalse%
\ simp\isanewline
\ \ \ \ \ \ \ \isacommand{from}\isamarkupfalse%
\ a{\isadigit{4}}\ \isakeyword{and}\ sg{\isadigit{7}}\ \isakeyword{and}\ sg{\isadigit{5}}\ \isakeyword{and}\ sgSuc\ \isakeyword{and}\ sg{\isadigit{8}}\ \isakeyword{and}\ sg{\isadigit{4}}\ \isakeyword{and}\ aa{\isadigit{0}}\ \isakeyword{and}\ aa{\isadigit{1}}\ \isanewline
\ \ \ \ \ \  \ \isacommand{show}\isamarkupfalse%
\ {\isacharquery}thesis\isanewline
\ \ \ \ \ \ \ \ \ \ \isacommand{by}\isamarkupfalse%
\ {\isacharparenleft}rule\ L{\isadigit{9}}{\isacharunderscore}Controller{\isacharparenright}\isanewline
\ \ \ \ \ \isacommand{next}\isamarkupfalse%
\isanewline
\ \ \ \ \ \ \ \isacommand{assume}\isamarkupfalse%
\ aaOne{\isacharcolon}{\isachardoublequoteopen}hd\ {\isacharparenleft}x\ i{\isacharparenright}\ {\isasymnoteq}\ Zero{\isachardoublequoteclose}\isanewline
\ \ \ \ \ \ \ \isacommand{from}\isamarkupfalse%
\ a{\isadigit{1}}\ \isakeyword{and}\ sg{\isadigit{4}}\ \isakeyword{and}\ this\ \isacommand{have}\isamarkupfalse%
\isanewline
\ \ \ \ \ \ \ \ \ sg{\isadigit{7}}{\isacharcolon}{\isachardoublequoteopen}{\isacharparenleft}fin{\isacharunderscore}make{\isacharunderscore}untimed\ {\isacharparenleft}inf{\isacharunderscore}truncate\ z\ i{\isacharparenright}{\isacharparenright}\ {\isacharbang}\ \isanewline
\ \ \ \ \ \ \ \ \ \ \ \ \ {\isacharparenleft}{\isacharparenleft}length\ {\isacharparenleft}fin{\isacharunderscore}make{\isacharunderscore}untimed\ {\isacharparenleft}inf{\isacharunderscore}truncate\ z\ i{\isacharparenright}{\isacharparenright}{\isacharparenright}\ {\isacharminus}\ Suc\ {\isadigit{0}}{\isacharparenright}\ {\isasymnoteq}\ Zero{\isachardoublequoteclose}\isanewline
\ \ \ \ \ \ \ \ \ \isacommand{by}\isamarkupfalse%
\ {\isacharparenleft}simp\ add{\isacharcolon}\ L{\isadigit{1}}{\isacharunderscore}Converter{\isacharparenright}\isanewline
\ \ \ \ \ \ \ \isacommand{from}\isamarkupfalse%
\ aa{\isadigit{2}}\ \isakeyword{and}\ aaOne\ \isacommand{have}\isamarkupfalse%
\ sg{\isadigit{9}}{\isacharcolon}{\isachardoublequoteopen}hd\ {\isacharparenleft}s\ {\isacharparenleft}Suc\ i{\isacharparenright}{\isacharparenright}\ {\isacharequal}\ hd\ {\isacharparenleft}s\ i{\isacharparenright}\ {\isacharplus}\ r{\isachardoublequoteclose}\ \isacommand{by}\isamarkupfalse%
\ simp\isanewline
\ \ \ \ \ \ \ \isacommand{from}\isamarkupfalse%
\ Suc\ \isakeyword{and}\ this\ \isacommand{show}\isamarkupfalse%
\ {\isacharquery}thesis\ \isacommand{by}\isamarkupfalse%
\ simp\isanewline
\ \ \ \ \ \isacommand{qed}\isamarkupfalse%
\isanewline
\ \ \isacommand{qed}\isamarkupfalse%
\isanewline
\isacommand{qed}\isamarkupfalse%
\endisatagproof
{\isafoldproof}%
\isadelimproof
\ \isanewline
\endisadelimproof
\isanewline
\isanewline
\isacommand{lemma}\isamarkupfalse%
\ L{\isadigit{3}}{\isacharunderscore}ControlSystem{\isacharcolon}\isanewline
\ \ \isakeyword{assumes}\ h{\isadigit{1}}{\isacharcolon}{\isachardoublequoteopen}ControlSystemArch\ s{\isachardoublequoteclose}\isanewline
\ \ \isakeyword{shows}\ {\isachardoublequoteopen}hd\ {\isacharparenleft}s\ i{\isacharparenright}\ {\isasymle}\ {\isacharparenleft}{\isadigit{8}}{\isadigit{0}}{\isadigit{0}}{\isacharcolon}{\isacharcolon}\ nat{\isacharparenright}{\isachardoublequoteclose}\isanewline
\isadelimproof
\endisadelimproof
\isatagproof
\isacommand{proof}\isamarkupfalse%
\ {\isacharminus}\ \isanewline
\ \ \isacommand{from}\isamarkupfalse%
\ h{\isadigit{1}}\ \isacommand{obtain}\isamarkupfalse%
\ x\ z\ y\ \isanewline
\ \ \ \ \isakeyword{where}\ a{\isadigit{1}}{\isacharcolon}{\isachardoublequoteopen}Converter\ z\ x{\isachardoublequoteclose}\ \isakeyword{and}\ a{\isadigit{2}}{\isacharcolon}\ {\isachardoublequoteopen}SteamBoiler\ x\ s\ y{\isachardoublequoteclose}\ \isakeyword{and}\ a{\isadigit{3}}{\isacharcolon}{\isachardoublequoteopen}Controller\ y\ z{\isachardoublequoteclose}\isanewline
\ \ \ \ \isacommand{by}\isamarkupfalse%
\ {\isacharparenleft}simp\ only{\isacharcolon}\ ControlSystemArch{\isacharunderscore}def{\isacharcomma}\ auto{\isacharparenright}\ \isanewline
\ \ \isacommand{from}\isamarkupfalse%
\ this\ \isacommand{have}\isamarkupfalse%
\ sg{\isadigit{1}}{\isacharcolon}{\isachardoublequoteopen}ts\ x{\isachardoublequoteclose}\ \ \isacommand{by}\isamarkupfalse%
\ {\isacharparenleft}simp\ add{\isacharcolon}\ Converter{\isacharunderscore}def{\isacharparenright}\isanewline
\ \ \isacommand{from}\isamarkupfalse%
\ sg{\isadigit{1}}\ \isakeyword{and}\ a{\isadigit{2}}\ \isacommand{have}\isamarkupfalse%
\ sg{\isadigit{2}}{\isacharcolon}{\isachardoublequoteopen}ts\ y{\isachardoublequoteclose}\ \ \isacommand{by}\isamarkupfalse%
\ {\isacharparenleft}simp\ add{\isacharcolon}\ L{\isadigit{2}}{\isacharunderscore}Boiler{\isacharparenright}\isanewline
\ \ \isacommand{from}\isamarkupfalse%
\ sg{\isadigit{1}}\ \isakeyword{and}\ a{\isadigit{2}}\ \isacommand{have}\isamarkupfalse%
\ sg{\isadigit{3}}{\isacharcolon}{\isachardoublequoteopen}y\ {\isacharequal}\ s{\isachardoublequoteclose}\ \isacommand{by}\isamarkupfalse%
\ {\isacharparenleft}simp\ add{\isacharcolon}\ SteamBoiler{\isacharunderscore}def{\isacharparenright}\isanewline
\ \ \isacommand{from}\isamarkupfalse%
\ a{\isadigit{1}}\ \isakeyword{and}\ a{\isadigit{2}}\ \isakeyword{and}\ a{\isadigit{3}}\ \isakeyword{and}\ sg{\isadigit{1}}\ \isakeyword{and}\ sg{\isadigit{2}}\ \isakeyword{and}\ sg{\isadigit{3}}\ \isacommand{show}\isamarkupfalse%
\ {\isachardoublequoteopen}hd\ {\isacharparenleft}s\ i{\isacharparenright}\ {\isasymle}\ {\isacharparenleft}{\isadigit{8}}{\isadigit{0}}{\isadigit{0}}{\isacharcolon}{\isacharcolon}\ nat{\isacharparenright}{\isachardoublequoteclose}\isanewline
\ \ \isacommand{proof}\isamarkupfalse%
\ {\isacharparenleft}induction\ i{\isacharparenright}\isanewline
\ \ \ \ \isacommand{case}\isamarkupfalse%
\ {\isadigit{0}}\isanewline
\ \ \ \ \isacommand{from}\isamarkupfalse%
\ this\ \isacommand{show}\isamarkupfalse%
\ {\isacharquery}case\ \ \isacommand{by}\isamarkupfalse%
\ {\isacharparenleft}simp\ add{\isacharcolon}\ L{\isadigit{4}}{\isacharunderscore}Boiler{\isacharparenright}\isanewline
\ \ \isacommand{next}\isamarkupfalse%
\isanewline
\ \ \ \ \isacommand{fix}\isamarkupfalse%
\ i\isanewline
\ \ \ \ \isacommand{case}\isamarkupfalse%
\ {\isacharparenleft}Suc\ i{\isacharparenright}\isanewline
\ \ \ \ \isacommand{from}\isamarkupfalse%
\ this\ \isacommand{obtain}\isamarkupfalse%
\ l\isanewline
\ \ \ \ \ \ \isakeyword{where}\ a{\isadigit{4}}{\isacharcolon}\ {\isachardoublequoteopen}Controller{\isacharunderscore}L\ s\ {\isacharparenleft}fin{\isacharunderscore}inf{\isacharunderscore}append\ {\isacharbrackleft}Zero{\isacharbrackright}\ l{\isacharparenright}\ l\ z{\isachardoublequoteclose}\isanewline
\ \ \ \ \ \ \isacommand{by}\isamarkupfalse%
\ {\isacharparenleft}simp\ add{\isacharcolon}\ Controller{\isacharunderscore}def{\isacharcomma}\ atomize{\isacharcomma}\ auto{\isacharparenright}\isanewline
\ \ \ \ \isacommand{from}\isamarkupfalse%
\ a{\isadigit{4}}\ \isacommand{have}\isamarkupfalse%
\ sg{\isadigit{4}}{\isacharcolon}{\isachardoublequoteopen}fin{\isacharunderscore}make{\isacharunderscore}untimed\ {\isacharparenleft}inf{\isacharunderscore}truncate\ z\ i{\isacharparenright}\ {\isasymnoteq}\ {\isacharbrackleft}{\isacharbrackright}{\isachardoublequoteclose}\isanewline
\ \ \ \ \ \ \isacommand{by}\isamarkupfalse%
\ {\isacharparenleft}simp\ add{\isacharcolon}\ \ L{\isadigit{1}}{\isacharunderscore}Controller{\isacharparenright}\isanewline
\ \ \ \ \isacommand{from}\isamarkupfalse%
\ a{\isadigit{2}}\ \isakeyword{and}\ sg{\isadigit{1}}\ \isacommand{have}\isamarkupfalse%
\ y{\isadigit{0}}asm{\isacharcolon}{\isachardoublequoteopen}y\ {\isadigit{0}}\ {\isacharequal}\ {\isacharbrackleft}{\isadigit{5}}{\isadigit{0}}{\isadigit{0}}{\isacharcolon}{\isacharcolon}nat{\isacharbrackright}{\isachardoublequoteclose}\ \ \isacommand{by}\isamarkupfalse%
\ {\isacharparenleft}simp\ add{\isacharcolon}\ SteamBoiler{\isacharunderscore}def{\isacharparenright}\isanewline
\ \ \ \ \isacommand{from}\isamarkupfalse%
\ Suc\ \isakeyword{and}\ a{\isadigit{4}}\ \isakeyword{and}\ sg{\isadigit{4}}\ \isakeyword{and}\ y{\isadigit{0}}asm\ \isacommand{have}\isamarkupfalse%
\ sg{\isadigit{5}}{\isacharcolon}\ {\isachardoublequoteopen}last\ {\isacharparenleft}fin{\isacharunderscore}make{\isacharunderscore}untimed\ {\isacharparenleft}inf{\isacharunderscore}truncate\ z\ i{\isacharparenright}{\isacharparenright}\ {\isacharequal}\ \ l\ i{\isachardoublequoteclose}\isanewline
\ \ \ \ \ \ \isacommand{by}\isamarkupfalse%
\ {\isacharparenleft}simp\ add{\isacharcolon}\ L{\isadigit{7}}{\isacharunderscore}Controller{\isacharparenright}\isanewline
\ \ \ \ \isacommand{from}\isamarkupfalse%
\ a{\isadigit{2}}\ \isakeyword{and}\ sg{\isadigit{1}}\ \isacommand{obtain}\isamarkupfalse%
\ r\ \isakeyword{where}\isanewline
\ \ \ \ \ \ \ aa{\isadigit{0}}{\isacharcolon}{\isachardoublequoteopen}{\isadigit{0}}\ {\isacharless}\ r{\isachardoublequoteclose}\ \isakeyword{and}\isanewline 
\ \ \ \ \ \ \ aa{\isadigit{1}}{\isacharcolon}{\isachardoublequoteopen}r\ {\isasymle}\ {\isadigit{1}}{\isadigit{0}}{\isachardoublequoteclose}\ \isakeyword{and}\ \isanewline
\ \ \ \ \ \ \ aa{\isadigit{2}}{\isacharcolon}{\isachardoublequoteopen}hd\ {\isacharparenleft}s\ {\isacharparenleft}Suc\ i{\isacharparenright}{\isacharparenright}\ {\isacharequal}\ {\isacharparenleft}if\ hd\ {\isacharparenleft}x\ i{\isacharparenright}\ {\isacharequal}\ Zero\ then\ hd\ {\isacharparenleft}s\ i{\isacharparenright}\ {\isacharminus}\ r\ else\ hd\ {\isacharparenleft}s\ i{\isacharparenright}\ {\isacharplus}\ r{\isacharparenright}{\isachardoublequoteclose}\isanewline
\ \ \ \ \ \ \ \ \ \isacommand{by}\isamarkupfalse%
\ {\isacharparenleft}simp\ add{\isacharcolon}\ SteamBoiler{\isacharunderscore}def{\isacharcomma}\ auto{\isacharparenright}\isanewline
\ \ \ \ \isacommand{from}\isamarkupfalse%
\ this\ \isakeyword{and}\ Suc\ \isakeyword{and}\ a{\isadigit{4}}\ \isakeyword{and}\ sg{\isadigit{4}}\ \isakeyword{and}\ sg{\isadigit{5}}\ \isacommand{show}\isamarkupfalse%
\ {\isacharquery}case\isanewline
\ \ \ \ \isacommand{proof}\isamarkupfalse%
\ {\isacharparenleft}cases\ {\isachardoublequoteopen}hd\ {\isacharparenleft}x\ i{\isacharparenright}\ {\isacharequal}\ Zero{\isachardoublequoteclose}{\isacharparenright}\isanewline
\ \ \ \ \ \ \ \isacommand{assume}\isamarkupfalse%
\ aaZero{\isacharcolon}{\isachardoublequoteopen}hd\ {\isacharparenleft}x\ i{\isacharparenright}\ {\isacharequal}\ Zero{\isachardoublequoteclose}\isanewline
\ \ \ \ \ \ \ \isacommand{from}\isamarkupfalse%
\ a{\isadigit{1}}\ \isakeyword{and}\ sg{\isadigit{4}}\ \isakeyword{and}\ this\ \isacommand{have}\isamarkupfalse%
\isanewline
\ \ \ \ \ \ \ \ \ sg{\isadigit{7}}{\isacharcolon}{\isachardoublequoteopen}{\isacharparenleft}fin{\isacharunderscore}make{\isacharunderscore}untimed\ {\isacharparenleft}inf{\isacharunderscore}truncate\ z\ i{\isacharparenright}{\isacharparenright}\ {\isacharbang}\ \isanewline
\ \ \ \ \ \ \ \ \ \ \ \ \ {\isacharparenleft}{\isacharparenleft}length\ {\isacharparenleft}fin{\isacharunderscore}make{\isacharunderscore}untimed\ {\isacharparenleft}inf{\isacharunderscore}truncate\ z\ i{\isacharparenright}{\isacharparenright}{\isacharparenright}\ {\isacharminus}\ Suc\ {\isadigit{0}}{\isacharparenright}\ \ {\isacharequal}\ Zero{\isachardoublequoteclose}\isanewline
\ \ \ \ \ \ \ \ \ \isacommand{by}\isamarkupfalse%
\ {\isacharparenleft}simp\ add{\isacharcolon}\ L{\isadigit{1}}{\isacharunderscore}Converter{\isacharparenright}\isanewline
\ \ \ \ \ \ \ \isacommand{from}\isamarkupfalse%
\ aa{\isadigit{2}}\ \isakeyword{and}\ aaZero\ \isacommand{have}\isamarkupfalse%
\ sg{\isadigit{8}}{\isacharcolon}{\isachardoublequoteopen}hd\ {\isacharparenleft}s\ {\isacharparenleft}Suc\ i{\isacharparenright}{\isacharparenright}\ {\isacharequal}\ hd\ {\isacharparenleft}s\ i{\isacharparenright}\ {\isacharminus}\ r{\isachardoublequoteclose}\ \isacommand{by}\isamarkupfalse%
\ simp\isanewline
\ \ \ \ \ \ \ \isacommand{from}\isamarkupfalse%
\ this\ \isakeyword{and}\ Suc\ \isacommand{show}\isamarkupfalse%
\ {\isacharquery}thesis\ \isacommand{by}\isamarkupfalse%
\ simp\isanewline
\ \ \ \ \ \isacommand{next}\isamarkupfalse%
\isanewline
\ \ \ \ \ \ \ \isacommand{assume}\isamarkupfalse%
\ aaOne{\isacharcolon}{\isachardoublequoteopen}hd\ {\isacharparenleft}x\ i{\isacharparenright}\ {\isasymnoteq}\ Zero{\isachardoublequoteclose}\isanewline
\ \ \ \ \ \ \ \isacommand{from}\isamarkupfalse%
\ a{\isadigit{1}}\ \isakeyword{and}\ sg{\isadigit{4}}\ \isakeyword{and}\ this\ \isacommand{have}\isamarkupfalse%
\isanewline
\ \ \ \ \ \ \ \ \ sg{\isadigit{7}}{\isacharcolon}{\isachardoublequoteopen}{\isacharparenleft}fin{\isacharunderscore}make{\isacharunderscore}untimed\ {\isacharparenleft}inf{\isacharunderscore}truncate\ z\ i{\isacharparenright}{\isacharparenright}\ {\isacharbang}\ \isanewline
\ \ \ \ \ \ \ \ \ \ \ \ \ {\isacharparenleft}{\isacharparenleft}length\ {\isacharparenleft}fin{\isacharunderscore}make{\isacharunderscore}untimed\ {\isacharparenleft}inf{\isacharunderscore}truncate\ z\ i{\isacharparenright}{\isacharparenright}{\isacharparenright}\ {\isacharminus}\ Suc\ {\isadigit{0}}{\isacharparenright}\ {\isasymnoteq}\ Zero{\isachardoublequoteclose}\isanewline
\ \ \ \ \ \ \ \ \ \isacommand{by}\isamarkupfalse%
\ {\isacharparenleft}simp\ add{\isacharcolon}\ L{\isadigit{1}}{\isacharunderscore}Converter{\isacharparenright}\isanewline
\ \ \ \ \ \ \ \isacommand{from}\isamarkupfalse%
\ aa{\isadigit{2}}\ \isakeyword{and}\ aaOne\ \isacommand{have}\isamarkupfalse%
\ sg{\isadigit{9}}{\isacharcolon}{\isachardoublequoteopen}hd\ {\isacharparenleft}s\ {\isacharparenleft}Suc\ i{\isacharparenright}{\isacharparenright}\ {\isacharequal}\ hd\ {\isacharparenleft}s\ i{\isacharparenright}\ {\isacharplus}\ r{\isachardoublequoteclose}\ \isacommand{by}\isamarkupfalse%
\ simp\isanewline
\ \ \ \ \ \ \ \isacommand{from}\isamarkupfalse%
\ Suc\ \isacommand{have}\isamarkupfalse%
\ sgSuc{\isacharcolon}{\isachardoublequoteopen}hd\ {\isacharparenleft}s\ i{\isacharparenright}\ {\isasymle}\ {\isadigit{8}}{\isadigit{0}}{\isadigit{0}}{\isachardoublequoteclose}\ \isacommand{by}\isamarkupfalse%
\ simp\isanewline
\ \ \ \ \ \ \ \isacommand{from}\isamarkupfalse%
\ a{\isadigit{4}}\ \isakeyword{and}\ sg{\isadigit{7}}\ \isakeyword{and}\ sg{\isadigit{5}}\ \isakeyword{and}\ sgSuc\ \isakeyword{and}\ sg{\isadigit{9}}\ \isakeyword{and}\ sg{\isadigit{4}}\ \isakeyword{and}\ aa{\isadigit{0}}\ \isakeyword{and}\ aa{\isadigit{1}}\ \isanewline
\ \ \ \ \ \ \   \isacommand{show}\isamarkupfalse%
\ {\isacharquery}thesis\isanewline
\ \ \ \ \ \ \ \ \ \ \isacommand{by}\isamarkupfalse%
\ {\isacharparenleft}rule\ L{\isadigit{1}}{\isadigit{0}}{\isacharunderscore}Controller{\isacharparenright}\isanewline
\ \ \ \ \ \isacommand{qed}\isamarkupfalse%
\isanewline
\ \ \isacommand{qed}\isamarkupfalse%
\isanewline
\isacommand{qed}\isamarkupfalse%
\endisatagproof
{\isafoldproof}%
\isadelimproof
\endisadelimproof
\isamarkupsubsection{Proof of the Refinement Relation%
}
\isamarkuptrue%
\isacommand{lemma}\isamarkupfalse%
\ L{\isadigit{0}}{\isacharunderscore}ControlSystem{\isacharcolon}\isanewline
\ \ \isakeyword{assumes}\ h{\isadigit{1}}{\isacharcolon}{\isachardoublequoteopen}ControlSystemArch\ s{\isachardoublequoteclose}\isanewline
\ \ \isakeyword{shows}\ \ \ {\isachardoublequoteopen}ControlSystem\ s{\isachardoublequoteclose}\isanewline
\isadelimproof
\ \ \ \ %
\endisadelimproof
\isatagproof
\isacommand{apply}\isamarkupfalse%
\ {\isacharparenleft}simp\ add{\isacharcolon}\ ControlSystem{\isacharunderscore}def{\isacharparenright}\isanewline
\ \ \ \ \isacommand{apply}\isamarkupfalse%
\ auto\ \isanewline
\isacommand{proof}\isamarkupfalse%
\ {\isacharminus}\isanewline
\ \ \isacommand{from}\isamarkupfalse%
\ h{\isadigit{1}}\ \isacommand{show}\isamarkupfalse%
\ sg{\isadigit{1}}{\isacharcolon}{\isachardoublequoteopen}ts\ s{\isachardoublequoteclose}\ \isacommand{by}\isamarkupfalse%
\ {\isacharparenleft}rule\ L{\isadigit{1}}{\isacharunderscore}ControlSystem{\isacharparenright}\ \isanewline
\isacommand{next}\isamarkupfalse%
\ \isanewline
\ \ \isacommand{fix}\isamarkupfalse%
\ j\ \ \ \isanewline
\ \ \isacommand{from}\isamarkupfalse%
\ h{\isadigit{1}}\ \isacommand{show}\isamarkupfalse%
\ sg{\isadigit{2}}{\isacharcolon}{\isachardoublequoteopen}{\isacharparenleft}{\isadigit{2}}{\isadigit{0}}{\isadigit{0}}{\isacharcolon}{\isacharcolon}nat{\isacharparenright}\ {\isasymle}\ hd\ {\isacharparenleft}s\ j{\isacharparenright}{\isachardoublequoteclose}\ \isacommand{by}\isamarkupfalse%
\ {\isacharparenleft}rule\ L{\isadigit{2}}{\isacharunderscore}ControlSystem{\isacharparenright}\isanewline
\isacommand{next}\isamarkupfalse%
\isanewline
\ \ \isacommand{fix}\isamarkupfalse%
\ j\isanewline
\ \ \isacommand{from}\isamarkupfalse%
\ h{\isadigit{1}}\ \isacommand{show}\isamarkupfalse%
\ sg{\isadigit{3}}{\isacharcolon}{\isachardoublequoteopen}hd\ {\isacharparenleft}s\ j{\isacharparenright}\ {\isasymle}\ {\isacharparenleft}{\isadigit{8}}{\isadigit{0}}{\isadigit{0}}{\isacharcolon}{\isacharcolon}\ nat{\isacharparenright}{\isachardoublequoteclose}\ \isacommand{by}\isamarkupfalse%
\ {\isacharparenleft}rule\ L{\isadigit{3}}{\isacharunderscore}ControlSystem{\isacharparenright}\isanewline
\isacommand{qed}\isamarkupfalse%
\endisatagproof
{\isafoldproof}%
\isadelimproof
\ \isanewline
\endisadelimproof
\isadelimtheory
\ \isanewline
\endisadelimtheory
\isatagtheory
\isacommand{end}\isamarkupfalse%
\endisatagtheory
{\isafoldtheory}%
\isadelimtheory
\endisadelimtheory
\ \end{isabellebody}%

%
\begin{isabellebody}%
\def\isabellecontext{FR{\isacharunderscore}types}%
\isamarkupheader{FlexRay: Types%
}
\isamarkuptrue%
\isadelimtheory
\endisadelimtheory
\isatagtheory
\isacommand{theory}\isamarkupfalse%
\ FR{\isacharunderscore}types\ \isanewline
\isakeyword{imports}\ stream\isanewline
\isakeyword{begin}%
\endisatagtheory
{\isafoldtheory}%
\isadelimtheory
\endisadelimtheory
\isanewline
\isanewline
\isacommand{record}\isamarkupfalse%
\ {\isacharprime}a\ Message\ {\isacharequal}\ \isanewline
\ \ \ message{\isacharunderscore}id\ {\isacharcolon}{\isacharcolon}\ nat\isanewline
\ \ \ ftcdata\ \ \ \ {\isacharcolon}{\isacharcolon}\ {\isacharprime}a\isanewline
\isanewline
\isacommand{record}\isamarkupfalse%
\ {\isacharprime}a\ Frame\ {\isacharequal}\ \isanewline
\ \ \ slot\ {\isacharcolon}{\isacharcolon}\ nat\isanewline
\ \ \ dataF\ {\isacharcolon}{\isacharcolon}\ {\isachardoublequoteopen}{\isacharparenleft}{\isacharprime}a\ Message{\isacharparenright}\ list{\isachardoublequoteclose}\isanewline
\isanewline
\isacommand{record}\isamarkupfalse%
\ Config\ {\isacharequal}\ \isanewline
\ \ \ schedule\ \ \ \ {\isacharcolon}{\isacharcolon}\ {\isachardoublequoteopen}nat\ list{\isachardoublequoteclose}\ \isanewline
\ \ \ cycleLength\ {\isacharcolon}{\isacharcolon}\ nat\isanewline
\isanewline
\isacommand{type{\isacharunderscore}synonym}\isamarkupfalse%
\ {\isacharprime}a\ nFrame\ {\isacharequal}\ {\isachardoublequoteopen}nat\ {\isasymRightarrow}\ {\isacharparenleft}{\isacharprime}a\ Frame{\isacharparenright}\ istream{\isachardoublequoteclose}\isanewline
\isacommand{type{\isacharunderscore}synonym}\isamarkupfalse%
\ nNat\ {\isacharequal}\ {\isachardoublequoteopen}nat\ {\isasymRightarrow}\ nat\ istream{\isachardoublequoteclose}
\isanewline
\isacommand{type{\isacharunderscore}synonym}\isamarkupfalse%
\ nConfig\ {\isacharequal}\ {\isachardoublequoteopen}nat\ {\isasymRightarrow}\ Config{\isachardoublequoteclose}\isanewline
\isanewline
\isacommand{consts}\isamarkupfalse%
\ sN\ {\isacharcolon}{\isacharcolon}\ {\isachardoublequoteopen}nat{\isachardoublequoteclose}\isanewline
\isanewline
\isacommand{definition}\isamarkupfalse%
\ \isanewline
\ \ sheafNumbers\ {\isacharcolon}{\isacharcolon}\ {\isachardoublequoteopen}nat\ list{\isachardoublequoteclose}\isanewline
\isakeyword{where} \ 
\ {\isachardoublequoteopen}sheafNumbers\ {\isasymequiv}\ {\isacharbrackleft}sN{\isacharbrackright}{\isachardoublequoteclose}\isanewline
\isadelimtheory
\isanewline
\endisadelimtheory
\isatagtheory
\isacommand{end}\isamarkupfalse%
\endisatagtheory
{\isafoldtheory}%
\isadelimtheory
\endisadelimtheory
\end{isabellebody}%

%
\begin{isabellebody}%
\def\isabellecontext{FR}%
\isamarkupheader{FlexRay: Specification%
}
\isamarkuptrue%
\isadelimtheory
\endisadelimtheory
\isatagtheory
\isacommand{theory}\isamarkupfalse%
\ \ FR\ \isanewline
\isakeyword{imports}\ FR{\isacharunderscore}types\isanewline
\isakeyword{begin}%
\endisatagtheory
{\isafoldtheory}%
\isadelimtheory
\endisadelimtheory
\isamarkupsubsection{Auxiliary predicates%
}
\isamarkuptrue%
\isamarkupcmt{The predicate DisjointSchedules is true  for sheaf of channels of type Config,%
}
\isanewline
\isamarkupcmt{if all bus configurations have disjoint scheduling tables.%
}
\isanewline
\isacommand{definition}\isamarkupfalse%
\isanewline
\ \ DisjointSchedules\ {\isacharcolon}{\isacharcolon}\ {\isachardoublequoteopen}nat\ {\isasymRightarrow}\ nConfig\ {\isasymRightarrow}\ bool{\isachardoublequoteclose}\ \isanewline
\isakeyword{where}\isanewline
\ {\isachardoublequoteopen}DisjointSchedules\ n\ nC\isanewline
\ \ {\isasymequiv}\isanewline
\ \ {\isasymforall}\ i\ j{\isachardot}\ i\ {\isacharless}\ n\ {\isasymand}\ j\ {\isacharless}\ n\ {\isasymand}\ i\ {\isasymnoteq}\ j\ {\isasymlongrightarrow}\ \isanewline
\ \ disjoint\ {\isacharparenleft}schedule\ {\isacharparenleft}nC\ i{\isacharparenright}{\isacharparenright}\ \ {\isacharparenleft}schedule\ {\isacharparenleft}nC\ j{\isacharparenright}{\isacharparenright}{\isachardoublequoteclose}\ \isanewline
\isanewline
\isamarkupcmt{The predicate IdenticCycleLength is true  for sheaf of channels of type Config,%
}
\ \isanewline
\isamarkupcmt{if all bus configurations have the equal length of the communication round.%
}
\isanewline
\isacommand{definition}\isamarkupfalse%
\isanewline
\ \ \ IdenticCycleLength\ {\isacharcolon}{\isacharcolon}\ {\isachardoublequoteopen}nat\ {\isasymRightarrow}\ nConfig\ {\isasymRightarrow}\ bool{\isachardoublequoteclose}\isanewline
\isakeyword{where}\isanewline
\ \ {\isachardoublequoteopen}IdenticCycleLength\ n\ nC\isanewline
\ \ \ {\isasymequiv}\isanewline
\ \ \ {\isasymforall}\ i\ j{\isachardot}\ i\ {\isacharless}\ n\ {\isasymand}\ j\ {\isacharless}\ n\ {\isasymlongrightarrow}\ \isanewline
\ \ \ cycleLength\ {\isacharparenleft}nC\ i{\isacharparenright}\ {\isacharequal}\ cycleLength\ {\isacharparenleft}nC\ j{\isacharparenright}{\isachardoublequoteclose}\isanewline
\ \ \ \isanewline
\isamarkupcmt{The predicate FrameTransmission defines the correct message transmission:%
}
\isanewline
\isamarkupcmt{if the time t is equal modulo the length of the cycle (Flexray communication round)%
}
\isanewline
\isamarkupcmt{to the element of the scheduler table of the node k, then this and only this node%
}
\isanewline
\isamarkupcmt{can send a data atn the $t$th time interval.%
}
\isanewline
\isacommand{definition}\isamarkupfalse%
\isanewline
\ \ \ FrameTransmission\ {\isacharcolon}{\isacharcolon}\ \isanewline
\ \ \ \ \ {\isachardoublequoteopen}nat\ {\isasymRightarrow}\ {\isacharprime}a\ nFrame\ {\isasymRightarrow}\ {\isacharprime}a\ nFrame\ {\isasymRightarrow}\ nNat\ {\isasymRightarrow}\ nConfig\ {\isasymRightarrow}\ bool{\isachardoublequoteclose}\isanewline
\isakeyword{where}\isanewline
\ \ {\isachardoublequoteopen}FrameTransmission\ n\ nStore\ nReturn\ nGet\ nC\isanewline
\ \ \ {\isasymequiv}\isanewline
\ \ \ {\isasymforall}\ {\isacharparenleft}t{\isacharcolon}{\isacharcolon}nat{\isacharparenright}\ {\isacharparenleft}k{\isacharcolon}{\isacharcolon}nat{\isacharparenright}{\isachardot}\ k\ {\isacharless}\ n\ {\isasymlongrightarrow}\isanewline
\ \ \ {\isacharparenleft}\ let\ s\ {\isacharequal}\ t\ mod\ {\isacharparenleft}cycleLength\ {\isacharparenleft}nC\ k{\isacharparenright}{\isacharparenright}\isanewline
\ \ \ \ \ in\ \isanewline
\ \ \ \ \ {\isacharparenleft}\ s\ mem\ {\isacharparenleft}schedule\ {\isacharparenleft}nC\ k{\isacharparenright}{\isacharparenright}\isanewline
\ \ \ \ \ \ \ {\isasymlongrightarrow}\isanewline
\ \ \ \ \ \ \ {\isacharparenleft}nGet\ k\ t{\isacharparenright}\ {\isacharequal}\ {\isacharbrackleft}s{\isacharbrackright}\ {\isasymand}\ \ \isanewline
\ \ \ \ \ \ \ {\isacharparenleft}{\isasymforall}\ j{\isachardot}\ j\ {\isacharless}\ n\ {\isasymand}\ j\ {\isasymnoteq}\ k\ {\isasymlongrightarrow}\ \isanewline
\ \ \ \ \ \ \ \ \ \ \ \ {\isacharparenleft}{\isacharparenleft}nStore\ j{\isacharparenright}\ t{\isacharparenright}\ {\isacharequal}\ \ {\isacharparenleft}{\isacharparenleft}nReturn\ k{\isacharparenright}\ t{\isacharparenright}{\isacharparenright}\ {\isacharparenright}{\isacharparenright}{\isachardoublequoteclose}\ \isanewline
\isanewline\isanewline
\isamarkupcmt{The   predicate Broadcast describes properties of FlexRay broadcast.%
}
\isanewline
\isacommand{definition}\isamarkupfalse%
\isanewline
\ \ \ Broadcast\ {\isacharcolon}{\isacharcolon}\ \isanewline
\ \ \ \ \ {\isachardoublequoteopen}nat\ {\isasymRightarrow}\ {\isacharprime}a\ nFrame\ {\isasymRightarrow}\ {\isacharprime}a\ Frame\ istream\ {\isasymRightarrow}\ bool{\isachardoublequoteclose}\isanewline
\isakeyword{where}\isanewline
\ \ {\isachardoublequoteopen}Broadcast\ n\ nSend\ recv\isanewline
\ \ \ {\isasymequiv}\ \isanewline
\ \ \ {\isasymforall}\ {\isacharparenleft}t{\isacharcolon}{\isacharcolon}nat{\isacharparenright}{\isachardot}\ \isanewline
\ \ \ \ {\isacharparenleft}\ if\ {\isasymexists}\ k{\isachardot}\ k\ {\isacharless}\ n\ {\isasymand}\ {\isacharparenleft}{\isacharparenleft}nSend\ k{\isacharparenright}\ t{\isacharparenright}\ {\isasymnoteq}\ {\isacharbrackleft}{\isacharbrackright}\isanewline
\ \ \ \ \ \ then\ {\isacharparenleft}recv\ t{\isacharparenright}\ {\isacharequal}\ {\isacharparenleft}{\isacharparenleft}nSend\ {\isacharparenleft}SOME\ k{\isachardot}\ k\ {\isacharless}\ n\ {\isasymand}\ {\isacharparenleft}{\isacharparenleft}nSend\ k{\isacharparenright}\ t{\isacharparenright}\ {\isasymnoteq}\ {\isacharbrackleft}{\isacharbrackright}{\isacharparenright}{\isacharparenright}\ t{\isacharparenright}\isanewline
\ \ \ \ \ \ else\ {\isacharparenleft}recv\ t{\isacharparenright}\ {\isacharequal}\ {\isacharbrackleft}{\isacharbrackright}\ {\isacharparenright}{\isachardoublequoteclose}\ \ \isanewline
\ \ \ \ \ \ \isanewline
\isamarkupcmt{The predicate Receive defines the  relations on the streams  to represent%
}
\ \isanewline
\isamarkupcmt{data receive by FlexRay controller.%
}
\isanewline
\isacommand{definition}\isamarkupfalse%
\isanewline
\ \ Receive\ {\isacharcolon}{\isacharcolon}\ \isanewline
\ \ {\isachardoublequoteopen}{\isacharprime}a\ Frame\ istream\ {\isasymRightarrow}\ {\isacharprime}a\ Frame\ istream\ {\isasymRightarrow}\ nat\ istream\ {\isasymRightarrow}\ bool{\isachardoublequoteclose}\ \ \ \ \ \isanewline
\isakeyword{where}\isanewline
\ \ {\isachardoublequoteopen}Receive\ recv\ store\ activation\isanewline
\ \ \ {\isasymequiv}\isanewline
\ \ \ {\isasymforall}\ {\isacharparenleft}t{\isacharcolon}{\isacharcolon}nat{\isacharparenright}{\isachardot}\isanewline
\ \ \ \ {\isacharparenleft}\ if\ \ {\isacharparenleft}activation\ t{\isacharparenright}\ {\isacharequal}\ {\isacharbrackleft}{\isacharbrackright}\isanewline
\ \ \ \ \ \ then\ {\isacharparenleft}store\ t{\isacharparenright}\ {\isacharequal}\ {\isacharparenleft}recv\ t{\isacharparenright}\isanewline
\ \ \ \ \ \ else\ {\isacharparenleft}store\ t{\isacharparenright}\ {\isacharequal}\ {\isacharbrackleft}{\isacharbrackright}{\isacharparenright}{\isachardoublequoteclose}\isanewline
\isanewline
\isamarkupcmt{The predicate Send defines the  relations on the streams  to represent%
}
\isanewline
\isamarkupcmt{sending data  by FlexRay controller.%
}
\isanewline
\isacommand{definition}\isamarkupfalse%
\isanewline
\ \ Send\ {\isacharcolon}{\isacharcolon}\ \isanewline
\ \ {\isachardoublequoteopen}{\isacharprime}a\ Frame\ istream\ {\isasymRightarrow}\ {\isacharprime}a\ Frame\ istream\ {\isasymRightarrow}\ nat\ istream\ {\isasymRightarrow}\ nat\ istream\ {\isasymRightarrow}\ bool{\isachardoublequoteclose}\isanewline
\isakeyword{where}\isanewline
\ \ {\isachardoublequoteopen}Send\ return\ send\ get\ activation\isanewline
\ \ \ {\isasymequiv}\isanewline
\ \ \ {\isasymforall}\ {\isacharparenleft}t{\isacharcolon}{\isacharcolon}nat{\isacharparenright}{\isachardot}\ \isanewline
\ \ \ {\isacharparenleft}\ if\ \ {\isacharparenleft}activation\ t{\isacharparenright}\ {\isacharequal}\ {\isacharbrackleft}{\isacharbrackright}\isanewline
\ \ \ \ \ then\ {\isacharparenleft}get\ t{\isacharparenright}\ {\isacharequal}\ {\isacharbrackleft}{\isacharbrackright}\ {\isasymand}\ {\isacharparenleft}send\ t{\isacharparenright}\ {\isacharequal}\ {\isacharbrackleft}{\isacharbrackright}\isanewline
\ \ \ \ \ else\ {\isacharparenleft}get\ t{\isacharparenright}\ {\isacharequal}\ {\isacharparenleft}activation\ t{\isacharparenright}\ {\isasymand}\ {\isacharparenleft}send\ t{\isacharparenright}\ {\isacharequal}\ {\isacharparenleft}return\ t{\isacharparenright}\ \ {\isacharparenright}{\isachardoublequoteclose}%
\isamarkupsubsection{Specifications of the FlexRay components%
}
\isamarkuptrue%
\isacommand{definition}\isamarkupfalse%
\isanewline
\ \ \ FlexRay\ {\isacharcolon}{\isacharcolon}\ \isanewline
\ \ {\isachardoublequoteopen}nat\ {\isasymRightarrow}\ {\isacharprime}a\ nFrame\ {\isasymRightarrow}\ nConfig\ {\isasymRightarrow}\ {\isacharprime}a\ nFrame\ {\isasymRightarrow}\ nNat\ {\isasymRightarrow}\ bool{\isachardoublequoteclose}\isanewline
\isakeyword{where}\isanewline
\ \ {\isachardoublequoteopen}FlexRay\ n\ nReturn\ nC\ nStore\ nGet\isanewline
\ \ \ {\isasymequiv}\ \isanewline
\ \ \ {\isacharparenleft}CorrectSheaf\ n{\isacharparenright}\ \ {\isasymand}\ \ \isanewline
\ \ \ {\isacharparenleft}{\isacharparenleft}{\isasymforall}\ {\isacharparenleft}i{\isacharcolon}{\isacharcolon}nat{\isacharparenright}{\isachardot}\ i\ {\isacharless}\ n\ {\isasymlongrightarrow}\ {\isacharparenleft}msg\ {\isadigit{1}}\ {\isacharparenleft}nReturn\ i{\isacharparenright}{\isacharparenright}{\isacharparenright}\ {\isasymand}\ \isanewline
\ \ \ \ {\isacharparenleft}DisjointSchedules\ n\ nC{\isacharparenright}\ {\isasymand}\ {\isacharparenleft}IdenticCycleLength\ n\ nC{\isacharparenright}\ \isanewline
\ \ \ {\isasymlongrightarrow}\isanewline
\ \ \ \ {\isacharparenleft}FrameTransmission\ n\ nStore\ nReturn\ nGet\ nC{\isacharparenright}\ {\isasymand}\ \isanewline
\ \ \ \ {\isacharparenleft}{\isasymforall}\ {\isacharparenleft}i{\isacharcolon}{\isacharcolon}nat{\isacharparenright}{\isachardot}\ i\ {\isacharless}\ n\ {\isasymlongrightarrow}\ {\isacharparenleft}msg\ {\isadigit{1}}\ {\isacharparenleft}nGet\ i{\isacharparenright}{\isacharparenright}\ {\isasymand}\ {\isacharparenleft}msg\ {\isadigit{1}}\ {\isacharparenleft}nStore\ i{\isacharparenright}{\isacharparenright}{\isacharparenright}\ {\isacharparenright}{\isachardoublequoteclose}\isanewline
\isanewline
\isanewline
\isacommand{definition}\isamarkupfalse%
\ \isanewline
\ \ \ Cable\ {\isacharcolon}{\isacharcolon}\ {\isachardoublequoteopen}nat\ {\isasymRightarrow}\ {\isacharprime}a\ nFrame\ {\isasymRightarrow}\ {\isacharprime}a\ Frame\ istream\ {\isasymRightarrow}\ bool{\isachardoublequoteclose}\ \isanewline
\isakeyword{where}\isanewline
\ \ {\isachardoublequoteopen}Cable\ n\ nSend\ recv\ \isanewline
\ \ \ {\isasymequiv}\isanewline
\ \ \ {\isacharparenleft}CorrectSheaf\ n{\isacharparenright}\ \isanewline
\ \ \ \ {\isasymand}\ \ \isanewline
\ \ \ {\isacharparenleft}{\isacharparenleft}inf{\isacharunderscore}disj\ n\ nSend{\isacharparenright}\ {\isasymlongrightarrow}\ {\isacharparenleft}Broadcast\ n\ nSend\ recv{\isacharparenright}{\isacharparenright}{\isachardoublequoteclose}\isanewline
\isanewline
\isacommand{definition}\isamarkupfalse%
\isanewline
\ \ \ Scheduler\ {\isacharcolon}{\isacharcolon}\ {\isachardoublequoteopen}Config\ {\isasymRightarrow}\ nat\ istream\ {\isasymRightarrow}\ bool{\isachardoublequoteclose}\isanewline
\isakeyword{where}\isanewline
\ \ {\isachardoublequoteopen}Scheduler\ c\ activation\isanewline
\ \ \ {\isasymequiv}\ \isanewline
\ \ \ {\isasymforall}\ {\isacharparenleft}t{\isacharcolon}{\isacharcolon}nat{\isacharparenright}{\isachardot}\isanewline
\ \ \ {\isacharparenleft}\ let\ s\ {\isacharequal}\ {\isacharparenleft}t\ mod\ {\isacharparenleft}cycleLength\ c{\isacharparenright}{\isacharparenright}\isanewline
\ \ \ \ \ \ in\isanewline
\ \ \ \ \ \ \ {\isacharparenleft}\ if\ \ {\isacharparenleft}s\ mem\ {\isacharparenleft}schedule\ c{\isacharparenright}{\isacharparenright}\isanewline
\ \ \ \ \ \ \ \ \ then\ {\isacharparenleft}activation\ t{\isacharparenright}\ {\isacharequal}\ {\isacharbrackleft}s{\isacharbrackright}\isanewline
\ \ \ \ \ \ \ \ \ else\ {\isacharparenleft}activation\ t{\isacharparenright}\ {\isacharequal}\ {\isacharbrackleft}{\isacharbrackright}{\isacharparenright}\ {\isacharparenright}{\isachardoublequoteclose}\isanewline
\ \ \ \ \ \ \ \ \ \isanewline
\isacommand{definition}\isamarkupfalse%
\isanewline
\ \ BusInterface\ {\isacharcolon}{\isacharcolon}\ \isanewline
\ \ \ \ {\isachardoublequoteopen}nat\ istream\ {\isasymRightarrow}\ {\isacharprime}a\ Frame\ istream\ {\isasymRightarrow}\ {\isacharprime}a\ Frame\ istream\ {\isasymRightarrow}\ \isanewline
\ \ \ \ \ {\isacharprime}a\ Frame\ istream\ {\isasymRightarrow}\ {\isacharprime}a\ Frame\ istream\ {\isasymRightarrow}\ nat\ istream\ {\isasymRightarrow}\ bool{\isachardoublequoteclose}\isanewline
\isakeyword{where}\isanewline
\ \ {\isachardoublequoteopen}BusInterface\ activation\ return\ recv\ \ store\ send\ get\isanewline
\ \ \ {\isasymequiv}\isanewline
\ \ \ {\isacharparenleft}Receive\ recv\ store\ activation{\isacharparenright}\ {\isasymand}\isanewline
\ \ \ {\isacharparenleft}Send\ return\ send\ get\ activation{\isacharparenright}{\isachardoublequoteclose}\isanewline
\isanewline
\isanewline
\isacommand{definition}\isamarkupfalse%
\isanewline
\ \ \ FlexRayController\ {\isacharcolon}{\isacharcolon}\ \isanewline
\ \ \ \ \ {\isachardoublequoteopen}{\isacharprime}a\ Frame\ istream\ {\isasymRightarrow}\ {\isacharprime}a\ Frame\ istream\ {\isasymRightarrow}\ \ Config\ {\isasymRightarrow}\isanewline
\ \ \ \ \ \ {\isacharprime}a\ Frame\ istream\ {\isasymRightarrow}\ {\isacharprime}a\ Frame\ istream\ {\isasymRightarrow}\ nat\ istream\ {\isasymRightarrow}\ bool{\isachardoublequoteclose}\isanewline
\isakeyword{where}\isanewline
\ \ {\isachardoublequoteopen}FlexRayController\ return\ recv\ c\ store\ send\ get\isanewline
\ \ \ {\isasymequiv}\isanewline
\ \ {\isacharparenleft}{\isasymexists}\ activation{\isachardot}\ \isanewline
\ \ \ \ \ {\isacharparenleft}Scheduler\ c\ activation{\isacharparenright}\ {\isasymand}\isanewline
\ \ \ \ \ {\isacharparenleft}BusInterface\ activation\ return\ recv\ store\ send\ get{\isacharparenright}{\isacharparenright}{\isachardoublequoteclose}\isanewline
\ \isanewline
\isanewline
\isacommand{definition}\isamarkupfalse%
\isanewline
\ \ FlexRayArchitecture\ {\isacharcolon}{\isacharcolon}\ \isanewline
\ \ \ {\isachardoublequoteopen}nat\ {\isasymRightarrow}\ {\isacharprime}a\ nFrame\ {\isasymRightarrow}\ nConfig\ {\isasymRightarrow}\ {\isacharprime}a\ nFrame\ {\isasymRightarrow}\ nNat\ {\isasymRightarrow}\ bool{\isachardoublequoteclose}\ \ \ \isanewline
\isakeyword{where}\isanewline
\ \ {\isachardoublequoteopen}FlexRayArchitecture\ n\ nReturn\ nC\ nStore\ nGet\isanewline
\ \ \ {\isasymequiv}\isanewline
\ \ \ {\isacharparenleft}CorrectSheaf\ n{\isacharparenright}\ {\isasymand}\isanewline
\ \ \ {\isacharparenleft}{\isasymexists}\ nSend\ recv{\isachardot}\isanewline
\ \ \ \ \ {\isacharparenleft}Cable\ n\ nSend\ recv{\isacharparenright}\ {\isasymand}\ \isanewline
\ \ \ \ \ {\isacharparenleft}{\isasymforall}\ {\isacharparenleft}i{\isacharcolon}{\isacharcolon}nat{\isacharparenright}{\isachardot}\ i\ {\isacharless}\ n\ {\isasymlongrightarrow}\ \ \isanewline
\ \ \ \ \ \ \ \ \ FlexRayController\ {\isacharparenleft}nReturn\ i{\isacharparenright}\ recv\ {\isacharparenleft}nC\ i{\isacharparenright}\ \isanewline
\ \ \ \ \ \ \ \ \ \ \ \ \ \ \ \ \ \ \ \ \ \ \ \ \ \ \ {\isacharparenleft}nStore\ i{\isacharparenright}\ {\isacharparenleft}nSend\ i{\isacharparenright}\ {\isacharparenleft}nGet\ i{\isacharparenright}{\isacharparenright}{\isacharparenright}{\isachardoublequoteclose}\ \ \isanewline
\ \ \ \isanewline
\isacommand{definition}\isamarkupfalse%
\isanewline
\ \ FlexRayArch\ {\isacharcolon}{\isacharcolon}\ \isanewline
\ \ \ {\isachardoublequoteopen}nat\ {\isasymRightarrow}\ {\isacharprime}a\ nFrame\ {\isasymRightarrow}\ nConfig\ {\isasymRightarrow}\ {\isacharprime}a\ nFrame\ {\isasymRightarrow}\ nNat\ {\isasymRightarrow}\ bool{\isachardoublequoteclose}\ \ \ \isanewline
\isakeyword{where}\isanewline
\ \ {\isachardoublequoteopen}FlexRayArch\ n\ nReturn\ nC\ nStore\ nGet\isanewline
\ \ \ {\isasymequiv}\isanewline
\ \ \ {\isacharparenleft}CorrectSheaf\ n{\isacharparenright}\ \ {\isasymand}\ \isanewline
\ \ \ {\isacharparenleft}{\isacharparenleft}{\isasymforall}\ {\isacharparenleft}i{\isacharcolon}{\isacharcolon}nat{\isacharparenright}{\isachardot}\ i\ {\isacharless}\ n\ {\isasymlongrightarrow}\ msg\ {\isadigit{1}}\ {\isacharparenleft}nReturn\ i{\isacharparenright}{\isacharparenright}\ {\isasymand}\ \isanewline
\ \ \ \ {\isacharparenleft}DisjointSchedules\ n\ nC{\isacharparenright}\ {\isasymand}\ {\isacharparenleft}IdenticCycleLength\ n\ nC{\isacharparenright}\ \ \ \ \isanewline
\ \ \ \ {\isasymlongrightarrow}\isanewline
\ \ \ \ {\isacharparenleft}FlexRayArchitecture\ n\ nReturn\ nC\ nStore\ nGet{\isacharparenright}{\isacharparenright}{\isachardoublequoteclose}\ \ \ \ \isanewline
\isadelimtheory
\isanewline
\endisadelimtheory
\isatagtheory
\isacommand{end}\isamarkupfalse%
\endisatagtheory
{\isafoldtheory}%
\isadelimtheory
\endisadelimtheory
\end{isabellebody}%

%
\begin{isabellebody}%
\def\isabellecontext{FR{\isacharunderscore}proof}%
\isamarkupheader{FlexRay: Verification%
}
\isamarkuptrue%
\isadelimtheory
\endisadelimtheory
\isatagtheory
\isacommand{theory}\isamarkupfalse%
\ \ FR{\isacharunderscore}proof\isanewline
\isakeyword{imports}\ FR\isanewline
\isakeyword{begin}%
\endisatagtheory
{\isafoldtheory}%
\isadelimtheory
\endisadelimtheory
\isamarkupsubsection{Properties of the function Send%
}
\isamarkuptrue%
\isacommand{lemma}\isamarkupfalse%
\ Send{\isacharunderscore}L{\isadigit{1}}{\isacharcolon}\isanewline
\ \isakeyword{assumes}\ h{\isadigit{1}}{\isacharcolon}{\isachardoublequoteopen}Send\ return\ send\ get\ activation{\isachardoublequoteclose}\isanewline
\ \ \ \ \ \isakeyword{and}\ h{\isadigit{2}}{\isacharcolon}{\isachardoublequoteopen}send\ t\ \ {\isasymnoteq}\ {\isacharbrackleft}{\isacharbrackright}{\isachardoublequoteclose}\isanewline
\ \isakeyword{shows}\ \ {\isachardoublequoteopen}{\isacharparenleft}activation\ t{\isacharparenright}\ {\isasymnoteq}\ {\isacharbrackleft}{\isacharbrackright}{\isachardoublequoteclose}\isanewline
\isadelimproof
\endisadelimproof
\isatagproof
\isacommand{using}\isamarkupfalse%
\ assms\ \isacommand{by}\isamarkupfalse%
\ {\isacharparenleft}simp\ add{\isacharcolon}\ Send{\isacharunderscore}def{\isacharcomma}\ auto{\isacharparenright}%
\endisatagproof
{\isafoldproof}%
\isadelimproof
\isanewline
\endisadelimproof
\isanewline
\isacommand{lemma}\isamarkupfalse%
\ Send{\isacharunderscore}L{\isadigit{2}}{\isacharcolon}\isanewline
\ \isakeyword{assumes}\ h{\isadigit{1}}{\isacharcolon}{\isachardoublequoteopen}Send\ return\ send\ get\ activation{\isachardoublequoteclose}\isanewline
\ \ \ \ \ \isakeyword{and}\ h{\isadigit{2}}{\isacharcolon}{\isachardoublequoteopen}{\isacharparenleft}activation\ t{\isacharparenright}\ \ {\isasymnoteq}\ {\isacharbrackleft}{\isacharbrackright}{\isachardoublequoteclose}\isanewline
\ \ \ \ \ \isakeyword{and}\ h{\isadigit{3}}{\isacharcolon}{\isachardoublequoteopen}return\ t\ {\isasymnoteq}\ {\isacharbrackleft}{\isacharbrackright}{\isachardoublequoteclose}\isanewline
\ \isakeyword{shows}\ \ {\isachardoublequoteopen}{\isacharparenleft}send\ t{\isacharparenright}\ {\isasymnoteq}\ {\isacharbrackleft}{\isacharbrackright}{\isachardoublequoteclose}\isanewline
\isadelimproof
\endisadelimproof
\isatagproof
\isacommand{using}\isamarkupfalse%
\ assms\ \ \isacommand{by}\isamarkupfalse%
\ {\isacharparenleft}simp\ add{\isacharcolon}\ Send{\isacharunderscore}def{\isacharparenright}%
\endisatagproof
{\isafoldproof}%
\isadelimproof
\endisadelimproof
\isamarkupsubsection{Properties of the component Scheduler%
}
\isamarkuptrue%
\isacommand{lemma}\isamarkupfalse%
\ Scheduler{\isacharunderscore}L{\isadigit{1}}{\isacharcolon}\isanewline
\ \isakeyword{assumes}\ h{\isadigit{1}}{\isacharcolon}{\isachardoublequoteopen}Scheduler\ C\ activation{\isachardoublequoteclose}\isanewline
\ \ \ \ \ \isakeyword{and}\ h{\isadigit{2}}{\isacharcolon}{\isachardoublequoteopen}{\isacharparenleft}activation\ t{\isacharparenright}\ {\isasymnoteq}\ {\isacharbrackleft}{\isacharbrackright}{\isachardoublequoteclose}\isanewline
\ \isakeyword{shows}\ {\isachardoublequoteopen}{\isacharparenleft}t\ mod\ {\isacharparenleft}cycleLength\ C{\isacharparenright}{\isacharparenright}\ mem\ {\isacharparenleft}schedule\ C{\isacharparenright}{\isachardoublequoteclose}\isanewline
\isadelimproof
\endisadelimproof
\isatagproof
\isacommand{using}\isamarkupfalse%
\ assms\isanewline
\isacommand{proof}\isamarkupfalse%
\ {\isacharminus}\ \isanewline
\ \ \ \isacommand{{\isacharbraceleft}}\isamarkupfalse%
\ \isacommand{assume}\isamarkupfalse%
\ \ a{\isadigit{1}}{\isacharcolon}{\isachardoublequoteopen}{\isasymnot}\ t\ mod\ cycleLength\ C\ mem\ schedule\ C{\isachardoublequoteclose}\isanewline
\ \ \ \ \ \isacommand{from}\isamarkupfalse%
\ h{\isadigit{1}}\ \isacommand{have}\isamarkupfalse%
\ sg{\isadigit{1}}{\isacharcolon}\isanewline
\ \ \ \ \ \ {\isachardoublequoteopen}if\ t\ mod\ cycleLength\ C\ mem\ schedule\ C\ \isanewline
\ \ \ \ \ \ \ then\ activation\ t\ {\isacharequal}\ {\isacharbrackleft}t\ mod\ cycleLength\ C{\isacharbrackright}\isanewline
\ \ \ \ \ \ \ else\ activation\ t\ {\isacharequal}\ {\isacharbrackleft}{\isacharbrackright}{\isachardoublequoteclose}\isanewline
\ \ \ \ \ \ \ \isacommand{by}\isamarkupfalse%
\ {\isacharparenleft}simp\ add{\isacharcolon}\ Scheduler{\isacharunderscore}def\ Let{\isacharunderscore}def{\isacharparenright}\isanewline
\ \ \ \ \ \isacommand{from}\isamarkupfalse%
\ a{\isadigit{1}}\ \isakeyword{and}\ sg{\isadigit{1}}\ \isacommand{have}\isamarkupfalse%
\ sg{\isadigit{2}}{\isacharcolon}{\isachardoublequoteopen}activation\ t\ {\isacharequal}\ {\isacharbrackleft}{\isacharbrackright}{\isachardoublequoteclose}\ \isacommand{by}\isamarkupfalse%
\ simp\isanewline
\ \ \ \ \ \isacommand{from}\isamarkupfalse%
\ sg{\isadigit{2}}\ \isakeyword{and}\ h{\isadigit{2}}\ \isacommand{have}\isamarkupfalse%
\ sg{\isadigit{3}}{\isacharcolon}{\isachardoublequoteopen}False{\isachardoublequoteclose}\ \isacommand{by}\isamarkupfalse%
\ simp\isanewline
\ \ \ \isacommand{{\isacharbraceright}}\isamarkupfalse%
\ \isacommand{from}\isamarkupfalse%
\ this\ \isacommand{have}\isamarkupfalse%
\ \ sg{\isadigit{4}}{\isacharcolon}{\isachardoublequoteopen}{\isacharparenleft}t\ mod\ {\isacharparenleft}cycleLength\ C{\isacharparenright}{\isacharparenright}\ mem\ {\isacharparenleft}schedule\ C{\isacharparenright}{\isachardoublequoteclose}\ \isacommand{by}\isamarkupfalse%
\ blast\isanewline
\ \ \isacommand{from}\isamarkupfalse%
\ this\ \isacommand{show}\isamarkupfalse%
\ {\isacharquery}thesis\ \isacommand{by}\isamarkupfalse%
\ simp\isanewline
\isacommand{qed}\isamarkupfalse%
\endisatagproof
{\isafoldproof}%
\isadelimproof
\isanewline
\endisadelimproof
\isanewline
\isanewline
\isacommand{lemma}\isamarkupfalse%
\ Scheduler{\isacharunderscore}L{\isadigit{2}}{\isacharcolon}\isanewline
\ \isakeyword{assumes}\ h{\isadigit{1}}{\isacharcolon}{\isachardoublequoteopen}Scheduler\ C\ activation{\isachardoublequoteclose}\isanewline
\ \ \ \ \ \isakeyword{and}\ h{\isadigit{2}}{\isacharcolon}{\isachardoublequoteopen}{\isasymnot}\ {\isacharparenleft}t\ mod\ cycleLength\ C{\isacharparenright}\ mem\ {\isacharparenleft}schedule\ C{\isacharparenright}{\isachardoublequoteclose}\isanewline
\ \isakeyword{shows}\ {\isachardoublequoteopen}activation\ t\ {\isacharequal}\ {\isacharbrackleft}{\isacharbrackright}{\isachardoublequoteclose}\isanewline
\isadelimproof
\endisadelimproof
\isatagproof
\isacommand{using}\isamarkupfalse%
\ assms\ \isacommand{by}\isamarkupfalse%
\ {\isacharparenleft}simp\ add{\isacharcolon}\ Scheduler{\isacharunderscore}def\ Let{\isacharunderscore}def{\isacharparenright}%
\endisatagproof
{\isafoldproof}%
\isadelimproof
\ \ \isanewline
\endisadelimproof
\isanewline
\isanewline
\isacommand{lemma}\isamarkupfalse%
\ Scheduler{\isacharunderscore}L{\isadigit{3}}{\isacharcolon}\isanewline
\ \isakeyword{assumes}\ h{\isadigit{1}}{\isacharcolon}{\isachardoublequoteopen}Scheduler\ C\ activation{\isachardoublequoteclose}\isanewline
\ \ \ \ \ \isakeyword{and}\ h{\isadigit{2}}{\isacharcolon}{\isachardoublequoteopen}{\isacharparenleft}t\ mod\ cycleLength\ C{\isacharparenright}\ mem\ {\isacharparenleft}schedule\ C{\isacharparenright}{\isachardoublequoteclose}\isanewline
\ \isakeyword{shows}\ {\isachardoublequoteopen}activation\ t\ {\isasymnoteq}\ {\isacharbrackleft}{\isacharbrackright}{\isachardoublequoteclose}\isanewline
\isadelimproof
\endisadelimproof
\isatagproof
\isacommand{using}\isamarkupfalse%
\ assms\ \isacommand{by}\isamarkupfalse%
\ {\isacharparenleft}simp\ add{\isacharcolon}\ Scheduler{\isacharunderscore}def\ Let{\isacharunderscore}def{\isacharparenright}%
\endisatagproof
{\isafoldproof}%
\isadelimproof
\isanewline
\endisadelimproof
\isanewline
\isanewline
\isacommand{lemma}\isamarkupfalse%
\ Scheduler{\isacharunderscore}L{\isadigit{4}}{\isacharcolon}\isanewline
\ \isakeyword{assumes}\ h{\isadigit{1}}{\isacharcolon}{\isachardoublequoteopen}Scheduler\ C\ activation{\isachardoublequoteclose}\isanewline
\ \ \ \ \ \isakeyword{and}\ h{\isadigit{2}}{\isacharcolon}{\isachardoublequoteopen}{\isacharparenleft}t\ mod\ cycleLength\ C{\isacharparenright}\ mem\ {\isacharparenleft}schedule\ C{\isacharparenright}{\isachardoublequoteclose}\isanewline
\ \isakeyword{shows}\ {\isachardoublequoteopen}activation\ t\ {\isacharequal}\ {\isacharbrackleft}t\ mod\ cycleLength\ C{\isacharbrackright}{\isachardoublequoteclose}\isanewline
\isadelimproof
\endisadelimproof
\isatagproof
\isacommand{using}\isamarkupfalse%
\ assms\ \isacommand{by}\isamarkupfalse%
\ {\isacharparenleft}simp\ add{\isacharcolon}\ Scheduler{\isacharunderscore}def\ Let{\isacharunderscore}def{\isacharparenright}%
\endisatagproof
{\isafoldproof}%
\isadelimproof
\isanewline
\endisadelimproof
\ \isanewline
\isanewline
\isacommand{lemma}\isamarkupfalse%
\ correct{\isacharunderscore}DisjointSchedules{\isadigit{1}}{\isacharcolon}\isanewline
\ \ \isakeyword{assumes}\ h{\isadigit{1}}{\isacharcolon}{\isachardoublequoteopen}DisjointSchedules\ n\ nC{\isachardoublequoteclose}\isanewline
\ \ \ \ \ \ \isakeyword{and}\ h{\isadigit{2}}{\isacharcolon}{\isachardoublequoteopen}IdenticCycleLength\ n\ nC{\isachardoublequoteclose}\isanewline
\ \ \ \ \ \ \isakeyword{and}\ h{\isadigit{3}}{\isacharcolon}{\isachardoublequoteopen}{\isacharparenleft}t\ mod\ cycleLength\ {\isacharparenleft}nC\ i{\isacharparenright}{\isacharparenright}\ mem\ schedule\ {\isacharparenleft}nC\ i{\isacharparenright}{\isachardoublequoteclose}\ \isanewline
\ \ \ \ \ \ \isakeyword{and}\ h{\isadigit{4}}{\isacharcolon}{\isachardoublequoteopen}i\ {\isacharless}\ n{\isachardoublequoteclose}\isanewline
\ \ \ \ \ \ \isakeyword{and}\ h{\isadigit{5}}{\isacharcolon}{\isachardoublequoteopen}j\ {\isacharless}\ n{\isachardoublequoteclose}\isanewline
\ \ \ \ \ \ \isakeyword{and}\ h{\isadigit{6}}{\isacharcolon}{\isachardoublequoteopen}i\ {\isasymnoteq}\ j{\isachardoublequoteclose}\isanewline
\ \ \isakeyword{shows}\ {\isachardoublequoteopen}{\isasymnot}\ {\isacharparenleft}t\ mod\ cycleLength\ {\isacharparenleft}nC\ j{\isacharparenright}\ mem\ schedule\ {\isacharparenleft}nC\ j{\isacharparenright}{\isacharparenright}{\isachardoublequoteclose}\isanewline
\isadelimproof
\endisadelimproof
\isatagproof
\isacommand{proof}\isamarkupfalse%
\ {\isacharminus}\isanewline
\ \ \isacommand{from}\isamarkupfalse%
\ h{\isadigit{1}}\ \isakeyword{and}\ h{\isadigit{4}}\ \isakeyword{and}\ h{\isadigit{5}}\ \isakeyword{and}\ h{\isadigit{6}}\ \isacommand{have}\isamarkupfalse%
\ sg{\isadigit{1}}{\isacharcolon}{\isachardoublequoteopen}disjoint\ {\isacharparenleft}schedule\ {\isacharparenleft}nC\ i{\isacharparenright}{\isacharparenright}\ {\isacharparenleft}schedule\ {\isacharparenleft}nC\ j{\isacharparenright}{\isacharparenright}{\isachardoublequoteclose}\isanewline
\ \ \ \ \isacommand{by}\isamarkupfalse%
\ {\isacharparenleft}simp\ add{\isacharcolon}\ DisjointSchedules{\isacharunderscore}def{\isacharparenright}\ \isanewline
\ \ \isacommand{from}\isamarkupfalse%
\ h{\isadigit{2}}\ \isakeyword{and}\ h{\isadigit{4}}\ \isakeyword{and}\ h{\isadigit{5}}\ \isacommand{have}\isamarkupfalse%
\ sg{\isadigit{2}}{\isacharcolon}{\isachardoublequoteopen}cycleLength\ {\isacharparenleft}nC\ i{\isacharparenright}\ {\isacharequal}\ cycleLength\ {\isacharparenleft}nC\ j{\isacharparenright}{\isachardoublequoteclose}\isanewline
\ \ \ \ \isacommand{by}\isamarkupfalse%
\ {\isacharparenleft}simp\ only{\isacharcolon}\ IdenticCycleLength{\isacharunderscore}def{\isacharcomma}\ blast{\isacharparenright}\ \isanewline
\ \ \isacommand{from}\isamarkupfalse%
\ sg{\isadigit{1}}\ \isakeyword{and}\ h{\isadigit{3}}\ \isacommand{have}\isamarkupfalse%
\ sg{\isadigit{3}}{\isacharcolon}{\isachardoublequoteopen}{\isasymnot}\ {\isacharparenleft}t\ mod\ {\isacharparenleft}cycleLength\ {\isacharparenleft}nC\ i{\isacharparenright}{\isacharparenright}{\isacharparenright}\ mem\ {\isacharparenleft}schedule\ {\isacharparenleft}nC\ j{\isacharparenright}{\isacharparenright}{\isachardoublequoteclose}\ \isanewline
\ \ \ \ \isacommand{by}\isamarkupfalse%
\ {\isacharparenleft}simp\ add{\isacharcolon}\ mem{\isacharunderscore}notdisjoint{\isadigit{2}}{\isacharparenright}\isanewline
\ \ \isacommand{from}\isamarkupfalse%
\ sg{\isadigit{2}}\ \isakeyword{and}\ sg{\isadigit{3}}\ \isacommand{show}\isamarkupfalse%
\ {\isacharquery}thesis\ \isacommand{by}\isamarkupfalse%
\ simp\isanewline
\isacommand{qed}\isamarkupfalse%
\endisatagproof
{\isafoldproof}%
\isadelimproof
\endisadelimproof
\isamarkupsubsection{Disjoint Frames%
}
\isamarkuptrue%
\isacommand{lemma}\isamarkupfalse%
\ disjointFrame{\isacharunderscore}L{\isadigit{1}}{\isacharcolon}\isanewline
\ \isakeyword{assumes}\ h{\isadigit{1}}{\isacharcolon}{\isachardoublequoteopen}DisjointSchedules\ n\ nC{\isachardoublequoteclose}\ \ \isanewline
\ \ \ \ \ \isakeyword{and}\ h{\isadigit{2}}{\isacharcolon}{\isachardoublequoteopen}IdenticCycleLength\ n\ nC{\isachardoublequoteclose}\isanewline
\ \ \ \ \ \isakeyword{and}\ h{\isadigit{3}}{\isacharcolon}{\isachardoublequoteopen}{\isasymforall}\ i\ {\isacharless}\ n{\isachardot}\ FlexRayController\ {\isacharparenleft}nReturn\ i{\isacharparenright}\ rcv\ \isanewline
\ \ \ \ \ \ \ \ \ \ \ \ \ \ \ \ \ \ \ \ \ \ {\isacharparenleft}nC\ i{\isacharparenright}\ {\isacharparenleft}nStore\ i{\isacharparenright}\ {\isacharparenleft}nSend\ i{\isacharparenright}\ {\isacharparenleft}nGet\ i{\isacharparenright}{\isachardoublequoteclose}\isanewline
\ \ \ \ \isakeyword{and}\ h{\isadigit{4}}{\isacharcolon}{\isachardoublequoteopen}nSend\ i\ t\ {\isasymnoteq}\ {\isacharbrackleft}{\isacharbrackright}{\isachardoublequoteclose}\isanewline
\ \ \ \ \isakeyword{and}\ h{\isadigit{5}}{\isacharcolon}{\isachardoublequoteopen}i\ {\isacharless}\ n{\isachardoublequoteclose}\isanewline
\ \ \ \ \isakeyword{and}\ h{\isadigit{6}}{\isacharcolon}{\isachardoublequoteopen}j\ {\isacharless}\ n{\isachardoublequoteclose}\isanewline
\ \ \ \ \isakeyword{and}\ h{\isadigit{7}}{\isacharcolon}{\isachardoublequoteopen}i\ {\isasymnoteq}\ j{\isachardoublequoteclose}\isanewline
\ \isakeyword{shows}\ {\isachardoublequoteopen}nSend\ j\ t\ {\isacharequal}\ {\isacharbrackleft}{\isacharbrackright}{\isachardoublequoteclose}\isanewline
\isadelimproof
\endisadelimproof
\isatagproof
\isacommand{proof}\isamarkupfalse%
\ {\isacharminus}\ \isanewline
\ \ \isacommand{from}\isamarkupfalse%
\ h{\isadigit{3}}\ \isakeyword{and}\ h{\isadigit{5}}\ \isacommand{have}\isamarkupfalse%
\ sg{\isadigit{1}}{\isacharcolon}\isanewline
\ \ \ {\isachardoublequoteopen}FlexRayController\ {\isacharparenleft}nReturn\ i{\isacharparenright}\ rcv\ {\isacharparenleft}nC\ i{\isacharparenright}\ {\isacharparenleft}nStore\ i{\isacharparenright}\ {\isacharparenleft}nSend\ i{\isacharparenright}\ {\isacharparenleft}nGet\ i{\isacharparenright}{\isachardoublequoteclose}\isanewline
\ \ \ \ \isacommand{by}\isamarkupfalse%
\ auto\isanewline
\ \ \isacommand{from}\isamarkupfalse%
\ h{\isadigit{3}}\ \isakeyword{and}\ h{\isadigit{6}}\ \isacommand{have}\isamarkupfalse%
\ sg{\isadigit{2}}{\isacharcolon}\isanewline
\ \ \ {\isachardoublequoteopen}FlexRayController\ {\isacharparenleft}nReturn\ j{\isacharparenright}\ rcv\ {\isacharparenleft}nC\ j{\isacharparenright}\ {\isacharparenleft}nStore\ j{\isacharparenright}\ {\isacharparenleft}nSend\ j{\isacharparenright}\ {\isacharparenleft}nGet\ j{\isacharparenright}{\isachardoublequoteclose}\isanewline
\ \ \ \ \isacommand{by}\isamarkupfalse%
\ auto\isanewline
\ \ \isacommand{from}\isamarkupfalse%
\ sg{\isadigit{1}}\ \isacommand{obtain}\isamarkupfalse%
\ activation{\isadigit{1}}\ \isakeyword{where}\isanewline
\ \ \ \ \ a{\isadigit{1}}{\isacharcolon}{\isachardoublequoteopen}Scheduler\ {\isacharparenleft}nC\ i{\isacharparenright}\ activation{\isadigit{1}}{\isachardoublequoteclose}\ \isakeyword{and}\ \isanewline
\ \ \ \ \ a{\isadigit{2}}{\isacharcolon}{\isachardoublequoteopen}BusInterface\ activation{\isadigit{1}}\ {\isacharparenleft}nReturn\ i{\isacharparenright}\ rcv\ {\isacharparenleft}nStore\ i{\isacharparenright}\ {\isacharparenleft}nSend\ i{\isacharparenright}\ {\isacharparenleft}nGet\ i{\isacharparenright}{\isachardoublequoteclose}\isanewline
\ \ \ \ \ \isacommand{by}\isamarkupfalse%
\ {\isacharparenleft}simp\ add{\isacharcolon}\ FlexRayController{\isacharunderscore}def{\isacharcomma}\ auto{\isacharparenright}\isanewline
\ \ \isacommand{from}\isamarkupfalse%
\ sg{\isadigit{2}}\ \isacommand{obtain}\isamarkupfalse%
\ activation{\isadigit{2}}\ \isakeyword{where}\isanewline
\ \ \ \ \ a{\isadigit{3}}{\isacharcolon}{\isachardoublequoteopen}Scheduler\ {\isacharparenleft}nC\ j{\isacharparenright}\ activation{\isadigit{2}}{\isachardoublequoteclose}\ \isakeyword{and}\ \isanewline
\ \ \ \ \ a{\isadigit{4}}{\isacharcolon}{\isachardoublequoteopen}BusInterface\ activation{\isadigit{2}}\ {\isacharparenleft}nReturn\ j{\isacharparenright}\ rcv\ {\isacharparenleft}nStore\ j{\isacharparenright}\ {\isacharparenleft}nSend\ j{\isacharparenright}\ {\isacharparenleft}nGet\ j{\isacharparenright}{\isachardoublequoteclose}\isanewline
\ \ \ \ \ \isacommand{by}\isamarkupfalse%
\ {\isacharparenleft}simp\ add{\isacharcolon}\ FlexRayController{\isacharunderscore}def{\isacharcomma}\ auto{\isacharparenright}\isanewline
\ \ \isacommand{from}\isamarkupfalse%
\ h{\isadigit{1}}\ \isakeyword{and}\ h{\isadigit{5}}\ \isakeyword{and}\ h{\isadigit{6}}\ \isakeyword{and}\ h{\isadigit{7}}\ \isacommand{have}\isamarkupfalse%
\ sg{\isadigit{3}}{\isacharcolon}{\isachardoublequoteopen}disjoint\ {\isacharparenleft}schedule\ {\isacharparenleft}nC\ i{\isacharparenright}{\isacharparenright}\ {\isacharparenleft}schedule\ {\isacharparenleft}nC\ j{\isacharparenright}{\isacharparenright}{\isachardoublequoteclose}\isanewline
\ \ \ \ \isacommand{by}\isamarkupfalse%
\ {\isacharparenleft}simp\ add{\isacharcolon}\ DisjointSchedules{\isacharunderscore}def{\isacharparenright}\ \isanewline
\ \ \isacommand{from}\isamarkupfalse%
\ a{\isadigit{2}}\ \isacommand{have}\isamarkupfalse%
\ \ sg{\isadigit{4}}a{\isacharcolon}{\isachardoublequoteopen}Send\ {\isacharparenleft}nReturn\ i{\isacharparenright}\ {\isacharparenleft}nSend\ i{\isacharparenright}\ {\isacharparenleft}nGet\ i{\isacharparenright}\ activation{\isadigit{1}}{\isachardoublequoteclose}\ \isanewline
\ \ \ \ \ \isacommand{by}\isamarkupfalse%
\ {\isacharparenleft}simp\ add{\isacharcolon}\ \ BusInterface{\isacharunderscore}def{\isacharparenright}\isanewline
\ \ \isacommand{from}\isamarkupfalse%
\ sg{\isadigit{4}}a\ \isakeyword{and}\ h{\isadigit{4}}\ \isacommand{have}\isamarkupfalse%
\ sg{\isadigit{5}}{\isacharcolon}{\isachardoublequoteopen}activation{\isadigit{1}}\ t\ \ {\isasymnoteq}\ {\isacharbrackleft}{\isacharbrackright}{\isachardoublequoteclose}\ \isacommand{by}\isamarkupfalse%
\ {\isacharparenleft}simp\ add{\isacharcolon}\ Send{\isacharunderscore}L{\isadigit{1}}{\isacharparenright}\isanewline
\ \ \isacommand{from}\isamarkupfalse%
\ a{\isadigit{1}}\ \isakeyword{and}\ sg{\isadigit{5}}\ \isacommand{have}\isamarkupfalse%
\ sg{\isadigit{6}}{\isacharcolon}{\isachardoublequoteopen}{\isacharparenleft}t\ mod\ {\isacharparenleft}cycleLength\ {\isacharparenleft}nC\ i{\isacharparenright}{\isacharparenright}{\isacharparenright}\ mem\ {\isacharparenleft}schedule\ {\isacharparenleft}nC\ i{\isacharparenright}{\isacharparenright}{\isachardoublequoteclose}\ \isanewline
\ \ \ \ \ \isacommand{by}\isamarkupfalse%
\ {\isacharparenleft}simp\ add{\isacharcolon}\ Scheduler{\isacharunderscore}L{\isadigit{1}}{\isacharparenright}\isanewline
\ \ \isacommand{from}\isamarkupfalse%
\ h{\isadigit{2}}\ \isakeyword{and}\ h{\isadigit{5}}\ \isakeyword{and}\ h{\isadigit{6}}\ \isacommand{have}\isamarkupfalse%
\ sg{\isadigit{7}}{\isacharcolon}{\isachardoublequoteopen}cycleLength\ {\isacharparenleft}nC\ i{\isacharparenright}\ {\isacharequal}\ cycleLength\ {\isacharparenleft}nC\ j{\isacharparenright}{\isachardoublequoteclose}\isanewline
\ \ \ \ \isacommand{by}\isamarkupfalse%
\ {\isacharparenleft}simp\ only{\isacharcolon}\ IdenticCycleLength{\isacharunderscore}def{\isacharcomma}\ blast{\isacharparenright}\ \ \isanewline
\ \ \isacommand{from}\isamarkupfalse%
\ sg{\isadigit{3}}\ \isakeyword{and}\ sg{\isadigit{6}}\ \isacommand{have}\isamarkupfalse%
\ sg{\isadigit{8}}{\isacharcolon}{\isachardoublequoteopen}{\isasymnot}\ {\isacharparenleft}t\ mod\ {\isacharparenleft}cycleLength\ {\isacharparenleft}nC\ i{\isacharparenright}{\isacharparenright}{\isacharparenright}\ mem\ {\isacharparenleft}schedule\ {\isacharparenleft}nC\ j{\isacharparenright}{\isacharparenright}{\isachardoublequoteclose}\ \isanewline
\ \ \ \ \isacommand{by}\isamarkupfalse%
\ {\isacharparenleft}simp\ add{\isacharcolon}\ mem{\isacharunderscore}notdisjoint{\isadigit{2}}{\isacharparenright}\isanewline
\ \ \isacommand{from}\isamarkupfalse%
\ sg{\isadigit{8}}\ \isakeyword{and}\ sg{\isadigit{7}}\ \isacommand{have}\isamarkupfalse%
\ sg{\isadigit{9}}{\isacharcolon}{\isachardoublequoteopen}{\isasymnot}\ {\isacharparenleft}t\ mod\ {\isacharparenleft}cycleLength\ {\isacharparenleft}nC\ j{\isacharparenright}{\isacharparenright}{\isacharparenright}\ mem\ {\isacharparenleft}schedule\ {\isacharparenleft}nC\ j{\isacharparenright}{\isacharparenright}{\isachardoublequoteclose}\ \isanewline
\ \ \ \ \isacommand{by}\isamarkupfalse%
\ simp\isanewline
\ \ \isacommand{from}\isamarkupfalse%
\ sg{\isadigit{9}}\ \isakeyword{and}\ a{\isadigit{3}}\ \isacommand{have}\isamarkupfalse%
\ sg{\isadigit{1}}{\isadigit{0}}{\isacharcolon}{\isachardoublequoteopen}activation{\isadigit{2}}\ t\ {\isacharequal}\ {\isacharbrackleft}{\isacharbrackright}{\isachardoublequoteclose}\ \isacommand{by}\isamarkupfalse%
\ {\isacharparenleft}simp\ add{\isacharcolon}\ Scheduler{\isacharunderscore}L{\isadigit{2}}{\isacharparenright}\ \isanewline
\ \ \isacommand{from}\isamarkupfalse%
\ a{\isadigit{4}}\ \isacommand{have}\isamarkupfalse%
\ sg{\isadigit{1}}{\isadigit{1}}{\isacharcolon}{\isachardoublequoteopen}Send\ {\isacharparenleft}nReturn\ j{\isacharparenright}\ {\isacharparenleft}nSend\ j{\isacharparenright}\ {\isacharparenleft}nGet\ j{\isacharparenright}\ activation{\isadigit{2}}{\isachardoublequoteclose}\ \isanewline
\ \ \ \ \ \isacommand{by}\isamarkupfalse%
\ {\isacharparenleft}simp\ add{\isacharcolon}\ \ BusInterface{\isacharunderscore}def{\isacharparenright}\isanewline
\ \ \isacommand{from}\isamarkupfalse%
\ sg{\isadigit{1}}{\isadigit{1}}\ \isakeyword{and}\ sg{\isadigit{1}}{\isadigit{0}}\ \isacommand{show}\isamarkupfalse%
\ {\isacharquery}thesis\ \isacommand{by}\isamarkupfalse%
\ {\isacharparenleft}simp\ add{\isacharcolon}\ Send{\isacharunderscore}def{\isacharparenright}\isanewline
\isacommand{qed}\isamarkupfalse%
\endisatagproof
{\isafoldproof}%
\isadelimproof
\isanewline
\endisadelimproof
\isanewline
\isanewline
\isacommand{lemma}\isamarkupfalse%
\ disjointFrame{\isacharunderscore}L{\isadigit{2}}{\isacharcolon}\isanewline
\ \isakeyword{assumes}\ h{\isadigit{1}}{\isacharcolon}{\isachardoublequoteopen}DisjointSchedules\ n\ nC{\isachardoublequoteclose}\ \ \isanewline
\ \ \ \ \ \isakeyword{and}\ h{\isadigit{2}}{\isacharcolon}{\isachardoublequoteopen}IdenticCycleLength\ n\ nC{\isachardoublequoteclose}\isanewline
\ \ \ \ \ \isakeyword{and}\ h{\isadigit{3}}{\isacharcolon}{\isachardoublequoteopen}{\isasymforall}\ i\ {\isacharless}\ n{\isachardot}\ FlexRayController\ {\isacharparenleft}nReturn\ i{\isacharparenright}\ rcv\ \isanewline
\ \ \ \ \ \ \ \ \ \ \ \ \ \ \ \ \ \ \ \ \ \ {\isacharparenleft}nC\ i{\isacharparenright}\ {\isacharparenleft}nStore\ i{\isacharparenright}\ {\isacharparenleft}nSend\ i{\isacharparenright}\ {\isacharparenleft}nGet\ i{\isacharparenright}{\isachardoublequoteclose}\isanewline
\ \isakeyword{shows}\ {\isachardoublequoteopen}inf{\isacharunderscore}disj\ n\ nSend{\isachardoublequoteclose}\isanewline
\isadelimproof
\endisadelimproof
\isatagproof
\isacommand{using}\isamarkupfalse%
\ assms\isanewline
\ \ \isacommand{apply}\isamarkupfalse%
\ {\isacharparenleft}simp\ add{\isacharcolon}\ inf{\isacharunderscore}disj{\isacharunderscore}def{\isacharcomma}\ clarify{\isacharparenright}\isanewline
\ \ \isacommand{by}\isamarkupfalse%
\ {\isacharparenleft}rule\ disjointFrame{\isacharunderscore}L{\isadigit{1}}{\isacharcomma}\ auto{\isacharparenright}%
\endisatagproof
{\isafoldproof}%
\isadelimproof
\isanewline
\endisadelimproof
\isanewline
\isanewline
\isacommand{lemma}\isamarkupfalse%
\ disjointFrame{\isacharunderscore}L{\isadigit{3}}{\isacharcolon}\isanewline
\ \isakeyword{assumes}\ h{\isadigit{1}}{\isacharcolon}{\isachardoublequoteopen}DisjointSchedules\ n\ nC{\isachardoublequoteclose}\ \ \isanewline
\ \ \ \ \ \isakeyword{and}\ h{\isadigit{2}}{\isacharcolon}{\isachardoublequoteopen}IdenticCycleLength\ n\ nC{\isachardoublequoteclose}\isanewline
\ \ \ \ \ \isakeyword{and}\ h{\isadigit{3}}{\isacharcolon}{\isachardoublequoteopen}{\isasymforall}\ i\ {\isacharless}\ n{\isachardot}\ FlexRayController\ {\isacharparenleft}nReturn\ i{\isacharparenright}\ rcv\ \isanewline
\ \ \ \ \ \ \ \ \ \ \ \ \ \ \ \ \ \ \ \ \ \ {\isacharparenleft}nC\ i{\isacharparenright}\ {\isacharparenleft}nStore\ i{\isacharparenright}\ {\isacharparenleft}nSend\ i{\isacharparenright}\ {\isacharparenleft}nGet\ i{\isacharparenright}{\isachardoublequoteclose}\isanewline
\ \ \ \ \isakeyword{and}\ h{\isadigit{4}}{\isacharcolon}{\isachardoublequoteopen}t\ mod\ cycleLength\ {\isacharparenleft}nC\ i{\isacharparenright}\ mem\ schedule\ {\isacharparenleft}nC\ i{\isacharparenright}{\isachardoublequoteclose}\isanewline
\ \ \ \ \isakeyword{and}\ h{\isadigit{5}}{\isacharcolon}{\isachardoublequoteopen}i\ {\isacharless}\ n{\isachardoublequoteclose}\isanewline
\ \ \ \ \isakeyword{and}\ h{\isadigit{6}}{\isacharcolon}{\isachardoublequoteopen}j\ {\isacharless}\ n{\isachardoublequoteclose}\isanewline
\ \ \ \ \isakeyword{and}\ h{\isadigit{7}}{\isacharcolon}{\isachardoublequoteopen}i\ {\isasymnoteq}\ j{\isachardoublequoteclose}\isanewline
\ \isakeyword{shows}\ {\isachardoublequoteopen}nSend\ j\ t\ {\isacharequal}\ {\isacharbrackleft}{\isacharbrackright}{\isachardoublequoteclose}\isanewline
\isadelimproof
\endisadelimproof
\isatagproof
\isacommand{proof}\isamarkupfalse%
\ {\isacharminus}\ \isanewline
\ \ \isacommand{from}\isamarkupfalse%
\ h{\isadigit{2}}\ \isakeyword{and}\ h{\isadigit{5}}\ \isakeyword{and}\ h{\isadigit{6}}\ \isacommand{have}\isamarkupfalse%
\ sg{\isadigit{1}}{\isacharcolon}{\isachardoublequoteopen}cycleLength\ {\isacharparenleft}nC\ i{\isacharparenright}\ {\isacharequal}\ cycleLength\ {\isacharparenleft}nC\ j{\isacharparenright}{\isachardoublequoteclose}\isanewline
\ \ \ \ \isacommand{by}\isamarkupfalse%
\ {\isacharparenleft}simp\ only{\isacharcolon}\ IdenticCycleLength{\isacharunderscore}def{\isacharcomma}\ blast{\isacharparenright}\ \ \isanewline
\ \ \isacommand{from}\isamarkupfalse%
\ h{\isadigit{1}}\ \isakeyword{and}\ h{\isadigit{5}}\ \isakeyword{and}\ h{\isadigit{6}}\ \isakeyword{and}\ h{\isadigit{7}}\ \isacommand{have}\isamarkupfalse%
\ sg{\isadigit{2}}{\isacharcolon}{\isachardoublequoteopen}disjoint\ {\isacharparenleft}schedule\ {\isacharparenleft}nC\ i{\isacharparenright}{\isacharparenright}\ {\isacharparenleft}schedule\ {\isacharparenleft}nC\ j{\isacharparenright}{\isacharparenright}{\isachardoublequoteclose}\isanewline
\ \ \ \ \isacommand{by}\isamarkupfalse%
\ {\isacharparenleft}simp\ add{\isacharcolon}\ DisjointSchedules{\isacharunderscore}def{\isacharparenright}\ \isanewline
\ \ \isacommand{from}\isamarkupfalse%
\ sg{\isadigit{2}}\ \isakeyword{and}\ h{\isadigit{4}}\ \isacommand{have}\isamarkupfalse%
\ sg{\isadigit{3}}{\isacharcolon}{\isachardoublequoteopen}{\isasymnot}\ {\isacharparenleft}t\ mod\ {\isacharparenleft}cycleLength\ {\isacharparenleft}nC\ i{\isacharparenright}{\isacharparenright}{\isacharparenright}\ mem\ {\isacharparenleft}schedule\ {\isacharparenleft}nC\ j{\isacharparenright}{\isacharparenright}{\isachardoublequoteclose}\ \isanewline
\ \ \ \ \isacommand{by}\isamarkupfalse%
\ {\isacharparenleft}simp\ add{\isacharcolon}\ mem{\isacharunderscore}notdisjoint{\isadigit{2}}{\isacharparenright}\isanewline
\ \ \isacommand{from}\isamarkupfalse%
\ sg{\isadigit{1}}\ \isakeyword{and}\ sg{\isadigit{3}}\ \isacommand{have}\isamarkupfalse%
\ sg{\isadigit{4}}{\isacharcolon}{\isachardoublequoteopen}{\isasymnot}\ {\isacharparenleft}t\ mod\ {\isacharparenleft}cycleLength\ {\isacharparenleft}nC\ j{\isacharparenright}{\isacharparenright}{\isacharparenright}\ mem\ {\isacharparenleft}schedule\ {\isacharparenleft}nC\ j{\isacharparenright}{\isacharparenright}{\isachardoublequoteclose}\ \isanewline
\ \ \ \ \isacommand{by}\isamarkupfalse%
\ simp\isanewline
\ \ \isacommand{from}\isamarkupfalse%
\ h{\isadigit{3}}\ \isakeyword{and}\ h{\isadigit{6}}\ \isacommand{have}\isamarkupfalse%
\ sg{\isadigit{5}}{\isacharcolon}\isanewline
\ \ \ {\isachardoublequoteopen}FlexRayController\ {\isacharparenleft}nReturn\ j{\isacharparenright}\ rcv\ {\isacharparenleft}nC\ j{\isacharparenright}\ {\isacharparenleft}nStore\ j{\isacharparenright}\ {\isacharparenleft}nSend\ j{\isacharparenright}\ {\isacharparenleft}nGet\ j{\isacharparenright}{\isachardoublequoteclose}\isanewline
\ \ \ \ \isacommand{by}\isamarkupfalse%
\ auto\isanewline
\ \ \isacommand{from}\isamarkupfalse%
\ sg{\isadigit{5}}\ \isacommand{obtain}\isamarkupfalse%
\ activation{\isadigit{2}}\ \isakeyword{where}\isanewline
\ \ \ \ \ a{\isadigit{1}}{\isacharcolon}{\isachardoublequoteopen}Scheduler\ {\isacharparenleft}nC\ j{\isacharparenright}\ activation{\isadigit{2}}{\isachardoublequoteclose}\ \isakeyword{and}\ \isanewline
\ \ \ \ \ a{\isadigit{2}}{\isacharcolon}{\isachardoublequoteopen}BusInterface\ activation{\isadigit{2}}\ {\isacharparenleft}nReturn\ j{\isacharparenright}\ rcv\ {\isacharparenleft}nStore\ j{\isacharparenright}\ {\isacharparenleft}nSend\ j{\isacharparenright}\ {\isacharparenleft}nGet\ j{\isacharparenright}{\isachardoublequoteclose}\isanewline
\ \ \ \ \ \isacommand{by}\isamarkupfalse%
\ {\isacharparenleft}simp\ add{\isacharcolon}\ FlexRayController{\isacharunderscore}def{\isacharcomma}\ auto{\isacharparenright}\isanewline
\ \ \isacommand{from}\isamarkupfalse%
\ sg{\isadigit{4}}\ \isakeyword{and}\ a{\isadigit{1}}\ \isacommand{have}\isamarkupfalse%
\ sg{\isadigit{6}}{\isacharcolon}{\isachardoublequoteopen}activation{\isadigit{2}}\ t\ {\isacharequal}\ {\isacharbrackleft}{\isacharbrackright}{\isachardoublequoteclose}\ \isacommand{by}\isamarkupfalse%
\ {\isacharparenleft}simp\ add{\isacharcolon}\ Scheduler{\isacharunderscore}L{\isadigit{2}}{\isacharparenright}\ \isanewline
\ \ \isacommand{from}\isamarkupfalse%
\ a{\isadigit{2}}\ \isacommand{have}\isamarkupfalse%
\ sg{\isadigit{7}}{\isacharcolon}{\isachardoublequoteopen}Send\ {\isacharparenleft}nReturn\ j{\isacharparenright}\ {\isacharparenleft}nSend\ j{\isacharparenright}\ {\isacharparenleft}nGet\ j{\isacharparenright}\ activation{\isadigit{2}}{\isachardoublequoteclose}\ \isanewline
\ \ \ \ \ \isacommand{by}\isamarkupfalse%
\ {\isacharparenleft}simp\ add{\isacharcolon}\ \ BusInterface{\isacharunderscore}def{\isacharparenright}\isanewline
\ \ \isacommand{from}\isamarkupfalse%
\ sg{\isadigit{7}}\ \isakeyword{and}\ sg{\isadigit{6}}\ \isacommand{show}\isamarkupfalse%
\ {\isacharquery}thesis\ \isacommand{by}\isamarkupfalse%
\ {\isacharparenleft}simp\ add{\isacharcolon}\ Send{\isacharunderscore}def{\isacharparenright}\isanewline
\isacommand{qed}\isamarkupfalse%
\endisatagproof
{\isafoldproof}%
\isadelimproof
\endisadelimproof
\isamarkupsubsection{Properties of the sheaf of channels nSend%
}
\isamarkuptrue%
\isacommand{lemma}\isamarkupfalse%
\ fr{\isacharunderscore}Send{\isadigit{1}}{\isacharcolon}\isanewline
\ \isakeyword{assumes}\ h{\isadigit{1}}{\isacharcolon}{\isachardoublequoteopen}FlexRayController\ {\isacharparenleft}nReturn\ i{\isacharparenright}\ recv\ {\isacharparenleft}nC\ i{\isacharparenright}\ {\isacharparenleft}nStore\ i{\isacharparenright}\ {\isacharparenleft}nSend\ i{\isacharparenright}\ {\isacharparenleft}nGet\ i{\isacharparenright}{\isachardoublequoteclose}\isanewline
\ \ \ \ \ \isakeyword{and}\ h{\isadigit{2}}{\isacharcolon}{\isachardoublequoteopen}{\isasymnot}\ {\isacharparenleft}t\ mod\ cycleLength\ {\isacharparenleft}nC\ i{\isacharparenright}\ mem\ schedule\ {\isacharparenleft}nC\ i{\isacharparenright}{\isacharparenright}{\isachardoublequoteclose}\isanewline
\ \isakeyword{shows}\ \ \ \ \ \ {\isachardoublequoteopen}{\isacharparenleft}nSend\ i{\isacharparenright}\ t\ {\isacharequal}\ {\isacharbrackleft}{\isacharbrackright}{\isachardoublequoteclose}\isanewline
\isadelimproof
\endisadelimproof
\isatagproof
\isacommand{proof}\isamarkupfalse%
\ {\isacharminus}\ \isanewline
\ \ \isacommand{from}\isamarkupfalse%
\ h{\isadigit{1}}\ \isacommand{obtain}\isamarkupfalse%
\ activation\ \isakeyword{where}\ \isanewline
\ \ \ \ a{\isadigit{1}}{\isacharcolon}{\isachardoublequoteopen}Scheduler\ {\isacharparenleft}nC\ i{\isacharparenright}\ activation{\isachardoublequoteclose}\ \ \isakeyword{and}\isanewline
\ \ \ \ a{\isadigit{2}}{\isacharcolon}{\isachardoublequoteopen}BusInterface\ activation\ {\isacharparenleft}nReturn\ i{\isacharparenright}\ recv\ {\isacharparenleft}nStore\ i{\isacharparenright}\ {\isacharparenleft}nSend\ i{\isacharparenright}\ {\isacharparenleft}nGet\ i{\isacharparenright}{\isachardoublequoteclose}\isanewline
\ \ \ \ \isacommand{by}\isamarkupfalse%
\ {\isacharparenleft}simp\ add{\isacharcolon}\ FlexRayController{\isacharunderscore}def{\isacharcomma}\ auto{\isacharparenright}\isanewline
\ \ \isacommand{from}\isamarkupfalse%
\ a{\isadigit{1}}\ \isakeyword{and}\ h{\isadigit{2}}\ \isacommand{have}\isamarkupfalse%
\ sg{\isadigit{1}}{\isacharcolon}{\isachardoublequoteopen}activation\ t\ {\isacharequal}\ {\isacharbrackleft}{\isacharbrackright}{\isachardoublequoteclose}\ \isacommand{by}\isamarkupfalse%
\ {\isacharparenleft}simp\ add{\isacharcolon}\ Scheduler{\isacharunderscore}L{\isadigit{2}}{\isacharparenright}\isanewline
\ \ \isacommand{from}\isamarkupfalse%
\ a{\isadigit{2}}\ \isacommand{have}\isamarkupfalse%
\ sg{\isadigit{2}}{\isacharcolon}{\isachardoublequoteopen}Send\ {\isacharparenleft}nReturn\ i{\isacharparenright}\ {\isacharparenleft}nSend\ i{\isacharparenright}\ {\isacharparenleft}nGet\ i{\isacharparenright}\ activation{\isachardoublequoteclose}\isanewline
\ \ \ \ \isacommand{by}\isamarkupfalse%
\ {\isacharparenleft}simp\ add{\isacharcolon}\ BusInterface{\isacharunderscore}def{\isacharparenright}\isanewline
\ \ \isacommand{from}\isamarkupfalse%
\ sg{\isadigit{2}}\ \isakeyword{and}\ sg{\isadigit{1}}\ \isacommand{show}\isamarkupfalse%
\ {\isacharquery}thesis\ \isacommand{by}\isamarkupfalse%
\ {\isacharparenleft}simp\ add{\isacharcolon}\ Send{\isacharunderscore}def{\isacharparenright}\isanewline
\isacommand{qed}\isamarkupfalse%
\endisatagproof
{\isafoldproof}%
\isadelimproof
\isanewline
\endisadelimproof
\isanewline
\isanewline
\isacommand{lemma}\isamarkupfalse%
\ fr{\isacharunderscore}Send{\isadigit{2}}{\isacharcolon}\isanewline
\ \isakeyword{assumes}\ h{\isadigit{1}}{\isacharcolon}{\isachardoublequoteopen}{\isasymforall}i{\isacharless}n{\isachardot}\ FlexRayController\ {\isacharparenleft}nReturn\ i{\isacharparenright}\ recv\ {\isacharparenleft}nC\ i{\isacharparenright}\ {\isacharparenleft}nStore\ i{\isacharparenright}\ {\isacharparenleft}nSend\ i{\isacharparenright}\ {\isacharparenleft}nGet\ i{\isacharparenright}{\isachardoublequoteclose}\isanewline
\ \ \ \ \ \isakeyword{and}\ h{\isadigit{2}}{\isacharcolon}{\isachardoublequoteopen}DisjointSchedules\ n\ nC{\isachardoublequoteclose}\isanewline
\ \ \ \ \ \isakeyword{and}\ h{\isadigit{3}}{\isacharcolon}{\isachardoublequoteopen}IdenticCycleLength\ n\ nC{\isachardoublequoteclose}\isanewline
\ \ \ \ \ \isakeyword{and}\ h{\isadigit{4}}{\isacharcolon}{\isachardoublequoteopen}t\ mod\ cycleLength\ {\isacharparenleft}nC\ k{\isacharparenright}\ mem\ schedule\ {\isacharparenleft}nC\ k{\isacharparenright}{\isachardoublequoteclose}\isanewline
\ \ \ \ \ \isakeyword{and}\ h{\isadigit{5}}{\isacharcolon}{\isachardoublequoteopen}k\ {\isacharless}\ n{\isachardoublequoteclose}\isanewline
\ \isakeyword{shows}\ {\isachardoublequoteopen}nSend\ k\ t\ {\isacharequal}\ nReturn\ k\ t{\isachardoublequoteclose}\isanewline
\isadelimproof
\endisadelimproof
\isatagproof
\isacommand{using}\isamarkupfalse%
\ assms\isanewline
\isacommand{proof}\isamarkupfalse%
\ {\isacharminus}\ \isanewline
\ \isacommand{from}\isamarkupfalse%
\ h{\isadigit{1}}\ \isakeyword{and}\ h{\isadigit{5}}\ \isacommand{have}\isamarkupfalse%
\ sg{\isadigit{1}}{\isacharcolon}\isanewline
\ \ \ {\isachardoublequoteopen}FlexRayController\ {\isacharparenleft}nReturn\ k{\isacharparenright}\ recv\ {\isacharparenleft}nC\ k{\isacharparenright}\ {\isacharparenleft}nStore\ k{\isacharparenright}\ {\isacharparenleft}nSend\ k{\isacharparenright}\ {\isacharparenleft}nGet\ k{\isacharparenright}{\isachardoublequoteclose}\isanewline
\ \ \ \ \isacommand{by}\isamarkupfalse%
\ auto\isanewline
\ \ \isacommand{from}\isamarkupfalse%
\ sg{\isadigit{1}}\ \isacommand{obtain}\isamarkupfalse%
\ activation\ \isakeyword{where}\isanewline
\ \ \ \ \ a{\isadigit{1}}{\isacharcolon}{\isachardoublequoteopen}Scheduler\ {\isacharparenleft}nC\ k{\isacharparenright}\ activation{\isachardoublequoteclose}\ \isakeyword{and}\ \isanewline
\ \ \ \ \ a{\isadigit{2}}{\isacharcolon}{\isachardoublequoteopen}BusInterface\ activation\ {\isacharparenleft}nReturn\ k{\isacharparenright}\ recv\ {\isacharparenleft}nStore\ k{\isacharparenright}\ {\isacharparenleft}nSend\ k{\isacharparenright}\ {\isacharparenleft}nGet\ k{\isacharparenright}{\isachardoublequoteclose}\isanewline
\ \ \ \ \ \isacommand{by}\isamarkupfalse%
\ {\isacharparenleft}simp\ add{\isacharcolon}\ FlexRayController{\isacharunderscore}def{\isacharcomma}\ auto{\isacharparenright}\isanewline
\ \ \isacommand{from}\isamarkupfalse%
\ a{\isadigit{1}}\ \isakeyword{and}\ h{\isadigit{4}}\ \isacommand{have}\isamarkupfalse%
\ sg{\isadigit{3}}{\isacharcolon}{\isachardoublequoteopen}activation\ t\ {\isasymnoteq}\ {\isacharbrackleft}{\isacharbrackright}{\isachardoublequoteclose}\ \isacommand{by}\isamarkupfalse%
\ {\isacharparenleft}simp\ add{\isacharcolon}\ Scheduler{\isacharunderscore}L{\isadigit{3}}{\isacharparenright}\isanewline
\ \ \isacommand{from}\isamarkupfalse%
\ a{\isadigit{2}}\ \isacommand{have}\isamarkupfalse%
\ \ sg{\isadigit{4}}{\isacharcolon}{\isachardoublequoteopen}Send\ {\isacharparenleft}nReturn\ k{\isacharparenright}\ {\isacharparenleft}nSend\ k{\isacharparenright}\ {\isacharparenleft}nGet\ k{\isacharparenright}\ activation{\isachardoublequoteclose}\ \isanewline
\ \ \ \ \ \isacommand{by}\isamarkupfalse%
\ {\isacharparenleft}simp\ add{\isacharcolon}\ \ BusInterface{\isacharunderscore}def{\isacharparenright}\isanewline
\ \ \isacommand{from}\isamarkupfalse%
\ sg{\isadigit{4}}\ \isakeyword{and}\ sg{\isadigit{3}}\ \isacommand{show}\isamarkupfalse%
\ {\isacharquery}thesis\ \isacommand{by}\isamarkupfalse%
\ {\isacharparenleft}simp\ add{\isacharcolon}\ Send{\isacharunderscore}def{\isacharparenright}\isanewline
\isacommand{qed}\isamarkupfalse%
\endisatagproof
{\isafoldproof}%
\isadelimproof
\isanewline
\endisadelimproof
\isanewline
\isanewline
\isacommand{lemma}\isamarkupfalse%
\ fr{\isacharunderscore}Send{\isadigit{3}}{\isacharcolon}\isanewline
\ \isakeyword{assumes}\ h{\isadigit{1}}{\isacharcolon}{\isachardoublequoteopen}{\isasymforall}i{\isacharless}n{\isachardot}\ FlexRayController\ {\isacharparenleft}nReturn\ i{\isacharparenright}\ recv\ {\isacharparenleft}nC\ i{\isacharparenright}\ {\isacharparenleft}nStore\ i{\isacharparenright}\ {\isacharparenleft}nSend\ i{\isacharparenright}\ {\isacharparenleft}nGet\ i{\isacharparenright}{\isachardoublequoteclose}\isanewline
\ \ \ \ \ \isakeyword{and}\ h{\isadigit{2}}{\isacharcolon}{\isachardoublequoteopen}DisjointSchedules\ n\ nC{\isachardoublequoteclose}\isanewline
\ \ \ \ \ \isakeyword{and}\ h{\isadigit{3}}{\isacharcolon}{\isachardoublequoteopen}IdenticCycleLength\ n\ nC{\isachardoublequoteclose}\isanewline
\ \ \ \ \ \isakeyword{and}\ h{\isadigit{4}}{\isacharcolon}{\isachardoublequoteopen}t\ mod\ cycleLength\ {\isacharparenleft}nC\ k{\isacharparenright}\ mem\ schedule\ {\isacharparenleft}nC\ k{\isacharparenright}{\isachardoublequoteclose}\isanewline
\ \ \ \ \ \isakeyword{and}\ h{\isadigit{5}}{\isacharcolon}{\isachardoublequoteopen}k\ {\isacharless}\ n{\isachardoublequoteclose}\isanewline
\ \ \ \ \ \isakeyword{and}\ h{\isadigit{6}}{\isacharcolon}{\isachardoublequoteopen}nReturn\ k\ t\ {\isasymnoteq}\ {\isacharbrackleft}{\isacharbrackright}{\isachardoublequoteclose}\isanewline
\ \isakeyword{shows}\ {\isachardoublequoteopen}nSend\ k\ t\ {\isasymnoteq}\ {\isacharbrackleft}{\isacharbrackright}{\isachardoublequoteclose}\isanewline
\isadelimproof
\endisadelimproof
\isatagproof
\isacommand{using}\isamarkupfalse%
\ assms\ \isacommand{by}\isamarkupfalse%
\ {\isacharparenleft}simp\ add{\isacharcolon}\ fr{\isacharunderscore}Send{\isadigit{2}}{\isacharparenright}%
\endisatagproof
{\isafoldproof}%
\isadelimproof
\isanewline
\endisadelimproof
\isanewline
\isanewline
\isacommand{lemma}\isamarkupfalse%
\ fr{\isacharunderscore}Send{\isadigit{4}}{\isacharcolon}\isanewline
\ \isakeyword{assumes}\ h{\isadigit{1}}{\isacharcolon}{\isachardoublequoteopen}{\isasymforall}i{\isacharless}n{\isachardot}\ FlexRayController\ {\isacharparenleft}nReturn\ i{\isacharparenright}\ recv\ {\isacharparenleft}nC\ i{\isacharparenright}\ {\isacharparenleft}nStore\ i{\isacharparenright}\ {\isacharparenleft}nSend\ i{\isacharparenright}\ {\isacharparenleft}nGet\ i{\isacharparenright}{\isachardoublequoteclose}\isanewline
\ \ \ \ \ \isakeyword{and}\ h{\isadigit{2}}{\isacharcolon}{\isachardoublequoteopen}DisjointSchedules\ n\ nC{\isachardoublequoteclose}\isanewline
\ \ \ \ \ \isakeyword{and}\ h{\isadigit{3}}{\isacharcolon}{\isachardoublequoteopen}IdenticCycleLength\ n\ nC{\isachardoublequoteclose}\isanewline
\ \ \ \ \ \isakeyword{and}\ h{\isadigit{4}}{\isacharcolon}{\isachardoublequoteopen}t\ mod\ cycleLength\ {\isacharparenleft}nC\ k{\isacharparenright}\ mem\ schedule\ {\isacharparenleft}nC\ k{\isacharparenright}{\isachardoublequoteclose}\isanewline
\ \ \ \ \ \isakeyword{and}\ h{\isadigit{5}}{\isacharcolon}{\isachardoublequoteopen}k\ {\isacharless}\ n{\isachardoublequoteclose}\isanewline
\ \ \ \ \ \isakeyword{and}\ h{\isadigit{6}}{\isacharcolon}{\isachardoublequoteopen}nReturn\ k\ t\ {\isasymnoteq}\ {\isacharbrackleft}{\isacharbrackright}{\isachardoublequoteclose}\isanewline
\ \isakeyword{shows}\ {\isachardoublequoteopen}{\isasymexists}k{\isachardot}\ k\ {\isacharless}\ n\ {\isasymlongrightarrow}\ nSend\ k\ t\ {\isasymnoteq}\ {\isacharbrackleft}{\isacharbrackright}{\isachardoublequoteclose}\isanewline
\isadelimproof
\endisadelimproof
\isatagproof
\isacommand{proof}\isamarkupfalse%
\ \isanewline
\ \isacommand{from}\isamarkupfalse%
\ assms\ \isacommand{show}\isamarkupfalse%
\ {\isachardoublequoteopen}k\ {\isacharless}\ n\ {\isasymlongrightarrow}\ nSend\ k\ t\ {\isasymnoteq}\ {\isacharbrackleft}{\isacharbrackright}{\isachardoublequoteclose}\ \ \isacommand{by}\isamarkupfalse%
\ {\isacharparenleft}simp\ add{\isacharcolon}\ fr{\isacharunderscore}Send{\isadigit{3}}{\isacharparenright}\isanewline
\isacommand{qed}\isamarkupfalse%
\endisatagproof
{\isafoldproof}%
\isadelimproof
\isanewline
\endisadelimproof
\isanewline
\isanewline
\isacommand{lemma}\isamarkupfalse%
\ fr{\isacharunderscore}Send{\isadigit{5}}{\isacharcolon}\isanewline
\ \isakeyword{assumes}\ h{\isadigit{1}}{\isacharcolon}{\isachardoublequoteopen}{\isasymforall}i{\isacharless}n{\isachardot}\ FlexRayController\ {\isacharparenleft}nReturn\ i{\isacharparenright}\ recv\ {\isacharparenleft}nC\ i{\isacharparenright}\ {\isacharparenleft}nStore\ i{\isacharparenright}\ {\isacharparenleft}nSend\ i{\isacharparenright}\ {\isacharparenleft}nGet\ i{\isacharparenright}{\isachardoublequoteclose}\isanewline
\ \ \ \ \ \isakeyword{and}\ h{\isadigit{2}}{\isacharcolon}{\isachardoublequoteopen}DisjointSchedules\ n\ nC{\isachardoublequoteclose}\isanewline
\ \ \ \ \ \isakeyword{and}\ h{\isadigit{3}}{\isacharcolon}{\isachardoublequoteopen}IdenticCycleLength\ n\ nC{\isachardoublequoteclose}\isanewline
\ \ \ \ \ \isakeyword{and}\ h{\isadigit{4}}{\isacharcolon}{\isachardoublequoteopen}t\ mod\ cycleLength\ {\isacharparenleft}nC\ k{\isacharparenright}\ mem\ schedule\ {\isacharparenleft}nC\ k{\isacharparenright}{\isachardoublequoteclose}\isanewline
\ \ \ \ \ \isakeyword{and}\ h{\isadigit{5}}{\isacharcolon}{\isachardoublequoteopen}k\ {\isacharless}\ n{\isachardoublequoteclose}\isanewline
\ \ \ \ \ \isakeyword{and}\ h{\isadigit{6}}{\isacharcolon}{\isachardoublequoteopen}nReturn\ k\ t\ {\isasymnoteq}\ {\isacharbrackleft}{\isacharbrackright}{\isachardoublequoteclose}\isanewline
\ \ \ \ \ \isakeyword{and}\ h{\isadigit{7}}{\isacharcolon}{\isachardoublequoteopen}{\isasymforall}k{\isacharless}n{\isachardot}\ nSend\ k\ t\ {\isacharequal}\ {\isacharbrackleft}{\isacharbrackright}{\isachardoublequoteclose}\isanewline
\ \isakeyword{shows}\ {\isachardoublequoteopen}False{\isachardoublequoteclose}\isanewline
\isadelimproof
\endisadelimproof
\isatagproof
\isacommand{proof}\isamarkupfalse%
\ {\isacharminus}\ \isanewline
\ \ \isacommand{from}\isamarkupfalse%
\ h{\isadigit{1}}\ \isakeyword{and}\ h{\isadigit{2}}\ \isakeyword{and}\ h{\isadigit{3}}\ \isakeyword{and}\ h{\isadigit{4}}\ \isakeyword{and}\ h{\isadigit{5}}\ \isakeyword{and}\ h{\isadigit{6}}\ \isacommand{have}\isamarkupfalse%
\ sg{\isadigit{1}}{\isacharcolon}{\isachardoublequoteopen}nSend\ k\ t\ {\isasymnoteq}\ {\isacharbrackleft}{\isacharbrackright}{\isachardoublequoteclose}\isanewline
\ \ \ \ \isacommand{by}\isamarkupfalse%
\ {\isacharparenleft}simp\ add{\isacharcolon}\ fr{\isacharunderscore}Send{\isadigit{2}}{\isacharparenright}\isanewline
\ \ \isacommand{from}\isamarkupfalse%
\ h{\isadigit{7}}\ \isakeyword{and}\ h{\isadigit{5}}\ \isacommand{have}\isamarkupfalse%
\ sg{\isadigit{2}}{\isacharcolon}{\isachardoublequoteopen}nSend\ k\ t\ {\isacharequal}\ {\isacharbrackleft}{\isacharbrackright}{\isachardoublequoteclose}\ \isacommand{by}\isamarkupfalse%
\ blast\ \isanewline
\ \ \isacommand{from}\isamarkupfalse%
\ sg{\isadigit{1}}\ \isakeyword{and}\ sg{\isadigit{2}}\ \isacommand{show}\isamarkupfalse%
\ {\isacharquery}thesis\ \isacommand{by}\isamarkupfalse%
\ simp\ \ \ \ \isanewline
\isacommand{qed}\isamarkupfalse%
\endisatagproof
{\isafoldproof}%
\isadelimproof
\isanewline
\endisadelimproof
\isanewline
\isanewline
\isacommand{lemma}\isamarkupfalse%
\ fr{\isacharunderscore}Send{\isadigit{6}}{\isacharcolon}\isanewline
\ \isakeyword{assumes}\ h{\isadigit{1}}{\isacharcolon}{\isachardoublequoteopen}{\isasymforall}i{\isacharless}n{\isachardot}\ FlexRayController\ {\isacharparenleft}nReturn\ i{\isacharparenright}\ recv\ {\isacharparenleft}nC\ i{\isacharparenright}\ {\isacharparenleft}nStore\ i{\isacharparenright}\ {\isacharparenleft}nSend\ i{\isacharparenright}\ {\isacharparenleft}nGet\ i{\isacharparenright}{\isachardoublequoteclose}\isanewline
\ \ \ \ \ \isakeyword{and}\ h{\isadigit{2}}{\isacharcolon}{\isachardoublequoteopen}DisjointSchedules\ n\ nC{\isachardoublequoteclose}\isanewline
\ \ \ \ \ \isakeyword{and}\ h{\isadigit{3}}{\isacharcolon}{\isachardoublequoteopen}IdenticCycleLength\ n\ nC{\isachardoublequoteclose}\isanewline
\ \ \ \ \ \isakeyword{and}\ h{\isadigit{4}}{\isacharcolon}{\isachardoublequoteopen}t\ mod\ cycleLength\ {\isacharparenleft}nC\ k{\isacharparenright}\ mem\ schedule\ {\isacharparenleft}nC\ k{\isacharparenright}{\isachardoublequoteclose}\isanewline
\ \ \ \ \ \isakeyword{and}\ h{\isadigit{5}}{\isacharcolon}{\isachardoublequoteopen}k\ {\isacharless}\ n{\isachardoublequoteclose}\isanewline
\ \ \ \ \ \isakeyword{and}\ h{\isadigit{6}}{\isacharcolon}{\isachardoublequoteopen}nReturn\ k\ t\ {\isasymnoteq}\ {\isacharbrackleft}{\isacharbrackright}{\isachardoublequoteclose}\isanewline
\ \isakeyword{shows}\ {\isachardoublequoteopen}{\isasymexists}k{\isacharless}n{\isachardot}\ nSend\ k\ t\ {\isasymnoteq}\ {\isacharbrackleft}{\isacharbrackright}{\isachardoublequoteclose}\isanewline
\isadelimproof
\endisadelimproof
\isatagproof
\isacommand{proof}\isamarkupfalse%
\ {\isacharparenleft}rule\ ccontr{\isacharparenright}\isanewline
\ \ \isacommand{assume}\isamarkupfalse%
\ {\isachardoublequoteopen}{\isasymnot}\ {\isacharparenleft}{\isasymexists}k{\isacharless}n{\isachardot}\ nSend\ k\ t\ {\isasymnoteq}\ {\isacharbrackleft}{\isacharbrackright}{\isacharparenright}{\isachardoublequoteclose}\ \isanewline
\ \ \isacommand{from}\isamarkupfalse%
\ this\ \isakeyword{and}\ assms\ \isacommand{show}\isamarkupfalse%
\ {\isachardoublequoteopen}False{\isachardoublequoteclose}\isanewline
\ \ \ \ \isacommand{apply}\isamarkupfalse%
\ auto\ \isanewline
\ \ \ \ \isacommand{by}\isamarkupfalse%
\ {\isacharparenleft}rule\ fr{\isacharunderscore}Send{\isadigit{5}}{\isacharcomma}\ auto{\isacharparenright}\isanewline
\isacommand{qed}\isamarkupfalse%
\endisatagproof
{\isafoldproof}%
\isadelimproof
\ \isanewline
\endisadelimproof
\isanewline
\isanewline
\isacommand{lemma}\isamarkupfalse%
\ fr{\isacharunderscore}Send{\isadigit{7}}{\isacharcolon}\isanewline
\ \isakeyword{assumes}\ h{\isadigit{1}}{\isacharcolon}{\isachardoublequoteopen}{\isasymforall}i{\isacharless}n{\isachardot}\ FlexRayController\ {\isacharparenleft}nReturn\ i{\isacharparenright}\ recv\ {\isacharparenleft}nC\ i{\isacharparenright}\ {\isacharparenleft}nStore\ i{\isacharparenright}\ {\isacharparenleft}nSend\ i{\isacharparenright}\ {\isacharparenleft}nGet\ i{\isacharparenright}{\isachardoublequoteclose}\isanewline
\ \ \ \ \ \isakeyword{and}\ h{\isadigit{2}}{\isacharcolon}{\isachardoublequoteopen}DisjointSchedules\ n\ nC{\isachardoublequoteclose}\isanewline
\ \ \ \ \ \isakeyword{and}\ h{\isadigit{3}}{\isacharcolon}{\isachardoublequoteopen}IdenticCycleLength\ n\ nC{\isachardoublequoteclose}\isanewline
\ \ \ \ \ \isakeyword{and}\ h{\isadigit{4}}{\isacharcolon}{\isachardoublequoteopen}t\ mod\ cycleLength\ {\isacharparenleft}nC\ k{\isacharparenright}\ mem\ schedule\ {\isacharparenleft}nC\ k{\isacharparenright}{\isachardoublequoteclose}\isanewline
\ \ \ \ \ \isakeyword{and}\ h{\isadigit{5}}{\isacharcolon}{\isachardoublequoteopen}k\ {\isacharless}\ n{\isachardoublequoteclose}\isanewline
\ \ \ \ \ \isakeyword{and}\ h{\isadigit{6}}{\isacharcolon}{\isachardoublequoteopen}j\ {\isacharless}\ n{\isachardoublequoteclose}\isanewline
\ \ \ \ \ \isakeyword{and}\ h{\isadigit{6}}{\isacharcolon}{\isachardoublequoteopen}nReturn\ k\ t\ {\isacharequal}\ {\isacharbrackleft}{\isacharbrackright}{\isachardoublequoteclose}\isanewline
\ \isakeyword{shows}\ {\isachardoublequoteopen}nSend\ j\ t\ {\isacharequal}\ {\isacharbrackleft}{\isacharbrackright}{\isachardoublequoteclose}\isanewline
\isadelimproof
\endisadelimproof
\isatagproof
\isacommand{using}\isamarkupfalse%
\ assms\isanewline
\isacommand{proof}\isamarkupfalse%
\ {\isacharparenleft}cases\ {\isachardoublequoteopen}j\ {\isacharequal}\ k{\isachardoublequoteclose}{\isacharparenright}\isanewline
\ \ \isacommand{assume}\isamarkupfalse%
\ a{\isadigit{1}}{\isacharcolon}\ {\isachardoublequoteopen}j\ {\isacharequal}\ k{\isachardoublequoteclose}\isanewline
\ \ \isacommand{from}\isamarkupfalse%
\ assms\ \isacommand{have}\isamarkupfalse%
\ sg{\isadigit{1}}{\isacharcolon}\ {\isachardoublequoteopen}nSend\ k\ t\ {\isacharequal}\ nReturn\ k\ t{\isachardoublequoteclose}\ \isacommand{by}\isamarkupfalse%
\ {\isacharparenleft}simp\ add{\isacharcolon}\ fr{\isacharunderscore}Send{\isadigit{2}}{\isacharparenright}\isanewline
\ \ \isacommand{from}\isamarkupfalse%
\ sg{\isadigit{1}}\ \isakeyword{and}\ a{\isadigit{1}}\ \isakeyword{and}\ h{\isadigit{6}}\ \isacommand{show}\isamarkupfalse%
\ {\isacharquery}thesis\ \isacommand{by}\isamarkupfalse%
\ simp\isanewline
\isacommand{next}\isamarkupfalse%
\ \isanewline
\ \ \isacommand{assume}\isamarkupfalse%
\ a{\isadigit{2}}{\isacharcolon}{\isachardoublequoteopen}j\ {\isasymnoteq}\ k{\isachardoublequoteclose}\isanewline
\ \ \isacommand{from}\isamarkupfalse%
\ assms\ \isakeyword{and}\ a{\isadigit{2}}\ \isacommand{show}\isamarkupfalse%
\ {\isacharquery}thesis\ \isacommand{by}\isamarkupfalse%
\ {\isacharparenleft}simp\ add{\isacharcolon}\ disjointFrame{\isacharunderscore}L{\isadigit{3}}{\isacharparenright}\isanewline
\isacommand{qed}\isamarkupfalse%
\endisatagproof
{\isafoldproof}%
\isadelimproof
\ \isanewline
\endisadelimproof
\isanewline
\isanewline
\isacommand{lemma}\isamarkupfalse%
\ fr{\isacharunderscore}Send{\isadigit{8}}{\isacharcolon}\isanewline
\ \isakeyword{assumes}\ h{\isadigit{1}}{\isacharcolon}{\isachardoublequoteopen}{\isasymforall}i{\isacharless}n{\isachardot}\ FlexRayController\ {\isacharparenleft}nReturn\ i{\isacharparenright}\ recv\ {\isacharparenleft}nC\ i{\isacharparenright}\ {\isacharparenleft}nStore\ i{\isacharparenright}\ {\isacharparenleft}nSend\ i{\isacharparenright}\ {\isacharparenleft}nGet\ i{\isacharparenright}{\isachardoublequoteclose}\isanewline
\ \ \ \ \ \isakeyword{and}\ h{\isadigit{2}}{\isacharcolon}{\isachardoublequoteopen}DisjointSchedules\ n\ nC{\isachardoublequoteclose}\isanewline
\ \ \ \ \ \isakeyword{and}\ h{\isadigit{3}}{\isacharcolon}{\isachardoublequoteopen}IdenticCycleLength\ n\ nC{\isachardoublequoteclose}\isanewline
\ \ \ \ \ \isakeyword{and}\ h{\isadigit{4}}{\isacharcolon}{\isachardoublequoteopen}t\ mod\ cycleLength\ {\isacharparenleft}nC\ k{\isacharparenright}\ mem\ schedule\ {\isacharparenleft}nC\ k{\isacharparenright}{\isachardoublequoteclose}\isanewline
\ \ \ \ \ \isakeyword{and}\ h{\isadigit{5}}{\isacharcolon}{\isachardoublequoteopen}k\ {\isacharless}\ n{\isachardoublequoteclose}\isanewline
\ \ \ \ \ \isakeyword{and}\ h{\isadigit{6}}{\isacharcolon}{\isachardoublequoteopen}nReturn\ k\ t\ {\isacharequal}\ {\isacharbrackleft}{\isacharbrackright}{\isachardoublequoteclose}\isanewline
\ \isakeyword{shows}\ {\isachardoublequoteopen}{\isasymnot}\ {\isacharparenleft}{\isasymexists}k{\isacharless}n{\isachardot}\ nSend\ k\ t\ {\isasymnoteq}\ {\isacharbrackleft}{\isacharbrackright}{\isacharparenright}{\isachardoublequoteclose}\isanewline
\isadelimproof
\endisadelimproof
\isatagproof
\isacommand{using}\isamarkupfalse%
\ assms\ \isacommand{by}\isamarkupfalse%
\ {\isacharparenleft}auto{\isacharcomma}\ simp\ add{\isacharcolon}\ fr{\isacharunderscore}Send{\isadigit{7}}{\isacharparenright}%
\endisatagproof
{\isafoldproof}%
\isadelimproof
\isanewline
\endisadelimproof
\isanewline
\isanewline
\isacommand{lemma}\isamarkupfalse%
\ fr{\isacharunderscore}nC{\isacharunderscore}Send{\isacharcolon}\isanewline
\ \isakeyword{assumes}\ h{\isadigit{1}}{\isacharcolon}{\isachardoublequoteopen}{\isasymforall}i{\isacharless}n{\isachardot}\ FlexRayController\ {\isacharparenleft}nReturn\ i{\isacharparenright}\ recv\ {\isacharparenleft}nC\ i{\isacharparenright}\ {\isacharparenleft}nStore\ i{\isacharparenright}\ {\isacharparenleft}nSend\ i{\isacharparenright}\ {\isacharparenleft}nGet\ i{\isacharparenright}{\isachardoublequoteclose}\isanewline
\ \ \ \ \ \isakeyword{and}\ h{\isadigit{2}}{\isacharcolon}{\isachardoublequoteopen}k\ {\isacharless}\ n{\isachardoublequoteclose}\isanewline
\ \ \ \ \ \isakeyword{and}\ h{\isadigit{3}}{\isacharcolon}{\isachardoublequoteopen}DisjointSchedules\ n\ nC{\isachardoublequoteclose}\isanewline
\ \ \ \ \ \isakeyword{and}\ h{\isadigit{4}}{\isacharcolon}{\isachardoublequoteopen}IdenticCycleLength\ n\ nC{\isachardoublequoteclose}\isanewline
\ \ \ \ \ \isakeyword{and}\ h{\isadigit{5}}{\isacharcolon}{\isachardoublequoteopen}t\ mod\ cycleLength\ {\isacharparenleft}nC\ k{\isacharparenright}\ mem\ schedule\ {\isacharparenleft}nC\ k{\isacharparenright}{\isachardoublequoteclose}\isanewline
\ \isakeyword{shows}\ {\isachardoublequoteopen}{\isasymforall}j{\isachardot}\ j\ {\isacharless}\ n\ {\isasymand}\ j\ {\isasymnoteq}\ k\ {\isasymlongrightarrow}\ {\isacharparenleft}nSend\ j{\isacharparenright}\ t\ {\isacharequal}\ {\isacharbrackleft}{\isacharbrackright}{\isachardoublequoteclose}\isanewline
\isadelimproof
\endisadelimproof
\isatagproof
\isacommand{using}\isamarkupfalse%
\ assms\ \isacommand{by}\isamarkupfalse%
\ {\isacharparenleft}clarify{\isacharcomma}\ simp\ add{\isacharcolon}\ disjointFrame{\isacharunderscore}L{\isadigit{3}}{\isacharparenright}%
\endisatagproof
{\isafoldproof}%
\isadelimproof
\isanewline
\endisadelimproof
\isanewline
\isanewline
\isacommand{lemma}\isamarkupfalse%
\ length{\isacharunderscore}nSend{\isacharcolon}\isanewline
\ \isakeyword{assumes}\ h{\isadigit{1}}{\isacharcolon}{\isachardoublequoteopen}BusInterface\ activation\ {\isacharparenleft}nReturn\ i{\isacharparenright}\ recv\ {\isacharparenleft}nStore\ i{\isacharparenright}\ {\isacharparenleft}nSend\ i{\isacharparenright}\ {\isacharparenleft}nGet\ i{\isacharparenright}{\isachardoublequoteclose}\isanewline
\ \ \ \ \ \isakeyword{and}\ h{\isadigit{2}}{\isacharcolon}{\isachardoublequoteopen}{\isasymforall}t{\isachardot}\ length\ {\isacharparenleft}nReturn\ i\ t{\isacharparenright}\ {\isasymle}\ Suc\ {\isadigit{0}}{\isachardoublequoteclose}\isanewline
\ \isakeyword{shows}\ \ \ {\isachardoublequoteopen}length\ {\isacharparenleft}nSend\ i\ t{\isacharparenright}\ {\isasymle}\ Suc\ {\isadigit{0}}{\isachardoublequoteclose}\isanewline
\isadelimproof
\endisadelimproof
\isatagproof
\isacommand{proof}\isamarkupfalse%
\ {\isacharminus}\ \isanewline
\ \ \isacommand{from}\isamarkupfalse%
\ h{\isadigit{1}}\ \isacommand{have}\isamarkupfalse%
\ sg{\isadigit{1}}{\isacharcolon}{\isachardoublequoteopen}Send\ {\isacharparenleft}nReturn\ i{\isacharparenright}\ {\isacharparenleft}nSend\ i{\isacharparenright}\ {\isacharparenleft}nGet\ i{\isacharparenright}\ activation{\isachardoublequoteclose}\isanewline
\ \ \ \ \isacommand{by}\isamarkupfalse%
\ {\isacharparenleft}simp\ add{\isacharcolon}\ BusInterface{\isacharunderscore}def{\isacharparenright}\isanewline
\ \ \isacommand{from}\isamarkupfalse%
\ sg{\isadigit{1}}\ \isacommand{have}\isamarkupfalse%
\ sg{\isadigit{2}}{\isacharcolon}\isanewline
\ \ \ {\isachardoublequoteopen}if\ activation\ t\ {\isacharequal}\ {\isacharbrackleft}{\isacharbrackright}\ then\ nGet\ i\ t\ {\isacharequal}\ {\isacharbrackleft}{\isacharbrackright}\ {\isasymand}\ nSend\ i\ t\ {\isacharequal}\ {\isacharbrackleft}{\isacharbrackright}\isanewline
\ \ \ \ else\ nGet\ i\ t\ {\isacharequal}\ activation\ t\ {\isasymand}\ nSend\ i\ t\ {\isacharequal}\ nReturn\ i\ t{\isachardoublequoteclose}\isanewline
\ \ \ \ \isacommand{by}\isamarkupfalse%
\ {\isacharparenleft}simp\ add{\isacharcolon}\ Send{\isacharunderscore}def{\isacharparenright}\isanewline
\ \ \isacommand{show}\isamarkupfalse%
\ {\isacharquery}thesis\isanewline
\ \ \isacommand{proof}\isamarkupfalse%
\ {\isacharparenleft}cases\ {\isachardoublequoteopen}activation\ t\ {\isacharequal}\ {\isacharbrackleft}{\isacharbrackright}{\isachardoublequoteclose}{\isacharparenright}\isanewline
\ \ \ \ \isacommand{assume}\isamarkupfalse%
\ a{\isadigit{1}}{\isacharcolon}{\isachardoublequoteopen}activation\ t\ {\isacharequal}\ {\isacharbrackleft}{\isacharbrackright}{\isachardoublequoteclose}\isanewline
\ \ \ \ \isacommand{from}\isamarkupfalse%
\ sg{\isadigit{2}}\ \isakeyword{and}\ a{\isadigit{1}}\ \isacommand{show}\isamarkupfalse%
\ {\isacharquery}thesis\ \isacommand{by}\isamarkupfalse%
\ simp\isanewline
\ \ \isacommand{next}\isamarkupfalse%
\isanewline
\ \ \ \ \isacommand{assume}\isamarkupfalse%
\ a{\isadigit{2}}{\isacharcolon}{\isachardoublequoteopen}activation\ t\ {\isasymnoteq}\ {\isacharbrackleft}{\isacharbrackright}{\isachardoublequoteclose}\isanewline
\ \ \ \ \isacommand{from}\isamarkupfalse%
\ h{\isadigit{2}}\ \isacommand{have}\isamarkupfalse%
\ sg{\isadigit{3}}{\isacharcolon}{\isachardoublequoteopen}length\ {\isacharparenleft}nReturn\ i\ t{\isacharparenright}\ {\isasymle}\ Suc\ {\isadigit{0}}{\isachardoublequoteclose}\ \isacommand{by}\isamarkupfalse%
\ auto\isanewline
\ \ \ \ \isacommand{from}\isamarkupfalse%
\ sg{\isadigit{2}}\ \isakeyword{and}\ a{\isadigit{2}}\ \isakeyword{and}\ sg{\isadigit{3}}\ \isacommand{show}\isamarkupfalse%
\ {\isacharquery}thesis\ \isacommand{by}\isamarkupfalse%
\ simp\isanewline
\ \ \isacommand{qed}\isamarkupfalse%
\isanewline
\isacommand{qed}\isamarkupfalse%
\endisatagproof
{\isafoldproof}%
\isadelimproof
\isanewline
\endisadelimproof
\isanewline
\isanewline
\isacommand{lemma}\isamarkupfalse%
\ msg{\isacharunderscore}nSend{\isacharcolon}\isanewline
\ \isakeyword{assumes}\ h{\isadigit{1}}{\isacharcolon}{\isachardoublequoteopen}BusInterface\ activation\ {\isacharparenleft}nReturn\ i{\isacharparenright}\ recv\ {\isacharparenleft}nStore\ i{\isacharparenright}\ {\isacharparenleft}nSend\ i{\isacharparenright}\ {\isacharparenleft}nGet\ i{\isacharparenright}{\isachardoublequoteclose}\isanewline
\ \ \ \ \ \isakeyword{and}\ h{\isadigit{2}}{\isacharcolon}{\isachardoublequoteopen}msg\ {\isacharparenleft}Suc\ {\isadigit{0}}{\isacharparenright}\ {\isacharparenleft}nReturn\ i{\isacharparenright}{\isachardoublequoteclose}\isanewline
\ \isakeyword{shows}\ {\isachardoublequoteopen}msg\ {\isacharparenleft}Suc\ {\isadigit{0}}{\isacharparenright}\ {\isacharparenleft}nSend\ i{\isacharparenright}{\isachardoublequoteclose}\isanewline
\isadelimproof
\endisadelimproof
\isatagproof
\isacommand{using}\isamarkupfalse%
\ assms\ \isacommand{by}\isamarkupfalse%
\ {\isacharparenleft}simp\ add{\isacharcolon}\ msg{\isacharunderscore}def{\isacharcomma}\ clarify{\isacharcomma}\ simp\ add{\isacharcolon}\ length{\isacharunderscore}nSend{\isacharparenright}%
\endisatagproof
{\isafoldproof}%
\isadelimproof
\isanewline
\endisadelimproof
\isanewline
\isanewline
\isacommand{lemma}\isamarkupfalse%
\ Broadcast{\isacharunderscore}nSend{\isacharunderscore}empty{\isadigit{1}}{\isacharcolon}\isanewline
\ \ \isakeyword{assumes}\ h{\isadigit{1}}{\isacharcolon}{\isachardoublequoteopen}Broadcast\ n\ nSend\ recv{\isachardoublequoteclose}\isanewline
\ \ \ \ \ \ \isakeyword{and}\ h{\isadigit{2}}{\isacharcolon}{\isachardoublequoteopen}{\isasymforall}k{\isacharless}n{\isachardot}\ nSend\ k\ t\ {\isacharequal}\ {\isacharbrackleft}{\isacharbrackright}{\isachardoublequoteclose}\isanewline
\ \ \isakeyword{shows}\ \ \ \ \ \ {\isachardoublequoteopen}recv\ t\ {\isacharequal}\ {\isacharbrackleft}{\isacharbrackright}{\isachardoublequoteclose}\isanewline
\isadelimproof
\endisadelimproof
\isatagproof
\isacommand{proof}\isamarkupfalse%
\ {\isacharminus}\ \isanewline
\ \ \isacommand{from}\isamarkupfalse%
\ h{\isadigit{1}}\ \isacommand{have}\isamarkupfalse%
\ sg{\isadigit{1}}{\isacharcolon}\isanewline
\ \ \ {\isachardoublequoteopen}if\ {\isasymexists}k{\isacharless}n{\isachardot}\ nSend\ k\ t\ {\isasymnoteq}\ {\isacharbrackleft}{\isacharbrackright}\ \isanewline
\ \ \ \ then\ recv\ t\ {\isacharequal}\ nSend\ {\isacharparenleft}SOME\ k{\isachardot}\ k\ {\isacharless}\ n\ {\isasymand}\ nSend\ k\ t\ {\isasymnoteq}\ {\isacharbrackleft}{\isacharbrackright}{\isacharparenright}\ t\ \isanewline
\ \ \ \ else\ recv\ t\ {\isacharequal}\ {\isacharbrackleft}{\isacharbrackright}{\isachardoublequoteclose}\isanewline
\ \ \ \isacommand{by}\isamarkupfalse%
\ {\isacharparenleft}simp\ add{\isacharcolon}\ Broadcast{\isacharunderscore}def{\isacharparenright}\isanewline
\ \ \isacommand{from}\isamarkupfalse%
\ sg{\isadigit{1}}\ \isakeyword{and}\ h{\isadigit{2}}\ \isacommand{show}\isamarkupfalse%
\ {\isacharquery}thesis\ \isacommand{by}\isamarkupfalse%
\ simp\isanewline
\isacommand{qed}\isamarkupfalse%
\endisatagproof
{\isafoldproof}%
\isadelimproof
\endisadelimproof
\isamarkupsubsection{Properties of the sheaf of channels  nGet%
}
\isamarkuptrue%
\isacommand{lemma}\isamarkupfalse%
\ fr{\isacharunderscore}nGet{\isadigit{1}}a{\isacharcolon}\isanewline
\ \isakeyword{assumes}\ h{\isadigit{1}}{\isacharcolon}{\isachardoublequoteopen}FlexRayController\ {\isacharparenleft}nReturn\ k{\isacharparenright}\ recv\ {\isacharparenleft}nC\ k{\isacharparenright}\ {\isacharparenleft}nStore\ k{\isacharparenright}\ {\isacharparenleft}nSend\ k{\isacharparenright}\ {\isacharparenleft}nGet\ k{\isacharparenright}{\isachardoublequoteclose}\ \isanewline
\ \ \ \ \ \isakeyword{and}\ h{\isadigit{2}}{\isacharcolon}{\isachardoublequoteopen}t\ mod\ cycleLength\ {\isacharparenleft}nC\ k{\isacharparenright}\ mem\ schedule\ {\isacharparenleft}nC\ k{\isacharparenright}{\isachardoublequoteclose}\isanewline
\ \isakeyword{shows}\ {\isachardoublequoteopen}nGet\ k\ t\ {\isacharequal}\ {\isacharbrackleft}t\ mod\ cycleLength\ {\isacharparenleft}nC\ k{\isacharparenright}{\isacharbrackright}{\isachardoublequoteclose}\isanewline
\isadelimproof
\endisadelimproof
\isatagproof
\isacommand{proof}\isamarkupfalse%
\ {\isacharminus}\isanewline
\ \ \isacommand{from}\isamarkupfalse%
\ h{\isadigit{1}}\ \ \isacommand{obtain}\isamarkupfalse%
\ activation{\isadigit{1}}\ \isakeyword{where}\isanewline
\ \ \ \ \ a{\isadigit{1}}{\isacharcolon}{\isachardoublequoteopen}Scheduler\ {\isacharparenleft}nC\ k{\isacharparenright}\ activation{\isadigit{1}}{\isachardoublequoteclose}\ \isakeyword{and}\ \isanewline
\ \ \ \ \ a{\isadigit{2}}{\isacharcolon}{\isachardoublequoteopen}BusInterface\ activation{\isadigit{1}}\ {\isacharparenleft}nReturn\ k{\isacharparenright}\ recv\ {\isacharparenleft}nStore\ k{\isacharparenright}\ {\isacharparenleft}nSend\ k{\isacharparenright}\ {\isacharparenleft}nGet\ k{\isacharparenright}{\isachardoublequoteclose}\isanewline
\ \ \ \ \ \isacommand{by}\isamarkupfalse%
\ {\isacharparenleft}simp\ add{\isacharcolon}\ FlexRayController{\isacharunderscore}def{\isacharcomma}\ auto{\isacharparenright}\isanewline
\ \ \ \isacommand{from}\isamarkupfalse%
\ a{\isadigit{2}}\ \isacommand{have}\isamarkupfalse%
\ sg{\isadigit{1}}{\isacharcolon}{\isachardoublequoteopen}Send\ {\isacharparenleft}nReturn\ k{\isacharparenright}\ {\isacharparenleft}nSend\ k{\isacharparenright}\ {\isacharparenleft}nGet\ k{\isacharparenright}\ activation{\isadigit{1}}{\isachardoublequoteclose}\isanewline
\ \ \ \ \ \isacommand{by}\isamarkupfalse%
\ {\isacharparenleft}simp\ add{\isacharcolon}\ BusInterface{\isacharunderscore}def{\isacharparenright}\isanewline
\ \ \ \isacommand{from}\isamarkupfalse%
\ sg{\isadigit{1}}\ \isacommand{have}\isamarkupfalse%
\ sg{\isadigit{2}}{\isacharcolon}\isanewline
\ \ \ \ {\isachardoublequoteopen}if\ activation{\isadigit{1}}\ t\ {\isacharequal}\ {\isacharbrackleft}{\isacharbrackright}\ then\ nGet\ k\ t\ {\isacharequal}\ {\isacharbrackleft}{\isacharbrackright}\ {\isasymand}\ nSend\ k\ t\ {\isacharequal}\ {\isacharbrackleft}{\isacharbrackright}\isanewline
\ \ \ \ \ else\ nGet\ k\ t\ {\isacharequal}\ activation{\isadigit{1}}\ t\ {\isasymand}\ nSend\ k\ t\ {\isacharequal}\ nReturn\ k\ t{\isachardoublequoteclose}\ \isanewline
\ \ \ \ \ \isacommand{by}\isamarkupfalse%
\ {\isacharparenleft}simp\ add{\isacharcolon}\ Send{\isacharunderscore}def{\isacharparenright}\isanewline
\ \ \isacommand{from}\isamarkupfalse%
\ a{\isadigit{1}}\ \isakeyword{and}\ h{\isadigit{2}}\ \isacommand{have}\isamarkupfalse%
\ sg{\isadigit{3}}{\isacharcolon}{\isachardoublequoteopen}activation{\isadigit{1}}\ t\ {\isacharequal}\ {\isacharbrackleft}t\ mod\ cycleLength\ {\isacharparenleft}nC\ k{\isacharparenright}{\isacharbrackright}{\isachardoublequoteclose}\isanewline
\ \ \ \ \ \isacommand{by}\isamarkupfalse%
\ {\isacharparenleft}simp\ add{\isacharcolon}\ Scheduler{\isacharunderscore}L{\isadigit{4}}{\isacharparenright}\isanewline
\ \ \isacommand{from}\isamarkupfalse%
\ sg{\isadigit{2}}\ \isakeyword{and}\ sg{\isadigit{3}}\ \isacommand{show}\isamarkupfalse%
\ {\isacharquery}thesis\ \ \isacommand{by}\isamarkupfalse%
\ simp\isanewline
\isacommand{qed}\isamarkupfalse%
\endisatagproof
{\isafoldproof}%
\isadelimproof
\isanewline
\endisadelimproof
\isanewline
\isanewline
\isacommand{lemma}\isamarkupfalse%
\ fr{\isacharunderscore}nGet{\isadigit{1}}{\isacharcolon}\isanewline
\ \isakeyword{assumes}\ h{\isadigit{1}}{\isacharcolon}{\isachardoublequoteopen}{\isasymforall}i{\isacharless}n{\isachardot}\ FlexRayController\ {\isacharparenleft}nReturn\ i{\isacharparenright}\ recv\ {\isacharparenleft}nC\ i{\isacharparenright}\ {\isacharparenleft}nStore\ i{\isacharparenright}\ {\isacharparenleft}nSend\ i{\isacharparenright}\ {\isacharparenleft}nGet\ i{\isacharparenright}{\isachardoublequoteclose}\ \isanewline
\ \ \ \ \ \isakeyword{and}\ h{\isadigit{2}}{\isacharcolon}{\isachardoublequoteopen}t\ mod\ cycleLength\ {\isacharparenleft}nC\ k{\isacharparenright}\ mem\ schedule\ {\isacharparenleft}nC\ k{\isacharparenright}{\isachardoublequoteclose}\isanewline
\ \ \ \ \ \isakeyword{and}\ h{\isadigit{3}}{\isacharcolon}{\isachardoublequoteopen}k\ {\isacharless}\ n{\isachardoublequoteclose}\isanewline
\ \isakeyword{shows}\ {\isachardoublequoteopen}nGet\ k\ t\ {\isacharequal}\ {\isacharbrackleft}t\ mod\ cycleLength\ {\isacharparenleft}nC\ k{\isacharparenright}{\isacharbrackright}{\isachardoublequoteclose}\isanewline
\isadelimproof
\endisadelimproof
\isatagproof
\isacommand{proof}\isamarkupfalse%
\ {\isacharminus}\isanewline
\ \ \isacommand{from}\isamarkupfalse%
\ h{\isadigit{1}}\ \isakeyword{and}\ h{\isadigit{3}}\ \isacommand{have}\isamarkupfalse%
\ sg{\isadigit{1}}{\isacharcolon}\isanewline
\ \ \ \ {\isachardoublequoteopen}FlexRayController\ {\isacharparenleft}nReturn\ k{\isacharparenright}\ recv\ {\isacharparenleft}nC\ k{\isacharparenright}\ {\isacharparenleft}nStore\ k{\isacharparenright}\ {\isacharparenleft}nSend\ k{\isacharparenright}\ {\isacharparenleft}nGet\ k{\isacharparenright}{\isachardoublequoteclose}\isanewline
\ \ \ \ \isacommand{by}\isamarkupfalse%
\ auto\isanewline
\ \ \isacommand{from}\isamarkupfalse%
\ sg{\isadigit{1}}\ \isakeyword{and}\ h{\isadigit{2}}\ \ \isacommand{show}\isamarkupfalse%
\ {\isacharquery}thesis\ \ \isacommand{by}\isamarkupfalse%
\ {\isacharparenleft}rule\ fr{\isacharunderscore}nGet{\isadigit{1}}a{\isacharparenright}\isanewline
\isacommand{qed}\isamarkupfalse%
\endisatagproof
{\isafoldproof}%
\isadelimproof
\isanewline
\endisadelimproof
\isanewline
\isanewline
\isacommand{lemma}\isamarkupfalse%
\ fr{\isacharunderscore}nGet{\isadigit{2}}a{\isacharcolon}\isanewline
\ \isakeyword{assumes}\ h{\isadigit{1}}{\isacharcolon}{\isachardoublequoteopen}FlexRayController\ {\isacharparenleft}nReturn\ k{\isacharparenright}\ recv\ {\isacharparenleft}nC\ k{\isacharparenright}\ {\isacharparenleft}nStore\ k{\isacharparenright}\ {\isacharparenleft}nSend\ k{\isacharparenright}\ {\isacharparenleft}nGet\ k{\isacharparenright}{\isachardoublequoteclose}\ \isanewline
\ \ \ \ \ \isakeyword{and}\ h{\isadigit{2}}{\isacharcolon}{\isachardoublequoteopen}{\isasymnot}\ {\isacharparenleft}t\ mod\ cycleLength\ {\isacharparenleft}nC\ k{\isacharparenright}\ mem\ schedule\ {\isacharparenleft}nC\ k{\isacharparenright}{\isacharparenright}{\isachardoublequoteclose}\isanewline
\ \isakeyword{shows}\ {\isachardoublequoteopen}nGet\ k\ t\ {\isacharequal}\ {\isacharbrackleft}{\isacharbrackright}{\isachardoublequoteclose}\isanewline
\isadelimproof
\endisadelimproof
\isatagproof
\isacommand{proof}\isamarkupfalse%
\ {\isacharminus}\isanewline
\ \ \ \isacommand{from}\isamarkupfalse%
\ h{\isadigit{1}}\ \isacommand{obtain}\isamarkupfalse%
\ activation{\isadigit{1}}\ \isakeyword{where}\isanewline
\ \ \ \ \ a{\isadigit{1}}{\isacharcolon}{\isachardoublequoteopen}Scheduler\ {\isacharparenleft}nC\ k{\isacharparenright}\ activation{\isadigit{1}}{\isachardoublequoteclose}\ \isakeyword{and}\ \isanewline
\ \ \ \ \ a{\isadigit{2}}{\isacharcolon}{\isachardoublequoteopen}BusInterface\ activation{\isadigit{1}}\ {\isacharparenleft}nReturn\ k{\isacharparenright}\ recv\ {\isacharparenleft}nStore\ k{\isacharparenright}\ {\isacharparenleft}nSend\ k{\isacharparenright}\ {\isacharparenleft}nGet\ k{\isacharparenright}{\isachardoublequoteclose}\isanewline
\ \ \ \ \ \isacommand{by}\isamarkupfalse%
\ {\isacharparenleft}simp\ add{\isacharcolon}\ FlexRayController{\isacharunderscore}def{\isacharcomma}\ auto{\isacharparenright}\isanewline
\ \ \ \isacommand{from}\isamarkupfalse%
\ a{\isadigit{2}}\ \isacommand{have}\isamarkupfalse%
\ sg{\isadigit{2}}{\isacharcolon}{\isachardoublequoteopen}Send\ {\isacharparenleft}nReturn\ k{\isacharparenright}\ {\isacharparenleft}nSend\ k{\isacharparenright}\ {\isacharparenleft}nGet\ k{\isacharparenright}\ activation{\isadigit{1}}{\isachardoublequoteclose}\isanewline
\ \ \ \ \ \isacommand{by}\isamarkupfalse%
\ {\isacharparenleft}simp\ add{\isacharcolon}\ BusInterface{\isacharunderscore}def{\isacharparenright}\isanewline
\ \ \ \isacommand{from}\isamarkupfalse%
\ sg{\isadigit{2}}\ \isacommand{have}\isamarkupfalse%
\ sg{\isadigit{3}}{\isacharcolon}\isanewline
\ \ \ \ {\isachardoublequoteopen}if\ activation{\isadigit{1}}\ t\ {\isacharequal}\ {\isacharbrackleft}{\isacharbrackright}\ then\ nGet\ k\ t\ {\isacharequal}\ {\isacharbrackleft}{\isacharbrackright}\ {\isasymand}\ nSend\ k\ t\ {\isacharequal}\ {\isacharbrackleft}{\isacharbrackright}\isanewline
\ \ \ \ \ else\ nGet\ k\ t\ {\isacharequal}\ activation{\isadigit{1}}\ t\ {\isasymand}\ nSend\ k\ t\ {\isacharequal}\ nReturn\ k\ t{\isachardoublequoteclose}\ \isanewline
\ \ \ \ \ \isacommand{by}\isamarkupfalse%
\ {\isacharparenleft}simp\ add{\isacharcolon}\ Send{\isacharunderscore}def{\isacharparenright}\isanewline
\ \ \isacommand{from}\isamarkupfalse%
\ a{\isadigit{1}}\ \isakeyword{and}\ h{\isadigit{2}}\ \isacommand{have}\isamarkupfalse%
\ sg{\isadigit{4}}{\isacharcolon}{\isachardoublequoteopen}activation{\isadigit{1}}\ t\ {\isacharequal}\ {\isacharbrackleft}{\isacharbrackright}{\isachardoublequoteclose}\isanewline
\ \ \ \ \ \isacommand{by}\isamarkupfalse%
\ {\isacharparenleft}simp\ add{\isacharcolon}\ Scheduler{\isacharunderscore}L{\isadigit{2}}{\isacharparenright}\isanewline
\ \ \isacommand{from}\isamarkupfalse%
\ sg{\isadigit{3}}\ \isakeyword{and}\ sg{\isadigit{4}}\ \isacommand{show}\isamarkupfalse%
\ {\isacharquery}thesis\ \ \isacommand{by}\isamarkupfalse%
\ simp\isanewline
\isacommand{qed}\isamarkupfalse%
\endisatagproof
{\isafoldproof}%
\isadelimproof
\isanewline
\endisadelimproof
\isanewline
\isanewline
\isacommand{lemma}\isamarkupfalse%
\ fr{\isacharunderscore}nGet{\isadigit{2}}{\isacharcolon}\isanewline
\ \isakeyword{assumes}\ h{\isadigit{1}}{\isacharcolon}{\isachardoublequoteopen}{\isasymforall}i{\isacharless}n{\isachardot}\ FlexRayController\ {\isacharparenleft}nReturn\ i{\isacharparenright}\ recv\ {\isacharparenleft}nC\ i{\isacharparenright}\ {\isacharparenleft}nStore\ i{\isacharparenright}\ {\isacharparenleft}nSend\ i{\isacharparenright}\ {\isacharparenleft}nGet\ i{\isacharparenright}{\isachardoublequoteclose}\ \isanewline
\ \ \ \ \ \isakeyword{and}\ h{\isadigit{2}}{\isacharcolon}{\isachardoublequoteopen}{\isasymnot}\ {\isacharparenleft}t\ mod\ cycleLength\ {\isacharparenleft}nC\ k{\isacharparenright}\ mem\ schedule\ {\isacharparenleft}nC\ k{\isacharparenright}{\isacharparenright}{\isachardoublequoteclose}\isanewline
\ \ \ \ \ \isakeyword{and}\ h{\isadigit{3}}{\isacharcolon}{\isachardoublequoteopen}k\ {\isacharless}\ n{\isachardoublequoteclose}\isanewline
\ \isakeyword{shows}\ {\isachardoublequoteopen}nGet\ k\ t\ {\isacharequal}\ {\isacharbrackleft}{\isacharbrackright}{\isachardoublequoteclose}\isanewline
\isadelimproof
\endisadelimproof
\isatagproof
\isacommand{proof}\isamarkupfalse%
\ {\isacharminus}\isanewline
\ \ \isacommand{from}\isamarkupfalse%
\ h{\isadigit{1}}\ \isakeyword{and}\ h{\isadigit{3}}\ \isacommand{have}\isamarkupfalse%
\ sg{\isadigit{1}}{\isacharcolon}\isanewline
\ \ \ \ {\isachardoublequoteopen}FlexRayController\ {\isacharparenleft}nReturn\ k{\isacharparenright}\ recv\ {\isacharparenleft}nC\ k{\isacharparenright}\ {\isacharparenleft}nStore\ k{\isacharparenright}\ {\isacharparenleft}nSend\ k{\isacharparenright}\ {\isacharparenleft}nGet\ k{\isacharparenright}{\isachardoublequoteclose}\isanewline
\ \ \ \ \isacommand{by}\isamarkupfalse%
\ auto\isanewline
\ \ \isacommand{from}\isamarkupfalse%
\ sg{\isadigit{1}}\ \isakeyword{and}\ h{\isadigit{2}}\ \isacommand{show}\isamarkupfalse%
\ {\isacharquery}thesis\ \isacommand{by}\isamarkupfalse%
\ {\isacharparenleft}rule\ fr{\isacharunderscore}nGet{\isadigit{2}}a{\isacharparenright}\isanewline
\isacommand{qed}\isamarkupfalse%
\endisatagproof
{\isafoldproof}%
\isadelimproof
\isanewline
\endisadelimproof
\isanewline
\isanewline
\isacommand{lemma}\isamarkupfalse%
\ length{\isacharunderscore}nGet{\isadigit{1}}{\isacharcolon}\isanewline
\ \isakeyword{assumes}\ h{\isadigit{1}}{\isacharcolon}{\isachardoublequoteopen}FlexRayController\ {\isacharparenleft}nReturn\ k{\isacharparenright}\ recv\ {\isacharparenleft}nC\ k{\isacharparenright}\ {\isacharparenleft}nStore\ k{\isacharparenright}\ {\isacharparenleft}nSend\ k{\isacharparenright}\ {\isacharparenleft}nGet\ k{\isacharparenright}{\isachardoublequoteclose}\isanewline
\ \isakeyword{shows}\ {\isachardoublequoteopen}length\ {\isacharparenleft}nGet\ k\ t{\isacharparenright}\ {\isasymle}\ Suc\ {\isadigit{0}}{\isachardoublequoteclose}\isanewline
\isadelimproof
\endisadelimproof
\isatagproof
\isacommand{proof}\isamarkupfalse%
\ {\isacharparenleft}cases\ {\isachardoublequoteopen}t\ mod\ cycleLength\ {\isacharparenleft}nC\ k{\isacharparenright}\ mem\ schedule\ {\isacharparenleft}nC\ k{\isacharparenright}{\isachardoublequoteclose}{\isacharparenright}\isanewline
\ \ \isacommand{assume}\isamarkupfalse%
\ a{\isadigit{1}}{\isacharcolon}{\isachardoublequoteopen}t\ mod\ cycleLength\ {\isacharparenleft}nC\ k{\isacharparenright}\ mem\ schedule\ {\isacharparenleft}nC\ k{\isacharparenright}{\isachardoublequoteclose}\isanewline
\ \ \isacommand{from}\isamarkupfalse%
\ h{\isadigit{1}}\ \isakeyword{and}\ a{\isadigit{1}}\ \isacommand{have}\isamarkupfalse%
\ sg{\isadigit{1}}{\isacharcolon}{\isachardoublequoteopen}nGet\ k\ t\ {\isacharequal}\ {\isacharbrackleft}t\ mod\ cycleLength\ {\isacharparenleft}nC\ k{\isacharparenright}{\isacharbrackright}{\isachardoublequoteclose}\ \isanewline
\ \ \ \ \isacommand{by}\isamarkupfalse%
\ {\isacharparenleft}rule\ fr{\isacharunderscore}nGet{\isadigit{1}}a{\isacharparenright}\isanewline
\ \ \isacommand{from}\isamarkupfalse%
\ sg{\isadigit{1}}\ \isacommand{show}\isamarkupfalse%
\ {\isacharquery}thesis\ \isacommand{by}\isamarkupfalse%
\ auto\isanewline
\isacommand{next}\isamarkupfalse%
\isanewline
\ \ \isacommand{assume}\isamarkupfalse%
\ a{\isadigit{2}}{\isacharcolon}{\isachardoublequoteopen}{\isasymnot}\ {\isacharparenleft}t\ mod\ cycleLength\ {\isacharparenleft}nC\ k{\isacharparenright}\ mem\ schedule\ {\isacharparenleft}nC\ k{\isacharparenright}{\isacharparenright}{\isachardoublequoteclose}\isanewline
\ \ \isacommand{from}\isamarkupfalse%
\ h{\isadigit{1}}\ \isakeyword{and}\ a{\isadigit{2}}\ \isacommand{have}\isamarkupfalse%
\ sg{\isadigit{2}}{\isacharcolon}{\isachardoublequoteopen}nGet\ k\ t\ {\isacharequal}\ {\isacharbrackleft}{\isacharbrackright}{\isachardoublequoteclose}\ \isacommand{by}\isamarkupfalse%
\ {\isacharparenleft}rule\ fr{\isacharunderscore}nGet{\isadigit{2}}a{\isacharparenright}\isanewline
\ \ \isacommand{from}\isamarkupfalse%
\ sg{\isadigit{2}}\ \isacommand{show}\isamarkupfalse%
\ {\isacharquery}thesis\ \isacommand{by}\isamarkupfalse%
\ auto\isanewline
\isacommand{qed}\isamarkupfalse%
\endisatagproof
{\isafoldproof}%
\isadelimproof
\isanewline
\endisadelimproof
\isanewline
\isanewline
\isacommand{lemma}\isamarkupfalse%
\ msg{\isacharunderscore}nGet{\isadigit{1}}{\isacharcolon}\isanewline
\ \isakeyword{assumes}\ h{\isadigit{1}}{\isacharcolon}{\isachardoublequoteopen}FlexRayController\ {\isacharparenleft}nReturn\ k{\isacharparenright}\ recv\ {\isacharparenleft}nC\ k{\isacharparenright}\ {\isacharparenleft}nStore\ k{\isacharparenright}\ {\isacharparenleft}nSend\ k{\isacharparenright}\ {\isacharparenleft}nGet\ k{\isacharparenright}{\isachardoublequoteclose}\isanewline
\ \isakeyword{shows}\ {\isachardoublequoteopen}msg\ {\isacharparenleft}Suc\ {\isadigit{0}}{\isacharparenright}\ {\isacharparenleft}nGet\ k{\isacharparenright}{\isachardoublequoteclose}\isanewline
\isadelimproof
\endisadelimproof
\isatagproof
\isacommand{using}\isamarkupfalse%
\ assms\ \isacommand{by}\isamarkupfalse%
\ {\isacharparenleft}simp\ add{\isacharcolon}\ msg{\isacharunderscore}def{\isacharcomma}\ auto{\isacharcomma}\ rule\ length{\isacharunderscore}nGet{\isadigit{1}}{\isacharparenright}%
\endisatagproof
{\isafoldproof}%
\isadelimproof
\isanewline
\endisadelimproof
\isanewline
\isanewline
\isacommand{lemma}\isamarkupfalse%
\ msg{\isacharunderscore}nGet{\isadigit{2}}{\isacharcolon}\isanewline
\ \isakeyword{assumes}\ h{\isadigit{1}}{\isacharcolon}{\isachardoublequoteopen}{\isasymforall}i{\isacharless}n{\isachardot}\ FlexRayController\ {\isacharparenleft}nReturn\ i{\isacharparenright}\ recv\ {\isacharparenleft}nC\ i{\isacharparenright}\ {\isacharparenleft}nStore\ i{\isacharparenright}\ {\isacharparenleft}nSend\ i{\isacharparenright}\ {\isacharparenleft}nGet\ i{\isacharparenright}{\isachardoublequoteclose}\isanewline
\ \ \ \ \ \isakeyword{and}\ h{\isadigit{2}}{\isacharcolon}{\isachardoublequoteopen}k\ {\isacharless}\ n{\isachardoublequoteclose}\isanewline
\ \isakeyword{shows}\ {\isachardoublequoteopen}msg\ {\isacharparenleft}Suc\ {\isadigit{0}}{\isacharparenright}\ {\isacharparenleft}nGet\ k{\isacharparenright}{\isachardoublequoteclose}\isanewline
\isadelimproof
\endisadelimproof
\isatagproof
\isacommand{proof}\isamarkupfalse%
\ {\isacharminus}\isanewline
\ \ \isacommand{from}\isamarkupfalse%
\ h{\isadigit{1}}\ \isakeyword{and}\ h{\isadigit{2}}\ \isacommand{have}\isamarkupfalse%
\ sg{\isadigit{1}}{\isacharcolon}\isanewline
\ \ \ {\isachardoublequoteopen}FlexRayController\ {\isacharparenleft}nReturn\ k{\isacharparenright}\ recv\ {\isacharparenleft}nC\ k{\isacharparenright}\ {\isacharparenleft}nStore\ k{\isacharparenright}\ {\isacharparenleft}nSend\ k{\isacharparenright}\ {\isacharparenleft}nGet\ k{\isacharparenright}{\isachardoublequoteclose}\isanewline
\ \ \ \ \isacommand{by}\isamarkupfalse%
\ auto\isanewline
\ \ \isacommand{from}\isamarkupfalse%
\ sg{\isadigit{1}}\ \isacommand{show}\isamarkupfalse%
\ {\isacharquery}thesis\ \isacommand{by}\isamarkupfalse%
\ {\isacharparenleft}rule\ msg{\isacharunderscore}nGet{\isadigit{1}}{\isacharparenright}\isanewline
\isacommand{qed}\isamarkupfalse%
\endisatagproof
{\isafoldproof}%
\isadelimproof
\endisadelimproof
\isamarkupsubsection{Properties of the sheaf of channels nStore%
}
\isamarkuptrue%
\isacommand{lemma}\isamarkupfalse%
\ fr{\isacharunderscore}nStore{\isacharunderscore}nReturn{\isadigit{1}}{\isacharcolon}\isanewline
\ \isakeyword{assumes}\ h{\isadigit{0}}{\isacharcolon}{\isachardoublequoteopen}Broadcast\ n\ nSend\ recv{\isachardoublequoteclose}\isanewline
\ \ \ \isakeyword{and}\ h{\isadigit{1}}{\isacharcolon}{\isachardoublequoteopen}inf{\isacharunderscore}disj\ n\ nSend{\isachardoublequoteclose}\isanewline
\ \ \ \isakeyword{and}\ h{\isadigit{2}}{\isacharcolon}{\isachardoublequoteopen}{\isasymforall}i{\isacharless}n{\isachardot}\ FlexRayController\ {\isacharparenleft}nReturn\ i{\isacharparenright}\ recv\ {\isacharparenleft}nC\ i{\isacharparenright}\ {\isacharparenleft}nStore\ i{\isacharparenright}\ {\isacharparenleft}nSend\ i{\isacharparenright}\ {\isacharparenleft}nGet\ i{\isacharparenright}{\isachardoublequoteclose}\ \isanewline
 \ \ \ \isakeyword{and}\ h{\isadigit{3}}{\isacharcolon}{\isachardoublequoteopen}DisjointSchedules\ n\ nC{\isachardoublequoteclose}\isanewline
\ \ \ \isakeyword{and}\ h{\isadigit{4}}{\isacharcolon}{\isachardoublequoteopen}IdenticCycleLength\ n\ nC{\isachardoublequoteclose}\isanewline
\ \ \ \isakeyword{and}\ h{\isadigit{5}}{\isacharcolon}{\isachardoublequoteopen}t\ mod\ cycleLength\ {\isacharparenleft}nC\ k{\isacharparenright}\ mem\ schedule\ {\isacharparenleft}nC\ k{\isacharparenright}{\isachardoublequoteclose}\isanewline
 \ \ \ \isakeyword{and}\ h{\isadigit{6}}{\isacharcolon}{\isachardoublequoteopen}k\ {\isacharless}\ n{\isachardoublequoteclose}\isanewline
 \ \ \ \isakeyword{and}\ h{\isadigit{7}}{\isacharcolon}{\isachardoublequoteopen}j\ {\isacharless}\ n{\isachardoublequoteclose}\isanewline
\ \ \ \isakeyword{and}\ h{\isadigit{8}}{\isacharcolon}{\isachardoublequoteopen}j\ {\isasymnoteq}\ k{\isachardoublequoteclose}\isanewline
\ \isakeyword{shows}\ \ {\isachardoublequoteopen}nStore\ j\ t\ {\isacharequal}\ nReturn\ k\ t{\isachardoublequoteclose}\isanewline
\isadelimproof
\endisadelimproof
\isatagproof
\isacommand{proof}\isamarkupfalse%
\ {\isacharminus}\ \isanewline
\ \ \isacommand{from}\isamarkupfalse%
\ h{\isadigit{2}}\ \isakeyword{and}\ h{\isadigit{6}}\ \isacommand{have}\isamarkupfalse%
\ sg{\isadigit{1}}{\isacharcolon}\isanewline
\ \ \ \ {\isachardoublequoteopen}FlexRayController\ {\isacharparenleft}nReturn\ k{\isacharparenright}\ recv\ {\isacharparenleft}nC\ k{\isacharparenright}\ {\isacharparenleft}nStore\ k{\isacharparenright}\ {\isacharparenleft}nSend\ k{\isacharparenright}\ {\isacharparenleft}nGet\ k{\isacharparenright}{\isachardoublequoteclose}\isanewline
\ \ \ \ \isacommand{by}\isamarkupfalse%
\ auto\isanewline
\ \ \isacommand{from}\isamarkupfalse%
\ h{\isadigit{2}}\ \isakeyword{and}\ h{\isadigit{7}}\ \isacommand{have}\isamarkupfalse%
\ sg{\isadigit{2}}{\isacharcolon}\isanewline
\ \ \ \ {\isachardoublequoteopen}FlexRayController\ {\isacharparenleft}nReturn\ j{\isacharparenright}\ recv\ {\isacharparenleft}nC\ j{\isacharparenright}\ {\isacharparenleft}nStore\ j{\isacharparenright}\ {\isacharparenleft}nSend\ j{\isacharparenright}\ {\isacharparenleft}nGet\ j{\isacharparenright}{\isachardoublequoteclose}\isanewline
\ \ \ \ \isacommand{by}\isamarkupfalse%
\ auto\isanewline
\ \ \ \isacommand{from}\isamarkupfalse%
\ sg{\isadigit{1}}\ \isacommand{obtain}\isamarkupfalse%
\ activation{\isadigit{1}}\ \isakeyword{where}\isanewline
\ \ \ \ \ a{\isadigit{1}}{\isacharcolon}{\isachardoublequoteopen}Scheduler\ {\isacharparenleft}nC\ k{\isacharparenright}\ activation{\isadigit{1}}{\isachardoublequoteclose}\ \isakeyword{and}\ \isanewline
\ \ \ \ \ a{\isadigit{2}}{\isacharcolon}{\isachardoublequoteopen}BusInterface\ activation{\isadigit{1}}\ {\isacharparenleft}nReturn\ k{\isacharparenright}\ recv\ {\isacharparenleft}nStore\ k{\isacharparenright}\ {\isacharparenleft}nSend\ k{\isacharparenright}\ {\isacharparenleft}nGet\ k{\isacharparenright}{\isachardoublequoteclose}\isanewline
\ \ \ \ \ \isacommand{by}\isamarkupfalse%
\ {\isacharparenleft}simp\ add{\isacharcolon}\ FlexRayController{\isacharunderscore}def{\isacharcomma}\ auto{\isacharparenright}\isanewline
\ \ \isacommand{from}\isamarkupfalse%
\ sg{\isadigit{2}}\ \isacommand{obtain}\isamarkupfalse%
\ activation{\isadigit{2}}\ \isakeyword{where}\isanewline
\ \ \ \ \ a{\isadigit{3}}{\isacharcolon}{\isachardoublequoteopen}Scheduler\ {\isacharparenleft}nC\ j{\isacharparenright}\ activation{\isadigit{2}}{\isachardoublequoteclose}\ \isakeyword{and}\ \isanewline
\ \ \ \ \ a{\isadigit{4}}{\isacharcolon}{\isachardoublequoteopen}BusInterface\ activation{\isadigit{2}}\ {\isacharparenleft}nReturn\ j{\isacharparenright}\ recv\ {\isacharparenleft}nStore\ j{\isacharparenright}\ {\isacharparenleft}nSend\ j{\isacharparenright}\ {\isacharparenleft}nGet\ j{\isacharparenright}{\isachardoublequoteclose}\isanewline
\ \ \ \ \ \isacommand{by}\isamarkupfalse%
\ {\isacharparenleft}simp\ add{\isacharcolon}\ FlexRayController{\isacharunderscore}def{\isacharcomma}\ auto{\isacharparenright}\isanewline
\ \ \isacommand{from}\isamarkupfalse%
\ a{\isadigit{4}}\ \isacommand{have}\isamarkupfalse%
\ sg{\isadigit{3}}{\isacharcolon}{\isachardoublequoteopen}Receive\ recv\ {\isacharparenleft}nStore\ j{\isacharparenright}\ activation{\isadigit{2}}{\isachardoublequoteclose}\isanewline
\ \ \ \ \ \isacommand{by}\isamarkupfalse%
\ {\isacharparenleft}simp\ add{\isacharcolon}\ BusInterface{\isacharunderscore}def{\isacharparenright}\isanewline
\ \ \isacommand{from}\isamarkupfalse%
\ this\ \isacommand{have}\isamarkupfalse%
\ sg{\isadigit{4}}{\isacharcolon}\isanewline
\ \ \ {\isachardoublequoteopen}if\ activation{\isadigit{2}}\ t\ {\isacharequal}\ {\isacharbrackleft}{\isacharbrackright}\ then\ nStore\ j\ t\ {\isacharequal}\ recv\ t\ else\ nStore\ j\ t\ {\isacharequal}\ {\isacharbrackleft}{\isacharbrackright}{\isachardoublequoteclose}\isanewline
\ \ \ \ \isacommand{by}\isamarkupfalse%
\ {\isacharparenleft}simp\ add{\isacharcolon}\ Receive{\isacharunderscore}def{\isacharparenright}\isanewline
\ \ \isacommand{from}\isamarkupfalse%
\ a{\isadigit{1}}\ \isakeyword{and}\ h{\isadigit{5}}\ \isacommand{have}\isamarkupfalse%
\ sg{\isadigit{5}}{\isacharcolon}{\isachardoublequoteopen}activation{\isadigit{1}}\ t\ {\isasymnoteq}\ {\isacharbrackleft}{\isacharbrackright}{\isachardoublequoteclose}\isanewline
\ \ \ \ \ \isacommand{by}\isamarkupfalse%
\ {\isacharparenleft}simp\ add{\isacharcolon}\ Scheduler{\isacharunderscore}L{\isadigit{3}}{\isacharparenright}\isanewline
\ \isacommand{from}\isamarkupfalse%
\ h{\isadigit{4}}\ \isakeyword{and}\ h{\isadigit{6}}\ \isakeyword{and}\ h{\isadigit{7}}\ \isacommand{have}\isamarkupfalse%
\ sg{\isadigit{6}}{\isacharcolon}{\isachardoublequoteopen}cycleLength\ {\isacharparenleft}nC\ k{\isacharparenright}\ {\isacharequal}\ cycleLength\ {\isacharparenleft}nC\ j{\isacharparenright}{\isachardoublequoteclose}\isanewline
\ \ \ \ \isacommand{by}\isamarkupfalse%
\ {\isacharparenleft}simp\ only{\isacharcolon}\ IdenticCycleLength{\isacharunderscore}def{\isacharcomma}\ blast{\isacharparenright}\ \ \isanewline
\ \ \isacommand{from}\isamarkupfalse%
\ h{\isadigit{3}}\ \isakeyword{and}\ h{\isadigit{6}}\ \isakeyword{and}\ h{\isadigit{7}}\ \isakeyword{and}\ h{\isadigit{8}}\ \isacommand{have}\isamarkupfalse%
\ sg{\isadigit{7}}{\isacharcolon}{\isachardoublequoteopen}disjoint\ {\isacharparenleft}schedule\ {\isacharparenleft}nC\ k{\isacharparenright}{\isacharparenright}\ {\isacharparenleft}schedule\ {\isacharparenleft}nC\ j{\isacharparenright}{\isacharparenright}{\isachardoublequoteclose}\isanewline
\ \ \ \ \isacommand{by}\isamarkupfalse%
\ {\isacharparenleft}simp\ add{\isacharcolon}\ DisjointSchedules{\isacharunderscore}def{\isacharparenright}\ \isanewline
\ \ \isacommand{from}\isamarkupfalse%
\ sg{\isadigit{7}}\ \isakeyword{and}\ h{\isadigit{5}}\ \isacommand{have}\isamarkupfalse%
\ sg{\isadigit{8}}{\isacharcolon}  \isanewline 
 \ \ \ {\isachardoublequoteopen}{\isasymnot}\ {\isacharparenleft}t\ mod\ {\isacharparenleft}cycleLength\ {\isacharparenleft}nC\ k{\isacharparenright}{\isacharparenright}{\isacharparenright}\ mem\ {\isacharparenleft}schedule\ {\isacharparenleft}nC\ j{\isacharparenright}{\isacharparenright}{\isachardoublequoteclose}\ \isanewline
\ \ \ \ \isacommand{by}\isamarkupfalse%
\ {\isacharparenleft}simp\ add{\isacharcolon}\ mem{\isacharunderscore}notdisjoint{\isadigit{2}}{\isacharparenright}\isanewline
\ \ \isacommand{from}\isamarkupfalse%
\ sg{\isadigit{6}}\ \isakeyword{and}\ sg{\isadigit{8}}\ \isacommand{have}\isamarkupfalse%
\ sg{\isadigit{9}}{\isacharcolon}\isanewline
\ \ \ {\isachardoublequoteopen}{\isasymnot}\ {\isacharparenleft}t\ mod\ {\isacharparenleft}cycleLength\ {\isacharparenleft}nC\ j{\isacharparenright}{\isacharparenright}{\isacharparenright}\ mem\ {\isacharparenleft}schedule\ {\isacharparenleft}nC\ j{\isacharparenright}{\isacharparenright}{\isachardoublequoteclose}\ \isanewline
\ \ \ \ \isacommand{by}\isamarkupfalse%
\ simp\isanewline
\ \ \isacommand{from}\isamarkupfalse%
\ sg{\isadigit{9}}\ \isakeyword{and}\ a{\isadigit{3}}\ \isacommand{have}\isamarkupfalse%
\ sg{\isadigit{1}}{\isadigit{0}}{\isacharcolon}{\isachardoublequoteopen}activation{\isadigit{2}}\ t\ {\isacharequal}\ {\isacharbrackleft}{\isacharbrackright}{\isachardoublequoteclose}\ \isacommand{by}\isamarkupfalse%
\ {\isacharparenleft}simp\ add{\isacharcolon}\ Scheduler{\isacharunderscore}L{\isadigit{2}}{\isacharparenright}\ \isanewline
\ \ \isacommand{from}\isamarkupfalse%
\ sg{\isadigit{1}}{\isadigit{0}}\ \isakeyword{and}\ sg{\isadigit{4}}\ \isacommand{have}\isamarkupfalse%
\ sg{\isadigit{1}}{\isadigit{1}}{\isacharcolon}{\isachardoublequoteopen}nStore\ j\ t\ {\isacharequal}\ recv\ t{\isachardoublequoteclose}\ \isacommand{by}\isamarkupfalse%
\ simp\isanewline
\ \ \isacommand{from}\isamarkupfalse%
\ h{\isadigit{0}}\ \isacommand{have}\isamarkupfalse%
\ sg{\isadigit{1}}{\isadigit{5}}{\isacharcolon}\isanewline
\ \ \ {\isachardoublequoteopen}if\ {\isasymexists}k{\isacharless}n{\isachardot}\ nSend\ k\ t\ {\isasymnoteq}\ {\isacharbrackleft}{\isacharbrackright}\ \isanewline
\ \ \ \ then\ recv\ t\ {\isacharequal}\ nSend\ {\isacharparenleft}SOME\ k{\isachardot}\ k\ {\isacharless}\ n\ {\isasymand}\ nSend\ k\ t\ {\isasymnoteq}\ {\isacharbrackleft}{\isacharbrackright}{\isacharparenright}\ t\isanewline
\ \ \ \ else\ recv\ t\ {\isacharequal}\ {\isacharbrackleft}{\isacharbrackright}{\isachardoublequoteclose}\isanewline
\ \ \ \ \isacommand{by}\isamarkupfalse%
\ {\isacharparenleft}simp\ add{\isacharcolon}\ Broadcast{\isacharunderscore}def{\isacharparenright}\isanewline
\ \ \isacommand{show}\isamarkupfalse%
\ {\isacharquery}thesis\isanewline
\ \ \isacommand{proof}\isamarkupfalse%
\ {\isacharparenleft}cases\ {\isachardoublequoteopen}nReturn\ k\ t\ {\isacharequal}\ {\isacharbrackleft}{\isacharbrackright}{\isachardoublequoteclose}{\isacharparenright}\ \isanewline
\ \ \ \ \isacommand{assume}\isamarkupfalse%
\ a{\isadigit{5}}{\isacharcolon}\ {\isachardoublequoteopen}nReturn\ k\ t\ {\isacharequal}\ {\isacharbrackleft}{\isacharbrackright}{\isachardoublequoteclose}\isanewline
\ \ \ \ \isacommand{from}\isamarkupfalse%
\ h{\isadigit{2}}\ \isakeyword{and}\ h{\isadigit{3}}\ \isakeyword{and}\ h{\isadigit{4}}\ \isakeyword{and}\ h{\isadigit{5}}\ \isakeyword{and}\ h{\isadigit{6}}\ \isakeyword{and}\ a{\isadigit{5}}\ \isacommand{have}\isamarkupfalse%
\ sg{\isadigit{1}}{\isadigit{6}}{\isacharcolon}\isanewline
\ \ \ \ {\isachardoublequoteopen}{\isasymnot}\ {\isacharparenleft}{\isasymexists}k{\isacharless}n{\isachardot}\ nSend\ k\ t\ {\isasymnoteq}\ {\isacharbrackleft}{\isacharbrackright}{\isacharparenright}{\isachardoublequoteclose}\isanewline
\ \ \ \ \ \ \isacommand{by}\isamarkupfalse%
\ {\isacharparenleft}simp\ add{\isacharcolon}\ fr{\isacharunderscore}Send{\isadigit{8}}{\isacharparenright}\isanewline
\ \ \ \ \isacommand{from}\isamarkupfalse%
\ sg{\isadigit{1}}{\isadigit{6}}\ \isakeyword{and}\ sg{\isadigit{1}}{\isadigit{5}}\ \isacommand{have}\isamarkupfalse%
\ sg{\isadigit{1}}{\isadigit{7}}{\isacharcolon}{\isachardoublequoteopen}recv\ t\ {\isacharequal}\ {\isacharbrackleft}{\isacharbrackright}{\isachardoublequoteclose}\ \isacommand{by}\isamarkupfalse%
\ simp\isanewline
\ \ \ \ \isacommand{from}\isamarkupfalse%
\ sg{\isadigit{1}}{\isadigit{1}}\ \isakeyword{and}\ sg{\isadigit{1}}{\isadigit{7}}\ \isacommand{have}\isamarkupfalse%
\ sg{\isadigit{1}}{\isadigit{8}}{\isacharcolon}{\isachardoublequoteopen}nStore\ j\ t\ {\isacharequal}\ {\isacharbrackleft}{\isacharbrackright}{\isachardoublequoteclose}\ \isacommand{by}\isamarkupfalse%
\ simp\isanewline
\ \ \ \ \isacommand{from}\isamarkupfalse%
\ this\ \isakeyword{and}\ a{\isadigit{5}}\ \isacommand{show}\isamarkupfalse%
\ {\isacharquery}thesis\ \isacommand{by}\isamarkupfalse%
\ simp\isanewline
\ \ \isacommand{next}\isamarkupfalse%
\isanewline
\ \ \ \ \isacommand{assume}\isamarkupfalse%
\ a{\isadigit{6}}{\isacharcolon}{\isachardoublequoteopen}nReturn\ k\ t\ {\isasymnoteq}\ {\isacharbrackleft}{\isacharbrackright}{\isachardoublequoteclose}\isanewline
\ \ \ \ \isacommand{from}\isamarkupfalse%
\ h{\isadigit{2}}\ \isakeyword{and}\ h{\isadigit{3}}\ \isakeyword{and}\ h{\isadigit{4}}\ \isakeyword{and}\ h{\isadigit{5}}\ \isakeyword{and}\ h{\isadigit{6}}\ \isakeyword{and}\ a{\isadigit{6}}\ \isacommand{have}\isamarkupfalse%
\ sg{\isadigit{1}}{\isadigit{9}}{\isacharcolon}\isanewline \ \ \ \ 
{\isachardoublequoteopen}{\isasymexists}k{\isacharless}n{\isachardot}\ nSend\ k\ t\ {\isasymnoteq}\ {\isacharbrackleft}{\isacharbrackright}{\isachardoublequoteclose}\isanewline
\ \ \ \ \ \ \isacommand{by}\isamarkupfalse%
\ {\isacharparenleft}simp\ add{\isacharcolon}\ fr{\isacharunderscore}Send{\isadigit{6}}{\isacharparenright}\isanewline
\ \ \ \ \isacommand{from}\isamarkupfalse%
\ h{\isadigit{2}}\ \isakeyword{and}\ h{\isadigit{3}}\ \isakeyword{and}\ h{\isadigit{4}}\ \isakeyword{and}\ h{\isadigit{5}}\ \isakeyword{and}\ h{\isadigit{6}}\ \isakeyword{and}\ a{\isadigit{6}}\ \isacommand{have}\isamarkupfalse%
\ sg{\isadigit{2}}{\isadigit{0}}{\isacharcolon}{\isachardoublequoteopen}nSend\ k\ t\ {\isasymnoteq}\ {\isacharbrackleft}{\isacharbrackright}{\isachardoublequoteclose}\isanewline
\ \ \ \ \ \ \isacommand{by}\isamarkupfalse%
\ {\isacharparenleft}simp\ add{\isacharcolon}\ fr{\isacharunderscore}Send{\isadigit{3}}{\isacharparenright}\isanewline
\ \ \ \ \isacommand{from}\isamarkupfalse%
\ h{\isadigit{1}}\ \isakeyword{and}\ sg{\isadigit{2}}{\isadigit{0}}\ \isakeyword{and}\ h{\isadigit{6}}\ \isacommand{have}\isamarkupfalse%
\ sg{\isadigit{2}}{\isadigit{1}}{\isacharcolon}{\isachardoublequoteopen}{\isacharparenleft}SOME\ k{\isachardot}\ k\ {\isacharless}\ n\ {\isasymand}\ nSend\ k\ t\ {\isasymnoteq}\ {\isacharbrackleft}{\isacharbrackright}{\isacharparenright}\ {\isacharequal}\ k{\isachardoublequoteclose}\isanewline
\ \ \ \ \ \ \isacommand{by}\isamarkupfalse%
\ {\isacharparenleft}simp\ add{\isacharcolon}\ inf{\isacharunderscore}disj{\isacharunderscore}index{\isacharparenright}\isanewline
\ \ \ \ \isacommand{from}\isamarkupfalse%
\ sg{\isadigit{1}}{\isadigit{5}}\ \isakeyword{and}\ sg{\isadigit{1}}{\isadigit{9}}\ \isacommand{have}\isamarkupfalse%
\ sg{\isadigit{2}}{\isadigit{2}}{\isacharcolon}\isanewline \ \ \ \ {\isachardoublequoteopen}recv\ t\ {\isacharequal}\ nSend\ {\isacharparenleft}SOME\ k{\isachardot}\ k\ {\isacharless}\ n\ {\isasymand}\ nSend\ k\ t\ {\isasymnoteq}\ {\isacharbrackleft}{\isacharbrackright}{\isacharparenright}\ t{\isachardoublequoteclose}\ \isanewline
\ \ \ \ \ \ \isacommand{by}\isamarkupfalse%
\ simp\isanewline
\ \ \ \ \isacommand{from}\isamarkupfalse%
\ sg{\isadigit{2}}{\isadigit{2}}\ \isakeyword{and}\ sg{\isadigit{2}}{\isadigit{1}}\ \isacommand{have}\isamarkupfalse%
\ sg{\isadigit{2}}{\isadigit{3}}{\isacharcolon}{\isachardoublequoteopen}recv\ t\ {\isacharequal}\ nSend\ k\ t{\isachardoublequoteclose}\ \isacommand{by}\isamarkupfalse%
\ simp\isanewline
\ \ \ \ \isacommand{from}\isamarkupfalse%
\ h{\isadigit{2}}\ \isakeyword{and}\ h{\isadigit{3}}\ \isakeyword{and}\ h{\isadigit{4}}\ \isakeyword{and}\ h{\isadigit{5}}\ \isakeyword{and}\ h{\isadigit{6}}\ \isacommand{have}\isamarkupfalse%
\ sg{\isadigit{2}}{\isadigit{4}}{\isacharcolon}{\isachardoublequoteopen}nSend\ k\ t\ {\isacharequal}\ \ nReturn\ k\ t{\isachardoublequoteclose}\isanewline
\ \ \ \ \ \ \isacommand{by}\isamarkupfalse%
\ {\isacharparenleft}simp\ add{\isacharcolon}\ fr{\isacharunderscore}Send{\isadigit{2}}{\isacharparenright}\isanewline
\ \ \ \ \isacommand{from}\isamarkupfalse%
\ sg{\isadigit{1}}{\isadigit{1}}\ \isakeyword{and}\ sg{\isadigit{2}}{\isadigit{3}}\ \isakeyword{and}\ sg{\isadigit{2}}{\isadigit{4}}\ \isacommand{show}\isamarkupfalse%
\ {\isacharquery}thesis\ \isacommand{by}\isamarkupfalse%
\ simp\isanewline
\ \ \isacommand{qed}\isamarkupfalse%
\isanewline
\isacommand{qed}\isamarkupfalse%
\endisatagproof
{\isafoldproof}%
\isadelimproof
\isanewline
\endisadelimproof
\isanewline
\isacommand{lemma}\isamarkupfalse%
\ fr{\isacharunderscore}nStore{\isacharunderscore}nReturn{\isadigit{2}}{\isacharcolon}\isanewline
\ \isakeyword{assumes}\ h{\isadigit{1}}{\isacharcolon}{\isachardoublequoteopen}Cable\ n\ nSend\ recv{\isachardoublequoteclose}\isanewline
\ \ \ \ \ \isakeyword{and}\ h{\isadigit{2}}{\isacharcolon}{\isachardoublequoteopen}{\isasymforall}i{\isacharless}n{\isachardot}\ FlexRayController\ {\isacharparenleft}nReturn\ i{\isacharparenright}\ recv\ {\isacharparenleft}nC\ i{\isacharparenright}\ {\isacharparenleft}nStore\ i{\isacharparenright}\ {\isacharparenleft}nSend\ i{\isacharparenright}\ {\isacharparenleft}nGet\ i{\isacharparenright}{\isachardoublequoteclose}\ \isanewline
\ \ \ \ \ \isakeyword{and}\ h{\isadigit{3}}{\isacharcolon}{\isachardoublequoteopen}DisjointSchedules\ n\ nC{\isachardoublequoteclose}\isanewline
\ \ \ \ \ \isakeyword{and}\ h{\isadigit{4}}{\isacharcolon}{\isachardoublequoteopen}IdenticCycleLength\ n\ nC{\isachardoublequoteclose}\isanewline
\ \ \ \ \ \isakeyword{and}\ h{\isadigit{5}}{\isacharcolon}{\isachardoublequoteopen}t\ mod\ cycleLength\ {\isacharparenleft}nC\ k{\isacharparenright}\ mem\ schedule\ {\isacharparenleft}nC\ k{\isacharparenright}{\isachardoublequoteclose}\isanewline
\ \ \ \ \ \isakeyword{and}\ h{\isadigit{6}}{\isacharcolon}{\isachardoublequoteopen}k\ {\isacharless}\ n{\isachardoublequoteclose}\isanewline
\ \ \ \ \ \isakeyword{and}\ h{\isadigit{7}}{\isacharcolon}{\isachardoublequoteopen}j\ {\isacharless}\ n{\isachardoublequoteclose}\isanewline
\ \ \ \ \ \isakeyword{and}\ h{\isadigit{8}}{\isacharcolon}{\isachardoublequoteopen}j\ {\isasymnoteq}\ k{\isachardoublequoteclose}\isanewline
\ \isakeyword{shows}\ \ {\isachardoublequoteopen}nStore\ j\ t\ {\isacharequal}\ nReturn\ k\ t{\isachardoublequoteclose}\isanewline
\isadelimproof
\endisadelimproof
\isatagproof
\isacommand{proof}\isamarkupfalse%
\ {\isacharminus}\ \isanewline
\ \ \isacommand{from}\isamarkupfalse%
\ h{\isadigit{1}}\ \isacommand{have}\isamarkupfalse%
\ sg{\isadigit{1}}{\isacharcolon}{\isachardoublequoteopen}inf{\isacharunderscore}disj\ n\ nSend\ {\isasymlongrightarrow}\ Broadcast\ n\ nSend\ recv{\isachardoublequoteclose}\isanewline
\ \ \ \ \isacommand{by}\isamarkupfalse%
\ {\isacharparenleft}simp\ add{\isacharcolon}\ Cable{\isacharunderscore}def{\isacharparenright}\ \isanewline
\ \ \isacommand{from}\isamarkupfalse%
\ \ h{\isadigit{3}}\ \isakeyword{and}\ h{\isadigit{4}}\ \isakeyword{and}\ h{\isadigit{2}}\ \isacommand{have}\isamarkupfalse%
\ sg{\isadigit{2}}{\isacharcolon}{\isachardoublequoteopen}inf{\isacharunderscore}disj\ n\ nSend{\isachardoublequoteclose}\ \isanewline
\ \ \ \ \isacommand{by}\isamarkupfalse%
\ {\isacharparenleft}simp\ add{\isacharcolon}\ disjointFrame{\isacharunderscore}L{\isadigit{2}}{\isacharparenright}\isanewline
\ \ \isacommand{from}\isamarkupfalse%
\ sg{\isadigit{1}}\ \isakeyword{and}\ sg{\isadigit{2}}\ \isacommand{have}\isamarkupfalse%
\ sg{\isadigit{3}}{\isacharcolon}{\isachardoublequoteopen}Broadcast\ n\ nSend\ recv{\isachardoublequoteclose}\ \isacommand{by}\isamarkupfalse%
\ simp\isanewline
\ \ \isacommand{from}\isamarkupfalse%
\ sg{\isadigit{3}}\ \isakeyword{and}\ sg{\isadigit{2}}\ \isakeyword{and}\ assms\ \isacommand{show}\isamarkupfalse%
\ {\isacharquery}thesis\ \ \isacommand{by}\isamarkupfalse%
\ {\isacharparenleft}simp\ add{\isacharcolon}\ fr{\isacharunderscore}nStore{\isacharunderscore}nReturn{\isadigit{1}}{\isacharparenright}\isanewline
\isacommand{qed}\isamarkupfalse%
\endisatagproof
{\isafoldproof}%
\isadelimproof
\isanewline
\endisadelimproof
\isanewline
\isacommand{lemma}\isamarkupfalse%
\ fr{\isacharunderscore}nStore{\isacharunderscore}empty{\isadigit{1}}{\isacharcolon}\isanewline
\ \isakeyword{assumes}\ h{\isadigit{1}}{\isacharcolon}{\isachardoublequoteopen}Cable\ n\ nSend\ recv{\isachardoublequoteclose}\isanewline
\ \ \ \ \ \isakeyword{and}\ h{\isadigit{2}}{\isacharcolon}{\isachardoublequoteopen}{\isasymforall}i{\isacharless}n{\isachardot}\ FlexRayController\ {\isacharparenleft}nReturn\ i{\isacharparenright}\ recv\ {\isacharparenleft}nC\ i{\isacharparenright}\ {\isacharparenleft}nStore\ i{\isacharparenright}\ {\isacharparenleft}nSend\ i{\isacharparenright}\ {\isacharparenleft}nGet\ i{\isacharparenright}{\isachardoublequoteclose}\ \isanewline
\ \ \ \ \ \isakeyword{and}\ h{\isadigit{3}}{\isacharcolon}{\isachardoublequoteopen}DisjointSchedules\ n\ nC{\isachardoublequoteclose}\isanewline
\ \ \ \ \ \isakeyword{and}\ h{\isadigit{4}}{\isacharcolon}{\isachardoublequoteopen}IdenticCycleLength\ n\ nC{\isachardoublequoteclose}\isanewline
\ \ \ \ \ \isakeyword{and}\ h{\isadigit{5}}{\isacharcolon}{\isachardoublequoteopen}{\isacharparenleft}t\ mod\ cycleLength\ {\isacharparenleft}nC\ k{\isacharparenright}\ mem\ schedule\ {\isacharparenleft}nC\ k{\isacharparenright}{\isacharparenright}{\isachardoublequoteclose}\isanewline
\ \ \ \ \ \isakeyword{and}\ h{\isadigit{6}}{\isacharcolon}{\isachardoublequoteopen}k\ {\isacharless}\ n{\isachardoublequoteclose}\isanewline
\ \isakeyword{shows}\ \ {\isachardoublequoteopen}nStore\ k\ t\ {\isacharequal}\ {\isacharbrackleft}{\isacharbrackright}{\isachardoublequoteclose}\isanewline
\isadelimproof
\endisadelimproof
\isatagproof
\isacommand{proof}\isamarkupfalse%
\ {\isacharminus}\ \isanewline
\ \ \isacommand{from}\isamarkupfalse%
\ h{\isadigit{2}}\ \isakeyword{and}\ h{\isadigit{6}}\ \isacommand{have}\isamarkupfalse%
\ sg{\isadigit{1}}{\isacharcolon}\isanewline
\ \ \ \ {\isachardoublequoteopen}FlexRayController\ {\isacharparenleft}nReturn\ k{\isacharparenright}\ recv\ {\isacharparenleft}nC\ k{\isacharparenright}\ {\isacharparenleft}nStore\ k{\isacharparenright}\ {\isacharparenleft}nSend\ k{\isacharparenright}\ {\isacharparenleft}nGet\ k{\isacharparenright}{\isachardoublequoteclose}\isanewline
\ \ \ \ \isacommand{by}\isamarkupfalse%
\ auto\isanewline
\ \ \ \isacommand{from}\isamarkupfalse%
\ sg{\isadigit{1}}\ \isacommand{obtain}\isamarkupfalse%
\ activation{\isadigit{1}}\ \isakeyword{where}\isanewline
\ \ \ \ \ a{\isadigit{1}}{\isacharcolon}{\isachardoublequoteopen}Scheduler\ {\isacharparenleft}nC\ k{\isacharparenright}\ activation{\isadigit{1}}{\isachardoublequoteclose}\ \isakeyword{and}\ \isanewline
\ \ \ \ \ a{\isadigit{2}}{\isacharcolon}{\isachardoublequoteopen}BusInterface\ activation{\isadigit{1}}\ {\isacharparenleft}nReturn\ k{\isacharparenright}\ recv\ {\isacharparenleft}nStore\ k{\isacharparenright}\ {\isacharparenleft}nSend\ k{\isacharparenright}\ {\isacharparenleft}nGet\ k{\isacharparenright}{\isachardoublequoteclose}\isanewline
\ \ \ \ \ \isacommand{by}\isamarkupfalse%
\ {\isacharparenleft}simp\ add{\isacharcolon}\ FlexRayController{\isacharunderscore}def{\isacharcomma}\ auto{\isacharparenright}\isanewline
\ \ \isacommand{from}\isamarkupfalse%
\ a{\isadigit{2}}\ \isacommand{have}\isamarkupfalse%
\ sg{\isadigit{2}}{\isacharcolon}{\isachardoublequoteopen}Receive\ recv\ {\isacharparenleft}nStore\ k{\isacharparenright}\ activation{\isadigit{1}}{\isachardoublequoteclose}\isanewline
\ \ \ \ \ \isacommand{by}\isamarkupfalse%
\ {\isacharparenleft}simp\ add{\isacharcolon}\ BusInterface{\isacharunderscore}def{\isacharparenright}\isanewline
\ \ \isacommand{from}\isamarkupfalse%
\ this\ \isacommand{have}\isamarkupfalse%
\ sg{\isadigit{3}}{\isacharcolon}\isanewline
\ \ \ {\isachardoublequoteopen}if\ activation{\isadigit{1}}\ t\ {\isacharequal}\ {\isacharbrackleft}{\isacharbrackright}\ then\ nStore\ k\ t\ {\isacharequal}\ recv\ t\ else\ nStore\ k\ t\ {\isacharequal}\ {\isacharbrackleft}{\isacharbrackright}{\isachardoublequoteclose}\isanewline
\ \ \ \ \isacommand{by}\isamarkupfalse%
\ {\isacharparenleft}simp\ add{\isacharcolon}\ Receive{\isacharunderscore}def{\isacharparenright}\isanewline
\ \ \isacommand{from}\isamarkupfalse%
\ a{\isadigit{1}}\ \isakeyword{and}\ h{\isadigit{5}}\ \isacommand{have}\isamarkupfalse%
\ sg{\isadigit{4}}{\isacharcolon}{\isachardoublequoteopen}activation{\isadigit{1}}\ t\ {\isasymnoteq}\ {\isacharbrackleft}{\isacharbrackright}{\isachardoublequoteclose}\isanewline
\ \ \ \ \ \isacommand{by}\isamarkupfalse%
\ {\isacharparenleft}simp\ add{\isacharcolon}\ Scheduler{\isacharunderscore}L{\isadigit{3}}{\isacharparenright}\isanewline
\ \isacommand{from}\isamarkupfalse%
\ sg{\isadigit{3}}\ \isakeyword{and}\ sg{\isadigit{4}}\ \isacommand{show}\isamarkupfalse%
\ {\isacharquery}thesis\ \isacommand{by}\isamarkupfalse%
\ simp\isanewline
\isacommand{qed}\isamarkupfalse%
\endisatagproof
{\isafoldproof}%
\isadelimproof
\isanewline
\endisadelimproof
\isanewline
\isanewline
\isacommand{lemma}\isamarkupfalse%
\ fr{\isacharunderscore}nStore{\isacharunderscore}nReturn{\isadigit{3}}{\isacharcolon}\isanewline
\ \isakeyword{assumes}\ h{\isadigit{1}}{\isacharcolon}{\isachardoublequoteopen}Cable\ n\ nSend\ recv{\isachardoublequoteclose}\isanewline
\ \ \ \ \ \isakeyword{and}\ h{\isadigit{2}}{\isacharcolon}{\isachardoublequoteopen}{\isasymforall}i{\isacharless}n{\isachardot}\ FlexRayController\ {\isacharparenleft}nReturn\ i{\isacharparenright}\ recv\ {\isacharparenleft}nC\ i{\isacharparenright}\ {\isacharparenleft}nStore\ i{\isacharparenright}\ {\isacharparenleft}nSend\ i{\isacharparenright}\ {\isacharparenleft}nGet\ i{\isacharparenright}{\isachardoublequoteclose}\ \isanewline
\ \ \ \ \ \isakeyword{and}\ h{\isadigit{3}}{\isacharcolon}{\isachardoublequoteopen}DisjointSchedules\ n\ nC{\isachardoublequoteclose}\isanewline
\ \ \ \ \ \isakeyword{and}\ h{\isadigit{4}}{\isacharcolon}{\isachardoublequoteopen}IdenticCycleLength\ n\ nC{\isachardoublequoteclose}\isanewline
\ \ \ \ \ \isakeyword{and}\ h{\isadigit{5}}{\isacharcolon}{\isachardoublequoteopen}t\ mod\ cycleLength\ {\isacharparenleft}nC\ k{\isacharparenright}\ mem\ schedule\ {\isacharparenleft}nC\ k{\isacharparenright}{\isachardoublequoteclose}\isanewline
\ \ \ \ \ \isakeyword{and}\ h{\isadigit{6}}{\isacharcolon}{\isachardoublequoteopen}k\ {\isacharless}\ n{\isachardoublequoteclose}\isanewline
\ \isakeyword{shows}\ \ {\isachardoublequoteopen}{\isasymforall}j{\isachardot}\ j\ {\isacharless}\ n\ {\isasymand}\ j\ {\isasymnoteq}\ k\ {\isasymlongrightarrow}\ nStore\ j\ t\ {\isacharequal}\ nReturn\ k\ t{\isachardoublequoteclose}\isanewline
\isadelimproof
\endisadelimproof
\isatagproof
\isacommand{using}\isamarkupfalse%
\ assms\isanewline
\ \ \isacommand{by}\isamarkupfalse%
\ {\isacharparenleft}clarify{\isacharcomma}\ simp\ add{\isacharcolon}\ fr{\isacharunderscore}nStore{\isacharunderscore}nReturn{\isadigit{2}}{\isacharparenright}%
\endisatagproof
{\isafoldproof}%
\isadelimproof
\isanewline
\endisadelimproof
\isanewline
\isanewline
\isacommand{lemma}\isamarkupfalse%
\ length{\isacharunderscore}nStore{\isacharcolon}\isanewline
\ \isakeyword{assumes}\ h{\isadigit{1}}{\isacharcolon}{\isachardoublequoteopen}{\isasymforall}i{\isacharless}n{\isachardot}\ FlexRayController\ {\isacharparenleft}nReturn\ i{\isacharparenright}\ recv\ {\isacharparenleft}nC\ i{\isacharparenright}\ {\isacharparenleft}nStore\ i{\isacharparenright}\ {\isacharparenleft}nSend\ i{\isacharparenright}\ {\isacharparenleft}nGet\ i{\isacharparenright}{\isachardoublequoteclose}\isanewline
\ \ \ \ \ \isakeyword{and}\ h{\isadigit{2}}{\isacharcolon}{\isachardoublequoteopen}DisjointSchedules\ n\ nC{\isachardoublequoteclose}\isanewline
\ \ \ \ \ \isakeyword{and}\ h{\isadigit{3}}{\isacharcolon}{\isachardoublequoteopen}IdenticCycleLength\ n\ nC{\isachardoublequoteclose}\isanewline
\ \ \ \ \ \isakeyword{and}\ h{\isadigit{4}}{\isacharcolon}{\isachardoublequoteopen}inf{\isacharunderscore}disj\ n\ nSend{\isachardoublequoteclose}\isanewline
\ \ \ \ \ \isakeyword{and}\ h{\isadigit{5}}{\isacharcolon}{\isachardoublequoteopen}i\ {\isacharless}\ n{\isachardoublequoteclose}\ \ \isanewline
\ \ \ \ \ \isakeyword{and}\ h{\isadigit{6}}{\isacharcolon}{\isachardoublequoteopen}{\isasymforall}\ i{\isacharless}n{\isachardot}\ msg\ {\isacharparenleft}Suc\ {\isadigit{0}}{\isacharparenright}\ {\isacharparenleft}nReturn\ i{\isacharparenright}{\isachardoublequoteclose}\isanewline
\ \ \ \ \ \isakeyword{and}\ h{\isadigit{7}}{\isacharcolon}{\isachardoublequoteopen}Broadcast\ n\ nSend\ recv{\isachardoublequoteclose}\isanewline
\ \isakeyword{shows}\ {\isachardoublequoteopen}length\ {\isacharparenleft}nStore\ i\ t{\isacharparenright}\ {\isasymle}\ Suc\ {\isadigit{0}}{\isachardoublequoteclose}\isanewline
\isadelimproof
\endisadelimproof
\isatagproof
\isacommand{proof}\isamarkupfalse%
\ {\isacharminus}\isanewline
\ \ \isacommand{from}\isamarkupfalse%
\ h{\isadigit{7}}\ \isacommand{have}\isamarkupfalse%
\ sg{\isadigit{1}}{\isacharcolon}\isanewline
\ \ \ {\isachardoublequoteopen}if\ {\isasymexists}k{\isacharless}n{\isachardot}\ nSend\ k\ t\ {\isasymnoteq}\ {\isacharbrackleft}{\isacharbrackright}\ \isanewline
\ \ \ \ then\ recv\ t\ {\isacharequal}\ nSend\ {\isacharparenleft}SOME\ k{\isachardot}\ k\ {\isacharless}\ n\ {\isasymand}\ nSend\ k\ t\ {\isasymnoteq}\ {\isacharbrackleft}{\isacharbrackright}{\isacharparenright}\ t\ \isanewline
\ \ \ \ else\ recv\ t\ {\isacharequal}\ {\isacharbrackleft}{\isacharbrackright}{\isachardoublequoteclose}\isanewline
\ \ \ \isacommand{by}\isamarkupfalse%
\ {\isacharparenleft}simp\ add{\isacharcolon}\ Broadcast{\isacharunderscore}def{\isacharparenright}\isanewline
\ \ \isacommand{show}\isamarkupfalse%
\ {\isacharquery}thesis\ \isanewline
\ \ \isacommand{proof}\isamarkupfalse%
\ {\isacharparenleft}cases\ {\isachardoublequoteopen}{\isasymexists}k{\isacharless}n{\isachardot}\ nSend\ k\ t\ {\isasymnoteq}\ {\isacharbrackleft}{\isacharbrackright}{\isachardoublequoteclose}{\isacharparenright}\isanewline
\ \ \ \ \isacommand{assume}\isamarkupfalse%
\ a{\isadigit{1}}{\isacharcolon}{\isachardoublequoteopen}{\isasymexists}k{\isacharless}n{\isachardot}\ nSend\ k\ t\ {\isasymnoteq}\ {\isacharbrackleft}{\isacharbrackright}{\isachardoublequoteclose}\isanewline
\ \ \ \ \isacommand{from}\isamarkupfalse%
\ a{\isadigit{1}}\ \isacommand{obtain}\isamarkupfalse%
\ k\ \isakeyword{where}\ a{\isadigit{2}}{\isacharcolon}{\isachardoublequoteopen}k{\isacharless}n{\isachardoublequoteclose}\ \isakeyword{and}\ a{\isadigit{3}}{\isacharcolon}{\isachardoublequoteopen}nSend\ k\ t\ {\isasymnoteq}\ {\isacharbrackleft}{\isacharbrackright}{\isachardoublequoteclose}\ \ \isacommand{by}\isamarkupfalse%
\ auto\isanewline
\ \ \ \ \isacommand{from}\isamarkupfalse%
\ h{\isadigit{1}}\ \isakeyword{and}\ a{\isadigit{2}}\ \isacommand{have}\isamarkupfalse%
\ sg{\isadigit{4}}{\isacharcolon}\isanewline
\ \ \ \ \ \ {\isachardoublequoteopen}FlexRayController\ {\isacharparenleft}nReturn\ k{\isacharparenright}\ recv\ {\isacharparenleft}nC\ k{\isacharparenright}\ {\isacharparenleft}nStore\ k{\isacharparenright}\ {\isacharparenleft}nSend\ k{\isacharparenright}\ {\isacharparenleft}nGet\ k{\isacharparenright}{\isachardoublequoteclose}\isanewline
\ \ \ \ \ \ \isacommand{by}\isamarkupfalse%
\ auto\isanewline
\ \ \ \ \isacommand{from}\isamarkupfalse%
\ sg{\isadigit{4}}\ \isacommand{obtain}\isamarkupfalse%
\ activation{\isadigit{1}}\ \isakeyword{where}\isanewline
\ \ \ \ \ \ a{\isadigit{4}}{\isacharcolon}{\isachardoublequoteopen}Scheduler\ {\isacharparenleft}nC\ k{\isacharparenright}\ activation{\isadigit{1}}{\isachardoublequoteclose}\ \isakeyword{and}\ \isanewline
\ \ \ \ \ \ a{\isadigit{5}}{\isacharcolon}{\isachardoublequoteopen}BusInterface\ activation{\isadigit{1}}\ {\isacharparenleft}nReturn\ k{\isacharparenright}\ recv\ {\isacharparenleft}nStore\ k{\isacharparenright}\ {\isacharparenleft}nSend\ k{\isacharparenright}\ {\isacharparenleft}nGet\ k{\isacharparenright}{\isachardoublequoteclose}\ \isanewline
\ \ \ \ \ \ \isacommand{by}\isamarkupfalse%
\ {\isacharparenleft}simp\ add{\isacharcolon}\ FlexRayController{\isacharunderscore}def{\isacharcomma}\ auto{\isacharparenright}\isanewline
\ \ \ \ \isacommand{from}\isamarkupfalse%
\ a{\isadigit{5}}\ \isacommand{have}\isamarkupfalse%
\ sg{\isadigit{5}}{\isacharcolon}{\isachardoublequoteopen}Send\ {\isacharparenleft}nReturn\ k{\isacharparenright}\ {\isacharparenleft}nSend\ k{\isacharparenright}\ {\isacharparenleft}nGet\ k{\isacharparenright}\ activation{\isadigit{1}}{\isachardoublequoteclose}\ \isanewline
\ \ \ \ \ \ \isacommand{by}\isamarkupfalse%
\ {\isacharparenleft}simp\ add{\isacharcolon}\ \ BusInterface{\isacharunderscore}def{\isacharparenright}\isanewline
\ \ \ \ \isacommand{from}\isamarkupfalse%
\ a{\isadigit{5}}\ \isacommand{have}\isamarkupfalse%
\ sg{\isadigit{6}}{\isacharcolon}{\isachardoublequoteopen}Receive\ recv\ {\isacharparenleft}nStore\ k{\isacharparenright}\ activation{\isadigit{1}}{\isachardoublequoteclose}\isanewline
\ \ \ \ \ \ \isacommand{by}\isamarkupfalse%
\ {\isacharparenleft}simp\ add{\isacharcolon}\ BusInterface{\isacharunderscore}def{\isacharparenright}\isanewline
\ \ \ \ \isacommand{from}\isamarkupfalse%
\ sg{\isadigit{5}}\ \isakeyword{and}\ a{\isadigit{3}}\ \isacommand{have}\isamarkupfalse%
\ sg{\isadigit{7}}{\isacharcolon}{\isachardoublequoteopen}{\isacharparenleft}activation{\isadigit{1}}\ t{\isacharparenright}\ {\isasymnoteq}\ {\isacharbrackleft}{\isacharbrackright}{\isachardoublequoteclose}\ \isacommand{by}\isamarkupfalse%
\ {\isacharparenleft}simp\ add{\isacharcolon}\ Send{\isacharunderscore}L{\isadigit{1}}{\isacharparenright}\isanewline
\ \ \ \ \isacommand{from}\isamarkupfalse%
\ sg{\isadigit{6}}\ \isacommand{have}\isamarkupfalse%
\ sg{\isadigit{8}}{\isacharcolon}\isanewline
\ \ \ \ \ {\isachardoublequoteopen}if\ activation{\isadigit{1}}\ t\ {\isacharequal}\ {\isacharbrackleft}{\isacharbrackright}\ \isanewline
\ \ \ \ \ \ then\ nStore\ k\ t\ {\isacharequal}\ recv\ t\ else\ nStore\ k\ t\ {\isacharequal}\ {\isacharbrackleft}{\isacharbrackright}{\isachardoublequoteclose}\isanewline
\ \ \ \ \ \ \isacommand{by}\isamarkupfalse%
\ {\isacharparenleft}simp\ add{\isacharcolon}\ Receive{\isacharunderscore}def{\isacharparenright}\isanewline
\ \ \ \ \isacommand{from}\isamarkupfalse%
\ sg{\isadigit{8}}\ \isakeyword{and}\ sg{\isadigit{7}}\ \isacommand{have}\isamarkupfalse%
\ sg{\isadigit{9}}{\isacharcolon}{\isachardoublequoteopen}nStore\ k\ t\ {\isacharequal}\ {\isacharbrackleft}{\isacharbrackright}{\isachardoublequoteclose}\ \isacommand{by}\isamarkupfalse%
\ simp\isanewline
\ \ \ \ \isacommand{from}\isamarkupfalse%
\ a{\isadigit{4}}\ \isakeyword{and}\ sg{\isadigit{7}}\ \isacommand{have}\isamarkupfalse%
\ sg{\isadigit{1}}{\isadigit{0}}{\isacharcolon}{\isachardoublequoteopen}{\isacharparenleft}t\ mod\ {\isacharparenleft}cycleLength\ {\isacharparenleft}nC\ k{\isacharparenright}{\isacharparenright}{\isacharparenright}\ mem\ {\isacharparenleft}schedule\ {\isacharparenleft}nC\ k{\isacharparenright}{\isacharparenright}{\isachardoublequoteclose}\isanewline
\ \ \ \ \ \ \isacommand{by}\isamarkupfalse%
\ {\isacharparenleft}simp\ add{\isacharcolon}\ Scheduler{\isacharunderscore}L{\isadigit{1}}{\isacharparenright}\isanewline
\ \ \ \ \isacommand{show}\isamarkupfalse%
\ {\isacharquery}thesis\ \isanewline
\ \ \ \ \isacommand{proof}\isamarkupfalse%
\ {\isacharparenleft}cases\ {\isachardoublequoteopen}i\ {\isacharequal}\ k{\isachardoublequoteclose}{\isacharparenright}\isanewline
\ \ \ \ \ \ \isacommand{assume}\isamarkupfalse%
\ aa{\isadigit{1}}{\isacharcolon}\ {\isachardoublequoteopen}i\ {\isacharequal}\ k{\isachardoublequoteclose}\isanewline
\ \ \ \ \ \ \isacommand{from}\isamarkupfalse%
\ sg{\isadigit{9}}\ \isakeyword{and}\ aa{\isadigit{1}}\ \isacommand{show}\isamarkupfalse%
\ {\isacharquery}thesis\ \isacommand{by}\isamarkupfalse%
\ simp\isanewline
\ \ \ \ \isacommand{next}\isamarkupfalse%
\isanewline
\ \ \ \ \ \ \isacommand{assume}\isamarkupfalse%
\ aa{\isadigit{2}}{\isacharcolon}{\isachardoublequoteopen}i\ {\isasymnoteq}\ k{\isachardoublequoteclose}\isanewline
\ \ \ \ \ \ \isacommand{from}\isamarkupfalse%
\ h{\isadigit{7}}\ \isakeyword{and}\ h{\isadigit{4}}\ \isakeyword{and}\ h{\isadigit{1}}\ \isakeyword{and}\ h{\isadigit{2}}\ \isakeyword{and}\ h{\isadigit{3}}\ \isakeyword{and}\ sg{\isadigit{1}}{\isadigit{0}}\ \isakeyword{and}\ a{\isadigit{2}}\ \isakeyword{and}\ h{\isadigit{5}}\ \isakeyword{and}\ aa{\isadigit{2}}\ \isacommand{have}\isamarkupfalse%
\ sg{\isadigit{1}}{\isadigit{1}}{\isacharcolon}\isanewline
\ \ \ \ \ \ \ {\isachardoublequoteopen}nStore\ i\ t\ {\isacharequal}\ nReturn\ k\ t{\isachardoublequoteclose}\isanewline
\ \ \ \ \ \ \ \ \isacommand{by}\isamarkupfalse%
\ {\isacharparenleft}simp\ add{\isacharcolon}\ fr{\isacharunderscore}nStore{\isacharunderscore}nReturn{\isadigit{1}}{\isacharparenright}\isanewline
\ \ \ \ \ \ \isacommand{from}\isamarkupfalse%
\ h{\isadigit{6}}\ \isakeyword{and}\ a{\isadigit{2}}\ \isacommand{have}\isamarkupfalse%
\ sg{\isadigit{1}}{\isadigit{2}}{\isacharcolon}{\isachardoublequoteopen}msg\ {\isacharparenleft}Suc\ {\isadigit{0}}{\isacharparenright}\ {\isacharparenleft}nReturn\ k{\isacharparenright}{\isachardoublequoteclose}\ \isacommand{by}\isamarkupfalse%
\ auto\isanewline
\ \ \ \ \ \ \isacommand{from}\isamarkupfalse%
\ a{\isadigit{2}}\ \isakeyword{and}\ h{\isadigit{6}}\ \isacommand{have}\isamarkupfalse%
\ sg{\isadigit{1}}{\isadigit{3}}{\isacharcolon}{\isachardoublequoteopen}length\ {\isacharparenleft}nReturn\ k\ t{\isacharparenright}\ {\isasymle}\ Suc\ {\isadigit{0}}{\isachardoublequoteclose}\ \isanewline
\ \ \ \ \ \ \ \ \isacommand{by}\isamarkupfalse%
\ {\isacharparenleft}simp\ add{\isacharcolon}\ msg{\isacharunderscore}def{\isacharparenright}\ \ \isanewline
\ \ \ \ \ \ \isacommand{from}\isamarkupfalse%
\ sg{\isadigit{1}}{\isadigit{1}}\ \isakeyword{and}\ sg{\isadigit{1}}{\isadigit{3}}\ \ \isacommand{show}\isamarkupfalse%
\ {\isacharquery}thesis\ \isacommand{by}\isamarkupfalse%
\ simp\isanewline
\ \ \ \ \isacommand{qed}\isamarkupfalse%
\isanewline
\ \ \isacommand{next}\isamarkupfalse%
\isanewline
\ \ \ \ \isacommand{assume}\isamarkupfalse%
\ a{\isadigit{1}}{\isadigit{0}}{\isacharcolon}{\isachardoublequoteopen}{\isasymnot}\ {\isacharparenleft}{\isasymexists}k{\isacharless}n{\isachardot}\ nSend\ k\ t\ {\isasymnoteq}\ {\isacharbrackleft}{\isacharbrackright}{\isacharparenright}{\isachardoublequoteclose}\isanewline
\ \ \ \ \isacommand{from}\isamarkupfalse%
\ h{\isadigit{7}}\ \isakeyword{and}\ a{\isadigit{1}}{\isadigit{0}}\ \isacommand{have}\isamarkupfalse%
\ sg{\isadigit{1}}{\isadigit{4}}{\isacharcolon}{\isachardoublequoteopen}recv\ t\ {\isacharequal}\ {\isacharbrackleft}{\isacharbrackright}{\isachardoublequoteclose}\ \isacommand{by}\isamarkupfalse%
\ {\isacharparenleft}simp\ add{\isacharcolon}\ Broadcast{\isacharunderscore}nSend{\isacharunderscore}empty{\isadigit{1}}{\isacharparenright} 
\ \ \ \ \ \ \isacommand{from}\isamarkupfalse%
\ h{\isadigit{1}}\ \isakeyword{and}\ h{\isadigit{5}}\ \isacommand{have}\isamarkupfalse%
\ sg{\isadigit{1}}{\isadigit{5}}{\isacharcolon}\isanewline
\ \ \ \ \ \ {\isachardoublequoteopen}FlexRayController\ {\isacharparenleft}nReturn\ i{\isacharparenright}\ recv\ {\isacharparenleft}nC\ i{\isacharparenright}\ {\isacharparenleft}nStore\ i{\isacharparenright}\ {\isacharparenleft}nSend\ i{\isacharparenright}\ {\isacharparenleft}nGet\ i{\isacharparenright}{\isachardoublequoteclose}\isanewline
\ \ \ \ \ \ \isacommand{by}\isamarkupfalse%
\ auto\isanewline
\ \ \ \ \isacommand{from}\isamarkupfalse%
\ sg{\isadigit{1}}{\isadigit{5}}\ \isacommand{obtain}\isamarkupfalse%
\ activation{\isadigit{2}}\ \isakeyword{where}\isanewline
\ \ \ \ \ \ a{\isadigit{1}}{\isadigit{1}}{\isacharcolon}{\isachardoublequoteopen}Scheduler\ {\isacharparenleft}nC\ i{\isacharparenright}\ activation{\isadigit{2}}{\isachardoublequoteclose}\ \isakeyword{and}\ \isanewline
\ \ \ \ \ \ a{\isadigit{1}}{\isadigit{2}}{\isacharcolon}{\isachardoublequoteopen}BusInterface\ activation{\isadigit{2}}\ {\isacharparenleft}nReturn\ i{\isacharparenright}\ recv\ {\isacharparenleft}nStore\ i{\isacharparenright}\ {\isacharparenleft}nSend\ i{\isacharparenright}\ {\isacharparenleft}nGet\ i{\isacharparenright}{\isachardoublequoteclose}\ \isanewline
\ \ \ \ \ \ \isacommand{by}\isamarkupfalse%
\ {\isacharparenleft}simp\ add{\isacharcolon}\ FlexRayController{\isacharunderscore}def{\isacharcomma}\ auto{\isacharparenright}\isanewline
\ \ \ \ \isacommand{from}\isamarkupfalse%
\ a{\isadigit{1}}{\isadigit{2}}\ \isacommand{have}\isamarkupfalse%
\ sg{\isadigit{1}}{\isadigit{6}}{\isacharcolon}{\isachardoublequoteopen}Receive\ recv\ {\isacharparenleft}nStore\ i{\isacharparenright}\ activation{\isadigit{2}}{\isachardoublequoteclose}\isanewline
\ \ \ \ \ \ \isacommand{by}\isamarkupfalse%
\ {\isacharparenleft}simp\ add{\isacharcolon}\ BusInterface{\isacharunderscore}def{\isacharparenright}\isanewline
\ \ \ \ \isacommand{from}\isamarkupfalse%
\ sg{\isadigit{1}}{\isadigit{6}}\ \isacommand{have}\isamarkupfalse%
\ sg{\isadigit{1}}{\isadigit{7}}{\isacharcolon}\isanewline
\ \ \ \ \ {\isachardoublequoteopen}if\ activation{\isadigit{2}}\ t\ {\isacharequal}\ {\isacharbrackleft}{\isacharbrackright}\ \isanewline
\ \ \ \ \ \ then\ nStore\ i\ t\ {\isacharequal}\ recv\ t\ else\ nStore\ i\ t\ {\isacharequal}\ {\isacharbrackleft}{\isacharbrackright}{\isachardoublequoteclose}\isanewline
\ \ \ \ \ \ \isacommand{by}\isamarkupfalse%
\ {\isacharparenleft}simp\ add{\isacharcolon}\ Receive{\isacharunderscore}def{\isacharparenright}\isanewline
\ \ \ \ \isacommand{show}\isamarkupfalse%
\ {\isacharquery}thesis\isanewline
\ \ \ \ \isacommand{proof}\isamarkupfalse%
\ {\isacharparenleft}cases\ {\isachardoublequoteopen}activation{\isadigit{2}}\ t\ {\isacharequal}\ {\isacharbrackleft}{\isacharbrackright}{\isachardoublequoteclose}{\isacharparenright}\isanewline
\ \ \ \ \ \ \isacommand{assume}\isamarkupfalse%
\ aa{\isadigit{3}}{\isacharcolon}{\isachardoublequoteopen}activation{\isadigit{2}}\ t\ {\isacharequal}\ {\isacharbrackleft}{\isacharbrackright}{\isachardoublequoteclose}\isanewline
\ \ \ \ \ \ \isacommand{from}\isamarkupfalse%
\ sg{\isadigit{1}}{\isadigit{7}}\ \isakeyword{and}\ aa{\isadigit{3}}\ \isakeyword{and}\ sg{\isadigit{1}}{\isadigit{4}}\ \isacommand{have}\isamarkupfalse%
\ sg{\isadigit{1}}{\isadigit{8}}{\isacharcolon}{\isachardoublequoteopen}nStore\ i\ t\ {\isacharequal}\ {\isacharbrackleft}{\isacharbrackright}{\isachardoublequoteclose}\ \isacommand{by}\isamarkupfalse%
\ simp\isanewline
\ \ \ \ \ \ \isacommand{from}\isamarkupfalse%
\ this\ \isacommand{show}\isamarkupfalse%
\ {\isacharquery}thesis\ \isacommand{by}\isamarkupfalse%
\ simp\isanewline
\ \ \ \ \isacommand{next}\isamarkupfalse%
\isanewline
\ \ \ \ \ \ \isacommand{assume}\isamarkupfalse%
\ aa{\isadigit{4}}{\isacharcolon}{\isachardoublequoteopen}activation{\isadigit{2}}\ t\ {\isasymnoteq}\ {\isacharbrackleft}{\isacharbrackright}{\isachardoublequoteclose}\isanewline
\ \ \ \ \ \ \isacommand{from}\isamarkupfalse%
\ sg{\isadigit{1}}{\isadigit{7}}\ \isakeyword{and}\ aa{\isadigit{4}}\ \isacommand{have}\isamarkupfalse%
\ sg{\isadigit{1}}{\isadigit{8}}{\isacharcolon}{\isachardoublequoteopen}nStore\ i\ t\ {\isacharequal}\ {\isacharbrackleft}{\isacharbrackright}{\isachardoublequoteclose}\ \isacommand{by}\isamarkupfalse%
\ simp\isanewline
\ \ \ \ \ \ \isacommand{from}\isamarkupfalse%
\ this\ \isacommand{show}\isamarkupfalse%
\ {\isacharquery}thesis\ \isacommand{by}\isamarkupfalse%
\ simp\isanewline
\ \ \ \ \isacommand{qed}\isamarkupfalse%
\isanewline
\ \ \isacommand{qed}\isamarkupfalse%
\isanewline
\isacommand{qed}\isamarkupfalse%
\endisatagproof
{\isafoldproof}%
\isadelimproof
\isanewline
\endisadelimproof
\isanewline
\isanewline
\isacommand{lemma}\isamarkupfalse%
\ msg{\isacharunderscore}nStore{\isacharcolon}\isanewline
\ \isakeyword{assumes}\ h{\isadigit{1}}{\isacharcolon}{\isachardoublequoteopen}\ {\isasymforall}i{\isacharless}n{\isachardot}\ FlexRayController\ {\isacharparenleft}nReturn\ i{\isacharparenright}\ recv\ {\isacharparenleft}nC\ i{\isacharparenright}\ {\isacharparenleft}nStore\ i{\isacharparenright}\ {\isacharparenleft}nSend\ i{\isacharparenright}\ {\isacharparenleft}nGet\ i{\isacharparenright}{\isachardoublequoteclose}\isanewline
\ \ \ \ \ \isakeyword{and}\ h{\isadigit{2}}{\isacharcolon}{\isachardoublequoteopen}DisjointSchedules\ n\ nC{\isachardoublequoteclose}\isanewline
\ \ \ \ \ \isakeyword{and}\ h{\isadigit{3}}{\isacharcolon}{\isachardoublequoteopen}IdenticCycleLength\ n\ nC{\isachardoublequoteclose}\isanewline
\ \ \ \ \ \isakeyword{and}\ h{\isadigit{4}}{\isacharcolon}{\isachardoublequoteopen}inf{\isacharunderscore}disj\ n\ nSend{\isachardoublequoteclose}\isanewline
\ \ \ \ \ \isakeyword{and}\ h{\isadigit{5}}{\isacharcolon}{\isachardoublequoteopen}i\ {\isacharless}\ n{\isachardoublequoteclose}\ \ \isanewline
\ \ \ \ \ \isakeyword{and}\ h{\isadigit{6}}{\isacharcolon}{\isachardoublequoteopen}{\isasymforall}\ i{\isacharless}n{\isachardot}\ msg\ {\isacharparenleft}Suc\ {\isadigit{0}}{\isacharparenright}\ {\isacharparenleft}nReturn\ i{\isacharparenright}{\isachardoublequoteclose}\isanewline
\ \ \ \ \ \isakeyword{and}\ h{\isadigit{7}}{\isacharcolon}{\isachardoublequoteopen}Cable\ n\ nSend\ recv{\isachardoublequoteclose}\isanewline
\ \isakeyword{shows}\ {\isachardoublequoteopen}msg\ {\isacharparenleft}Suc\ {\isadigit{0}}{\isacharparenright}\ {\isacharparenleft}nStore\ i{\isacharparenright}{\isachardoublequoteclose}\isanewline
\isadelimproof
\endisadelimproof
\isatagproof
\isacommand{using}\isamarkupfalse%
\ assms\isanewline
\ \ \isacommand{apply}\isamarkupfalse%
\ {\isacharparenleft}simp\ {\isacharparenleft}no{\isacharunderscore}asm{\isacharparenright}\ add{\isacharcolon}\ msg{\isacharunderscore}def{\isacharcomma}\ simp\ add{\isacharcolon}\ Cable{\isacharunderscore}def{\isacharcomma}\ clarify{\isacharparenright}\isanewline
\ \ \isacommand{by}\isamarkupfalse%
\ {\isacharparenleft}simp\ add{\isacharcolon}\ length{\isacharunderscore}nStore{\isacharparenright}%
\endisatagproof
{\isafoldproof}%
\isadelimproof
\endisadelimproof
\isamarkupsubsection{Refinement Properties%
}
\isamarkuptrue%
\isacommand{lemma}\isamarkupfalse%
\ fr{\isacharunderscore}refinement{\isacharunderscore}FrameTransmission{\isacharcolon}\isanewline
\ \isakeyword{assumes}\ h{\isadigit{1}}{\isacharcolon}{\isachardoublequoteopen}Cable\ n\ nSend\ recv{\isachardoublequoteclose}\isanewline
 \ \ \isakeyword{and}\ h{\isadigit{2}}{\isacharcolon}{\isachardoublequoteopen}{\isasymforall}i{\isacharless}n{\isachardot}\ FlexRayController\ {\isacharparenleft}nReturn\ i{\isacharparenright}\ recv\ {\isacharparenleft}nC\ i{\isacharparenright}\ {\isacharparenleft}nStore\ i{\isacharparenright}\ {\isacharparenleft}nSend\ i{\isacharparenright}\ {\isacharparenleft}nGet\ i{\isacharparenright}{\isachardoublequoteclose}\ \isanewline
\ \ \isakeyword{and}\ h{\isadigit{3}}{\isacharcolon}{\isachardoublequoteopen}DisjointSchedules\ n\ nC{\isachardoublequoteclose}\isanewline
\ \ \isakeyword{and}\ h{\isadigit{4}}{\isacharcolon}{\isachardoublequoteopen}IdenticCycleLength\ n\ nC{\isachardoublequoteclose}\ \isanewline
\ \isakeyword{shows}\ {\isachardoublequoteopen}FrameTransmission\ n\ nStore\ nReturn\ nGet\ nC{\isachardoublequoteclose}\isanewline
\isadelimproof
\endisadelimproof
\isatagproof
\isacommand{using}\isamarkupfalse%
\ assms\isanewline
\ \ \isacommand{apply}\isamarkupfalse%
\ {\isacharparenleft}simp\ add{\isacharcolon}\ FrameTransmission{\isacharunderscore}def\ Let{\isacharunderscore}def{\isacharcomma}\ auto{\isacharparenright}\isanewline
\ \ \isacommand{apply}\isamarkupfalse%
\ {\isacharparenleft}simp\ add{\isacharcolon}\ fr{\isacharunderscore}nGet{\isadigit{1}}{\isacharparenright}\isanewline
\ \ \isacommand{by}\isamarkupfalse%
\ {\isacharparenleft}simp\ add{\isacharcolon}\ fr{\isacharunderscore}nStore{\isacharunderscore}nReturn{\isadigit{3}}{\isacharparenright}%
\endisatagproof
{\isafoldproof}%
\isadelimproof
\isanewline
\endisadelimproof
\isanewline
\isacommand{lemma}\isamarkupfalse%
\ FlexRayArch{\isacharunderscore}CorrectSheaf{\isacharcolon}\isanewline
\ \isakeyword{assumes}\ h{\isadigit{1}}{\isacharcolon}{\isachardoublequoteopen}FlexRayArch\ n\ nReturn\ nC\ nStore\ nGet{\isachardoublequoteclose}\isanewline
\ \ \ \isakeyword{shows}\ \ \ \ {\isachardoublequoteopen}CorrectSheaf\ n{\isachardoublequoteclose}\isanewline
\isadelimproof
\endisadelimproof
\isatagproof
\isacommand{using}\isamarkupfalse%
\ assms\ \isacommand{by}\isamarkupfalse%
\ {\isacharparenleft}simp\ add{\isacharcolon}\ FlexRayArch{\isacharunderscore}def{\isacharparenright}%
\endisatagproof
{\isafoldproof}%
\isadelimproof
\isanewline
\endisadelimproof
\isanewline
\isacommand{lemma}\isamarkupfalse%
\ FlexRayArch{\isacharunderscore}FrameTransmission{\isacharcolon}\isanewline
\ \isakeyword{assumes}\ h{\isadigit{1}}{\isacharcolon}{\isachardoublequoteopen}FlexRayArch\ n\ nReturn\ nC\ nStore\ nGet{\isachardoublequoteclose}\isanewline
\ \ \isakeyword{and}\ h{\isadigit{2}}{\isacharcolon}{\isachardoublequoteopen}{\isasymforall}i{\isacharless}n{\isachardot}\ msg\ {\isacharparenleft}Suc\ {\isadigit{0}}{\isacharparenright}\ {\isacharparenleft}nReturn\ i{\isacharparenright}{\isachardoublequoteclose}\isanewline
\ \ \isakeyword{and}\ h{\isadigit{3}}{\isacharcolon}{\isachardoublequoteopen}DisjointSchedules\ n\ nC{\isachardoublequoteclose}\isanewline
\ \ \isakeyword{and}\ h{\isadigit{4}}{\isacharcolon}{\isachardoublequoteopen}IdenticCycleLength\ n\ nC{\isachardoublequoteclose}\isanewline
\ \isakeyword{shows}\ \ \ \ \ \ {\isachardoublequoteopen}FrameTransmission\ n\ nStore\ nReturn\ nGet\ nC{\isachardoublequoteclose}\isanewline
\isadelimproof
\endisadelimproof
\isatagproof
\isacommand{proof}\isamarkupfalse%
\ {\isacharminus}\ \isanewline
\isacommand{from}\isamarkupfalse%
\ assms\ \isacommand{obtain}\isamarkupfalse%
\ nSend\ recv\ \isakeyword{where}\isanewline
a{\isadigit{1}}{\isacharcolon}{\isachardoublequoteopen}Cable\ n\ nSend\ recv{\isachardoublequoteclose}\ \isakeyword{and}\ \isanewline
a{\isadigit{2}}{\isacharcolon}{\isachardoublequoteopen}{\isasymforall}i{\isacharless}n{\isachardot}\ FlexRayController\ {\isacharparenleft}nReturn\ i{\isacharparenright}\ recv\ {\isacharparenleft}nC\ i{\isacharparenright} {\isacharparenleft}nStore\ i{\isacharparenright} {\isacharparenleft}nSend\ i{\isacharparenright} {\isacharparenleft}nGet\ i{\isacharparenright}{\isachardoublequoteclose}\ \isacommand{by}\isamarkupfalse%
\ {\isacharparenleft}simp\ add{\isacharcolon}\ FlexRayArch{\isacharunderscore}def\ \ FlexRayArchitecture{\isacharunderscore}def{\isacharcomma}\ auto{\isacharparenright}\isanewline
\ \ \isacommand{from}\isamarkupfalse%
\ a{\isadigit{1}}\ \isakeyword{and}\ a{\isadigit{2}}\ \isakeyword{and}\ h{\isadigit{3}}\ \isakeyword{and}\ h{\isadigit{4}}\ \isacommand{show}\isamarkupfalse%
\ {\isacharquery}thesis\ \isanewline
\ \ \ \ \isacommand{by}\isamarkupfalse%
\ {\isacharparenleft}rule\ fr{\isacharunderscore}refinement{\isacharunderscore}FrameTransmission{\isacharparenright}\isanewline
\isacommand{qed}\isamarkupfalse%
\endisatagproof
{\isafoldproof}%
\isadelimproof
\isanewline
\endisadelimproof
\isanewline
\isacommand{lemma}\isamarkupfalse%
\ FlexRayArch{\isacharunderscore}nGet{\isacharcolon}\isanewline
\ \isakeyword{assumes}\ h{\isadigit{1}}{\isacharcolon}{\isachardoublequoteopen}FlexRayArch\ n\ nReturn\ nC\ nStore\ nGet{\isachardoublequoteclose}\isanewline
\ \ \ \ \ \isakeyword{and}\ h{\isadigit{2}}{\isacharcolon}{\isachardoublequoteopen}{\isasymforall}i{\isacharless}n{\isachardot}\ msg\ {\isacharparenleft}Suc\ {\isadigit{0}}{\isacharparenright}\ {\isacharparenleft}nReturn\ i{\isacharparenright}{\isachardoublequoteclose}\isanewline
\ \ \ \ \ \isakeyword{and}\ h{\isadigit{3}}{\isacharcolon}{\isachardoublequoteopen}DisjointSchedules\ n\ nC{\isachardoublequoteclose}\isanewline
\ \ \ \ \ \isakeyword{and}\ h{\isadigit{4}}{\isacharcolon}{\isachardoublequoteopen}IdenticCycleLength\ n\ nC{\isachardoublequoteclose}\isanewline
\ \ \ \ \ \isakeyword{and}\ h{\isadigit{5}}{\isacharcolon}{\isachardoublequoteopen}i\ {\isacharless}\ n{\isachardoublequoteclose}\isanewline
\ \isakeyword{shows}\ \ \ \ \ \ {\isachardoublequoteopen}msg\ {\isacharparenleft}Suc\ {\isadigit{0}}{\isacharparenright}\ {\isacharparenleft}nGet\ i{\isacharparenright}{\isachardoublequoteclose}\isanewline
\isadelimproof
\endisadelimproof
\isatagproof
\isacommand{proof}\isamarkupfalse%
\ {\isacharminus}\ \isanewline
\ \isacommand{from}\isamarkupfalse%
\ assms\ \isacommand{obtain}\isamarkupfalse%
\ nSend\ recv\ \isakeyword{where}\isanewline
\ \ \ \ a{\isadigit{1}}{\isacharcolon}{\isachardoublequoteopen}Cable\ n\ nSend\ recv{\isachardoublequoteclose}\ \isakeyword{and}\ \isanewline
\ \ \ \ a{\isadigit{2}}{\isacharcolon}{\isachardoublequoteopen}{\isasymforall}i{\isacharless}n{\isachardot}\ FlexRayController\ {\isacharparenleft}nReturn\ i{\isacharparenright}\ recv\ {\isacharparenleft}nC\ i{\isacharparenright}\ {\isacharparenleft}nStore\ i{\isacharparenright}\ {\isacharparenleft}nSend\ i{\isacharparenright}\ {\isacharparenleft}nGet\ i{\isacharparenright}{\isachardoublequoteclose}\isanewline
\ \ \ \ \isacommand{by}\isamarkupfalse%
\ {\isacharparenleft}simp\ add{\isacharcolon}\ FlexRayArch{\isacharunderscore}def\ \ FlexRayArchitecture{\isacharunderscore}def{\isacharcomma}\ auto{\isacharparenright}\isanewline
\ \ \isacommand{from}\isamarkupfalse%
\ a{\isadigit{2}}\ \isakeyword{and}\ h{\isadigit{5}}\ \isacommand{show}\isamarkupfalse%
\ {\isacharquery}thesis\ \isacommand{by}\isamarkupfalse%
\ {\isacharparenleft}rule\ msg{\isacharunderscore}nGet{\isadigit{2}}{\isacharparenright}\isanewline
\isacommand{qed}\isamarkupfalse%
\endisatagproof
{\isafoldproof}%
\isadelimproof
\isanewline
\endisadelimproof
\isanewline
\isacommand{lemma}\isamarkupfalse%
\ FlexRayArch{\isacharunderscore}nStore{\isacharcolon}\isanewline
\ \isakeyword{assumes}\ h{\isadigit{1}}{\isacharcolon}{\isachardoublequoteopen}FlexRayArch\ n\ nReturn\ nC\ nStore\ nGet{\isachardoublequoteclose}\isanewline
\ \ \ \ \ \isakeyword{and}\ h{\isadigit{2}}{\isacharcolon}{\isachardoublequoteopen}{\isasymforall}i{\isacharless}n{\isachardot}\ msg\ {\isacharparenleft}Suc\ {\isadigit{0}}{\isacharparenright}\ {\isacharparenleft}nReturn\ i{\isacharparenright}{\isachardoublequoteclose}\isanewline
\ \ \ \ \ \isakeyword{and}\ h{\isadigit{3}}{\isacharcolon}{\isachardoublequoteopen}DisjointSchedules\ n\ nC{\isachardoublequoteclose}\isanewline
\ \ \ \ \ \isakeyword{and}\ h{\isadigit{4}}{\isacharcolon}{\isachardoublequoteopen}IdenticCycleLength\ n\ nC{\isachardoublequoteclose}\isanewline
\ \ \ \ \ \isakeyword{and}\ h{\isadigit{5}}{\isacharcolon}{\isachardoublequoteopen}i\ {\isacharless}\ n{\isachardoublequoteclose}\isanewline
\ \isakeyword{shows}\ \ \ \ \ \ {\isachardoublequoteopen}msg\ {\isacharparenleft}Suc\ {\isadigit{0}}{\isacharparenright}\ {\isacharparenleft}nStore\ i{\isacharparenright}{\isachardoublequoteclose}\isanewline
\isadelimproof
\endisadelimproof
\isatagproof
\isacommand{proof}\isamarkupfalse%
\ {\isacharminus}\ \isanewline
\ \isacommand{from}\isamarkupfalse%
\ assms\ \isacommand{obtain}\isamarkupfalse%
\ nSend\ recv\ \isakeyword{where}\isanewline
\ \ \ \ a{\isadigit{1}}{\isacharcolon}{\isachardoublequoteopen}Cable\ n\ nSend\ recv{\isachardoublequoteclose}\ \isakeyword{and}\ \isanewline
\ \ \ \ a{\isadigit{2}}{\isacharcolon}{\isachardoublequoteopen}{\isasymforall}i{\isacharless}n{\isachardot}\ FlexRayController\ {\isacharparenleft}nReturn\ i{\isacharparenright}\ recv\ {\isacharparenleft}nC\ i{\isacharparenright}\ {\isacharparenleft}nStore\ i{\isacharparenright}\ {\isacharparenleft}nSend\ i{\isacharparenright}\ {\isacharparenleft}nGet\ i{\isacharparenright}{\isachardoublequoteclose}\isanewline
\ \ \ \ \isacommand{by}\isamarkupfalse%
\ {\isacharparenleft}simp\ add{\isacharcolon}\ FlexRayArch{\isacharunderscore}def\ \ FlexRayArchitecture{\isacharunderscore}def{\isacharcomma}\ auto{\isacharparenright}\isanewline
\ \ \isacommand{from}\isamarkupfalse%
\ h{\isadigit{3}}\ \isakeyword{and}\ h{\isadigit{4}}\ \isakeyword{and}\ a{\isadigit{2}}\ \isacommand{have}\isamarkupfalse%
\ sg{\isadigit{1}}{\isacharcolon}{\isachardoublequoteopen}inf{\isacharunderscore}disj\ n\ nSend{\isachardoublequoteclose}\ \isacommand{by}\isamarkupfalse%
\ {\isacharparenleft}simp\ add{\isacharcolon}\ disjointFrame{\isacharunderscore}L{\isadigit{2}}{\isacharparenright}\isanewline
\ \ \isacommand{from}\isamarkupfalse%
\ a{\isadigit{2}}\ \isakeyword{and}\ h{\isadigit{3}}\ \isakeyword{and}\ h{\isadigit{4}}\ \isakeyword{and}\ sg{\isadigit{1}}\ \isakeyword{and}\ h{\isadigit{5}}\ \isakeyword{and}\ h{\isadigit{2}}\ \isakeyword{and}\ a{\isadigit{1}}\ \ \isacommand{show}\isamarkupfalse%
\ {\isacharquery}thesis\ \isanewline
\ \ \ \ \isacommand{by}\isamarkupfalse%
\ {\isacharparenleft}rule\ msg{\isacharunderscore}nStore{\isacharparenright}\isanewline
\isacommand{qed}\isamarkupfalse%
\endisatagproof
{\isafoldproof}%
\isadelimproof
\isanewline
\endisadelimproof
\isanewline
\isanewline
\isacommand{theorem}\isamarkupfalse%
\ main{\isacharunderscore}fr{\isacharunderscore}refinement{\isacharcolon}\isanewline
\ \isakeyword{assumes}\ h{\isadigit{1}}{\isacharcolon}{\isachardoublequoteopen}FlexRayArch\ n\ nReturn\ nC\ nStore\ nGet{\isachardoublequoteclose}\isanewline
\ \isakeyword{shows}\ \ \ \ \ \ {\isachardoublequoteopen}FlexRay\ n\ nReturn\ nC\ nStore\ nGet{\isachardoublequoteclose}\isanewline
\isadelimproof
\endisadelimproof
\isatagproof
\isacommand{using}\isamarkupfalse%
\ assms\isanewline
\ \ \isacommand{by}\isamarkupfalse%
\ {\isacharparenleft}simp\ add{\isacharcolon}\ FlexRay{\isacharunderscore}def\isanewline
\ \ \ \ \ \ \ \ \ \ \ \ \ \ \ \ FlexRayArch{\isacharunderscore}CorrectSheaf\ \isanewline
\ \ \ \ \ \ \ \ \ \ \ \ \ \ \ \ FlexRayArch{\isacharunderscore}FrameTransmission\ \isanewline
\ \ \ \ \ \ \ \ \ \ \ \ \ \ \ \ FlexRayArch{\isacharunderscore}nGet\ \isanewline
\ \ \ \ \ \ \ \ \ \ \ \ \ \ \ \ FlexRayArch{\isacharunderscore}nStore{\isacharparenright}%
\endisatagproof
{\isafoldproof}%
\isadelimproof
\isanewline
\endisadelimproof
\isadelimtheory
\ \ \ \ \ \isanewline
\endisadelimtheory
\isatagtheory
\isacommand{end}\isamarkupfalse%
\endisatagtheory
{\isafoldtheory}%
\isadelimtheory
\endisadelimtheory
\end{isabellebody}%

\newpage
\begin{isabellebody}%
\def\isabellecontext{Gateway{\isacharunderscore}types}%
\isamarkupheader{Gateway: Types%
}
\isamarkuptrue%
\isadelimtheory
\endisadelimtheory
\isatagtheory
\isacommand{theory}\isamarkupfalse%
\ Gateway{\isacharunderscore}types\ \isanewline
\isakeyword{imports}\ stream\isanewline
\isakeyword{begin}%
\endisatagtheory
{\isafoldtheory}%
\isadelimtheory
\endisadelimtheory
\isanewline
\isanewline
\isacommand{type{\isacharunderscore}synonym}\isamarkupfalse%
\isanewline
\ \ \ Coordinates\ {\isacharequal}\ {\isachardoublequoteopen}nat\ {\isasymtimes}\ nat{\isachardoublequoteclose}\isanewline
\isacommand{type{\isacharunderscore}synonym}\isamarkupfalse%
\ \isanewline
\ \ \ CollisionSpeed\ {\isacharequal}\ {\isachardoublequoteopen}nat{\isachardoublequoteclose}\isanewline
\isanewline
\isacommand{record}\isamarkupfalse%
\ ECall{\isacharunderscore}Info\ {\isacharequal}\ \isanewline
\ \ \ coord\ {\isacharcolon}{\isacharcolon}\ Coordinates\isanewline
\ \ \ speed\ {\isacharcolon}{\isacharcolon}\ CollisionSpeed\isanewline
\isanewline
\isacommand{datatype}\isamarkupfalse%
\ GatewayStatus\ {\isacharequal}\ \isanewline
\ \ \ \ init{\isacharunderscore}state\isanewline
\ \ {\isacharbar}\ call\isanewline
\ \ {\isacharbar}\ connection{\isacharunderscore}ok\isanewline
\ \ {\isacharbar}\ sending{\isacharunderscore}data\isanewline
\ \ {\isacharbar}\ voice{\isacharunderscore}com\isanewline
\isanewline
\isacommand{datatype}\isamarkupfalse%
\ reqType\ {\isacharequal}\ init\ {\isacharbar}\ send\isanewline
\isacommand{datatype}\isamarkupfalse%
\ stopType\ {\isacharequal}\ stop{\isacharunderscore}vc\isanewline
\isacommand{datatype}\isamarkupfalse%
\ vcType\ {\isacharequal}\ vc{\isacharunderscore}com\isanewline
\isacommand{datatype}\isamarkupfalse%
\ aType\ {\isacharequal}\ sc{\isacharunderscore}ack\isanewline
\isadelimtheory
\isanewline
\endisadelimtheory
\isatagtheory
\isacommand{end}\isamarkupfalse%
\endisatagtheory
{\isafoldtheory}%
\isadelimtheory
\endisadelimtheory
\end{isabellebody}%

%
\begin{isabellebody}%
\def\isabellecontext{Gateway}%
\isamarkupheader{Gateway: Specification%
}
\isamarkuptrue%
\isadelimtheory
\endisadelimtheory
\isatagtheory
\isacommand{theory}\isamarkupfalse%
\ Gateway\ \isanewline
\isakeyword{imports}\ Gateway{\isacharunderscore}types\isanewline
\isakeyword{begin}%
\endisatagtheory
{\isafoldtheory}%
\isadelimtheory
\endisadelimtheory
\isanewline
\isanewline
\isacommand{definition}\isamarkupfalse%
\isanewline
\ ServiceCenter\ {\isacharcolon}{\isacharcolon}\ \isanewline
\ \ \ {\isachardoublequoteopen}ECall{\isacharunderscore}Info\ istream\ {\isasymRightarrow}\ aType\ istream\ {\isasymRightarrow}\ bool\ {\isachardoublequoteclose}\isanewline
\isakeyword{where}\isanewline
\ {\isachardoublequoteopen}ServiceCenter\ i\ a\ \isanewline
\ \ {\isasymequiv}\ \isanewline
\ \ {\isasymforall}\ {\isacharparenleft}t{\isacharcolon}{\isacharcolon}nat{\isacharparenright}{\isachardot}\ \isanewline
\ \ a\ {\isadigit{0}}\ {\isacharequal}\ {\isacharbrackleft}{\isacharbrackright}\ {\isasymand}\ \ a\ {\isacharparenleft}Suc\ t{\isacharparenright}\ {\isacharequal}\ \ {\isacharparenleft}if\ {\isacharparenleft}i\ t{\isacharparenright}\ {\isacharequal}\ {\isacharbrackleft}{\isacharbrackright}\ then\ {\isacharbrackleft}{\isacharbrackright}\ \ else\ {\isacharbrackleft}sc{\isacharunderscore}ack{\isacharbrackright}{\isacharparenright}{\isachardoublequoteclose}\isanewline
\isanewline
\isacommand{definition}\isamarkupfalse%
\isanewline
\ \ Loss\ {\isacharcolon}{\isacharcolon}\isanewline
\ \ \ {\isachardoublequoteopen}bool\ istream\ {\isasymRightarrow}\ aType\ istream\ {\isasymRightarrow}\ ECall{\isacharunderscore}Info\ istream\ {\isasymRightarrow}\ \isanewline
\ \ \ \ aType\ istream\ {\isasymRightarrow}\ ECall{\isacharunderscore}Info\ istream\ {\isasymRightarrow}\ bool{\isachardoublequoteclose}\isanewline
\isakeyword{where}\isanewline
\ {\isachardoublequoteopen}Loss\ lose\ a\ i{\isadigit{2}}\ a{\isadigit{2}}\ i\ \isanewline
\ \ {\isasymequiv}\ \isanewline
\ \ {\isasymforall}\ {\isacharparenleft}t{\isacharcolon}{\isacharcolon}nat{\isacharparenright}{\isachardot}\ \isanewline
\ \ {\isacharparenleft}\ if\ lose\ t\ {\isacharequal}\ \ {\isacharbrackleft}False{\isacharbrackright}\isanewline
\ \ \ \ then\ a{\isadigit{2}}\ t\ {\isacharequal}\ a\ t\ {\isasymand}\ i\ t\ {\isacharequal}\ i{\isadigit{2}}\ t\ \isanewline
\ \ \ \ else\ a{\isadigit{2}}\ t\ {\isacharequal}\ {\isacharbrackleft}{\isacharbrackright}\ \ {\isasymand}\ i\ t\ {\isacharequal}\ {\isacharbrackleft}{\isacharbrackright}\ \ \ {\isacharparenright}{\isachardoublequoteclose}\isanewline
\isanewline
\isacommand{definition}\isamarkupfalse%
\isanewline
\ Delay\ {\isacharcolon}{\isacharcolon}\isanewline
\ \ {\isachardoublequoteopen}aType\ istream\ {\isasymRightarrow}\ ECall{\isacharunderscore}Info\ istream\ {\isasymRightarrow}\ nat\ {\isasymRightarrow}\ \isanewline
\ \ \ aType\ istream\ {\isasymRightarrow}\ ECall{\isacharunderscore}Info\ istream\ {\isasymRightarrow}\ bool{\isachardoublequoteclose}\isanewline
\isakeyword{where}\isanewline
\ {\isachardoublequoteopen}Delay\ a{\isadigit{2}}\ i{\isadigit{1}}\ d\ a{\isadigit{1}}\ i{\isadigit{2}}\ \isanewline
\ \ {\isasymequiv}\ \isanewline
\ \ {\isasymforall}\ {\isacharparenleft}t{\isacharcolon}{\isacharcolon}nat{\isacharparenright}{\isachardot}\ \ \isanewline
\ \ \ {\isacharparenleft}t\ {\isacharless}\ d\ {\isasymlongrightarrow}\ a{\isadigit{1}}\ t\ {\isacharequal}\ {\isacharbrackleft}{\isacharbrackright}\ {\isasymand}\ i{\isadigit{2}}\ t\ {\isacharequal}\ {\isacharbrackleft}{\isacharbrackright}{\isacharparenright}\ {\isasymand}\ \isanewline
\ \ \ {\isacharparenleft}t\ {\isasymge}\ d\ {\isasymlongrightarrow}\ {\isacharparenleft}a{\isadigit{1}}\ t\ {\isacharequal}\ a{\isadigit{2}}\ {\isacharparenleft}t{\isacharminus}d{\isacharparenright}{\isacharparenright}\ {\isasymand}\ {\isacharparenleft}i{\isadigit{2}}\ t\ {\isacharequal}\ i{\isadigit{1}}\ {\isacharparenleft}t{\isacharminus}d{\isacharparenright}{\isacharparenright}{\isacharparenright}{\isachardoublequoteclose}\isanewline
\isanewline
\isacommand{definition}\isamarkupfalse%
\isanewline
\ tiTable{\isacharunderscore}SampleT\ {\isacharcolon}{\isacharcolon}\isanewline
\ {\isachardoublequoteopen}reqType\ istream\ \ {\isasymRightarrow}\ aType\ istream\ {\isasymRightarrow}\ \isanewline
\ \ stopType\ istream\ {\isasymRightarrow}\ bool\ istream\ {\isasymRightarrow}\ \isanewline
\ \ {\isacharparenleft}nat\ {\isasymRightarrow}\ GatewayStatus{\isacharparenright}\ {\isasymRightarrow}\ \ {\isacharparenleft}nat\ {\isasymRightarrow}\ ECall{\isacharunderscore}Info\ list{\isacharparenright}\ {\isasymRightarrow}\isanewline
\ \ GatewayStatus\ istream\ {\isasymRightarrow}\ ECall{\isacharunderscore}Info\ istream\ {\isasymRightarrow}\ vcType\ istream\ \isanewline
\ \ {\isasymRightarrow}\ {\isacharparenleft}nat\ {\isasymRightarrow}\ GatewayStatus{\isacharparenright}\ {\isasymRightarrow}\ bool\ {\isachardoublequoteclose}\isanewline
\isakeyword{where}\isanewline
\ {\isachardoublequoteopen}tiTable{\isacharunderscore}SampleT\ req\ a{\isadigit{1}}\ stop\ lose\ st{\isacharunderscore}in\ buffer{\isacharunderscore}in\ \isanewline
\ \ \ \ \ \ \ \ \ ack\ i{\isadigit{1}}\ vc\ st{\isacharunderscore}out\isanewline
\ \ {\isasymequiv}\isanewline
\ \ {\isasymforall}\ {\isacharparenleft}t{\isacharcolon}{\isacharcolon}nat{\isacharparenright}\ \isanewline
\ \ \ \ \ {\isacharparenleft}r{\isacharcolon}{\isacharcolon}reqType\ list{\isacharparenright}\ {\isacharparenleft}x{\isacharcolon}{\isacharcolon}aType\ list{\isacharparenright}\ \isanewline
\ \ \ \ \ {\isacharparenleft}y{\isacharcolon}{\isacharcolon}stopType\ list{\isacharparenright}\ {\isacharparenleft}z{\isacharcolon}{\isacharcolon}bool\ list{\isacharparenright}{\isachardot}\isanewline
\ \ \ {\isacharparenleft}{\isacharasterisk}{\isadigit{1}}{\isacharasterisk}{\isacharparenright}\isanewline
\ \ \ {\isacharparenleft}\ st{\isacharunderscore}in\ t\ {\isacharequal}\ init{\isacharunderscore}state\ {\isasymand}\ req\ t\ {\isacharequal}\ {\isacharbrackleft}init{\isacharbrackright}\isanewline
\ \ \ \ \ {\isasymlongrightarrow}\ ack\ t\ {\isacharequal}\ {\isacharbrackleft}call{\isacharbrackright}\ {\isasymand}\ i{\isadigit{1}}\ t\ {\isacharequal}\ {\isacharbrackleft}{\isacharbrackright}\ {\isasymand}\ vc\ t\ {\isacharequal}\ {\isacharbrackleft}{\isacharbrackright}\ \isanewline
\ \ \ \ \ \ \ \ \ {\isasymand}\ st{\isacharunderscore}out\ t\ {\isacharequal}\ call\ {\isacharparenright}\ \isanewline
\ \ \ {\isasymand}\ \ \ \ \ \isanewline
\ \ \ {\isacharparenleft}{\isacharasterisk}{\isadigit{2}}{\isacharasterisk}{\isacharparenright}\isanewline
\ \ \ {\isacharparenleft}\ st{\isacharunderscore}in\ t\ {\isacharequal}\ init{\isacharunderscore}state\ {\isasymand}\ req\ t\ {\isasymnoteq}\ {\isacharbrackleft}init{\isacharbrackright}\isanewline
\ \ \ \ \ {\isasymlongrightarrow}\ \ ack\ t\ {\isacharequal}\ {\isacharbrackleft}init{\isacharunderscore}state{\isacharbrackright}\ {\isasymand}\ i{\isadigit{1}}\ t\ {\isacharequal}\ {\isacharbrackleft}{\isacharbrackright}\ {\isasymand}\ vc\ t\ {\isacharequal}\ {\isacharbrackleft}{\isacharbrackright}\ \isanewline
\ \ \ \ \ \ \ \ \ {\isasymand}\ st{\isacharunderscore}out\ t\ {\isacharequal}\ init{\isacharunderscore}state\ {\isacharparenright}\ \isanewline
\ \ \ {\isasymand}\isanewline
\ \ \ {\isacharparenleft}{\isacharasterisk}{\isadigit{3}}{\isacharasterisk}{\isacharparenright}\isanewline
\ \ \ {\isacharparenleft}\ {\isacharparenleft}st{\isacharunderscore}in\ t\ {\isacharequal}\ call\ {\isasymor}\ {\isacharparenleft}st{\isacharunderscore}in\ t\ {\isacharequal}\ connection{\isacharunderscore}ok\ {\isasymand}\ r\ {\isasymnoteq}\ {\isacharbrackleft}send{\isacharbrackright}{\isacharparenright}{\isacharparenright}\ {\isasymand}\isanewline
\ \ \ \ \ req\ t\ {\isacharequal}\ r\ {\isasymand}\ lose\ t\ {\isacharequal}\ {\isacharbrackleft}False{\isacharbrackright}\isanewline
\ \ \ \ \ {\isasymlongrightarrow}\ \ ack\ t\ {\isacharequal}\ {\isacharbrackleft}connection{\isacharunderscore}ok{\isacharbrackright}\ {\isasymand}\ i{\isadigit{1}}\ t\ {\isacharequal}\ {\isacharbrackleft}{\isacharbrackright}\ {\isasymand}\ vc\ t\ {\isacharequal}\ {\isacharbrackleft}{\isacharbrackright}\ \isanewline
\ \ \ \ \ \ \ \ \ {\isasymand}\ st{\isacharunderscore}out\ t\ {\isacharequal}\ connection{\isacharunderscore}ok\ {\isacharparenright}\ \isanewline
\ \ \ {\isasymand}\isanewline
\ \ \ {\isacharparenleft}{\isacharasterisk}{\isadigit{4}}{\isacharasterisk}{\isacharparenright}\isanewline
\ \ \ {\isacharparenleft}\ {\isacharparenleft}st{\isacharunderscore}in\ t\ {\isacharequal}\ call\ {\isasymor}\ st{\isacharunderscore}in\ t\ {\isacharequal}\ connection{\isacharunderscore}ok\ {\isasymor}\ st{\isacharunderscore}in\ t\ {\isacharequal}\ sending{\isacharunderscore}data{\isacharparenright}\isanewline
\ \ \ \ \ {\isasymand}\ lose\ t\ {\isacharequal}\ {\isacharbrackleft}True{\isacharbrackright}\isanewline
\ \ \ \ \ {\isasymlongrightarrow}\ \ ack\ t\ {\isacharequal}\ {\isacharbrackleft}init{\isacharunderscore}state{\isacharbrackright}\ {\isasymand}\ i{\isadigit{1}}\ t\ {\isacharequal}\ {\isacharbrackleft}{\isacharbrackright}\ {\isasymand}\ vc\ t\ {\isacharequal}\ {\isacharbrackleft}{\isacharbrackright}\ \isanewline
\ \ \ \ \ \ \ \ \ {\isasymand}\ st{\isacharunderscore}out\ t\ {\isacharequal}\ init{\isacharunderscore}state\ {\isacharparenright}\ \isanewline
\ \ \ {\isasymand}\isanewline
\ \ \ {\isacharparenleft}{\isacharasterisk}{\isadigit{5}}{\isacharasterisk}{\isacharparenright}\isanewline
\ \ \ {\isacharparenleft}\ st{\isacharunderscore}in\ t\ {\isacharequal}\ connection{\isacharunderscore}ok\ {\isasymand}\ req\ t\ {\isacharequal}\ {\isacharbrackleft}send{\isacharbrackright}\ {\isasymand}\ lose\ t\ {\isacharequal}\ {\isacharbrackleft}False{\isacharbrackright}\isanewline
\ \ \ \ \ {\isasymlongrightarrow}\ ack\ t\ {\isacharequal}\ {\isacharbrackleft}sending{\isacharunderscore}data{\isacharbrackright}\ {\isasymand}\ i{\isadigit{1}}\ t\ {\isacharequal}\ buffer{\isacharunderscore}in\ t\ {\isasymand}\ vc\ t\ {\isacharequal}\ {\isacharbrackleft}{\isacharbrackright}\ \isanewline
\ \ \ \ \ \ \ \ \ {\isasymand}\ st{\isacharunderscore}out\ t\ {\isacharequal}\ sending{\isacharunderscore}data\ {\isacharparenright}\ \isanewline
\ \ \ {\isasymand}\isanewline
\ \ \ {\isacharparenleft}{\isacharasterisk}{\isadigit{6}}{\isacharasterisk}{\isacharparenright}\isanewline
\ \ \ {\isacharparenleft}\ st{\isacharunderscore}in\ t\ {\isacharequal}\ sending{\isacharunderscore}data\ {\isasymand}\ a{\isadigit{1}}\ t\ {\isacharequal}\ {\isacharbrackleft}{\isacharbrackright}\ {\isasymand}\ lose\ t\ {\isacharequal}\ {\isacharbrackleft}False{\isacharbrackright}\isanewline
\ \ \ \ \ {\isasymlongrightarrow}\ ack\ t\ {\isacharequal}\ {\isacharbrackleft}sending{\isacharunderscore}data{\isacharbrackright}\ {\isasymand}\ i{\isadigit{1}}\ t\ {\isacharequal}\ {\isacharbrackleft}{\isacharbrackright}\ {\isasymand}\ vc\ t\ {\isacharequal}\ {\isacharbrackleft}{\isacharbrackright}\ \isanewline
\ \ \ \ \ \ \ \ \ {\isasymand}\ st{\isacharunderscore}out\ t\ {\isacharequal}\ sending{\isacharunderscore}data\ {\isacharparenright}\ \isanewline
\ \ \ {\isasymand}\isanewline
\ \ \ {\isacharparenleft}{\isacharasterisk}{\isadigit{7}}{\isacharasterisk}{\isacharparenright}\isanewline
\ \ \ {\isacharparenleft}\ st{\isacharunderscore}in\ t\ {\isacharequal}\ sending{\isacharunderscore}data\ {\isasymand}\ a{\isadigit{1}}\ t\ {\isacharequal}\ {\isacharbrackleft}sc{\isacharunderscore}ack{\isacharbrackright}\ {\isasymand}\ lose\ t\ {\isacharequal}\ {\isacharbrackleft}False{\isacharbrackright}\isanewline
\ \ \ \ \ {\isasymlongrightarrow}\ ack\ t\ {\isacharequal}\ {\isacharbrackleft}voice{\isacharunderscore}com{\isacharbrackright}\ {\isasymand}\ i{\isadigit{1}}\ t\ {\isacharequal}\ {\isacharbrackleft}{\isacharbrackright}\ {\isasymand}\ vc\ t\ {\isacharequal}\ {\isacharbrackleft}vc{\isacharunderscore}com{\isacharbrackright}\ \isanewline
\ \ \ \ \ \ \ \ \ {\isasymand}\ st{\isacharunderscore}out\ t\ {\isacharequal}\ voice{\isacharunderscore}com\ {\isacharparenright}\ \isanewline
\ \ \ {\isasymand}\isanewline
\ \ \ {\isacharparenleft}{\isacharasterisk}{\isadigit{8}}{\isacharasterisk}{\isacharparenright}\isanewline
\ \ \ {\isacharparenleft}\ st{\isacharunderscore}in\ t\ {\isacharequal}\ voice{\isacharunderscore}com\ {\isasymand}\ stop\ t\ {\isacharequal}\ {\isacharbrackleft}{\isacharbrackright}\ {\isasymand}\ lose\ t\ {\isacharequal}\ {\isacharbrackleft}False{\isacharbrackright}\isanewline
\ \ \ \ \ {\isasymlongrightarrow}\ ack\ t\ {\isacharequal}\ {\isacharbrackleft}voice{\isacharunderscore}com{\isacharbrackright}\ {\isasymand}\ i{\isadigit{1}}\ t\ {\isacharequal}\ {\isacharbrackleft}{\isacharbrackright}\ {\isasymand}\ vc\ t\ {\isacharequal}\ {\isacharbrackleft}vc{\isacharunderscore}com{\isacharbrackright}\ \isanewline
\ \ \ \ \ \ \ \ \ {\isasymand}\ st{\isacharunderscore}out\ t\ {\isacharequal}\ voice{\isacharunderscore}com\ {\isacharparenright}\ \isanewline
\ \ \ {\isasymand}\isanewline
\ \ \ {\isacharparenleft}{\isacharasterisk}{\isadigit{9}}{\isacharasterisk}{\isacharparenright}\isanewline
\ \ \ {\isacharparenleft}\ st{\isacharunderscore}in\ t\ {\isacharequal}\ voice{\isacharunderscore}com\ {\isasymand}\ stop\ t\ {\isacharequal}\ {\isacharbrackleft}{\isacharbrackright}\ {\isasymand}\ lose\ t\ {\isacharequal}\ {\isacharbrackleft}True{\isacharbrackright}\isanewline
\ \ \ \ \ {\isasymlongrightarrow}\ ack\ t\ {\isacharequal}\ {\isacharbrackleft}voice{\isacharunderscore}com{\isacharbrackright}\ {\isasymand}\ i{\isadigit{1}}\ t\ {\isacharequal}\ {\isacharbrackleft}{\isacharbrackright}\ {\isasymand}\ vc\ t\ {\isacharequal}\ {\isacharbrackleft}{\isacharbrackright}\ \isanewline
\ \ \ \ \ \ \ \ \ {\isasymand}\ st{\isacharunderscore}out\ t\ {\isacharequal}\ voice{\isacharunderscore}com\ {\isacharparenright}\ \isanewline
\ \ \ {\isasymand}\isanewline
\ \ \ {\isacharparenleft}{\isacharasterisk}{\isadigit{1}}{\isadigit{0}}{\isacharasterisk}{\isacharparenright}\isanewline
\ \ \ {\isacharparenleft}\ st{\isacharunderscore}in\ t\ {\isacharequal}\ voice{\isacharunderscore}com\ {\isasymand}\ stop\ t\ {\isacharequal}\ {\isacharbrackleft}stop{\isacharunderscore}vc{\isacharbrackright}\ \isanewline
\ \ \ \ \ {\isasymlongrightarrow}\ ack\ t\ {\isacharequal}\ {\isacharbrackleft}init{\isacharunderscore}state{\isacharbrackright}\ {\isasymand}\ i{\isadigit{1}}\ t\ {\isacharequal}\ {\isacharbrackleft}{\isacharbrackright}\ {\isasymand}\ vc\ t\ {\isacharequal}\ {\isacharbrackleft}{\isacharbrackright}\ \isanewline
\ \ \ \ \ \ \ \ \ {\isasymand}\ st{\isacharunderscore}out\ t\ {\isacharequal}\ init{\isacharunderscore}state\ {\isacharparenright}{\isachardoublequoteclose}\ \isanewline
\isanewline
\isacommand{definition}\isamarkupfalse%
\isanewline
\ Sample{\isacharunderscore}L\ {\isacharcolon}{\isacharcolon}\isanewline
\ {\isachardoublequoteopen}reqType\ istream\ \ {\isasymRightarrow}\ ECall{\isacharunderscore}Info\ istream\ {\isasymRightarrow}\ aType\ istream\ {\isasymRightarrow}\ \isanewline
\ \ stopType\ istream\ {\isasymRightarrow}\ bool\ istream\ {\isasymRightarrow}\ \isanewline
\ \ {\isacharparenleft}nat\ {\isasymRightarrow}\ GatewayStatus{\isacharparenright}\ {\isasymRightarrow}\ \ {\isacharparenleft}nat\ {\isasymRightarrow}\ ECall{\isacharunderscore}Info\ list{\isacharparenright}\ {\isasymRightarrow}\isanewline
\ \ GatewayStatus\ istream\ {\isasymRightarrow}\ ECall{\isacharunderscore}Info\ istream\ {\isasymRightarrow}\ vcType\ istream\ \isanewline
\ \ {\isasymRightarrow}\ {\isacharparenleft}nat\ {\isasymRightarrow}\ GatewayStatus{\isacharparenright}\ \ {\isasymRightarrow}\ {\isacharparenleft}nat\ {\isasymRightarrow}\ ECall{\isacharunderscore}Info\ list{\isacharparenright}\isanewline
\ \ {\isasymRightarrow}\ bool\ {\isachardoublequoteclose}\isanewline
\isakeyword{where}\isanewline
\ {\isachardoublequoteopen}Sample{\isacharunderscore}L\ req\ dt\ a{\isadigit{1}}\ stop\ lose\ st{\isacharunderscore}in\ buffer{\isacharunderscore}in\ \isanewline
\ \ \ \ \ \ \ \ \ ack\ i{\isadigit{1}}\ vc\ st{\isacharunderscore}out\ buffer{\isacharunderscore}out\isanewline
\ \ {\isasymequiv}\isanewline
\ \ {\isacharparenleft}{\isasymforall}\ {\isacharparenleft}t{\isacharcolon}{\isacharcolon}nat{\isacharparenright}{\isachardot}\ \isanewline
\ \ \ buffer{\isacharunderscore}out\ t\ {\isacharequal}\ \isanewline
\ \ \ \ {\isacharparenleft}if\ dt\ t\ {\isacharequal}\ {\isacharbrackleft}{\isacharbrackright}\ then\ buffer{\isacharunderscore}in\ t\ \ else\ dt\ t{\isacharparenright}\ {\isacharparenright}\isanewline
\ \ {\isasymand}\ \isanewline
\ \ {\isacharparenleft}tiTable{\isacharunderscore}SampleT\ req\ a{\isadigit{1}}\ stop\ lose\ st{\isacharunderscore}in\ buffer{\isacharunderscore}in\ \isanewline
\ \ \ \ \ \ \ \ \ \ \ \ \ \ \ \ \ \ \ ack\ i{\isadigit{1}}\ vc\ st{\isacharunderscore}out{\isacharparenright}{\isachardoublequoteclose}\ \isanewline
\isanewline
\isacommand{definition}\isamarkupfalse%
\isanewline
\ Sample\ {\isacharcolon}{\isacharcolon}\isanewline
\ {\isachardoublequoteopen}reqType\ istream\ \ {\isasymRightarrow}\ ECall{\isacharunderscore}Info\ istream\ {\isasymRightarrow}\ aType\ istream\ {\isasymRightarrow}\ \isanewline
\ \ stopType\ istream\ {\isasymRightarrow}\ bool\ istream\ {\isasymRightarrow}\ \isanewline
\ \ GatewayStatus\ istream\ {\isasymRightarrow}\ ECall{\isacharunderscore}Info\ istream\ {\isasymRightarrow}\ vcType\ istream\isanewline
\ \ {\isasymRightarrow}\ bool\ {\isachardoublequoteclose}\isanewline
\isakeyword{where}\isanewline
\ {\isachardoublequoteopen}Sample\ req\ dt\ a{\isadigit{1}}\ stop\ lose\ \ ack\ i{\isadigit{1}}\ vc\ \isanewline
\ \ {\isasymequiv}\isanewline
\ \ {\isacharparenleft}{\isacharparenleft}msg\ {\isacharparenleft}{\isadigit{1}}{\isacharcolon}{\isacharcolon}nat{\isacharparenright}\ req{\isacharparenright}\ {\isasymand}\ \isanewline
\ \ \ {\isacharparenleft}msg\ {\isacharparenleft}{\isadigit{1}}{\isacharcolon}{\isacharcolon}nat{\isacharparenright}\ a{\isadigit{1}}{\isacharparenright}\ \ {\isasymand}\ \isanewline
\ \ \ {\isacharparenleft}msg\ {\isacharparenleft}{\isadigit{1}}{\isacharcolon}{\isacharcolon}nat{\isacharparenright}\ stop{\isacharparenright}{\isacharparenright}\ \isanewline
\ \ {\isasymlongrightarrow}\ \isanewline
\ \ {\isacharparenleft}{\isasymexists}\ st\ buffer{\isachardot}\ \isanewline
\ \ {\isacharparenleft}Sample{\isacharunderscore}L\ req\ dt\ a{\isadigit{1}}\ stop\ lose\ \isanewline
\ \ \ \ \ \ \ \ \ \ \ \ {\isacharparenleft}fin{\isacharunderscore}inf{\isacharunderscore}append\ {\isacharbrackleft}init{\isacharunderscore}state{\isacharbrackright}\ st{\isacharparenright}\isanewline
\ \ \ \ \ \ \ \ \ \ \ \ {\isacharparenleft}fin{\isacharunderscore}inf{\isacharunderscore}append\ {\isacharbrackleft}{\isacharbrackleft}{\isacharbrackright}{\isacharbrackright}\ buffer{\isacharparenright}\isanewline
\ \ \ \ \ \ \ \ \ \ \ \ ack\ i{\isadigit{1}}\ vc\ st\ buffer{\isacharparenright}\ {\isacharparenright}{\isachardoublequoteclose}\isanewline
\isanewline
\isacommand{definition}\isamarkupfalse%
\isanewline
\ Gateway\ {\isacharcolon}{\isacharcolon}\ \isanewline
\ \ \ {\isachardoublequoteopen}reqType\ istream\ {\isasymRightarrow}\ ECall{\isacharunderscore}Info\ istream\ {\isasymRightarrow}\ aType\ istream\ {\isasymRightarrow}\ \isanewline
\ \ \ \ stopType\ istream\ {\isasymRightarrow}\ bool\ istream\ {\isasymRightarrow}\ nat\ {\isasymRightarrow}\isanewline
\ \ \ \ GatewayStatus\ istream\ {\isasymRightarrow}\ ECall{\isacharunderscore}Info\ istream\ {\isasymRightarrow}\ vcType\ istream\isanewline
\ \ \ \ {\isasymRightarrow}\ bool{\isachardoublequoteclose}\isanewline
\isakeyword{where}\isanewline
\ {\isachardoublequoteopen}Gateway\ req\ dt\ a\ stop\ lose\ d\ ack\ i\ vc\ \isanewline
\ \ {\isasymequiv}\ {\isasymexists}\ i{\isadigit{1}}\ i{\isadigit{2}}\ x\ y{\isachardot}\ \isanewline
\ \ \ \ {\isacharparenleft}Sample\ req\ dt\ x\ stop\ lose\ ack\ i{\isadigit{1}}\ vc{\isacharparenright}\ {\isasymand}\ \isanewline
\ \ \ \ {\isacharparenleft}Delay\ y\ i{\isadigit{1}}\ d\ x\ i{\isadigit{2}}{\isacharparenright}\ {\isasymand}\ \isanewline
\ \ \ \ {\isacharparenleft}Loss\ lose\ a\ i{\isadigit{2}}\ y\ i{\isacharparenright}{\isachardoublequoteclose}\isanewline
\isanewline
\isacommand{definition}\isamarkupfalse%
\isanewline
\ \ GatewaySystem\ {\isacharcolon}{\isacharcolon}\ \isanewline
\ \ \ {\isachardoublequoteopen}reqType\ istream\ {\isasymRightarrow}\ ECall{\isacharunderscore}Info\ istream\ {\isasymRightarrow}\ \isanewline
\ \ \ \ stopType\ istream\ {\isasymRightarrow}\ bool\ istream\ {\isasymRightarrow}\ nat\ {\isasymRightarrow}\isanewline
\ \ \ \ GatewayStatus\ istream\ {\isasymRightarrow}\ vcType\ istream\isanewline
\ \ \ \ {\isasymRightarrow}\ bool{\isachardoublequoteclose}\isanewline
\isakeyword{where}\isanewline
\ {\isachardoublequoteopen}GatewaySystem\ req\ dt\ stop\ lose\ d\ ack\ vc\ \isanewline
\ \ {\isasymequiv}\isanewline
\ \ {\isasymexists}\ a\ i{\isachardot}\ \isanewline
\ {\isacharparenleft}Gateway\ req\ dt\ a\ stop\ lose\ d\ ack\ i\ vc{\isacharparenright}\ {\isasymand}\ \isanewline
\ {\isacharparenleft}ServiceCenter\ i\ a{\isacharparenright}{\isachardoublequoteclose}\isanewline
\isanewline
\isacommand{definition}\isamarkupfalse%
\isanewline
\ GatewayReq\ {\isacharcolon}{\isacharcolon}\ \isanewline
\ \ \ {\isachardoublequoteopen}reqType\ istream\ {\isasymRightarrow}\ ECall{\isacharunderscore}Info\ istream\ {\isasymRightarrow}\ aType\ istream\ {\isasymRightarrow}\ \isanewline
\ \ \ \ stopType\ istream\ {\isasymRightarrow}\ bool\ istream\ {\isasymRightarrow}\ nat\ {\isasymRightarrow}\isanewline
\ \ \ \ GatewayStatus\ istream\ {\isasymRightarrow}\ ECall{\isacharunderscore}Info\ istream\ {\isasymRightarrow}\ vcType\ istream\isanewline
\ \ \ \ {\isasymRightarrow}\ bool{\isachardoublequoteclose}\isanewline
\isakeyword{where}\isanewline
\ {\isachardoublequoteopen}GatewayReq\ req\ dt\ a\ stop\ lose\ d\ ack\ i\ vc\ \isanewline
\ \ {\isasymequiv}\ \ \isanewline
\ {\isacharparenleft}{\isacharparenleft}msg\ {\isacharparenleft}{\isadigit{1}}{\isacharcolon}{\isacharcolon}nat{\isacharparenright}\ req{\isacharparenright}\ {\isasymand}\ \ {\isacharparenleft}msg\ {\isacharparenleft}{\isadigit{1}}{\isacharcolon}{\isacharcolon}nat{\isacharparenright}\ a{\isacharparenright}\ \ {\isasymand}\ \isanewline
\ \ {\isacharparenleft}msg\ {\isacharparenleft}{\isadigit{1}}{\isacharcolon}{\isacharcolon}nat{\isacharparenright}\ stop{\isacharparenright}\ {\isasymand}\ \ {\isacharparenleft}ts\ lose{\isacharparenright}{\isacharparenright}\ \isanewline
\ \ \ {\isasymlongrightarrow}\isanewline
\ \ {\isacharparenleft}{\isasymforall}\ {\isacharparenleft}t{\isacharcolon}{\isacharcolon}nat{\isacharparenright}{\isachardot}\isanewline
\ \ {\isacharparenleft}\ ack\ t\ {\isacharequal}\ {\isacharbrackleft}init{\isacharunderscore}state{\isacharbrackright}\ {\isasymand}\ req\ {\isacharparenleft}Suc\ t{\isacharparenright}\ {\isacharequal}\ {\isacharbrackleft}init{\isacharbrackright}\ {\isasymand}\ \isanewline
\ \ \ \ lose\ {\isacharparenleft}t{\isacharplus}{\isadigit{1}}{\isacharparenright}\ {\isacharequal}\ {\isacharbrackleft}False{\isacharbrackright}\ {\isasymand}\ lose\ {\isacharparenleft}t{\isacharplus}{\isadigit{2}}{\isacharparenright}\ {\isacharequal}\ {\isacharbrackleft}False{\isacharbrackright}\isanewline
\ \ \ \ {\isasymlongrightarrow}\ ack\ {\isacharparenleft}t{\isacharplus}{\isadigit{2}}{\isacharparenright}\ {\isacharequal}\ {\isacharbrackleft}connection{\isacharunderscore}ok{\isacharbrackright}{\isacharparenright}\ \isanewline
\ \ {\isasymand}\ \isanewline
\ \ {\isacharparenleft}\ ack\ t\ {\isacharequal}\ {\isacharbrackleft}connection{\isacharunderscore}ok{\isacharbrackright}\ {\isasymand}\ req\ {\isacharparenleft}Suc\ t{\isacharparenright}\ \ {\isacharequal}\ {\isacharbrackleft}send{\isacharbrackright}\ {\isasymand}\ \ \isanewline
\ \ \ \ {\isacharparenleft}{\isasymforall}\ {\isacharparenleft}k{\isacharcolon}{\isacharcolon}nat{\isacharparenright}{\isachardot}\ k\ {\isasymle}\ {\isacharparenleft}d{\isacharplus}{\isadigit{1}}{\isacharparenright}\ {\isasymlongrightarrow}\ lose\ {\isacharparenleft}t{\isacharplus}k{\isacharparenright}\ {\isacharequal}\ {\isacharbrackleft}False{\isacharbrackright}{\isacharparenright}\ \isanewline
\ \ \ \ {\isasymlongrightarrow}\ i\ {\isacharparenleft}{\isacharparenleft}Suc\ t{\isacharparenright}\ {\isacharplus}\ d{\isacharparenright}\ {\isacharequal}\ inf{\isacharunderscore}last{\isacharunderscore}ti\ dt\ t\ \isanewline
\ \ \ \ \ \ \ \ {\isasymand}\ \ ack\ {\isacharparenleft}Suc\ t{\isacharparenright}\ {\isacharequal}\ {\isacharbrackleft}sending{\isacharunderscore}data{\isacharbrackright}{\isacharparenright}\isanewline
\ \ {\isasymand}\ \isanewline
\ \ {\isacharparenleft}\ ack\ {\isacharparenleft}t{\isacharplus}d{\isacharparenright}\ {\isacharequal}\ {\isacharbrackleft}sending{\isacharunderscore}data{\isacharbrackright}\ {\isasymand}\ a\ {\isacharparenleft}Suc\ t{\isacharparenright}\ {\isacharequal}\ {\isacharbrackleft}sc{\isacharunderscore}ack{\isacharbrackright}\ {\isasymand}\ \ \isanewline
\ \ \ \ {\isacharparenleft}{\isasymforall}\ {\isacharparenleft}k{\isacharcolon}{\isacharcolon}nat{\isacharparenright}{\isachardot}\ k\ {\isasymle}\ {\isacharparenleft}d{\isacharplus}{\isadigit{1}}{\isacharparenright}\ {\isasymlongrightarrow}\ lose\ {\isacharparenleft}t{\isacharplus}k{\isacharparenright}\ {\isacharequal}\ {\isacharbrackleft}False{\isacharbrackright}{\isacharparenright}\ \isanewline
\ \ \ \ {\isasymlongrightarrow}\ vc\ {\isacharparenleft}{\isacharparenleft}Suc\ t{\isacharparenright}\ {\isacharplus}\ d{\isacharparenright}\ {\isacharequal}\ {\isacharbrackleft}vc{\isacharunderscore}com{\isacharbrackright}{\isacharparenright}\ {\isacharparenright}{\isachardoublequoteclose}\isanewline
\isanewline
\isacommand{definition}\isamarkupfalse%
\isanewline
\ \ GatewaySystemReq\ {\isacharcolon}{\isacharcolon}\ \isanewline
\ \ \ {\isachardoublequoteopen}reqType\ istream\ {\isasymRightarrow}\ ECall{\isacharunderscore}Info\ istream\ {\isasymRightarrow}\ \isanewline
\ \ \ \ stopType\ istream\ {\isasymRightarrow}\ bool\ istream\ {\isasymRightarrow}\ nat\ {\isasymRightarrow}\isanewline
\ \ \ \ GatewayStatus\ istream\ {\isasymRightarrow}\ vcType\ istream\isanewline
\ \ \ \ {\isasymRightarrow}\ bool{\isachardoublequoteclose}\isanewline
\isakeyword{where}\isanewline
\ {\isachardoublequoteopen}GatewaySystemReq\ req\ dt\ \ stop\ lose\ d\ ack\ vc\ \isanewline
\ \ {\isasymequiv}\isanewline
\ {\isacharparenleft}{\isacharparenleft}msg\ {\isacharparenleft}{\isadigit{1}}{\isacharcolon}{\isacharcolon}nat{\isacharparenright}\ req{\isacharparenright}\ {\isasymand}\ {\isacharparenleft}msg\ {\isacharparenleft}{\isadigit{1}}{\isacharcolon}{\isacharcolon}nat{\isacharparenright}\ stop{\isacharparenright}\ {\isasymand}\ {\isacharparenleft}ts\ lose{\isacharparenright}{\isacharparenright}\isanewline
\ \ \ {\isasymlongrightarrow}\isanewline
\ \ {\isacharparenleft}{\isasymforall}\ {\isacharparenleft}t{\isacharcolon}{\isacharcolon}nat{\isacharparenright}\ {\isacharparenleft}k{\isacharcolon}{\isacharcolon}nat{\isacharparenright}{\isachardot}\ \isanewline
\ \ {\isacharparenleft}\ ack\ t\ {\isacharequal}\ {\isacharbrackleft}init{\isacharunderscore}state{\isacharbrackright}\ {\isasymand}\ req\ {\isacharparenleft}Suc\ t{\isacharparenright}\ {\isacharequal}\ {\isacharbrackleft}init{\isacharbrackright}\isanewline
\ \ {\isasymand}\ {\isacharparenleft}{\isasymforall}\ t{\isadigit{1}}{\isachardot}\ t{\isadigit{1}}\ {\isasymle}\ t\ {\isasymlongrightarrow}\ req\ t{\isadigit{1}}\ {\isacharequal}\ {\isacharbrackleft}{\isacharbrackright}{\isacharparenright}\isanewline
\ \ {\isasymand}\ req\ {\isacharparenleft}t{\isacharplus}{\isadigit{2}}{\isacharparenright}\ {\isacharequal}\ {\isacharbrackleft}{\isacharbrackright}\isanewline
\ \ {\isasymand}\ {\isacharparenleft}{\isasymforall}\ m{\isachardot}\ m\ {\isacharless}\ k\ {\isacharplus}\ {\isadigit{3}}\ {\isasymlongrightarrow}\ req\ {\isacharparenleft}t\ {\isacharplus}\ m{\isacharparenright}\ {\isasymnoteq}\ {\isacharbrackleft}send{\isacharbrackright}{\isacharparenright}\isanewline
\ \ {\isasymand}\ \ req\ {\isacharparenleft}t{\isacharplus}{\isadigit{3}}{\isacharplus}k{\isacharparenright}\ \ {\isacharequal}\ {\isacharbrackleft}send{\isacharbrackright}\ {\isasymand}\ \ inf{\isacharunderscore}last{\isacharunderscore}ti\ dt\ {\isacharparenleft}t{\isacharplus}{\isadigit{2}}{\isacharparenright}\ {\isasymnoteq}\ {\isacharbrackleft}{\isacharbrackright}\isanewline
\ \ {\isasymand}\ {\isacharparenleft}{\isasymforall}\ {\isacharparenleft}j{\isacharcolon}{\isacharcolon}nat{\isacharparenright}{\isachardot}\ \isanewline
\ \ \ \ \ j\ {\isasymle}\ {\isacharparenleft}{\isadigit{4}}\ {\isacharplus}\ k\ {\isacharplus}\ d\ {\isacharplus}\ d{\isacharparenright}\ {\isasymlongrightarrow}\ lose\ {\isacharparenleft}t{\isacharplus}j{\isacharparenright}\ {\isacharequal}\ {\isacharbrackleft}False{\isacharbrackright}{\isacharparenright}\ \isanewline
\ \ {\isasymlongrightarrow}\ vc\ {\isacharparenleft}t\ {\isacharplus}\ {\isadigit{4}}\ {\isacharplus}\ k\ {\isacharplus}\ d\ {\isacharplus}\ d{\isacharparenright}\ {\isacharequal}\ {\isacharbrackleft}vc{\isacharunderscore}com{\isacharbrackright}{\isacharparenright}\ {\isacharparenright}{\isachardoublequoteclose}\isanewline
\isadelimtheory
\isanewline
\endisadelimtheory
\isatagtheory
\isacommand{end}\isamarkupfalse%
\endisatagtheory
{\isafoldtheory}%
\isadelimtheory
\endisadelimtheory
\end{isabellebody}%

%
\begin{isabellebody}%
\def\isabellecontext{Gateway{\isacharunderscore}proof{\isacharunderscore}aux}%
\isamarkupheader{Gateway: Verification%
}
\isamarkuptrue%
\isadelimtheory
\endisadelimtheory
\isatagtheory
\isacommand{theory}\isamarkupfalse%
\ Gateway{\isacharunderscore}proof{\isacharunderscore}aux\ \isanewline
\isakeyword{imports}\ Gateway\ BitBoolTS\isanewline
\isakeyword{begin}%
\endisatagtheory
{\isafoldtheory}%
\isadelimtheory
\endisadelimtheory
\isamarkupsubsection{Properties of the defined data types%
}
\isamarkuptrue%
\isacommand{lemma}\isamarkupfalse%
\ aType{\isacharunderscore}empty{\isacharcolon}\ \isanewline
\ \ \isakeyword{assumes}\ h{\isadigit{1}}{\isacharcolon}{\isachardoublequoteopen}msg\ {\isacharparenleft}Suc\ {\isadigit{0}}{\isacharparenright}\ a{\isachardoublequoteclose}\isanewline
\ \ \ \ \ \ \isakeyword{and}\ h{\isadigit{2}}{\isacharcolon}\ {\isachardoublequoteopen}a\ t\ {\isasymnoteq}\ {\isacharbrackleft}sc{\isacharunderscore}ack{\isacharbrackright}{\isachardoublequoteclose}\isanewline
\ \ \isakeyword{shows}\ \ \ \ \ \ \ {\isachardoublequoteopen}a\ t\ {\isacharequal}\ {\isacharbrackleft}{\isacharbrackright}{\isachardoublequoteclose}\isanewline
\isadelimproof
\endisadelimproof
\isatagproof
\isacommand{proof}\isamarkupfalse%
\ {\isacharparenleft}cases\ {\isachardoublequoteopen}a\ t{\isachardoublequoteclose}{\isacharparenright}\isanewline
\ \ \isacommand{assume}\isamarkupfalse%
\ a{\isadigit{1}}{\isacharcolon}{\isachardoublequoteopen}a\ t\ {\isacharequal}\ {\isacharbrackleft}{\isacharbrackright}{\isachardoublequoteclose}\isanewline
\ \ \isacommand{from}\isamarkupfalse%
\ this\ \isacommand{show}\isamarkupfalse%
\ {\isacharquery}thesis\ \isacommand{by}\isamarkupfalse%
\ simp\isanewline
\isacommand{next}\isamarkupfalse%
\isanewline
\ \ \isacommand{fix}\isamarkupfalse%
\ aa\ l\isanewline
\ \ \isacommand{assume}\isamarkupfalse%
\ a{\isadigit{2}}{\isacharcolon}{\isachardoublequoteopen}a\ t\ {\isacharequal}\ aa\ {\isacharhash}\ l{\isachardoublequoteclose}\isanewline
\ \ \isacommand{show}\isamarkupfalse%
\ {\isacharquery}thesis\isanewline
\ \ \ \ \isacommand{proof}\isamarkupfalse%
\ {\isacharparenleft}cases\ {\isachardoublequoteopen}aa{\isachardoublequoteclose}{\isacharparenright}\ \isanewline
\ \ \ \ \ \ \isacommand{assume}\isamarkupfalse%
\ a{\isadigit{3}}{\isacharcolon}{\isachardoublequoteopen}aa\ {\isacharequal}\ sc{\isacharunderscore}ack{\isachardoublequoteclose}\isanewline
\ \ \ \ \ \ \isacommand{from}\isamarkupfalse%
\ h{\isadigit{1}}\ \isacommand{have}\isamarkupfalse%
\ sg{\isadigit{1}}{\isacharcolon}{\isachardoublequoteopen}length\ {\isacharparenleft}a\ t{\isacharparenright}\ {\isasymle}\ Suc\ {\isadigit{0}}{\isachardoublequoteclose}\ \isacommand{by}\isamarkupfalse%
\ {\isacharparenleft}simp\ add{\isacharcolon}\ msg{\isacharunderscore}def{\isacharparenright}\isanewline
\ \ \ \ \ \ \isacommand{from}\isamarkupfalse%
\ this\ \isakeyword{and}\ h{\isadigit{1}}\ \isakeyword{and}\ h{\isadigit{2}}\ \isakeyword{and}\ a{\isadigit{2}}\ \isakeyword{and}\ a{\isadigit{3}}\ \isacommand{show}\isamarkupfalse%
\ {\isacharquery}thesis\ \isacommand{by}\isamarkupfalse%
\ auto\ \isanewline
\ \ \ \ \isacommand{qed}\isamarkupfalse%
\isanewline
\isacommand{qed}\isamarkupfalse%
\endisatagproof
{\isafoldproof}%
\isadelimproof
\isanewline
\endisadelimproof
\isanewline
\isacommand{lemma}\isamarkupfalse%
\ aType{\isacharunderscore}nonempty{\isacharcolon}\ \isanewline
\ \ \isakeyword{assumes}\ h{\isadigit{1}}{\isacharcolon}{\isachardoublequoteopen}msg\ {\isacharparenleft}Suc\ {\isadigit{0}}{\isacharparenright}\ a{\isachardoublequoteclose}\isanewline
\ \ \ \ \ \ \isakeyword{and}\ h{\isadigit{2}}{\isacharcolon}\ {\isachardoublequoteopen}a\ t\ {\isasymnoteq}\ {\isacharbrackleft}{\isacharbrackright}{\isachardoublequoteclose}\isanewline
\ \ \isakeyword{shows}\ \ \ \ \ \ \ {\isachardoublequoteopen}a\ t\ {\isacharequal}\ {\isacharbrackleft}sc{\isacharunderscore}ack{\isacharbrackright}{\isachardoublequoteclose}\isanewline
\isadelimproof
\endisadelimproof
\isatagproof
\isacommand{proof}\isamarkupfalse%
\ {\isacharparenleft}cases\ {\isachardoublequoteopen}a\ t{\isachardoublequoteclose}{\isacharparenright}\isanewline
\ \ \isacommand{assume}\isamarkupfalse%
\ a{\isadigit{1}}{\isacharcolon}{\isachardoublequoteopen}a\ t\ {\isacharequal}\ {\isacharbrackleft}{\isacharbrackright}{\isachardoublequoteclose}\isanewline
\ \ \isacommand{from}\isamarkupfalse%
\ this\ \isakeyword{and}\ h{\isadigit{2}}\ \isacommand{show}\isamarkupfalse%
\ {\isacharquery}thesis\ \isacommand{by}\isamarkupfalse%
\ simp\isanewline
\isacommand{next}\isamarkupfalse%
\isanewline
\ \ \isacommand{fix}\isamarkupfalse%
\ aa\ l\isanewline
\ \ \isacommand{assume}\isamarkupfalse%
\ a{\isadigit{2}}{\isacharcolon}{\isachardoublequoteopen}a\ t\ {\isacharequal}\ aa\ {\isacharhash}\ l{\isachardoublequoteclose}\isanewline
\ \ \isacommand{from}\isamarkupfalse%
\ a{\isadigit{2}}\ \isakeyword{and}\ h{\isadigit{1}}\ \isacommand{have}\isamarkupfalse%
\ sg{\isadigit{1}}{\isacharcolon}\ {\isachardoublequoteopen}l\ {\isacharequal}\ {\isacharbrackleft}{\isacharbrackright}{\isachardoublequoteclose}\ \isacommand{by}\isamarkupfalse%
\ {\isacharparenleft}simp\ add{\isacharcolon}\ msg{\isacharunderscore}nonempty{\isadigit{1}}{\isacharparenright}\isanewline
\ \ \isacommand{from}\isamarkupfalse%
\ a{\isadigit{2}}\ \isakeyword{and}\ h{\isadigit{1}}\ \isakeyword{and}\ sg{\isadigit{1}}\ \isacommand{show}\isamarkupfalse%
\ {\isacharquery}thesis\isanewline
\ \ \ \ \isacommand{proof}\isamarkupfalse%
\ {\isacharparenleft}cases\ {\isachardoublequoteopen}aa{\isachardoublequoteclose}{\isacharparenright}\ \isanewline
\ \ \ \ \ \ \isacommand{assume}\isamarkupfalse%
\ a{\isadigit{3}}{\isacharcolon}{\isachardoublequoteopen}aa\ {\isacharequal}\ sc{\isacharunderscore}ack{\isachardoublequoteclose}\ \isanewline
\ \ \ \ \ \ \isacommand{from}\isamarkupfalse%
\ this\ \isakeyword{and}\ sg{\isadigit{1}}\ \isakeyword{and}\ h{\isadigit{2}}\ \isakeyword{and}\ a{\isadigit{2}}\ \isacommand{show}\isamarkupfalse%
\ {\isacharquery}thesis\ \ \isacommand{by}\isamarkupfalse%
\ simp\isanewline
\ \ \ \ \isacommand{qed}\isamarkupfalse%
\isanewline
\isacommand{qed}\isamarkupfalse%
\endisatagproof
{\isafoldproof}%
\isadelimproof
\isanewline
\endisadelimproof
\isanewline
\isacommand{lemma}\isamarkupfalse%
\ aType{\isacharunderscore}lemma{\isacharcolon}\ \isanewline
\ \ \isakeyword{assumes}\ h{\isadigit{1}}{\isacharcolon}{\isachardoublequoteopen}msg\ {\isacharparenleft}Suc\ {\isadigit{0}}{\isacharparenright}\ a{\isachardoublequoteclose}\ \isanewline
\ \ \isakeyword{shows}\ \ \ \ \ \ {\isachardoublequoteopen}a\ t\ {\isacharequal}\ {\isacharbrackleft}{\isacharbrackright}\ {\isasymor}\ a\ t\ {\isacharequal}\ {\isacharbrackleft}sc{\isacharunderscore}ack{\isacharbrackright}{\isachardoublequoteclose}\isanewline
\isadelimproof
\endisadelimproof
\isatagproof
\isacommand{using}\isamarkupfalse%
\ assms\isanewline
\ \ \isacommand{apply}\isamarkupfalse%
\ auto\isanewline
\ \ \isacommand{by}\isamarkupfalse%
\ {\isacharparenleft}simp\ add{\isacharcolon}\ aType{\isacharunderscore}empty{\isacharparenright}%
\endisatagproof
{\isafoldproof}%
\isadelimproof
\isanewline
\endisadelimproof
\isanewline
\isacommand{lemma}\isamarkupfalse%
\ stopType{\isacharunderscore}empty{\isacharcolon}\ \isanewline
\ \ \isakeyword{assumes}\ h{\isadigit{1}}{\isacharcolon}{\isachardoublequoteopen}msg\ {\isacharparenleft}Suc\ {\isadigit{0}}{\isacharparenright}\ a{\isachardoublequoteclose}\isanewline
\ \ \ \ \ \ \isakeyword{and}\ h{\isadigit{2}}{\isacharcolon}{\isachardoublequoteopen}a\ t\ {\isasymnoteq}\ {\isacharbrackleft}stop{\isacharunderscore}vc{\isacharbrackright}{\isachardoublequoteclose}\isanewline
\ \ \isakeyword{shows}\ {\isachardoublequoteopen}a\ t\ {\isacharequal}\ {\isacharbrackleft}{\isacharbrackright}{\isachardoublequoteclose}\isanewline
\isadelimproof
\endisadelimproof
\isatagproof
\isacommand{proof}\isamarkupfalse%
\ {\isacharparenleft}cases\ {\isachardoublequoteopen}a\ t{\isachardoublequoteclose}{\isacharparenright}\isanewline
\ \ \isacommand{assume}\isamarkupfalse%
\ a{\isadigit{1}}{\isacharcolon}{\isachardoublequoteopen}a\ t\ {\isacharequal}\ {\isacharbrackleft}{\isacharbrackright}{\isachardoublequoteclose}\isanewline
\ \ \isacommand{from}\isamarkupfalse%
\ this\ \isacommand{show}\isamarkupfalse%
\ {\isacharquery}thesis\ \isacommand{by}\isamarkupfalse%
\ simp\isanewline
\isacommand{next}\isamarkupfalse%
\isanewline
\ \ \isacommand{fix}\isamarkupfalse%
\ aa\ l\isanewline
\ \ \isacommand{assume}\isamarkupfalse%
\ a{\isadigit{2}}{\isacharcolon}{\isachardoublequoteopen}a\ t\ {\isacharequal}\ aa\ {\isacharhash}\ l{\isachardoublequoteclose}\isanewline
\ \ \isacommand{show}\isamarkupfalse%
\ {\isacharquery}thesis\isanewline
\ \ \ \ \isacommand{proof}\isamarkupfalse%
\ {\isacharparenleft}cases\ {\isachardoublequoteopen}aa{\isachardoublequoteclose}{\isacharparenright}\ \isanewline
\ \ \ \ \ \ \isacommand{assume}\isamarkupfalse%
\ a{\isadigit{3}}{\isacharcolon}{\isachardoublequoteopen}aa\ {\isacharequal}\ stop{\isacharunderscore}vc{\isachardoublequoteclose}\isanewline
\ \ \ \ \ \ \isacommand{from}\isamarkupfalse%
\ h{\isadigit{1}}\ \isacommand{have}\isamarkupfalse%
\ sg{\isadigit{1}}{\isacharcolon}{\isachardoublequoteopen}length\ {\isacharparenleft}a\ t{\isacharparenright}\ {\isasymle}\ Suc\ {\isadigit{0}}{\isachardoublequoteclose}\ \isacommand{by}\isamarkupfalse%
\ {\isacharparenleft}simp\ add{\isacharcolon}\ msg{\isacharunderscore}def{\isacharparenright}\isanewline
\ \ \ \ \ \ \isacommand{from}\isamarkupfalse%
\ this\ \isakeyword{and}\ h{\isadigit{1}}\ \isakeyword{and}\ h{\isadigit{2}}\ \isakeyword{and}\ a{\isadigit{2}}\ \isakeyword{and}\ a{\isadigit{3}}\ \isacommand{show}\isamarkupfalse%
\ {\isacharquery}thesis\ \isacommand{by}\isamarkupfalse%
\ auto\ \isanewline
\ \ \ \ \isacommand{qed}\isamarkupfalse%
\isanewline
\isacommand{qed}\isamarkupfalse%
\endisatagproof
{\isafoldproof}%
\isadelimproof
\isanewline
\endisadelimproof
\isanewline
\isacommand{lemma}\isamarkupfalse%
\ stopType{\isacharunderscore}nonempty{\isacharcolon}\ \isanewline
\ \ \isakeyword{assumes}\ h{\isadigit{1}}{\isacharcolon}{\isachardoublequoteopen}msg\ {\isacharparenleft}Suc\ {\isadigit{0}}{\isacharparenright}\ a{\isachardoublequoteclose}\isanewline
\ \ \ \ \ \ \isakeyword{and}\ h{\isadigit{2}}{\isacharcolon}{\isachardoublequoteopen}a\ t\ {\isasymnoteq}\ {\isacharbrackleft}{\isacharbrackright}{\isachardoublequoteclose}\isanewline
\ \ \isakeyword{shows}\ {\isachardoublequoteopen}a\ t\ {\isacharequal}\ {\isacharbrackleft}stop{\isacharunderscore}vc{\isacharbrackright}{\isachardoublequoteclose}\isanewline
\isadelimproof
\endisadelimproof
\isatagproof
\isacommand{proof}\isamarkupfalse%
\ {\isacharparenleft}cases\ {\isachardoublequoteopen}a\ t{\isachardoublequoteclose}{\isacharparenright}\isanewline
\ \ \isacommand{assume}\isamarkupfalse%
\ a{\isadigit{1}}{\isacharcolon}{\isachardoublequoteopen}a\ t\ {\isacharequal}\ {\isacharbrackleft}{\isacharbrackright}{\isachardoublequoteclose}\isanewline
\ \ \isacommand{from}\isamarkupfalse%
\ this\ \isakeyword{and}\ h{\isadigit{2}}\ \isacommand{show}\isamarkupfalse%
\ {\isacharquery}thesis\ \isacommand{by}\isamarkupfalse%
\ simp\isanewline
\isacommand{next}\isamarkupfalse%
\isanewline
\ \ \isacommand{fix}\isamarkupfalse%
\ aa\ l\isanewline
\ \ \isacommand{assume}\isamarkupfalse%
\ a{\isadigit{2}}{\isacharcolon}{\isachardoublequoteopen}a\ t\ {\isacharequal}\ aa\ {\isacharhash}\ l{\isachardoublequoteclose}\isanewline
\ \ \isacommand{show}\isamarkupfalse%
\ {\isacharquery}thesis\isanewline
\ \ \ \ \isacommand{proof}\isamarkupfalse%
\ {\isacharparenleft}cases\ {\isachardoublequoteopen}aa{\isachardoublequoteclose}{\isacharparenright}\ \isanewline
\ \ \ \ \ \ \isacommand{assume}\isamarkupfalse%
\ a{\isadigit{3}}{\isacharcolon}{\isachardoublequoteopen}aa\ {\isacharequal}\ stop{\isacharunderscore}vc{\isachardoublequoteclose}\isanewline
\ \ \ \ \ \ \isacommand{from}\isamarkupfalse%
\ h{\isadigit{1}}\ \isacommand{have}\isamarkupfalse%
\ sg{\isadigit{1}}{\isacharcolon}{\isachardoublequoteopen}length\ {\isacharparenleft}a\ t{\isacharparenright}\ {\isasymle}\ Suc\ {\isadigit{0}}{\isachardoublequoteclose}\ \isacommand{by}\isamarkupfalse%
\ {\isacharparenleft}simp\ add{\isacharcolon}\ msg{\isacharunderscore}def{\isacharparenright}\isanewline
\ \ \ \ \ \ \isacommand{from}\isamarkupfalse%
\ this\ \isakeyword{and}\ h{\isadigit{1}}\ \isakeyword{and}\ h{\isadigit{2}}\ \isakeyword{and}\ a{\isadigit{2}}\ \isakeyword{and}\ a{\isadigit{3}}\ \isacommand{show}\isamarkupfalse%
\ {\isacharquery}thesis\ \isacommand{by}\isamarkupfalse%
\ auto\ \isanewline
\ \ \ \ \isacommand{qed}\isamarkupfalse%
\isanewline
\isacommand{qed}\isamarkupfalse%
\endisatagproof
{\isafoldproof}%
\isadelimproof
\isanewline
\endisadelimproof
\isanewline
\isacommand{lemma}\isamarkupfalse%
\ stopType{\isacharunderscore}lemma{\isacharcolon}\ \isanewline
\ \ \isakeyword{assumes}\ h{\isadigit{1}}{\isacharcolon}{\isachardoublequoteopen}msg\ {\isacharparenleft}Suc\ {\isadigit{0}}{\isacharparenright}\ a{\isachardoublequoteclose}\ \isanewline
\ \ \isakeyword{shows}\ \ \ \ \ \ {\isachardoublequoteopen}a\ t\ {\isacharequal}\ {\isacharbrackleft}{\isacharbrackright}\ {\isasymor}\ a\ t\ {\isacharequal}\ {\isacharbrackleft}stop{\isacharunderscore}vc{\isacharbrackright}{\isachardoublequoteclose}\isanewline
\isadelimproof
\endisadelimproof
\isatagproof
\isacommand{using}\isamarkupfalse%
\ assms\isanewline
\ \ \isacommand{apply}\isamarkupfalse%
\ auto\isanewline
\ \ \isacommand{by}\isamarkupfalse%
\ {\isacharparenleft}simp\ add{\isacharcolon}\ stopType{\isacharunderscore}empty{\isacharparenright}%
\endisatagproof
{\isafoldproof}%
\isadelimproof
\isanewline
\endisadelimproof
\isanewline
\isacommand{lemma}\isamarkupfalse%
\ vcType{\isacharunderscore}empty{\isacharcolon}\ \isanewline
\ \ \isakeyword{assumes}\ h{\isadigit{1}}{\isacharcolon}{\isachardoublequoteopen}msg\ {\isacharparenleft}Suc\ {\isadigit{0}}{\isacharparenright}\ a{\isachardoublequoteclose}\isanewline
\ \ \ \ \ \ \isakeyword{and}\ h{\isadigit{2}}{\isacharcolon}{\isachardoublequoteopen}a\ t\ {\isasymnoteq}\ {\isacharbrackleft}vc{\isacharunderscore}com{\isacharbrackright}{\isachardoublequoteclose}\isanewline
\ \ \isakeyword{shows}{\isachardoublequoteopen}a\ t\ {\isacharequal}\ {\isacharbrackleft}{\isacharbrackright}{\isachardoublequoteclose}\isanewline
\isadelimproof
\endisadelimproof
\isatagproof
\isacommand{proof}\isamarkupfalse%
\ {\isacharparenleft}cases\ {\isachardoublequoteopen}a\ t{\isachardoublequoteclose}{\isacharparenright}\isanewline
\ \ \isacommand{assume}\isamarkupfalse%
\ a{\isadigit{1}}{\isacharcolon}{\isachardoublequoteopen}a\ t\ {\isacharequal}\ {\isacharbrackleft}{\isacharbrackright}{\isachardoublequoteclose}\isanewline
\ \ \isacommand{from}\isamarkupfalse%
\ this\ \isacommand{show}\isamarkupfalse%
\ {\isacharquery}thesis\ \isacommand{by}\isamarkupfalse%
\ simp\isanewline
\isacommand{next}\isamarkupfalse%
\isanewline
\ \ \isacommand{fix}\isamarkupfalse%
\ aa\ l\isanewline
\ \ \isacommand{assume}\isamarkupfalse%
\ a{\isadigit{2}}{\isacharcolon}{\isachardoublequoteopen}a\ t\ {\isacharequal}\ aa\ {\isacharhash}\ l{\isachardoublequoteclose}\isanewline
\ \ \isacommand{show}\isamarkupfalse%
\ {\isacharquery}thesis\isanewline
\ \ \ \ \isacommand{proof}\isamarkupfalse%
\ {\isacharparenleft}cases\ {\isachardoublequoteopen}aa{\isachardoublequoteclose}{\isacharparenright}\ \isanewline
\ \ \ \ \ \ \isacommand{assume}\isamarkupfalse%
\ a{\isadigit{3}}{\isacharcolon}{\isachardoublequoteopen}aa\ {\isacharequal}\ vc{\isacharunderscore}com{\isachardoublequoteclose}\isanewline
\ \ \ \ \ \ \isacommand{from}\isamarkupfalse%
\ h{\isadigit{1}}\ \isacommand{have}\isamarkupfalse%
\ sg{\isadigit{1}}{\isacharcolon}{\isachardoublequoteopen}length\ {\isacharparenleft}a\ t{\isacharparenright}\ {\isasymle}\ Suc\ {\isadigit{0}}{\isachardoublequoteclose}\ \isacommand{by}\isamarkupfalse%
\ {\isacharparenleft}simp\ add{\isacharcolon}\ msg{\isacharunderscore}def{\isacharparenright}\isanewline
\ \ \ \ \ \ \isacommand{from}\isamarkupfalse%
\ this\ \isakeyword{and}\ h{\isadigit{1}}\ \isakeyword{and}\ h{\isadigit{2}}\ \isakeyword{and}\ a{\isadigit{2}}\ \isakeyword{and}\ a{\isadigit{3}}\ \isacommand{show}\isamarkupfalse%
\ {\isacharquery}thesis\ \isacommand{by}\isamarkupfalse%
\ auto\ \isanewline
\ \ \ \ \isacommand{qed}\isamarkupfalse%
\isanewline
\isacommand{qed}\isamarkupfalse%
\endisatagproof
{\isafoldproof}%
\isadelimproof
\isanewline
\endisadelimproof
\isanewline
\isacommand{lemma}\isamarkupfalse%
\ vcType{\isacharunderscore}lemma{\isacharcolon}\ \isanewline
\ \ \isakeyword{assumes}\ h{\isadigit{1}}{\isacharcolon}{\isachardoublequoteopen}msg\ {\isacharparenleft}Suc\ {\isadigit{0}}{\isacharparenright}\ a{\isachardoublequoteclose}\ \isanewline
\ \ \isakeyword{shows}\ \ \ \ \ \ {\isachardoublequoteopen}a\ t\ {\isacharequal}\ {\isacharbrackleft}{\isacharbrackright}\ {\isasymor}\ a\ t\ {\isacharequal}\ {\isacharbrackleft}vc{\isacharunderscore}com{\isacharbrackright}{\isachardoublequoteclose}\isanewline
\isadelimproof
\endisadelimproof
\isatagproof
\isacommand{using}\isamarkupfalse%
\ assms\isanewline
\ \ \isacommand{apply}\isamarkupfalse%
\ auto\isanewline
\ \ \isacommand{by}\isamarkupfalse%
\ {\isacharparenleft}simp\ add{\isacharcolon}\ vcType{\isacharunderscore}empty{\isacharparenright}%
\endisatagproof
{\isafoldproof}%
\isadelimproof
\endisadelimproof
\isamarkupsubsection{Properties of the Delay component%
}
\isamarkuptrue%
\isacommand{lemma}\isamarkupfalse%
\ Delay{\isacharunderscore}L{\isadigit{1}}{\isacharcolon}\isanewline
\ \isakeyword{assumes}\ h{\isadigit{1}}{\isacharcolon}{\isachardoublequoteopen}{\isasymforall}t{\isadigit{1}}\ {\isacharless}\ t{\isachardot}\ i{\isadigit{1}}\ t{\isadigit{1}}\ {\isacharequal}\ {\isacharbrackleft}{\isacharbrackright}{\isachardoublequoteclose}\isanewline
\ \ \ \ \ \isakeyword{and}\ h{\isadigit{2}}{\isacharcolon}{\isachardoublequoteopen}Delay\ y\ i{\isadigit{1}}\ d\ x\ i{\isadigit{2}}{\isachardoublequoteclose}\isanewline
\ \ \ \ \ \isakeyword{and}\ h{\isadigit{3}}{\isacharcolon}{\isachardoublequoteopen}t{\isadigit{2}}\ {\isacharless}\ t\ {\isacharplus}\ d{\isachardoublequoteclose}\isanewline
\ \ \ \isakeyword{shows}\ {\isachardoublequoteopen}i{\isadigit{2}}\ t{\isadigit{2}}\ {\isacharequal}\ {\isacharbrackleft}{\isacharbrackright}{\isachardoublequoteclose}\isanewline
\isadelimproof
\endisadelimproof
\isatagproof
\isacommand{proof}\isamarkupfalse%
\ {\isacharparenleft}cases\ {\isachardoublequoteopen}t{\isadigit{2}}\ {\isacharless}\ d{\isachardoublequoteclose}{\isacharparenright}\isanewline
\ \ \isacommand{assume}\isamarkupfalse%
\ a{\isadigit{1}}{\isacharcolon}{\isachardoublequoteopen}t{\isadigit{2}}\ {\isacharless}\ d{\isachardoublequoteclose}\isanewline
\ \ \isacommand{from}\isamarkupfalse%
\ h{\isadigit{2}}\ \isacommand{have}\isamarkupfalse%
\ sg{\isadigit{1}}{\isacharcolon}{\isachardoublequoteopen}t{\isadigit{2}}\ {\isacharless}\ d\ {\isasymlongrightarrow}\ i{\isadigit{2}}\ t{\isadigit{2}}\ {\isacharequal}\ {\isacharbrackleft}{\isacharbrackright}{\isachardoublequoteclose}\isanewline
\ \ \ \ \isacommand{by}\isamarkupfalse%
\ {\isacharparenleft}simp\ add{\isacharcolon}\ Delay{\isacharunderscore}def{\isacharparenright}\isanewline
\ \ \isacommand{from}\isamarkupfalse%
\ sg{\isadigit{1}}\ \isakeyword{and}\ a{\isadigit{1}}\ \isacommand{show}\isamarkupfalse%
\ {\isacharquery}thesis\ \isacommand{by}\isamarkupfalse%
\ simp\isanewline
\isacommand{next}\isamarkupfalse%
\isanewline
\ \ \isacommand{assume}\isamarkupfalse%
\ a{\isadigit{2}}{\isacharcolon}{\isachardoublequoteopen}{\isasymnot}\ t{\isadigit{2}}\ {\isacharless}\ d{\isachardoublequoteclose}\isanewline
\ \ \isacommand{from}\isamarkupfalse%
\ h{\isadigit{2}}\ \isacommand{have}\isamarkupfalse%
\ sg{\isadigit{2}}{\isacharcolon}{\isachardoublequoteopen}d\ {\isasymle}\ t{\isadigit{2}}\ {\isasymlongrightarrow}\ i{\isadigit{2}}\ t{\isadigit{2}}\ {\isacharequal}\ i{\isadigit{1}}\ {\isacharparenleft}t{\isadigit{2}}\ {\isacharminus}\ d{\isacharparenright}{\isachardoublequoteclose}\isanewline
\ \ \ \ \isacommand{by}\isamarkupfalse%
\ {\isacharparenleft}simp\ add{\isacharcolon}\ Delay{\isacharunderscore}def{\isacharparenright}\isanewline
\ \ \isacommand{from}\isamarkupfalse%
\ a{\isadigit{2}}\ \isakeyword{and}\ sg{\isadigit{2}}\ \isacommand{have}\isamarkupfalse%
\ sg{\isadigit{3}}{\isacharcolon}{\isachardoublequoteopen}i{\isadigit{2}}\ t{\isadigit{2}}\ {\isacharequal}\ i{\isadigit{1}}\ {\isacharparenleft}t{\isadigit{2}}\ {\isacharminus}\ d{\isacharparenright}{\isachardoublequoteclose}\ \isacommand{by}\isamarkupfalse%
\ simp\isanewline
\ \ \isacommand{from}\isamarkupfalse%
\ h{\isadigit{1}}\ \isakeyword{and}\ a{\isadigit{2}}\ \isakeyword{and}\ h{\isadigit{3}}\ \isakeyword{and}\ sg{\isadigit{3}}\ \isacommand{show}\isamarkupfalse%
\ {\isacharquery}thesis\ \isacommand{by}\isamarkupfalse%
\ auto\isanewline
\isacommand{qed}\isamarkupfalse%
\endisatagproof
{\isafoldproof}%
\isadelimproof
\isanewline
\endisadelimproof
\isanewline
\isanewline
\isacommand{lemma}\isamarkupfalse%
\ Delay{\isacharunderscore}L{\isadigit{2}}{\isacharcolon}\isanewline
\ \isakeyword{assumes}\ h{\isadigit{1}}{\isacharcolon}{\isachardoublequoteopen}{\isasymforall}t{\isadigit{1}}\ {\isacharless}\ t{\isachardot}\ i{\isadigit{1}}\ t{\isadigit{1}}\ {\isacharequal}\ {\isacharbrackleft}{\isacharbrackright}{\isachardoublequoteclose}\isanewline
\ \ \ \ \ \isakeyword{and}\ h{\isadigit{2}}{\isacharcolon}{\isachardoublequoteopen}Delay\ y\ i{\isadigit{1}}\ d\ x\ i{\isadigit{2}}{\isachardoublequoteclose}\isanewline
\ \ \ \isakeyword{shows}\ {\isachardoublequoteopen}{\isasymforall}t{\isadigit{2}}\ {\isacharless}\ t\ {\isacharplus}\ d{\isachardot}\ i{\isadigit{2}}\ t{\isadigit{2}}\ {\isacharequal}\ {\isacharbrackleft}{\isacharbrackright}{\isachardoublequoteclose}\isanewline
\isadelimproof
\endisadelimproof
\isatagproof
\isacommand{using}\isamarkupfalse%
\ assms\ \isacommand{by}\isamarkupfalse%
\ {\isacharparenleft}clarify{\isacharcomma}\ rule\ Delay{\isacharunderscore}L{\isadigit{1}}{\isacharcomma}\ auto{\isacharparenright}%
\endisatagproof
{\isafoldproof}%
\isadelimproof
\isanewline
\endisadelimproof
\isanewline
\isanewline
\isacommand{lemma}\isamarkupfalse%
\ Delay{\isacharunderscore}L{\isadigit{3}}{\isacharcolon}\isanewline
\ \isakeyword{assumes}\ h{\isadigit{1}}{\isacharcolon}{\isachardoublequoteopen}{\isasymforall}t{\isadigit{1}}\ {\isasymle}\ t{\isachardot}\ y\ t{\isadigit{1}}\ {\isacharequal}\ {\isacharbrackleft}{\isacharbrackright}{\isachardoublequoteclose}\isanewline
\ \ \ \ \ \isakeyword{and}\ h{\isadigit{2}}{\isacharcolon}{\isachardoublequoteopen}Delay\ y\ i{\isadigit{1}}\ d\ x\ i{\isadigit{2}}{\isachardoublequoteclose}\isanewline
\ \ \ \ \ \isakeyword{and}\ h{\isadigit{3}}{\isacharcolon}{\isachardoublequoteopen}t{\isadigit{2}}\ {\isasymle}\ t\ {\isacharplus}\ d{\isachardoublequoteclose}\isanewline
\ \ \ \isakeyword{shows}\ {\isachardoublequoteopen}x\ t{\isadigit{2}}\ {\isacharequal}\ {\isacharbrackleft}{\isacharbrackright}{\isachardoublequoteclose}\isanewline
\isadelimproof
\endisadelimproof
\isatagproof
\isacommand{proof}\isamarkupfalse%
\ {\isacharparenleft}cases\ {\isachardoublequoteopen}t{\isadigit{2}}\ {\isacharless}\ d{\isachardoublequoteclose}{\isacharparenright}\isanewline
\ \ \isacommand{assume}\isamarkupfalse%
\ a{\isadigit{1}}{\isacharcolon}{\isachardoublequoteopen}t{\isadigit{2}}\ {\isacharless}\ d{\isachardoublequoteclose}\isanewline
\ \ \isacommand{from}\isamarkupfalse%
\ h{\isadigit{2}}\ \isacommand{have}\isamarkupfalse%
\ sg{\isadigit{1}}{\isacharcolon}{\isachardoublequoteopen}t{\isadigit{2}}\ {\isacharless}\ d\ {\isasymlongrightarrow}\ x\ t{\isadigit{2}}\ {\isacharequal}\ {\isacharbrackleft}{\isacharbrackright}{\isachardoublequoteclose}\isanewline
\ \ \ \ \isacommand{by}\isamarkupfalse%
\ {\isacharparenleft}simp\ add{\isacharcolon}\ Delay{\isacharunderscore}def{\isacharparenright}\isanewline
\ \ \isacommand{from}\isamarkupfalse%
\ sg{\isadigit{1}}\ \isakeyword{and}\ a{\isadigit{1}}\ \isacommand{show}\isamarkupfalse%
\ {\isacharquery}thesis\ \isacommand{by}\isamarkupfalse%
\ simp\isanewline
\isacommand{next}\isamarkupfalse%
\isanewline
\ \ \isacommand{assume}\isamarkupfalse%
\ a{\isadigit{2}}{\isacharcolon}{\isachardoublequoteopen}{\isasymnot}\ t{\isadigit{2}}\ {\isacharless}\ d{\isachardoublequoteclose}\isanewline
\ \ \isacommand{from}\isamarkupfalse%
\ h{\isadigit{2}}\ \isacommand{have}\isamarkupfalse%
\ sg{\isadigit{2}}{\isacharcolon}{\isachardoublequoteopen}d\ {\isasymle}\ t{\isadigit{2}}\ {\isasymlongrightarrow}\ x\ t{\isadigit{2}}\ {\isacharequal}\ y\ {\isacharparenleft}t{\isadigit{2}}\ {\isacharminus}\ d{\isacharparenright}{\isachardoublequoteclose}\isanewline
\ \ \ \ \isacommand{by}\isamarkupfalse%
\ {\isacharparenleft}simp\ add{\isacharcolon}\ Delay{\isacharunderscore}def{\isacharparenright}\isanewline
\ \ \isacommand{from}\isamarkupfalse%
\ a{\isadigit{2}}\ \isakeyword{and}\ sg{\isadigit{2}}\ \isacommand{have}\isamarkupfalse%
\ sg{\isadigit{3}}{\isacharcolon}{\isachardoublequoteopen}x\ t{\isadigit{2}}\ {\isacharequal}\ y\ {\isacharparenleft}t{\isadigit{2}}\ {\isacharminus}\ d{\isacharparenright}{\isachardoublequoteclose}\ \isacommand{by}\isamarkupfalse%
\ simp\isanewline
\ \ \isacommand{from}\isamarkupfalse%
\ h{\isadigit{1}}\ \isakeyword{and}\ a{\isadigit{2}}\ \isakeyword{and}\ h{\isadigit{3}}\ \isakeyword{and}\ sg{\isadigit{3}}\ \isacommand{show}\isamarkupfalse%
\ {\isacharquery}thesis\ \isacommand{by}\isamarkupfalse%
\ auto\isanewline
\isacommand{qed}\isamarkupfalse%
\endisatagproof
{\isafoldproof}%
\isadelimproof
\isanewline
\endisadelimproof
\isanewline
\isanewline
\isacommand{lemma}\isamarkupfalse%
\ Delay{\isacharunderscore}L{\isadigit{4}}{\isacharcolon}\isanewline
\ \isakeyword{assumes}\ h{\isadigit{1}}{\isacharcolon}{\isachardoublequoteopen}{\isasymforall}t{\isadigit{1}}\ {\isasymle}\ t{\isachardot}\ y\ t{\isadigit{1}}\ {\isacharequal}\ {\isacharbrackleft}{\isacharbrackright}{\isachardoublequoteclose}\isanewline
\ \ \ \ \ \isakeyword{and}\ h{\isadigit{2}}{\isacharcolon}{\isachardoublequoteopen}Delay\ y\ i{\isadigit{1}}\ d\ x\ i{\isadigit{2}}{\isachardoublequoteclose}\isanewline
\ \ \ \isakeyword{shows}\ {\isachardoublequoteopen}{\isasymforall}t{\isadigit{2}}\ {\isasymle}\ t\ {\isacharplus}\ d{\isachardot}\ x\ t{\isadigit{2}}\ {\isacharequal}\ {\isacharbrackleft}{\isacharbrackright}{\isachardoublequoteclose}\isanewline
\isadelimproof
\endisadelimproof
\isatagproof
\isacommand{using}\isamarkupfalse%
\ assms\ \isacommand{by}\isamarkupfalse%
\ {\isacharparenleft}clarify{\isacharcomma}\ rule\ Delay{\isacharunderscore}L{\isadigit{3}}{\isacharcomma}\ auto{\isacharparenright}%
\endisatagproof
{\isafoldproof}%
\isadelimproof
\isanewline
\endisadelimproof
\isanewline
\isanewline
\isacommand{lemma}\isamarkupfalse%
\ Delay{\isacharunderscore}lengthOut{\isadigit{1}}{\isacharcolon}\isanewline
\ \ \isakeyword{assumes}\ h{\isadigit{1}}{\isacharcolon}{\isachardoublequoteopen}{\isasymforall}t{\isachardot}\ length\ {\isacharparenleft}x\ t{\isacharparenright}\ {\isasymle}\ Suc\ {\isadigit{0}}{\isachardoublequoteclose}\isanewline
\ \ \ \ \ \ \isakeyword{and}\ h{\isadigit{2}}{\isacharcolon}{\isachardoublequoteopen}Delay\ x\ i{\isadigit{1}}\ d\ y\ i{\isadigit{2}}{\isachardoublequoteclose}\isanewline
\ \ \isakeyword{shows}\ {\isachardoublequoteopen}length\ {\isacharparenleft}y\ t{\isacharparenright}\ {\isasymle}\ Suc\ {\isadigit{0}}{\isachardoublequoteclose}\isanewline
\isadelimproof
\endisadelimproof
\isatagproof
\isacommand{proof}\isamarkupfalse%
\ {\isacharparenleft}cases\ {\isachardoublequoteopen}t\ {\isacharless}\ d{\isachardoublequoteclose}{\isacharparenright}\isanewline
\ \ \isacommand{assume}\isamarkupfalse%
\ a{\isadigit{1}}{\isacharcolon}{\isachardoublequoteopen}t\ {\isacharless}\ d{\isachardoublequoteclose}\isanewline
\ \ \isacommand{from}\isamarkupfalse%
\ h{\isadigit{2}}\ \isacommand{have}\isamarkupfalse%
\ sg{\isadigit{1}}{\isacharcolon}{\isachardoublequoteopen}t\ {\isacharless}\ d\ {\isasymlongrightarrow}\ y\ t\ {\isacharequal}\ {\isacharbrackleft}{\isacharbrackright}{\isachardoublequoteclose}\isanewline
\ \ \ \ \isacommand{by}\isamarkupfalse%
\ {\isacharparenleft}simp\ add{\isacharcolon}\ Delay{\isacharunderscore}def{\isacharparenright}\isanewline
\ \ \isacommand{from}\isamarkupfalse%
\ a{\isadigit{1}}\ \isakeyword{and}\ sg{\isadigit{1}}\ \isacommand{show}\isamarkupfalse%
\ {\isacharquery}thesis\ \isacommand{by}\isamarkupfalse%
\ auto\isanewline
\isacommand{next}\isamarkupfalse%
\isanewline
\ \ \isacommand{assume}\isamarkupfalse%
\ a{\isadigit{2}}{\isacharcolon}{\isachardoublequoteopen}{\isasymnot}\ t\ {\isacharless}\ d{\isachardoublequoteclose}\isanewline
\ \ \isacommand{from}\isamarkupfalse%
\ h{\isadigit{2}}\ \isacommand{have}\isamarkupfalse%
\ sg{\isadigit{2}}{\isacharcolon}{\isachardoublequoteopen}t\ {\isasymge}\ d\ {\isasymlongrightarrow}\ {\isacharparenleft}y\ t\ {\isacharequal}\ x\ {\isacharparenleft}t{\isacharminus}d{\isacharparenright}{\isacharparenright}{\isachardoublequoteclose}\isanewline
\ \ \ \ \isacommand{by}\isamarkupfalse%
\ {\isacharparenleft}simp\ add{\isacharcolon}\ Delay{\isacharunderscore}def{\isacharparenright}\isanewline
\ \ \isacommand{from}\isamarkupfalse%
\ a{\isadigit{2}}\ \isakeyword{and}\ sg{\isadigit{2}}\ \isakeyword{and}\ h{\isadigit{1}}\ \isacommand{show}\isamarkupfalse%
\ {\isacharquery}thesis\ \isacommand{by}\isamarkupfalse%
\ auto\ \isanewline
\isacommand{qed}\isamarkupfalse%
\endisatagproof
{\isafoldproof}%
\isadelimproof
\ \isanewline
\endisadelimproof
\isanewline
\isanewline
\isacommand{lemma}\isamarkupfalse%
\ Delay{\isacharunderscore}msg{\isadigit{1}}{\isacharcolon}\isanewline
\ \ \isakeyword{assumes}\ h{\isadigit{1}}{\isacharcolon}{\isachardoublequoteopen}msg\ {\isacharparenleft}Suc\ {\isadigit{0}}{\isacharparenright}\ x{\isachardoublequoteclose}\isanewline
\ \ \ \ \ \ \isakeyword{and}\ h{\isadigit{2}}{\isacharcolon}{\isachardoublequoteopen}Delay\ x\ i{\isadigit{1}}\ d\ y\ i{\isadigit{2}}{\isachardoublequoteclose}\ \isanewline
\ \ \isakeyword{shows}\ \ \ \ \ \ {\isachardoublequoteopen}msg\ {\isacharparenleft}Suc\ {\isadigit{0}}{\isacharparenright}\ y{\isachardoublequoteclose}\isanewline
\isadelimproof
\endisadelimproof
\isatagproof
\isacommand{using}\isamarkupfalse%
\ assms\isanewline
\ \ \isacommand{by}\isamarkupfalse%
\ {\isacharparenleft}simp\ add{\isacharcolon}\ msg{\isacharunderscore}def\ Delay{\isacharunderscore}lengthOut{\isadigit{1}}{\isacharparenright}%
\endisatagproof
{\isafoldproof}%
\isadelimproof
\endisadelimproof
\isamarkupsubsection{Properties of the Loss component%
}
\isamarkuptrue%
\isacommand{lemma}\isamarkupfalse%
\ Loss{\isacharunderscore}L{\isadigit{1}}{\isacharcolon}\isanewline
\ \isakeyword{assumes}\ h{\isadigit{1}}{\isacharcolon}{\isachardoublequoteopen}{\isasymforall}t{\isadigit{2}}{\isacharless}t{\isachardot}\ i{\isadigit{2}}\ t{\isadigit{2}}\ {\isacharequal}\ {\isacharbrackleft}{\isacharbrackright}{\isachardoublequoteclose}\isanewline
\ \ \ \ \ \isakeyword{and}\ h{\isadigit{2}}{\isacharcolon}{\isachardoublequoteopen}Loss\ lose\ a\ i{\isadigit{2}}\ y\ i{\isachardoublequoteclose}\isanewline
\ \ \ \ \ \isakeyword{and}\ h{\isadigit{3}}{\isacharcolon}{\isachardoublequoteopen}t{\isadigit{2}}\ {\isacharless}\ t{\isachardoublequoteclose}\isanewline
\ \ \ \ \ \isakeyword{and}\ h{\isadigit{4}}{\isacharcolon}{\isachardoublequoteopen}ts\ lose{\isachardoublequoteclose}\isanewline
\ \isakeyword{shows}\ {\isachardoublequoteopen}i\ t{\isadigit{2}}\ {\isacharequal}\ {\isacharbrackleft}{\isacharbrackright}{\isachardoublequoteclose}\isanewline
\isadelimproof
\endisadelimproof
\isatagproof
\isacommand{proof}\isamarkupfalse%
\ {\isacharparenleft}cases\ {\isachardoublequoteopen}lose\ t{\isadigit{2}}\ {\isacharequal}\ {\isacharbrackleft}False{\isacharbrackright}{\isachardoublequoteclose}{\isacharparenright}\isanewline
\ \ \isacommand{assume}\isamarkupfalse%
\ a{\isadigit{1}}{\isacharcolon}{\isachardoublequoteopen}lose\ t{\isadigit{2}}\ {\isacharequal}\ {\isacharbrackleft}False{\isacharbrackright}{\isachardoublequoteclose}\isanewline
\ \ \isacommand{from}\isamarkupfalse%
\ assms\ \isakeyword{and}\ a{\isadigit{1}}\ \isacommand{show}\isamarkupfalse%
\ {\isacharquery}thesis\ \isacommand{by}\isamarkupfalse%
\ {\isacharparenleft}simp\ add{\isacharcolon}\ Loss{\isacharunderscore}def{\isacharparenright}\isanewline
\isacommand{next}\isamarkupfalse%
\isanewline
\ \ \isacommand{assume}\isamarkupfalse%
\ a{\isadigit{2}}{\isacharcolon}{\isachardoublequoteopen}lose\ t{\isadigit{2}}\ {\isasymnoteq}\ {\isacharbrackleft}False{\isacharbrackright}{\isachardoublequoteclose}\isanewline
\ \ \isacommand{from}\isamarkupfalse%
\ a{\isadigit{2}}\ \isakeyword{and}\ h{\isadigit{4}}\ \isacommand{have}\isamarkupfalse%
\ sg{\isadigit{1}}{\isacharcolon}{\isachardoublequoteopen}lose\ t{\isadigit{2}}\ {\isacharequal}\ {\isacharbrackleft}True{\isacharbrackright}{\isachardoublequoteclose}\ \isacommand{by}\isamarkupfalse%
\ {\isacharparenleft}simp\ add{\isacharcolon}\ ts{\isacharunderscore}bool{\isacharunderscore}True{\isacharparenright}\isanewline
\ \ \isacommand{from}\isamarkupfalse%
\ assms\ \isakeyword{and}\ sg{\isadigit{1}}\ \isacommand{show}\isamarkupfalse%
\ {\isacharquery}thesis\ \isacommand{by}\isamarkupfalse%
\ {\isacharparenleft}simp\ add{\isacharcolon}\ Loss{\isacharunderscore}def{\isacharparenright}\isanewline
\isacommand{qed}\isamarkupfalse%
\endisatagproof
{\isafoldproof}%
\isadelimproof
\isanewline
\endisadelimproof
\isanewline
\isacommand{lemma}\isamarkupfalse%
\ Loss{\isacharunderscore}L{\isadigit{2}}{\isacharcolon}\isanewline
\ \isakeyword{assumes}\ h{\isadigit{1}}{\isacharcolon}{\isachardoublequoteopen}{\isasymforall}t{\isadigit{2}}{\isacharless}t{\isachardot}\ i{\isadigit{2}}\ t{\isadigit{2}}\ {\isacharequal}\ {\isacharbrackleft}{\isacharbrackright}{\isachardoublequoteclose}\isanewline
\ \ \ \ \ \isakeyword{and}\ h{\isadigit{2}}{\isacharcolon}{\isachardoublequoteopen}Loss\ lose\ a\ i{\isadigit{2}}\ y\ i{\isachardoublequoteclose}\isanewline
\ \ \ \ \ \isakeyword{and}\ h{\isadigit{3}}{\isacharcolon}{\isachardoublequoteopen}ts\ lose{\isachardoublequoteclose}\isanewline
\ \isakeyword{shows}\ \ {\isachardoublequoteopen}{\isasymforall}t{\isadigit{2}}{\isacharless}t{\isachardot}\ i\ t{\isadigit{2}}\ {\isacharequal}\ {\isacharbrackleft}{\isacharbrackright}{\isachardoublequoteclose}\isanewline
\isadelimproof
\endisadelimproof
\isatagproof
\isacommand{using}\isamarkupfalse%
\ assms\isanewline
\ \ \isacommand{apply}\isamarkupfalse%
\ clarify\ \isanewline
\ \ \isacommand{by}\isamarkupfalse%
\ {\isacharparenleft}rule\ Loss{\isacharunderscore}L{\isadigit{1}}{\isacharcomma}\ auto{\isacharparenright}%
\endisatagproof
{\isafoldproof}%
\isadelimproof
\isanewline
\endisadelimproof
\isanewline
\isacommand{lemma}\isamarkupfalse%
\ Loss{\isacharunderscore}L{\isadigit{3}}{\isacharcolon}\isanewline
\ \isakeyword{assumes}\ h{\isadigit{1}}{\isacharcolon}{\isachardoublequoteopen}{\isasymforall}t{\isadigit{2}}{\isacharless}t{\isachardot}\ a\ t{\isadigit{2}}\ {\isacharequal}\ {\isacharbrackleft}{\isacharbrackright}{\isachardoublequoteclose}\isanewline
\ \ \ \ \ \isakeyword{and}\ h{\isadigit{2}}{\isacharcolon}{\isachardoublequoteopen}Loss\ lose\ a\ i{\isadigit{2}}\ y\ i{\isachardoublequoteclose}\isanewline
\ \ \ \ \ \isakeyword{and}\ h{\isadigit{3}}{\isacharcolon}{\isachardoublequoteopen}t{\isadigit{2}}\ {\isacharless}\ t{\isachardoublequoteclose}\isanewline
\ \ \ \ \ \isakeyword{and}\ h{\isadigit{4}}{\isacharcolon}{\isachardoublequoteopen}ts\ lose{\isachardoublequoteclose}\isanewline
\ \isakeyword{shows}\ {\isachardoublequoteopen}y\ t{\isadigit{2}}\ {\isacharequal}\ {\isacharbrackleft}{\isacharbrackright}{\isachardoublequoteclose}\isanewline
\isadelimproof
\endisadelimproof
\isatagproof
\isacommand{proof}\isamarkupfalse%
\ {\isacharparenleft}cases\ {\isachardoublequoteopen}lose\ t{\isadigit{2}}\ {\isacharequal}\ {\isacharbrackleft}False{\isacharbrackright}{\isachardoublequoteclose}{\isacharparenright}\isanewline
\ \ \isacommand{assume}\isamarkupfalse%
\ a{\isadigit{1}}{\isacharcolon}{\isachardoublequoteopen}lose\ t{\isadigit{2}}\ {\isacharequal}\ {\isacharbrackleft}False{\isacharbrackright}{\isachardoublequoteclose}\isanewline
\ \ \isacommand{from}\isamarkupfalse%
\ assms\ \isakeyword{and}\ a{\isadigit{1}}\ \isacommand{show}\isamarkupfalse%
\ {\isacharquery}thesis\ \isacommand{by}\isamarkupfalse%
\ {\isacharparenleft}simp\ add{\isacharcolon}\ Loss{\isacharunderscore}def{\isacharparenright}\isanewline
\isacommand{next}\isamarkupfalse%
\isanewline
\ \ \isacommand{assume}\isamarkupfalse%
\ a{\isadigit{2}}{\isacharcolon}{\isachardoublequoteopen}lose\ t{\isadigit{2}}\ {\isasymnoteq}\ {\isacharbrackleft}False{\isacharbrackright}{\isachardoublequoteclose}\isanewline
\ \ \isacommand{from}\isamarkupfalse%
\ a{\isadigit{2}}\ \isakeyword{and}\ h{\isadigit{4}}\ \isacommand{have}\isamarkupfalse%
\ sg{\isadigit{1}}{\isacharcolon}{\isachardoublequoteopen}lose\ t{\isadigit{2}}\ {\isacharequal}\ {\isacharbrackleft}True{\isacharbrackright}{\isachardoublequoteclose}\ \isacommand{by}\isamarkupfalse%
\ {\isacharparenleft}simp\ add{\isacharcolon}\ ts{\isacharunderscore}bool{\isacharunderscore}True{\isacharparenright}\isanewline
\ \ \isacommand{from}\isamarkupfalse%
\ assms\ \isakeyword{and}\ sg{\isadigit{1}}\ \isacommand{show}\isamarkupfalse%
\ {\isacharquery}thesis\ \isacommand{by}\isamarkupfalse%
\ {\isacharparenleft}simp\ add{\isacharcolon}\ Loss{\isacharunderscore}def{\isacharparenright}\isanewline
\isacommand{qed}\isamarkupfalse%
\endisatagproof
{\isafoldproof}%
\isadelimproof
\isanewline
\endisadelimproof
\isanewline
\isacommand{lemma}\isamarkupfalse%
\ Loss{\isacharunderscore}L{\isadigit{4}}{\isacharcolon}\isanewline
\ \isakeyword{assumes}\ h{\isadigit{1}}{\isacharcolon}{\isachardoublequoteopen}{\isasymforall}t{\isadigit{2}}{\isacharless}t{\isachardot}\ a\ t{\isadigit{2}}\ {\isacharequal}\ {\isacharbrackleft}{\isacharbrackright}{\isachardoublequoteclose}\isanewline
\ \ \ \ \ \isakeyword{and}\ h{\isadigit{2}}{\isacharcolon}{\isachardoublequoteopen}Loss\ lose\ a\ i{\isadigit{2}}\ y\ i{\isachardoublequoteclose}\isanewline
\ \ \ \ \ \isakeyword{and}\ h{\isadigit{3}}{\isacharcolon}{\isachardoublequoteopen}ts\ lose{\isachardoublequoteclose}\isanewline
\ \isakeyword{shows}\ \ {\isachardoublequoteopen}{\isasymforall}t{\isadigit{2}}{\isacharless}t{\isachardot}\ y\ t{\isadigit{2}}\ {\isacharequal}\ {\isacharbrackleft}{\isacharbrackright}{\isachardoublequoteclose}\isanewline
\isadelimproof
\endisadelimproof
\isatagproof
\isacommand{using}\isamarkupfalse%
\ assms\isanewline
\ \ \isacommand{apply}\isamarkupfalse%
\ clarify\ \isanewline
\ \ \isacommand{by}\isamarkupfalse%
\ {\isacharparenleft}rule\ Loss{\isacharunderscore}L{\isadigit{3}}{\isacharcomma}\ auto{\isacharparenright}%
\endisatagproof
{\isafoldproof}%
\isadelimproof
\isanewline
\endisadelimproof
\isanewline
\isacommand{lemma}\isamarkupfalse%
\ Loss{\isacharunderscore}L{\isadigit{5}}{\isacharcolon}\isanewline
\ \isakeyword{assumes}\ h{\isadigit{1}}{\isacharcolon}{\isachardoublequoteopen}{\isasymforall}t{\isadigit{1}}\ {\isasymle}\ t{\isachardot}\ a\ t{\isadigit{1}}\ {\isacharequal}\ {\isacharbrackleft}{\isacharbrackright}{\isachardoublequoteclose}\isanewline
\ \ \ \ \ \isakeyword{and}\ h{\isadigit{2}}{\isacharcolon}{\isachardoublequoteopen}Loss\ lose\ a\ i{\isadigit{2}}\ y\ i{\isachardoublequoteclose}\isanewline
\ \ \ \ \ \isakeyword{and}\ h{\isadigit{3}}{\isacharcolon}{\isachardoublequoteopen}t{\isadigit{2}}\ {\isasymle}\ t{\isachardoublequoteclose}\isanewline
\ \ \ \ \ \isakeyword{and}\ h{\isadigit{4}}{\isacharcolon}{\isachardoublequoteopen}ts\ lose{\isachardoublequoteclose}\isanewline
\ \isakeyword{shows}\ {\isachardoublequoteopen}y\ t{\isadigit{2}}\ {\isacharequal}\ {\isacharbrackleft}{\isacharbrackright}{\isachardoublequoteclose}\isanewline
\isadelimproof
\endisadelimproof
\isatagproof
\isacommand{proof}\isamarkupfalse%
\ {\isacharparenleft}cases\ {\isachardoublequoteopen}lose\ t{\isadigit{2}}\ {\isacharequal}\ {\isacharbrackleft}False{\isacharbrackright}{\isachardoublequoteclose}{\isacharparenright}\isanewline
\ \ \isacommand{assume}\isamarkupfalse%
\ a{\isadigit{1}}{\isacharcolon}{\isachardoublequoteopen}lose\ t{\isadigit{2}}\ {\isacharequal}\ {\isacharbrackleft}False{\isacharbrackright}{\isachardoublequoteclose}\isanewline
\ \ \isacommand{from}\isamarkupfalse%
\ assms\ \isakeyword{and}\ a{\isadigit{1}}\ \isacommand{show}\isamarkupfalse%
\ {\isacharquery}thesis\ \isacommand{by}\isamarkupfalse%
\ {\isacharparenleft}simp\ add{\isacharcolon}\ Loss{\isacharunderscore}def{\isacharparenright}\isanewline
\isacommand{next}\isamarkupfalse%
\isanewline
\ \ \isacommand{assume}\isamarkupfalse%
\ a{\isadigit{2}}{\isacharcolon}{\isachardoublequoteopen}lose\ t{\isadigit{2}}\ {\isasymnoteq}\ {\isacharbrackleft}False{\isacharbrackright}{\isachardoublequoteclose}\isanewline
\ \ \isacommand{from}\isamarkupfalse%
\ a{\isadigit{2}}\ \isakeyword{and}\ h{\isadigit{4}}\ \isacommand{have}\isamarkupfalse%
\ sg{\isadigit{1}}{\isacharcolon}{\isachardoublequoteopen}lose\ t{\isadigit{2}}\ {\isacharequal}\ {\isacharbrackleft}True{\isacharbrackright}{\isachardoublequoteclose}\ \isacommand{by}\isamarkupfalse%
\ {\isacharparenleft}simp\ add{\isacharcolon}\ ts{\isacharunderscore}bool{\isacharunderscore}True{\isacharparenright}\isanewline
\ \ \isacommand{from}\isamarkupfalse%
\ assms\ \isakeyword{and}\ sg{\isadigit{1}}\ \isacommand{show}\isamarkupfalse%
\ {\isacharquery}thesis\ \isacommand{by}\isamarkupfalse%
\ {\isacharparenleft}simp\ add{\isacharcolon}\ Loss{\isacharunderscore}def{\isacharparenright}\isanewline
\isacommand{qed}\isamarkupfalse%
\endisatagproof
{\isafoldproof}%
\isadelimproof
\isanewline
\endisadelimproof
\isanewline
\isacommand{lemma}\isamarkupfalse%
\ Loss{\isacharunderscore}L{\isadigit{5}}Suc{\isacharcolon}\isanewline
\ \isakeyword{assumes}\ h{\isadigit{1}}{\isacharcolon}{\isachardoublequoteopen}{\isasymforall}j\ {\isasymle}\ d{\isachardot}\ a\ {\isacharparenleft}t\ {\isacharplus}\ Suc\ j{\isacharparenright}\ {\isacharequal}\ {\isacharbrackleft}{\isacharbrackright}{\isachardoublequoteclose}\isanewline
\ \ \ \ \ \isakeyword{and}\ h{\isadigit{2}}{\isacharcolon}{\isachardoublequoteopen}Loss\ lose\ a\ i{\isadigit{2}}\ y\ i{\isachardoublequoteclose}\isanewline
\ \ \ \ \ \isakeyword{and}\ h{\isadigit{3}}{\isacharcolon}{\isachardoublequoteopen}Suc\ j\ {\isasymle}\ d{\isachardoublequoteclose}\isanewline
\ \ \ \ \ \isakeyword{and}\ h{\isadigit{4}}{\isacharcolon}{\isachardoublequoteopen}ts\ lose{\isachardoublequoteclose}\isanewline
\ \isakeyword{shows}\ {\isachardoublequoteopen}y\ {\isacharparenleft}t\ {\isacharplus}\ Suc\ j{\isacharparenright}\ {\isacharequal}\ {\isacharbrackleft}{\isacharbrackright}{\isachardoublequoteclose}\isanewline
\isadelimproof
\endisadelimproof
\isatagproof
\isacommand{proof}\isamarkupfalse%
\ {\isacharparenleft}cases\ {\isachardoublequoteopen}lose\ {\isacharparenleft}t\ {\isacharplus}\ Suc\ j{\isacharparenright}\ {\isacharequal}\ {\isacharbrackleft}False{\isacharbrackright}{\isachardoublequoteclose}{\isacharparenright}\isanewline
\ \ \isacommand{assume}\isamarkupfalse%
\ a{\isadigit{1}}{\isacharcolon}{\isachardoublequoteopen}lose\ {\isacharparenleft}t\ {\isacharplus}\ Suc\ j{\isacharparenright}\ {\isacharequal}\ {\isacharbrackleft}False{\isacharbrackright}{\isachardoublequoteclose}\isanewline
\ \ \isacommand{from}\isamarkupfalse%
\ assms\ \isakeyword{and}\ a{\isadigit{1}}\ \isacommand{show}\isamarkupfalse%
\ {\isacharquery}thesis\ \isacommand{by}\isamarkupfalse%
\ {\isacharparenleft}simp\ add{\isacharcolon}\ Loss{\isacharunderscore}def{\isacharparenright}\isanewline
\isacommand{next}\isamarkupfalse%
\isanewline
\ \ \isacommand{assume}\isamarkupfalse%
\ a{\isadigit{2}}{\isacharcolon}{\isachardoublequoteopen}lose\ {\isacharparenleft}t\ {\isacharplus}\ Suc\ j{\isacharparenright}\ {\isasymnoteq}\ {\isacharbrackleft}False{\isacharbrackright}{\isachardoublequoteclose}\isanewline
\ \ \isacommand{from}\isamarkupfalse%
\ a{\isadigit{2}}\ \isakeyword{and}\ h{\isadigit{4}}\ \isacommand{have}\isamarkupfalse%
\ sg{\isadigit{1}}{\isacharcolon}{\isachardoublequoteopen}lose\ {\isacharparenleft}t\ {\isacharplus}\ Suc\ j{\isacharparenright}\ {\isacharequal}\ {\isacharbrackleft}True{\isacharbrackright}{\isachardoublequoteclose}\ \isacommand{by}\isamarkupfalse%
\ {\isacharparenleft}simp\ add{\isacharcolon}\ ts{\isacharunderscore}bool{\isacharunderscore}True{\isacharparenright}\isanewline
\ \ \isacommand{from}\isamarkupfalse%
\ assms\ \isakeyword{and}\ sg{\isadigit{1}}\ \isacommand{show}\isamarkupfalse%
\ {\isacharquery}thesis\ \isacommand{by}\isamarkupfalse%
\ {\isacharparenleft}simp\ add{\isacharcolon}\ Loss{\isacharunderscore}def{\isacharparenright}\isanewline
\isacommand{qed}\isamarkupfalse%
\endisatagproof
{\isafoldproof}%
\isadelimproof
\isanewline
\endisadelimproof
\isanewline
\isacommand{lemma}\isamarkupfalse%
\ Loss{\isacharunderscore}L{\isadigit{6}}{\isacharcolon}\isanewline
\ \isakeyword{assumes}\ h{\isadigit{1}}{\isacharcolon}{\isachardoublequoteopen}{\isasymforall}t{\isadigit{2}}\ {\isasymle}\ t{\isachardot}\ a\ t{\isadigit{2}}\ {\isacharequal}\ {\isacharbrackleft}{\isacharbrackright}{\isachardoublequoteclose}\isanewline
\ \ \ \ \ \isakeyword{and}\ h{\isadigit{2}}{\isacharcolon}{\isachardoublequoteopen}Loss\ lose\ a\ i{\isadigit{2}}\ y\ i{\isachardoublequoteclose}\isanewline
\ \ \ \ \ \isakeyword{and}\ h{\isadigit{3}}{\isacharcolon}{\isachardoublequoteopen}ts\ lose{\isachardoublequoteclose}\isanewline
\ \isakeyword{shows}\ \ {\isachardoublequoteopen}{\isasymforall}t{\isadigit{2}}\ {\isasymle}\ t{\isachardot}\ y\ t{\isadigit{2}}\ {\isacharequal}\ {\isacharbrackleft}{\isacharbrackright}{\isachardoublequoteclose}\isanewline
\isadelimproof
\endisadelimproof
\isatagproof
\isacommand{using}\isamarkupfalse%
\ assms\isanewline
\ \ \isacommand{apply}\isamarkupfalse%
\ clarify\ \isanewline
\ \ \isacommand{by}\isamarkupfalse%
\ {\isacharparenleft}rule\ Loss{\isacharunderscore}L{\isadigit{5}}{\isacharcomma}\ auto{\isacharparenright}%
\endisatagproof
{\isafoldproof}%
\isadelimproof
\isanewline
\endisadelimproof
\isanewline
\isacommand{lemma}\isamarkupfalse%
\ Loss{\isacharunderscore}lengthOut{\isadigit{1}}{\isacharcolon}\isanewline
\ \ \isakeyword{assumes}\ h{\isadigit{1}}{\isacharcolon}{\isachardoublequoteopen}{\isasymforall}t{\isachardot}\ length\ {\isacharparenleft}a\ t{\isacharparenright}\ {\isasymle}\ Suc\ {\isadigit{0}}{\isachardoublequoteclose}\isanewline
\ \ \ \ \ \ \isakeyword{and}\ h{\isadigit{2}}{\isacharcolon}{\isachardoublequoteopen}Loss\ lose\ a\ i{\isadigit{2}}\ x\ i{\isachardoublequoteclose}\isanewline
\ \ \isakeyword{shows}\ {\isachardoublequoteopen}length\ {\isacharparenleft}x\ t{\isacharparenright}\ {\isasymle}\ Suc\ {\isadigit{0}}{\isachardoublequoteclose}\isanewline
\isadelimproof
\endisadelimproof
\isatagproof
\isacommand{proof}\isamarkupfalse%
\ {\isacharparenleft}cases\ {\isachardoublequoteopen}lose\ t\ {\isacharequal}\ \ {\isacharbrackleft}False{\isacharbrackright}{\isachardoublequoteclose}{\isacharparenright}\isanewline
\ \ \isacommand{assume}\isamarkupfalse%
\ a{\isadigit{1}}{\isacharcolon}{\isachardoublequoteopen}lose\ t\ {\isacharequal}\ \ {\isacharbrackleft}False{\isacharbrackright}{\isachardoublequoteclose}\isanewline
\ \ \isacommand{from}\isamarkupfalse%
\ a{\isadigit{1}}\ \isakeyword{and}\ h{\isadigit{2}}\ \isacommand{have}\isamarkupfalse%
\ sg{\isadigit{1}}{\isacharcolon}{\isachardoublequoteopen}x\ t\ {\isacharequal}\ a\ t{\isachardoublequoteclose}\ \isacommand{by}\isamarkupfalse%
\ {\isacharparenleft}simp\ add{\isacharcolon}\ Loss{\isacharunderscore}def{\isacharparenright}\isanewline
\ \ \isacommand{from}\isamarkupfalse%
\ h{\isadigit{1}}\ \isacommand{have}\isamarkupfalse%
\ sg{\isadigit{2}}{\isacharcolon}{\isachardoublequoteopen}length\ {\isacharparenleft}a\ t{\isacharparenright}\ {\isasymle}\ Suc\ {\isadigit{0}}{\isachardoublequoteclose}\ \isacommand{by}\isamarkupfalse%
\ auto\isanewline
\ \ \isacommand{from}\isamarkupfalse%
\ sg{\isadigit{1}}\ \isakeyword{and}\ sg{\isadigit{2}}\ \isacommand{show}\isamarkupfalse%
\ {\isacharquery}thesis\ \isacommand{by}\isamarkupfalse%
\ simp\isanewline
\isacommand{next}\isamarkupfalse%
\isanewline
\ \ \isacommand{assume}\isamarkupfalse%
\ a{\isadigit{2}}{\isacharcolon}{\isachardoublequoteopen}lose\ t\ {\isasymnoteq}\ {\isacharbrackleft}False{\isacharbrackright}{\isachardoublequoteclose}\isanewline
\ \ \isacommand{from}\isamarkupfalse%
\ a{\isadigit{2}}\ \isakeyword{and}\ h{\isadigit{2}}\ \isacommand{have}\isamarkupfalse%
\ sg{\isadigit{2}}{\isacharcolon}{\isachardoublequoteopen}x\ t\ {\isacharequal}\ {\isacharbrackleft}{\isacharbrackright}{\isachardoublequoteclose}\ \isacommand{by}\isamarkupfalse%
\ {\isacharparenleft}simp\ add{\isacharcolon}\ Loss{\isacharunderscore}def{\isacharparenright}\isanewline
\ \ \isacommand{from}\isamarkupfalse%
\ sg{\isadigit{2}}\ \isacommand{show}\isamarkupfalse%
\ {\isacharquery}thesis\ \isacommand{by}\isamarkupfalse%
\ simp\isanewline
\isacommand{qed}\isamarkupfalse%
\endisatagproof
{\isafoldproof}%
\isadelimproof
\isanewline
\endisadelimproof
\isanewline
\isacommand{lemma}\isamarkupfalse%
\ Loss{\isacharunderscore}lengthOut{\isadigit{2}}{\isacharcolon}\isanewline
\ \ \isakeyword{assumes}\ h{\isadigit{1}}{\isacharcolon}{\isachardoublequoteopen}{\isasymforall}t{\isachardot}\ length\ {\isacharparenleft}a\ t{\isacharparenright}\ {\isasymle}\ Suc\ {\isadigit{0}}{\isachardoublequoteclose}\isanewline
\ \ \ \ \ \ \isakeyword{and}\ h{\isadigit{2}}{\isacharcolon}{\isachardoublequoteopen}Loss\ lose\ a\ i{\isadigit{2}}\ x\ i{\isachardoublequoteclose}\isanewline
\ \ \isakeyword{shows}\ {\isachardoublequoteopen}{\isasymforall}t{\isachardot}\ length\ {\isacharparenleft}x\ t{\isacharparenright}\ {\isasymle}\ Suc\ {\isadigit{0}}{\isachardoublequoteclose}\isanewline
\isadelimproof
\endisadelimproof
\isatagproof
\isacommand{using}\isamarkupfalse%
\ assms\isanewline
\ \ \isacommand{by}\isamarkupfalse%
\ {\isacharparenleft}simp\ add{\isacharcolon}\ Loss{\isacharunderscore}lengthOut{\isadigit{1}}{\isacharparenright}%
\endisatagproof
{\isafoldproof}%
\isadelimproof
\isanewline
\endisadelimproof
\isanewline
\isacommand{lemma}\isamarkupfalse%
\ Loss{\isacharunderscore}msg{\isadigit{1}}{\isacharcolon}\isanewline
\ \ \isakeyword{assumes}\ h{\isadigit{1}}{\isacharcolon}{\isachardoublequoteopen}msg\ {\isacharparenleft}Suc\ {\isadigit{0}}{\isacharparenright}\ a{\isachardoublequoteclose}\ \isanewline
\ \ \ \ \ \ \isakeyword{and}\ h{\isadigit{2}}{\isacharcolon}{\isachardoublequoteopen}Loss\ lose\ a\ i{\isadigit{2}}\ x\ i{\isachardoublequoteclose}\isanewline
\ \ \isakeyword{shows}\ \ \ \ \ \ {\isachardoublequoteopen}msg\ {\isacharparenleft}Suc\ {\isadigit{0}}{\isacharparenright}\ x{\isachardoublequoteclose}\isanewline
\isadelimproof
\endisadelimproof
\isatagproof
\isacommand{using}\isamarkupfalse%
\ assms\isanewline
\ \ \isacommand{by}\isamarkupfalse%
\ {\isacharparenleft}simp\ add{\isacharcolon}\ msg{\isacharunderscore}def\ Loss{\isacharunderscore}def\ Loss{\isacharunderscore}lengthOut{\isadigit{1}}{\isacharparenright}%
\endisatagproof
{\isafoldproof}%
\isadelimproof
\endisadelimproof
\isamarkupsubsection{Properties of the composition of Delay and Loss components%
}
\isamarkuptrue%
\isacommand{lemma}\isamarkupfalse%
\ Loss{\isacharunderscore}Delay{\isacharunderscore}length{\isacharunderscore}y{\isacharcolon}\isanewline
\ \ \isakeyword{assumes}\ h{\isadigit{1}}{\isacharcolon}{\isachardoublequoteopen}{\isasymforall}t{\isachardot}\ length\ {\isacharparenleft}a\ t{\isacharparenright}\ {\isasymle}\ Suc\ {\isadigit{0}}{\isachardoublequoteclose}\isanewline
\ \ \ \ \ \ \isakeyword{and}\ h{\isadigit{2}}{\isacharcolon}{\isachardoublequoteopen}Delay\ x\ i{\isadigit{1}}\ d\ y\ i{\isadigit{2}}{\isachardoublequoteclose}\isanewline
\ \ \ \ \ \ \isakeyword{and}\ h{\isadigit{3}}{\isacharcolon}{\isachardoublequoteopen}Loss\ lose\ a\ i{\isadigit{2}}\ x\ i{\isachardoublequoteclose}\isanewline
\ \ \isakeyword{shows}\ {\isachardoublequoteopen}length\ {\isacharparenleft}y\ t{\isacharparenright}\ {\isasymle}\ Suc\ {\isadigit{0}}{\isachardoublequoteclose}\isanewline
\isadelimproof
\endisadelimproof
\isatagproof
\isacommand{proof}\isamarkupfalse%
\ {\isacharminus}\ \isanewline
\ \ \isacommand{from}\isamarkupfalse%
\ h{\isadigit{1}}\ \isakeyword{and}\ h{\isadigit{3}}\ \isacommand{have}\isamarkupfalse%
\ sg{\isadigit{1}}{\isacharcolon}{\isachardoublequoteopen}{\isasymforall}t{\isachardot}\ length\ {\isacharparenleft}x\ t{\isacharparenright}\ {\isasymle}\ Suc\ {\isadigit{0}}{\isachardoublequoteclose}\isanewline
\ \ \ \ \isacommand{by}\isamarkupfalse%
\ {\isacharparenleft}simp\ add{\isacharcolon}\ Loss{\isacharunderscore}lengthOut{\isadigit{2}}{\isacharparenright}\isanewline
\ \ \isacommand{from}\isamarkupfalse%
\ this\ \isakeyword{and}\ h{\isadigit{2}}\ \isacommand{show}\isamarkupfalse%
\ {\isacharquery}thesis\ \isanewline
\ \ \ \ \isacommand{by}\isamarkupfalse%
\ {\isacharparenleft}simp\ add{\isacharcolon}\ Delay{\isacharunderscore}lengthOut{\isadigit{1}}{\isacharparenright}\isanewline
\isacommand{qed}\isamarkupfalse%
\endisatagproof
{\isafoldproof}%
\isadelimproof
\isanewline
\endisadelimproof
\isanewline
\isacommand{lemma}\isamarkupfalse%
\ Loss{\isacharunderscore}Delay{\isacharunderscore}msg{\isacharunderscore}a{\isacharcolon}\isanewline
\ \ \isakeyword{assumes}\ h{\isadigit{1}}{\isacharcolon}{\isachardoublequoteopen}msg\ {\isacharparenleft}Suc\ {\isadigit{0}}{\isacharparenright}\ a{\isachardoublequoteclose}\isanewline
\ \ \ \ \ \ \isakeyword{and}\ h{\isadigit{2}}{\isacharcolon}{\isachardoublequoteopen}Delay\ x\ i{\isadigit{1}}\ d\ y\ i{\isadigit{2}}{\isachardoublequoteclose}\isanewline
\ \ \ \ \ \ \isakeyword{and}\ h{\isadigit{3}}{\isacharcolon}{\isachardoublequoteopen}Loss\ lose\ a\ i{\isadigit{2}}\ x\ i{\isachardoublequoteclose}\isanewline
\ \ \isakeyword{shows}\ \ \ \ \ \ {\isachardoublequoteopen}msg\ {\isacharparenleft}Suc\ {\isadigit{0}}{\isacharparenright}\ y{\isachardoublequoteclose}\isanewline
\isadelimproof
\endisadelimproof
\isatagproof
\isacommand{using}\isamarkupfalse%
\ assms\isanewline
\ \ \isacommand{by}\isamarkupfalse%
\ {\isacharparenleft}simp\ add{\isacharcolon}\ msg{\isacharunderscore}def\ Loss{\isacharunderscore}Delay{\isacharunderscore}length{\isacharunderscore}y{\isacharparenright}%
\endisatagproof
{\isafoldproof}%
\isadelimproof
\endisadelimproof
\isamarkupsubsection{Auxiliary Lemmas%
}
\isamarkuptrue%
\isacommand{lemma}\isamarkupfalse%
\ inf{\isacharunderscore}last{\isacharunderscore}ti{\isadigit{2}}{\isacharcolon}\isanewline
\ \ \isakeyword{assumes}\ h{\isadigit{1}}{\isacharcolon}{\isachardoublequoteopen}inf{\isacharunderscore}last{\isacharunderscore}ti\ dt\ {\isacharparenleft}Suc\ {\isacharparenleft}Suc\ t{\isacharparenright}{\isacharparenright}\ {\isasymnoteq}\ {\isacharbrackleft}{\isacharbrackright}{\isachardoublequoteclose}\isanewline
\ \ \isakeyword{shows}\ \ \ \ \ \ {\isachardoublequoteopen}inf{\isacharunderscore}last{\isacharunderscore}ti\ dt\ {\isacharparenleft}Suc\ {\isacharparenleft}Suc\ {\isacharparenleft}t\ {\isacharplus}\ k{\isacharparenright}{\isacharparenright}{\isacharparenright}\ {\isasymnoteq}\ {\isacharbrackleft}{\isacharbrackright}{\isachardoublequoteclose}\isanewline
\isadelimproof
\endisadelimproof
\isatagproof
\isacommand{using}\isamarkupfalse%
\ assms\isanewline
\isacommand{proof}\isamarkupfalse%
\ {\isacharparenleft}induct\ k{\isacharparenright}\isanewline
\ \ \isacommand{case}\isamarkupfalse%
\ {\isadigit{0}}\isanewline
\ \ \isacommand{from}\isamarkupfalse%
\ this\ \isacommand{show}\isamarkupfalse%
\ {\isacharquery}case\ \isacommand{by}\isamarkupfalse%
\ auto\isanewline
\isacommand{next}\isamarkupfalse%
\isanewline
\ \ \isacommand{case}\isamarkupfalse%
\ Suc\isanewline
\ \ \isacommand{from}\isamarkupfalse%
\ this\ \isacommand{show}\isamarkupfalse%
\ {\isacharquery}case\ \isacommand{by}\isamarkupfalse%
\ auto\isanewline
\isacommand{qed}\isamarkupfalse%
\endisatagproof
{\isafoldproof}%
\isadelimproof
\isanewline
\endisadelimproof
\ \ \isanewline
\isanewline
\isacommand{lemma}\isamarkupfalse%
\ aux{\isacharunderscore}ack{\isacharunderscore}t{\isadigit{2}}{\isacharcolon}\isanewline
\ \ \isakeyword{assumes}\ h{\isadigit{1}}{\isacharcolon}{\isachardoublequoteopen}{\isasymforall}m{\isasymle}k{\isachardot}\ ack\ {\isacharparenleft}Suc\ {\isacharparenleft}Suc\ {\isacharparenleft}t\ {\isacharplus}\ m{\isacharparenright}{\isacharparenright}{\isacharparenright}\ {\isacharequal}\ {\isacharbrackleft}connection{\isacharunderscore}ok{\isacharbrackright}{\isachardoublequoteclose}\isanewline
\ \ \ \ \ \ \isakeyword{and}\ h{\isadigit{2}}{\isacharcolon}{\isachardoublequoteopen}Suc\ {\isacharparenleft}Suc\ t{\isacharparenright}\ {\isacharless}\ t{\isadigit{2}}{\isachardoublequoteclose}\isanewline
\ \ \ \ \ \ \isakeyword{and}\ h{\isadigit{3}}{\isacharcolon}{\isachardoublequoteopen}t{\isadigit{2}}\ {\isacharless}\ t\ {\isacharplus}\ {\isadigit{3}}\ {\isacharplus}\ k{\isachardoublequoteclose}\isanewline
\ \ \isakeyword{shows}\ \ \ \ \ \ {\isachardoublequoteopen}ack\ t{\isadigit{2}}\ {\isacharequal}\ {\isacharbrackleft}connection{\isacharunderscore}ok{\isacharbrackright}{\isachardoublequoteclose}\isanewline
\isadelimproof
\endisadelimproof
\isatagproof
\isacommand{proof}\isamarkupfalse%
\ {\isacharminus}\isanewline
\ \ \isacommand{from}\isamarkupfalse%
\ h{\isadigit{3}}\ \isacommand{have}\isamarkupfalse%
\ sg{\isadigit{1}}{\isacharcolon}{\isachardoublequoteopen}t{\isadigit{2}}\ {\isacharminus}\ Suc\ {\isacharparenleft}Suc\ t{\isacharparenright}\ {\isasymle}\ k{\isachardoublequoteclose}\ \isacommand{by}\isamarkupfalse%
\ arith\isanewline
\ \ \isacommand{from}\isamarkupfalse%
\ h{\isadigit{1}}\ \isakeyword{and}\ sg{\isadigit{1}}\ \isanewline
\ \ \ \ \isacommand{obtain}\isamarkupfalse%
\ m\ \isakeyword{where}\ a{\isadigit{1}}{\isacharcolon}{\isachardoublequoteopen}m\ {\isacharequal}\ t{\isadigit{2}}\ {\isacharminus}\ Suc\ {\isacharparenleft}Suc\ t{\isacharparenright}{\isachardoublequoteclose}\ \isanewline
\ \ \ \ \ \ \ \ \ \ \ \ \ \ \ \isakeyword{and}\ a{\isadigit{2}}{\isacharcolon}{\isachardoublequoteopen}ack\ {\isacharparenleft}Suc\ {\isacharparenleft}Suc\ {\isacharparenleft}t\ {\isacharplus}\ m{\isacharparenright}{\isacharparenright}{\isacharparenright}\ {\isacharequal}\ {\isacharbrackleft}connection{\isacharunderscore}ok{\isacharbrackright}{\isachardoublequoteclose}\ \isanewline
\ \ \ \ \isacommand{by}\isamarkupfalse%
\ auto\ \isanewline
\ \ \isacommand{from}\isamarkupfalse%
\ h{\isadigit{2}}\ \isacommand{have}\isamarkupfalse%
\ sg{\isadigit{2}}{\isacharcolon}{\isachardoublequoteopen}{\isacharparenleft}Suc\ {\isacharparenleft}Suc\ {\isacharparenleft}t{\isadigit{2}}\ {\isacharminus}\ {\isadigit{2}}{\isacharparenright}{\isacharparenright}{\isacharparenright}\ {\isacharequal}\ \ t{\isadigit{2}}{\isachardoublequoteclose}\ \isacommand{by}\isamarkupfalse%
\ arith\isanewline
\ \ \isacommand{from}\isamarkupfalse%
\ h{\isadigit{2}}\ \isacommand{have}\isamarkupfalse%
\ sg{\isadigit{3}}{\isacharcolon}{\isachardoublequoteopen}Suc\ {\isacharparenleft}Suc\ {\isacharparenleft}t\ {\isacharplus}\ {\isacharparenleft}t{\isadigit{2}}\ {\isacharminus}\ Suc\ {\isacharparenleft}Suc\ t{\isacharparenright}{\isacharparenright}{\isacharparenright}{\isacharparenright}\ {\isacharequal}\ \ t{\isadigit{2}}{\isachardoublequoteclose}\ \isacommand{by}\isamarkupfalse%
\ arith\isanewline
\ \ \isacommand{from}\isamarkupfalse%
\ sg{\isadigit{1}}\ \isakeyword{and}\ a{\isadigit{1}}\ \isakeyword{and}\ a{\isadigit{2}}\ \isakeyword{and}\ sg{\isadigit{2}}\ \isakeyword{and}\ sg{\isadigit{3}}\ \isacommand{show}\isamarkupfalse%
\ {\isacharquery}thesis\ \isacommand{by}\isamarkupfalse%
\ simp\isanewline
\isacommand{qed}\isamarkupfalse%
\endisatagproof
{\isafoldproof}%
\isadelimproof
\isanewline
\endisadelimproof
\isanewline
\isanewline
\isacommand{lemma}\isamarkupfalse%
\ aux{\isacharunderscore}lemma{\isacharunderscore}lose{\isacharunderscore}{\isadigit{1}}{\isacharcolon}\isanewline
\ \ \isakeyword{assumes}\ h{\isadigit{1}}{\isacharcolon}{\isachardoublequoteopen}{\isasymforall}j{\isasymle}{\isacharparenleft}{\isacharparenleft}{\isadigit{2}}{\isacharcolon}{\isacharcolon}nat{\isacharparenright}\ {\isacharasterisk}\ d\ {\isacharplus}\ {\isacharparenleft}{\isacharparenleft}{\isadigit{4}}{\isacharcolon}{\isacharcolon}nat{\isacharparenright}\ {\isacharplus}\ k{\isacharparenright}{\isacharparenright}{\isachardot}\ {\isacharparenleft}lose\ {\isacharparenleft}t\ {\isacharplus}\ j{\isacharparenright}\ {\isacharequal}\ x{\isacharparenright}{\isachardoublequoteclose}\isanewline
\ \ \ \ \ \ \isakeyword{and}\ h{\isadigit{2}}{\isacharcolon}{\isachardoublequoteopen}ka{\isasymle}Suc\ d{\isachardoublequoteclose}\isanewline
\ \ \isakeyword{shows}\ \ \ \ \ \ {\isachardoublequoteopen}lose\ {\isacharparenleft}Suc\ {\isacharparenleft}Suc\ {\isacharparenleft}t\ {\isacharplus}\ k\ {\isacharplus}\ ka{\isacharparenright}{\isacharparenright}{\isacharparenright}\ {\isacharequal}\ x{\isachardoublequoteclose}\isanewline
\isadelimproof
\endisadelimproof
\isatagproof
\isacommand{proof}\isamarkupfalse%
\ {\isacharminus}\isanewline
\ \ \isacommand{from}\isamarkupfalse%
\ h{\isadigit{2}}\ \isacommand{have}\isamarkupfalse%
\ sg{\isadigit{1}}{\isacharcolon}{\isachardoublequoteopen}k\ {\isacharplus}\ {\isacharparenleft}{\isadigit{2}}{\isacharcolon}{\isacharcolon}nat{\isacharparenright}\ {\isacharplus}\ ka\ {\isasymle}\ {\isacharparenleft}{\isadigit{2}}{\isacharcolon}{\isacharcolon}nat{\isacharparenright}\ {\isacharasterisk}\ d\ {\isacharplus}\ {\isacharparenleft}{\isacharparenleft}{\isadigit{4}}{\isacharcolon}{\isacharcolon}nat{\isacharparenright}\ {\isacharplus}\ k{\isacharparenright}{\isachardoublequoteclose}\ \isacommand{by}\isamarkupfalse%
\ arith\isanewline
\ \ \isacommand{from}\isamarkupfalse%
\ h{\isadigit{2}}\ \isakeyword{and}\ sg{\isadigit{1}}\ \isacommand{have}\isamarkupfalse%
\ sg{\isadigit{2}}{\isacharcolon}{\isachardoublequoteopen}Suc\ {\isacharparenleft}Suc\ {\isacharparenleft}k\ {\isacharplus}\ ka{\isacharparenright}{\isacharparenright}\ {\isasymle}{\isadigit{2}}\ {\isacharasterisk}\ d\ {\isacharplus}\ {\isacharparenleft}{\isadigit{4}}\ {\isacharplus}\ k{\isacharparenright}{\isachardoublequoteclose}\ \isacommand{by}\isamarkupfalse%
\ arith\isanewline
\ \ \isacommand{from}\isamarkupfalse%
\ sg{\isadigit{1}}\ \isakeyword{and}\ sg{\isadigit{2}}\ \isakeyword{and}\ h{\isadigit{1}}\ \isakeyword{and}\ h{\isadigit{2}}\ \ \isacommand{obtain}\isamarkupfalse%
\ j\ \isakeyword{where}\ a{\isadigit{1}}{\isacharcolon}{\isachardoublequoteopen}j\ {\isacharequal}\ k\ {\isacharplus}\ {\isacharparenleft}{\isadigit{2}}{\isacharcolon}{\isacharcolon}nat{\isacharparenright}\ {\isacharplus}\ ka{\isachardoublequoteclose}\isanewline
\ \ \ \ \ \ \ \ \ \ \ \ \ \ \ \ \ \ \ \ \ \ \ \ \ \ \ \ \ \ \ \ \ \ \ \ \ \isakeyword{and}\ a{\isadigit{2}}{\isacharcolon}{\isachardoublequoteopen}lose\ {\isacharparenleft}t\ {\isacharplus}\ j{\isacharparenright}\ {\isacharequal}\ x{\isachardoublequoteclose}\isanewline
\ \ \ \ \isacommand{by}\isamarkupfalse%
\ arith\isanewline
\ \ \isacommand{have}\isamarkupfalse%
\ sg{\isadigit{3}}{\isacharcolon}{\isachardoublequoteopen}Suc\ {\isacharparenleft}Suc\ {\isacharparenleft}t\ {\isacharplus}\ {\isacharparenleft}k\ {\isacharplus}\ ka{\isacharparenright}{\isacharparenright}{\isacharparenright}\ {\isacharequal}\ Suc\ {\isacharparenleft}Suc\ {\isacharparenleft}t\ {\isacharplus}\ k\ {\isacharplus}\ ka{\isacharparenright}{\isacharparenright}{\isachardoublequoteclose}\ \isacommand{by}\isamarkupfalse%
\ arith\isanewline
\ \ \isacommand{from}\isamarkupfalse%
\ a{\isadigit{1}}\ \isakeyword{and}\ a{\isadigit{2}}\ \isakeyword{and}\ sg{\isadigit{3}}\ \isacommand{show}\isamarkupfalse%
\ {\isacharquery}thesis\ \ \isacommand{by}\isamarkupfalse%
\ simp\isanewline
\isacommand{qed}\isamarkupfalse%
\endisatagproof
{\isafoldproof}%
\isadelimproof
\isanewline
\endisadelimproof
\isanewline
\isanewline
\isacommand{lemma}\isamarkupfalse%
\ aux{\isacharunderscore}lemma{\isacharunderscore}lose{\isacharunderscore}{\isadigit{2}}{\isacharcolon}\isanewline
\ \ \isakeyword{assumes}\ h{\isadigit{1}}{\isacharcolon}{\isachardoublequoteopen}{\isasymforall}j{\isasymle}{\isacharparenleft}{\isadigit{2}}{\isacharcolon}{\isacharcolon}nat{\isacharparenright}\ {\isacharasterisk}\ d\ {\isacharplus}\ {\isacharparenleft}{\isacharparenleft}{\isadigit{4}}{\isacharcolon}{\isacharcolon}nat{\isacharparenright}\ {\isacharplus}\ k{\isacharparenright}{\isachardot}\ lose\ {\isacharparenleft}t\ {\isacharplus}\ j{\isacharparenright}\ {\isacharequal}\ {\isacharbrackleft}False{\isacharbrackright}{\isachardoublequoteclose}\isanewline
\ \ \isakeyword{shows}\ \ \ {\isachardoublequoteopen}{\isasymforall}x{\isasymle}d\ {\isacharplus}\ {\isacharparenleft}{\isadigit{1}}{\isacharcolon}{\isacharcolon}nat{\isacharparenright}{\isachardot}\ lose\ {\isacharparenleft}t\ {\isacharplus}\ x{\isacharparenright}\ {\isacharequal}\ {\isacharbrackleft}False{\isacharbrackright}{\isachardoublequoteclose}\isanewline
\isadelimproof
\endisadelimproof
\isatagproof
\isacommand{using}\isamarkupfalse%
\ assms\ \isacommand{by}\isamarkupfalse%
\ auto%
\endisatagproof
{\isafoldproof}%
\isadelimproof
\isanewline
\endisadelimproof
\isanewline
\isanewline
\isacommand{lemma}\isamarkupfalse%
\ aux{\isacharunderscore}lemma{\isacharunderscore}lose{\isacharunderscore}{\isadigit{3}}a{\isacharcolon}\isanewline
\ \ \isakeyword{assumes}\ h{\isadigit{1}}{\isacharcolon}{\isachardoublequoteopen}{\isasymforall}j{\isasymle}{\isadigit{2}}\ {\isacharasterisk}\ d\ {\isacharplus}\ {\isacharparenleft}{\isadigit{4}}\ {\isacharplus}\ k{\isacharparenright}{\isachardot}\ lose\ {\isacharparenleft}t\ {\isacharplus}\ j{\isacharparenright}\ {\isacharequal}\ {\isacharbrackleft}False{\isacharbrackright}{\isachardoublequoteclose}\ \isanewline
\ \ \ \ \ \ \isakeyword{and}\ h{\isadigit{2}}{\isacharcolon}{\isachardoublequoteopen}ka\ {\isasymle}\ Suc\ d{\isachardoublequoteclose}\isanewline
\ \ \isakeyword{shows}\ {\isachardoublequoteopen}lose\ {\isacharparenleft}d\ {\isacharplus}\ {\isacharparenleft}t\ {\isacharplus}\ {\isacharparenleft}{\isadigit{3}}\ {\isacharplus}\ k{\isacharparenright}{\isacharparenright}\ {\isacharplus}\ ka{\isacharparenright}\ {\isacharequal}\ {\isacharbrackleft}False{\isacharbrackright}{\isachardoublequoteclose}\isanewline
\isadelimproof
\endisadelimproof
\isatagproof
\isacommand{proof}\isamarkupfalse%
\ {\isacharminus}\ \isanewline
\ \ \isacommand{from}\isamarkupfalse%
\ h{\isadigit{2}}\ \isacommand{have}\isamarkupfalse%
\ sg{\isadigit{1}}{\isacharcolon}{\isachardoublequoteopen}{\isacharparenleft}d\ {\isacharplus}\ {\isadigit{3}}\ {\isacharplus}\ k\ {\isacharplus}\ ka{\isacharparenright}\ {\isasymle}{\isadigit{2}}\ {\isacharasterisk}\ d\ {\isacharplus}\ {\isacharparenleft}{\isadigit{4}}\ {\isacharplus}\ k{\isacharparenright}{\isachardoublequoteclose}\isanewline
\ \ \ \ \isacommand{by}\isamarkupfalse%
\ arith\isanewline
\ \ \isacommand{from}\isamarkupfalse%
\ h{\isadigit{1}}\ \isakeyword{and}\ h{\isadigit{2}}\ \isakeyword{and}\ sg{\isadigit{1}}\ \ \isacommand{obtain}\isamarkupfalse%
\ j\ \isakeyword{where}\ a{\isadigit{1}}{\isacharcolon}{\isachardoublequoteopen}j\ {\isacharequal}\ {\isacharparenleft}d\ {\isacharplus}\ {\isadigit{3}}\ {\isacharplus}\ k\ {\isacharplus}\ ka{\isacharparenright}{\isachardoublequoteclose}\ \isakeyword{and}\ \isanewline
\ \ \ \ \ \ \ \ \ \ \ \ \ \ \ \ \ \ \ \ \ \ \ \ \ \ \ \ \ \ \ \ \ \ \ \ \ \ \ \ \ a{\isadigit{2}}{\isacharcolon}{\isachardoublequoteopen}lose\ {\isacharparenleft}t\ {\isacharplus}\ j{\isacharparenright}\ {\isacharequal}\ {\isacharbrackleft}False{\isacharbrackright}{\isachardoublequoteclose}\ \isanewline
\ \ \ \ \isacommand{by}\isamarkupfalse%
\ simp\isanewline
\ \ \isacommand{from}\isamarkupfalse%
\ h{\isadigit{2}}\ \isakeyword{and}\ sg{\isadigit{1}}\ \isacommand{have}\isamarkupfalse%
\ sg{\isadigit{2}}{\isacharcolon}{\isachardoublequoteopen}{\isacharparenleft}t\ {\isacharplus}\ {\isacharparenleft}d\ {\isacharplus}\ {\isadigit{3}}\ {\isacharplus}\ k\ {\isacharplus}\ ka{\isacharparenright}{\isacharparenright}\ {\isacharequal}\ {\isacharparenleft}d\ {\isacharplus}\ {\isacharparenleft}t\ {\isacharplus}\ {\isacharparenleft}{\isadigit{3}}\ {\isacharplus}\ k{\isacharparenright}{\isacharparenright}\ {\isacharplus}\ ka{\isacharparenright}{\isachardoublequoteclose}\ \isanewline
\ \ \ \ \isacommand{by}\isamarkupfalse%
\ arith\isanewline
\ \ \isacommand{from}\isamarkupfalse%
\ h{\isadigit{1}}\ \isakeyword{and}\ h{\isadigit{2}}\ \isakeyword{and}\ a{\isadigit{1}}\ \isakeyword{and}\ a{\isadigit{2}}\ \isakeyword{and}\ sg{\isadigit{2}}\ \isacommand{show}\isamarkupfalse%
\ {\isacharquery}thesis\isanewline
\ \ \ \ \isacommand{by}\isamarkupfalse%
\ simp\ \isanewline
\isacommand{qed}\isamarkupfalse%
\endisatagproof
{\isafoldproof}%
\isadelimproof
\isanewline
\endisadelimproof
\isanewline
\isanewline
\isacommand{lemma}\isamarkupfalse%
\ aux{\isacharunderscore}lemma{\isacharunderscore}lose{\isacharunderscore}{\isadigit{3}}{\isacharcolon}\isanewline
\ \ \isakeyword{assumes}\ h{\isadigit{1}}{\isacharcolon}{\isachardoublequoteopen}{\isasymforall}j{\isasymle}{\isadigit{2}}\ {\isacharasterisk}\ d\ {\isacharplus}\ {\isacharparenleft}{\isadigit{4}}\ {\isacharplus}\ k{\isacharparenright}{\isachardot}\ lose\ {\isacharparenleft}t\ {\isacharplus}\ j{\isacharparenright}\ {\isacharequal}\ {\isacharbrackleft}False{\isacharbrackright}{\isachardoublequoteclose}\isanewline
\ \ \isakeyword{shows}\ \ \ \ \ \ {\isachardoublequoteopen}{\isasymforall}ka{\isasymle}Suc\ d{\isachardot}\ lose\ {\isacharparenleft}d\ {\isacharplus}\ {\isacharparenleft}t\ {\isacharplus}\ {\isacharparenleft}{\isadigit{3}}\ {\isacharplus}\ k{\isacharparenright}{\isacharparenright}\ {\isacharplus}\ ka{\isacharparenright}\ {\isacharequal}\ {\isacharbrackleft}False{\isacharbrackright}{\isachardoublequoteclose}\isanewline
\isadelimproof
\endisadelimproof
\isatagproof
\isacommand{using}\isamarkupfalse%
\ assms\isanewline
\ \ \isacommand{by}\isamarkupfalse%
\ {\isacharparenleft}auto{\isacharcomma}\ simp\ add{\isacharcolon}\ aux{\isacharunderscore}lemma{\isacharunderscore}lose{\isacharunderscore}{\isadigit{3}}a{\isacharparenright}%
\endisatagproof
{\isafoldproof}%
\isadelimproof
\isanewline
\endisadelimproof
\isanewline
\isanewline
\isacommand{lemma}\isamarkupfalse%
\ aux{\isacharunderscore}arith{\isadigit{1}}{\isacharunderscore}Gateway{\isadigit{7}}{\isacharcolon}\isanewline
\ \ \isakeyword{assumes}\ h{\isadigit{1}}{\isacharcolon}{\isachardoublequoteopen}t{\isadigit{2}}\ {\isacharminus}\ t\ {\isasymle}\ {\isacharparenleft}{\isadigit{2}}{\isacharcolon}{\isacharcolon}nat{\isacharparenright}\ {\isacharasterisk}\ d\ {\isacharplus}\ {\isacharparenleft}t\ {\isacharplus}\ {\isacharparenleft}{\isacharparenleft}{\isadigit{4}}{\isacharcolon}{\isacharcolon}nat{\isacharparenright}\ {\isacharplus}\ k{\isacharparenright}{\isacharparenright}{\isachardoublequoteclose}\isanewline
\ \ \ \ \ \ \isakeyword{and}\ h{\isadigit{2}}{\isacharcolon}{\isachardoublequoteopen}t{\isadigit{2}}\ {\isacharless}\ t\ {\isacharplus}\ {\isacharparenleft}{\isadigit{3}}{\isacharcolon}{\isacharcolon}nat{\isacharparenright}\ {\isacharplus}\ k\ {\isacharplus}\ d{\isachardoublequoteclose}\isanewline
\ \ \ \ \ \ \isakeyword{and}\ h{\isadigit{3}}{\isacharcolon}{\isachardoublequoteopen}{\isasymnot}\ t{\isadigit{2}}\ {\isacharminus}\ d\ {\isacharless}\ {\isacharparenleft}{\isadigit{0}}{\isacharcolon}{\isacharcolon}nat{\isacharparenright}{\isachardoublequoteclose}\isanewline
\ \ \isakeyword{shows}\ {\isachardoublequoteopen}t{\isadigit{2}}\ {\isacharminus}\ d\ {\isacharless}\ t\ {\isacharplus}\ {\isacharparenleft}{\isadigit{3}}{\isacharcolon}{\isacharcolon}nat{\isacharparenright}\ {\isacharplus}\ k{\isachardoublequoteclose}\isanewline
\isadelimproof
\endisadelimproof
\isatagproof
\isacommand{using}\isamarkupfalse%
\ assms\ \ \isacommand{by}\isamarkupfalse%
\ arith%
\endisatagproof
{\isafoldproof}%
\isadelimproof
\isanewline
\endisadelimproof
\isanewline
\isanewline
\isacommand{lemma}\isamarkupfalse%
\ ts{\isacharunderscore}lose{\isacharunderscore}ack{\isacharunderscore}st{\isadigit{1}}ts{\isacharcolon}\isanewline
\ \ \isakeyword{assumes}\ h{\isadigit{1}}{\isacharcolon}{\isachardoublequoteopen}ts\ lose{\isachardoublequoteclose}\ \isanewline
\ \ \isakeyword{and}\ h{\isadigit{2}}{\isacharcolon}{\isachardoublequoteopen}lose\ t\ {\isacharequal}\ {\isacharbrackleft}True{\isacharbrackright}\ \ {\isasymlongrightarrow}\ \ ack\ t\ {\isacharequal}\ {\isacharbrackleft}x{\isacharbrackright}\ \ {\isasymand}\ st{\isacharunderscore}out\ t\ {\isacharequal}\ x{\isachardoublequoteclose}\isanewline
\ \ \isakeyword{and}\ h{\isadigit{3}}{\isacharcolon}{\isachardoublequoteopen}lose\ t\ {\isacharequal}\ {\isacharbrackleft}False{\isacharbrackright}\ {\isasymlongrightarrow}\ \ ack\ t\ {\isacharequal}\ {\isacharbrackleft}y{\isacharbrackright}\ \ {\isasymand}\ st{\isacharunderscore}out\ t\ {\isacharequal}\ y{\isachardoublequoteclose}\isanewline
\ \ \isakeyword{shows}\ {\isachardoublequoteopen}ack\ t\ {\isacharequal}\ {\isacharbrackleft}st{\isacharunderscore}out\ t{\isacharbrackright}{\isachardoublequoteclose}\isanewline
\isadelimproof
\endisadelimproof
\isatagproof
\isacommand{proof}\isamarkupfalse%
\ {\isacharparenleft}cases\ {\isachardoublequoteopen}lose\ t\ {\isacharequal}\ {\isacharbrackleft}False{\isacharbrackright}{\isachardoublequoteclose}{\isacharparenright}\isanewline
\ \ \isacommand{assume}\isamarkupfalse%
\ a{\isadigit{1}}{\isacharcolon}{\isachardoublequoteopen}lose\ t\ {\isacharequal}\ {\isacharbrackleft}False{\isacharbrackright}{\isachardoublequoteclose}\isanewline
\ \ \isacommand{from}\isamarkupfalse%
\ this\ \isakeyword{and}\ h{\isadigit{3}}\ \isacommand{show}\isamarkupfalse%
\ {\isacharquery}thesis\ \isacommand{by}\isamarkupfalse%
\ simp\isanewline
\isacommand{next}\isamarkupfalse%
\ \isanewline
\ \ \isacommand{assume}\isamarkupfalse%
\ a{\isadigit{2}}{\isacharcolon}{\isachardoublequoteopen}lose\ t\ {\isasymnoteq}\ {\isacharbrackleft}False{\isacharbrackright}{\isachardoublequoteclose}\isanewline
\ \ \isacommand{from}\isamarkupfalse%
\ this\ \isakeyword{and}\ h{\isadigit{1}}\ \isacommand{have}\isamarkupfalse%
\ ag{\isadigit{1}}{\isacharcolon}{\isachardoublequoteopen}lose\ t\ {\isacharequal}\ {\isacharbrackleft}True{\isacharbrackright}{\isachardoublequoteclose}\ \isacommand{by}\isamarkupfalse%
\ {\isacharparenleft}simp\ add{\isacharcolon}\ ts{\isacharunderscore}bool{\isacharunderscore}True{\isacharparenright}\isanewline
\ \ \isacommand{from}\isamarkupfalse%
\ this\ \isakeyword{and}\ a{\isadigit{2}}\ \isakeyword{and}\ h{\isadigit{2}}\ \isacommand{show}\isamarkupfalse%
\ {\isacharquery}thesis\ \isacommand{by}\isamarkupfalse%
\ simp\isanewline
\isacommand{qed}\isamarkupfalse%
\endisatagproof
{\isafoldproof}%
\isadelimproof
\isanewline
\endisadelimproof
\isanewline
\isanewline
\isacommand{lemma}\isamarkupfalse%
\ ts{\isacharunderscore}lose{\isacharunderscore}ack{\isacharunderscore}st{\isadigit{1}}{\isacharcolon}\isanewline
\ \ \isakeyword{assumes}\ h{\isadigit{1}}{\isacharcolon}{\isachardoublequoteopen}lose\ t\ {\isacharequal}\ {\isacharbrackleft}True{\isacharbrackright}\ {\isasymor}\ lose\ t\ {\isacharequal}\ {\isacharbrackleft}False{\isacharbrackright}{\isachardoublequoteclose}\ \isanewline
\ \ \isakeyword{and}\ h{\isadigit{2}}{\isacharcolon}{\isachardoublequoteopen}lose\ t\ {\isacharequal}\ {\isacharbrackleft}True{\isacharbrackright}\ \ {\isasymlongrightarrow}\ \ ack\ t\ {\isacharequal}\ {\isacharbrackleft}x{\isacharbrackright}\ \ {\isasymand}\ st{\isacharunderscore}out\ t\ {\isacharequal}\ x{\isachardoublequoteclose}\isanewline
\ \ \isakeyword{and}\ h{\isadigit{3}}{\isacharcolon}{\isachardoublequoteopen}lose\ t\ {\isacharequal}\ {\isacharbrackleft}False{\isacharbrackright}\ {\isasymlongrightarrow}\ \ ack\ t\ {\isacharequal}\ {\isacharbrackleft}y{\isacharbrackright}\ \ {\isasymand}\ st{\isacharunderscore}out\ t\ {\isacharequal}\ y{\isachardoublequoteclose}\isanewline
\ \ \isakeyword{shows}\ {\isachardoublequoteopen}ack\ t\ {\isacharequal}\ {\isacharbrackleft}st{\isacharunderscore}out\ t{\isacharbrackright}{\isachardoublequoteclose}\isanewline
\isadelimproof
\endisadelimproof
\isatagproof
\isacommand{proof}\isamarkupfalse%
\ {\isacharparenleft}cases\ {\isachardoublequoteopen}lose\ t\ {\isacharequal}\ {\isacharbrackleft}False{\isacharbrackright}{\isachardoublequoteclose}{\isacharparenright}\isanewline
\ \ \isacommand{assume}\isamarkupfalse%
\ a{\isadigit{1}}{\isacharcolon}{\isachardoublequoteopen}lose\ t\ {\isacharequal}\ {\isacharbrackleft}False{\isacharbrackright}{\isachardoublequoteclose}\isanewline
\ \ \isacommand{from}\isamarkupfalse%
\ this\ \isakeyword{and}\ h{\isadigit{3}}\ \isacommand{show}\isamarkupfalse%
\ {\isacharquery}thesis\ \isacommand{by}\isamarkupfalse%
\ simp\isanewline
\isacommand{next}\isamarkupfalse%
\ \isanewline
\ \ \isacommand{assume}\isamarkupfalse%
\ a{\isadigit{2}}{\isacharcolon}{\isachardoublequoteopen}lose\ t\ {\isasymnoteq}\ {\isacharbrackleft}False{\isacharbrackright}{\isachardoublequoteclose}\isanewline
\ \ \isacommand{from}\isamarkupfalse%
\ this\ \isakeyword{and}\ h{\isadigit{1}}\ \isacommand{have}\isamarkupfalse%
\ ag{\isadigit{1}}{\isacharcolon}{\isachardoublequoteopen}lose\ t\ {\isacharequal}\ {\isacharbrackleft}True{\isacharbrackright}{\isachardoublequoteclose}\ \isacommand{by}\isamarkupfalse%
\ {\isacharparenleft}simp\ add{\isacharcolon}\ ts{\isacharunderscore}bool{\isacharunderscore}True{\isacharparenright}\isanewline
\ \ \isacommand{from}\isamarkupfalse%
\ this\ \isakeyword{and}\ a{\isadigit{2}}\ \isakeyword{and}\ h{\isadigit{2}}\ \isacommand{show}\isamarkupfalse%
\ {\isacharquery}thesis\ \isacommand{by}\isamarkupfalse%
\ simp\isanewline
\isacommand{qed}\isamarkupfalse%
\endisatagproof
{\isafoldproof}%
\isadelimproof
\isanewline
\endisadelimproof
\isanewline
\isanewline
\isacommand{lemma}\isamarkupfalse%
\ ts{\isacharunderscore}lose{\isacharunderscore}ack{\isacharunderscore}st{\isadigit{2}}ts{\isacharcolon}\isanewline
\ \ \isakeyword{assumes}\ h{\isadigit{1}}{\isacharcolon}{\isachardoublequoteopen}ts\ lose{\isachardoublequoteclose}\ \isanewline
\ \ \isakeyword{and}\ h{\isadigit{2}}{\isacharcolon}{\isachardoublequoteopen}lose\ t\ {\isacharequal}\ {\isacharbrackleft}True{\isacharbrackright}\ {\isasymlongrightarrow}\ \isanewline
\ \ \ \ \ \ ack\ t\ {\isacharequal}\ {\isacharbrackleft}x{\isacharbrackright}\ \ {\isasymand}\ i{\isadigit{1}}\ t\ {\isacharequal}\ {\isacharbrackleft}{\isacharbrackright}\ {\isasymand}\ vc\ t\ {\isacharequal}\ {\isacharbrackleft}{\isacharbrackright}\ {\isasymand}\ st{\isacharunderscore}out\ t\ {\isacharequal}\ x{\isachardoublequoteclose}\isanewline
\ \ \isakeyword{and}\ h{\isadigit{3}}{\isacharcolon}{\isachardoublequoteopen}lose\ t\ {\isacharequal}\ {\isacharbrackleft}False{\isacharbrackright}\ {\isasymlongrightarrow}\ \isanewline
\ \ \ \ \ \ ack\ t\ {\isacharequal}\ {\isacharbrackleft}y{\isacharbrackright}\ {\isasymand}\ i{\isadigit{1}}\ t\ {\isacharequal}\ {\isacharbrackleft}{\isacharbrackright}\ {\isasymand}\ vc\ t\ {\isacharequal}\ {\isacharbrackleft}{\isacharbrackright}\ {\isasymand}\ st{\isacharunderscore}out\ t\ {\isacharequal}\ y{\isachardoublequoteclose}\isanewline
\ \ \isakeyword{shows}\ {\isachardoublequoteopen}ack\ t\ {\isacharequal}\ {\isacharbrackleft}st{\isacharunderscore}out\ t{\isacharbrackright}{\isachardoublequoteclose}\isanewline
\isadelimproof
\endisadelimproof
\isatagproof
\isacommand{proof}\isamarkupfalse%
\ {\isacharparenleft}cases\ {\isachardoublequoteopen}lose\ t\ {\isacharequal}\ {\isacharbrackleft}False{\isacharbrackright}{\isachardoublequoteclose}{\isacharparenright}\isanewline
\ \ \isacommand{assume}\isamarkupfalse%
\ a{\isadigit{1}}{\isacharcolon}{\isachardoublequoteopen}lose\ t\ {\isacharequal}\ {\isacharbrackleft}False{\isacharbrackright}{\isachardoublequoteclose}\isanewline
\ \ \isacommand{from}\isamarkupfalse%
\ this\ \isakeyword{and}\ h{\isadigit{3}}\ \isacommand{show}\isamarkupfalse%
\ {\isacharquery}thesis\ \isacommand{by}\isamarkupfalse%
\ simp\isanewline
\isacommand{next}\isamarkupfalse%
\ \isanewline
\ \ \isacommand{assume}\isamarkupfalse%
\ a{\isadigit{2}}{\isacharcolon}{\isachardoublequoteopen}lose\ t\ {\isasymnoteq}\ {\isacharbrackleft}False{\isacharbrackright}{\isachardoublequoteclose}\isanewline
\ \ \isacommand{from}\isamarkupfalse%
\ this\ \isakeyword{and}\ h{\isadigit{1}}\ \isacommand{have}\isamarkupfalse%
\ ag{\isadigit{1}}{\isacharcolon}{\isachardoublequoteopen}lose\ t\ {\isacharequal}\ {\isacharbrackleft}True{\isacharbrackright}{\isachardoublequoteclose}\ \isacommand{by}\isamarkupfalse%
\ {\isacharparenleft}simp\ add{\isacharcolon}\ ts{\isacharunderscore}bool{\isacharunderscore}True{\isacharparenright}\isanewline
\ \ \isacommand{from}\isamarkupfalse%
\ this\ \isakeyword{and}\ a{\isadigit{2}}\ \isakeyword{and}\ h{\isadigit{2}}\ \isacommand{show}\isamarkupfalse%
\ {\isacharquery}thesis\ \isacommand{by}\isamarkupfalse%
\ simp\isanewline
\isacommand{qed}\isamarkupfalse%
\endisatagproof
{\isafoldproof}%
\isadelimproof
\isanewline
\endisadelimproof
\isanewline
\isanewline
\isacommand{lemma}\isamarkupfalse%
\ ts{\isacharunderscore}lose{\isacharunderscore}ack{\isacharunderscore}st{\isadigit{2}}{\isacharcolon}\isanewline
\ \ \isakeyword{assumes}\ h{\isadigit{1}}{\isacharcolon}{\isachardoublequoteopen}lose\ t\ {\isacharequal}\ {\isacharbrackleft}True{\isacharbrackright}\ {\isasymor}\ lose\ t\ {\isacharequal}\ {\isacharbrackleft}False{\isacharbrackright}{\isachardoublequoteclose}\ \isanewline
\ \ \isakeyword{and}\ h{\isadigit{2}}{\isacharcolon}{\isachardoublequoteopen}lose\ t\ {\isacharequal}\ {\isacharbrackleft}True{\isacharbrackright}\ {\isasymlongrightarrow}\ \isanewline
\ \ \ \ \ \ ack\ t\ {\isacharequal}\ {\isacharbrackleft}x{\isacharbrackright}\ \ {\isasymand}\ i{\isadigit{1}}\ t\ {\isacharequal}\ {\isacharbrackleft}{\isacharbrackright}\ {\isasymand}\ vc\ t\ {\isacharequal}\ {\isacharbrackleft}{\isacharbrackright}\ {\isasymand}\ st{\isacharunderscore}out\ t\ {\isacharequal}\ x{\isachardoublequoteclose}\isanewline
\ \ \isakeyword{and}\ h{\isadigit{3}}{\isacharcolon}{\isachardoublequoteopen}lose\ t\ {\isacharequal}\ {\isacharbrackleft}False{\isacharbrackright}\ {\isasymlongrightarrow}\ \isanewline
\ \ \ \ \ \ ack\ t\ {\isacharequal}\ {\isacharbrackleft}y{\isacharbrackright}\ {\isasymand}\ i{\isadigit{1}}\ t\ {\isacharequal}\ {\isacharbrackleft}{\isacharbrackright}\ {\isasymand}\ vc\ t\ {\isacharequal}\ {\isacharbrackleft}{\isacharbrackright}\ {\isasymand}\ st{\isacharunderscore}out\ t\ {\isacharequal}\ y{\isachardoublequoteclose}\isanewline
\ \ \isakeyword{shows}\ {\isachardoublequoteopen}ack\ t\ {\isacharequal}\ {\isacharbrackleft}st{\isacharunderscore}out\ t{\isacharbrackright}{\isachardoublequoteclose}\isanewline
\isadelimproof
\endisadelimproof
\isatagproof
\isacommand{proof}\isamarkupfalse%
\ {\isacharparenleft}cases\ {\isachardoublequoteopen}lose\ t\ {\isacharequal}\ {\isacharbrackleft}False{\isacharbrackright}{\isachardoublequoteclose}{\isacharparenright}\isanewline
\ \ \isacommand{assume}\isamarkupfalse%
\ a{\isadigit{1}}{\isacharcolon}{\isachardoublequoteopen}lose\ t\ {\isacharequal}\ {\isacharbrackleft}False{\isacharbrackright}{\isachardoublequoteclose}\isanewline
\ \ \isacommand{from}\isamarkupfalse%
\ this\ \isakeyword{and}\ h{\isadigit{3}}\ \isacommand{show}\isamarkupfalse%
\ {\isacharquery}thesis\ \isacommand{by}\isamarkupfalse%
\ simp\isanewline
\isacommand{next}\isamarkupfalse%
\ \isanewline
\ \ \isacommand{assume}\isamarkupfalse%
\ a{\isadigit{2}}{\isacharcolon}{\isachardoublequoteopen}lose\ t\ {\isasymnoteq}\ {\isacharbrackleft}False{\isacharbrackright}{\isachardoublequoteclose}\isanewline
\ \ \isacommand{from}\isamarkupfalse%
\ this\ \isakeyword{and}\ h{\isadigit{1}}\ \isacommand{have}\isamarkupfalse%
\ ag{\isadigit{1}}{\isacharcolon}{\isachardoublequoteopen}lose\ t\ {\isacharequal}\ {\isacharbrackleft}True{\isacharbrackright}{\isachardoublequoteclose}\ \isacommand{by}\isamarkupfalse%
\ {\isacharparenleft}simp\ add{\isacharcolon}\ ts{\isacharunderscore}bool{\isacharunderscore}True{\isacharparenright}\isanewline
\ \ \isacommand{from}\isamarkupfalse%
\ this\ \isakeyword{and}\ a{\isadigit{2}}\ \isakeyword{and}\ h{\isadigit{2}}\ \isacommand{show}\isamarkupfalse%
\ {\isacharquery}thesis\ \isacommand{by}\isamarkupfalse%
\ simp\isanewline
\isacommand{qed}\isamarkupfalse%
\endisatagproof
{\isafoldproof}%
\isadelimproof
\isanewline
\endisadelimproof
\isanewline
\isacommand{lemma}\isamarkupfalse%
\ ts{\isacharunderscore}lose{\isacharunderscore}ack{\isacharunderscore}st{\isadigit{2}}vc{\isacharunderscore}com{\isacharcolon}\isanewline
\ \ \isakeyword{assumes}\ h{\isadigit{1}}{\isacharcolon}{\isachardoublequoteopen}lose\ t\ {\isacharequal}\ {\isacharbrackleft}True{\isacharbrackright}\ {\isasymor}\ lose\ t\ {\isacharequal}\ {\isacharbrackleft}False{\isacharbrackright}{\isachardoublequoteclose}\ \isanewline
\ \ \isakeyword{and}\ h{\isadigit{2}}{\isacharcolon}{\isachardoublequoteopen}lose\ t\ {\isacharequal}\ {\isacharbrackleft}True{\isacharbrackright}\ {\isasymlongrightarrow}\ \isanewline
\ \ \ \ \ \ ack\ t\ {\isacharequal}\ {\isacharbrackleft}x{\isacharbrackright}\ \ {\isasymand}\ i{\isadigit{1}}\ t\ {\isacharequal}\ {\isacharbrackleft}{\isacharbrackright}\ {\isasymand}\ vc\ t\ {\isacharequal}\ {\isacharbrackleft}{\isacharbrackright}\ {\isasymand}\ st{\isacharunderscore}out\ t\ {\isacharequal}\ x{\isachardoublequoteclose}\isanewline
\ \ \isakeyword{and}\ h{\isadigit{3}}{\isacharcolon}{\isachardoublequoteopen}lose\ t\ {\isacharequal}\ {\isacharbrackleft}False{\isacharbrackright}\ {\isasymlongrightarrow}\ \isanewline
\ \ \ \ \ \ ack\ t\ {\isacharequal}\ {\isacharbrackleft}y{\isacharbrackright}\ {\isasymand}\ i{\isadigit{1}}\ t\ {\isacharequal}\ {\isacharbrackleft}{\isacharbrackright}\ {\isasymand}\ vc\ t\ {\isacharequal}\ {\isacharbrackleft}vc{\isacharunderscore}com{\isacharbrackright}\ {\isasymand}\ st{\isacharunderscore}out\ t\ {\isacharequal}\ y{\isachardoublequoteclose}\isanewline
\ \ \isakeyword{shows}\ {\isachardoublequoteopen}ack\ t\ {\isacharequal}\ {\isacharbrackleft}st{\isacharunderscore}out\ t{\isacharbrackright}{\isachardoublequoteclose}\isanewline
\isadelimproof
\endisadelimproof
\isatagproof
\isacommand{proof}\isamarkupfalse%
\ {\isacharparenleft}cases\ {\isachardoublequoteopen}lose\ t\ {\isacharequal}\ {\isacharbrackleft}False{\isacharbrackright}{\isachardoublequoteclose}{\isacharparenright}\isanewline
\ \ \isacommand{assume}\isamarkupfalse%
\ a{\isadigit{1}}{\isacharcolon}{\isachardoublequoteopen}lose\ t\ {\isacharequal}\ {\isacharbrackleft}False{\isacharbrackright}{\isachardoublequoteclose}\isanewline
\ \ \isacommand{from}\isamarkupfalse%
\ this\ \isakeyword{and}\ h{\isadigit{3}}\ \isacommand{show}\isamarkupfalse%
\ {\isacharquery}thesis\ \isacommand{by}\isamarkupfalse%
\ simp\isanewline
\isacommand{next}\isamarkupfalse%
\ \isanewline
\ \ \isacommand{assume}\isamarkupfalse%
\ a{\isadigit{2}}{\isacharcolon}{\isachardoublequoteopen}lose\ t\ {\isasymnoteq}\ {\isacharbrackleft}False{\isacharbrackright}{\isachardoublequoteclose}\isanewline
\ \ \isacommand{from}\isamarkupfalse%
\ this\ \isakeyword{and}\ h{\isadigit{1}}\ \isacommand{have}\isamarkupfalse%
\ ag{\isadigit{1}}{\isacharcolon}{\isachardoublequoteopen}lose\ t\ {\isacharequal}\ {\isacharbrackleft}True{\isacharbrackright}{\isachardoublequoteclose}\ \isacommand{by}\isamarkupfalse%
\ {\isacharparenleft}simp\ add{\isacharcolon}\ ts{\isacharunderscore}bool{\isacharunderscore}True{\isacharparenright}\isanewline
\ \ \isacommand{from}\isamarkupfalse%
\ this\ \isakeyword{and}\ a{\isadigit{2}}\ \isakeyword{and}\ h{\isadigit{2}}\ \isacommand{show}\isamarkupfalse%
\ {\isacharquery}thesis\ \isacommand{by}\isamarkupfalse%
\ simp\isanewline
\isacommand{qed}\isamarkupfalse%
\endisatagproof
{\isafoldproof}%
\isadelimproof
\isanewline
\endisadelimproof
\isanewline
\isanewline
\isacommand{lemma}\isamarkupfalse%
\ ts{\isacharunderscore}lose{\isacharunderscore}ack{\isacharunderscore}st{\isadigit{2}}send{\isacharcolon}\isanewline
\ \ \isakeyword{assumes}\ h{\isadigit{1}}{\isacharcolon}{\isachardoublequoteopen}lose\ t\ {\isacharequal}\ {\isacharbrackleft}True{\isacharbrackright}\ {\isasymor}\ lose\ t\ {\isacharequal}\ {\isacharbrackleft}False{\isacharbrackright}{\isachardoublequoteclose}\ \isanewline
\ \ \isakeyword{and}\ h{\isadigit{2}}{\isacharcolon}{\isachardoublequoteopen}lose\ t\ {\isacharequal}\ {\isacharbrackleft}True{\isacharbrackright}\ {\isasymlongrightarrow}\ \isanewline
\ \ \ \ \ \ ack\ t\ {\isacharequal}\ {\isacharbrackleft}x{\isacharbrackright}\ \ {\isasymand}\ i{\isadigit{1}}\ t\ {\isacharequal}\ {\isacharbrackleft}{\isacharbrackright}\ {\isasymand}\ vc\ t\ {\isacharequal}\ {\isacharbrackleft}{\isacharbrackright}\ {\isasymand}\ st{\isacharunderscore}out\ t\ {\isacharequal}\ x{\isachardoublequoteclose}\isanewline
\ \ \isakeyword{and}\ h{\isadigit{3}}{\isacharcolon}{\isachardoublequoteopen}lose\ t\ {\isacharequal}\ {\isacharbrackleft}False{\isacharbrackright}\ {\isasymlongrightarrow}\ \isanewline
\ \ \ \ \ \ ack\ t\ {\isacharequal}\ {\isacharbrackleft}y{\isacharbrackright}\ {\isasymand}\ i{\isadigit{1}}\ t\ {\isacharequal}\ b\ t\ {\isasymand}\ vc\ t\ {\isacharequal}\ {\isacharbrackleft}{\isacharbrackright}\ {\isasymand}\ st{\isacharunderscore}out\ t\ {\isacharequal}\ y{\isachardoublequoteclose}\isanewline
\ \ \isakeyword{shows}\ {\isachardoublequoteopen}ack\ t\ {\isacharequal}\ {\isacharbrackleft}st{\isacharunderscore}out\ t{\isacharbrackright}{\isachardoublequoteclose}\isanewline
\isadelimproof
\endisadelimproof
\isatagproof
\isacommand{proof}\isamarkupfalse%
\ {\isacharparenleft}cases\ {\isachardoublequoteopen}lose\ t\ {\isacharequal}\ {\isacharbrackleft}False{\isacharbrackright}{\isachardoublequoteclose}{\isacharparenright}\isanewline
\ \ \isacommand{assume}\isamarkupfalse%
\ a{\isadigit{1}}{\isacharcolon}{\isachardoublequoteopen}lose\ t\ {\isacharequal}\ {\isacharbrackleft}False{\isacharbrackright}{\isachardoublequoteclose}\isanewline
\ \ \isacommand{from}\isamarkupfalse%
\ this\ \isakeyword{and}\ h{\isadigit{3}}\ \isacommand{show}\isamarkupfalse%
\ {\isacharquery}thesis\ \isacommand{by}\isamarkupfalse%
\ simp\isanewline
\isacommand{next}\isamarkupfalse%
\ \isanewline
\ \ \isacommand{assume}\isamarkupfalse%
\ a{\isadigit{2}}{\isacharcolon}{\isachardoublequoteopen}lose\ t\ {\isasymnoteq}\ {\isacharbrackleft}False{\isacharbrackright}{\isachardoublequoteclose}\isanewline
\ \ \isacommand{from}\isamarkupfalse%
\ this\ \isakeyword{and}\ h{\isadigit{1}}\ \isacommand{have}\isamarkupfalse%
\ ag{\isadigit{1}}{\isacharcolon}{\isachardoublequoteopen}lose\ t\ {\isacharequal}\ {\isacharbrackleft}True{\isacharbrackright}{\isachardoublequoteclose}\ \isacommand{by}\isamarkupfalse%
\ {\isacharparenleft}simp\ add{\isacharcolon}\ ts{\isacharunderscore}bool{\isacharunderscore}True{\isacharparenright}\isanewline
\ \ \isacommand{from}\isamarkupfalse%
\ this\ \isakeyword{and}\ a{\isadigit{2}}\ \isakeyword{and}\ h{\isadigit{2}}\ \isacommand{show}\isamarkupfalse%
\ {\isacharquery}thesis\ \isacommand{by}\isamarkupfalse%
\ simp\isanewline
\isacommand{qed}\isamarkupfalse%
\endisatagproof
{\isafoldproof}%
\isadelimproof
\isanewline
\endisadelimproof
\isanewline
\isanewline
\isacommand{lemma}\isamarkupfalse%
\ tiTable{\isacharunderscore}ack{\isacharunderscore}st{\isacharunderscore}splitten{\isacharcolon}\isanewline
\ \ \isakeyword{assumes}\ h{\isadigit{1}}{\isacharcolon}{\isachardoublequoteopen}ts\ lose{\isachardoublequoteclose}\isanewline
\ \ \ \ \ \ \isakeyword{and}\ h{\isadigit{2}}{\isacharcolon}{\isachardoublequoteopen}msg\ {\isacharparenleft}Suc\ {\isadigit{0}}{\isacharparenright}\ a{\isadigit{1}}{\isachardoublequoteclose}\isanewline
\ \ \ \ \ \ \isakeyword{and}\ h{\isadigit{3}}{\isacharcolon}{\isachardoublequoteopen}msg\ {\isacharparenleft}Suc\ {\isadigit{0}}{\isacharparenright}\ stop{\isachardoublequoteclose}\isanewline
\ \ \ \ \ \ \isakeyword{and}\ h{\isadigit{4}}{\isacharcolon}{\isachardoublequoteopen}st{\isacharunderscore}in\ t\ {\isacharequal}\ init{\isacharunderscore}state\ {\isasymand}\ req\ t\ {\isacharequal}\ {\isacharbrackleft}init{\isacharbrackright}\ {\isasymlongrightarrow}\ \isanewline
\ \ \ \ \ \ \ \ \ \ ack\ t\ {\isacharequal}\ {\isacharbrackleft}call{\isacharbrackright}\ {\isasymand}\ i{\isadigit{1}}\ t\ {\isacharequal}\ {\isacharbrackleft}{\isacharbrackright}\ {\isasymand}\ vc\ t\ {\isacharequal}\ {\isacharbrackleft}{\isacharbrackright}\ {\isasymand}\ st{\isacharunderscore}out\ t\ {\isacharequal}\ call{\isachardoublequoteclose}\isanewline
\ \ \ \ \ \ \isakeyword{and}\ h{\isadigit{5}}{\isacharcolon}{\isachardoublequoteopen}st{\isacharunderscore}in\ t\ {\isacharequal}\ init{\isacharunderscore}state\ {\isasymand}\ req\ t\ {\isasymnoteq}\ {\isacharbrackleft}init{\isacharbrackright}\ {\isasymlongrightarrow}\isanewline
\ \ \ \ \ \ \ \ \ \ ack\ t\ {\isacharequal}\ {\isacharbrackleft}init{\isacharunderscore}state{\isacharbrackright}\ {\isasymand}\ i{\isadigit{1}}\ t\ {\isacharequal}\ {\isacharbrackleft}{\isacharbrackright}\ {\isasymand}\ vc\ t\ {\isacharequal}\ {\isacharbrackleft}{\isacharbrackright}\ {\isasymand}\ st{\isacharunderscore}out\ t\ {\isacharequal}\ init{\isacharunderscore}state{\isachardoublequoteclose}\isanewline
\ \ \ \ \ \ \isakeyword{and}\ h{\isadigit{6}}{\isacharcolon}{\isachardoublequoteopen}{\isacharparenleft}st{\isacharunderscore}in\ t\ {\isacharequal}\ call\ {\isasymor}\ st{\isacharunderscore}in\ t\ {\isacharequal}\ connection{\isacharunderscore}ok\ {\isasymand}\ req\ t\ {\isasymnoteq}\ {\isacharbrackleft}send{\isacharbrackright}{\isacharparenright}\ {\isasymand}\ lose\ t\ {\isacharequal}\ {\isacharbrackleft}False{\isacharbrackright}\ {\isasymlongrightarrow}\isanewline
\ \ \ \ \ \ \ \ \ \ ack\ t\ {\isacharequal}\ {\isacharbrackleft}connection{\isacharunderscore}ok{\isacharbrackright}\ {\isasymand}\ i{\isadigit{1}}\ t\ {\isacharequal}\ {\isacharbrackleft}{\isacharbrackright}\ {\isasymand}\ vc\ t\ {\isacharequal}\ {\isacharbrackleft}{\isacharbrackright}\ {\isasymand}\ st{\isacharunderscore}out\ t\ {\isacharequal}\ connection{\isacharunderscore}ok{\isachardoublequoteclose}\isanewline
\ \ \ \ \ \ \isakeyword{and}\ h{\isadigit{7}}{\isacharcolon}{\isachardoublequoteopen}{\isacharparenleft}st{\isacharunderscore}in\ t\ {\isacharequal}\ call\ {\isasymor}\ st{\isacharunderscore}in\ t\ {\isacharequal}\ connection{\isacharunderscore}ok\ {\isasymor}\ st{\isacharunderscore}in\ t\ {\isacharequal}\ sending{\isacharunderscore}data{\isacharparenright}\ {\isasymand}\ lose\ t\ {\isacharequal}\ {\isacharbrackleft}True{\isacharbrackright}\ {\isasymlongrightarrow}\isanewline
\ \ \ \ \ \ \ \ \ \ ack\ t\ {\isacharequal}\ {\isacharbrackleft}init{\isacharunderscore}state{\isacharbrackright}\ {\isasymand}\ i{\isadigit{1}}\ t\ {\isacharequal}\ {\isacharbrackleft}{\isacharbrackright}\ {\isasymand}\ vc\ t\ {\isacharequal}\ {\isacharbrackleft}{\isacharbrackright}\ {\isasymand}\ st{\isacharunderscore}out\ t\ {\isacharequal}\ init{\isacharunderscore}state{\isachardoublequoteclose}\isanewline
\ \ \ \ \ \ \isakeyword{and}\ h{\isadigit{8}}{\isacharcolon}{\isachardoublequoteopen}st{\isacharunderscore}in\ t\ {\isacharequal}\ connection{\isacharunderscore}ok\ {\isasymand}\ req\ t\ {\isacharequal}\ {\isacharbrackleft}send{\isacharbrackright}\ {\isasymand}\ lose\ t\ {\isacharequal}\ {\isacharbrackleft}False{\isacharbrackright}\ {\isasymlongrightarrow}\isanewline
\ \ \ \ \ \ \ \ \ \ ack\ t\ {\isacharequal}\ {\isacharbrackleft}sending{\isacharunderscore}data{\isacharbrackright}\ {\isasymand}\ i{\isadigit{1}}\ t\ {\isacharequal}\ b\ t\ {\isasymand}\ vc\ t\ {\isacharequal}\ {\isacharbrackleft}{\isacharbrackright}\ {\isasymand}\ st{\isacharunderscore}out\ t\ {\isacharequal}\ sending{\isacharunderscore}data{\isachardoublequoteclose}\isanewline
\ \ \ \ \ \ \isakeyword{and}\ h{\isadigit{9}}{\isacharcolon}{\isachardoublequoteopen}st{\isacharunderscore}in\ t\ {\isacharequal}\ sending{\isacharunderscore}data\ {\isasymand}\ a{\isadigit{1}}\ t\ {\isacharequal}\ {\isacharbrackleft}{\isacharbrackright}\ {\isasymand}\ lose\ t\ {\isacharequal}\ {\isacharbrackleft}False{\isacharbrackright}\ {\isasymlongrightarrow}\isanewline
\ \ \ \ \ \ \ \ \ \ ack\ t\ {\isacharequal}\ {\isacharbrackleft}sending{\isacharunderscore}data{\isacharbrackright}\ {\isasymand}\ i{\isadigit{1}}\ t\ {\isacharequal}\ {\isacharbrackleft}{\isacharbrackright}\ {\isasymand}\ vc\ t\ {\isacharequal}\ {\isacharbrackleft}{\isacharbrackright}\ {\isasymand}\ st{\isacharunderscore}out\ t\ {\isacharequal}\ sending{\isacharunderscore}data{\isachardoublequoteclose}\isanewline
\ \ \ \ \ \ \isakeyword{and}\ h{\isadigit{1}}{\isadigit{0}}{\isacharcolon}{\isachardoublequoteopen}st{\isacharunderscore}in\ t\ {\isacharequal}\ sending{\isacharunderscore}data\ {\isasymand}\ a{\isadigit{1}}\ t\ {\isacharequal}\ {\isacharbrackleft}sc{\isacharunderscore}ack{\isacharbrackright}\ {\isasymand}\ lose\ t\ {\isacharequal}\ {\isacharbrackleft}False{\isacharbrackright}\ {\isasymlongrightarrow}\isanewline
\ \ \ \ \ \ \ \ \ \ ack\ t\ {\isacharequal}\ {\isacharbrackleft}voice{\isacharunderscore}com{\isacharbrackright}\ {\isasymand}\ i{\isadigit{1}}\ t\ {\isacharequal}\ {\isacharbrackleft}{\isacharbrackright}\ {\isasymand}\ vc\ t\ {\isacharequal}\ {\isacharbrackleft}vc{\isacharunderscore}com{\isacharbrackright}\ {\isasymand}\ st{\isacharunderscore}out\ t\ {\isacharequal}\ voice{\isacharunderscore}com{\isachardoublequoteclose}\isanewline
\ \ \ \ \ \ \isakeyword{and}\ h{\isadigit{1}}{\isadigit{1}}{\isacharcolon}{\isachardoublequoteopen}st{\isacharunderscore}in\ t\ {\isacharequal}\ voice{\isacharunderscore}com\ {\isasymand}\ stop\ t\ {\isacharequal}\ {\isacharbrackleft}{\isacharbrackright}\ {\isasymand}\ lose\ t\ {\isacharequal}\ {\isacharbrackleft}False{\isacharbrackright}\ {\isasymlongrightarrow}\isanewline
\ \ \ \ \ \ \ \ \ \ ack\ t\ {\isacharequal}\ {\isacharbrackleft}voice{\isacharunderscore}com{\isacharbrackright}\ {\isasymand}\ i{\isadigit{1}}\ t\ {\isacharequal}\ {\isacharbrackleft}{\isacharbrackright}\ {\isasymand}\ vc\ t\ {\isacharequal}\ {\isacharbrackleft}vc{\isacharunderscore}com{\isacharbrackright}\ {\isasymand}\ st{\isacharunderscore}out\ t\ {\isacharequal}\ voice{\isacharunderscore}com{\isachardoublequoteclose}\isanewline
\ \ \ \ \ \ \isakeyword{and}\ h{\isadigit{1}}{\isadigit{2}}{\isacharcolon}{\isachardoublequoteopen}st{\isacharunderscore}in\ t\ {\isacharequal}\ voice{\isacharunderscore}com\ {\isasymand}\ stop\ t\ {\isacharequal}\ {\isacharbrackleft}{\isacharbrackright}\ {\isasymand}\ lose\ t\ {\isacharequal}\ {\isacharbrackleft}True{\isacharbrackright}\ {\isasymlongrightarrow}\isanewline
\ \ \ \ \ \ \ \ \ \ ack\ t\ {\isacharequal}\ {\isacharbrackleft}voice{\isacharunderscore}com{\isacharbrackright}\ {\isasymand}\ i{\isadigit{1}}\ t\ {\isacharequal}\ {\isacharbrackleft}{\isacharbrackright}\ {\isasymand}\ vc\ t\ {\isacharequal}\ {\isacharbrackleft}{\isacharbrackright}\ {\isasymand}\ st{\isacharunderscore}out\ t\ {\isacharequal}\ voice{\isacharunderscore}com{\isachardoublequoteclose}\isanewline
\ \ \ \ \ \ \isakeyword{and}\ h{\isadigit{1}}{\isadigit{3}}{\isacharcolon}{\isachardoublequoteopen}st{\isacharunderscore}in\ t\ {\isacharequal}\ voice{\isacharunderscore}com\ {\isasymand}\ stop\ t\ {\isacharequal}\ {\isacharbrackleft}stop{\isacharunderscore}vc{\isacharbrackright}\ {\isasymlongrightarrow}\isanewline
\ \ \ \ \ \ \ \ \ \ ack\ t\ {\isacharequal}\ {\isacharbrackleft}init{\isacharunderscore}state{\isacharbrackright}\ {\isasymand}\ i{\isadigit{1}}\ t\ {\isacharequal}\ {\isacharbrackleft}{\isacharbrackright}\ {\isasymand}\ vc\ t\ {\isacharequal}\ {\isacharbrackleft}{\isacharbrackright}\ {\isasymand}\ st{\isacharunderscore}out\ t\ {\isacharequal}\ init{\isacharunderscore}state{\isachardoublequoteclose}\isanewline
\ \ \isakeyword{shows}\ {\isachardoublequoteopen}ack\ t\ {\isacharequal}\ {\isacharbrackleft}st{\isacharunderscore}out\ t{\isacharbrackright}{\isachardoublequoteclose}\isanewline
\isadelimproof
\endisadelimproof
\isatagproof
\isacommand{proof}\isamarkupfalse%
\ {\isacharminus}\ \isanewline
\ \ \isacommand{from}\isamarkupfalse%
\ h{\isadigit{1}}\ \isakeyword{and}\ h{\isadigit{6}}\ \isakeyword{and}\ h{\isadigit{7}}\ \isacommand{have}\isamarkupfalse%
\ sg{\isadigit{1}}{\isacharcolon}{\isachardoublequoteopen}lose\ t\ {\isacharequal}\ {\isacharbrackleft}True{\isacharbrackright}\ {\isasymor}\ lose\ t\ {\isacharequal}\ {\isacharbrackleft}False{\isacharbrackright}{\isachardoublequoteclose}\isanewline
\ \ \ \ \isacommand{by}\isamarkupfalse%
\ {\isacharparenleft}simp\ add{\isacharcolon}\ ts{\isacharunderscore}bool{\isacharunderscore}True{\isacharunderscore}False{\isacharparenright}\isanewline
\isacommand{show}\isamarkupfalse%
\ {\isacharquery}thesis\ \isanewline
\isacommand{proof}\isamarkupfalse%
\ {\isacharparenleft}cases\ {\isachardoublequoteopen}st{\isacharunderscore}in\ t{\isachardoublequoteclose}{\isacharparenright}\isanewline
\ \ \isacommand{assume}\isamarkupfalse%
\ a{\isadigit{1}}{\isacharcolon}{\isachardoublequoteopen}st{\isacharunderscore}in\ t\ {\isacharequal}\ init{\isacharunderscore}state{\isachardoublequoteclose}\isanewline
\ \ \isacommand{from}\isamarkupfalse%
\ a{\isadigit{1}}\ \isakeyword{and}\ h{\isadigit{4}}\ \isakeyword{and}\ h{\isadigit{5}}\ \isacommand{show}\isamarkupfalse%
\ {\isacharquery}thesis\ \isanewline
\ \ \isacommand{proof}\isamarkupfalse%
\ {\isacharparenleft}cases\ {\isachardoublequoteopen}req\ t\ {\isacharequal}\ {\isacharbrackleft}init{\isacharbrackright}{\isachardoublequoteclose}{\isacharparenright}\isanewline
\ \ \ \ \isacommand{assume}\isamarkupfalse%
\ a{\isadigit{1}}{\isadigit{1}}{\isacharcolon}{\isachardoublequoteopen}req\ t\ {\isacharequal}\ {\isacharbrackleft}init{\isacharbrackright}{\isachardoublequoteclose}\isanewline
\ \ \ \ \isacommand{from}\isamarkupfalse%
\ a{\isadigit{1}}{\isadigit{1}}\ \isakeyword{and}\ a{\isadigit{1}}\ \isakeyword{and}\ h{\isadigit{4}}\ \isakeyword{and}\ h{\isadigit{5}}\ \isacommand{show}\isamarkupfalse%
\ {\isacharquery}thesis\ \isacommand{by}\isamarkupfalse%
\ simp\isanewline
\ \ \isacommand{next}\isamarkupfalse%
\isanewline
\ \ \ \ \isacommand{assume}\isamarkupfalse%
\ a{\isadigit{1}}{\isadigit{2}}{\isacharcolon}{\isachardoublequoteopen}req\ t\ {\isasymnoteq}\ {\isacharbrackleft}init{\isacharbrackright}{\isachardoublequoteclose}\isanewline
\ \ \ \ \isacommand{from}\isamarkupfalse%
\ a{\isadigit{1}}{\isadigit{2}}\ \isakeyword{and}\ a{\isadigit{1}}\ \isakeyword{and}\ h{\isadigit{4}}\ \isakeyword{and}\ h{\isadigit{5}}\ \isacommand{show}\isamarkupfalse%
\ {\isacharquery}thesis\ \isacommand{by}\isamarkupfalse%
\ simp\isanewline
\ \ \isacommand{qed}\isamarkupfalse%
\isanewline
\isacommand{next}\isamarkupfalse%
\isanewline
\ \ \isacommand{assume}\isamarkupfalse%
\ a{\isadigit{2}}{\isacharcolon}{\isachardoublequoteopen}st{\isacharunderscore}in\ t\ {\isacharequal}\ call{\isachardoublequoteclose}\isanewline
\ \ \isacommand{from}\isamarkupfalse%
\ a{\isadigit{2}}\ \isakeyword{and}\ sg{\isadigit{1}}\ \isakeyword{and}\ h{\isadigit{6}}\ \isakeyword{and}\ h{\isadigit{7}}\ \isacommand{show}\isamarkupfalse%
\ {\isacharquery}thesis\ \isanewline
\ \ \ \ \ \isacommand{apply}\isamarkupfalse%
\ simp\isanewline
\ \ \ \ \ \isacommand{by}\isamarkupfalse%
\ {\isacharparenleft}rule\ ts{\isacharunderscore}lose{\isacharunderscore}ack{\isacharunderscore}st{\isadigit{2}}{\isacharcomma}\ assumption{\isacharplus}{\isacharparenright}\isanewline
\isacommand{next}\isamarkupfalse%
\isanewline
\ \ \isacommand{assume}\isamarkupfalse%
\ a{\isadigit{3}}{\isacharcolon}{\isachardoublequoteopen}st{\isacharunderscore}in\ t\ {\isacharequal}\ connection{\isacharunderscore}ok{\isachardoublequoteclose}\isanewline
\ \ \isacommand{from}\isamarkupfalse%
\ a{\isadigit{3}}\ \isakeyword{and}\ h{\isadigit{6}}\ \isakeyword{and}\ h{\isadigit{7}}\ \isakeyword{and}\ h{\isadigit{8}}\ \isacommand{show}\isamarkupfalse%
\ {\isacharquery}thesis\ \isacommand{apply}\isamarkupfalse%
\ simp\isanewline
\ \ \isacommand{proof}\isamarkupfalse%
\ {\isacharparenleft}cases\ {\isachardoublequoteopen}req\ t\ {\isacharequal}\ {\isacharbrackleft}send{\isacharbrackright}{\isachardoublequoteclose}{\isacharparenright}\isanewline
\ \ \ \ \isacommand{assume}\isamarkupfalse%
\ a{\isadigit{3}}{\isadigit{1}}{\isacharcolon}{\isachardoublequoteopen}req\ t\ {\isacharequal}\ {\isacharbrackleft}send{\isacharbrackright}{\isachardoublequoteclose}\isanewline
\ \ \ \ \isacommand{from}\isamarkupfalse%
\ this\ \isakeyword{and}\ a{\isadigit{3}}\ \isakeyword{and}\ h{\isadigit{6}}\ \isakeyword{and}\ h{\isadigit{7}}\ \isakeyword{and}\ h{\isadigit{8}}\ \isakeyword{and}\ sg{\isadigit{1}}\ \isacommand{show}\isamarkupfalse%
\ {\isacharquery}thesis\ \isanewline
\ \ \ \ \ \ \isacommand{apply}\isamarkupfalse%
\ simp\isanewline
\ \ \ \ \ \ \isacommand{by}\isamarkupfalse%
\ {\isacharparenleft}rule\ ts{\isacharunderscore}lose{\isacharunderscore}ack{\isacharunderscore}st{\isadigit{2}}send{\isacharcomma}\ assumption{\isacharplus}{\isacharparenright}\ \isanewline
\ \ \isacommand{next}\isamarkupfalse%
\isanewline
\ \ \ \ \isacommand{assume}\isamarkupfalse%
\ a{\isadigit{3}}{\isadigit{2}}{\isacharcolon}{\isachardoublequoteopen}req\ t\ {\isasymnoteq}\ {\isacharbrackleft}send{\isacharbrackright}{\isachardoublequoteclose}\isanewline
\ \ \ \ \isacommand{from}\isamarkupfalse%
\ this\ \isakeyword{and}\ a{\isadigit{3}}\ \isakeyword{and}\ h{\isadigit{6}}\ \isakeyword{and}\ h{\isadigit{7}}\ \isakeyword{and}\ h{\isadigit{8}}\ \isakeyword{and}\ sg{\isadigit{1}}\ \isacommand{show}\isamarkupfalse%
\ {\isacharquery}thesis\ \isanewline
\ \ \ \ \ \ \isacommand{apply}\isamarkupfalse%
\ simp\isanewline
\ \ \ \ \ \ \isacommand{by}\isamarkupfalse%
\ {\isacharparenleft}rule\ ts{\isacharunderscore}lose{\isacharunderscore}ack{\isacharunderscore}st{\isadigit{2}}{\isacharcomma}\ assumption{\isacharplus}{\isacharparenright}\ \isanewline
\ \ \isacommand{qed}\isamarkupfalse%
\isanewline
\isacommand{next}\isamarkupfalse%
\isanewline
\ \ \isacommand{assume}\isamarkupfalse%
\ a{\isadigit{4}}{\isacharcolon}{\isachardoublequoteopen}st{\isacharunderscore}in\ t\ {\isacharequal}\ sending{\isacharunderscore}data{\isachardoublequoteclose}\ \isanewline
\ \ \isacommand{from}\isamarkupfalse%
\ sg{\isadigit{1}}\ \isakeyword{and}\ a{\isadigit{4}}\ \isakeyword{and}\ h{\isadigit{7}}\ \isakeyword{and}\ h{\isadigit{9}}\ \isakeyword{and}\ h{\isadigit{1}}{\isadigit{0}}\ \ \isacommand{show}\isamarkupfalse%
\ {\isacharquery}thesis\ \isacommand{apply}\isamarkupfalse%
\ simp\ \isanewline
\ \ \isacommand{proof}\isamarkupfalse%
\ {\isacharparenleft}cases\ {\isachardoublequoteopen}a{\isadigit{1}}\ t\ {\isacharequal}\ {\isacharbrackleft}{\isacharbrackright}{\isachardoublequoteclose}{\isacharparenright}\isanewline
\ \ \ \ \isacommand{assume}\isamarkupfalse%
\ a{\isadigit{4}}{\isadigit{1}}{\isacharcolon}{\isachardoublequoteopen}a{\isadigit{1}}\ t\ {\isacharequal}\ {\isacharbrackleft}{\isacharbrackright}{\isachardoublequoteclose}\isanewline
\ \ \ \ \isacommand{from}\isamarkupfalse%
\ this\ \isakeyword{and}\ a{\isadigit{4}}\ \isakeyword{and}\ sg{\isadigit{1}}\ \isakeyword{and}\ h{\isadigit{7}}\ \isakeyword{and}\ h{\isadigit{9}}\ \isakeyword{and}\ h{\isadigit{1}}{\isadigit{0}}\ \isacommand{show}\isamarkupfalse%
\ {\isacharquery}thesis\isanewline
\ \ \ \ \ \ \isacommand{apply}\isamarkupfalse%
\ simp\isanewline
\ \ \ \ \ \ \isacommand{by}\isamarkupfalse%
\ {\isacharparenleft}rule\ ts{\isacharunderscore}lose{\isacharunderscore}ack{\isacharunderscore}st{\isadigit{2}}{\isacharcomma}\ assumption{\isacharplus}{\isacharparenright}\ \isanewline
\ \ \isacommand{next}\isamarkupfalse%
\isanewline
\ \ \ \ \isacommand{assume}\isamarkupfalse%
\ a{\isadigit{4}}{\isadigit{2}}{\isacharcolon}{\isachardoublequoteopen}a{\isadigit{1}}\ t\ {\isasymnoteq}\ {\isacharbrackleft}{\isacharbrackright}{\isachardoublequoteclose}\isanewline
\ \ \ \ \isacommand{from}\isamarkupfalse%
\ this\ \isakeyword{and}\ h{\isadigit{2}}\ \ \isacommand{have}\isamarkupfalse%
\ {\isachardoublequoteopen}a{\isadigit{1}}\ t\ {\isacharequal}\ {\isacharbrackleft}sc{\isacharunderscore}ack{\isacharbrackright}{\isachardoublequoteclose}\ \ \isacommand{by}\isamarkupfalse%
\ {\isacharparenleft}simp\ add{\isacharcolon}\ aType{\isacharunderscore}nonempty{\isacharparenright}\isanewline
\ \ \ \ \isacommand{from}\isamarkupfalse%
\ this\ \isakeyword{and}\ a{\isadigit{4}}\ \isakeyword{and}\ a{\isadigit{4}}{\isadigit{2}}\ \isakeyword{and}\ sg{\isadigit{1}}\ \isakeyword{and}\ h{\isadigit{7}}\ \isakeyword{and}\ h{\isadigit{9}}\ \isakeyword{and}\ h{\isadigit{1}}{\isadigit{0}}\ \isacommand{show}\isamarkupfalse%
\ {\isacharquery}thesis\isanewline
\ \ \ \ \ \ \isacommand{apply}\isamarkupfalse%
\ simp\isanewline
\ \ \ \ \ \ \isacommand{by}\isamarkupfalse%
\ {\isacharparenleft}rule\ ts{\isacharunderscore}lose{\isacharunderscore}ack{\isacharunderscore}st{\isadigit{2}}vc{\isacharunderscore}com{\isacharcomma}\ assumption{\isacharplus}{\isacharparenright}\isanewline
\ \ \isacommand{qed}\isamarkupfalse%
\ \ \isanewline
\isacommand{next}\isamarkupfalse%
\isanewline
\ \ \isacommand{assume}\isamarkupfalse%
\ a{\isadigit{5}}{\isacharcolon}{\isachardoublequoteopen}st{\isacharunderscore}in\ t\ {\isacharequal}\ voice{\isacharunderscore}com{\isachardoublequoteclose}\isanewline
\ \ \isacommand{from}\isamarkupfalse%
\ a{\isadigit{5}}\ \isakeyword{and}\ h{\isadigit{1}}{\isadigit{1}}\ \isakeyword{and}\ h{\isadigit{1}}{\isadigit{2}}\ \isakeyword{and}\ h{\isadigit{1}}{\isadigit{3}}\ \isacommand{show}\isamarkupfalse%
\ {\isacharquery}thesis\ \isacommand{apply}\isamarkupfalse%
\ simp\isanewline
\ \ \isacommand{proof}\isamarkupfalse%
\ {\isacharparenleft}cases\ {\isachardoublequoteopen}stop\ t\ {\isacharequal}\ {\isacharbrackleft}{\isacharbrackright}{\isachardoublequoteclose}{\isacharparenright}\isanewline
\ \ \ \ \isacommand{assume}\isamarkupfalse%
\ a{\isadigit{5}}{\isadigit{1}}{\isacharcolon}{\isachardoublequoteopen}stop\ t\ {\isacharequal}\ {\isacharbrackleft}{\isacharbrackright}{\isachardoublequoteclose}\isanewline
\ \ \ \ \isacommand{from}\isamarkupfalse%
\ this\ \isakeyword{and}\ a{\isadigit{5}}\ \isakeyword{and}\ h{\isadigit{1}}{\isadigit{1}}\ \isakeyword{and}\ h{\isadigit{1}}{\isadigit{2}}\ \isakeyword{and}\ h{\isadigit{1}}{\isadigit{3}}\ \isakeyword{and}\ sg{\isadigit{1}}\ \isacommand{show}\isamarkupfalse%
\ {\isacharquery}thesis\isanewline
\ \ \ \ \ \ \isacommand{apply}\isamarkupfalse%
\ simp\isanewline
\ \ \ \ \ \ \isacommand{by}\isamarkupfalse%
\ {\isacharparenleft}rule\ ts{\isacharunderscore}lose{\isacharunderscore}ack{\isacharunderscore}st{\isadigit{2}}vc{\isacharunderscore}com{\isacharcomma}\ assumption{\isacharplus}{\isacharparenright}\isanewline
\ \ \isacommand{next}\isamarkupfalse%
\isanewline
\ \ \ \ \isacommand{assume}\isamarkupfalse%
\ a{\isadigit{5}}{\isadigit{2}}{\isacharcolon}{\isachardoublequoteopen}stop\ t\ {\isasymnoteq}\ {\isacharbrackleft}{\isacharbrackright}{\isachardoublequoteclose}\isanewline
\ \ \ \ \isacommand{from}\isamarkupfalse%
\ this\ \isakeyword{and}\ h{\isadigit{3}}\ \isacommand{have}\isamarkupfalse%
\ sg{\isadigit{7}}{\isacharcolon}{\isachardoublequoteopen}stop\ t\ {\isacharequal}\ {\isacharbrackleft}stop{\isacharunderscore}vc{\isacharbrackright}{\isachardoublequoteclose}\ \ \isanewline
\ \ \ \ \ \ \isacommand{by}\isamarkupfalse%
\ {\isacharparenleft}simp\ add{\isacharcolon}\ stopType{\isacharunderscore}nonempty{\isacharparenright}\isanewline
\ \ \ \ \isacommand{from}\isamarkupfalse%
\ this\ \isakeyword{and}\ a{\isadigit{5}}\ \isakeyword{and}\ a{\isadigit{5}}{\isadigit{2}}\ \isakeyword{and}\ h{\isadigit{1}}{\isadigit{3}}\ \isacommand{show}\isamarkupfalse%
\ {\isacharquery}thesis\ \isacommand{by}\isamarkupfalse%
\ simp\isanewline
\ \ \isacommand{qed}\isamarkupfalse%
\isanewline
\isacommand{qed}\isamarkupfalse%
\isanewline
\isacommand{qed}\isamarkupfalse%
\endisatagproof
{\isafoldproof}%
\isadelimproof
\isanewline
\endisadelimproof
\isanewline
\isacommand{lemma}\isamarkupfalse%
\ tiTable{\isacharunderscore}ack{\isacharunderscore}st{\isacharcolon}\isanewline
\ \ \isakeyword{assumes}\ h{\isadigit{1}}{\isacharcolon}{\isachardoublequoteopen}tiTable{\isacharunderscore}SampleT\ req\ a{\isadigit{1}}\ stop\ lose\ st{\isacharunderscore}in\ b\ ack\ i{\isadigit{1}}\ vc\ st{\isacharunderscore}out{\isachardoublequoteclose}\isanewline
\ \ \ \ \ \ \isakeyword{and}\ h{\isadigit{2}}{\isacharcolon}{\isachardoublequoteopen}ts\ lose{\isachardoublequoteclose}\isanewline
\ \ \ \ \ \ \isakeyword{and}\ h{\isadigit{3}}{\isacharcolon}{\isachardoublequoteopen}msg\ {\isacharparenleft}Suc\ {\isadigit{0}}{\isacharparenright}\ a{\isadigit{1}}{\isachardoublequoteclose}\ \ \ \ \ \ \isanewline
\ \ \ \ \ \ \isakeyword{and}\ h{\isadigit{4}}{\isacharcolon}{\isachardoublequoteopen}msg\ {\isacharparenleft}Suc\ {\isadigit{0}}{\isacharparenright}\ stop{\isachardoublequoteclose}\isanewline
\ \ \isakeyword{shows}\ \ \ \ \ \ {\isachardoublequoteopen}ack\ t\ {\isacharequal}\ {\isacharbrackleft}st{\isacharunderscore}out\ t{\isacharbrackright}\ {\isachardoublequoteclose}\isanewline
\isadelimproof
\endisadelimproof
\isatagproof
\isacommand{proof}\isamarkupfalse%
\ {\isacharminus}\isanewline
\ \ \isacommand{from}\isamarkupfalse%
\ assms\ \isacommand{have}\isamarkupfalse%
\ sg{\isadigit{1}}{\isacharcolon}\isanewline
\ \ \ {\isachardoublequoteopen}st{\isacharunderscore}in\ t\ {\isacharequal}\ init{\isacharunderscore}state\ {\isasymand}\ req\ t\ {\isacharequal}\ {\isacharbrackleft}init{\isacharbrackright}\ {\isasymlongrightarrow}\isanewline
\ \ \ \ ack\ t\ {\isacharequal}\ {\isacharbrackleft}call{\isacharbrackright}\ {\isasymand}\ i{\isadigit{1}}\ t\ {\isacharequal}\ {\isacharbrackleft}{\isacharbrackright}\ {\isasymand}\ vc\ t\ {\isacharequal}\ {\isacharbrackleft}{\isacharbrackright}\ {\isasymand}\ st{\isacharunderscore}out\ t\ {\isacharequal}\ call{\isachardoublequoteclose}\isanewline
\ \ \ \ \ \isacommand{by}\isamarkupfalse%
\ {\isacharparenleft}simp\ add{\isacharcolon}\ tiTable{\isacharunderscore}SampleT{\isacharunderscore}def{\isacharparenright}\isanewline
\ \ \isacommand{from}\isamarkupfalse%
\ assms\ \isacommand{have}\isamarkupfalse%
\ sg{\isadigit{2}}{\isacharcolon}\isanewline
\ \ \ {\isachardoublequoteopen}st{\isacharunderscore}in\ t\ {\isacharequal}\ init{\isacharunderscore}state\ {\isasymand}\ req\ t\ {\isasymnoteq}\ {\isacharbrackleft}init{\isacharbrackright}\ {\isasymlongrightarrow}\isanewline
\ \ \ \ ack\ t\ {\isacharequal}\ {\isacharbrackleft}init{\isacharunderscore}state{\isacharbrackright}\ {\isasymand}\ i{\isadigit{1}}\ t\ {\isacharequal}\ {\isacharbrackleft}{\isacharbrackright}\ {\isasymand}\ vc\ t\ {\isacharequal}\ {\isacharbrackleft}{\isacharbrackright}\ {\isasymand}\ st{\isacharunderscore}out\ t\ {\isacharequal}\ init{\isacharunderscore}state{\isachardoublequoteclose}\isanewline
\ \ \ \ \ \isacommand{by}\isamarkupfalse%
\ {\isacharparenleft}simp\ add{\isacharcolon}\ tiTable{\isacharunderscore}SampleT{\isacharunderscore}def{\isacharparenright}\isanewline
\ \ \isacommand{from}\isamarkupfalse%
\ assms\ \isacommand{have}\isamarkupfalse%
\ sg{\isadigit{3}}{\isacharcolon}\isanewline
\ \ \ {\isachardoublequoteopen}{\isacharparenleft}st{\isacharunderscore}in\ t\ {\isacharequal}\ call\ {\isasymor}\ st{\isacharunderscore}in\ t\ {\isacharequal}\ connection{\isacharunderscore}ok\ {\isasymand}\ req\ t\ {\isasymnoteq}\ {\isacharbrackleft}send{\isacharbrackright}{\isacharparenright}\ {\isasymand}\ \isanewline
\ \ \ \ \ lose\ t\ {\isacharequal}\ {\isacharbrackleft}False{\isacharbrackright}\ {\isasymlongrightarrow}\isanewline
\ \ \ \ \ ack\ t\ {\isacharequal}\ {\isacharbrackleft}connection{\isacharunderscore}ok{\isacharbrackright}\ {\isasymand}\ i{\isadigit{1}}\ t\ {\isacharequal}\ {\isacharbrackleft}{\isacharbrackright}\ {\isasymand}\ vc\ t\ {\isacharequal}\ {\isacharbrackleft}{\isacharbrackright}\ {\isasymand}\ st{\isacharunderscore}out\ t\ {\isacharequal}\ connection{\isacharunderscore}ok{\isachardoublequoteclose}\isanewline
\ \ \ \ \ \isacommand{by}\isamarkupfalse%
\ {\isacharparenleft}simp\ add{\isacharcolon}\ tiTable{\isacharunderscore}SampleT{\isacharunderscore}def{\isacharparenright}\isanewline
\ \ \isacommand{from}\isamarkupfalse%
\ assms\ \isacommand{have}\isamarkupfalse%
\ sg{\isadigit{4}}{\isacharcolon}\isanewline
\ \ \ {\isachardoublequoteopen}{\isacharparenleft}st{\isacharunderscore}in\ t\ {\isacharequal}\ call\ {\isasymor}\ st{\isacharunderscore}in\ t\ {\isacharequal}\ connection{\isacharunderscore}ok\ {\isasymor}\ st{\isacharunderscore}in\ t\ {\isacharequal}\ sending{\isacharunderscore}data{\isacharparenright}\ {\isasymand}\ \isanewline
\ \ \ \ \ lose\ t\ {\isacharequal}\ {\isacharbrackleft}True{\isacharbrackright}\ {\isasymlongrightarrow}\isanewline
\ \ \ \ \ ack\ t\ {\isacharequal}\ {\isacharbrackleft}init{\isacharunderscore}state{\isacharbrackright}\ {\isasymand}\ i{\isadigit{1}}\ t\ {\isacharequal}\ {\isacharbrackleft}{\isacharbrackright}\ {\isasymand}\ vc\ t\ {\isacharequal}\ {\isacharbrackleft}{\isacharbrackright}\ {\isasymand}\ st{\isacharunderscore}out\ t\ {\isacharequal}\ init{\isacharunderscore}state{\isachardoublequoteclose}\isanewline
\ \ \ \ \ \isacommand{by}\isamarkupfalse%
\ {\isacharparenleft}simp\ add{\isacharcolon}\ tiTable{\isacharunderscore}SampleT{\isacharunderscore}def{\isacharparenright}\isanewline
\ \ \isacommand{from}\isamarkupfalse%
\ assms\ \isacommand{have}\isamarkupfalse%
\ sg{\isadigit{5}}{\isacharcolon}\isanewline
\ \ \ {\isachardoublequoteopen}st{\isacharunderscore}in\ t\ {\isacharequal}\ connection{\isacharunderscore}ok\ {\isasymand}\ req\ t\ {\isacharequal}\ {\isacharbrackleft}send{\isacharbrackright}\ {\isasymand}\ lose\ t\ {\isacharequal}\ {\isacharbrackleft}False{\isacharbrackright}\ {\isasymlongrightarrow}\isanewline
\ \ \ \ ack\ t\ {\isacharequal}\ {\isacharbrackleft}sending{\isacharunderscore}data{\isacharbrackright}\ {\isasymand}\ i{\isadigit{1}}\ t\ {\isacharequal}\ b\ t\ {\isasymand}\ vc\ t\ {\isacharequal}\ {\isacharbrackleft}{\isacharbrackright}\ {\isasymand}\ st{\isacharunderscore}out\ t\ {\isacharequal}\ sending{\isacharunderscore}data{\isachardoublequoteclose}\isanewline
\ \ \ \ \isacommand{by}\isamarkupfalse%
\ {\isacharparenleft}simp\ add{\isacharcolon}\ tiTable{\isacharunderscore}SampleT{\isacharunderscore}def{\isacharparenright}\isanewline
\ \ \isacommand{from}\isamarkupfalse%
\ assms\ \isacommand{have}\isamarkupfalse%
\ sg{\isadigit{6}}{\isacharcolon}\isanewline
\ \ \ {\isachardoublequoteopen}st{\isacharunderscore}in\ t\ {\isacharequal}\ sending{\isacharunderscore}data\ {\isasymand}\ a{\isadigit{1}}\ t\ {\isacharequal}\ {\isacharbrackleft}{\isacharbrackright}\ {\isasymand}\ lose\ t\ {\isacharequal}\ {\isacharbrackleft}False{\isacharbrackright}\ {\isasymlongrightarrow}\isanewline
\ \ \ \ ack\ t\ {\isacharequal}\ {\isacharbrackleft}sending{\isacharunderscore}data{\isacharbrackright}\ {\isasymand}\ i{\isadigit{1}}\ t\ {\isacharequal}\ {\isacharbrackleft}{\isacharbrackright}\ {\isasymand}\ vc\ t\ {\isacharequal}\ {\isacharbrackleft}{\isacharbrackright}\ {\isasymand}\ st{\isacharunderscore}out\ t\ {\isacharequal}\ sending{\isacharunderscore}data{\isachardoublequoteclose}\isanewline
\ \ \ \ \ \isacommand{by}\isamarkupfalse%
\ {\isacharparenleft}simp\ add{\isacharcolon}\ tiTable{\isacharunderscore}SampleT{\isacharunderscore}def{\isacharparenright}\isanewline
\ \ \isacommand{from}\isamarkupfalse%
\ assms\ \isacommand{have}\isamarkupfalse%
\ sg{\isadigit{7}}{\isacharcolon}\isanewline
\ \ \ {\isachardoublequoteopen}st{\isacharunderscore}in\ t\ {\isacharequal}\ sending{\isacharunderscore}data\ {\isasymand}\ a{\isadigit{1}}\ t\ {\isacharequal}\ {\isacharbrackleft}sc{\isacharunderscore}ack{\isacharbrackright}\ {\isasymand}\ lose\ t\ {\isacharequal}\ {\isacharbrackleft}False{\isacharbrackright}\ {\isasymlongrightarrow}\isanewline
\ \ \ \ ack\ t\ {\isacharequal}\ {\isacharbrackleft}voice{\isacharunderscore}com{\isacharbrackright}\ {\isasymand}\ i{\isadigit{1}}\ t\ {\isacharequal}\ {\isacharbrackleft}{\isacharbrackright}\ {\isasymand}\ vc\ t\ {\isacharequal}\ {\isacharbrackleft}vc{\isacharunderscore}com{\isacharbrackright}\ {\isasymand}\ st{\isacharunderscore}out\ t\ {\isacharequal}\ voice{\isacharunderscore}com{\isachardoublequoteclose}\isanewline
\ \ \ \ \isacommand{by}\isamarkupfalse%
\ {\isacharparenleft}simp\ add{\isacharcolon}\ tiTable{\isacharunderscore}SampleT{\isacharunderscore}def{\isacharparenright}\isanewline
\ \ \isacommand{from}\isamarkupfalse%
\ assms\ \isacommand{have}\isamarkupfalse%
\ sg{\isadigit{8}}{\isacharcolon}\isanewline
\ \ \ {\isachardoublequoteopen}st{\isacharunderscore}in\ t\ {\isacharequal}\ voice{\isacharunderscore}com\ {\isasymand}\ stop\ t\ {\isacharequal}\ {\isacharbrackleft}{\isacharbrackright}\ {\isasymand}\ lose\ t\ {\isacharequal}\ {\isacharbrackleft}False{\isacharbrackright}\ {\isasymlongrightarrow}\isanewline
\ \ \ \ ack\ t\ {\isacharequal}\ {\isacharbrackleft}voice{\isacharunderscore}com{\isacharbrackright}\ {\isasymand}\ i{\isadigit{1}}\ t\ {\isacharequal}\ {\isacharbrackleft}{\isacharbrackright}\ {\isasymand}\ vc\ t\ {\isacharequal}\ {\isacharbrackleft}vc{\isacharunderscore}com{\isacharbrackright}\ {\isasymand}\ st{\isacharunderscore}out\ t\ {\isacharequal}\ voice{\isacharunderscore}com{\isachardoublequoteclose}\isanewline
\ \ \ \ \isacommand{by}\isamarkupfalse%
\ {\isacharparenleft}simp\ add{\isacharcolon}\ tiTable{\isacharunderscore}SampleT{\isacharunderscore}def{\isacharparenright}\isanewline
\ \ \isacommand{from}\isamarkupfalse%
\ assms\ \isacommand{have}\isamarkupfalse%
\ sg{\isadigit{9}}{\isacharcolon}\isanewline
\ \ \ {\isachardoublequoteopen}st{\isacharunderscore}in\ t\ {\isacharequal}\ voice{\isacharunderscore}com\ {\isasymand}\ stop\ t\ {\isacharequal}\ {\isacharbrackleft}{\isacharbrackright}\ {\isasymand}\ lose\ t\ {\isacharequal}\ {\isacharbrackleft}True{\isacharbrackright}\ {\isasymlongrightarrow}\isanewline
\ \ \ \ ack\ t\ {\isacharequal}\ {\isacharbrackleft}voice{\isacharunderscore}com{\isacharbrackright}\ {\isasymand}\ i{\isadigit{1}}\ t\ {\isacharequal}\ {\isacharbrackleft}{\isacharbrackright}\ {\isasymand}\ vc\ t\ {\isacharequal}\ {\isacharbrackleft}{\isacharbrackright}\ {\isasymand}\ st{\isacharunderscore}out\ t\ {\isacharequal}\ voice{\isacharunderscore}com{\isachardoublequoteclose}\isanewline
\ \ \ \ \isacommand{by}\isamarkupfalse%
\ {\isacharparenleft}simp\ add{\isacharcolon}\ tiTable{\isacharunderscore}SampleT{\isacharunderscore}def{\isacharparenright}\isanewline
\ \ \isacommand{from}\isamarkupfalse%
\ assms\ \isacommand{have}\isamarkupfalse%
\ sg{\isadigit{1}}{\isadigit{0}}{\isacharcolon}\isanewline
\ \ \ {\isachardoublequoteopen}st{\isacharunderscore}in\ t\ {\isacharequal}\ voice{\isacharunderscore}com\ {\isasymand}\ stop\ t\ {\isacharequal}\ {\isacharbrackleft}stop{\isacharunderscore}vc{\isacharbrackright}\ {\isasymlongrightarrow}\isanewline
\ \ \ \ ack\ t\ {\isacharequal}\ {\isacharbrackleft}init{\isacharunderscore}state{\isacharbrackright}\ {\isasymand}\ i{\isadigit{1}}\ t\ {\isacharequal}\ {\isacharbrackleft}{\isacharbrackright}\ {\isasymand}\ vc\ t\ {\isacharequal}\ {\isacharbrackleft}{\isacharbrackright}\ {\isasymand}\ st{\isacharunderscore}out\ t\ {\isacharequal}\ init{\isacharunderscore}state{\isachardoublequoteclose}\isanewline
\ \ \ \ \isacommand{by}\isamarkupfalse%
\ {\isacharparenleft}simp\ add{\isacharcolon}\ tiTable{\isacharunderscore}SampleT{\isacharunderscore}def{\isacharparenright}\isanewline
\ \ \isacommand{from}\isamarkupfalse%
\ h{\isadigit{2}}\ \isakeyword{and}\ h{\isadigit{3}}\ \isakeyword{and}\ h{\isadigit{4}}\ \isakeyword{and}\ sg{\isadigit{1}}\ \isakeyword{and}\ sg{\isadigit{2}}\ \isakeyword{and}\ sg{\isadigit{3}}\ \isakeyword{and}\ sg{\isadigit{4}}\ \isakeyword{and}\ sg{\isadigit{5}}\ \isakeyword{and}\ \isanewline
\ \ sg{\isadigit{6}}\ \isakeyword{and}\ sg{\isadigit{7}}\ \isakeyword{and}\ sg{\isadigit{8}}\ \isakeyword{and}\ sg{\isadigit{9}}\ \isakeyword{and}\ sg{\isadigit{1}}{\isadigit{0}}\ \isacommand{show}\isamarkupfalse%
\ {\isacharquery}thesis\ \isanewline
\ \ \ \ \isacommand{by}\isamarkupfalse%
\ {\isacharparenleft}rule\ tiTable{\isacharunderscore}ack{\isacharunderscore}st{\isacharunderscore}splitten{\isacharparenright}\isanewline
\isacommand{qed}\isamarkupfalse%
\endisatagproof
{\isafoldproof}%
\isadelimproof
\isanewline
\endisadelimproof
\isanewline 
\isacommand{lemma}\isamarkupfalse%
\ tiTable{\isacharunderscore}ack{\isacharunderscore}st{\isacharunderscore}hd{\isacharcolon}\isanewline
\ \ \isakeyword{assumes}\ h{\isadigit{1}}{\isacharcolon}{\isachardoublequoteopen}tiTable{\isacharunderscore}SampleT\ req\ a{\isadigit{1}}\ stop\ lose\ st{\isacharunderscore}in\ b\ ack\ i{\isadigit{1}}\ vc\ st{\isacharunderscore}out{\isachardoublequoteclose}\isanewline
\ \ \ \ \ \ \isakeyword{and}\ h{\isadigit{2}}{\isacharcolon}{\isachardoublequoteopen}ts\ lose{\isachardoublequoteclose}\isanewline
\ \ \ \ \ \ \isakeyword{and}\ h{\isadigit{3}}{\isacharcolon}{\isachardoublequoteopen}msg\ {\isacharparenleft}Suc\ {\isadigit{0}}{\isacharparenright}\ a{\isadigit{1}}{\isachardoublequoteclose}\isanewline
\ \ \ \ \ \ \isakeyword{and}\ h{\isadigit{4}}{\isacharcolon}{\isachardoublequoteopen}msg\ {\isacharparenleft}Suc\ {\isadigit{0}}{\isacharparenright}\ stop{\isachardoublequoteclose}\isanewline
\ \ \isakeyword{shows}\ {\isachardoublequoteopen}st{\isacharunderscore}out\ t\ {\isacharequal}\ \ hd\ {\isacharparenleft}ack\ t{\isacharparenright}{\isachardoublequoteclose}\isanewline
\isadelimproof
\endisadelimproof
\isatagproof
\isacommand{using}\isamarkupfalse%
\ assms\ \isacommand{by}\isamarkupfalse%
\ {\isacharparenleft}simp\ add{\isacharcolon}\ \ tiTable{\isacharunderscore}ack{\isacharunderscore}st{\isacharparenright}%
\endisatagproof
{\isafoldproof}%
\isadelimproof
\isanewline
\endisadelimproof
\isanewline 
\isacommand{lemma}\isamarkupfalse%
\ tiTable{\isacharunderscore}ack{\isacharunderscore}connection{\isacharunderscore}ok{\isacharcolon}\isanewline
\ \ \isakeyword{assumes}\ h{\isadigit{1}}{\isacharcolon}{\isachardoublequoteopen}tiTable{\isacharunderscore}SampleT\ req\ x\ stop\ lose\ st{\isacharunderscore}in\ b\ ack\ i{\isadigit{1}}\ vc\ st{\isacharunderscore}out{\isachardoublequoteclose}\isanewline
\ \ \ \ \ \ \isakeyword{and}\ h{\isadigit{2}}{\isacharcolon}{\isachardoublequoteopen}ack\ t\ {\isacharequal}\ {\isacharbrackleft}connection{\isacharunderscore}ok{\isacharbrackright}{\isachardoublequoteclose}\isanewline
\ \ \ \ \ \ \isakeyword{and}\ h{\isadigit{3}}{\isacharcolon}{\isachardoublequoteopen}msg\ {\isacharparenleft}Suc\ {\isadigit{0}}{\isacharparenright}\ x{\isachardoublequoteclose}\isanewline
\ \ \ \ \ \ \isakeyword{and}\ h{\isadigit{4}}{\isacharcolon}{\isachardoublequoteopen}ts\ lose{\isachardoublequoteclose}\isanewline
\ \ \ \ \ \ \isakeyword{and}\ h{\isadigit{5}}{\isacharcolon}{\isachardoublequoteopen}msg\ {\isacharparenleft}Suc\ {\isadigit{0}}{\isacharparenright}\ stop{\isachardoublequoteclose}\isanewline
\ \ \isakeyword{shows}\ {\isachardoublequoteopen}{\isacharparenleft}st{\isacharunderscore}in\ t\ {\isacharequal}\ call\ {\isasymor}\ st{\isacharunderscore}in\ t\ {\isacharequal}\ connection{\isacharunderscore}ok\ {\isasymand}\ req\ t\ {\isasymnoteq}\ {\isacharbrackleft}send{\isacharbrackright}{\isacharparenright}\ {\isasymand}\ \isanewline
\ \ \ \ \ \ \ \ \ lose\ t\ {\isacharequal}\ {\isacharbrackleft}False{\isacharbrackright}{\isachardoublequoteclose}\isanewline
\isadelimproof
\endisadelimproof
\isatagproof
\isacommand{proof}\isamarkupfalse%
\ {\isacharminus}\ \isanewline
\ \ \isacommand{from}\isamarkupfalse%
\ h{\isadigit{1}}\ \isakeyword{and}\ h{\isadigit{4}}\ \isacommand{have}\isamarkupfalse%
\ sg{\isadigit{1}}{\isacharcolon}{\isachardoublequoteopen}lose\ t\ {\isacharequal}\ {\isacharbrackleft}True{\isacharbrackright}\ {\isasymor}\ lose\ t\ {\isacharequal}\ {\isacharbrackleft}False{\isacharbrackright}{\isachardoublequoteclose}\isanewline
\ \ \ \ \isacommand{by}\isamarkupfalse%
\ {\isacharparenleft}simp\ add{\isacharcolon}\ ts{\isacharunderscore}bool{\isacharunderscore}True{\isacharunderscore}False{\isacharparenright}\isanewline
\ \ \isacommand{from}\isamarkupfalse%
\ h{\isadigit{1}}\ \isakeyword{and}\ h{\isadigit{3}}\ \isacommand{have}\isamarkupfalse%
\ sg{\isadigit{2}}{\isacharcolon}{\isachardoublequoteopen}x\ t\ {\isacharequal}\ {\isacharbrackleft}{\isacharbrackright}\ {\isasymor}\ x\ t\ {\isacharequal}\ {\isacharbrackleft}sc{\isacharunderscore}ack{\isacharbrackright}{\isachardoublequoteclose}\isanewline
\ \ \ \ \isacommand{by}\isamarkupfalse%
\ {\isacharparenleft}simp\ add{\isacharcolon}\ aType{\isacharunderscore}lemma{\isacharparenright}\ \isanewline
\ \ \isacommand{from}\isamarkupfalse%
\ h{\isadigit{1}}\ \isakeyword{and}\ h{\isadigit{5}}\ \isacommand{have}\isamarkupfalse%
\ sg{\isadigit{3}}{\isacharcolon}{\isachardoublequoteopen}stop\ t\ {\isacharequal}\ {\isacharbrackleft}{\isacharbrackright}\ {\isasymor}\ stop\ t\ {\isacharequal}\ {\isacharbrackleft}stop{\isacharunderscore}vc{\isacharbrackright}{\isachardoublequoteclose}\isanewline
\ \ \ \ \isacommand{by}\isamarkupfalse%
\ {\isacharparenleft}simp\ add{\isacharcolon}\ stopType{\isacharunderscore}lemma{\isacharparenright}\ 
\ \ \isacommand{show}\isamarkupfalse%
\ {\isacharquery}thesis\isanewline
\ \ \isacommand{proof}\isamarkupfalse%
\ {\isacharparenleft}cases\ {\isachardoublequoteopen}st{\isacharunderscore}in\ t{\isachardoublequoteclose}{\isacharparenright}\isanewline
\ \ \ \ \isacommand{assume}\isamarkupfalse%
\ a{\isadigit{1}}{\isacharcolon}{\isachardoublequoteopen}st{\isacharunderscore}in\ t\ {\isacharequal}\ init{\isacharunderscore}state{\isachardoublequoteclose}\isanewline
\ \ \ \ \isacommand{show}\isamarkupfalse%
\ {\isacharquery}thesis\isanewline
\ \ \ \ \isacommand{proof}\isamarkupfalse%
\ {\isacharparenleft}cases\ {\isachardoublequoteopen}req\ t\ {\isacharequal}\ {\isacharbrackleft}init{\isacharbrackright}{\isachardoublequoteclose}{\isacharparenright}\isanewline
\ \ \ \ \ \ \isacommand{assume}\isamarkupfalse%
\ a{\isadigit{1}}{\isadigit{1}}{\isacharcolon}{\isachardoublequoteopen}req\ t\ {\isacharequal}\ {\isacharbrackleft}init{\isacharbrackright}{\isachardoublequoteclose}\isanewline
\ \ \ \ \ \ \isacommand{from}\isamarkupfalse%
\ h{\isadigit{1}}\ \isakeyword{and}\ a{\isadigit{1}}\ \isakeyword{and}\ a{\isadigit{1}}{\isadigit{1}}\ \isakeyword{and}\ h{\isadigit{2}}\ \isacommand{show}\isamarkupfalse%
\ {\isacharquery}thesis\ \isacommand{by}\isamarkupfalse%
\ {\isacharparenleft}simp\ add{\isacharcolon}\ tiTable{\isacharunderscore}SampleT{\isacharunderscore}def{\isacharparenright}\isanewline
\ \ \ \ \isacommand{next}\isamarkupfalse%
\isanewline
\ \ \ \ \ \ \isacommand{assume}\isamarkupfalse%
\ a{\isadigit{1}}{\isadigit{2}}{\isacharcolon}{\isachardoublequoteopen}req\ t\ {\isasymnoteq}\ {\isacharbrackleft}init{\isacharbrackright}{\isachardoublequoteclose}\isanewline
\ \ \ \ \ \ \isacommand{from}\isamarkupfalse%
\ h{\isadigit{1}}\ \isakeyword{and}\ a{\isadigit{1}}\ \isakeyword{and}\ a{\isadigit{1}}{\isadigit{2}}\ \isakeyword{and}\ h{\isadigit{2}}\ \isacommand{show}\isamarkupfalse%
\ {\isacharquery}thesis\ \isacommand{by}\isamarkupfalse%
\ {\isacharparenleft}simp\ add{\isacharcolon}\ tiTable{\isacharunderscore}SampleT{\isacharunderscore}def{\isacharparenright}\isanewline
\ \ \ \ \isacommand{qed}\isamarkupfalse%
\ \isanewline
\ \ \isacommand{next}\isamarkupfalse%
\isanewline
\ \ \ \ \isacommand{assume}\isamarkupfalse%
\ a{\isadigit{2}}{\isacharcolon}{\isachardoublequoteopen}st{\isacharunderscore}in\ t\ {\isacharequal}\ call{\isachardoublequoteclose}\isanewline
\ \ \ \ \isacommand{show}\isamarkupfalse%
\ {\isacharquery}thesis\isanewline
\ \ \ \ \isacommand{proof}\isamarkupfalse%
\ {\isacharparenleft}cases\ {\isachardoublequoteopen}lose\ t\ {\isacharequal}\ {\isacharbrackleft}True{\isacharbrackright}{\isachardoublequoteclose}{\isacharparenright}\isanewline
\ \ \ \ \ \ \isacommand{assume}\isamarkupfalse%
\ a{\isadigit{2}}{\isadigit{1}}{\isacharcolon}{\isachardoublequoteopen}lose\ t\ {\isacharequal}\ {\isacharbrackleft}True{\isacharbrackright}{\isachardoublequoteclose}\isanewline
\ \ \ \ \ \ \isacommand{from}\isamarkupfalse%
\ h{\isadigit{1}}\ \isakeyword{and}\ a{\isadigit{2}}\ \isakeyword{and}\ a{\isadigit{2}}{\isadigit{1}}\ \isakeyword{and}\ h{\isadigit{2}}\ \isacommand{show}\isamarkupfalse%
\ {\isacharquery}thesis\ \isacommand{by}\isamarkupfalse%
\ {\isacharparenleft}simp\ add{\isacharcolon}\ tiTable{\isacharunderscore}SampleT{\isacharunderscore}def{\isacharparenright}\isanewline
\ \ \ \ \isacommand{next}\isamarkupfalse%
\isanewline
\ \ \ \ \ \ \isacommand{assume}\isamarkupfalse%
\ a{\isadigit{2}}{\isadigit{2}}{\isacharcolon}{\isachardoublequoteopen}lose\ t\ {\isasymnoteq}\ {\isacharbrackleft}True{\isacharbrackright}{\isachardoublequoteclose}\isanewline
\ \ \ \ \ \ \isacommand{from}\isamarkupfalse%
\ this\ \isakeyword{and}\ h{\isadigit{4}}\ \isacommand{have}\isamarkupfalse%
\ a{\isadigit{2}}{\isadigit{2}}a{\isacharcolon}{\isachardoublequoteopen}lose\ t\ {\isacharequal}\ {\isacharbrackleft}False{\isacharbrackright}{\isachardoublequoteclose}\ \isacommand{by}\isamarkupfalse%
\ {\isacharparenleft}simp\ add{\isacharcolon}\ ts{\isacharunderscore}bool{\isacharunderscore}False{\isacharparenright}\isanewline
\ \ \ \ \ \ \isacommand{from}\isamarkupfalse%
\ h{\isadigit{1}}\ \isacommand{have}\isamarkupfalse%
\ \isanewline
\ \ \ \ \ \ \ {\isachardoublequoteopen}{\isacharparenleft}st{\isacharunderscore}in\ t\ {\isacharequal}\ call\ {\isasymor}\ st{\isacharunderscore}in\ t\ {\isacharequal}\ connection{\isacharunderscore}ok\ {\isasymand}\ req\ t\ {\isasymnoteq}\ {\isacharbrackleft}send{\isacharbrackright}{\isacharparenright}\ {\isasymand}\ \isanewline
\ \ \ \ \ \ \ \ \ lose\ t\ {\isacharequal}\ {\isacharbrackleft}False{\isacharbrackright}\ {\isasymlongrightarrow}\isanewline
\ \ \ \ \ \ \ \ \ ack\ t\ {\isacharequal}\ {\isacharbrackleft}connection{\isacharunderscore}ok{\isacharbrackright}\ {\isasymand}\ i{\isadigit{1}}\ t\ {\isacharequal}\ {\isacharbrackleft}{\isacharbrackright}\ {\isasymand}\ vc\ t\ {\isacharequal}\ {\isacharbrackleft}{\isacharbrackright}\ {\isasymand}\ st{\isacharunderscore}out\ t\ {\isacharequal}\ connection{\isacharunderscore}ok{\isachardoublequoteclose}\isanewline
\ \ \ \ \ \ \ \ \ \isacommand{by}\isamarkupfalse%
\ {\isacharparenleft}simp\ add{\isacharcolon}\ tiTable{\isacharunderscore}SampleT{\isacharunderscore}def{\isacharparenright}\isanewline
\ \ \ \ \ \ \isacommand{from}\isamarkupfalse%
\ this\ \isakeyword{and}\ a{\isadigit{2}}\ \isakeyword{and}\ a{\isadigit{2}}{\isadigit{2}}a\ \isakeyword{and}\ h{\isadigit{2}}\ \isacommand{show}\isamarkupfalse%
\ {\isacharquery}thesis\ \isacommand{by}\isamarkupfalse%
\ simp\isanewline
\ \ \ \ \isacommand{qed}\isamarkupfalse%
\ \isanewline
\ \ \isacommand{next}\isamarkupfalse%
\isanewline
\ \ \ \ \isacommand{assume}\isamarkupfalse%
\ a{\isadigit{3}}{\isacharcolon}{\isachardoublequoteopen}st{\isacharunderscore}in\ t\ {\isacharequal}\ connection{\isacharunderscore}ok{\isachardoublequoteclose}\isanewline
\ \ \ \ \isacommand{show}\isamarkupfalse%
\ {\isacharquery}thesis\isanewline
\ \ \ \ \isacommand{proof}\isamarkupfalse%
\ {\isacharparenleft}cases\ {\isachardoublequoteopen}lose\ t\ {\isacharequal}\ {\isacharbrackleft}True{\isacharbrackright}{\isachardoublequoteclose}{\isacharparenright}\isanewline
\ \ \ \ \ \ \isacommand{assume}\isamarkupfalse%
\ a{\isadigit{3}}{\isadigit{1}}{\isacharcolon}{\isachardoublequoteopen}lose\ t\ {\isacharequal}\ {\isacharbrackleft}True{\isacharbrackright}{\isachardoublequoteclose}\isanewline
\ \ \ \ \ \ \isacommand{from}\isamarkupfalse%
\ h{\isadigit{1}}\ \isacommand{have}\isamarkupfalse%
\ \isanewline
\ \ \ \ \ \ \ {\isachardoublequoteopen}{\isacharparenleft}st{\isacharunderscore}in\ t\ {\isacharequal}\ call\ {\isasymor}\ st{\isacharunderscore}in\ t\ {\isacharequal}\ connection{\isacharunderscore}ok\ {\isasymor}\ st{\isacharunderscore}in\ t\ {\isacharequal}\ sending{\isacharunderscore}data{\isacharparenright}\ {\isasymand}\ \isanewline
\ \ \ \ \ \ \ \ \ lose\ t\ {\isacharequal}\ {\isacharbrackleft}True{\isacharbrackright}\ {\isasymlongrightarrow}\isanewline
\ \ \ \ \ \ \ \ \ ack\ t\ {\isacharequal}\ {\isacharbrackleft}init{\isacharunderscore}state{\isacharbrackright}\ {\isasymand}\ i{\isadigit{1}}\ t\ {\isacharequal}\ {\isacharbrackleft}{\isacharbrackright}\ {\isasymand}\ vc\ t\ {\isacharequal}\ {\isacharbrackleft}{\isacharbrackright}\ {\isasymand}\ st{\isacharunderscore}out\ t\ {\isacharequal}\ init{\isacharunderscore}state{\isachardoublequoteclose}\isanewline
\ \ \ \ \ \ \ \ \isacommand{by}\isamarkupfalse%
\ {\isacharparenleft}simp\ add{\isacharcolon}\ tiTable{\isacharunderscore}SampleT{\isacharunderscore}def{\isacharparenright}\isanewline
\ \ \ \ \ \ \isacommand{from}\isamarkupfalse%
\ this\ \isakeyword{and}\ a{\isadigit{3}}\ \isakeyword{and}\ a{\isadigit{3}}{\isadigit{1}}\ \isakeyword{and}\ h{\isadigit{2}}\ \isacommand{show}\isamarkupfalse%
\ {\isacharquery}thesis\ \isacommand{by}\isamarkupfalse%
\ simp\isanewline
\ \ \ \ \isacommand{next}\isamarkupfalse%
\isanewline
\ \ \ \ \ \ \isacommand{assume}\isamarkupfalse%
\ a{\isadigit{3}}{\isadigit{2}}{\isacharcolon}{\isachardoublequoteopen}lose\ t\ {\isasymnoteq}\ {\isacharbrackleft}True{\isacharbrackright}{\isachardoublequoteclose}\isanewline
\ \ \ \ \ \ \isacommand{from}\isamarkupfalse%
\ this\ \isakeyword{and}\ h{\isadigit{4}}\ \isacommand{have}\isamarkupfalse%
\ a{\isadigit{3}}{\isadigit{2}}a{\isacharcolon}{\isachardoublequoteopen}lose\ t\ {\isacharequal}\ {\isacharbrackleft}False{\isacharbrackright}{\isachardoublequoteclose}\ \isacommand{by}\isamarkupfalse%
\ {\isacharparenleft}simp\ add{\isacharcolon}\ ts{\isacharunderscore}bool{\isacharunderscore}False{\isacharparenright}\isanewline
\ \ \ \ \ \ \isacommand{show}\isamarkupfalse%
\ {\isacharquery}thesis\isanewline
\ \ \ \ \ \ \isacommand{proof}\isamarkupfalse%
\ {\isacharparenleft}cases\ {\isachardoublequoteopen}req\ t\ {\isacharequal}\ {\isacharbrackleft}send{\isacharbrackright}{\isachardoublequoteclose}{\isacharparenright}\isanewline
\ \ \ \ \ \ \ \ \isacommand{assume}\isamarkupfalse%
\ a{\isadigit{3}}{\isadigit{2}}{\isadigit{1}}{\isacharcolon}{\isachardoublequoteopen}req\ t\ {\isacharequal}\ {\isacharbrackleft}send{\isacharbrackright}{\isachardoublequoteclose}\isanewline
\ \ \ \ \ \ \ \ \isacommand{from}\isamarkupfalse%
\ h{\isadigit{1}}\ \isakeyword{and}\ a{\isadigit{3}}\ \isakeyword{and}\ a{\isadigit{3}}{\isadigit{2}}a\ \isakeyword{and}\ a{\isadigit{3}}{\isadigit{2}}{\isadigit{1}}\ \isakeyword{and}\ h{\isadigit{2}}\ \isacommand{show}\isamarkupfalse%
\ {\isacharquery}thesis\ \isanewline
\ \ \ \ \ \ \ \ \ \ \isacommand{by}\isamarkupfalse%
\ {\isacharparenleft}simp\ add{\isacharcolon}\ tiTable{\isacharunderscore}SampleT{\isacharunderscore}def{\isacharparenright}\isanewline
\ \ \ \ \ \ \isacommand{next}\isamarkupfalse%
\isanewline
\ \ \ \ \ \ \ \ \isacommand{assume}\isamarkupfalse%
\ a{\isadigit{3}}{\isadigit{2}}{\isadigit{2}}{\isacharcolon}{\isachardoublequoteopen}req\ t\ {\isasymnoteq}\ {\isacharbrackleft}send{\isacharbrackright}{\isachardoublequoteclose}\isanewline
\ \ \ \ \ \ \ \ \isacommand{from}\isamarkupfalse%
\ h{\isadigit{1}}\ \isakeyword{and}\ a{\isadigit{3}}\ \isakeyword{and}\ a{\isadigit{3}}{\isadigit{2}}a\ \isakeyword{and}\ a{\isadigit{3}}{\isadigit{2}}{\isadigit{2}}\ \isakeyword{and}\ h{\isadigit{2}}\ \isacommand{show}\isamarkupfalse%
\ {\isacharquery}thesis\isanewline
\ \ \ \ \ \ \ \ \ \ \ \isacommand{by}\isamarkupfalse%
\ {\isacharparenleft}simp\ add{\isacharcolon}\ tiTable{\isacharunderscore}SampleT{\isacharunderscore}def{\isacharparenright}\isanewline
\ \ \ \ \ \ \isacommand{qed}\isamarkupfalse%
\isanewline
\ \ \ \ \isacommand{qed}\isamarkupfalse%
\ \isanewline
\ \ \isacommand{next}\isamarkupfalse%
\isanewline
\ \ \ \ \isacommand{assume}\isamarkupfalse%
\ a{\isadigit{4}}{\isacharcolon}{\isachardoublequoteopen}st{\isacharunderscore}in\ t\ {\isacharequal}\ sending{\isacharunderscore}data{\isachardoublequoteclose}\isanewline
\ \ \ \ \isacommand{show}\isamarkupfalse%
\ {\isacharquery}thesis\isanewline
\ \ \ \ \isacommand{proof}\isamarkupfalse%
\ {\isacharparenleft}cases\ {\isachardoublequoteopen}lose\ t\ {\isacharequal}\ {\isacharbrackleft}True{\isacharbrackright}{\isachardoublequoteclose}{\isacharparenright}\isanewline
\ \ \ \ \ \ \isacommand{assume}\isamarkupfalse%
\ a{\isadigit{4}}{\isadigit{1}}{\isacharcolon}{\isachardoublequoteopen}lose\ t\ {\isacharequal}\ {\isacharbrackleft}True{\isacharbrackright}{\isachardoublequoteclose}\isanewline
\ \ \ \ \ \ \isacommand{from}\isamarkupfalse%
\ h{\isadigit{1}}\ \isakeyword{and}\ a{\isadigit{4}}\ \isakeyword{and}\ a{\isadigit{4}}{\isadigit{1}}\ \isakeyword{and}\ h{\isadigit{2}}\ \isacommand{show}\isamarkupfalse%
\ {\isacharquery}thesis\ \isacommand{by}\isamarkupfalse%
\ {\isacharparenleft}simp\ add{\isacharcolon}\ tiTable{\isacharunderscore}SampleT{\isacharunderscore}def{\isacharparenright}\isanewline
\ \ \ \ \isacommand{next}\isamarkupfalse%
\isanewline
\ \ \ \ \ \ \isacommand{assume}\isamarkupfalse%
\ a{\isadigit{4}}{\isadigit{2}}{\isacharcolon}{\isachardoublequoteopen}lose\ t\ {\isasymnoteq}\ {\isacharbrackleft}True{\isacharbrackright}{\isachardoublequoteclose}\isanewline
\ \ \ \ \ \ \isacommand{from}\isamarkupfalse%
\ this\ \isakeyword{and}\ h{\isadigit{4}}\ \isacommand{have}\isamarkupfalse%
\ a{\isadigit{4}}{\isadigit{2}}a{\isacharcolon}{\isachardoublequoteopen}lose\ t\ {\isacharequal}\ {\isacharbrackleft}False{\isacharbrackright}{\isachardoublequoteclose}\ \isacommand{by}\isamarkupfalse%
\ {\isacharparenleft}simp\ add{\isacharcolon}\ ts{\isacharunderscore}bool{\isacharunderscore}False{\isacharparenright}\isanewline
\ \ \ \ \ \ \isacommand{show}\isamarkupfalse%
\ {\isacharquery}thesis\isanewline
\ \ \ \ \ \ \isacommand{proof}\isamarkupfalse%
\ {\isacharparenleft}cases\ {\isachardoublequoteopen}x\ t\ {\isacharequal}\ {\isacharbrackleft}sc{\isacharunderscore}ack{\isacharbrackright}{\isachardoublequoteclose}{\isacharparenright}\isanewline
\ \ \ \ \ \ \ \ \isacommand{assume}\isamarkupfalse%
\ a{\isadigit{4}}{\isadigit{2}}{\isadigit{1}}{\isacharcolon}{\isachardoublequoteopen}x\ t\ {\isacharequal}\ {\isacharbrackleft}sc{\isacharunderscore}ack{\isacharbrackright}{\isachardoublequoteclose}\isanewline
\ \ \ \ \ \ \ \ \isacommand{from}\isamarkupfalse%
\ h{\isadigit{1}}\ \isakeyword{and}\ a{\isadigit{4}}\ \isakeyword{and}\ a{\isadigit{4}}{\isadigit{2}}a\ \isakeyword{and}\ a{\isadigit{4}}{\isadigit{2}}{\isadigit{1}}\ \isakeyword{and}\ h{\isadigit{2}}\ \isacommand{show}\isamarkupfalse%
\ {\isacharquery}thesis\ \isanewline
\ \ \ \ \ \ \ \ \ \ \isacommand{by}\isamarkupfalse%
\ {\isacharparenleft}simp\ add{\isacharcolon}\ tiTable{\isacharunderscore}SampleT{\isacharunderscore}def{\isacharparenright}\isanewline
\ \ \ \ \ \ \isacommand{next}\isamarkupfalse%
\isanewline
\ \ \ \ \ \ \ \ \isacommand{assume}\isamarkupfalse%
\ a{\isadigit{4}}{\isadigit{2}}{\isadigit{2}}{\isacharcolon}{\isachardoublequoteopen}\ x\ t\ {\isasymnoteq}\ {\isacharbrackleft}sc{\isacharunderscore}ack{\isacharbrackright}{\isachardoublequoteclose}\isanewline
\ \ \ \ \ \ \ \ \isacommand{from}\isamarkupfalse%
\ this\ \isakeyword{and}\ h{\isadigit{3}}\ \ \isacommand{have}\isamarkupfalse%
\ a{\isadigit{4}}{\isadigit{2}}{\isadigit{2}}a{\isacharcolon}{\isachardoublequoteopen}x\ t\ {\isacharequal}\ {\isacharbrackleft}{\isacharbrackright}{\isachardoublequoteclose}\ \isacommand{by}\isamarkupfalse%
\ {\isacharparenleft}simp\ add{\isacharcolon}\ aType{\isacharunderscore}empty{\isacharparenright}\isanewline
\ \ \ \ \ \ \ \ \isacommand{from}\isamarkupfalse%
\ h{\isadigit{1}}\ \isakeyword{and}\ a{\isadigit{4}}\ \isakeyword{and}\ a{\isadigit{4}}{\isadigit{2}}a\ \isakeyword{and}\ a{\isadigit{4}}{\isadigit{2}}{\isadigit{2}}a\ \isakeyword{and}\ h{\isadigit{2}}\ \isacommand{show}\isamarkupfalse%
\ {\isacharquery}thesis\ \isanewline
\ \ \ \ \ \ \ \ \ \ \isacommand{by}\isamarkupfalse%
\ {\isacharparenleft}simp\ add{\isacharcolon}\ tiTable{\isacharunderscore}SampleT{\isacharunderscore}def{\isacharparenright}\isanewline
\ \ \ \ \ \ \isacommand{qed}\isamarkupfalse%
\isanewline
\ \ \ \ \isacommand{qed}\isamarkupfalse%
\isanewline
\ \ \isacommand{next}\isamarkupfalse%
\ \ \isanewline
\ \ \ \ \isacommand{assume}\isamarkupfalse%
\ a{\isadigit{5}}{\isacharcolon}{\isachardoublequoteopen}st{\isacharunderscore}in\ t\ {\isacharequal}\ voice{\isacharunderscore}com{\isachardoublequoteclose}\isanewline
\ \ \ \ \isacommand{show}\isamarkupfalse%
\ {\isacharquery}thesis\isanewline
\ \ \ \ \isacommand{proof}\isamarkupfalse%
\ {\isacharparenleft}cases\ {\isachardoublequoteopen}stop\ t\ {\isacharequal}\ {\isacharbrackleft}stop{\isacharunderscore}vc{\isacharbrackright}{\isachardoublequoteclose}{\isacharparenright}\isanewline
\ \ \ \ \ \ \isacommand{assume}\isamarkupfalse%
\ a{\isadigit{5}}{\isadigit{1}}{\isacharcolon}{\isachardoublequoteopen}stop\ t\ {\isacharequal}\ {\isacharbrackleft}stop{\isacharunderscore}vc{\isacharbrackright}{\isachardoublequoteclose}\isanewline
\ \ \ \ \ \ \isacommand{from}\isamarkupfalse%
\ h{\isadigit{1}}\ \isakeyword{and}\ a{\isadigit{5}}\ \isakeyword{and}\ a{\isadigit{5}}{\isadigit{1}}\ \isakeyword{and}\ h{\isadigit{2}}\ \isacommand{show}\isamarkupfalse%
\ {\isacharquery}thesis\ \isanewline
\ \ \ \ \ \ \ \ \ \isacommand{by}\isamarkupfalse%
\ {\isacharparenleft}simp\ add{\isacharcolon}\ tiTable{\isacharunderscore}SampleT{\isacharunderscore}def{\isacharparenright}\isanewline
\ \ \ \ \isacommand{next}\isamarkupfalse%
\isanewline
\ \ \ \ \ \ \isacommand{assume}\isamarkupfalse%
\ a{\isadigit{5}}{\isadigit{2}}{\isacharcolon}{\isachardoublequoteopen}stop\ t\ {\isasymnoteq}\ {\isacharbrackleft}stop{\isacharunderscore}vc{\isacharbrackright}{\isachardoublequoteclose}\isanewline
\ \ \ \ \ \ \isacommand{from}\isamarkupfalse%
\ this\ \isakeyword{and}\ h{\isadigit{5}}\ \isacommand{have}\isamarkupfalse%
\ a{\isadigit{5}}{\isadigit{2}}a{\isacharcolon}{\isachardoublequoteopen}stop\ t\ {\isacharequal}\ {\isacharbrackleft}{\isacharbrackright}{\isachardoublequoteclose}\ \isacommand{by}\isamarkupfalse%
\ {\isacharparenleft}simp\ add{\isacharcolon}\ stopType{\isacharunderscore}empty{\isacharparenright}\isanewline
\ \ \ \ \ \ \isacommand{show}\isamarkupfalse%
\ {\isacharquery}thesis\isanewline
\ \ \ \ \ \ \isacommand{proof}\isamarkupfalse%
\ {\isacharparenleft}cases\ {\isachardoublequoteopen}lose\ t\ {\isacharequal}\ {\isacharbrackleft}True{\isacharbrackright}{\isachardoublequoteclose}{\isacharparenright}\isanewline
\ \ \ \ \ \ \ \ \isacommand{assume}\isamarkupfalse%
\ a{\isadigit{5}}{\isadigit{2}}{\isadigit{1}}{\isacharcolon}{\isachardoublequoteopen}lose\ t\ {\isacharequal}\ {\isacharbrackleft}True{\isacharbrackright}{\isachardoublequoteclose}\isanewline
\ \ \ \ \ \ \ \ \isacommand{from}\isamarkupfalse%
\ h{\isadigit{1}}\ \isakeyword{and}\ a{\isadigit{5}}\ \isakeyword{and}\ a{\isadigit{5}}{\isadigit{2}}a\ \isakeyword{and}\ a{\isadigit{5}}{\isadigit{2}}{\isadigit{1}}\ \isakeyword{and}\ h{\isadigit{2}}\ \isacommand{show}\isamarkupfalse%
\ {\isacharquery}thesis\ \isanewline
\ \ \ \ \ \ \ \ \ \ \isacommand{by}\isamarkupfalse%
\ {\isacharparenleft}simp\ add{\isacharcolon}\ tiTable{\isacharunderscore}SampleT{\isacharunderscore}def{\isacharparenright}\isanewline
\ \ \ \ \ \ \isacommand{next}\isamarkupfalse%
\isanewline
\ \ \ \ \ \ \ \ \isacommand{assume}\isamarkupfalse%
\ a{\isadigit{5}}{\isadigit{2}}{\isadigit{2}}{\isacharcolon}{\isachardoublequoteopen}lose\ t\ {\isasymnoteq}\ {\isacharbrackleft}True{\isacharbrackright}{\isachardoublequoteclose}\isanewline
\ \ \ \ \ \ \ \ \isacommand{from}\isamarkupfalse%
\ this\ \isakeyword{and}\ h{\isadigit{4}}\ \isacommand{have}\isamarkupfalse%
\ a{\isadigit{5}}{\isadigit{2}}{\isadigit{2}}a{\isacharcolon}{\isachardoublequoteopen}lose\ t\ {\isacharequal}\ {\isacharbrackleft}False{\isacharbrackright}{\isachardoublequoteclose}\ \isacommand{by}\isamarkupfalse%
\ {\isacharparenleft}simp\ add{\isacharcolon}\ ts{\isacharunderscore}bool{\isacharunderscore}False{\isacharparenright}\isanewline
\ \ \ \ \ \ \ \ \isacommand{from}\isamarkupfalse%
\ h{\isadigit{1}}\ \isakeyword{and}\ a{\isadigit{5}}\ \isakeyword{and}\ a{\isadigit{5}}{\isadigit{2}}a\ \isakeyword{and}\ a{\isadigit{5}}{\isadigit{2}}{\isadigit{2}}a\ \isakeyword{and}\ h{\isadigit{2}}\ \isacommand{show}\isamarkupfalse%
\ {\isacharquery}thesis\ \isanewline
\ \ \ \ \ \ \ \ \ \ \isacommand{by}\isamarkupfalse%
\ {\isacharparenleft}simp\ add{\isacharcolon}\ tiTable{\isacharunderscore}SampleT{\isacharunderscore}def{\isacharparenright}\isanewline
\ \ \ \ \ \ \isacommand{qed}\isamarkupfalse%
\isanewline
\ \ \ \ \isacommand{qed}\isamarkupfalse%
\isanewline
\ \ \isacommand{qed}\isamarkupfalse%
\isanewline
\isacommand{qed}\isamarkupfalse%
\endisatagproof
{\isafoldproof}%
\isadelimproof
\isanewline
\endisadelimproof
\isanewline
\isacommand{lemma}\isamarkupfalse%
\ tiTable{\isacharunderscore}i{\isadigit{1}}{\isacharunderscore}{\isadigit{1}}{\isacharcolon}\isanewline
\ \ \isakeyword{assumes}\ h{\isadigit{1}}{\isacharcolon}{\isachardoublequoteopen}tiTable{\isacharunderscore}SampleT\ req\ x\ stop\ lose\ st{\isacharunderscore}in\ b\ ack\ i{\isadigit{1}}\ vc\ st{\isacharunderscore}out{\isachardoublequoteclose}\isanewline
\ \ \ \ \ \ \isakeyword{and}\ h{\isadigit{2}}{\isacharcolon}{\isachardoublequoteopen}ts\ lose{\isachardoublequoteclose}\isanewline
\ \ \ \ \ \ \isakeyword{and}\ h{\isadigit{3}}{\isacharcolon}{\isachardoublequoteopen}msg\ {\isacharparenleft}Suc\ {\isadigit{0}}{\isacharparenright}\ x{\isachardoublequoteclose}\isanewline
\ \ \ \ \ \ \isakeyword{and}\ h{\isadigit{4}}{\isacharcolon}{\isachardoublequoteopen}msg\ {\isacharparenleft}Suc\ {\isadigit{0}}{\isacharparenright}\ stop{\isachardoublequoteclose}\isanewline
\ \ \ \ \ \ \isakeyword{and}\ h{\isadigit{5}}{\isacharcolon}{\isachardoublequoteopen}ack\ t\ {\isacharequal}\ {\isacharbrackleft}connection{\isacharunderscore}ok{\isacharbrackright}{\isachardoublequoteclose}\isanewline
\ \ \isakeyword{shows}\ {\isachardoublequoteopen}i{\isadigit{1}}\ t\ {\isacharequal}\ {\isacharbrackleft}{\isacharbrackright}{\isachardoublequoteclose}\isanewline
\isadelimproof
\endisadelimproof
\isatagproof
\isacommand{proof}\isamarkupfalse%
\ {\isacharminus}\isanewline
\ \ \isacommand{from}\isamarkupfalse%
\ assms\ \isacommand{have}\isamarkupfalse%
\ sg{\isadigit{1}}{\isacharcolon}\isanewline
\ \ \ {\isachardoublequoteopen}{\isacharparenleft}st{\isacharunderscore}in\ t\ {\isacharequal}\ call\ {\isasymor}\ st{\isacharunderscore}in\ t\ {\isacharequal}\ connection{\isacharunderscore}ok\ {\isasymand}\ req\ t\ {\isasymnoteq}\ {\isacharbrackleft}send{\isacharbrackright}{\isacharparenright}\ {\isasymand}\ \isanewline
\ \ \ \ lose\ t\ {\isacharequal}\ {\isacharbrackleft}False{\isacharbrackright}{\isachardoublequoteclose}\isanewline
\ \ \ \ \isacommand{by}\isamarkupfalse%
\ {\isacharparenleft}simp\ add{\isacharcolon}\ tiTable{\isacharunderscore}ack{\isacharunderscore}connection{\isacharunderscore}ok{\isacharparenright}\isanewline
\ \ \isacommand{from}\isamarkupfalse%
\ this\ \isakeyword{and}\ h{\isadigit{1}}\ \isacommand{show}\isamarkupfalse%
\ {\isacharquery}thesis\ \isacommand{by}\isamarkupfalse%
\ {\isacharparenleft}simp\ add{\isacharcolon}\ tiTable{\isacharunderscore}SampleT{\isacharunderscore}def{\isacharparenright}\isanewline
\isacommand{qed}\isamarkupfalse%
\endisatagproof
{\isafoldproof}%
\isadelimproof
\isanewline
\endisadelimproof
\isanewline
\isacommand{lemma}\isamarkupfalse%
\ tiTable{\isacharunderscore}ack{\isacharunderscore}call{\isacharcolon}\isanewline
\ \ \isakeyword{assumes}\ h{\isadigit{1}}{\isacharcolon}{\isachardoublequoteopen}tiTable{\isacharunderscore}SampleT\ req\ x\ stop\ lose\ st{\isacharunderscore}in\ b\ ack\ i{\isadigit{1}}\ vc\ st{\isacharunderscore}out{\isachardoublequoteclose}\isanewline
\ \ \ \ \ \ \isakeyword{and}\ h{\isadigit{2}}{\isacharcolon}{\isachardoublequoteopen}ack\ t\ {\isacharequal}\ {\isacharbrackleft}call{\isacharbrackright}{\isachardoublequoteclose}\isanewline
\ \ \ \ \ \ \isakeyword{and}\ h{\isadigit{3}}{\isacharcolon}{\isachardoublequoteopen}msg\ {\isacharparenleft}Suc\ {\isadigit{0}}{\isacharparenright}\ x{\isachardoublequoteclose}\isanewline
\ \ \ \ \ \ \isakeyword{and}\ h{\isadigit{4}}{\isacharcolon}{\isachardoublequoteopen}ts\ lose{\isachardoublequoteclose}\isanewline
\ \ \ \ \ \ \isakeyword{and}\ h{\isadigit{5}}{\isacharcolon}{\isachardoublequoteopen}msg\ {\isacharparenleft}Suc\ {\isadigit{0}}{\isacharparenright}\ stop{\isachardoublequoteclose}\isanewline
\ \ \isakeyword{shows}\ {\isachardoublequoteopen}st{\isacharunderscore}in\ t\ {\isacharequal}\ init{\isacharunderscore}state\ {\isasymand}\ req\ t\ {\isacharequal}\ {\isacharbrackleft}init{\isacharbrackright}{\isachardoublequoteclose}\isanewline
\isadelimproof
\endisadelimproof
\isatagproof
\isacommand{proof}\isamarkupfalse%
\ {\isacharminus}\ \isanewline
\ \ \isacommand{from}\isamarkupfalse%
\ h{\isadigit{1}}\ \isakeyword{and}\ h{\isadigit{4}}\ \isacommand{have}\isamarkupfalse%
\ sg{\isadigit{1}}{\isacharcolon}{\isachardoublequoteopen}lose\ t\ {\isacharequal}\ {\isacharbrackleft}True{\isacharbrackright}\ {\isasymor}\ lose\ t\ {\isacharequal}\ {\isacharbrackleft}False{\isacharbrackright}{\isachardoublequoteclose}\isanewline
\ \ \ \ \isacommand{by}\isamarkupfalse%
\ {\isacharparenleft}simp\ add{\isacharcolon}\ ts{\isacharunderscore}bool{\isacharunderscore}True{\isacharunderscore}False{\isacharparenright}\isanewline
\ \ \isacommand{from}\isamarkupfalse%
\ h{\isadigit{1}}\ \isakeyword{and}\ h{\isadigit{3}}\ \isacommand{have}\isamarkupfalse%
\ sg{\isadigit{2}}{\isacharcolon}{\isachardoublequoteopen}x\ t\ {\isacharequal}\ {\isacharbrackleft}{\isacharbrackright}\ {\isasymor}\ x\ t\ {\isacharequal}\ {\isacharbrackleft}sc{\isacharunderscore}ack{\isacharbrackright}{\isachardoublequoteclose}\isanewline
\ \ \ \ \isacommand{by}\isamarkupfalse%
\ {\isacharparenleft}simp\ add{\isacharcolon}\ aType{\isacharunderscore}lemma{\isacharparenright}\ \isanewline
\ \ \isacommand{from}\isamarkupfalse%
\ h{\isadigit{1}}\ \isakeyword{and}\ h{\isadigit{5}}\ \isacommand{have}\isamarkupfalse%
\ sg{\isadigit{3}}{\isacharcolon}{\isachardoublequoteopen}stop\ t\ {\isacharequal}\ {\isacharbrackleft}{\isacharbrackright}\ {\isasymor}\ stop\ t\ {\isacharequal}\ {\isacharbrackleft}stop{\isacharunderscore}vc{\isacharbrackright}{\isachardoublequoteclose}\isanewline
\ \ \ \ \isacommand{by}\isamarkupfalse%
\ {\isacharparenleft}simp\ add{\isacharcolon}\ stopType{\isacharunderscore}lemma{\isacharparenright}\ \isanewline
\ \ \isacommand{show}\isamarkupfalse%
\ {\isacharquery}thesis\isanewline
\ \ \isacommand{proof}\isamarkupfalse%
\ {\isacharparenleft}cases\ {\isachardoublequoteopen}st{\isacharunderscore}in\ t{\isachardoublequoteclose}{\isacharparenright}\isanewline
\ \ \ \ \isacommand{assume}\isamarkupfalse%
\ a{\isadigit{1}}{\isacharcolon}{\isachardoublequoteopen}st{\isacharunderscore}in\ t\ {\isacharequal}\ init{\isacharunderscore}state{\isachardoublequoteclose}\isanewline
\ \ \ \ \isacommand{show}\isamarkupfalse%
\ {\isacharquery}thesis\isanewline
\ \ \ \ \isacommand{proof}\isamarkupfalse%
\ {\isacharparenleft}cases\ {\isachardoublequoteopen}req\ t\ {\isacharequal}\ {\isacharbrackleft}init{\isacharbrackright}{\isachardoublequoteclose}{\isacharparenright}\isanewline
\ \ \ \ \ \ \isacommand{assume}\isamarkupfalse%
\ a{\isadigit{1}}{\isadigit{1}}{\isacharcolon}{\isachardoublequoteopen}req\ t\ {\isacharequal}\ {\isacharbrackleft}init{\isacharbrackright}{\isachardoublequoteclose}\isanewline
\ \ \ \ \ \ \isacommand{from}\isamarkupfalse%
\ h{\isadigit{1}}\ \isakeyword{and}\ a{\isadigit{1}}\ \isakeyword{and}\ a{\isadigit{1}}{\isadigit{1}}\ \isakeyword{and}\ h{\isadigit{2}}\ \isacommand{show}\isamarkupfalse%
\ {\isacharquery}thesis\ \isanewline
\ \ \ \ \ \ \ \ \isacommand{by}\isamarkupfalse%
\ {\isacharparenleft}simp\ add{\isacharcolon}\ tiTable{\isacharunderscore}SampleT{\isacharunderscore}def{\isacharparenright}\isanewline
\ \ \ \ \isacommand{next}\isamarkupfalse%
\isanewline
\ \ \ \ \ \ \isacommand{assume}\isamarkupfalse%
\ a{\isadigit{1}}{\isadigit{2}}{\isacharcolon}{\isachardoublequoteopen}req\ t\ {\isasymnoteq}\ {\isacharbrackleft}init{\isacharbrackright}{\isachardoublequoteclose}\isanewline
\ \ \ \ \ \ \isacommand{from}\isamarkupfalse%
\ h{\isadigit{1}}\ \isakeyword{and}\ a{\isadigit{1}}\ \isakeyword{and}\ a{\isadigit{1}}{\isadigit{2}}\ \isakeyword{and}\ h{\isadigit{2}}\ \isacommand{show}\isamarkupfalse%
\ {\isacharquery}thesis\ \isanewline
\ \ \ \ \ \ \ \ \ \isacommand{by}\isamarkupfalse%
\ {\isacharparenleft}simp\ add{\isacharcolon}\ tiTable{\isacharunderscore}SampleT{\isacharunderscore}def{\isacharparenright}\isanewline
\ \ \ \ \isacommand{qed}\isamarkupfalse%
\ \isanewline
\ \ \isacommand{next}\isamarkupfalse%
\isanewline
\ \ \ \ \isacommand{assume}\isamarkupfalse%
\ a{\isadigit{2}}{\isacharcolon}{\isachardoublequoteopen}st{\isacharunderscore}in\ t\ {\isacharequal}\ call{\isachardoublequoteclose}\isanewline
\ \ \ \ \isacommand{show}\isamarkupfalse%
\ {\isacharquery}thesis\isanewline
\ \ \ \ \isacommand{proof}\isamarkupfalse%
\ {\isacharparenleft}cases\ {\isachardoublequoteopen}lose\ t\ {\isacharequal}\ {\isacharbrackleft}True{\isacharbrackright}{\isachardoublequoteclose}{\isacharparenright}\isanewline
\ \ \ \ \ \ \isacommand{assume}\isamarkupfalse%
\ a{\isadigit{2}}{\isadigit{1}}{\isacharcolon}{\isachardoublequoteopen}lose\ t\ {\isacharequal}\ {\isacharbrackleft}True{\isacharbrackright}{\isachardoublequoteclose}\isanewline
\ \ \ \ \ \ \isacommand{from}\isamarkupfalse%
\ h{\isadigit{1}}\ \isakeyword{and}\ a{\isadigit{2}}\ \isakeyword{and}\ a{\isadigit{2}}{\isadigit{1}}\ \isakeyword{and}\ h{\isadigit{2}}\ \isacommand{show}\isamarkupfalse%
\ {\isacharquery}thesis\ \isanewline
\ \ \ \ \ \ \ \ \isacommand{by}\isamarkupfalse%
\ {\isacharparenleft}simp\ add{\isacharcolon}\ tiTable{\isacharunderscore}SampleT{\isacharunderscore}def{\isacharparenright}\isanewline
\ \ \ \ \isacommand{next}\isamarkupfalse%
\isanewline
\ \ \ \ \ \ \isacommand{assume}\isamarkupfalse%
\ a{\isadigit{2}}{\isadigit{2}}{\isacharcolon}{\isachardoublequoteopen}lose\ t\ {\isasymnoteq}\ {\isacharbrackleft}True{\isacharbrackright}{\isachardoublequoteclose}\isanewline
\ \ \ \ \ \ \isacommand{from}\isamarkupfalse%
\ this\ \isakeyword{and}\ h{\isadigit{4}}\ \isacommand{have}\isamarkupfalse%
\ a{\isadigit{2}}{\isadigit{2}}a{\isacharcolon}{\isachardoublequoteopen}lose\ t\ {\isacharequal}\ {\isacharbrackleft}False{\isacharbrackright}{\isachardoublequoteclose}\ \isacommand{by}\isamarkupfalse%
\ {\isacharparenleft}simp\ add{\isacharcolon}\ ts{\isacharunderscore}bool{\isacharunderscore}False{\isacharparenright}\isanewline
\ \ \ \ \ \ \isacommand{from}\isamarkupfalse%
\ h{\isadigit{1}}\ \isakeyword{and}\ a{\isadigit{2}}\ \isakeyword{and}\ a{\isadigit{2}}{\isadigit{2}}a\ \isakeyword{and}\ h{\isadigit{2}}\ \isacommand{show}\isamarkupfalse%
\ {\isacharquery}thesis\isanewline
\ \ \ \ \ \ \ \ \ \isacommand{by}\isamarkupfalse%
\ {\isacharparenleft}simp\ add{\isacharcolon}\ tiTable{\isacharunderscore}SampleT{\isacharunderscore}def{\isacharparenright}\isanewline
\ \ \ \ \isacommand{qed}\isamarkupfalse%
\ \isanewline
\ \ \isacommand{next}\isamarkupfalse%
\isanewline
\ \ \ \ \isacommand{assume}\isamarkupfalse%
\ a{\isadigit{3}}{\isacharcolon}{\isachardoublequoteopen}st{\isacharunderscore}in\ t\ {\isacharequal}\ connection{\isacharunderscore}ok{\isachardoublequoteclose}\isanewline
\ \ \ \ \isacommand{show}\isamarkupfalse%
\ {\isacharquery}thesis\isanewline
\ \ \ \ \isacommand{proof}\isamarkupfalse%
\ {\isacharparenleft}cases\ {\isachardoublequoteopen}lose\ t\ {\isacharequal}\ {\isacharbrackleft}True{\isacharbrackright}{\isachardoublequoteclose}{\isacharparenright}\isanewline
\ \ \ \ \ \ \isacommand{assume}\isamarkupfalse%
\ a{\isadigit{3}}{\isadigit{1}}{\isacharcolon}{\isachardoublequoteopen}lose\ t\ {\isacharequal}\ {\isacharbrackleft}True{\isacharbrackright}{\isachardoublequoteclose}\isanewline
\ \ \ \ \ \ \isacommand{from}\isamarkupfalse%
\ h{\isadigit{1}}\ \isakeyword{and}\ a{\isadigit{3}}\ \isakeyword{and}\ a{\isadigit{3}}{\isadigit{1}}\ \isakeyword{and}\ h{\isadigit{2}}\ \isacommand{show}\isamarkupfalse%
\ {\isacharquery}thesis\ \isacommand{by}\isamarkupfalse%
\ {\isacharparenleft}simp\ add{\isacharcolon}\ tiTable{\isacharunderscore}SampleT{\isacharunderscore}def{\isacharparenright}\isanewline
\ \ \ \ \isacommand{next}\isamarkupfalse%
\isanewline
\ \ \ \ \ \ \isacommand{assume}\isamarkupfalse%
\ a{\isadigit{3}}{\isadigit{2}}{\isacharcolon}{\isachardoublequoteopen}lose\ t\ {\isasymnoteq}\ {\isacharbrackleft}True{\isacharbrackright}{\isachardoublequoteclose}\isanewline
\ \ \ \ \ \ \isacommand{from}\isamarkupfalse%
\ this\ \isakeyword{and}\ h{\isadigit{4}}\ \isacommand{have}\isamarkupfalse%
\ a{\isadigit{3}}{\isadigit{2}}a{\isacharcolon}{\isachardoublequoteopen}lose\ t\ {\isacharequal}\ {\isacharbrackleft}False{\isacharbrackright}{\isachardoublequoteclose}\ \isacommand{by}\isamarkupfalse%
\ {\isacharparenleft}simp\ add{\isacharcolon}\ ts{\isacharunderscore}bool{\isacharunderscore}False{\isacharparenright}\isanewline
\ \ \ \ \ \ \isacommand{show}\isamarkupfalse%
\ {\isacharquery}thesis\isanewline
\ \ \ \ \ \ \isacommand{proof}\isamarkupfalse%
\ {\isacharparenleft}cases\ {\isachardoublequoteopen}req\ t\ {\isacharequal}\ {\isacharbrackleft}send{\isacharbrackright}{\isachardoublequoteclose}{\isacharparenright}\isanewline
\ \ \ \ \ \ \ \ \isacommand{assume}\isamarkupfalse%
\ a{\isadigit{3}}{\isadigit{2}}{\isadigit{1}}{\isacharcolon}{\isachardoublequoteopen}req\ t\ {\isacharequal}\ {\isacharbrackleft}send{\isacharbrackright}{\isachardoublequoteclose}\isanewline
\ \ \ \ \ \ \ \ \isacommand{from}\isamarkupfalse%
\ h{\isadigit{1}}\ \isakeyword{and}\ a{\isadigit{3}}\ \isakeyword{and}\ a{\isadigit{3}}{\isadigit{2}}a\ \isakeyword{and}\ a{\isadigit{3}}{\isadigit{2}}{\isadigit{1}}\ \isakeyword{and}\ h{\isadigit{2}}\ \isacommand{show}\isamarkupfalse%
\ {\isacharquery}thesis\ \isanewline
\ \ \ \ \ \ \ \ \ \ \isacommand{by}\isamarkupfalse%
\ {\isacharparenleft}simp\ add{\isacharcolon}\ tiTable{\isacharunderscore}SampleT{\isacharunderscore}def{\isacharparenright}\isanewline
\ \ \ \ \ \ \isacommand{next}\isamarkupfalse%
\isanewline
\ \ \ \ \ \ \ \ \isacommand{assume}\isamarkupfalse%
\ a{\isadigit{3}}{\isadigit{2}}{\isadigit{2}}{\isacharcolon}{\isachardoublequoteopen}req\ t\ {\isasymnoteq}\ {\isacharbrackleft}send{\isacharbrackright}{\isachardoublequoteclose}\isanewline
\ \ \ \ \ \ \ \ \isacommand{from}\isamarkupfalse%
\ h{\isadigit{1}}\ \isakeyword{and}\ a{\isadigit{3}}\ \isakeyword{and}\ a{\isadigit{3}}{\isadigit{2}}a\ \isakeyword{and}\ a{\isadigit{3}}{\isadigit{2}}{\isadigit{2}}\ \isakeyword{and}\ h{\isadigit{2}}\ \isacommand{show}\isamarkupfalse%
\ {\isacharquery}thesis\isanewline
\ \ \ \ \ \ \ \ \ \ \ \isacommand{by}\isamarkupfalse%
\ {\isacharparenleft}simp\ add{\isacharcolon}\ tiTable{\isacharunderscore}SampleT{\isacharunderscore}def{\isacharparenright}\isanewline
\ \ \ \ \ \ \isacommand{qed}\isamarkupfalse%
\isanewline
\ \ \ \ \isacommand{qed}\isamarkupfalse%
\ \isanewline
\ \ \isacommand{next}\isamarkupfalse%
\isanewline
\ \ \ \ \isacommand{assume}\isamarkupfalse%
\ a{\isadigit{4}}{\isacharcolon}{\isachardoublequoteopen}st{\isacharunderscore}in\ t\ {\isacharequal}\ sending{\isacharunderscore}data{\isachardoublequoteclose}\isanewline
\ \ \ \ \isacommand{show}\isamarkupfalse%
\ {\isacharquery}thesis\isanewline
\ \ \ \ \isacommand{proof}\isamarkupfalse%
\ {\isacharparenleft}cases\ {\isachardoublequoteopen}lose\ t\ {\isacharequal}\ {\isacharbrackleft}True{\isacharbrackright}{\isachardoublequoteclose}{\isacharparenright}\isanewline
\ \ \ \ \ \ \isacommand{assume}\isamarkupfalse%
\ a{\isadigit{4}}{\isadigit{1}}{\isacharcolon}{\isachardoublequoteopen}lose\ t\ {\isacharequal}\ {\isacharbrackleft}True{\isacharbrackright}{\isachardoublequoteclose}\isanewline
\ \ \ \ \ \ \isacommand{from}\isamarkupfalse%
\ h{\isadigit{1}}\ \isakeyword{and}\ a{\isadigit{4}}\ \isakeyword{and}\ a{\isadigit{4}}{\isadigit{1}}\ \isakeyword{and}\ h{\isadigit{2}}\ \isacommand{show}\isamarkupfalse%
\ {\isacharquery}thesis\isanewline
\ \ \ \ \ \ \ \ \isacommand{by}\isamarkupfalse%
\ {\isacharparenleft}simp\ add{\isacharcolon}\ tiTable{\isacharunderscore}SampleT{\isacharunderscore}def{\isacharparenright}\isanewline
\ \ \ \ \isacommand{next}\isamarkupfalse%
\isanewline
\ \ \ \ \ \ \isacommand{assume}\isamarkupfalse%
\ a{\isadigit{4}}{\isadigit{2}}{\isacharcolon}{\isachardoublequoteopen}lose\ t\ {\isasymnoteq}\ {\isacharbrackleft}True{\isacharbrackright}{\isachardoublequoteclose}\isanewline
\ \ \ \ \ \ \isacommand{from}\isamarkupfalse%
\ this\ \isakeyword{and}\ h{\isadigit{4}}\ \isacommand{have}\isamarkupfalse%
\ a{\isadigit{4}}{\isadigit{2}}a{\isacharcolon}{\isachardoublequoteopen}lose\ t\ {\isacharequal}\ {\isacharbrackleft}False{\isacharbrackright}{\isachardoublequoteclose}\ \isacommand{by}\isamarkupfalse%
\ {\isacharparenleft}simp\ add{\isacharcolon}\ ts{\isacharunderscore}bool{\isacharunderscore}False{\isacharparenright}\isanewline
\ \ \ \ \ \ \isacommand{show}\isamarkupfalse%
\ {\isacharquery}thesis\isanewline
\ \ \ \ \ \ \isacommand{proof}\isamarkupfalse%
\ {\isacharparenleft}cases\ {\isachardoublequoteopen}x\ t\ {\isacharequal}\ {\isacharbrackleft}sc{\isacharunderscore}ack{\isacharbrackright}{\isachardoublequoteclose}{\isacharparenright}\isanewline
\ \ \ \ \ \ \ \ \isacommand{assume}\isamarkupfalse%
\ a{\isadigit{4}}{\isadigit{2}}{\isadigit{1}}{\isacharcolon}{\isachardoublequoteopen}x\ t\ {\isacharequal}\ {\isacharbrackleft}sc{\isacharunderscore}ack{\isacharbrackright}{\isachardoublequoteclose}\isanewline
\ \ \ \ \ \ \ \ \isacommand{from}\isamarkupfalse%
\ h{\isadigit{1}}\ \isakeyword{and}\ a{\isadigit{4}}\ \isakeyword{and}\ a{\isadigit{4}}{\isadigit{2}}a\ \isakeyword{and}\ a{\isadigit{4}}{\isadigit{2}}{\isadigit{1}}\ \isakeyword{and}\ h{\isadigit{2}}\ \isacommand{show}\isamarkupfalse%
\ {\isacharquery}thesis\isanewline
\ \ \ \ \ \ \ \ \ \ \isacommand{by}\isamarkupfalse%
\ {\isacharparenleft}simp\ add{\isacharcolon}\ tiTable{\isacharunderscore}SampleT{\isacharunderscore}def{\isacharparenright}\isanewline
\ \ \ \ \ \ \isacommand{next}\isamarkupfalse%
\isanewline
\ \ \ \ \ \ \ \ \isacommand{assume}\isamarkupfalse%
\ a{\isadigit{4}}{\isadigit{2}}{\isadigit{2}}{\isacharcolon}{\isachardoublequoteopen}\ x\ t\ {\isasymnoteq}\ {\isacharbrackleft}sc{\isacharunderscore}ack{\isacharbrackright}{\isachardoublequoteclose}\isanewline
\ \ \ \ \ \ \ \ \isacommand{from}\isamarkupfalse%
\ this\ \isakeyword{and}\ h{\isadigit{3}}\ \ \isacommand{have}\isamarkupfalse%
\ a{\isadigit{4}}{\isadigit{2}}{\isadigit{2}}a{\isacharcolon}{\isachardoublequoteopen}x\ t\ {\isacharequal}\ {\isacharbrackleft}{\isacharbrackright}{\isachardoublequoteclose}\ \isacommand{by}\isamarkupfalse%
\ {\isacharparenleft}simp\ add{\isacharcolon}\ aType{\isacharunderscore}empty{\isacharparenright}\isanewline
\ \ \ \ \ \ \ \ \isacommand{from}\isamarkupfalse%
\ h{\isadigit{1}}\ \isakeyword{and}\ a{\isadigit{4}}\ \isakeyword{and}\ a{\isadigit{4}}{\isadigit{2}}a\ \isakeyword{and}\ a{\isadigit{4}}{\isadigit{2}}{\isadigit{2}}a\ \isakeyword{and}\ h{\isadigit{2}}\ \isacommand{show}\isamarkupfalse%
\ {\isacharquery}thesis\ \isanewline
\ \ \ \ \ \ \ \ \ \ \isacommand{by}\isamarkupfalse%
\ {\isacharparenleft}simp\ add{\isacharcolon}\ tiTable{\isacharunderscore}SampleT{\isacharunderscore}def{\isacharparenright}\isanewline
\ \ \ \ \ \ \isacommand{qed}\isamarkupfalse%
\isanewline
\ \ \ \ \isacommand{qed}\isamarkupfalse%
\isanewline
\ \ \isacommand{next}\isamarkupfalse%
\ \ \isanewline
\ \ \ \ \isacommand{assume}\isamarkupfalse%
\ a{\isadigit{5}}{\isacharcolon}{\isachardoublequoteopen}st{\isacharunderscore}in\ t\ {\isacharequal}\ voice{\isacharunderscore}com{\isachardoublequoteclose}\isanewline
\ \ \ \ \isacommand{show}\isamarkupfalse%
\ {\isacharquery}thesis\isanewline
\ \ \ \ \isacommand{proof}\isamarkupfalse%
\ {\isacharparenleft}cases\ {\isachardoublequoteopen}stop\ t\ {\isacharequal}\ {\isacharbrackleft}stop{\isacharunderscore}vc{\isacharbrackright}{\isachardoublequoteclose}{\isacharparenright}\isanewline
\ \ \ \ \ \ \isacommand{assume}\isamarkupfalse%
\ a{\isadigit{5}}{\isadigit{1}}{\isacharcolon}{\isachardoublequoteopen}stop\ t\ {\isacharequal}\ {\isacharbrackleft}stop{\isacharunderscore}vc{\isacharbrackright}{\isachardoublequoteclose}\isanewline
\ \ \ \ \ \ \isacommand{from}\isamarkupfalse%
\ h{\isadigit{1}}\ \isakeyword{and}\ a{\isadigit{5}}\ \isakeyword{and}\ a{\isadigit{5}}{\isadigit{1}}\ \isakeyword{and}\ h{\isadigit{2}}\ \isacommand{show}\isamarkupfalse%
\ {\isacharquery}thesis\ \isanewline
\ \ \ \ \ \ \ \ \isacommand{by}\isamarkupfalse%
\ {\isacharparenleft}simp\ add{\isacharcolon}\ tiTable{\isacharunderscore}SampleT{\isacharunderscore}def{\isacharparenright}\isanewline
\ \ \ \ \isacommand{next}\isamarkupfalse%
\isanewline
\ \ \ \ \ \ \isacommand{assume}\isamarkupfalse%
\ a{\isadigit{5}}{\isadigit{2}}{\isacharcolon}{\isachardoublequoteopen}stop\ t\ {\isasymnoteq}\ {\isacharbrackleft}stop{\isacharunderscore}vc{\isacharbrackright}{\isachardoublequoteclose}\isanewline
\ \ \ \ \ \ \isacommand{from}\isamarkupfalse%
\ this\ \isakeyword{and}\ h{\isadigit{5}}\ \isacommand{have}\isamarkupfalse%
\ a{\isadigit{5}}{\isadigit{2}}a{\isacharcolon}{\isachardoublequoteopen}stop\ t\ {\isacharequal}\ {\isacharbrackleft}{\isacharbrackright}{\isachardoublequoteclose}\ \isacommand{by}\isamarkupfalse%
\ {\isacharparenleft}simp\ add{\isacharcolon}\ stopType{\isacharunderscore}empty{\isacharparenright}\isanewline
\ \ \ \ \ \ \isacommand{show}\isamarkupfalse%
\ {\isacharquery}thesis\isanewline
\ \ \ \ \ \ \isacommand{proof}\isamarkupfalse%
\ {\isacharparenleft}cases\ {\isachardoublequoteopen}lose\ t\ {\isacharequal}\ {\isacharbrackleft}True{\isacharbrackright}{\isachardoublequoteclose}{\isacharparenright}\isanewline
\ \ \ \ \ \ \ \ \isacommand{assume}\isamarkupfalse%
\ a{\isadigit{5}}{\isadigit{2}}{\isadigit{1}}{\isacharcolon}{\isachardoublequoteopen}lose\ t\ {\isacharequal}\ {\isacharbrackleft}True{\isacharbrackright}{\isachardoublequoteclose}\isanewline
\ \ \ \ \ \ \ \ \isacommand{from}\isamarkupfalse%
\ h{\isadigit{1}}\ \isakeyword{and}\ a{\isadigit{5}}\ \isakeyword{and}\ a{\isadigit{5}}{\isadigit{2}}a\ \isakeyword{and}\ a{\isadigit{5}}{\isadigit{2}}{\isadigit{1}}\ \isakeyword{and}\ h{\isadigit{2}}\ \isacommand{show}\isamarkupfalse%
\ {\isacharquery}thesis\ \isanewline
\ \ \ \ \ \ \ \ \ \ \isacommand{by}\isamarkupfalse%
\ {\isacharparenleft}simp\ add{\isacharcolon}\ tiTable{\isacharunderscore}SampleT{\isacharunderscore}def{\isacharparenright}\isanewline
\ \ \ \ \ \ \isacommand{next}\isamarkupfalse%
\isanewline
\ \ \ \ \ \ \ \ \isacommand{assume}\isamarkupfalse%
\ a{\isadigit{5}}{\isadigit{2}}{\isadigit{2}}{\isacharcolon}{\isachardoublequoteopen}lose\ t\ {\isasymnoteq}\ {\isacharbrackleft}True{\isacharbrackright}{\isachardoublequoteclose}\isanewline
\ \ \ \ \ \ \ \ \isacommand{from}\isamarkupfalse%
\ this\ \isakeyword{and}\ h{\isadigit{4}}\ \isacommand{have}\isamarkupfalse%
\ a{\isadigit{5}}{\isadigit{2}}{\isadigit{2}}a{\isacharcolon}{\isachardoublequoteopen}lose\ t\ {\isacharequal}\ {\isacharbrackleft}False{\isacharbrackright}{\isachardoublequoteclose}\ \isacommand{by}\isamarkupfalse%
\ {\isacharparenleft}simp\ add{\isacharcolon}\ ts{\isacharunderscore}bool{\isacharunderscore}False{\isacharparenright}\isanewline
\ \ \ \ \ \ \ \ \isacommand{from}\isamarkupfalse%
\ h{\isadigit{1}}\ \isakeyword{and}\ a{\isadigit{5}}\ \isakeyword{and}\ a{\isadigit{5}}{\isadigit{2}}a\ \isakeyword{and}\ a{\isadigit{5}}{\isadigit{2}}{\isadigit{2}}a\ \isakeyword{and}\ h{\isadigit{2}}\ \isacommand{show}\isamarkupfalse%
\ {\isacharquery}thesis\ \isanewline
\ \ \ \ \ \ \ \ \ \ \isacommand{by}\isamarkupfalse%
\ {\isacharparenleft}simp\ add{\isacharcolon}\ tiTable{\isacharunderscore}SampleT{\isacharunderscore}def{\isacharparenright}\isanewline
\ \ \ \ \ \ \isacommand{qed}\isamarkupfalse%
\isanewline
\ \ \ \ \isacommand{qed}\isamarkupfalse%
\isanewline
\ \ \isacommand{qed}\isamarkupfalse%
\isanewline
\isacommand{qed}\isamarkupfalse%
\endisatagproof
{\isafoldproof}%
\isadelimproof
\isanewline
\endisadelimproof
\isanewline
\isacommand{lemma}\isamarkupfalse%
\ tiTable{\isacharunderscore}i{\isadigit{1}}{\isacharunderscore}{\isadigit{2}}{\isacharcolon}\isanewline
\ \ \isakeyword{assumes}\ h{\isadigit{1}}{\isacharcolon}{\isachardoublequoteopen}tiTable{\isacharunderscore}SampleT\ req\ a{\isadigit{1}}\ stop\ lose\ st{\isacharunderscore}in\ b\ ack\ i{\isadigit{1}}\ vc\ st{\isacharunderscore}out{\isachardoublequoteclose}\ \isanewline
\ \ \ \ \ \ \isakeyword{and}\ h{\isadigit{2}}{\isacharcolon}{\isachardoublequoteopen}ts\ lose{\isachardoublequoteclose}\isanewline
\ \ \ \ \ \ \isakeyword{and}\ h{\isadigit{3}}{\isacharcolon}{\isachardoublequoteopen}msg\ {\isacharparenleft}Suc\ {\isadigit{0}}{\isacharparenright}\ a{\isadigit{1}}{\isachardoublequoteclose} 
\ \isakeyword{and}\ h{\isadigit{4}}{\isacharcolon}{\isachardoublequoteopen}msg\ {\isacharparenleft}Suc\ {\isadigit{0}}{\isacharparenright}\ stop{\isachardoublequoteclose}\ \isanewline
\ \ \ \ \ \ \isakeyword{and}\ h{\isadigit{5}}{\isacharcolon}{\isachardoublequoteopen}ack\ t\ {\isacharequal}\ {\isacharbrackleft}call{\isacharbrackright}{\isachardoublequoteclose}\isanewline
\ \ \isakeyword{shows}\ {\isachardoublequoteopen}i{\isadigit{1}}\ t\ {\isacharequal}\ {\isacharbrackleft}{\isacharbrackright}{\isachardoublequoteclose}\isanewline
\isadelimproof
\endisadelimproof
\isatagproof
\isacommand{proof}\isamarkupfalse%
\ {\isacharminus}\isanewline
\ \ \isacommand{from}\isamarkupfalse%
\ assms\ \isacommand{have}\isamarkupfalse%
\ sg{\isadigit{1}}{\isacharcolon}{\isachardoublequoteopen}st{\isacharunderscore}in\ t\ {\isacharequal}\ init{\isacharunderscore}state\ {\isasymand}\ req\ t\ {\isacharequal}\ {\isacharbrackleft}init{\isacharbrackright}{\isachardoublequoteclose}\isanewline
\ \ \ \ \isacommand{by}\isamarkupfalse%
\ {\isacharparenleft}simp\ add{\isacharcolon}\ tiTable{\isacharunderscore}ack{\isacharunderscore}call{\isacharparenright}\isanewline
\ \ \isacommand{from}\isamarkupfalse%
\ this\ \isakeyword{and}\ h{\isadigit{1}}\ \isacommand{show}\isamarkupfalse%
\ {\isacharquery}thesis\isanewline
\ \ \ \ \isacommand{by}\isamarkupfalse%
\ {\isacharparenleft}simp\ add{\isacharcolon}\ tiTable{\isacharunderscore}SampleT{\isacharunderscore}def{\isacharparenright}\isanewline
\isacommand{qed}\isamarkupfalse%
\endisatagproof
{\isafoldproof}%
\isadelimproof
\ \isanewline
\endisadelimproof
\isanewline
\isacommand{lemma}\isamarkupfalse%
\ tiTable{\isacharunderscore}ack{\isacharunderscore}init{\isadigit{0}}{\isacharcolon}\isanewline
\ \ \isakeyword{assumes}\ h{\isadigit{1}}{\isacharcolon}{\isachardoublequoteopen}tiTable{\isacharunderscore}SampleT\ req\ a{\isadigit{1}}\ stop\ lose\ \isanewline
\ \ \ \ \ \ \ \ \ \ \ \ \ \ \ \ \ \ {\isacharparenleft}fin{\isacharunderscore}inf{\isacharunderscore}append\ {\isacharbrackleft}init{\isacharunderscore}state{\isacharbrackright}\ st{\isacharparenright}\ \isanewline
\ \ \ \ \ \ \ \ \ \ \ \ \ \ \ \ \ \ \ b\ ack\ i{\isadigit{1}}\ vc\ st{\isachardoublequoteclose}\ \isanewline
\ \ \ \ \ \ \isakeyword{and}\ h{\isadigit{2}}{\isacharcolon}{\isachardoublequoteopen}req\ {\isadigit{0}}\ {\isacharequal}\ {\isacharbrackleft}{\isacharbrackright}{\isachardoublequoteclose}\isanewline
\ \ \isakeyword{shows}\ {\isachardoublequoteopen}ack\ {\isadigit{0}}\ {\isacharequal}\ {\isacharbrackleft}init{\isacharunderscore}state{\isacharbrackright}{\isachardoublequoteclose}\isanewline
\isadelimproof
\endisadelimproof
\isatagproof
\isacommand{proof}\isamarkupfalse%
\ {\isacharminus}\isanewline
\ \ \isacommand{have}\isamarkupfalse%
\ sg{\isadigit{1}}{\isacharcolon}{\isachardoublequoteopen}{\isacharparenleft}fin{\isacharunderscore}inf{\isacharunderscore}append\ {\isacharbrackleft}init{\isacharunderscore}state{\isacharbrackright}\ st{\isacharparenright}\ {\isacharparenleft}{\isadigit{0}}{\isacharcolon}{\isacharcolon}nat{\isacharparenright}\ {\isacharequal}\ init{\isacharunderscore}state{\isachardoublequoteclose}\ \isanewline
\ \ \ \ \isacommand{by}\isamarkupfalse%
\ {\isacharparenleft}simp\ add{\isacharcolon}\ fin{\isacharunderscore}inf{\isacharunderscore}append{\isacharunderscore}def{\isacharparenright}\isanewline
\ \ \isacommand{from}\isamarkupfalse%
\ h{\isadigit{1}}\ \isakeyword{and}\ sg{\isadigit{1}}\ \isakeyword{and}\ h{\isadigit{2}}\ \isacommand{show}\isamarkupfalse%
\ {\isacharquery}thesis\ \isacommand{by}\isamarkupfalse%
\ {\isacharparenleft}simp\ add{\isacharcolon}\ tiTable{\isacharunderscore}SampleT{\isacharunderscore}def{\isacharparenright}\isanewline
\isacommand{qed}\isamarkupfalse%
\endisatagproof
{\isafoldproof}%
\isadelimproof
\isanewline
\endisadelimproof
\isanewline
\isanewline
\isacommand{lemma}\isamarkupfalse%
\ tiTable{\isacharunderscore}ack{\isacharunderscore}init{\isacharcolon}\isanewline
\ \ \isakeyword{assumes}\ h{\isadigit{1}}{\isacharcolon}{\isachardoublequoteopen}tiTable{\isacharunderscore}SampleT\ req\ a{\isadigit{1}}\ stop\ lose\ \isanewline
\ \ \ \ \ \ \ \ \ \ \ \ \ \ \ \ \ \ {\isacharparenleft}fin{\isacharunderscore}inf{\isacharunderscore}append\ {\isacharbrackleft}init{\isacharunderscore}state{\isacharbrackright}\ st{\isacharparenright}\ \isanewline
\ \ \ \ \ \ \ \ \ \ \ \ \ \ \ \ \ \ \ b\ ack\ i{\isadigit{1}}\ vc\ st{\isachardoublequoteclose}\isanewline
\ \ \ \ \ \ \isakeyword{and}\ h{\isadigit{2}}{\isacharcolon}{\isachardoublequoteopen}ts\ lose{\isachardoublequoteclose}\isanewline
\ \ \ \ \ \ \isakeyword{and}\ h{\isadigit{3}}{\isacharcolon}{\isachardoublequoteopen}msg\ {\isacharparenleft}Suc\ {\isadigit{0}}{\isacharparenright}\ a{\isadigit{1}}{\isachardoublequoteclose}\isanewline
\ \ \ \ \ \ \isakeyword{and}\ h{\isadigit{4}}{\isacharcolon}{\isachardoublequoteopen}msg\ {\isacharparenleft}Suc\ {\isadigit{0}}{\isacharparenright}\ stop{\isachardoublequoteclose}\isanewline
\ \ \ \ \ \ \isakeyword{and}\ h{\isadigit{5}}{\isacharcolon}{\isachardoublequoteopen}{\isasymforall}\ t{\isadigit{1}}\ {\isasymle}\ t{\isachardot}\ req\ t{\isadigit{1}}\ {\isacharequal}\ {\isacharbrackleft}{\isacharbrackright}{\isachardoublequoteclose}\isanewline
\ \ \isakeyword{shows}\ {\isachardoublequoteopen}ack\ t\ {\isacharequal}\ {\isacharbrackleft}init{\isacharunderscore}state{\isacharbrackright}{\isachardoublequoteclose}\isanewline
\isadelimproof
\endisadelimproof
\isatagproof
\isacommand{using}\isamarkupfalse%
\ assms\isanewline
\isacommand{proof}\isamarkupfalse%
\ {\isacharparenleft}induction\ t{\isacharparenright}\isanewline
\ \ \isacommand{case}\isamarkupfalse%
\ {\isadigit{0}}\isanewline
\ \ \isacommand{from}\isamarkupfalse%
\ this\ \isacommand{show}\isamarkupfalse%
\ {\isacharquery}case\isanewline
\ \ \ \ \isacommand{by}\isamarkupfalse%
\ {\isacharparenleft}simp\ add{\isacharcolon}\ tiTable{\isacharunderscore}ack{\isacharunderscore}init{\isadigit{0}}{\isacharparenright}\isanewline
\isacommand{next}\isamarkupfalse%
\ \isanewline
\ \ \isacommand{case}\isamarkupfalse%
\ {\isacharparenleft}Suc\ t{\isacharparenright}\isanewline
\ \ \isacommand{from}\isamarkupfalse%
\ Suc\ \isacommand{have}\isamarkupfalse%
\ sg{\isadigit{1}}{\isacharcolon}\ {\isachardoublequoteopen}st\ t\ {\isacharequal}\ \ hd\ {\isacharparenleft}ack\ t{\isacharparenright}{\isachardoublequoteclose}\isanewline
\ \ \ \ \isacommand{by}\isamarkupfalse%
\ {\isacharparenleft}simp\ add{\isacharcolon}\ tiTable{\isacharunderscore}ack{\isacharunderscore}st{\isacharunderscore}hd{\isacharparenright}\ \ \isanewline
\ \ \isacommand{from}\isamarkupfalse%
\ Suc\ \isakeyword{and}\ sg{\isadigit{1}}\ \isacommand{have}\isamarkupfalse%
\ sg{\isadigit{2}}{\isacharcolon}\ \isanewline
\ \ \ {\isachardoublequoteopen}{\isacharparenleft}fin{\isacharunderscore}inf{\isacharunderscore}append\ {\isacharbrackleft}init{\isacharunderscore}state{\isacharbrackright}\ st{\isacharparenright}\ {\isacharparenleft}Suc\ t{\isacharparenright}\ {\isacharequal}\ init{\isacharunderscore}state{\isachardoublequoteclose}\isanewline
\ \ \ \ \isacommand{by}\isamarkupfalse%
\ {\isacharparenleft}simp\ add{\isacharcolon}\ correct{\isacharunderscore}fin{\isacharunderscore}inf{\isacharunderscore}append{\isadigit{2}}{\isacharparenright}\isanewline
\ \ \isacommand{from}\isamarkupfalse%
\ Suc\ \isakeyword{and}\ sg{\isadigit{1}}\ \isakeyword{and}\ sg{\isadigit{2}}\ \isacommand{show}\isamarkupfalse%
\ {\isacharquery}case\isanewline
\ \ \ \ \isacommand{by}\isamarkupfalse%
\ {\isacharparenleft}simp\ add{\isacharcolon}\ tiTable{\isacharunderscore}SampleT{\isacharunderscore}def{\isacharparenright}\isanewline
\isacommand{qed}\isamarkupfalse%
\endisatagproof
{\isafoldproof}%
\isadelimproof
\isanewline
\endisadelimproof
\isanewline
\isanewline
\isacommand{lemma}\isamarkupfalse%
\ tiTable{\isacharunderscore}i{\isadigit{1}}{\isacharunderscore}{\isadigit{3}}{\isacharcolon}\isanewline
\ \ \isakeyword{assumes}\ h{\isadigit{1}}{\isacharcolon}{\isachardoublequoteopen}tiTable{\isacharunderscore}SampleT\ req\ x\ stop\ lose\ \isanewline
\ \ \ \ \ \ \ \ \ \ \ \ \ \ \ \ \ \ {\isacharparenleft}fin{\isacharunderscore}inf{\isacharunderscore}append\ {\isacharbrackleft}init{\isacharunderscore}state{\isacharbrackright}\ st{\isacharparenright}\ \isanewline
\ \ \ \ \ \ \ \ \ \ \ \ \ \ \ \ \ \ \ b\ ack\ i{\isadigit{1}}\ vc\ st{\isachardoublequoteclose}\ \isanewline
\ \ \ \ \ \ \isakeyword{and}\ h{\isadigit{2}}{\isacharcolon}{\isachardoublequoteopen}ts\ lose{\isachardoublequoteclose}\isanewline
\ \ \ \ \ \ \isakeyword{and}\ h{\isadigit{3}}{\isacharcolon}{\isachardoublequoteopen}msg\ {\isacharparenleft}Suc\ {\isadigit{0}}{\isacharparenright}\ x{\isachardoublequoteclose}\isanewline
\ \ \ \ \ \ \isakeyword{and}\ h{\isadigit{4}}{\isacharcolon}{\isachardoublequoteopen}msg\ {\isacharparenleft}Suc\ {\isadigit{0}}{\isacharparenright}\ stop{\isachardoublequoteclose}\isanewline
\ \ \ \ \ \ \isakeyword{and}\ h{\isadigit{5}}{\isacharcolon}{\isachardoublequoteopen}{\isasymforall}\ t{\isadigit{1}}\ {\isasymle}\ t{\isachardot}\ req\ t{\isadigit{1}}\ {\isacharequal}\ {\isacharbrackleft}{\isacharbrackright}{\isachardoublequoteclose}\ \isanewline
\ \isakeyword{shows}\ {\isachardoublequoteopen}i{\isadigit{1}}\ t\ {\isacharequal}\ {\isacharbrackleft}{\isacharbrackright}{\isachardoublequoteclose}\isanewline
\isadelimproof
\endisadelimproof
\isatagproof
\isacommand{proof}\isamarkupfalse%
\ {\isacharminus}\ \isanewline
\ \ \isacommand{from}\isamarkupfalse%
\ assms\ \isacommand{have}\isamarkupfalse%
\ sg{\isadigit{1}}{\isacharcolon}{\isachardoublequoteopen}ack\ t\ {\isacharequal}\ {\isacharbrackleft}init{\isacharunderscore}state{\isacharbrackright}{\isachardoublequoteclose}\isanewline
\ \ \ \ \isacommand{by}\isamarkupfalse%
\ {\isacharparenleft}simp\ add{\isacharcolon}\ tiTable{\isacharunderscore}ack{\isacharunderscore}init{\isacharparenright}\isanewline
\ \ \isacommand{from}\isamarkupfalse%
\ assms\ \isacommand{have}\isamarkupfalse%
\ sg{\isadigit{2}}{\isacharcolon}{\isachardoublequoteopen}st\ t\ {\isacharequal}\ \ hd\ {\isacharparenleft}ack\ t{\isacharparenright}{\isachardoublequoteclose}\isanewline
\ \ \ \ \isacommand{by}\isamarkupfalse%
\ {\isacharparenleft}simp\ add{\isacharcolon}\ tiTable{\isacharunderscore}ack{\isacharunderscore}st{\isacharunderscore}hd{\isacharparenright}\ \ \isanewline
\ \ \isacommand{from}\isamarkupfalse%
\ sg{\isadigit{1}}\ \isakeyword{and}\ sg{\isadigit{2}}\ \isacommand{have}\isamarkupfalse%
\ sg{\isadigit{3}}{\isacharcolon}\isanewline
\ \ \ {\isachardoublequoteopen}{\isacharparenleft}fin{\isacharunderscore}inf{\isacharunderscore}append\ {\isacharbrackleft}init{\isacharunderscore}state{\isacharbrackright}\ st{\isacharparenright}\ {\isacharparenleft}Suc\ t{\isacharparenright}\ {\isacharequal}\ init{\isacharunderscore}state{\isachardoublequoteclose}\isanewline
\ \ \ \ \isacommand{by}\isamarkupfalse%
\ {\isacharparenleft}simp\ add{\isacharcolon}\ correct{\isacharunderscore}fin{\isacharunderscore}inf{\isacharunderscore}append{\isadigit{2}}{\isacharparenright}\isanewline
\ \ \isacommand{from}\isamarkupfalse%
\ h{\isadigit{1}}\ \isakeyword{and}\ h{\isadigit{2}}\ \isacommand{have}\isamarkupfalse%
\ sg{\isadigit{4}}{\isacharcolon}{\isachardoublequoteopen}lose\ t\ {\isacharequal}\ {\isacharbrackleft}True{\isacharbrackright}\ {\isasymor}\ lose\ t\ {\isacharequal}\ {\isacharbrackleft}False{\isacharbrackright}{\isachardoublequoteclose}\isanewline
\ \ \ \ \isacommand{by}\isamarkupfalse%
\ {\isacharparenleft}simp\ add{\isacharcolon}\ ts{\isacharunderscore}bool{\isacharunderscore}True{\isacharunderscore}False{\isacharparenright}\isanewline
\ \ \isacommand{from}\isamarkupfalse%
\ h{\isadigit{1}}\ \isakeyword{and}\ h{\isadigit{3}}\ \isacommand{have}\isamarkupfalse%
\ sg{\isadigit{5}}{\isacharcolon}{\isachardoublequoteopen}x\ t\ {\isacharequal}\ {\isacharbrackleft}{\isacharbrackright}\ {\isasymor}\ x\ t\ {\isacharequal}\ {\isacharbrackleft}sc{\isacharunderscore}ack{\isacharbrackright}{\isachardoublequoteclose}\isanewline
\ \ \ \ \isacommand{by}\isamarkupfalse%
\ {\isacharparenleft}simp\ add{\isacharcolon}\ aType{\isacharunderscore}lemma{\isacharparenright}\ \isanewline
\ \ \isacommand{from}\isamarkupfalse%
\ h{\isadigit{1}}\ \isakeyword{and}\ h{\isadigit{4}}\ \isacommand{have}\isamarkupfalse%
\ sg{\isadigit{6}}{\isacharcolon}{\isachardoublequoteopen}stop\ t\ {\isacharequal}\ {\isacharbrackleft}{\isacharbrackright}\ {\isasymor}\ stop\ t\ {\isacharequal}\ {\isacharbrackleft}stop{\isacharunderscore}vc{\isacharbrackright}{\isachardoublequoteclose}\isanewline
\ \ \ \ \isacommand{by}\isamarkupfalse%
\ {\isacharparenleft}simp\ add{\isacharcolon}\ stopType{\isacharunderscore}lemma{\isacharparenright}\ \isanewline
\ \ \isacommand{show}\isamarkupfalse%
\ {\isacharquery}thesis\isanewline
\ \ \isacommand{proof}\isamarkupfalse%
\ {\isacharparenleft}cases\ {\isachardoublequoteopen}fin{\isacharunderscore}inf{\isacharunderscore}append\ {\isacharbrackleft}init{\isacharunderscore}state{\isacharbrackright}\ st\ t{\isachardoublequoteclose}{\isacharparenright}\isanewline
\ \ \ \ \isacommand{assume}\isamarkupfalse%
\ a{\isadigit{1}}{\isacharcolon}{\isachardoublequoteopen}fin{\isacharunderscore}inf{\isacharunderscore}append\ {\isacharbrackleft}init{\isacharunderscore}state{\isacharbrackright}\ st\ t\ {\isacharequal}\ init{\isacharunderscore}state{\isachardoublequoteclose}\isanewline
\ \ \ \ \isacommand{from}\isamarkupfalse%
\ assms\ \isakeyword{and}\ sg{\isadigit{1}}\ \isakeyword{and}\ sg{\isadigit{2}}\ \isakeyword{and}\ sg{\isadigit{3}}\ \isakeyword{and}\ a{\isadigit{1}}\ \isacommand{show}\isamarkupfalse%
\ {\isacharquery}thesis\isanewline
\ \ \ \ \ \ \isacommand{by}\isamarkupfalse%
\ {\isacharparenleft}simp\ add{\isacharcolon}\ \ tiTable{\isacharunderscore}SampleT{\isacharunderscore}def{\isacharparenright}\isanewline
\ \ \isacommand{next}\isamarkupfalse%
\isanewline
\ \ \ \ \isacommand{assume}\isamarkupfalse%
\ a{\isadigit{2}}{\isacharcolon}{\isachardoublequoteopen}fin{\isacharunderscore}inf{\isacharunderscore}append\ {\isacharbrackleft}init{\isacharunderscore}state{\isacharbrackright}\ st\ t\ {\isacharequal}\ call{\isachardoublequoteclose}\isanewline
\ \ \ \ \isacommand{show}\isamarkupfalse%
\ {\isacharquery}thesis\isanewline
\ \ \ \ \isacommand{proof}\isamarkupfalse%
\ {\isacharparenleft}cases\ {\isachardoublequoteopen}lose\ t\ {\isacharequal}\ {\isacharbrackleft}True{\isacharbrackright}{\isachardoublequoteclose}{\isacharparenright}\isanewline
\ \ \ \ \ \ \isacommand{assume}\isamarkupfalse%
\ a{\isadigit{2}}{\isadigit{1}}{\isacharcolon}{\isachardoublequoteopen}lose\ t\ {\isacharequal}\ {\isacharbrackleft}True{\isacharbrackright}{\isachardoublequoteclose}\isanewline
\ \ \ \ \ \ \isacommand{from}\isamarkupfalse%
\ h{\isadigit{1}}\ \isakeyword{and}\ a{\isadigit{2}}\ \isakeyword{and}\ a{\isadigit{2}}{\isadigit{1}}\ \isacommand{show}\isamarkupfalse%
\ {\isacharquery}thesis\ \isacommand{by}\isamarkupfalse%
\ {\isacharparenleft}simp\ add{\isacharcolon}\ tiTable{\isacharunderscore}SampleT{\isacharunderscore}def{\isacharparenright}\isanewline
\ \ \ \ \isacommand{next}\isamarkupfalse%
\isanewline
\ \ \ \ \ \ \isacommand{assume}\isamarkupfalse%
\ a{\isadigit{2}}{\isadigit{2}}{\isacharcolon}{\isachardoublequoteopen}lose\ t\ {\isasymnoteq}\ {\isacharbrackleft}True{\isacharbrackright}{\isachardoublequoteclose}\isanewline
\ \ \ \ \ \ \isacommand{from}\isamarkupfalse%
\ this\ \isakeyword{and}\ h{\isadigit{2}}\ \isacommand{have}\isamarkupfalse%
\ a{\isadigit{2}}{\isadigit{2}}a{\isacharcolon}{\isachardoublequoteopen}lose\ t\ {\isacharequal}\ {\isacharbrackleft}False{\isacharbrackright}{\isachardoublequoteclose}\ \isacommand{by}\isamarkupfalse%
\ {\isacharparenleft}simp\ add{\isacharcolon}\ ts{\isacharunderscore}bool{\isacharunderscore}False{\isacharparenright}\isanewline
\ \ \ \ \ \ \isacommand{from}\isamarkupfalse%
\ h{\isadigit{1}}\ \isakeyword{and}\ a{\isadigit{2}}\ \isakeyword{and}\ a{\isadigit{2}}{\isadigit{2}}a\ \isacommand{show}\isamarkupfalse%
\ {\isacharquery}thesis\ \isacommand{by}\isamarkupfalse%
\ {\isacharparenleft}simp\ add{\isacharcolon}\ tiTable{\isacharunderscore}SampleT{\isacharunderscore}def{\isacharparenright}\isanewline
\ \ \ \ \isacommand{qed}\isamarkupfalse%
\ \isanewline
\ \ \isacommand{next}\isamarkupfalse%
\isanewline
\ \ \ \ \isacommand{assume}\isamarkupfalse%
\ a{\isadigit{3}}{\isacharcolon}{\isachardoublequoteopen}fin{\isacharunderscore}inf{\isacharunderscore}append\ {\isacharbrackleft}init{\isacharunderscore}state{\isacharbrackright}\ st\ t\ {\isacharequal}\ connection{\isacharunderscore}ok{\isachardoublequoteclose}\isanewline
\ \ \ \ \isacommand{show}\isamarkupfalse%
\ {\isacharquery}thesis\isanewline
\ \ \ \ \isacommand{proof}\isamarkupfalse%
\ {\isacharparenleft}cases\ {\isachardoublequoteopen}lose\ t\ {\isacharequal}\ {\isacharbrackleft}True{\isacharbrackright}{\isachardoublequoteclose}{\isacharparenright}\isanewline
\ \ \ \ \ \ \isacommand{assume}\isamarkupfalse%
\ a{\isadigit{3}}{\isadigit{1}}{\isacharcolon}{\isachardoublequoteopen}lose\ t\ {\isacharequal}\ {\isacharbrackleft}True{\isacharbrackright}{\isachardoublequoteclose}\isanewline
\ \ \ \ \ \ \isacommand{from}\isamarkupfalse%
\ h{\isadigit{1}}\ \isakeyword{and}\ a{\isadigit{3}}\ \isakeyword{and}\ a{\isadigit{3}}{\isadigit{1}}\ \isacommand{show}\isamarkupfalse%
\ {\isacharquery}thesis\ \isacommand{by}\isamarkupfalse%
\ {\isacharparenleft}simp\ add{\isacharcolon}\ tiTable{\isacharunderscore}SampleT{\isacharunderscore}def{\isacharparenright}\isanewline
\ \ \ \ \isacommand{next}\isamarkupfalse%
\isanewline
\ \ \ \ \ \ \isacommand{assume}\isamarkupfalse%
\ a{\isadigit{3}}{\isadigit{2}}{\isacharcolon}{\isachardoublequoteopen}lose\ t\ {\isasymnoteq}\ {\isacharbrackleft}True{\isacharbrackright}{\isachardoublequoteclose}\isanewline
\ \ \ \ \ \ \isacommand{from}\isamarkupfalse%
\ this\ \isakeyword{and}\ h{\isadigit{2}}\ \isacommand{have}\isamarkupfalse%
\ a{\isadigit{3}}{\isadigit{2}}a{\isacharcolon}{\isachardoublequoteopen}lose\ t\ {\isacharequal}\ {\isacharbrackleft}False{\isacharbrackright}{\isachardoublequoteclose}\ \isacommand{by}\isamarkupfalse%
\ {\isacharparenleft}simp\ add{\isacharcolon}\ ts{\isacharunderscore}bool{\isacharunderscore}False{\isacharparenright}\isanewline
\ \ \ \ \ \ \isacommand{from}\isamarkupfalse%
\ h{\isadigit{5}}\ \isacommand{have}\isamarkupfalse%
\ a{\isadigit{3}}{\isadigit{2}}{\isadigit{2}}{\isacharcolon}{\isachardoublequoteopen}req\ t\ {\isasymnoteq}\ {\isacharbrackleft}send{\isacharbrackright}{\isachardoublequoteclose}\ \isacommand{by}\isamarkupfalse%
\ auto\isanewline
\ \ \ \ \ \ \isacommand{from}\isamarkupfalse%
\ h{\isadigit{1}}\ \isakeyword{and}\ a{\isadigit{3}}\ \isakeyword{and}\ a{\isadigit{3}}{\isadigit{2}}a\ \isakeyword{and}\ a{\isadigit{3}}{\isadigit{2}}{\isadigit{2}}\ \isacommand{show}\isamarkupfalse%
\ {\isacharquery}thesis\ \isanewline
\ \ \ \ \ \ \ \ \isacommand{by}\isamarkupfalse%
\ {\isacharparenleft}simp\ add{\isacharcolon}\ tiTable{\isacharunderscore}SampleT{\isacharunderscore}def{\isacharparenright}\isanewline
\ \ \ \ \isacommand{qed}\isamarkupfalse%
\ \isanewline
\ \ \isacommand{next}\isamarkupfalse%
\isanewline
\ \ \ \ \isacommand{assume}\isamarkupfalse%
\ a{\isadigit{4}}{\isacharcolon}{\isachardoublequoteopen}fin{\isacharunderscore}inf{\isacharunderscore}append\ {\isacharbrackleft}init{\isacharunderscore}state{\isacharbrackright}\ st\ t\ {\isacharequal}\ sending{\isacharunderscore}data{\isachardoublequoteclose}\isanewline
\ \ \ \ \isacommand{show}\isamarkupfalse%
\ {\isacharquery}thesis\isanewline
\ \ \ \ \isacommand{proof}\isamarkupfalse%
\ {\isacharparenleft}cases\ {\isachardoublequoteopen}lose\ t\ {\isacharequal}\ {\isacharbrackleft}True{\isacharbrackright}{\isachardoublequoteclose}{\isacharparenright}\isanewline
\ \ \ \ \ \ \isacommand{assume}\isamarkupfalse%
\ a{\isadigit{4}}{\isadigit{1}}{\isacharcolon}{\isachardoublequoteopen}lose\ t\ {\isacharequal}\ {\isacharbrackleft}True{\isacharbrackright}{\isachardoublequoteclose}\isanewline
\ \ \ \ \ \ \isacommand{from}\isamarkupfalse%
\ h{\isadigit{1}}\ \isakeyword{and}\ a{\isadigit{4}}\ \isakeyword{and}\ a{\isadigit{4}}{\isadigit{1}}\ \isacommand{show}\isamarkupfalse%
\ {\isacharquery}thesis\ \isacommand{by}\isamarkupfalse%
\ {\isacharparenleft}simp\ add{\isacharcolon}\ tiTable{\isacharunderscore}SampleT{\isacharunderscore}def{\isacharparenright}\ \isanewline
\ \ \ \ \isacommand{next}\isamarkupfalse%
\isanewline
\ \ \ \ \ \ \isacommand{assume}\isamarkupfalse%
\ a{\isadigit{4}}{\isadigit{2}}{\isacharcolon}{\isachardoublequoteopen}lose\ t\ {\isasymnoteq}\ {\isacharbrackleft}True{\isacharbrackright}{\isachardoublequoteclose}\isanewline
\ \ \ \ \ \ \isacommand{from}\isamarkupfalse%
\ this\ \isakeyword{and}\ h{\isadigit{2}}\ \isacommand{have}\isamarkupfalse%
\ a{\isadigit{4}}{\isadigit{2}}a{\isacharcolon}{\isachardoublequoteopen}lose\ t\ {\isacharequal}\ {\isacharbrackleft}False{\isacharbrackright}{\isachardoublequoteclose}\ \isacommand{by}\isamarkupfalse%
\ {\isacharparenleft}simp\ add{\isacharcolon}\ ts{\isacharunderscore}bool{\isacharunderscore}False{\isacharparenright}\isanewline
\ \ \ \ \ \ \isacommand{show}\isamarkupfalse%
\ {\isacharquery}thesis\isanewline
\ \ \ \ \ \ \isacommand{proof}\isamarkupfalse%
\ {\isacharparenleft}cases\ {\isachardoublequoteopen}x\ t\ {\isacharequal}\ {\isacharbrackleft}sc{\isacharunderscore}ack{\isacharbrackright}{\isachardoublequoteclose}{\isacharparenright}\isanewline
\ \ \ \ \ \ \ \ \isacommand{assume}\isamarkupfalse%
\ a{\isadigit{4}}{\isadigit{2}}{\isadigit{1}}{\isacharcolon}{\isachardoublequoteopen}x\ t\ {\isacharequal}\ {\isacharbrackleft}sc{\isacharunderscore}ack{\isacharbrackright}{\isachardoublequoteclose}\isanewline
\ \ \ \ \ \ \ \ \isacommand{from}\isamarkupfalse%
\ h{\isadigit{1}}\ \isakeyword{and}\ a{\isadigit{4}}\ \isakeyword{and}\ a{\isadigit{4}}{\isadigit{2}}a\ \isakeyword{and}\ a{\isadigit{4}}{\isadigit{2}}{\isadigit{1}}\ \isakeyword{and}\ h{\isadigit{2}}\ \isacommand{show}\isamarkupfalse%
\ {\isacharquery}thesis\ \isanewline
\ \ \ \ \ \ \ \ \ \ \isacommand{by}\isamarkupfalse%
\ {\isacharparenleft}simp\ add{\isacharcolon}\ tiTable{\isacharunderscore}SampleT{\isacharunderscore}def{\isacharparenright}\isanewline
\ \ \ \ \ \ \isacommand{next}\isamarkupfalse%
\isanewline
\ \ \ \ \ \ \ \ \isacommand{assume}\isamarkupfalse%
\ a{\isadigit{4}}{\isadigit{2}}{\isadigit{2}}{\isacharcolon}{\isachardoublequoteopen}\ x\ t\ {\isasymnoteq}\ {\isacharbrackleft}sc{\isacharunderscore}ack{\isacharbrackright}{\isachardoublequoteclose}\isanewline
\ \ \ \ \ \ \ \ \isacommand{from}\isamarkupfalse%
\ this\ \isakeyword{and}\ h{\isadigit{3}}\ \ \isacommand{have}\isamarkupfalse%
\ a{\isadigit{4}}{\isadigit{2}}{\isadigit{2}}a{\isacharcolon}{\isachardoublequoteopen}x\ t\ {\isacharequal}\ {\isacharbrackleft}{\isacharbrackright}{\isachardoublequoteclose}\ \isacommand{by}\isamarkupfalse%
\ {\isacharparenleft}simp\ add{\isacharcolon}\ aType{\isacharunderscore}empty{\isacharparenright}\isanewline
\ \ \ \ \ \ \ \ \isacommand{from}\isamarkupfalse%
\ h{\isadigit{1}}\ \isakeyword{and}\ a{\isadigit{4}}\ \isakeyword{and}\ a{\isadigit{4}}{\isadigit{2}}a\ \isakeyword{and}\ a{\isadigit{4}}{\isadigit{2}}{\isadigit{2}}a\ \isakeyword{and}\ h{\isadigit{2}}\ \isacommand{show}\isamarkupfalse%
\ {\isacharquery}thesis\ \isanewline
\ \ \ \ \ \ \ \ \ \ \isacommand{by}\isamarkupfalse%
\ {\isacharparenleft}simp\ add{\isacharcolon}\ tiTable{\isacharunderscore}SampleT{\isacharunderscore}def{\isacharparenright}\isanewline
\ \ \ \ \ \ \isacommand{qed}\isamarkupfalse%
\isanewline
\ \ \ \ \isacommand{qed}\isamarkupfalse%
\isanewline
\ \ \isacommand{next}\isamarkupfalse%
\isanewline
\ \ \ \ \isacommand{assume}\isamarkupfalse%
\ a{\isadigit{5}}{\isacharcolon}{\isachardoublequoteopen}fin{\isacharunderscore}inf{\isacharunderscore}append\ {\isacharbrackleft}init{\isacharunderscore}state{\isacharbrackright}\ st\ t\ {\isacharequal}\ voice{\isacharunderscore}com{\isachardoublequoteclose}\isanewline
\ \ \ \ \isacommand{show}\isamarkupfalse%
\ {\isacharquery}thesis\isanewline
\ \ \ \ \isacommand{proof}\isamarkupfalse%
\ {\isacharparenleft}cases\ {\isachardoublequoteopen}stop\ t\ {\isacharequal}\ {\isacharbrackleft}stop{\isacharunderscore}vc{\isacharbrackright}{\isachardoublequoteclose}{\isacharparenright}\isanewline
\ \ \ \ \ \ \isacommand{assume}\isamarkupfalse%
\ a{\isadigit{5}}{\isadigit{1}}{\isacharcolon}{\isachardoublequoteopen}stop\ t\ {\isacharequal}\ {\isacharbrackleft}stop{\isacharunderscore}vc{\isacharbrackright}{\isachardoublequoteclose}\isanewline
\ \ \ \ \ \ \isacommand{from}\isamarkupfalse%
\ h{\isadigit{1}}\ \isakeyword{and}\ a{\isadigit{5}}\ \isakeyword{and}\ a{\isadigit{5}}{\isadigit{1}}\ \isakeyword{and}\ h{\isadigit{2}}\ \isacommand{show}\isamarkupfalse%
\ {\isacharquery}thesis\ \isacommand{by}\isamarkupfalse%
\ {\isacharparenleft}simp\ add{\isacharcolon}\ tiTable{\isacharunderscore}SampleT{\isacharunderscore}def{\isacharparenright}\isanewline
\ \ \ \ \isacommand{next}\isamarkupfalse%
\isanewline
\ \ \ \ \ \ \isacommand{assume}\isamarkupfalse%
\ a{\isadigit{5}}{\isadigit{2}}{\isacharcolon}{\isachardoublequoteopen}stop\ t\ {\isasymnoteq}\ {\isacharbrackleft}stop{\isacharunderscore}vc{\isacharbrackright}{\isachardoublequoteclose}\isanewline
\ \ \ \ \ \ \isacommand{from}\isamarkupfalse%
\ this\ \isakeyword{and}\ h{\isadigit{4}}\ \isacommand{have}\isamarkupfalse%
\ a{\isadigit{5}}{\isadigit{2}}a{\isacharcolon}{\isachardoublequoteopen}stop\ t\ {\isacharequal}\ {\isacharbrackleft}{\isacharbrackright}{\isachardoublequoteclose}\ \isacommand{by}\isamarkupfalse%
\ {\isacharparenleft}simp\ add{\isacharcolon}\ stopType{\isacharunderscore}empty{\isacharparenright}\isanewline
\ \ \ \ \ \ \isacommand{show}\isamarkupfalse%
\ {\isacharquery}thesis\isanewline
\ \ \ \ \ \ \isacommand{proof}\isamarkupfalse%
\ {\isacharparenleft}cases\ {\isachardoublequoteopen}lose\ t\ {\isacharequal}\ {\isacharbrackleft}True{\isacharbrackright}{\isachardoublequoteclose}{\isacharparenright}\isanewline
\ \ \ \ \ \ \ \ \isacommand{assume}\isamarkupfalse%
\ a{\isadigit{5}}{\isadigit{2}}{\isadigit{1}}{\isacharcolon}{\isachardoublequoteopen}lose\ t\ {\isacharequal}\ {\isacharbrackleft}True{\isacharbrackright}{\isachardoublequoteclose}\isanewline
\ \ \ \ \ \ \ \ \isacommand{from}\isamarkupfalse%
\ h{\isadigit{1}}\ \isakeyword{and}\ a{\isadigit{5}}\ \isakeyword{and}\ a{\isadigit{5}}{\isadigit{2}}a\ \isakeyword{and}\ a{\isadigit{5}}{\isadigit{2}}{\isadigit{1}}\ \isakeyword{and}\ h{\isadigit{2}}\ \isacommand{show}\isamarkupfalse%
\ {\isacharquery}thesis\ \isanewline
\ \ \ \ \ \ \ \ \ \ \isacommand{by}\isamarkupfalse%
\ {\isacharparenleft}simp\ add{\isacharcolon}\ tiTable{\isacharunderscore}SampleT{\isacharunderscore}def{\isacharparenright}\isanewline
\ \ \ \ \ \ \isacommand{next}\isamarkupfalse%
\isanewline
\ \ \ \ \ \ \ \ \isacommand{assume}\isamarkupfalse%
\ a{\isadigit{5}}{\isadigit{2}}{\isadigit{2}}{\isacharcolon}{\isachardoublequoteopen}lose\ t\ {\isasymnoteq}\ {\isacharbrackleft}True{\isacharbrackright}{\isachardoublequoteclose}\isanewline
\ \ \ \ \ \ \ \ \isacommand{from}\isamarkupfalse%
\ this\ \isakeyword{and}\ h{\isadigit{2}}\ \isacommand{have}\isamarkupfalse%
\ a{\isadigit{5}}{\isadigit{2}}{\isadigit{2}}a{\isacharcolon}{\isachardoublequoteopen}lose\ t\ {\isacharequal}\ {\isacharbrackleft}False{\isacharbrackright}{\isachardoublequoteclose}\ \isacommand{by}\isamarkupfalse%
\ {\isacharparenleft}simp\ add{\isacharcolon}\ ts{\isacharunderscore}bool{\isacharunderscore}False{\isacharparenright}\isanewline
\ \ \ \ \ \ \ \ \isacommand{from}\isamarkupfalse%
\ h{\isadigit{1}}\ \isakeyword{and}\ a{\isadigit{5}}\ \isakeyword{and}\ a{\isadigit{5}}{\isadigit{2}}a\ \isakeyword{and}\ a{\isadigit{5}}{\isadigit{2}}{\isadigit{2}}a\ \isakeyword{and}\ h{\isadigit{2}}\ \isacommand{show}\isamarkupfalse%
\ {\isacharquery}thesis\ \isanewline
\ \ \ \ \ \ \ \ \ \ \isacommand{by}\isamarkupfalse%
\ {\isacharparenleft}simp\ add{\isacharcolon}\ tiTable{\isacharunderscore}SampleT{\isacharunderscore}def{\isacharparenright}\isanewline
\ \ \ \ \ \ \isacommand{qed}\isamarkupfalse%
\isanewline
\ \ \ \ \isacommand{qed}\isamarkupfalse%
\isanewline
\ \ \isacommand{qed}\isamarkupfalse%
\isanewline
\isacommand{qed}\isamarkupfalse%
\endisatagproof
{\isafoldproof}%
\isadelimproof
\isanewline
\endisadelimproof
\isanewline
\isanewline
\isacommand{lemma}\isamarkupfalse%
\ tiTable{\isacharunderscore}st{\isacharunderscore}call{\isacharunderscore}ok{\isacharcolon}\isanewline
\ \ \isakeyword{assumes}\ h{\isadigit{1}}{\isacharcolon}{\isachardoublequoteopen}tiTable{\isacharunderscore}SampleT\ req\ x\ stop\ lose\ \isanewline
\ \ \ \ \ \ \ \ \ \ \ \ \ \ \ \ \ \ {\isacharparenleft}fin{\isacharunderscore}inf{\isacharunderscore}append\ {\isacharbrackleft}init{\isacharunderscore}state{\isacharbrackright}\ st{\isacharparenright}\ \isanewline
\ \ \ \ \ \ \ \ \ \ \ \ \ \ \ \ \ \ \ b\ ack\ i{\isadigit{1}}\ vc\ st{\isachardoublequoteclose}\isanewline
\ \ \ \ \ \ \isakeyword{and}\ h{\isadigit{2}}{\isacharcolon}{\isachardoublequoteopen}ts\ lose{\isachardoublequoteclose}\isanewline
\ \ \ \ \ \ \isakeyword{and}\ h{\isadigit{3}}{\isacharcolon}{\isachardoublequoteopen}{\isasymforall}m\ {\isasymle}\ k{\isachardot}\ ack\ {\isacharparenleft}Suc\ {\isacharparenleft}Suc\ {\isacharparenleft}t\ {\isacharplus}\ m{\isacharparenright}{\isacharparenright}{\isacharparenright}\ {\isacharequal}\ {\isacharbrackleft}connection{\isacharunderscore}ok{\isacharbrackright}{\isachardoublequoteclose}\isanewline
\ \ \ \ \ \ \isakeyword{and}\ h{\isadigit{4}}{\isacharcolon}{\isachardoublequoteopen}st\ {\isacharparenleft}Suc\ t{\isacharparenright}\ {\isacharequal}\ call{\isachardoublequoteclose}\isanewline
\ \ \isakeyword{shows}\ {\isachardoublequoteopen}st\ {\isacharparenleft}Suc\ {\isacharparenleft}Suc\ t{\isacharparenright}{\isacharparenright}\ {\isacharequal}\ connection{\isacharunderscore}ok{\isachardoublequoteclose}\isanewline
\isadelimproof
\endisadelimproof
\isatagproof
\isacommand{proof}\isamarkupfalse%
\ {\isacharminus}\ \isanewline
\ \ \ \ \isacommand{from}\isamarkupfalse%
\ h{\isadigit{4}}\ \isacommand{have}\isamarkupfalse%
\ sg{\isadigit{1}}{\isacharcolon}\isanewline
\ \ \ \ \ {\isachardoublequoteopen}{\isacharparenleft}fin{\isacharunderscore}inf{\isacharunderscore}append\ {\isacharbrackleft}init{\isacharunderscore}state{\isacharbrackright}\ st{\isacharparenright}\ {\isacharparenleft}Suc\ {\isacharparenleft}Suc\ t{\isacharparenright}{\isacharparenright}\ {\isacharequal}\ call{\isachardoublequoteclose}\isanewline
\ \ \ \ \ \ \isacommand{by}\isamarkupfalse%
\ {\isacharparenleft}simp\ add{\isacharcolon}\ correct{\isacharunderscore}fin{\isacharunderscore}inf{\isacharunderscore}append{\isadigit{2}}{\isacharparenright}\isanewline
\ \ \ \isacommand{from}\isamarkupfalse%
\ h{\isadigit{1}}\ \isakeyword{and}\ h{\isadigit{2}}\ \isacommand{have}\isamarkupfalse%
\ sg{\isadigit{2}}{\isacharcolon}{\isachardoublequoteopen}lose\ {\isacharparenleft}Suc\ {\isacharparenleft}Suc\ t{\isacharparenright}{\isacharparenright}\ {\isacharequal}\ {\isacharbrackleft}True{\isacharbrackright}\ {\isasymor}\ lose\ {\isacharparenleft}Suc\ {\isacharparenleft}Suc\ t{\isacharparenright}{\isacharparenright}\ {\isacharequal}\ {\isacharbrackleft}False{\isacharbrackright}{\isachardoublequoteclose}\isanewline
\ \ \ \ \isacommand{by}\isamarkupfalse%
\ {\isacharparenleft}simp\ add{\isacharcolon}\ ts{\isacharunderscore}bool{\isacharunderscore}True{\isacharunderscore}False{\isacharparenright}\ \isanewline
\ \ \ \isacommand{show}\isamarkupfalse%
\ {\isacharquery}thesis\isanewline
\ \ \ \isacommand{proof}\isamarkupfalse%
\ {\isacharparenleft}cases\ {\isachardoublequoteopen}lose\ {\isacharparenleft}Suc\ {\isacharparenleft}Suc\ t{\isacharparenright}{\isacharparenright}\ {\isacharequal}\ {\isacharbrackleft}False{\isacharbrackright}{\isachardoublequoteclose}{\isacharparenright}\isanewline
\ \ \ \ \ \isacommand{assume}\isamarkupfalse%
\ a{\isadigit{1}}{\isacharcolon}{\isachardoublequoteopen}lose\ {\isacharparenleft}Suc\ {\isacharparenleft}Suc\ t{\isacharparenright}{\isacharparenright}\ {\isacharequal}\ {\isacharbrackleft}False{\isacharbrackright}{\isachardoublequoteclose}\isanewline
\ \ \ \ \ \isacommand{from}\isamarkupfalse%
\ h{\isadigit{1}}\ \isakeyword{and}\ a{\isadigit{1}}\ \isakeyword{and}\ sg{\isadigit{1}}\ \isacommand{show}\isamarkupfalse%
\ {\isacharquery}thesis\ \ \isanewline
\ \ \ \ \ \ \ \isacommand{by}\isamarkupfalse%
\ {\isacharparenleft}simp\ add{\isacharcolon}\ tiTable{\isacharunderscore}SampleT{\isacharunderscore}def{\isacharparenright}\isanewline
\ \ \ \isacommand{next}\isamarkupfalse%
\isanewline
\ \ \ \ \ \isacommand{assume}\isamarkupfalse%
\ a{\isadigit{2}}{\isacharcolon}{\isachardoublequoteopen}lose\ {\isacharparenleft}Suc\ {\isacharparenleft}Suc\ t{\isacharparenright}{\isacharparenright}\ {\isasymnoteq}\ {\isacharbrackleft}False{\isacharbrackright}{\isachardoublequoteclose}\isanewline
\ \ \ \ \ \isacommand{from}\isamarkupfalse%
\ h{\isadigit{3}}\ \isacommand{have}\isamarkupfalse%
\ sg{\isadigit{3}}{\isacharcolon}{\isachardoublequoteopen}ack\ {\isacharparenleft}Suc\ {\isacharparenleft}Suc\ t{\isacharparenright}{\isacharparenright}\ {\isacharequal}\ {\isacharbrackleft}connection{\isacharunderscore}ok{\isacharbrackright}{\isachardoublequoteclose}\ \isacommand{by}\isamarkupfalse%
\ auto\isanewline
\ \ \ \ \ \isacommand{from}\isamarkupfalse%
\ h{\isadigit{1}}\ \isakeyword{and}\ a{\isadigit{2}}\ \isakeyword{and}\ sg{\isadigit{1}}\ \isakeyword{and}\ sg{\isadigit{2}}\ \isakeyword{and}\ sg{\isadigit{3}}\ \isacommand{show}\isamarkupfalse%
\ {\isacharquery}thesis\isanewline
\ \ \ \ \ \ \ \isacommand{by}\isamarkupfalse%
\ {\isacharparenleft}simp\ add{\isacharcolon}\ tiTable{\isacharunderscore}SampleT{\isacharunderscore}def{\isacharparenright}\ \ \ \isanewline
\ \ \ \isacommand{qed}\isamarkupfalse%
\isanewline
\isacommand{qed}\isamarkupfalse%
\endisatagproof
{\isafoldproof}%
\isadelimproof
\isanewline
\endisadelimproof
\isanewline
\isanewline
\isacommand{lemma}\isamarkupfalse%
\ tiTable{\isacharunderscore}i{\isadigit{1}}{\isacharunderscore}{\isadigit{4}}b{\isacharcolon}\isanewline
\ \ \isakeyword{assumes}\ h{\isadigit{1}}{\isacharcolon}{\isachardoublequoteopen}tiTable{\isacharunderscore}SampleT\ req\ x\ stop\ lose\ \isanewline
\ \ \ \ \ \ \ \ \ \ \ \ \ \ \ \ \ \ {\isacharparenleft}fin{\isacharunderscore}inf{\isacharunderscore}append\ {\isacharbrackleft}init{\isacharunderscore}state{\isacharbrackright}\ st{\isacharparenright}\ \isanewline
\ \ \ \ \ \ \ \ \ \ \ \ \ \ \ \ \ \ \ b\ ack\ i{\isadigit{1}}\ vc\ st{\isachardoublequoteclose}\isanewline
\ \ \ \ \ \ \isakeyword{and}\ h{\isadigit{2}}{\isacharcolon}{\isachardoublequoteopen}ts\ lose{\isachardoublequoteclose}\isanewline
\ \ \ \ \ \ \isakeyword{and}\ h{\isadigit{3}}{\isacharcolon}{\isachardoublequoteopen}msg\ {\isacharparenleft}Suc\ {\isadigit{0}}{\isacharparenright}\ x{\isachardoublequoteclose}\isanewline
\ \ \ \ \ \ \isakeyword{and}\ h{\isadigit{4}}{\isacharcolon}{\isachardoublequoteopen}msg\ {\isacharparenleft}Suc\ {\isadigit{0}}{\isacharparenright}\ stop{\isachardoublequoteclose}\ \isanewline
\ \ \ \ \ \ \isakeyword{and}\ h{\isadigit{5}}{\isacharcolon}{\isachardoublequoteopen}{\isasymforall}\ t{\isadigit{1}}\ {\isasymle}\ t{\isachardot}\ req\ t{\isadigit{1}}\ {\isacharequal}\ {\isacharbrackleft}{\isacharbrackright}{\isachardoublequoteclose}\isanewline
\ \ \ \ \ \ \isakeyword{and}\ h{\isadigit{6}}{\isacharcolon}{\isachardoublequoteopen}req\ {\isacharparenleft}Suc\ t{\isacharparenright}\ {\isacharequal}\ {\isacharbrackleft}init{\isacharbrackright}{\isachardoublequoteclose}\isanewline
\ \ \ \ \ \ \isakeyword{and}\ h{\isadigit{7}}{\isacharcolon}{\isachardoublequoteopen}{\isasymforall}m\ {\isacharless}\ k\ {\isacharplus}\ {\isadigit{3}}{\isachardot}\ req\ {\isacharparenleft}t\ {\isacharplus}\ m{\isacharparenright}\ {\isasymnoteq}\ {\isacharbrackleft}send{\isacharbrackright}{\isachardoublequoteclose}\isanewline
\ \ \ \ \ \ \isakeyword{and}\ h{\isadigit{7}}{\isacharcolon}{\isachardoublequoteopen}{\isasymforall}m\ {\isasymle}\ k{\isachardot}\ ack\ {\isacharparenleft}Suc\ {\isacharparenleft}Suc\ {\isacharparenleft}t\ {\isacharplus}\ m{\isacharparenright}{\isacharparenright}{\isacharparenright}\ {\isacharequal}\ {\isacharbrackleft}connection{\isacharunderscore}ok{\isacharbrackright}{\isachardoublequoteclose}\isanewline
\ \ \ \ \ \ \isakeyword{and}\ h{\isadigit{8}}{\isacharcolon}{\isachardoublequoteopen}{\isasymforall}j\ {\isasymle}\ k\ {\isacharplus}\ {\isadigit{3}}{\isachardot}\ lose\ {\isacharparenleft}t\ {\isacharplus}\ j{\isacharparenright}\ {\isacharequal}\ {\isacharbrackleft}False{\isacharbrackright}{\isachardoublequoteclose}\isanewline
\ \ \ \ \ \ \isakeyword{and}\ h{\isadigit{9}}{\isacharcolon}{\isachardoublequoteopen}t{\isadigit{2}}\ {\isacharless}\ {\isacharparenleft}t\ {\isacharplus}\ {\isadigit{3}}\ {\isacharplus}\ k{\isacharparenright}{\isachardoublequoteclose}\isanewline
\ \ \isakeyword{shows}\ {\isachardoublequoteopen}i{\isadigit{1}}\ t{\isadigit{2}}\ {\isacharequal}\ {\isacharbrackleft}{\isacharbrackright}{\isachardoublequoteclose}\isanewline
\isadelimproof
\endisadelimproof
\isatagproof
\isacommand{proof}\isamarkupfalse%
\ {\isacharparenleft}cases\ {\isachardoublequoteopen}t{\isadigit{2}}\ {\isasymle}\ t{\isachardoublequoteclose}{\isacharparenright}\isanewline
\ \ \isacommand{assume}\isamarkupfalse%
\ a{\isadigit{1}}{\isacharcolon}{\isachardoublequoteopen}t{\isadigit{2}}\ {\isasymle}\ t{\isachardoublequoteclose}\isanewline
\ \ \isacommand{from}\isamarkupfalse%
\ assms\ \isakeyword{and}\ a{\isadigit{1}}\ \isacommand{show}\isamarkupfalse%
\ {\isacharquery}thesis\ \isacommand{by}\isamarkupfalse%
\ {\isacharparenleft}simp\ add{\isacharcolon}\ tiTable{\isacharunderscore}i{\isadigit{1}}{\isacharunderscore}{\isadigit{3}}{\isacharparenright}\isanewline
\isacommand{next}\isamarkupfalse%
\ \isanewline
\ \ \isacommand{assume}\isamarkupfalse%
\ a{\isadigit{2}}{\isacharcolon}{\isachardoublequoteopen}{\isasymnot}\ t{\isadigit{2}}\ {\isasymle}\ t{\isachardoublequoteclose}\isanewline
\ \ \isacommand{from}\isamarkupfalse%
\ assms\ \isacommand{have}\isamarkupfalse%
\ sg{\isadigit{1}}{\isacharcolon}{\isachardoublequoteopen}ack\ t\ {\isacharequal}\ {\isacharbrackleft}init{\isacharunderscore}state{\isacharbrackright}{\isachardoublequoteclose}\ \isacommand{by}\isamarkupfalse%
\ {\isacharparenleft}simp\ add{\isacharcolon}\ tiTable{\isacharunderscore}ack{\isacharunderscore}init{\isacharparenright}\isanewline
\ \ \isacommand{from}\isamarkupfalse%
\ assms\ \isacommand{have}\isamarkupfalse%
\ sg{\isadigit{2}}{\isacharcolon}{\isachardoublequoteopen}st\ t\ {\isacharequal}\ \ hd\ {\isacharparenleft}ack\ t{\isacharparenright}{\isachardoublequoteclose}\ \isacommand{by}\isamarkupfalse%
\ {\isacharparenleft}simp\ add{\isacharcolon}\ tiTable{\isacharunderscore}ack{\isacharunderscore}st{\isacharunderscore}hd{\isacharparenright}\ \ \isanewline
\ \ \isacommand{from}\isamarkupfalse%
\ sg{\isadigit{1}}\ \isakeyword{and}\ sg{\isadigit{2}}\ \isacommand{have}\isamarkupfalse%
\ sg{\isadigit{3}}{\isacharcolon}\isanewline
\ \ \ {\isachardoublequoteopen}{\isacharparenleft}fin{\isacharunderscore}inf{\isacharunderscore}append\ {\isacharbrackleft}init{\isacharunderscore}state{\isacharbrackright}\ st{\isacharparenright}\ {\isacharparenleft}Suc\ t{\isacharparenright}\ {\isacharequal}\ init{\isacharunderscore}state{\isachardoublequoteclose}\isanewline
\ \ \ \ \isacommand{by}\isamarkupfalse%
\ {\isacharparenleft}simp\ add{\isacharcolon}\ correct{\isacharunderscore}fin{\isacharunderscore}inf{\isacharunderscore}append{\isadigit{2}}{\isacharparenright}\isanewline
\ \ \isacommand{from}\isamarkupfalse%
\ assms\ \isakeyword{and}\ sg{\isadigit{3}}\ \isacommand{have}\isamarkupfalse%
\ sg{\isadigit{4}}{\isacharcolon}{\isachardoublequoteopen}st\ {\isacharparenleft}Suc\ t{\isacharparenright}\ {\isacharequal}\ call{\isachardoublequoteclose}\isanewline
\ \ \ \ \isacommand{by}\isamarkupfalse%
\ {\isacharparenleft}simp\ add{\isacharcolon}\ tiTable{\isacharunderscore}SampleT{\isacharunderscore}def{\isacharparenright}\isanewline
\ \ \isacommand{show}\isamarkupfalse%
\ {\isacharquery}thesis\isanewline
\ \ \isacommand{proof}\isamarkupfalse%
\ {\isacharparenleft}cases\ {\isachardoublequoteopen}t{\isadigit{2}}\ {\isacharequal}\ Suc\ t{\isachardoublequoteclose}{\isacharparenright}\isanewline
\ \ \ \ \isacommand{assume}\isamarkupfalse%
\ a{\isadigit{3}}{\isacharcolon}{\isachardoublequoteopen}t{\isadigit{2}}\ {\isacharequal}\ Suc\ t{\isachardoublequoteclose}\isanewline
\ \ \ \ \isacommand{from}\isamarkupfalse%
\ assms\ \isakeyword{and}\ sg{\isadigit{3}}\ \isakeyword{and}\ a{\isadigit{3}}\ \isacommand{show}\isamarkupfalse%
\ {\isacharquery}thesis\isanewline
\ \ \ \ \ \ \isacommand{by}\isamarkupfalse%
\ {\isacharparenleft}simp\ add{\isacharcolon}\ tiTable{\isacharunderscore}SampleT{\isacharunderscore}def{\isacharparenright}\ \ \isanewline
\ \ \isacommand{next}\isamarkupfalse%
\isanewline
\ \ \ \ \isacommand{assume}\isamarkupfalse%
\ a{\isadigit{4}}{\isacharcolon}{\isachardoublequoteopen}t{\isadigit{2}}\ {\isasymnoteq}\ Suc\ t{\isachardoublequoteclose}\ \isanewline
\ \ \ \ \isacommand{from}\isamarkupfalse%
\ assms\ \isakeyword{and}\ sg{\isadigit{4}}\ \isakeyword{and}\ a{\isadigit{4}}\ \isakeyword{and}\ a{\isadigit{2}}\ \isacommand{have}\isamarkupfalse%
\ sg{\isadigit{7}}{\isacharcolon}{\isachardoublequoteopen}st\ {\isacharparenleft}Suc\ {\isacharparenleft}Suc\ t{\isacharparenright}{\isacharparenright}\ {\isacharequal}\ connection{\isacharunderscore}ok{\isachardoublequoteclose}\isanewline
\ \ \ \ \ \ \isacommand{by}\isamarkupfalse%
\ {\isacharparenleft}simp\ add{\isacharcolon}\ tiTable{\isacharunderscore}st{\isacharunderscore}call{\isacharunderscore}ok{\isacharparenright}\isanewline
\ \ \ \ \isacommand{from}\isamarkupfalse%
\ assms\ \isacommand{have}\isamarkupfalse%
\ sg{\isadigit{8}}{\isacharcolon}{\isachardoublequoteopen}ack\ {\isacharparenleft}Suc\ {\isacharparenleft}Suc\ t{\isacharparenright}{\isacharparenright}\ {\isacharequal}\ {\isacharbrackleft}st\ {\isacharparenleft}Suc\ {\isacharparenleft}Suc\ t{\isacharparenright}{\isacharparenright}{\isacharbrackright}{\isachardoublequoteclose}\isanewline
\ \ \ \ \ \ \isacommand{by}\isamarkupfalse%
\ {\isacharparenleft}simp\ add{\isacharcolon}\ tiTable{\isacharunderscore}ack{\isacharunderscore}st{\isacharparenright}\isanewline
\ \ \ \ \isacommand{show}\isamarkupfalse%
\ {\isacharquery}thesis\isanewline
\ \ \ \ \isacommand{proof}\isamarkupfalse%
\ {\isacharparenleft}cases\ {\isachardoublequoteopen}t{\isadigit{2}}\ {\isacharequal}\ \ Suc\ {\isacharparenleft}Suc\ t{\isacharparenright}{\isachardoublequoteclose}{\isacharparenright}\isanewline
\ \ \ \ \ \ \isacommand{assume}\isamarkupfalse%
\ a{\isadigit{5}}{\isacharcolon}{\isachardoublequoteopen}t{\isadigit{2}}\ {\isacharequal}\ \ Suc\ {\isacharparenleft}Suc\ t{\isacharparenright}{\isachardoublequoteclose}\isanewline
\ \ \ \ \ \ \isacommand{from}\isamarkupfalse%
\ h{\isadigit{7}}\ \isakeyword{and}\ h{\isadigit{9}}\ \isakeyword{and}\ a{\isadigit{5}}\ \isacommand{have}\isamarkupfalse%
\ sg{\isadigit{9}}{\isacharcolon}{\isachardoublequoteopen}ack\ t{\isadigit{2}}\ {\isacharequal}\ {\isacharbrackleft}connection{\isacharunderscore}ok{\isacharbrackright}{\isachardoublequoteclose}\ \isacommand{by}\isamarkupfalse%
\ auto\isanewline
\ \ \ \ \ \ \isacommand{from}\isamarkupfalse%
\ assms\ \isakeyword{and}\ sg{\isadigit{9}}\ \isacommand{show}\isamarkupfalse%
\ {\isacharquery}thesis\ \isacommand{by}\isamarkupfalse%
\ {\isacharparenleft}simp\ add{\isacharcolon}\ \ tiTable{\isacharunderscore}i{\isadigit{1}}{\isacharunderscore}{\isadigit{1}}{\isacharparenright}\isanewline
\ \ \ \ \isacommand{next}\isamarkupfalse%
\ \isanewline
\ \ \ \ \ \ \isacommand{assume}\isamarkupfalse%
\ a{\isadigit{6}}{\isacharcolon}{\isachardoublequoteopen}t{\isadigit{2}}\ {\isasymnoteq}\ Suc\ {\isacharparenleft}Suc\ t{\isacharparenright}{\isachardoublequoteclose}\isanewline
\ \ \ \ \ \ \isacommand{from}\isamarkupfalse%
\ a{\isadigit{6}}\ \isakeyword{and}\ a{\isadigit{4}}\ \isakeyword{and}\ a{\isadigit{2}}\ \isacommand{have}\isamarkupfalse%
\ sg{\isadigit{1}}{\isadigit{0}}{\isacharcolon}{\isachardoublequoteopen}Suc\ {\isacharparenleft}Suc\ t{\isacharparenright}\ {\isacharless}\ t{\isadigit{2}}{\isachardoublequoteclose}\ \isacommand{by}\isamarkupfalse%
\ arith\isanewline
\ \ \ \ \ \ \isacommand{from}\isamarkupfalse%
\ h{\isadigit{7}}\ \isakeyword{and}\ h{\isadigit{9}}\ \isakeyword{and}\ sg{\isadigit{1}}{\isadigit{0}}\ \isacommand{have}\isamarkupfalse%
\ sg{\isadigit{1}}{\isadigit{1}}{\isacharcolon}{\isachardoublequoteopen}ack\ t{\isadigit{2}}\ {\isacharequal}\ {\isacharbrackleft}connection{\isacharunderscore}ok{\isacharbrackright}{\isachardoublequoteclose}\ \isanewline
\ \ \ \ \ \ \ \ \isacommand{by}\isamarkupfalse%
\ {\isacharparenleft}simp\ add{\isacharcolon}\ aux{\isacharunderscore}ack{\isacharunderscore}t{\isadigit{2}}{\isacharparenright}\isanewline
\ \ \ \ \ \ \isacommand{from}\isamarkupfalse%
\ assms\ \isakeyword{and}\ a{\isadigit{6}}\ \isakeyword{and}\ sg{\isadigit{7}}\ \isakeyword{and}\ sg{\isadigit{8}}\ \isakeyword{and}\ sg{\isadigit{1}}{\isadigit{1}}\ \isacommand{show}\isamarkupfalse%
\ {\isacharquery}thesis\ \isanewline
\ \ \ \ \ \ \ \ \isacommand{by}\isamarkupfalse%
\ {\isacharparenleft}simp\ add{\isacharcolon}\ \ tiTable{\isacharunderscore}i{\isadigit{1}}{\isacharunderscore}{\isadigit{1}}{\isacharparenright}\isanewline
\ \ \ \ \isacommand{qed}\isamarkupfalse%
\isanewline
\ \ \isacommand{qed}\isamarkupfalse%
\isanewline
\isacommand{qed}\isamarkupfalse%
\endisatagproof
{\isafoldproof}%
\isadelimproof
\isanewline
\endisadelimproof
\ \isanewline
\isanewline
\isacommand{lemma}\isamarkupfalse%
\ tiTable{\isacharunderscore}i{\isadigit{1}}{\isacharunderscore}{\isadigit{4}}{\isacharcolon}\isanewline
\ \ \isakeyword{assumes}\ h{\isadigit{1}}{\isacharcolon}{\isachardoublequoteopen}tiTable{\isacharunderscore}SampleT\ req\ a{\isadigit{1}}\ stop\ lose\ \isanewline
\ \ \ \ \ \ \ \ \ \ \ \ \ \ \ \ \ \ {\isacharparenleft}fin{\isacharunderscore}inf{\isacharunderscore}append\ {\isacharbrackleft}init{\isacharunderscore}state{\isacharbrackright}\ st{\isacharparenright}\ \isanewline
\ \ \ \ \ \ \ \ \ \ \ \ \ \ \ \ \ \ \ b\ ack\ i{\isadigit{1}}\ vc\ st{\isachardoublequoteclose}\isanewline
\ \ \ \ \ \ \isakeyword{and}\ h{\isadigit{2}}{\isacharcolon}{\isachardoublequoteopen}ts\ lose{\isachardoublequoteclose}\isanewline
\ \ \ \ \ \ \isakeyword{and}\ h{\isadigit{3}}{\isacharcolon}{\isachardoublequoteopen}msg\ {\isacharparenleft}Suc\ {\isadigit{0}}{\isacharparenright}\ a{\isadigit{1}}{\isachardoublequoteclose}\isanewline
\ \ \ \ \ \ \isakeyword{and}\ h{\isadigit{4}}{\isacharcolon}{\isachardoublequoteopen}msg\ {\isacharparenleft}Suc\ {\isadigit{0}}{\isacharparenright}\ stop{\isachardoublequoteclose}\ \isanewline
\ \ \ \ \ \ \isakeyword{and}\ h{\isadigit{5}}{\isacharcolon}{\isachardoublequoteopen}{\isasymforall}\ t{\isadigit{1}}\ {\isasymle}\ t{\isachardot}\ req\ t{\isadigit{1}}\ {\isacharequal}\ {\isacharbrackleft}{\isacharbrackright}{\isachardoublequoteclose}\isanewline
\ \ \ \ \ \ \isakeyword{and}\ h{\isadigit{6}}{\isacharcolon}{\isachardoublequoteopen}req\ {\isacharparenleft}Suc\ t{\isacharparenright}\ {\isacharequal}\ {\isacharbrackleft}init{\isacharbrackright}{\isachardoublequoteclose}\isanewline
\ \ \ \ \ \ \isakeyword{and}\ h{\isadigit{7}}{\isacharcolon}{\isachardoublequoteopen}{\isasymforall}m\ {\isacharless}\ k\ {\isacharplus}\ {\isadigit{3}}{\isachardot}\ req\ {\isacharparenleft}t\ {\isacharplus}\ m{\isacharparenright}\ {\isasymnoteq}\ {\isacharbrackleft}send{\isacharbrackright}{\isachardoublequoteclose}\isanewline
\ \ \ \ \ \ \isakeyword{and}\ h{\isadigit{7}}{\isacharcolon}{\isachardoublequoteopen}{\isasymforall}m\ {\isasymle}\ k{\isachardot}\ ack\ {\isacharparenleft}Suc\ {\isacharparenleft}Suc\ {\isacharparenleft}t\ {\isacharplus}\ m{\isacharparenright}{\isacharparenright}{\isacharparenright}\ {\isacharequal}\ {\isacharbrackleft}connection{\isacharunderscore}ok{\isacharbrackright}{\isachardoublequoteclose}\isanewline
\ \ \ \ \ \ \isakeyword{and}\ h{\isadigit{8}}{\isacharcolon}{\isachardoublequoteopen}{\isasymforall}j\ {\isasymle}\ k\ {\isacharplus}\ {\isadigit{3}}{\isachardot}\ lose\ {\isacharparenleft}t\ {\isacharplus}\ j{\isacharparenright}\ {\isacharequal}\ {\isacharbrackleft}False{\isacharbrackright}{\isachardoublequoteclose}\isanewline
\ \ \isakeyword{shows}\ {\isachardoublequoteopen}{\isasymforall}\ t{\isadigit{2}}\ {\isacharless}\ {\isacharparenleft}t\ {\isacharplus}\ {\isadigit{3}}\ {\isacharplus}\ k{\isacharparenright}{\isachardot}\ i{\isadigit{1}}\ t{\isadigit{2}}\ {\isacharequal}\ {\isacharbrackleft}{\isacharbrackright}{\isachardoublequoteclose}\isanewline
\isadelimproof
\endisadelimproof
\isatagproof
\isacommand{using}\isamarkupfalse%
\ assms\ \isacommand{by}\isamarkupfalse%
\ {\isacharparenleft}simp\ add{\isacharcolon}\ tiTable{\isacharunderscore}i{\isadigit{1}}{\isacharunderscore}{\isadigit{4}}b{\isacharparenright}%
\endisatagproof
{\isafoldproof}%
\isadelimproof
\isanewline
\endisadelimproof
\isanewline
\isanewline
\isacommand{lemma}\isamarkupfalse%
\ tiTable{\isacharunderscore}ack{\isacharunderscore}ok{\isacharcolon}\isanewline
\ \ \isakeyword{assumes}\ h{\isadigit{1}}{\isacharcolon}{\isachardoublequoteopen}{\isasymforall}j{\isasymle}\ d\ {\isacharplus}\ {\isadigit{2}}{\isachardot}\ lose\ {\isacharparenleft}t\ {\isacharplus}\ j{\isacharparenright}\ {\isacharequal}\ {\isacharbrackleft}False{\isacharbrackright}{\isachardoublequoteclose}\isanewline
\ \ \ \ \ \ \isakeyword{and}\ h{\isadigit{2}}{\isacharcolon}{\isachardoublequoteopen}ts\ lose{\isachardoublequoteclose}\isanewline
\ \ \ \ \ \ \isakeyword{and}\ h{\isadigit{4}}{\isacharcolon}{\isachardoublequoteopen}msg\ {\isacharparenleft}Suc\ {\isadigit{0}}{\isacharparenright}\ stop{\isachardoublequoteclose}\isanewline
\ \ \ \ \ \ \isakeyword{and}\ h{\isadigit{5}}{\isacharcolon}{\isachardoublequoteopen}msg\ {\isacharparenleft}Suc\ {\isadigit{0}}{\isacharparenright}\ a{\isadigit{1}}{\isachardoublequoteclose}\isanewline
\ \ \ \ \ \ \isakeyword{and}\ h{\isadigit{6}}{\isacharcolon}{\isachardoublequoteopen}req\ {\isacharparenleft}Suc\ t{\isacharparenright}\ {\isasymnoteq}\ {\isacharbrackleft}send{\isacharbrackright}{\isachardoublequoteclose}\isanewline
\ \ \ \ \ \ \isakeyword{and}\ h{\isadigit{7}}{\isacharcolon}{\isachardoublequoteopen}ack\ t\ {\isacharequal}\ {\isacharbrackleft}connection{\isacharunderscore}ok{\isacharbrackright}{\isachardoublequoteclose}\isanewline
\ \ \ \ \ \ \isakeyword{and}\ h{\isadigit{8}}{\isacharcolon}{\isachardoublequoteopen}tiTable{\isacharunderscore}SampleT\ req\ a{\isadigit{1}}\ stop\ lose\ {\isacharparenleft}fin{\isacharunderscore}inf{\isacharunderscore}append\ {\isacharbrackleft}init{\isacharunderscore}state{\isacharbrackright}\ st{\isacharparenright}\ b\ ack\ i{\isadigit{1}}\ vc\ st{\isachardoublequoteclose}\isanewline
\ \ \isakeyword{shows}\ {\isachardoublequoteopen}ack\ {\isacharparenleft}Suc\ t{\isacharparenright}\ {\isacharequal}\ {\isacharbrackleft}connection{\isacharunderscore}ok{\isacharbrackright}{\isachardoublequoteclose}\isanewline
\isadelimproof
\endisadelimproof
\isatagproof
\isacommand{proof}\isamarkupfalse%
\ {\isacharminus}\isanewline
\ \ \isacommand{from}\isamarkupfalse%
\ h{\isadigit{8}}\ \isakeyword{and}\ h{\isadigit{2}}\ \isakeyword{and}\ h{\isadigit{5}}\ \isakeyword{and}\ h{\isadigit{4}}\ \isacommand{have}\isamarkupfalse%
\ sg{\isadigit{1}}{\isacharcolon}{\isachardoublequoteopen}st\ t\ {\isacharequal}\ \ hd\ {\isacharparenleft}ack\ t{\isacharparenright}{\isachardoublequoteclose}\isanewline
\ \ \ \ \isacommand{by}\isamarkupfalse%
\ {\isacharparenleft}simp\ add{\isacharcolon}\ tiTable{\isacharunderscore}ack{\isacharunderscore}st{\isacharunderscore}hd{\isacharparenright}\ \ \isanewline
\ \ \isacommand{from}\isamarkupfalse%
\ sg{\isadigit{1}}\ \isakeyword{and}\ h{\isadigit{7}}\ \isacommand{have}\isamarkupfalse%
\ sg{\isadigit{2}}{\isacharcolon}\isanewline
\ \ \ {\isachardoublequoteopen}{\isacharparenleft}fin{\isacharunderscore}inf{\isacharunderscore}append\ {\isacharbrackleft}init{\isacharunderscore}state{\isacharbrackright}\ st{\isacharparenright}\ {\isacharparenleft}Suc\ t{\isacharparenright}\ {\isacharequal}\ \ connection{\isacharunderscore}ok{\isachardoublequoteclose}\isanewline
\ \ \ \ \isacommand{by}\isamarkupfalse%
\ {\isacharparenleft}simp\ add{\isacharcolon}\ correct{\isacharunderscore}fin{\isacharunderscore}inf{\isacharunderscore}append{\isadigit{2}}{\isacharparenright}\isanewline
\ \ \isacommand{have}\isamarkupfalse%
\ sg{\isadigit{3}}a{\isacharcolon}{\isachardoublequoteopen}Suc\ {\isadigit{0}}\ {\isasymle}\ d\ {\isacharplus}\ {\isadigit{2}}{\isachardoublequoteclose}\ \isacommand{by}\isamarkupfalse%
\ arith\isanewline
\ \ \isacommand{from}\isamarkupfalse%
\ h{\isadigit{1}}\ \isakeyword{and}\ sg{\isadigit{3}}a\ \isacommand{have}\isamarkupfalse%
\ sg{\isadigit{3}}{\isacharcolon}{\isachardoublequoteopen}lose\ {\isacharparenleft}t\ {\isacharplus}\ Suc\ {\isadigit{0}}{\isacharparenright}\ {\isacharequal}\ {\isacharbrackleft}False{\isacharbrackright}{\isachardoublequoteclose}\ \isacommand{by}\isamarkupfalse%
\ auto\ \isanewline
\ \ \isacommand{from}\isamarkupfalse%
\ sg{\isadigit{2}}\ \isakeyword{and}\ sg{\isadigit{3}}\ \isakeyword{and}\ h{\isadigit{6}}\ \isakeyword{and}\ h{\isadigit{8}}\ \isacommand{show}\isamarkupfalse%
\ {\isacharquery}thesis\isanewline
\ \ \ \ \isacommand{by}\isamarkupfalse%
\ {\isacharparenleft}simp\ add{\isacharcolon}\ tiTable{\isacharunderscore}SampleT{\isacharunderscore}def{\isacharparenright}\ \isanewline
\isacommand{qed}\isamarkupfalse%
\endisatagproof
{\isafoldproof}%
\isadelimproof
\isanewline
\endisadelimproof
\isanewline
\isanewline
\isacommand{lemma}\isamarkupfalse%
\ Gateway{\isacharunderscore}L{\isadigit{7}}a{\isacharcolon}\isanewline
\ \ \isakeyword{assumes}\ h{\isadigit{1}}{\isacharcolon}{\isachardoublequoteopen}Gateway\ req\ dt\ a\ stop\ lose\ d\ ack\ i\ vc{\isachardoublequoteclose}\isanewline
\ \ \ \ \ \ \isakeyword{and}\ h{\isadigit{2}}{\isacharcolon}{\isachardoublequoteopen}msg\ {\isacharparenleft}Suc\ {\isadigit{0}}{\isacharparenright}\ a{\isachardoublequoteclose}\isanewline
\ \ \ \ \ \ \isakeyword{and}\ h{\isadigit{3}}{\isacharcolon}{\isachardoublequoteopen}msg\ {\isacharparenleft}Suc\ {\isadigit{0}}{\isacharparenright}\ stop{\isachardoublequoteclose}\isanewline
\ \ \ \ \ \ \isakeyword{and}\ h{\isadigit{4}}{\isacharcolon}{\isachardoublequoteopen}msg\ {\isacharparenleft}Suc\ {\isadigit{0}}{\isacharparenright}\ req{\isachardoublequoteclose}\isanewline
\ \ \ \ \ \ \isakeyword{and}\ h{\isadigit{5}}{\isacharcolon}{\isachardoublequoteopen}ts\ lose{\isachardoublequoteclose}\isanewline
\ \ \ \ \ \ \isakeyword{and}\ h{\isadigit{6}}{\isacharcolon}{\isachardoublequoteopen}{\isasymforall}j{\isasymle}\ d\ {\isacharplus}\ {\isadigit{2}}{\isachardot}\ lose\ {\isacharparenleft}t\ {\isacharplus}\ j{\isacharparenright}\ {\isacharequal}\ {\isacharbrackleft}False{\isacharbrackright}{\isachardoublequoteclose}\isanewline
\ \ \ \ \ \ \isakeyword{and}\ h{\isadigit{7}}{\isacharcolon}{\isachardoublequoteopen}req\ {\isacharparenleft}Suc\ t{\isacharparenright}\ {\isasymnoteq}\ {\isacharbrackleft}send{\isacharbrackright}{\isachardoublequoteclose}\isanewline
\ \ \ \ \ \ \isakeyword{and}\ h{\isadigit{8}}{\isacharcolon}{\isachardoublequoteopen}ack\ {\isacharparenleft}t{\isacharparenright}\ {\isacharequal}\ {\isacharbrackleft}connection{\isacharunderscore}ok{\isacharbrackright}{\isachardoublequoteclose}\isanewline
\ \ \isakeyword{shows}\ {\isachardoublequoteopen}ack\ {\isacharparenleft}Suc\ t{\isacharparenright}\ {\isacharequal}\ {\isacharbrackleft}connection{\isacharunderscore}ok{\isacharbrackright}{\isachardoublequoteclose}\isanewline
\isadelimproof
\endisadelimproof
\isatagproof
\isacommand{proof}\isamarkupfalse%
\ {\isacharminus}\isanewline
\ \ \isacommand{from}\isamarkupfalse%
\ h{\isadigit{1}}\ \isakeyword{and}\ h{\isadigit{3}}\ \isakeyword{and}\ h{\isadigit{4}}\ \isakeyword{and}\ h{\isadigit{7}}\ \isacommand{obtain}\isamarkupfalse%
\ i{\isadigit{1}}\ i{\isadigit{2}}\ a{\isadigit{1}}\ a{\isadigit{2}}\ \isakeyword{where}\ \isanewline
\ \ \ \ ah{\isadigit{1}}{\isacharcolon}{\isachardoublequoteopen}Sample\ req\ dt\ a{\isadigit{1}}\ stop\ lose\ ack\ i{\isadigit{1}}\ vc{\isachardoublequoteclose}\ \isakeyword{and}\isanewline
\ \ \ \ ah{\isadigit{2}}{\isacharcolon}{\isachardoublequoteopen}Delay\ a{\isadigit{2}}\ i{\isadigit{1}}\ d\ a{\isadigit{1}}\ i{\isadigit{2}}{\isachardoublequoteclose}\ \isakeyword{and}\isanewline
\ \ \ \ ah{\isadigit{3}}{\isacharcolon}{\isachardoublequoteopen}Loss\ lose\ a\ i{\isadigit{2}}\ a{\isadigit{2}}\ i{\isachardoublequoteclose}\isanewline
\ \ \ \ \isacommand{by}\isamarkupfalse%
\ {\isacharparenleft}simp\ add{\isacharcolon}\ Gateway{\isacharunderscore}def{\isacharcomma}\ auto{\isacharparenright}\isanewline
\ \ \isacommand{from}\isamarkupfalse%
\ ah{\isadigit{2}}\ \isakeyword{and}\ ah{\isadigit{3}}\ \isakeyword{and}\ h{\isadigit{2}}\ \isacommand{have}\isamarkupfalse%
\ sg{\isadigit{1}}{\isacharcolon}{\isachardoublequoteopen}msg\ {\isacharparenleft}Suc\ {\isadigit{0}}{\isacharparenright}\ a{\isadigit{1}}{\isachardoublequoteclose}\isanewline
\ \ \ \ \isacommand{by}\isamarkupfalse%
\ {\isacharparenleft}simp\ add{\isacharcolon}\ Loss{\isacharunderscore}Delay{\isacharunderscore}msg{\isacharunderscore}a{\isacharparenright}\ \isanewline
\ \ \isacommand{from}\isamarkupfalse%
\ ah{\isadigit{1}}\ \isakeyword{and}\ sg{\isadigit{1}}\ \isakeyword{and}\ h{\isadigit{3}}\ \isakeyword{and}\ h{\isadigit{4}}\ \isacommand{obtain}\isamarkupfalse%
\ st\ buffer\ \isakeyword{where}\isanewline
\ \ \ \ ah{\isadigit{4}}{\isacharcolon}{\isachardoublequoteopen}Sample{\isacharunderscore}L\ req\ dt\ a{\isadigit{1}}\ stop\ lose\ {\isacharparenleft}fin{\isacharunderscore}inf{\isacharunderscore}append\ {\isacharbrackleft}init{\isacharunderscore}state{\isacharbrackright}\ st{\isacharparenright}\ \isanewline
\ \ \ \ \ \ \ \ \ \ \ \ \ {\isacharparenleft}fin{\isacharunderscore}inf{\isacharunderscore}append\ {\isacharbrackleft}{\isacharbrackleft}{\isacharbrackright}{\isacharbrackright}\ buffer{\isacharparenright}\isanewline
\ \ \ \ \ \ \ \ \ \ \ \ \ ack\ i{\isadigit{1}}\ vc\ st\ buffer{\isachardoublequoteclose}\isanewline
\ \ \ \ \isacommand{by}\isamarkupfalse%
\ {\isacharparenleft}simp\ add{\isacharcolon}\ Sample{\isacharunderscore}def{\isacharcomma}\ auto{\isacharparenright}\isanewline
\ \ \isacommand{from}\isamarkupfalse%
\ ah{\isadigit{4}}\ \isacommand{have}\isamarkupfalse%
\ sg{\isadigit{2}}{\isacharcolon}\isanewline
\ \ \ \ {\isachardoublequoteopen}tiTable{\isacharunderscore}SampleT\ req\ a{\isadigit{1}}\ stop\ lose\ {\isacharparenleft}fin{\isacharunderscore}inf{\isacharunderscore}append\ {\isacharbrackleft}init{\isacharunderscore}state{\isacharbrackright}\ st{\isacharparenright}\ \isanewline
\ \ \ \ \ \ \ \ \ {\isacharparenleft}fin{\isacharunderscore}inf{\isacharunderscore}append\ {\isacharbrackleft}{\isacharbrackleft}{\isacharbrackright}{\isacharbrackright}\ buffer{\isacharparenright}\isanewline
\ \ \ \ \ \ \ \ \ ack\ i{\isadigit{1}}\ vc\ st{\isachardoublequoteclose}\isanewline
\ \ \ \ \isacommand{by}\isamarkupfalse%
\ {\isacharparenleft}simp\ add{\isacharcolon}\ Sample{\isacharunderscore}L{\isacharunderscore}def{\isacharparenright}\isanewline
\ \ \isacommand{from}\isamarkupfalse%
\ h{\isadigit{6}}\ \isakeyword{and}\ h{\isadigit{5}}\ \isakeyword{and}\ h{\isadigit{3}}\ \isakeyword{and}\ sg{\isadigit{1}}\ \isakeyword{and}\ h{\isadigit{7}}\ \isakeyword{and}\ h{\isadigit{8}}\ \isakeyword{and}\ sg{\isadigit{2}}\ \isacommand{show}\isamarkupfalse%
\ {\isacharquery}thesis\isanewline
\ \ \ \ \isacommand{by}\isamarkupfalse%
\ {\isacharparenleft}simp\ add{\isacharcolon}\ tiTable{\isacharunderscore}ack{\isacharunderscore}ok{\isacharparenright}\isanewline
\isacommand{qed}\isamarkupfalse%
\endisatagproof
{\isafoldproof}%
\isadelimproof
\isanewline
\endisadelimproof
\isanewline
\isanewline
\isacommand{lemma}\isamarkupfalse%
\ Sample{\isacharunderscore}L{\isacharunderscore}buffer{\isacharcolon}\isanewline
\ \ \isakeyword{assumes}\ h{\isadigit{1}}{\isacharcolon}\ \isanewline
\ \ \ \ {\isachardoublequoteopen}Sample{\isacharunderscore}L\ req\ dt\ a{\isadigit{1}}\ stop\ lose\ {\isacharparenleft}fin{\isacharunderscore}inf{\isacharunderscore}append\ {\isacharbrackleft}init{\isacharunderscore}state{\isacharbrackright}\ st{\isacharparenright}\isanewline
\ \ \ \ \ \ \ \ \ \ {\isacharparenleft}fin{\isacharunderscore}inf{\isacharunderscore}append\ {\isacharbrackleft}{\isacharbrackleft}{\isacharbrackright}{\isacharbrackright}\ buffer{\isacharparenright}\isanewline
\ \ \ \ \ \ \ \ \ \ \ ack\ i{\isadigit{1}}\ vc\ st\ buffer{\isachardoublequoteclose}\isanewline
\ \ \isakeyword{shows}\ {\isachardoublequoteopen}buffer\ t\ {\isacharequal}\ inf{\isacharunderscore}last{\isacharunderscore}ti\ dt\ t{\isachardoublequoteclose}\isanewline
\isadelimproof
\endisadelimproof
\isatagproof
\isacommand{proof}\isamarkupfalse%
\ {\isacharminus}\ \isanewline
\ \ \isacommand{from}\isamarkupfalse%
\ h{\isadigit{1}}\ \isacommand{have}\isamarkupfalse%
\ sg{\isadigit{1}}{\isacharcolon}\isanewline
\ \ \ {\isachardoublequoteopen}{\isasymforall}t{\isachardot}\ buffer\ t\ {\isacharequal}\ \isanewline
\ \ \ \ {\isacharparenleft}if\ dt\ t\ {\isacharequal}\ {\isacharbrackleft}{\isacharbrackright}\ then\ fin{\isacharunderscore}inf{\isacharunderscore}append\ {\isacharbrackleft}{\isacharbrackleft}{\isacharbrackright}{\isacharbrackright}\ buffer\ t\ else\ dt\ t{\isacharparenright}{\isachardoublequoteclose}\isanewline
\ \ \ \ \isacommand{by}\isamarkupfalse%
\ {\isacharparenleft}simp\ add{\isacharcolon}\ Sample{\isacharunderscore}L{\isacharunderscore}def{\isacharparenright}\ \isanewline
\ \ \isacommand{from}\isamarkupfalse%
\ sg{\isadigit{1}}\ \isacommand{show}\isamarkupfalse%
\ {\isacharquery}thesis\ \isanewline
\ \ \isacommand{proof}\isamarkupfalse%
\ {\isacharparenleft}induct\ t{\isacharparenright}\isanewline
\ \ \ \ \isacommand{case}\isamarkupfalse%
\ {\isadigit{0}}\ \isanewline
\ \ \ \ \isacommand{from}\isamarkupfalse%
\ this\ \isacommand{show}\isamarkupfalse%
\ {\isacharquery}case\isanewline
\ \ \ \ \ \ \isacommand{by}\isamarkupfalse%
\ {\isacharparenleft}simp\ add{\isacharcolon}\ fin{\isacharunderscore}inf{\isacharunderscore}append{\isacharunderscore}def{\isacharparenright}\isanewline
\ \ \isacommand{next}\isamarkupfalse%
\isanewline
\ \ \ \ \isacommand{fix}\isamarkupfalse%
\ t\isanewline
\ \ \ \ \isacommand{case}\isamarkupfalse%
\ {\isacharparenleft}Suc\ t{\isacharparenright}\ \ \isanewline
\ \ \ \ \isacommand{from}\isamarkupfalse%
\ this\ \isacommand{show}\isamarkupfalse%
\ {\isacharquery}case\isanewline
\ \ \ \ \isacommand{proof}\isamarkupfalse%
\ {\isacharparenleft}cases\ {\isachardoublequoteopen}dt\ t\ {\isacharequal}\ {\isacharbrackleft}{\isacharbrackright}{\isachardoublequoteclose}{\isacharparenright}\isanewline
\ \ \ \ \ \ \isacommand{assume}\isamarkupfalse%
\ a{\isadigit{1}}{\isacharcolon}{\isachardoublequoteopen}dt\ t\ {\isacharequal}\ {\isacharbrackleft}{\isacharbrackright}{\isachardoublequoteclose}\isanewline
\ \ \ \ \ \ \isacommand{from}\isamarkupfalse%
\ a{\isadigit{1}}\ \isakeyword{and}\ Suc\ \isacommand{show}\isamarkupfalse%
\ {\isacharquery}thesis\isanewline
\ \ \ \ \ \ \ \ \isacommand{by}\isamarkupfalse%
\ {\isacharparenleft}simp\ add{\isacharcolon}\ correct{\isacharunderscore}fin{\isacharunderscore}inf{\isacharunderscore}append{\isadigit{1}}{\isacharparenright}\isanewline
\ \ \ \ \isacommand{next}\isamarkupfalse%
\isanewline
\ \ \ \ \ \ \isacommand{assume}\isamarkupfalse%
\ a{\isadigit{2}}{\isacharcolon}{\isachardoublequoteopen}dt\ t\ {\isasymnoteq}\ {\isacharbrackleft}{\isacharbrackright}{\isachardoublequoteclose}\isanewline
\ \ \ \ \ \ \isacommand{from}\isamarkupfalse%
\ a{\isadigit{2}}\ \isakeyword{and}\ Suc\ \isacommand{show}\isamarkupfalse%
\ {\isacharquery}thesis\isanewline
\ \ \ \ \ \ \ \ \isacommand{by}\isamarkupfalse%
\ {\isacharparenleft}simp\ add{\isacharcolon}\ correct{\isacharunderscore}fin{\isacharunderscore}inf{\isacharunderscore}append{\isadigit{1}}{\isacharparenright}\isanewline
\ \ \ \ \isacommand{qed}\isamarkupfalse%
\isanewline
\ \ \isacommand{qed}\isamarkupfalse%
\isanewline
\isacommand{qed}\isamarkupfalse%
\endisatagproof
{\isafoldproof}%
\isadelimproof
\isanewline
\endisadelimproof
\ \isanewline
\ \isanewline
\isacommand{lemma}\isamarkupfalse%
\ \ tiTable{\isacharunderscore}SampleT{\isacharunderscore}i{\isadigit{1}}{\isacharunderscore}buffer{\isacharcolon}\isanewline
\ \ \isakeyword{assumes}\ h{\isadigit{1}}{\isacharcolon}{\isachardoublequoteopen}ack\ t\ {\isacharequal}\ {\isacharbrackleft}connection{\isacharunderscore}ok{\isacharbrackright}{\isachardoublequoteclose}\isanewline
\ \ \ \ \ \ \isakeyword{and}\ h{\isadigit{2}}{\isacharcolon}{\isachardoublequoteopen}req\ {\isacharparenleft}Suc\ t{\isacharparenright}\ {\isacharequal}\ {\isacharbrackleft}send{\isacharbrackright}{\isachardoublequoteclose}\ \isanewline
\ \ \ \ \ \ \isakeyword{and}\ h{\isadigit{3}}{\isacharcolon}{\isachardoublequoteopen}{\isasymforall}k{\isasymle}Suc\ d{\isachardot}\ lose\ {\isacharparenleft}t\ {\isacharplus}\ k{\isacharparenright}\ {\isacharequal}\ {\isacharbrackleft}False{\isacharbrackright}{\isachardoublequoteclose}\ \isanewline
\ \ \ \ \ \ \isakeyword{and}\ h{\isadigit{4}}{\isacharcolon}\ {\isachardoublequoteopen}buffer\ t\ {\isacharequal}\ inf{\isacharunderscore}last{\isacharunderscore}ti\ dt\ t{\isachardoublequoteclose}\isanewline
\ \ \ \ \ \isakeyword{and}\ h{\isadigit{6}}{\isacharcolon}{\isachardoublequoteopen}tiTable{\isacharunderscore}SampleT\ req\ a{\isadigit{1}}\ stop\ lose\ {\isacharparenleft}fin{\isacharunderscore}inf{\isacharunderscore}append\ {\isacharbrackleft}init{\isacharunderscore}state{\isacharbrackright}\ st{\isacharparenright}\ \isanewline
\ \ \ \ \ \ {\isacharparenleft}fin{\isacharunderscore}inf{\isacharunderscore}append\ {\isacharbrackleft}{\isacharbrackleft}{\isacharbrackright}{\isacharbrackright}\ buffer{\isacharparenright}\ ack\isanewline
\ \ \ \ \ \ i{\isadigit{1}}\ vc\ st{\isachardoublequoteclose}\isanewline
\ \ \ \ \ \isakeyword{and}\ h{\isadigit{7}}{\isacharcolon}{\isachardoublequoteopen}st\ t\ {\isacharequal}\ hd\ {\isacharparenleft}ack\ t{\isacharparenright}{\isachardoublequoteclose}\isanewline
\ \ \ \ \ \isakeyword{and}\ h{\isadigit{8}}{\isacharcolon}{\isachardoublequoteopen}fin{\isacharunderscore}inf{\isacharunderscore}append\ {\isacharbrackleft}init{\isacharunderscore}state{\isacharbrackright}\ st\ {\isacharparenleft}Suc\ t{\isacharparenright}\ {\isacharequal}\ connection{\isacharunderscore}ok{\isachardoublequoteclose}\isanewline
\ \ \isakeyword{shows}\ {\isachardoublequoteopen}i{\isadigit{1}}\ {\isacharparenleft}Suc\ t{\isacharparenright}\ {\isacharequal}\ inf{\isacharunderscore}last{\isacharunderscore}ti\ dt\ t{\isachardoublequoteclose}\isanewline
\isadelimproof
\endisadelimproof
\isatagproof
\isacommand{proof}\isamarkupfalse%
\ {\isacharminus}\ \ \isanewline
\ \ \isacommand{have}\isamarkupfalse%
\ sg{\isadigit{1}}{\isacharcolon}{\isachardoublequoteopen}Suc\ {\isadigit{0}}\ {\isasymle}Suc\ d{\isachardoublequoteclose}\ \isacommand{by}\isamarkupfalse%
\ arith\isanewline
\ \ \isacommand{from}\isamarkupfalse%
\ h{\isadigit{3}}\ \isakeyword{and}\ sg{\isadigit{1}}\ \isacommand{have}\isamarkupfalse%
\ sg{\isadigit{2}}{\isacharcolon}{\isachardoublequoteopen}lose\ {\isacharparenleft}Suc\ t{\isacharparenright}\ {\isacharequal}\ {\isacharbrackleft}False{\isacharbrackright}{\isachardoublequoteclose}\ \isacommand{by}\isamarkupfalse%
\ auto\isanewline
\ \ \isacommand{from}\isamarkupfalse%
\ h{\isadigit{6}}\ \isacommand{have}\isamarkupfalse%
\isanewline
\ \ \ {\isachardoublequoteopen}fin{\isacharunderscore}inf{\isacharunderscore}append\ {\isacharbrackleft}init{\isacharunderscore}state{\isacharbrackright}\ st\ {\isacharparenleft}Suc\ t{\isacharparenright}\ {\isacharequal}\ connection{\isacharunderscore}ok\ {\isasymand}\ \isanewline
\ \ \ \ req\ {\isacharparenleft}Suc\ t{\isacharparenright}\ {\isacharequal}\ {\isacharbrackleft}send{\isacharbrackright}\ {\isasymand}\ \isanewline
\ \ \ \ lose\ {\isacharparenleft}Suc\ t{\isacharparenright}\ {\isacharequal}\ {\isacharbrackleft}False{\isacharbrackright}\ {\isasymlongrightarrow}\isanewline
\ \ \ \ ack\ {\isacharparenleft}Suc\ t{\isacharparenright}\ {\isacharequal}\ {\isacharbrackleft}sending{\isacharunderscore}data{\isacharbrackright}\ {\isasymand}\ \isanewline
\ \ \ \ i{\isadigit{1}}\ {\isacharparenleft}Suc\ t{\isacharparenright}\ {\isacharequal}\ {\isacharparenleft}fin{\isacharunderscore}inf{\isacharunderscore}append\ {\isacharbrackleft}{\isacharbrackleft}{\isacharbrackright}{\isacharbrackright}\ buffer{\isacharparenright}\ {\isacharparenleft}Suc\ t{\isacharparenright}\ {\isasymand}\ \isanewline
\ \ \ \ vc\ {\isacharparenleft}Suc\ t{\isacharparenright}\ {\isacharequal}\ {\isacharbrackleft}{\isacharbrackright}\ {\isasymand}\ st\ {\isacharparenleft}Suc\ t{\isacharparenright}\ {\isacharequal}\ sending{\isacharunderscore}data{\isachardoublequoteclose}\isanewline
\ \ \ \ \isacommand{by}\isamarkupfalse%
\ {\isacharparenleft}simp\ add{\isacharcolon}\ tiTable{\isacharunderscore}SampleT{\isacharunderscore}def{\isacharparenright}\ \ \isanewline
\ \ \isacommand{from}\isamarkupfalse%
\ this\ \isakeyword{and}\ h{\isadigit{8}}\ \isakeyword{and}\ h{\isadigit{2}}\ \isakeyword{and}\ sg{\isadigit{2}}\ \isacommand{have}\isamarkupfalse%
\ \isanewline
\ \ \ {\isachardoublequoteopen}i{\isadigit{1}}\ {\isacharparenleft}Suc\ t{\isacharparenright}\ {\isacharequal}\ {\isacharparenleft}fin{\isacharunderscore}inf{\isacharunderscore}append\ {\isacharbrackleft}{\isacharbrackleft}{\isacharbrackright}{\isacharbrackright}\ buffer{\isacharparenright}\ {\isacharparenleft}Suc\ t{\isacharparenright}{\isachardoublequoteclose}\ \isacommand{by}\isamarkupfalse%
\ simp\isanewline
\ \ \isacommand{from}\isamarkupfalse%
\ this\ \isakeyword{and}\ h{\isadigit{4}}\ \isacommand{show}\isamarkupfalse%
\ {\isacharquery}thesis\ \isacommand{by}\isamarkupfalse%
\ {\isacharparenleft}simp\ add{\isacharcolon}\ correct{\isacharunderscore}fin{\isacharunderscore}inf{\isacharunderscore}append{\isadigit{1}}{\isacharparenright}\ \isanewline
\isacommand{qed}\isamarkupfalse%
\endisatagproof
{\isafoldproof}%
\isadelimproof
\ \ \isanewline
\endisadelimproof
\isanewline
\isanewline
\isacommand{lemma}\isamarkupfalse%
\ Sample{\isacharunderscore}L{\isacharunderscore}i{\isadigit{1}}{\isacharunderscore}buffer{\isacharcolon}\isanewline
\ \ \isakeyword{assumes}\ h{\isadigit{1}}{\isacharcolon}{\isachardoublequoteopen}msg\ {\isacharparenleft}Suc\ {\isadigit{0}}{\isacharparenright}\ req{\isachardoublequoteclose}\isanewline
\ \ \ \ \ \ \isakeyword{and}\ h{\isadigit{2}}{\isacharcolon}{\isachardoublequoteopen}msg\ {\isacharparenleft}Suc\ {\isadigit{0}}{\isacharparenright}\ a{\isachardoublequoteclose}\isanewline
\ \ \ \ \ \ \isakeyword{and}\ h{\isadigit{3}}{\isacharcolon}{\isachardoublequoteopen}msg\ {\isacharparenleft}Suc\ {\isadigit{0}}{\isacharparenright}\ stop{\isachardoublequoteclose}\isanewline
\ \ \ \ \ \ \isakeyword{and}\ h{\isadigit{4}}{\isacharcolon}{\isachardoublequoteopen}msg\ {\isacharparenleft}Suc\ {\isadigit{0}}{\isacharparenright}\ a{\isadigit{1}}{\isachardoublequoteclose}\isanewline
\ \ \ \ \ \ \isakeyword{and}\ h{\isadigit{5}}{\isacharcolon}{\isachardoublequoteopen}ts\ lose{\isachardoublequoteclose}\isanewline
\ \ \ \ \ \ \isakeyword{and}\ h{\isadigit{6}}{\isacharcolon}{\isachardoublequoteopen}ack\ t\ {\isacharequal}\ {\isacharbrackleft}connection{\isacharunderscore}ok{\isacharbrackright}{\isachardoublequoteclose}\isanewline
\ \ \ \ \ \ \isakeyword{and}\ h{\isadigit{7}}{\isacharcolon}{\isachardoublequoteopen}req\ {\isacharparenleft}Suc\ t{\isacharparenright}\ {\isacharequal}\ {\isacharbrackleft}send{\isacharbrackright}{\isachardoublequoteclose}\isanewline
\ \ \ \ \ \ \isakeyword{and}\ h{\isadigit{8}}{\isacharcolon}{\isachardoublequoteopen}{\isasymforall}k{\isasymle}Suc\ d{\isachardot}\ lose\ {\isacharparenleft}t\ {\isacharplus}\ k{\isacharparenright}\ {\isacharequal}\ {\isacharbrackleft}False{\isacharbrackright}{\isachardoublequoteclose}\isanewline
\ \ \ \ \ \ \isakeyword{and}\ h{\isadigit{9}}{\isacharcolon}{\isachardoublequoteopen}Sample{\isacharunderscore}L\ req\ dt\ a{\isadigit{1}}\ stop\ lose\ \isanewline
\ \ \ \ \ \ \ \ \ \ \ \ \ \ \ \ {\isacharparenleft}fin{\isacharunderscore}inf{\isacharunderscore}append\ {\isacharbrackleft}init{\isacharunderscore}state{\isacharbrackright}\ st{\isacharparenright}\ \isanewline
\ \ \ \ \ \ \ \ \ \ \ \ \ \ \ \ {\isacharparenleft}fin{\isacharunderscore}inf{\isacharunderscore}append\ {\isacharbrackleft}{\isacharbrackleft}{\isacharbrackright}{\isacharbrackright}\ buffer{\isacharparenright}\ ack\ i{\isadigit{1}}\ vc\ st\ buffer{\isachardoublequoteclose}\isanewline
\ \ \isakeyword{shows}\ {\isachardoublequoteopen}i{\isadigit{1}}\ {\isacharparenleft}Suc\ t{\isacharparenright}\ {\isacharequal}\ \ buffer\ t{\isachardoublequoteclose}\isanewline
\isadelimproof
\endisadelimproof
\isatagproof
\isacommand{proof}\isamarkupfalse%
\ {\isacharminus}\ \isanewline
\ \ \isacommand{from}\isamarkupfalse%
\ h{\isadigit{9}}\ \isacommand{have}\isamarkupfalse%
\ sg{\isadigit{1}}{\isacharcolon}{\isachardoublequoteopen}buffer\ t\ {\isacharequal}\ inf{\isacharunderscore}last{\isacharunderscore}ti\ dt\ t{\isachardoublequoteclose}\isanewline
\ \ \ \ \isacommand{by}\isamarkupfalse%
\ {\isacharparenleft}simp\ add{\isacharcolon}\ Sample{\isacharunderscore}L{\isacharunderscore}buffer{\isacharparenright}\isanewline
\ \ \isacommand{from}\isamarkupfalse%
\ h{\isadigit{9}}\ \isacommand{have}\isamarkupfalse%
\ sg{\isadigit{2}}{\isacharcolon}\isanewline
\ \ \ \ {\isachardoublequoteopen}{\isasymforall}t{\isachardot}\ buffer\ t\ {\isacharequal}\ {\isacharparenleft}if\ dt\ t\ {\isacharequal}\ {\isacharbrackleft}{\isacharbrackright}\ then\ fin{\isacharunderscore}inf{\isacharunderscore}append\ {\isacharbrackleft}{\isacharbrackleft}{\isacharbrackright}{\isacharbrackright}\ buffer\ t\ else\ dt\ t{\isacharparenright}{\isachardoublequoteclose}\isanewline
\ \ \ \ \isacommand{by}\isamarkupfalse%
\ {\isacharparenleft}simp\ add{\isacharcolon}\ Sample{\isacharunderscore}L{\isacharunderscore}def{\isacharparenright}\isanewline
\ \ \isacommand{from}\isamarkupfalse%
\ h{\isadigit{9}}\ \isacommand{have}\isamarkupfalse%
\ sg{\isadigit{3}}{\isacharcolon}\ \isanewline
\ \ \ \ {\isachardoublequoteopen}tiTable{\isacharunderscore}SampleT\ req\ a{\isadigit{1}}\ stop\ lose\ {\isacharparenleft}fin{\isacharunderscore}inf{\isacharunderscore}append\ {\isacharbrackleft}init{\isacharunderscore}state{\isacharbrackright}\ st{\isacharparenright}\ \isanewline
\ \ \ \ \ \ {\isacharparenleft}fin{\isacharunderscore}inf{\isacharunderscore}append\ {\isacharbrackleft}{\isacharbrackleft}{\isacharbrackright}{\isacharbrackright}\ buffer{\isacharparenright}\ ack\isanewline
\ \ \ \ \ \ i{\isadigit{1}}\ vc\ st{\isachardoublequoteclose}\ \ \ \isanewline
\ \ \ \ \isacommand{by}\isamarkupfalse%
\ {\isacharparenleft}simp\ add{\isacharcolon}\ Sample{\isacharunderscore}L{\isacharunderscore}def{\isacharparenright}\ \isanewline
\ \ \isacommand{from}\isamarkupfalse%
\ sg{\isadigit{3}}\ \isakeyword{and}\ h{\isadigit{5}}\ \isakeyword{and}\ h{\isadigit{4}}\ \isakeyword{and}\ h{\isadigit{3}}\ \isacommand{have}\isamarkupfalse%
\ sg{\isadigit{4}}{\isacharcolon}{\isachardoublequoteopen}st\ t\ {\isacharequal}\ \ hd\ {\isacharparenleft}ack\ t{\isacharparenright}{\isachardoublequoteclose}\isanewline
\ \ \ \ \isacommand{by}\isamarkupfalse%
\ {\isacharparenleft}simp\ add{\isacharcolon}\ tiTable{\isacharunderscore}ack{\isacharunderscore}st{\isacharunderscore}hd{\isacharparenright}\ \ \isanewline
\ \ \isacommand{from}\isamarkupfalse%
\ h{\isadigit{6}}\ \isakeyword{and}\ sg{\isadigit{4}}\ \isacommand{have}\isamarkupfalse%
\ sg{\isadigit{5}}{\isacharcolon}\isanewline
\ \ \ \ {\isachardoublequoteopen}{\isacharparenleft}fin{\isacharunderscore}inf{\isacharunderscore}append\ {\isacharbrackleft}init{\isacharunderscore}state{\isacharbrackright}\ st{\isacharparenright}\ {\isacharparenleft}Suc\ t{\isacharparenright}\ {\isacharequal}\ connection{\isacharunderscore}ok{\isachardoublequoteclose}\isanewline
\ \ \ \ \isacommand{by}\isamarkupfalse%
\ {\isacharparenleft}simp\ add{\isacharcolon}\ correct{\isacharunderscore}fin{\isacharunderscore}inf{\isacharunderscore}append{\isadigit{1}}{\isacharparenright}\isanewline
\ \ \isacommand{from}\isamarkupfalse%
\ h{\isadigit{6}}\ \isakeyword{and}\ h{\isadigit{7}}\ \isakeyword{and}\ h{\isadigit{8}}\ \isakeyword{and}\ sg{\isadigit{1}}\ \isakeyword{and}\ sg{\isadigit{3}}\ \isakeyword{and}\ sg{\isadigit{4}}\ \isakeyword{and}\ sg{\isadigit{5}}\ \isacommand{have}\isamarkupfalse%
\ sg{\isadigit{6}}{\isacharcolon}\isanewline
\ \ \ \ {\isachardoublequoteopen}i{\isadigit{1}}\ {\isacharparenleft}Suc\ t{\isacharparenright}\ {\isacharequal}\ inf{\isacharunderscore}last{\isacharunderscore}ti\ dt\ t{\isachardoublequoteclose}\isanewline
\ \ \ \ \ \isacommand{by}\isamarkupfalse%
\ {\isacharparenleft}simp\ add{\isacharcolon}\ tiTable{\isacharunderscore}SampleT{\isacharunderscore}i{\isadigit{1}}{\isacharunderscore}buffer{\isacharparenright}\isanewline
\ \ \isacommand{from}\isamarkupfalse%
\ this\ \isakeyword{and}\ sg{\isadigit{1}}\ \isacommand{show}\isamarkupfalse%
\ {\isacharquery}thesis\ \isacommand{by}\isamarkupfalse%
\ simp\isanewline
\isacommand{qed}\isamarkupfalse%
\endisatagproof
{\isafoldproof}%
\isadelimproof
\isanewline
\endisadelimproof
\isanewline
\isanewline
\isacommand{lemma}\isamarkupfalse%
\ tiTable{\isacharunderscore}SampleT{\isacharunderscore}sending{\isacharunderscore}data{\isacharcolon}\isanewline
\ \ \isakeyword{assumes}\ h{\isadigit{1}}{\isacharcolon}\ {\isachardoublequoteopen}tiTable{\isacharunderscore}SampleT\ req\ a{\isadigit{1}}\ stop\ lose\ {\isacharparenleft}fin{\isacharunderscore}inf{\isacharunderscore}append\ {\isacharbrackleft}init{\isacharunderscore}state{\isacharbrackright}\ st{\isacharparenright}\ \isanewline
\ \ \ \ \ \ \ \ \ {\isacharparenleft}fin{\isacharunderscore}inf{\isacharunderscore}append\ {\isacharbrackleft}{\isacharbrackleft}{\isacharbrackright}{\isacharbrackright}\ buffer{\isacharparenright}\isanewline
\ \ \ \ \ \ \ \ \ ack\ i{\isadigit{1}}\ vc\ st{\isachardoublequoteclose}\isanewline
\ \ \ \ \ \ \isakeyword{and}\ h{\isadigit{2}}{\isacharcolon}{\isachardoublequoteopen}{\isasymforall}j{\isasymle}{\isadigit{2}}\ {\isacharasterisk}\ d{\isachardot}\ lose\ {\isacharparenleft}t\ {\isacharplus}\ j{\isacharparenright}\ {\isacharequal}\ {\isacharbrackleft}False{\isacharbrackright}{\isachardoublequoteclose}\isanewline
\ \ \ \ \ \ \isakeyword{and}\ h{\isadigit{3}}{\isacharcolon}{\isachardoublequoteopen}{\isasymforall}t{\isadigit{4}}{\isasymle}t\ {\isacharplus}\ d\ {\isacharplus}\ d{\isachardot}\ a{\isadigit{1}}\ t{\isadigit{4}}\ {\isacharequal}\ {\isacharbrackleft}{\isacharbrackright}{\isachardoublequoteclose}\isanewline
\ \ \ \ \ \ \isakeyword{and}\ h{\isadigit{4}}{\isacharcolon}{\isachardoublequoteopen}ack\ {\isacharparenleft}t\ {\isacharplus}\ x{\isacharparenright}\ {\isacharequal}\ {\isacharbrackleft}sending{\isacharunderscore}data{\isacharbrackright}{\isachardoublequoteclose}\isanewline
\ \ \ \ \ \ \isakeyword{and}\ h{\isadigit{5}}{\isacharcolon}{\isachardoublequoteopen}fin{\isacharunderscore}inf{\isacharunderscore}append\ {\isacharbrackleft}init{\isacharunderscore}state{\isacharbrackright}\ st\ {\isacharparenleft}Suc\ {\isacharparenleft}t\ {\isacharplus}\ x{\isacharparenright}{\isacharparenright}\ {\isacharequal}\ sending{\isacharunderscore}data{\isachardoublequoteclose}\isanewline
\ \ \ \ \ \ \isakeyword{and}\ h{\isadigit{6}}{\isacharcolon}{\isachardoublequoteopen}Suc\ {\isacharparenleft}t\ {\isacharplus}\ x{\isacharparenright}\ {\isasymle}\ {\isadigit{2}}\ {\isacharasterisk}\ d\ {\isacharplus}\ t{\isachardoublequoteclose}\isanewline
\ \ \isakeyword{shows}\ {\isachardoublequoteopen}ack\ {\isacharparenleft}Suc\ {\isacharparenleft}t\ {\isacharplus}\ x{\isacharparenright}{\isacharparenright}\ {\isacharequal}\ {\isacharbrackleft}sending{\isacharunderscore}data{\isacharbrackright}{\isachardoublequoteclose}\isanewline
\isadelimproof
\endisadelimproof
\isatagproof
\isacommand{proof}\isamarkupfalse%
\ {\isacharminus}\isanewline
\ \ \isacommand{from}\isamarkupfalse%
\ h{\isadigit{6}}\ \isacommand{have}\isamarkupfalse%
\ {\isachardoublequoteopen}Suc\ x\ {\isasymle}\ {\isadigit{2}}\ {\isacharasterisk}\ d{\isachardoublequoteclose}\ \isacommand{by}\isamarkupfalse%
\ arith\isanewline
\ \ \isacommand{from}\isamarkupfalse%
\ this\ \isakeyword{and}\ h{\isadigit{2}}\ \isacommand{have}\isamarkupfalse%
\ sg{\isadigit{1}}{\isacharcolon}{\isachardoublequoteopen}lose\ {\isacharparenleft}t\ {\isacharplus}\ Suc\ x{\isacharparenright}\ {\isacharequal}\ {\isacharbrackleft}False{\isacharbrackright}{\isachardoublequoteclose}\ \isacommand{by}\isamarkupfalse%
\ auto\isanewline
\ \ \isacommand{from}\isamarkupfalse%
\ h{\isadigit{6}}\ \isacommand{have}\isamarkupfalse%
\ {\isachardoublequoteopen}Suc\ {\isacharparenleft}t\ {\isacharplus}\ x{\isacharparenright}\ {\isasymle}t\ {\isacharplus}\ d\ {\isacharplus}\ d{\isachardoublequoteclose}\ \isacommand{by}\isamarkupfalse%
\ arith\isanewline
\ \ \isacommand{from}\isamarkupfalse%
\ this\ \isakeyword{and}\ h{\isadigit{3}}\ \isacommand{have}\isamarkupfalse%
\ sg{\isadigit{2}}{\isacharcolon}{\isachardoublequoteopen}a{\isadigit{1}}\ {\isacharparenleft}Suc\ {\isacharparenleft}t\ {\isacharplus}\ x{\isacharparenright}{\isacharparenright}\ {\isacharequal}\ {\isacharbrackleft}{\isacharbrackright}{\isachardoublequoteclose}\ \isacommand{by}\isamarkupfalse%
\ auto\isanewline
\ \ \isacommand{from}\isamarkupfalse%
\ h{\isadigit{1}}\ \isakeyword{and}\ sg{\isadigit{1}}\ \isakeyword{and}\ sg{\isadigit{2}}\ \isakeyword{and}\ h{\isadigit{5}}\ \isacommand{show}\isamarkupfalse%
\ {\isacharquery}thesis\ \isanewline
\ \ \ \ \isacommand{by}\isamarkupfalse%
\ {\isacharparenleft}simp\ add{\isacharcolon}\ tiTable{\isacharunderscore}SampleT{\isacharunderscore}def{\isacharparenright}\ \isanewline
\isacommand{qed}\isamarkupfalse%
\endisatagproof
{\isafoldproof}%
\isadelimproof
\isanewline
\endisadelimproof
\isanewline
\isanewline
\isacommand{lemma}\isamarkupfalse%
\ Sample{\isacharunderscore}sending{\isacharunderscore}data{\isacharcolon}\isanewline
\ \ \isakeyword{assumes}\ h{\isadigit{1}}{\isacharcolon}{\isachardoublequoteopen}msg\ {\isacharparenleft}Suc\ {\isadigit{0}}{\isacharparenright}\ stop{\isachardoublequoteclose}\isanewline
\ \ \ \ \ \ \isakeyword{and}\ h{\isadigit{2}}{\isacharcolon}{\isachardoublequoteopen}ts\ lose{\isachardoublequoteclose}\isanewline
\ \ \ \ \ \ \isakeyword{and}\ h{\isadigit{3}}{\isacharcolon}{\isachardoublequoteopen}msg\ {\isacharparenleft}Suc\ {\isadigit{0}}{\isacharparenright}\ req{\isachardoublequoteclose}\isanewline
\ \ \ \ \ \ \isakeyword{and}\ h{\isadigit{4}}{\isacharcolon}{\isachardoublequoteopen}msg\ {\isacharparenleft}Suc\ {\isadigit{0}}{\isacharparenright}\ a{\isadigit{1}}{\isachardoublequoteclose}\isanewline
\ \ \ \ \ \ \isakeyword{and}\ h{\isadigit{5}}{\isacharcolon}{\isachardoublequoteopen}{\isasymforall}j{\isasymle}{\isadigit{2}}\ {\isacharasterisk}\ d{\isachardot}\ lose\ {\isacharparenleft}t\ {\isacharplus}\ j{\isacharparenright}\ {\isacharequal}\ {\isacharbrackleft}False{\isacharbrackright}{\isachardoublequoteclose}\isanewline
\ \ \ \ \ \ \isakeyword{and}\ h{\isadigit{6}}{\isacharcolon}{\isachardoublequoteopen}ack\ t\ {\isacharequal}\ {\isacharbrackleft}sending{\isacharunderscore}data{\isacharbrackright}{\isachardoublequoteclose}\isanewline
\ \ \ \ \ \ \isakeyword{and}\ h{\isadigit{7}}{\isacharcolon}{\isachardoublequoteopen}Sample\ req\ dt\ a{\isadigit{1}}\ stop\ lose\ ack\ i{\isadigit{1}}\ vc{\isachardoublequoteclose}\isanewline
\ \ \ \ \ \ \isakeyword{and}\ h{\isadigit{8}}{\isacharcolon}{\isachardoublequoteopen}x\ {\isasymle}\ d\ {\isacharplus}\ d{\isachardoublequoteclose}\isanewline
\ \ \ \ \ \ \isakeyword{and}\ h{\isadigit{9}}{\isacharcolon}{\isachardoublequoteopen}{\isasymforall}t{\isadigit{4}}\ {\isasymle}\ t\ {\isacharplus}\ d\ {\isacharplus}\ d{\isachardot}\ a{\isadigit{1}}\ t{\isadigit{4}}\ {\isacharequal}\ {\isacharbrackleft}{\isacharbrackright}{\isachardoublequoteclose}\isanewline
\ \isakeyword{shows}\ {\isachardoublequoteopen}ack\ {\isacharparenleft}t\ {\isacharplus}\ x{\isacharparenright}\ {\isacharequal}\ {\isacharbrackleft}sending{\isacharunderscore}data{\isacharbrackright}{\isachardoublequoteclose}\isanewline
\isadelimproof
\endisadelimproof
\isatagproof
\isacommand{using}\isamarkupfalse%
\ assms\isanewline
\isacommand{proof}\isamarkupfalse%
\ {\isacharminus}\isanewline
\ \ \isacommand{from}\isamarkupfalse%
\ h{\isadigit{1}}\ \isakeyword{and}\ h{\isadigit{3}}\ \isakeyword{and}\ h{\isadigit{4}}\ \isakeyword{and}\ h{\isadigit{7}}\ \isacommand{obtain}\isamarkupfalse%
\ st\ buffer\ \isakeyword{where}\ a{\isadigit{1}}{\isacharcolon}\ \isanewline
\ \ \ {\isachardoublequoteopen}Sample{\isacharunderscore}L\ req\ dt\ a{\isadigit{1}}\ stop\ lose\ {\isacharparenleft}fin{\isacharunderscore}inf{\isacharunderscore}append\ {\isacharbrackleft}init{\isacharunderscore}state{\isacharbrackright}\ st{\isacharparenright}\ \isanewline
\ \ \ \ \ \ \ \ \ \ \ \ \ {\isacharparenleft}fin{\isacharunderscore}inf{\isacharunderscore}append\ {\isacharbrackleft}{\isacharbrackleft}{\isacharbrackright}{\isacharbrackright}\ buffer{\isacharparenright}\ ack\isanewline
\ \ \ \ \ \ \ \ \ \ \ \ \ i{\isadigit{1}}\ vc\ st\ buffer{\isachardoublequoteclose}\isanewline
\ \ \ \ \isacommand{by}\isamarkupfalse%
\ {\isacharparenleft}simp\ add{\isacharcolon}\ Sample{\isacharunderscore}def{\isacharcomma}\ auto{\isacharparenright}\isanewline
\ \ \isacommand{from}\isamarkupfalse%
\ a{\isadigit{1}}\ \isacommand{have}\isamarkupfalse%
\ sg{\isadigit{1}}{\isacharcolon}\isanewline
\ \ \ \ {\isachardoublequoteopen}tiTable{\isacharunderscore}SampleT\ req\ a{\isadigit{1}}\ stop\ lose\ {\isacharparenleft}fin{\isacharunderscore}inf{\isacharunderscore}append\ {\isacharbrackleft}init{\isacharunderscore}state{\isacharbrackright}\ st{\isacharparenright}\ \isanewline
\ \ \ \ \ \ \ \ {\isacharparenleft}fin{\isacharunderscore}inf{\isacharunderscore}append\ {\isacharbrackleft}{\isacharbrackleft}{\isacharbrackright}{\isacharbrackright}\ buffer{\isacharparenright}\isanewline
\ \ \ \ \ \ \ \ \ ack\ i{\isadigit{1}}\ vc\ st{\isachardoublequoteclose}\ \isanewline
\ \ \ \ \ \isacommand{by}\isamarkupfalse%
\ {\isacharparenleft}simp\ add{\isacharcolon}\ Sample{\isacharunderscore}L{\isacharunderscore}def{\isacharparenright}\isanewline
\ \ \isacommand{from}\isamarkupfalse%
\ a{\isadigit{1}}\ \isacommand{have}\isamarkupfalse%
\ sg{\isadigit{2}}{\isacharcolon}\isanewline
\ \ \ \ {\isachardoublequoteopen}{\isasymforall}t{\isachardot}\ buffer\ t\ {\isacharequal}\ {\isacharparenleft}if\ dt\ t\ {\isacharequal}\ {\isacharbrackleft}{\isacharbrackright}\ then\ fin{\isacharunderscore}inf{\isacharunderscore}append\ {\isacharbrackleft}{\isacharbrackleft}{\isacharbrackright}{\isacharbrackright}\ buffer\ t\ else\ dt\ t{\isacharparenright}{\isachardoublequoteclose}\isanewline
\ \ \ \ \ \isacommand{by}\isamarkupfalse%
\ {\isacharparenleft}simp\ add{\isacharcolon}\ Sample{\isacharunderscore}L{\isacharunderscore}def{\isacharparenright}\isanewline
\ \ \isacommand{from}\isamarkupfalse%
\ h{\isadigit{1}}\ \isakeyword{and}\ h{\isadigit{2}}\ \isakeyword{and}\ h{\isadigit{4}}\ \isakeyword{and}\ h{\isadigit{6}}\ \isakeyword{and}\ h{\isadigit{8}}\ \isakeyword{and}\ sg{\isadigit{1}}\ \isakeyword{and}\ sg{\isadigit{2}}\ \isacommand{show}\isamarkupfalse%
\ {\isacharquery}thesis\isanewline
\ \ \isacommand{proof}\isamarkupfalse%
\ {\isacharparenleft}induct\ {\isachardoublequoteopen}x{\isachardoublequoteclose}{\isacharparenright}\isanewline
\ \ \ \ \isacommand{case}\isamarkupfalse%
\ {\isadigit{0}}\isanewline
\ \ \ \ \isacommand{from}\isamarkupfalse%
\ this\ \isacommand{show}\isamarkupfalse%
\ {\isacharquery}case\ \isacommand{by}\isamarkupfalse%
\ simp\isanewline
\ \ \isacommand{next}\isamarkupfalse%
\isanewline
\ \ \ \ \isacommand{fix}\isamarkupfalse%
\ x\isanewline
\ \ \ \ \isacommand{case}\isamarkupfalse%
\ {\isacharparenleft}Suc\ x{\isacharparenright}\ \ \isanewline
\ \ \ \ \isacommand{from}\isamarkupfalse%
\ this\ \isacommand{have}\isamarkupfalse%
\ sg{\isadigit{3}}{\isacharcolon}{\isachardoublequoteopen}st\ {\isacharparenleft}t\ {\isacharplus}\ x{\isacharparenright}\ {\isacharequal}\ hd\ {\isacharparenleft}ack\ {\isacharparenleft}t\ {\isacharplus}\ x{\isacharparenright}{\isacharparenright}{\isachardoublequoteclose}\isanewline
\ \ \ \ \ \ \isacommand{by}\isamarkupfalse%
\ {\isacharparenleft}simp\ add{\isacharcolon}\ tiTable{\isacharunderscore}ack{\isacharunderscore}st{\isacharunderscore}hd{\isacharparenright}\ \isanewline
\ \ \ \ \isacommand{from}\isamarkupfalse%
\ Suc\ \isacommand{have}\isamarkupfalse%
\ sg{\isadigit{4}}{\isacharcolon}{\isachardoublequoteopen}x\ {\isasymle}\ d\ {\isacharplus}\ d{\isachardoublequoteclose}\ \isacommand{by}\isamarkupfalse%
\ arith\ \isanewline
\ \ \ \ \isacommand{from}\isamarkupfalse%
\ Suc\ \isakeyword{and}\ sg{\isadigit{3}}\ \isakeyword{and}\ sg{\isadigit{4}}\ \isacommand{have}\isamarkupfalse%
\ sg{\isadigit{5}}{\isacharcolon}\ \ \isanewline
\ \ \ \ \ {\isachardoublequoteopen}{\isacharparenleft}fin{\isacharunderscore}inf{\isacharunderscore}append\ {\isacharbrackleft}init{\isacharunderscore}state{\isacharbrackright}\ st{\isacharparenright}\ {\isacharparenleft}Suc\ {\isacharparenleft}t\ {\isacharplus}\ x{\isacharparenright}{\isacharparenright}\ {\isacharequal}\ sending{\isacharunderscore}data{\isachardoublequoteclose}\isanewline
\ \ \ \ \ \ \isacommand{by}\isamarkupfalse%
\ {\isacharparenleft}simp\ add{\isacharcolon}\ fin{\isacharunderscore}inf{\isacharunderscore}append{\isacharunderscore}def{\isacharparenright}\isanewline
\ \ \ \ \isacommand{from}\isamarkupfalse%
\ Suc\ \isacommand{have}\isamarkupfalse%
\ sg{\isadigit{6}}{\isacharcolon}{\isachardoublequoteopen}Suc\ {\isacharparenleft}t\ {\isacharplus}\ x{\isacharparenright}\ {\isasymle}\ {\isadigit{2}}\ {\isacharasterisk}\ d\ {\isacharplus}\ t{\isachardoublequoteclose}\ \isacommand{by}\isamarkupfalse%
\ simp\isanewline
\ \ \ \ \isacommand{from}\isamarkupfalse%
\ Suc\ \isacommand{have}\isamarkupfalse%
\ sg{\isadigit{7}}{\isacharcolon}{\isachardoublequoteopen}ack\ {\isacharparenleft}t\ {\isacharplus}\ x{\isacharparenright}\ {\isacharequal}\ {\isacharbrackleft}sending{\isacharunderscore}data{\isacharbrackright}{\isachardoublequoteclose}\ \isacommand{by}\isamarkupfalse%
\ simp\isanewline
\ \ \ \ \isacommand{from}\isamarkupfalse%
\ sg{\isadigit{1}}\ \isakeyword{and}\ h{\isadigit{5}}\ \isakeyword{and}\ h{\isadigit{9}}\ \isakeyword{and}\ sg{\isadigit{7}}\ \isakeyword{and}\ sg{\isadigit{5}}\ \isakeyword{and}\ sg{\isadigit{6}}\ \isacommand{have}\isamarkupfalse%
\ sg{\isadigit{7}}{\isacharcolon}\isanewline
\ \ \ \ \ {\isachardoublequoteopen}ack\ {\isacharparenleft}Suc\ {\isacharparenleft}t\ {\isacharplus}\ x{\isacharparenright}{\isacharparenright}\ {\isacharequal}\ {\isacharbrackleft}sending{\isacharunderscore}data{\isacharbrackright}{\isachardoublequoteclose}\isanewline
\ \ \ \ \ \ \isacommand{by}\isamarkupfalse%
\ {\isacharparenleft}simp\ add{\isacharcolon}\ tiTable{\isacharunderscore}SampleT{\isacharunderscore}sending{\isacharunderscore}data{\isacharparenright}\isanewline
\ \ \ \ \isacommand{from}\isamarkupfalse%
\ this\ \isacommand{show}\isamarkupfalse%
\ {\isacharquery}case\ \isacommand{by}\isamarkupfalse%
\ simp\isanewline
\ \ \isacommand{qed}\isamarkupfalse%
\isanewline
\isacommand{qed}\isamarkupfalse%
\endisatagproof
{\isafoldproof}%
\isadelimproof
\endisadelimproof
\isamarkupsubsection{Properties of the ServiceCenter component%
}
\isamarkuptrue%
\isacommand{lemma}\isamarkupfalse%
\ ServiceCenter{\isacharunderscore}a{\isacharunderscore}l{\isacharcolon}\isanewline
\ \ \isakeyword{assumes}\ h{\isadigit{1}}{\isacharcolon}{\isachardoublequoteopen}ServiceCenter\ i\ a{\isachardoublequoteclose}\isanewline
\ \ \isakeyword{shows}\ \ \ \ \ \ {\isachardoublequoteopen}length\ {\isacharparenleft}a\ t{\isacharparenright}\ {\isasymle}\ {\isacharparenleft}Suc\ {\isadigit{0}}{\isacharparenright}{\isachardoublequoteclose}\ \isanewline
\isadelimproof
\endisadelimproof
\isatagproof
\isacommand{proof}\isamarkupfalse%
\ {\isacharparenleft}cases\ {\isachardoublequoteopen}t{\isachardoublequoteclose}{\isacharparenright}\isanewline
\ \ \isacommand{case}\isamarkupfalse%
\ {\isadigit{0}}\ \isanewline
\ \ \isacommand{from}\isamarkupfalse%
\ this\ \isakeyword{and}\ h{\isadigit{1}}\ \isacommand{show}\isamarkupfalse%
\ {\isacharquery}thesis\ \isacommand{by}\isamarkupfalse%
\ {\isacharparenleft}simp\ add{\isacharcolon}\ ServiceCenter{\isacharunderscore}def{\isacharparenright}\isanewline
\isacommand{next}\isamarkupfalse%
\isanewline
\ \ \isacommand{fix}\isamarkupfalse%
\ m\ \isacommand{assume}\isamarkupfalse%
\ Suc{\isacharcolon}{\isachardoublequoteopen}t\ {\isacharequal}\ Suc\ m{\isachardoublequoteclose}\isanewline
\ \ \isacommand{from}\isamarkupfalse%
\ this\ \isakeyword{and}\ h{\isadigit{1}}\ \isacommand{show}\isamarkupfalse%
\ {\isacharquery}thesis\ \isacommand{by}\isamarkupfalse%
\ {\isacharparenleft}simp\ add{\isacharcolon}\ ServiceCenter{\isacharunderscore}def{\isacharparenright}\isanewline
\isacommand{qed}\isamarkupfalse%
\endisatagproof
{\isafoldproof}%
\isadelimproof
\isanewline
\endisadelimproof
\isanewline
\isacommand{lemma}\isamarkupfalse%
\ ServiceCenter{\isacharunderscore}a{\isacharunderscore}msg{\isacharcolon}\isanewline
\ \ \isakeyword{assumes}\ h{\isadigit{1}}{\isacharcolon}{\isachardoublequoteopen}ServiceCenter\ i\ a{\isachardoublequoteclose}\isanewline
\ \ \isakeyword{shows}\ \ \ \ \ \ {\isachardoublequoteopen}msg\ {\isacharparenleft}Suc\ {\isadigit{0}}{\isacharparenright}\ a{\isachardoublequoteclose}\isanewline
\isadelimproof
\endisadelimproof
\isatagproof
\isacommand{using}\isamarkupfalse%
\ assms\ \ \isacommand{by}\isamarkupfalse%
\ {\isacharparenleft}simp\ add{\isacharcolon}\ msg{\isacharunderscore}def\ ServiceCenter{\isacharunderscore}a{\isacharunderscore}l{\isacharparenright}%
\endisatagproof
{\isafoldproof}%
\isadelimproof
\isanewline
\endisadelimproof
\isanewline
\isacommand{lemma}\isamarkupfalse%
\ ServiceCenter{\isacharunderscore}L{\isadigit{1}}{\isacharcolon}\isanewline
\ \ \isakeyword{assumes}\ h{\isadigit{1}}{\isacharcolon}{\isachardoublequoteopen}{\isasymforall}\ t{\isadigit{2}}\ {\isacharless}\ x{\isachardot}\ i\ t{\isadigit{2}}\ {\isacharequal}\ {\isacharbrackleft}{\isacharbrackright}{\isachardoublequoteclose}\isanewline
\ \ \ \ \ \ \isakeyword{and}\ h{\isadigit{2}}{\isacharcolon}{\isachardoublequoteopen}ServiceCenter\ i\ a{\isachardoublequoteclose}\isanewline
\ \ \ \ \ \ \isakeyword{and}\ h{\isadigit{3}}{\isacharcolon}{\isachardoublequoteopen}t\ {\isasymle}\ x{\isachardoublequoteclose}\isanewline
\ \ \isakeyword{shows}\ {\isachardoublequoteopen}a\ t\ {\isacharequal}\ {\isacharbrackleft}{\isacharbrackright}{\isachardoublequoteclose}\isanewline
\isadelimproof
\endisadelimproof
\isatagproof
\isacommand{using}\isamarkupfalse%
\ assms\isanewline
\isacommand{proof}\isamarkupfalse%
\ {\isacharparenleft}induct\ t{\isacharparenright}\isanewline
\ \ \ \isacommand{case}\isamarkupfalse%
\ {\isadigit{0}}\ \isanewline
\ \ \ \isacommand{from}\isamarkupfalse%
\ this\ \isacommand{show}\isamarkupfalse%
\ {\isacharquery}case\ \isacommand{by}\isamarkupfalse%
\ {\isacharparenleft}simp\ add{\isacharcolon}\ ServiceCenter{\isacharunderscore}def{\isacharparenright}\isanewline
\isacommand{next}\isamarkupfalse%
\isanewline
\ \ \ \isacommand{case}\isamarkupfalse%
\ {\isacharparenleft}Suc\ t{\isacharparenright}\isanewline
\ \ \ \isacommand{from}\isamarkupfalse%
\ this\ \isacommand{show}\isamarkupfalse%
\ {\isacharquery}case\ \isacommand{by}\isamarkupfalse%
\ {\isacharparenleft}simp\ add{\isacharcolon}\ ServiceCenter{\isacharunderscore}def{\isacharparenright}\isanewline
\isacommand{qed}\isamarkupfalse%
\endisatagproof
{\isafoldproof}%
\isadelimproof
\isanewline
\endisadelimproof
\isanewline
\isacommand{lemma}\isamarkupfalse%
\ ServiceCenter{\isacharunderscore}L{\isadigit{2}}{\isacharcolon}\isanewline
\ \ \isakeyword{assumes}\ h{\isadigit{1}}{\isacharcolon}{\isachardoublequoteopen}{\isasymforall}\ t{\isadigit{2}}\ {\isacharless}\ x{\isachardot}\ i\ t{\isadigit{2}}\ {\isacharequal}\ {\isacharbrackleft}{\isacharbrackright}{\isachardoublequoteclose}\isanewline
\ \ \ \ \ \ \isakeyword{and}\ h{\isadigit{2}}{\isacharcolon}{\isachardoublequoteopen}ServiceCenter\ i\ a{\isachardoublequoteclose}\isanewline
\ \ \isakeyword{shows}\ {\isachardoublequoteopen}{\isasymforall}\ t{\isadigit{3}}\ {\isasymle}\ x{\isachardot}\ a\ t{\isadigit{3}}\ {\isacharequal}\ {\isacharbrackleft}{\isacharbrackright}{\isachardoublequoteclose}\isanewline
\isadelimproof
\endisadelimproof
\isatagproof
\isacommand{using}\isamarkupfalse%
\ assms\ \isacommand{by}\isamarkupfalse%
\ {\isacharparenleft}clarify{\isacharcomma}\ simp\ add{\isacharcolon}\ ServiceCenter{\isacharunderscore}L{\isadigit{1}}{\isacharparenright}%
\endisatagproof
{\isafoldproof}%
\isadelimproof
\endisadelimproof
\isamarkupsubsection{General properties of stream values%
}
\isamarkuptrue%
\isacommand{lemma}\isamarkupfalse%
\ streamValue{\isadigit{1}}{\isacharcolon}\ \isanewline
\ \ \isakeyword{assumes}\ h{\isadigit{1}}{\isacharcolon}{\isachardoublequoteopen}{\isasymforall}j{\isasymle}\ D\ {\isacharplus}\ {\isacharparenleft}z{\isacharcolon}{\isacharcolon}nat{\isacharparenright}{\isachardot}\ str\ {\isacharparenleft}t\ {\isacharplus}\ j{\isacharparenright}\ {\isacharequal}\ x{\isachardoublequoteclose}\isanewline
\ \ \ \ \ \ \isakeyword{and}\ h{\isadigit{2}}{\isacharcolon}\ {\isachardoublequoteopen}j{\isasymle}\ D{\isachardoublequoteclose}\isanewline
\ \ \isakeyword{shows}\ \ \ \ \ \ {\isachardoublequoteopen}str\ {\isacharparenleft}t\ {\isacharplus}\ j\ {\isacharplus}\ z{\isacharparenright}\ {\isacharequal}\ x{\isachardoublequoteclose}\isanewline
\isadelimproof
\endisadelimproof
\isatagproof
\isacommand{proof}\isamarkupfalse%
\ {\isacharminus}\ \isanewline
\ \ \ \ \isacommand{from}\isamarkupfalse%
\ h{\isadigit{2}}\ \isacommand{have}\isamarkupfalse%
\ sg{\isadigit{1}}{\isacharcolon}{\isachardoublequoteopen}\ j\ {\isacharplus}\ z\ {\isasymle}\ D\ {\isacharplus}\ z{\isachardoublequoteclose}\ \isacommand{by}\isamarkupfalse%
\ arith\isanewline
\ \ \ \ \isacommand{have}\isamarkupfalse%
\ sg{\isadigit{2}}{\isacharcolon}{\isachardoublequoteopen}t\ {\isacharplus}\ j\ {\isacharplus}\ z\ {\isacharequal}\ t\ {\isacharplus}\ {\isacharparenleft}j\ {\isacharplus}\ z{\isacharparenright}{\isachardoublequoteclose}\ \isacommand{by}\isamarkupfalse%
\ arith\ \isanewline
\ \ \ \ \isacommand{from}\isamarkupfalse%
\ h{\isadigit{1}}\ \isakeyword{and}\ sg{\isadigit{1}}\ \isakeyword{and}\ sg{\isadigit{2}}\ \isacommand{show}\isamarkupfalse%
\ {\isacharquery}thesis\ \isacommand{by}\isamarkupfalse%
\ {\isacharparenleft}simp\ {\isacharparenleft}no{\isacharunderscore}asm{\isacharunderscore}simp{\isacharparenright}{\isacharparenright}\isanewline
\isacommand{qed}\isamarkupfalse%
\endisatagproof
{\isafoldproof}%
\isadelimproof
\isanewline
\endisadelimproof
\isanewline
\isacommand{lemma}\isamarkupfalse%
\ streamValue{\isadigit{2}}{\isacharcolon}\isanewline
\ \ \isakeyword{assumes}\ h{\isadigit{1}}{\isacharcolon}{\isachardoublequoteopen}{\isasymforall}j{\isasymle}\ D\ {\isacharplus}\ {\isacharparenleft}z{\isacharcolon}{\isacharcolon}nat{\isacharparenright}{\isachardot}\ str\ {\isacharparenleft}t\ {\isacharplus}\ j{\isacharparenright}\ {\isacharequal}\ x{\isachardoublequoteclose}\isanewline
\ \ \isakeyword{shows}\ \ \ \ \ \ {\isachardoublequoteopen}{\isasymforall}j{\isasymle}\ D{\isachardot}\ str\ {\isacharparenleft}t\ {\isacharplus}\ j\ {\isacharplus}\ z{\isacharparenright}\ {\isacharequal}\ x{\isachardoublequoteclose}\isanewline
\isadelimproof
\endisadelimproof
\isatagproof
\isacommand{using}\isamarkupfalse%
\ assms\ \isacommand{by}\isamarkupfalse%
\ {\isacharparenleft}clarify{\isacharcomma}\ simp\ add{\isacharcolon}\ streamValue{\isadigit{1}}{\isacharparenright}%
\endisatagproof
{\isafoldproof}%
\isadelimproof
\isanewline
\endisadelimproof
\isanewline
\isacommand{lemma}\isamarkupfalse%
\ streamValue{\isadigit{3}}{\isacharcolon}\isanewline
\ \ \isakeyword{assumes}\ h{\isadigit{1}}{\isacharcolon}{\isachardoublequoteopen}{\isasymforall}j{\isasymle}\ D{\isachardot}\ str\ {\isacharparenleft}t\ {\isacharplus}\ j\ {\isacharplus}\ {\isacharparenleft}Suc\ y{\isacharparenright}{\isacharparenright}\ {\isacharequal}\ x{\isachardoublequoteclose}\isanewline
\ \ \ \ \ \ \isakeyword{and}\ h{\isadigit{2}}{\isacharcolon}{\isachardoublequoteopen}j\ {\isasymle}\ D{\isachardoublequoteclose}\isanewline
\ \ \ \ \ \ \isakeyword{and}\ h{\isadigit{3}}{\isacharcolon}{\isachardoublequoteopen}str\ {\isacharparenleft}t\ {\isacharplus}\ y{\isacharparenright}\ {\isacharequal}\ x{\isachardoublequoteclose}\isanewline
\ \ \ \ \isakeyword{shows}\ \ \ \ {\isachardoublequoteopen}str\ {\isacharparenleft}t\ {\isacharplus}\ j\ {\isacharplus}\ y{\isacharparenright}\ {\isacharequal}\ x{\isachardoublequoteclose}\isanewline
\isadelimproof
\endisadelimproof
\isatagproof
\isacommand{using}\isamarkupfalse%
\ assms\isanewline
\isacommand{proof}\isamarkupfalse%
\ {\isacharparenleft}induct\ j{\isacharparenright}\ \isanewline
\ \ \isacommand{case}\isamarkupfalse%
\ {\isadigit{0}}\isanewline
\ \ \isacommand{from}\isamarkupfalse%
\ h{\isadigit{3}}\ \isacommand{show}\isamarkupfalse%
\ {\isacharquery}case\ \isacommand{by}\isamarkupfalse%
\ simp\isanewline
\isacommand{next}\isamarkupfalse%
\isanewline
\ \ \isacommand{case}\isamarkupfalse%
\ {\isacharparenleft}Suc\ j{\isacharparenright}\ \isanewline
\ \ \isacommand{from}\isamarkupfalse%
\ this\ \isacommand{show}\isamarkupfalse%
\ {\isacharquery}case\ \isacommand{by}\isamarkupfalse%
\ auto\isanewline
\isacommand{qed}\isamarkupfalse%
\endisatagproof
{\isafoldproof}%
\isadelimproof
\isanewline
\endisadelimproof
\isanewline
\isacommand{lemma}\isamarkupfalse%
\ streamValue{\isadigit{4}}{\isacharcolon}\isanewline
\ \ \isakeyword{assumes}\ h{\isadigit{1}}{\isacharcolon}{\isachardoublequoteopen}{\isasymforall}j{\isasymle}\ D{\isachardot}\ str\ {\isacharparenleft}t\ {\isacharplus}\ j\ {\isacharplus}\ {\isacharparenleft}Suc\ y{\isacharparenright}{\isacharparenright}\ {\isacharequal}\ x{\isachardoublequoteclose}\isanewline
\ \ \ \ \ \ \isakeyword{and}\ h{\isadigit{3}}{\isacharcolon}{\isachardoublequoteopen}str\ {\isacharparenleft}t\ {\isacharplus}\ y{\isacharparenright}\ {\isacharequal}\ x{\isachardoublequoteclose}\isanewline
\ \ \ \ \isakeyword{shows}\ \ \ \ \ {\isachardoublequoteopen}{\isasymforall}j{\isasymle}\ D{\isachardot}\ str\ {\isacharparenleft}t\ {\isacharplus}\ j\ {\isacharplus}\ y{\isacharparenright}\ {\isacharequal}\ x{\isachardoublequoteclose}\isanewline
\isadelimproof
\endisadelimproof
\isatagproof
\isacommand{using}\isamarkupfalse%
\ assms\ \isanewline
\ \ \isacommand{by}\isamarkupfalse%
\ {\isacharparenleft}clarify{\isacharcomma}\ \ simp\ add{\isacharcolon}\ streamValue{\isadigit{3}}{\isacharparenright}%
\endisatagproof
{\isafoldproof}%
\isadelimproof
\isanewline
\endisadelimproof
\isanewline
\isacommand{lemma}\isamarkupfalse%
\ streamValue{\isadigit{5}}{\isacharcolon}\isanewline
\ \ \isakeyword{assumes}\ h{\isadigit{1}}{\isacharcolon}{\isachardoublequoteopen}{\isasymforall}j{\isasymle}\ D{\isachardot}\ str\ {\isacharparenleft}t\ {\isacharplus}\ j\ {\isacharplus}\ {\isacharparenleft}{\isacharparenleft}i{\isacharcolon}{\isacharcolon}nat{\isacharparenright}\ {\isacharplus}\ k{\isacharparenright}{\isacharparenright}\ {\isacharequal}\ x{\isachardoublequoteclose}\isanewline
\ \ \ \ \ \ \isakeyword{and}\ h{\isadigit{2}}{\isacharcolon}{\isachardoublequoteopen}j{\isasymle}\ D{\isachardoublequoteclose}\isanewline
\ \ \isakeyword{shows}\ \ \ \ \ \ {\isachardoublequoteopen}str\ {\isacharparenleft}t\ {\isacharplus}\ i\ {\isacharplus}\ k\ {\isacharplus}\ j{\isacharparenright}\ {\isacharequal}\ x{\isachardoublequoteclose}\isanewline
\isadelimproof
\endisadelimproof
\isatagproof
\isacommand{proof}\isamarkupfalse%
\ {\isacharminus}\ \isanewline
\ \ \ \isacommand{have}\isamarkupfalse%
\ sg{\isadigit{1}}{\isacharcolon}{\isachardoublequoteopen}t\ {\isacharplus}\ i\ {\isacharplus}\ k\ {\isacharplus}\ j\ {\isacharequal}\ t\ {\isacharplus}\ j\ {\isacharplus}\ {\isacharparenleft}i\ {\isacharplus}\ k{\isacharparenright}{\isachardoublequoteclose}\ \isacommand{by}\isamarkupfalse%
\ arith\isanewline
\ \ \ \isacommand{from}\isamarkupfalse%
\ assms\ \isakeyword{and}\ sg{\isadigit{1}}\ \isacommand{show}\isamarkupfalse%
\ {\isacharquery}thesis\ \isacommand{by}\isamarkupfalse%
\ {\isacharparenleft}simp\ {\isacharparenleft}no{\isacharunderscore}asm{\isacharunderscore}simp{\isacharparenright}{\isacharparenright}\isanewline
\isacommand{qed}\isamarkupfalse%
\endisatagproof
{\isafoldproof}%
\isadelimproof
\isanewline
\endisadelimproof
\isanewline
\isacommand{lemma}\isamarkupfalse%
\ streamValue{\isadigit{6}}{\isacharcolon}\isanewline
\ \ \isakeyword{assumes}\ h{\isadigit{1}}{\isacharcolon}{\isachardoublequoteopen}{\isasymforall}j{\isasymle}\ D{\isachardot}\ str\ {\isacharparenleft}t\ {\isacharplus}\ j\ {\isacharplus}\ {\isacharparenleft}{\isacharparenleft}i{\isacharcolon}{\isacharcolon}nat{\isacharparenright}\ {\isacharplus}\ k{\isacharparenright}{\isacharparenright}\ {\isacharequal}\ x{\isachardoublequoteclose}\isanewline
\ \ \isakeyword{shows}\ \ \ \ \ \ {\isachardoublequoteopen}{\isasymforall}j{\isasymle}\ D{\isachardot}\ str\ {\isacharparenleft}t\ {\isacharplus}\ {\isacharparenleft}i{\isacharcolon}{\isacharcolon}nat{\isacharparenright}\ {\isacharplus}\ k\ {\isacharplus}\ j{\isacharparenright}\ {\isacharequal}\ x{\isachardoublequoteclose}\isanewline
\isadelimproof
\endisadelimproof
\isatagproof
\isacommand{using}\isamarkupfalse%
\ assms\ \isacommand{by}\isamarkupfalse%
\ {\isacharparenleft}clarify{\isacharcomma}\ simp\ add{\isacharcolon}\ streamValue{\isadigit{5}}{\isacharparenright}%
\endisatagproof
{\isafoldproof}%
\isadelimproof
\isanewline
\endisadelimproof
\isanewline
\isacommand{lemma}\isamarkupfalse%
\ streamValue{\isadigit{7}}{\isacharcolon}\isanewline
\ \ \isakeyword{assumes}\ h{\isadigit{1}}{\isacharcolon}{\isachardoublequoteopen}{\isasymforall}j{\isasymle}d{\isachardot}\ str\ {\isacharparenleft}t\ {\isacharplus}\ i\ {\isacharplus}\ k\ {\isacharplus}\ d\ {\isacharplus}\ Suc\ j{\isacharparenright}\ {\isacharequal}\ x{\isachardoublequoteclose}\isanewline
\ \ \ \ \ \ \isakeyword{and}\ h{\isadigit{2}}{\isacharcolon}{\isachardoublequoteopen}str\ {\isacharparenleft}t\ {\isacharplus}\ i\ {\isacharplus}\ k\ {\isacharplus}\ d{\isacharparenright}\ {\isacharequal}\ x{\isachardoublequoteclose}\isanewline
\ \ \ \ \ \ \isakeyword{and}\ h{\isadigit{3}}{\isacharcolon}{\isachardoublequoteopen}j{\isasymle}\ Suc\ d{\isachardoublequoteclose}\isanewline
\ \ \isakeyword{shows}\ \ \ \ \ \ {\isachardoublequoteopen}str\ {\isacharparenleft}t\ {\isacharplus}\ i\ {\isacharplus}\ k\ {\isacharplus}\ d\ {\isacharplus}\ j{\isacharparenright}\ {\isacharequal}\ x{\isachardoublequoteclose}\isanewline
\isadelimproof
\endisadelimproof
\isatagproof
\isacommand{proof}\isamarkupfalse%
\ {\isacharminus}\ \isanewline
\ \ \isacommand{from}\isamarkupfalse%
\ h{\isadigit{1}}\ \isacommand{have}\isamarkupfalse%
\ sg{\isadigit{1}}{\isacharcolon}{\isachardoublequoteopen}str\ {\isacharparenleft}t\ {\isacharplus}\ i\ {\isacharplus}\ k\ {\isacharplus}\ d\ {\isacharplus}\ Suc\ d{\isacharparenright}\ {\isacharequal}\ x{\isachardoublequoteclose}\ \isanewline
\ \ \ \ \ \isacommand{by}\isamarkupfalse%
\ {\isacharparenleft}simp\ {\isacharparenleft}no{\isacharunderscore}asm{\isacharunderscore}simp{\isacharparenright}{\isacharcomma}\ simp{\isacharparenright}\ \isanewline
\ \ \isacommand{from}\isamarkupfalse%
\ assms\ \isacommand{show}\isamarkupfalse%
\ {\isacharquery}thesis\ \isanewline
\ \ \isacommand{proof}\isamarkupfalse%
\ {\isacharparenleft}cases\ {\isachardoublequoteopen}j\ {\isacharequal}\ Suc\ d{\isachardoublequoteclose}{\isacharparenright}\isanewline
\ \ \ \ \isacommand{assume}\isamarkupfalse%
\ a{\isadigit{1}}{\isacharcolon}{\isachardoublequoteopen}j\ {\isacharequal}\ Suc\ d{\isachardoublequoteclose}\isanewline
\ \ \ \ \isacommand{from}\isamarkupfalse%
\ a{\isadigit{1}}\ \isakeyword{and}\ sg{\isadigit{1}}\ \isacommand{show}\isamarkupfalse%
\ {\isacharquery}thesis\ \isacommand{by}\isamarkupfalse%
\ simp\isanewline
\ \ \isacommand{next}\isamarkupfalse%
\isanewline
\ \ \ \ \isacommand{assume}\isamarkupfalse%
\ a{\isadigit{2}}{\isacharcolon}{\isachardoublequoteopen}j\ {\isasymnoteq}\ Suc\ d{\isachardoublequoteclose}\isanewline
\ \ \ \ \isacommand{from}\isamarkupfalse%
\ a{\isadigit{2}}\ \isakeyword{and}\ h{\isadigit{3}}\ \isacommand{have}\isamarkupfalse%
\ sg{\isadigit{2}}{\isacharcolon}{\isachardoublequoteopen}j{\isasymle}d{\isachardoublequoteclose}\ \isacommand{by}\isamarkupfalse%
\ auto\isanewline
\ \ \ \ \isacommand{from}\isamarkupfalse%
\ assms\ \isakeyword{and}\ sg{\isadigit{2}}\ \isacommand{show}\isamarkupfalse%
\ {\isacharquery}thesis\isanewline
\ \ \ \ \isacommand{proof}\isamarkupfalse%
\ {\isacharparenleft}cases\ {\isachardoublequoteopen}j\ {\isachargreater}\ {\isadigit{0}}{\isachardoublequoteclose}{\isacharparenright}\isanewline
\ \ \ \ \ \ \isacommand{assume}\isamarkupfalse%
\ a{\isadigit{3}}{\isacharcolon}{\isachardoublequoteopen}{\isadigit{0}}\ {\isacharless}\ j{\isachardoublequoteclose}\isanewline
\ \ \ \ \ \ \isacommand{from}\isamarkupfalse%
\ a{\isadigit{3}}\ \isakeyword{and}\ h{\isadigit{3}}\ \isacommand{have}\isamarkupfalse%
\ sg{\isadigit{3}}{\isacharcolon}{\isachardoublequoteopen}j\ {\isacharminus}\ {\isacharparenleft}{\isadigit{1}}{\isacharcolon}{\isacharcolon}nat{\isacharparenright}\ {\isasymle}\ d{\isachardoublequoteclose}\ \isacommand{by}\isamarkupfalse%
\ simp\isanewline
\ \ \ \ \ \ \isacommand{from}\isamarkupfalse%
\ a{\isadigit{3}}\ \isacommand{have}\isamarkupfalse%
\ sg{\isadigit{4}}{\isacharcolon}{\isachardoublequoteopen}Suc\ {\isacharparenleft}j\ {\isacharminus}\ {\isacharparenleft}{\isadigit{1}}{\isacharcolon}{\isacharcolon}nat{\isacharparenright}{\isacharparenright}\ {\isacharequal}\ j{\isachardoublequoteclose}\ \ \isacommand{by}\isamarkupfalse%
\ arith\isanewline
\ \ \ \ \ \ \isacommand{from}\isamarkupfalse%
\ sg{\isadigit{3}}\ \isakeyword{and}\ h{\isadigit{1}}\ \isakeyword{and}\ sg{\isadigit{4}}\ \isacommand{have}\isamarkupfalse%
\ sg{\isadigit{5}}{\isacharcolon}{\isachardoublequoteopen}str\ {\isacharparenleft}t\ {\isacharplus}\ i\ {\isacharplus}\ k\ {\isacharplus}\ d\ {\isacharplus}\ j{\isacharparenright}\ {\isacharequal}\ x{\isachardoublequoteclose}\ \isacommand{by}\isamarkupfalse%
\ auto\isanewline
\ \ \ \ \ \ \isacommand{from}\isamarkupfalse%
\ sg{\isadigit{5}}\ \isacommand{show}\isamarkupfalse%
\ {\isacharquery}thesis\ \isacommand{by}\isamarkupfalse%
\ simp\isanewline
\ \ \ \ \isacommand{next}\isamarkupfalse%
\isanewline
\ \ \ \ \ \ \isacommand{assume}\isamarkupfalse%
\ a{\isadigit{4}}{\isacharcolon}{\isachardoublequoteopen}{\isasymnot}\ {\isadigit{0}}\ {\isacharless}\ j{\isachardoublequoteclose}\isanewline
\ \ \ \ \ \ \isacommand{from}\isamarkupfalse%
\ a{\isadigit{4}}\ \isacommand{have}\isamarkupfalse%
\ sg{\isadigit{6}}{\isacharcolon}{\isachardoublequoteopen}j\ {\isacharequal}\ {\isadigit{0}}{\isachardoublequoteclose}\ \isacommand{by}\isamarkupfalse%
\ simp\isanewline
\ \ \ \ \ \ \isacommand{from}\isamarkupfalse%
\ h{\isadigit{2}}\ \isakeyword{and}\ sg{\isadigit{6}}\ \isacommand{show}\isamarkupfalse%
\ {\isacharquery}thesis\ \isacommand{by}\isamarkupfalse%
\ simp\isanewline
\ \ \ \ \isacommand{qed}\isamarkupfalse%
\isanewline
\ \ \isacommand{qed}\isamarkupfalse%
\isanewline
\isacommand{qed}\isamarkupfalse%
\endisatagproof
{\isafoldproof}%
\isadelimproof
\isanewline
\endisadelimproof
\isanewline
\isacommand{lemma}\isamarkupfalse%
\ streamValue{\isadigit{8}}{\isacharcolon}\isanewline
\ \ \isakeyword{assumes}\ h{\isadigit{1}}{\isacharcolon}{\isachardoublequoteopen}{\isasymforall}j{\isasymle}d{\isachardot}\ str\ {\isacharparenleft}t\ {\isacharplus}\ i\ {\isacharplus}\ k\ {\isacharplus}\ d\ {\isacharplus}\ Suc\ j{\isacharparenright}\ {\isacharequal}\ x{\isachardoublequoteclose}\isanewline
\ \ \ \ \ \ \isakeyword{and}\ h{\isadigit{2}}{\isacharcolon}{\isachardoublequoteopen}str\ {\isacharparenleft}t\ {\isacharplus}\ i\ {\isacharplus}\ k\ {\isacharplus}\ d{\isacharparenright}\ {\isacharequal}\ x{\isachardoublequoteclose}\ \isanewline
\ \ \isakeyword{shows}\ \ \ \ \ \ {\isachardoublequoteopen}{\isasymforall}\ j{\isasymle}\ Suc\ d{\isachardot}\ str\ {\isacharparenleft}t\ {\isacharplus}\ i\ {\isacharplus}\ k\ {\isacharplus}\ d\ {\isacharplus}\ j{\isacharparenright}\ {\isacharequal}\ x{\isachardoublequoteclose}\isanewline
\isadelimproof
\endisadelimproof
\isatagproof
\isacommand{using}\isamarkupfalse%
\ assms\ \isacommand{by}\isamarkupfalse%
\ {\isacharparenleft}clarify{\isacharcomma}\ simp\ add{\isacharcolon}\ streamValue{\isadigit{7}}{\isacharparenright}%
\endisatagproof
{\isafoldproof}%
\isadelimproof
\isanewline
\endisadelimproof
\isanewline
\isanewline
\isacommand{lemma}\isamarkupfalse%
\ arith{\isacharunderscore}streamValue{\isadigit{9}}aux{\isacharcolon}\isanewline
{\isachardoublequoteopen}Suc\ {\isacharparenleft}t\ {\isacharplus}\ {\isacharparenleft}j\ {\isacharplus}\ d{\isacharparenright}\ {\isacharplus}\ {\isacharparenleft}i\ {\isacharplus}\ k{\isacharparenright}{\isacharparenright}\ {\isacharequal}\ \ Suc\ {\isacharparenleft}t\ {\isacharplus}\ i\ {\isacharplus}\ k\ {\isacharplus}\ d\ {\isacharplus}\ j{\isacharparenright}{\isachardoublequoteclose}\ \isanewline
\isadelimproof
\endisadelimproof
\isatagproof
\isacommand{by}\isamarkupfalse%
\ arith%
\endisatagproof
{\isafoldproof}%
\isadelimproof
\isanewline
\endisadelimproof
\isanewline
\isacommand{lemma}\isamarkupfalse%
\ streamValue{\isadigit{9}}{\isacharcolon}\isanewline
\ \ \isakeyword{assumes}\ h{\isadigit{1}}{\isacharcolon}{\isachardoublequoteopen}{\isasymforall}j{\isasymle}{\isadigit{2}}\ {\isacharasterisk}\ d{\isachardot}\ str\ {\isacharparenleft}t\ {\isacharplus}\ j\ {\isacharplus}\ Suc\ {\isacharparenleft}i\ {\isacharplus}\ k{\isacharparenright}{\isacharparenright}\ {\isacharequal}\ x{\isachardoublequoteclose}\isanewline
\ \ \ \ \ \ \isakeyword{and}\ h{\isadigit{2}}{\isacharcolon}{\isachardoublequoteopen}j{\isasymle}d{\isachardoublequoteclose}\isanewline
\ \ \isakeyword{shows}\ \ \ \ \ \ {\isachardoublequoteopen}str\ {\isacharparenleft}t\ {\isacharplus}\ i\ {\isacharplus}\ k\ {\isacharplus}\ d\ {\isacharplus}\ Suc\ j{\isacharparenright}\ {\isacharequal}\ x{\isachardoublequoteclose}\isanewline
\isadelimproof
\endisadelimproof
\isatagproof
\isacommand{proof}\isamarkupfalse%
\ {\isacharminus}\isanewline
\ \ \isacommand{from}\isamarkupfalse%
\ h{\isadigit{2}}\ \isacommand{have}\isamarkupfalse%
\ {\isachardoublequoteopen}{\isacharparenleft}j{\isacharplus}d{\isacharparenright}\ {\isasymle}{\isadigit{2}}\ {\isacharasterisk}\ d{\isachardoublequoteclose}\ \isacommand{by}\isamarkupfalse%
\ arith\isanewline
\ \ \isacommand{from}\isamarkupfalse%
\ h{\isadigit{1}}\ \isakeyword{and}\ this\ \isacommand{have}\isamarkupfalse%
\ {\isachardoublequoteopen}str\ {\isacharparenleft}t\ {\isacharplus}\ {\isacharparenleft}j\ {\isacharplus}\ d{\isacharparenright}\ {\isacharplus}\ Suc\ {\isacharparenleft}i\ {\isacharplus}\ k{\isacharparenright}{\isacharparenright}\ {\isacharequal}\ x{\isachardoublequoteclose}\ \isacommand{by}\isamarkupfalse%
\ auto\isanewline
\ \ \isacommand{from}\isamarkupfalse%
\ this\ \isacommand{show}\isamarkupfalse%
\ {\isacharquery}thesis\ \ \isacommand{by}\isamarkupfalse%
\ {\isacharparenleft}simp\ add{\isacharcolon}\ arith{\isacharunderscore}streamValue{\isadigit{9}}aux{\isacharparenright}\ \ \isanewline
\isacommand{qed}\isamarkupfalse%
\endisatagproof
{\isafoldproof}%
\isadelimproof
\ \ \ \ \ \isanewline
\endisadelimproof
\isanewline
\isanewline
\isacommand{lemma}\isamarkupfalse%
\ streamValue{\isadigit{1}}{\isadigit{0}}{\isacharcolon}\isanewline
\ \ \isakeyword{assumes}\ h{\isadigit{1}}{\isacharcolon}{\isachardoublequoteopen}{\isasymforall}j{\isasymle}{\isadigit{2}}\ {\isacharasterisk}\ d{\isachardot}\ str\ {\isacharparenleft}t\ {\isacharplus}\ j\ {\isacharplus}\ Suc\ {\isacharparenleft}i\ {\isacharplus}\ k{\isacharparenright}{\isacharparenright}\ {\isacharequal}\ x{\isachardoublequoteclose}\isanewline
\ \ \isakeyword{shows}\ \ \ \ \ \ {\isachardoublequoteopen}{\isasymforall}j{\isasymle}d{\isachardot}\ str\ {\isacharparenleft}t\ {\isacharplus}\ i\ {\isacharplus}\ k\ {\isacharplus}\ d\ {\isacharplus}\ Suc\ j{\isacharparenright}\ {\isacharequal}\ x{\isachardoublequoteclose}\isanewline
\isadelimproof
\endisadelimproof
\isatagproof
\isacommand{using}\isamarkupfalse%
\ assms\ \isanewline
\ \ \isacommand{apply}\isamarkupfalse%
\ clarify\isanewline
\ \ \isacommand{by}\isamarkupfalse%
\ {\isacharparenleft}rule\ streamValue{\isadigit{9}}{\isacharcomma}\ auto{\isacharparenright}%
\endisatagproof
{\isafoldproof}%
\isadelimproof
\isanewline
\endisadelimproof
\isanewline
\isacommand{lemma}\isamarkupfalse%
\ arith{\isacharunderscore}sum{\isadigit{1}}{\isacharcolon}{\isachardoublequoteopen}{\isacharparenleft}t{\isacharcolon}{\isacharcolon}nat{\isacharparenright}\ {\isacharplus}\ {\isacharparenleft}i\ {\isacharplus}\ k\ {\isacharplus}\ d{\isacharparenright}\ {\isacharequal}\ \ t\ {\isacharplus}\ i\ {\isacharplus}\ k\ {\isacharplus}\ d{\isachardoublequoteclose}\isanewline
\isadelimproof
\endisadelimproof
\isatagproof
\isacommand{by}\isamarkupfalse%
\ arith%
\endisatagproof
{\isafoldproof}%
\isadelimproof
\isanewline
\endisadelimproof
\isanewline
\isacommand{lemma}\isamarkupfalse%
\ arith{\isacharunderscore}sum{\isadigit{2}}{\isacharcolon}{\isachardoublequoteopen}Suc\ {\isacharparenleft}Suc\ {\isacharparenleft}t\ {\isacharplus}\ k\ {\isacharplus}\ j{\isacharparenright}{\isacharparenright}\ {\isacharequal}\ Suc\ {\isacharparenleft}Suc\ {\isacharparenleft}t\ {\isacharplus}\ {\isacharparenleft}k\ {\isacharplus}\ j{\isacharparenright}{\isacharparenright}{\isacharparenright}{\isachardoublequoteclose}\isanewline
\isadelimproof
\endisadelimproof
\isatagproof
\isacommand{by}\isamarkupfalse%
\ arith%
\endisatagproof
{\isafoldproof}%
\isadelimproof
\isanewline
\endisadelimproof
\isanewline
\isacommand{lemma}\isamarkupfalse%
\ arith{\isacharunderscore}sum{\isadigit{4}}{\isacharcolon}{\isachardoublequoteopen}t\ {\isacharplus}\ {\isadigit{3}}\ {\isacharplus}\ k\ {\isacharplus}\ d\ {\isacharequal}\ Suc\ {\isacharparenleft}t\ {\isacharplus}\ {\isacharparenleft}{\isadigit{2}}{\isacharcolon}{\isacharcolon}nat{\isacharparenright}\ {\isacharplus}\ k\ {\isacharplus}\ d{\isacharparenright}{\isachardoublequoteclose}\isanewline
\isadelimproof
\endisadelimproof
\isatagproof
\isacommand{by}\isamarkupfalse%
\ arith%
\endisatagproof
{\isafoldproof}%
\isadelimproof
\isanewline
\endisadelimproof
\isanewline 
\isacommand{lemma}\isamarkupfalse%
\ streamValue{\isadigit{1}}{\isadigit{1}}{\isacharcolon}\isanewline
\ \isakeyword{assumes}\ h{\isadigit{1}}{\isacharcolon}{\isachardoublequoteopen}{\isasymforall}j{\isasymle}{\isadigit{2}}\ {\isacharasterisk}\ d\ {\isacharplus}\ {\isacharparenleft}{\isadigit{4}}\ {\isacharplus}\ k{\isacharparenright}{\isachardot}\ lose\ {\isacharparenleft}t\ {\isacharplus}\ j{\isacharparenright}\ {\isacharequal}\ x{\isachardoublequoteclose}\isanewline
\ \ \ \ \ \isakeyword{and}\ h{\isadigit{2}}{\isacharcolon}{\isachardoublequoteopen}j{\isasymle}Suc\ d{\isachardoublequoteclose}\isanewline
\ \isakeyword{shows}\ \ \ \ \ \ {\isachardoublequoteopen}lose\ {\isacharparenleft}t\ {\isacharplus}\ {\isadigit{2}}\ {\isacharplus}\ k\ {\isacharplus}\ j{\isacharparenright}\ {\isacharequal}\ x{\isachardoublequoteclose}\isanewline
\isadelimproof
\endisadelimproof
\isatagproof
\isacommand{proof}\isamarkupfalse%
\ {\isacharminus}\isanewline
\ \ \isacommand{from}\isamarkupfalse%
\ h{\isadigit{2}}\ \isacommand{have}\isamarkupfalse%
\ sg{\isadigit{1}}{\isacharcolon}{\isachardoublequoteopen}{\isadigit{2}}\ {\isacharplus}\ k\ {\isacharplus}\ j\ {\isasymle}{\isadigit{2}}\ {\isacharasterisk}\ d\ {\isacharplus}\ {\isacharparenleft}{\isadigit{4}}\ {\isacharplus}\ k{\isacharparenright}{\isachardoublequoteclose}\ \isacommand{by}\isamarkupfalse%
\ arith\isanewline
\ \ \isacommand{have}\isamarkupfalse%
\ sg{\isadigit{2}}{\isacharcolon}{\isachardoublequoteopen}Suc\ {\isacharparenleft}Suc\ {\isacharparenleft}t\ {\isacharplus}\ k\ {\isacharplus}\ j{\isacharparenright}{\isacharparenright}\ {\isacharequal}\ Suc\ {\isacharparenleft}Suc\ {\isacharparenleft}t\ {\isacharplus}\ {\isacharparenleft}k\ {\isacharplus}\ j{\isacharparenright}{\isacharparenright}{\isacharparenright}{\isachardoublequoteclose}\ \isacommand{by}\isamarkupfalse%
\ arith\isanewline
\ \ \isacommand{from}\isamarkupfalse%
\ sg{\isadigit{1}}\ \isakeyword{and}\ h{\isadigit{1}}\ \isacommand{have}\isamarkupfalse%
\ {\isachardoublequoteopen}lose\ {\isacharparenleft}t\ {\isacharplus}\ {\isacharparenleft}{\isadigit{2}}\ {\isacharplus}\ k\ {\isacharplus}\ j{\isacharparenright}{\isacharparenright}\ {\isacharequal}\ x{\isachardoublequoteclose}\ \isacommand{by}\isamarkupfalse%
\ blast\isanewline
\ \ \isacommand{from}\isamarkupfalse%
\ this\ \isakeyword{and}\ sg{\isadigit{2}}\ \isacommand{show}\isamarkupfalse%
\ {\isacharquery}thesis\ \isacommand{by}\isamarkupfalse%
\ {\isacharparenleft}simp\ add{\isacharcolon}\ arith{\isacharunderscore}sum{\isadigit{2}}{\isacharparenright}\isanewline
\isacommand{qed}\isamarkupfalse%
\endisatagproof
{\isafoldproof}%
\isadelimproof
\ \isanewline
\endisadelimproof
\isanewline 
\isacommand{lemma}\isamarkupfalse%
\ streamValue{\isadigit{1}}{\isadigit{2}}{\isacharcolon}\isanewline
\ \isakeyword{assumes}\ h{\isadigit{1}}{\isacharcolon}{\isachardoublequoteopen}{\isasymforall}j{\isasymle}{\isadigit{2}}\ {\isacharasterisk}\ d\ {\isacharplus}\ {\isacharparenleft}{\isadigit{4}}\ {\isacharplus}\ k{\isacharparenright}{\isachardot}\ lose\ {\isacharparenleft}t\ {\isacharplus}\ j{\isacharparenright}\ {\isacharequal}\ x{\isachardoublequoteclose}\isanewline
\ \isakeyword{shows}\ \ \ \ \ \ {\isachardoublequoteopen}{\isasymforall}j{\isasymle}Suc\ d{\isachardot}\ lose\ {\isacharparenleft}t\ {\isacharplus}\ {\isadigit{2}}\ {\isacharplus}\ k\ {\isacharplus}\ j{\isacharparenright}\ {\isacharequal}\ x{\isachardoublequoteclose}\isanewline
\isadelimproof
\endisadelimproof
\isatagproof
\isacommand{using}\isamarkupfalse%
\ assms\isanewline
\ \ \isacommand{apply}\isamarkupfalse%
\ clarify\
\isacommand{by}\isamarkupfalse%
\ {\isacharparenleft}rule\ streamValue{\isadigit{1}}{\isadigit{1}}{\isacharcomma}\ auto{\isacharparenright}%
\endisatagproof
{\isafoldproof}%
\isadelimproof
\isanewline
\endisadelimproof
\isanewline 
\isacommand{lemma}\isamarkupfalse%
\ streamValue{\isadigit{4}}{\isadigit{3}}{\isacharcolon}\isanewline
\ \ \isakeyword{assumes}\ h{\isadigit{1}}{\isacharcolon}{\isachardoublequoteopen}{\isasymforall}j{\isasymle}{\isadigit{2}}\ {\isacharasterisk}\ d\ {\isacharplus}\ {\isacharparenleft}{\isacharparenleft}{\isadigit{4}}{\isacharcolon}{\isacharcolon}nat{\isacharparenright}\ {\isacharplus}\ k{\isacharparenright}{\isachardot}\ lose\ {\isacharparenleft}t\ {\isacharplus}\ j{\isacharparenright}\ {\isacharequal}\ {\isacharbrackleft}False{\isacharbrackright}{\isachardoublequoteclose}\isanewline
\ \ \isakeyword{shows}\ \ {\isachardoublequoteopen}{\isasymforall}j{\isasymle}{\isadigit{2}}\ {\isacharasterisk}\ d{\isachardot}\ lose\ {\isacharparenleft}{\isacharparenleft}t\ {\isacharplus}\ {\isacharparenleft}{\isadigit{3}}{\isacharcolon}{\isacharcolon}nat{\isacharparenright}\ {\isacharplus}\ k{\isacharparenright}\ {\isacharplus}\ j{\isacharparenright}\ {\isacharequal}\ {\isacharbrackleft}False{\isacharbrackright}{\isachardoublequoteclose}\isanewline
\isadelimproof
\endisadelimproof
\isatagproof
\isacommand{proof}\isamarkupfalse%
\ {\isacharminus}\isanewline
\ \ \isacommand{from}\isamarkupfalse%
\ h{\isadigit{1}}\ \isacommand{have}\isamarkupfalse%
\ sg{\isadigit{1}}{\isacharcolon}{\isachardoublequoteopen}{\isasymforall}j{\isasymle}{\isadigit{2}}\ {\isacharasterisk}\ d{\isachardot}\ lose\ {\isacharparenleft}t\ {\isacharplus}\ j\ {\isacharplus}\ {\isacharparenleft}{\isadigit{4}}\ {\isacharplus}\ k{\isacharparenright}{\isacharparenright}\ {\isacharequal}\ {\isacharbrackleft}False{\isacharbrackright}{\isachardoublequoteclose}\ \isanewline
\ \ \ \ \isacommand{by}\isamarkupfalse%
\ {\isacharparenleft}simp\ add{\isacharcolon}\ streamValue{\isadigit{2}}{\isacharparenright}\isanewline
\ \ \isacommand{have}\isamarkupfalse%
\ sg{\isadigit{2}}{\isacharcolon}{\isachardoublequoteopen}Suc\ {\isacharparenleft}{\isadigit{3}}\ {\isacharplus}\ k{\isacharparenright}\ {\isacharequal}\ {\isacharparenleft}{\isadigit{4}}\ {\isacharplus}\ k{\isacharparenright}{\isachardoublequoteclose}\ \isacommand{by}\isamarkupfalse%
\ arith\isanewline
\ \ \isacommand{from}\isamarkupfalse%
\ sg{\isadigit{1}}\ \isakeyword{and}\ sg{\isadigit{2}}\ \isacommand{have}\isamarkupfalse%
\ sg{\isadigit{3}}{\isacharcolon}{\isachardoublequoteopen}{\isasymforall}j{\isasymle}{\isadigit{2}}\ {\isacharasterisk}\ d{\isachardot}\ lose\ {\isacharparenleft}t\ {\isacharplus}\ j\ {\isacharplus}\ Suc\ {\isacharparenleft}{\isadigit{3}}\ {\isacharplus}\ k{\isacharparenright}{\isacharparenright}\ {\isacharequal}\ {\isacharbrackleft}False{\isacharbrackright}{\isachardoublequoteclose}\ \isanewline
\ \ \ \ \isacommand{by}\isamarkupfalse%
\ {\isacharparenleft}simp\ {\isacharparenleft}no{\isacharunderscore}asm{\isacharunderscore}simp{\isacharparenright}{\isacharparenright}\ \ \isanewline
\ \ \isacommand{from}\isamarkupfalse%
\ h{\isadigit{1}}\ \isacommand{have}\isamarkupfalse%
\ sg{\isadigit{4}}{\isacharcolon}{\isachardoublequoteopen}lose\ {\isacharparenleft}t\ {\isacharplus}\ {\isacharparenleft}{\isadigit{3}}\ {\isacharplus}\ k{\isacharparenright}{\isacharparenright}\ {\isacharequal}\ {\isacharbrackleft}False{\isacharbrackright}{\isachardoublequoteclose}\ \isacommand{by}\isamarkupfalse%
\ auto\isanewline
\ \ \isacommand{from}\isamarkupfalse%
\ sg{\isadigit{3}}\ \isakeyword{and}\ sg{\isadigit{4}}\ \isacommand{have}\isamarkupfalse%
\ sg{\isadigit{5}}{\isacharcolon}{\isachardoublequoteopen}{\isasymforall}j{\isasymle}{\isadigit{2}}\ {\isacharasterisk}\ d{\isachardot}\ lose\ {\isacharparenleft}t\ {\isacharplus}\ j\ {\isacharplus}\ {\isacharparenleft}{\isadigit{3}}\ {\isacharplus}\ k{\isacharparenright}{\isacharparenright}\ {\isacharequal}\ {\isacharbrackleft}False{\isacharbrackright}{\isachardoublequoteclose}\ \isanewline
\ \ \ \ \isacommand{by}\isamarkupfalse%
\ {\isacharparenleft}rule\ streamValue{\isadigit{4}}{\isacharparenright}\ \isanewline
\ \ \isacommand{from}\isamarkupfalse%
\ sg{\isadigit{5}}\ \isacommand{show}\isamarkupfalse%
\ {\isacharquery}thesis\ \isacommand{by}\isamarkupfalse%
\ {\isacharparenleft}rule\ streamValue{\isadigit{6}}{\isacharparenright}\ \isanewline
\isacommand{qed}\isamarkupfalse%
\endisatagproof
{\isafoldproof}%
\isadelimproof
\isanewline
\endisadelimproof
\isadelimtheory
\isanewline
\endisadelimtheory
\isatagtheory
\isacommand{end}\isamarkupfalse%
\endisatagtheory
{\isafoldtheory}%
\isadelimtheory
\endisadelimtheory
\end{isabellebody}%

%
\begin{isabellebody}%
\def\isabellecontext{Gateway{\isacharunderscore}proof}%
\isadelimtheory
\endisadelimtheory
\isatagtheory
\isanewline
\isanewline
\isanewline
\isacommand{theory}\isamarkupfalse%
\ Gateway{\isacharunderscore}proof\ \isanewline
\isakeyword{imports}\ Gateway{\isacharunderscore}proof{\isacharunderscore}aux\isanewline
\isakeyword{begin}%
\endisatagtheory
{\isafoldtheory}%
\isadelimtheory
\endisadelimtheory
\isamarkupsubsection{Properties of the Gateway%
}
\isamarkuptrue%
\isacommand{lemma}\isamarkupfalse%
\ Gateway{\isacharunderscore}L{\isadigit{1}}{\isacharcolon}\isanewline
\ \ \isakeyword{assumes}\ h{\isadigit{1}}{\isacharcolon}{\isachardoublequoteopen}Gateway\ req\ dt\ a\ stop\ lose\ d\ ack\ i\ vc{\isachardoublequoteclose}\isanewline
\ \ \ \ \ \ \isakeyword{and}\ h{\isadigit{2}}{\isacharcolon}{\isachardoublequoteopen}msg\ {\isacharparenleft}Suc\ {\isadigit{0}}{\isacharparenright}\ req{\isachardoublequoteclose}\isanewline
\ \ \ \ \ \ \isakeyword{and}\ h{\isadigit{3}}{\isacharcolon}{\isachardoublequoteopen}msg\ {\isacharparenleft}Suc\ {\isadigit{0}}{\isacharparenright}\ a{\isachardoublequoteclose}\isanewline
\ \ \ \ \ \ \isakeyword{and}\ h{\isadigit{4}}{\isacharcolon}{\isachardoublequoteopen}msg\ {\isacharparenleft}Suc\ {\isadigit{0}}{\isacharparenright}\ stop{\isachardoublequoteclose}\ \isanewline
\ \ \ \ \ \ \isakeyword{and}\ h{\isadigit{5}}{\isacharcolon}{\isachardoublequoteopen}ts\ lose{\isachardoublequoteclose}\isanewline
\ \ \ \ \ \ \isakeyword{and}\ h{\isadigit{6}}{\isacharcolon}{\isachardoublequoteopen}ack\ t\ {\isacharequal}\ {\isacharbrackleft}init{\isacharunderscore}state{\isacharbrackright}{\isachardoublequoteclose}\isanewline
\ \ \ \ \ \ \isakeyword{and}\ h{\isadigit{7}}{\isacharcolon}{\isachardoublequoteopen}req\ {\isacharparenleft}Suc\ t{\isacharparenright}\ {\isacharequal}\ {\isacharbrackleft}init{\isacharbrackright}{\isachardoublequoteclose}\isanewline
\ \ \ \ \ \ \isakeyword{and}\ h{\isadigit{8}}{\isacharcolon}{\isachardoublequoteopen}lose\ {\isacharparenleft}Suc\ t{\isacharparenright}\ {\isacharequal}\ {\isacharbrackleft}False{\isacharbrackright}{\isachardoublequoteclose}\isanewline
\ \ \ \ \ \ \isakeyword{and}\ h{\isadigit{9}}{\isacharcolon}{\isachardoublequoteopen}lose\ {\isacharparenleft}Suc\ {\isacharparenleft}Suc\ t{\isacharparenright}{\isacharparenright}\ {\isacharequal}\ {\isacharbrackleft}False{\isacharbrackright}{\isachardoublequoteclose}\isanewline
\ \ \isakeyword{shows}\ {\isachardoublequoteopen}ack\ {\isacharparenleft}Suc\ {\isacharparenleft}Suc\ t{\isacharparenright}{\isacharparenright}\ {\isacharequal}\ {\isacharbrackleft}connection{\isacharunderscore}ok{\isacharbrackright}{\isachardoublequoteclose}\ \isanewline
\isadelimproof
\endisadelimproof
\isatagproof
\isacommand{proof}\isamarkupfalse%
\ {\isacharminus}\ \isanewline
\ \ \isacommand{from}\isamarkupfalse%
\ h{\isadigit{1}}\ \isacommand{obtain}\isamarkupfalse%
\ i{\isadigit{1}}\ i{\isadigit{2}}\ x\ y\isanewline
\ \ \ \ \isakeyword{where}\ a{\isadigit{1}}{\isacharcolon}{\isachardoublequoteopen}Sample\ req\ dt\ x\ stop\ lose\ ack\ i{\isadigit{1}}\ vc{\isachardoublequoteclose}\isanewline
\ \ \ \ \ \ \isakeyword{and}\ a{\isadigit{2}}{\isacharcolon}{\isachardoublequoteopen}Delay\ y\ i{\isadigit{1}}\ d\ x\ i{\isadigit{2}}{\isachardoublequoteclose}\isanewline
\ \ \ \ \ \ \isakeyword{and}\ a{\isadigit{3}}{\isacharcolon}{\isachardoublequoteopen}Loss\ lose\ a\ i{\isadigit{2}}\ y\ i{\isachardoublequoteclose}\isanewline
\ \ \ \ \isacommand{by}\isamarkupfalse%
\ {\isacharparenleft}simp\ only{\isacharcolon}\ Gateway{\isacharunderscore}def{\isacharcomma}\ auto{\isacharparenright}\ \isanewline
\ \ \isacommand{from}\isamarkupfalse%
\ a{\isadigit{2}}\ \isakeyword{and}\ a{\isadigit{3}}\ \isakeyword{and}\ h{\isadigit{3}}\ \isacommand{have}\isamarkupfalse%
\ sg{\isadigit{1}}{\isacharcolon}{\isachardoublequoteopen}msg\ {\isacharparenleft}Suc\ {\isadigit{0}}{\isacharparenright}\ x{\isachardoublequoteclose}\ \ \ \isanewline
\ \ \ \ \isacommand{by}\isamarkupfalse%
\ {\isacharparenleft}simp\ add{\isacharcolon}\ Loss{\isacharunderscore}Delay{\isacharunderscore}msg{\isacharunderscore}a{\isacharparenright}\ \isanewline
\ \ \isacommand{from}\isamarkupfalse%
\ a{\isadigit{1}}\ \isakeyword{and}\ h{\isadigit{2}}\ \isakeyword{and}\ h{\isadigit{4}}\ \isakeyword{and}\ sg{\isadigit{1}}\ \isacommand{obtain}\isamarkupfalse%
\ st\ buffer\ \isakeyword{where}\ a{\isadigit{4}}{\isacharcolon}\isanewline
\ \ \ \ {\isachardoublequoteopen}tiTable{\isacharunderscore}SampleT\ req\ x\ stop\ lose\ \isanewline
\ \ \ \ \ \ \ \ {\isacharparenleft}fin{\isacharunderscore}inf{\isacharunderscore}append\ {\isacharbrackleft}init{\isacharunderscore}state{\isacharbrackright}\ st{\isacharparenright}\ {\isacharparenleft}fin{\isacharunderscore}inf{\isacharunderscore}append\ {\isacharbrackleft}{\isacharbrackleft}{\isacharbrackright}{\isacharbrackright}\ buffer{\isacharparenright}\ ack\isanewline
\ \ \ \ \ \ \ \ \ i{\isadigit{1}}\ vc\ st{\isachardoublequoteclose}\isanewline
\ \ \ \ \ \isacommand{by}\isamarkupfalse%
\ {\isacharparenleft}simp\ add{\isacharcolon}\ Sample{\isacharunderscore}def\ Sample{\isacharunderscore}L{\isacharunderscore}def{\isacharcomma}\ auto{\isacharparenright}\isanewline
\ \ \isacommand{from}\isamarkupfalse%
\ a{\isadigit{4}}\ \isakeyword{and}\ h{\isadigit{5}}\ \isakeyword{and}\ sg{\isadigit{1}}\ \isakeyword{and}\ h{\isadigit{4}}\ \isacommand{have}\isamarkupfalse%
\ sg{\isadigit{2}}{\isacharcolon}{\isachardoublequoteopen}st\ t\ {\isacharequal}\ \ hd\ {\isacharparenleft}ack\ t{\isacharparenright}{\isachardoublequoteclose}\isanewline
\ \ \ \ \isacommand{by}\isamarkupfalse%
\ {\isacharparenleft}simp\ add{\isacharcolon}\ tiTable{\isacharunderscore}ack{\isacharunderscore}st{\isacharunderscore}hd{\isacharparenright}\ \ \isanewline
\ \ \isacommand{from}\isamarkupfalse%
\ h{\isadigit{6}}\ \isakeyword{and}\ sg{\isadigit{1}}\ \isakeyword{and}\ sg{\isadigit{2}}\ \isakeyword{and}\ h{\isadigit{4}}\ \isacommand{have}\isamarkupfalse%
\ sg{\isadigit{3}}{\isacharcolon}\isanewline
\ \ \ {\isachardoublequoteopen}{\isacharparenleft}fin{\isacharunderscore}inf{\isacharunderscore}append\ {\isacharbrackleft}init{\isacharunderscore}state{\isacharbrackright}\ st{\isacharparenright}\ {\isacharparenleft}Suc\ t{\isacharparenright}\ {\isacharequal}\ init{\isacharunderscore}state{\isachardoublequoteclose}\isanewline
\ \ \ \ \isacommand{by}\isamarkupfalse%
\ {\isacharparenleft}simp\ add{\isacharcolon}\ correct{\isacharunderscore}fin{\isacharunderscore}inf{\isacharunderscore}append{\isadigit{1}}{\isacharparenright}\isanewline
\ \ \isacommand{from}\isamarkupfalse%
\ a{\isadigit{4}}\ \isakeyword{and}\ h{\isadigit{7}}\ \isakeyword{and}\ sg{\isadigit{3}}\ \isacommand{have}\isamarkupfalse%
\ sg{\isadigit{4}}{\isacharcolon}{\isachardoublequoteopen}st\ {\isacharparenleft}Suc\ t{\isacharparenright}\ {\isacharequal}\ call{\isachardoublequoteclose}\isanewline
\ \ \ \ \isacommand{by}\isamarkupfalse%
\ {\isacharparenleft}simp\ add{\isacharcolon}\ tiTable{\isacharunderscore}SampleT{\isacharunderscore}def{\isacharparenright}\isanewline
\ \ \isacommand{from}\isamarkupfalse%
\ sg{\isadigit{4}}\ \isacommand{have}\isamarkupfalse%
\ sg{\isadigit{5}}{\isacharcolon}{\isachardoublequoteopen}{\isacharparenleft}fin{\isacharunderscore}inf{\isacharunderscore}append\ {\isacharbrackleft}init{\isacharunderscore}state{\isacharbrackright}\ st{\isacharparenright}\ {\isacharparenleft}Suc\ {\isacharparenleft}Suc\ t{\isacharparenright}{\isacharparenright}\ {\isacharequal}\ call{\isachardoublequoteclose}\isanewline
\ \ \ \ \isacommand{by}\isamarkupfalse%
\ {\isacharparenleft}simp\ add{\isacharcolon}\ correct{\isacharunderscore}fin{\isacharunderscore}inf{\isacharunderscore}append{\isadigit{1}}{\isacharparenright}\isanewline
\ \ \isacommand{from}\isamarkupfalse%
\ a{\isadigit{4}}\ \isakeyword{and}\ sg{\isadigit{5}}\ \isakeyword{and}\ assms\ \isacommand{show}\isamarkupfalse%
\ {\isacharquery}thesis\ \isanewline
\ \ \ \ \isacommand{by}\isamarkupfalse%
\ {\isacharparenleft}simp\ add{\isacharcolon}\ tiTable{\isacharunderscore}SampleT{\isacharunderscore}def{\isacharparenright}\isanewline
\isacommand{qed}\isamarkupfalse%
\endisatagproof
{\isafoldproof}%
\isadelimproof
\isanewline
\endisadelimproof
\isanewline
\isanewline
\isanewline
\isanewline
\ \isacommand{lemma}\isamarkupfalse%
\ Gateway{\isacharunderscore}L{\isadigit{2}}{\isacharcolon}\isanewline
\ \ \isakeyword{assumes}\ h{\isadigit{1}}{\isacharcolon}{\isachardoublequoteopen}Gateway\ req\ dt\ a\ stop\ lose\ d\ ack\ i\ vc{\isachardoublequoteclose}\isanewline
\ \ \ \ \ \ \isakeyword{and}\ h{\isadigit{2}}{\isacharcolon}{\isachardoublequoteopen}msg\ {\isacharparenleft}Suc\ {\isadigit{0}}{\isacharparenright}\ req{\isachardoublequoteclose}\isanewline
\ \ \ \ \ \ \isakeyword{and}\ h{\isadigit{3}}{\isacharcolon}{\isachardoublequoteopen}msg\ {\isacharparenleft}Suc\ {\isadigit{0}}{\isacharparenright}\ a{\isachardoublequoteclose}\isanewline
\ \ \ \ \ \ \isakeyword{and}\ h{\isadigit{4}}{\isacharcolon}{\isachardoublequoteopen}msg\ {\isacharparenleft}Suc\ {\isadigit{0}}{\isacharparenright}\ stop{\isachardoublequoteclose}\ \isanewline
\ \ \ \ \ \ \isakeyword{and}\ h{\isadigit{5}}{\isacharcolon}{\isachardoublequoteopen}ts\ lose{\isachardoublequoteclose}\isanewline
\ \ \ \ \ \ \isakeyword{and}\ h{\isadigit{6}}{\isacharcolon}{\isachardoublequoteopen}ack\ t\ {\isacharequal}\ {\isacharbrackleft}connection{\isacharunderscore}ok{\isacharbrackright}{\isachardoublequoteclose}\isanewline
\ \ \ \ \ \ \isakeyword{and}\ h{\isadigit{7}}{\isacharcolon}{\isachardoublequoteopen}req\ {\isacharparenleft}Suc\ t{\isacharparenright}\ {\isacharequal}\ {\isacharbrackleft}send{\isacharbrackright}{\isachardoublequoteclose}\isanewline
\ \ \ \ \ \ \isakeyword{and}\ h{\isadigit{8}}{\isacharcolon}{\isachardoublequoteopen}{\isasymforall}k{\isasymle}Suc\ d{\isachardot}\ lose\ {\isacharparenleft}t\ {\isacharplus}\ k{\isacharparenright}\ {\isacharequal}\ {\isacharbrackleft}False{\isacharbrackright}{\isachardoublequoteclose}\ \isanewline
\ \ \isakeyword{shows}\ {\isachardoublequoteopen}i\ {\isacharparenleft}Suc\ {\isacharparenleft}t\ {\isacharplus}\ d{\isacharparenright}{\isacharparenright}\ \ {\isacharequal}\ inf{\isacharunderscore}last{\isacharunderscore}ti\ dt\ t{\isachardoublequoteclose}\isanewline
\isadelimproof
\endisadelimproof
\isatagproof
\isacommand{proof}\isamarkupfalse%
\ {\isacharminus}\ \isanewline
\ \ \isacommand{from}\isamarkupfalse%
\ h{\isadigit{1}}\ \isacommand{obtain}\isamarkupfalse%
\ i{\isadigit{1}}\ i{\isadigit{2}}\ x\ y\isanewline
\ \ \ \ \isakeyword{where}\ a{\isadigit{1}}{\isacharcolon}{\isachardoublequoteopen}Sample\ req\ dt\ x\ stop\ lose\ ack\ i{\isadigit{1}}\ vc{\isachardoublequoteclose}\isanewline
\ \ \ \ \ \ \isakeyword{and}\ a{\isadigit{2}}{\isacharcolon}{\isachardoublequoteopen}Delay\ y\ i{\isadigit{1}}\ d\ x\ i{\isadigit{2}}{\isachardoublequoteclose}\isanewline
\ \ \ \ \ \ \isakeyword{and}\ a{\isadigit{3}}{\isacharcolon}{\isachardoublequoteopen}Loss\ lose\ a\ i{\isadigit{2}}\ y\ i{\isachardoublequoteclose}\isanewline
\ \ \ \ \isacommand{by}\isamarkupfalse%
\ {\isacharparenleft}simp\ only{\isacharcolon}\ Gateway{\isacharunderscore}def{\isacharcomma}\ auto{\isacharparenright}\ \isanewline
\ \ \isacommand{from}\isamarkupfalse%
\ a{\isadigit{2}}\ \isakeyword{and}\ a{\isadigit{3}}\ \isakeyword{and}\ h{\isadigit{3}}\ \isacommand{have}\isamarkupfalse%
\ sg{\isadigit{1}}{\isacharcolon}{\isachardoublequoteopen}msg\ {\isacharparenleft}Suc\ {\isadigit{0}}{\isacharparenright}\ x{\isachardoublequoteclose}\ \ \ \isanewline
\ \ \ \ \isacommand{by}\isamarkupfalse%
\ {\isacharparenleft}simp\ add{\isacharcolon}\ Loss{\isacharunderscore}Delay{\isacharunderscore}msg{\isacharunderscore}a{\isacharparenright}\ \isanewline
\ \ \isacommand{from}\isamarkupfalse%
\ a{\isadigit{1}}\ \isakeyword{and}\ h{\isadigit{2}}\ \isakeyword{and}\ h{\isadigit{4}}\ \isakeyword{and}\ sg{\isadigit{1}}\ \isacommand{obtain}\isamarkupfalse%
\ st\ buffer\ \isakeyword{where}\ a{\isadigit{4}}{\isacharcolon}\isanewline
\ \ \ \ {\isachardoublequoteopen}Sample{\isacharunderscore}L\ req\ dt\ x\ stop\ lose\ {\isacharparenleft}fin{\isacharunderscore}inf{\isacharunderscore}append\ {\isacharbrackleft}init{\isacharunderscore}state{\isacharbrackright}\ st{\isacharparenright}\ \isanewline
\ \ \ \ \ \ \ \ \ {\isacharparenleft}fin{\isacharunderscore}inf{\isacharunderscore}append\ {\isacharbrackleft}{\isacharbrackleft}{\isacharbrackright}{\isacharbrackright}\ buffer{\isacharparenright}\ ack\ i{\isadigit{1}}\ vc\ st\ buffer{\isachardoublequoteclose}\isanewline
\ \ \ \ \isacommand{by}\isamarkupfalse%
\ {\isacharparenleft}simp\ add{\isacharcolon}\ Sample{\isacharunderscore}def{\isacharcomma}\ auto{\isacharparenright}\isanewline
\ \ \isacommand{from}\isamarkupfalse%
\ a{\isadigit{4}}\ \isacommand{have}\isamarkupfalse%
\ sg{\isadigit{2}}{\isacharcolon}{\isachardoublequoteopen}buffer\ t\ {\isacharequal}\ inf{\isacharunderscore}last{\isacharunderscore}ti\ dt\ t{\isachardoublequoteclose}\isanewline
\ \ \ \ \isacommand{by}\isamarkupfalse%
\ {\isacharparenleft}simp\ add{\isacharcolon}\ Sample{\isacharunderscore}L{\isacharunderscore}buffer{\isacharparenright}\ \isanewline
\ \ \isacommand{from}\isamarkupfalse%
\ assms\ \isakeyword{and}\ a{\isadigit{1}}\ \isakeyword{and}\ a{\isadigit{4}}\ \isakeyword{and}\ sg{\isadigit{1}}\ \isakeyword{and}\ sg{\isadigit{2}}\ \isacommand{have}\isamarkupfalse%
\ sg{\isadigit{3}}{\isacharcolon}{\isachardoublequoteopen}i{\isadigit{1}}\ {\isacharparenleft}Suc\ t{\isacharparenright}\ {\isacharequal}\ \ buffer\ t{\isachardoublequoteclose}\isanewline
\ \ \ \ \isacommand{by}\isamarkupfalse%
\ {\isacharparenleft}simp\ add{\isacharcolon}\ Sample{\isacharunderscore}L{\isacharunderscore}i{\isadigit{1}}{\isacharunderscore}buffer{\isacharparenright}\isanewline
\ \ \isacommand{from}\isamarkupfalse%
\ a{\isadigit{2}}\ \isakeyword{and}\ sg{\isadigit{1}}\ \isacommand{have}\isamarkupfalse%
\ sg{\isadigit{4}}{\isacharcolon}{\isachardoublequoteopen}i{\isadigit{2}}\ {\isacharparenleft}{\isacharparenleft}Suc\ t{\isacharparenright}\ {\isacharplus}\ d{\isacharparenright}\ \ {\isacharequal}\ \ i{\isadigit{1}}\ {\isacharparenleft}Suc\ t{\isacharparenright}{\isachardoublequoteclose}\isanewline
\ \ \ \ \isacommand{by}\isamarkupfalse%
\ {\isacharparenleft}simp\ add{\isacharcolon}\ Delay{\isacharunderscore}def{\isacharparenright}\isanewline
\ \ \isacommand{from}\isamarkupfalse%
\ a{\isadigit{3}}\ \isakeyword{and}\ h{\isadigit{8}}\ \isacommand{have}\isamarkupfalse%
\ sg{\isadigit{5}}{\isacharcolon}{\isachardoublequoteopen}i\ {\isacharparenleft}{\isacharparenleft}Suc\ t{\isacharparenright}\ {\isacharplus}\ d{\isacharparenright}\ \ {\isacharequal}\ \ i{\isadigit{2}}\ {\isacharparenleft}{\isacharparenleft}Suc\ t{\isacharparenright}\ {\isacharplus}\ d{\isacharparenright}{\isachardoublequoteclose}\isanewline
\ \ \ \ \isacommand{by}\isamarkupfalse%
\ {\isacharparenleft}simp\ add{\isacharcolon}\ Loss{\isacharunderscore}def{\isacharcomma}\ auto{\isacharparenright}\isanewline
\ \ \isacommand{from}\isamarkupfalse%
\ sg{\isadigit{5}}\ \isakeyword{and}\ sg{\isadigit{4}}\ \isakeyword{and}\ sg{\isadigit{3}}\ \isakeyword{and}\ sg{\isadigit{2}}\ \isacommand{show}\isamarkupfalse%
\ {\isacharquery}thesis\ \isacommand{by}\isamarkupfalse%
\ simp\isanewline
\isacommand{qed}\isamarkupfalse%
\endisatagproof
{\isafoldproof}%
\isadelimproof
\isanewline
\endisadelimproof
\isanewline
\isacommand{lemma}\isamarkupfalse%
\ Gateway{\isacharunderscore}L{\isadigit{3}}{\isacharcolon}\isanewline
\ \ \isakeyword{assumes}\ h{\isadigit{1}}{\isacharcolon}{\isachardoublequoteopen}Gateway\ req\ dt\ a\ stop\ lose\ d\ ack\ i\ vc{\isachardoublequoteclose}\isanewline
\ \ \ \ \ \ \isakeyword{and}\ h{\isadigit{2}}{\isacharcolon}{\isachardoublequoteopen}msg\ {\isacharparenleft}Suc\ {\isadigit{0}}{\isacharparenright}\ req{\isachardoublequoteclose}\isanewline
\ \ \ \ \ \ \isakeyword{and}\ h{\isadigit{3}}{\isacharcolon}{\isachardoublequoteopen}msg\ {\isacharparenleft}Suc\ {\isadigit{0}}{\isacharparenright}\ a{\isachardoublequoteclose}\isanewline
\ \ \ \ \ \ \isakeyword{and}\ h{\isadigit{4}}{\isacharcolon}{\isachardoublequoteopen}msg\ {\isacharparenleft}Suc\ {\isadigit{0}}{\isacharparenright}\ stop{\isachardoublequoteclose}\ \isanewline
\ \ \ \ \ \ \isakeyword{and}\ h{\isadigit{5}}{\isacharcolon}{\isachardoublequoteopen}ts\ lose{\isachardoublequoteclose}\isanewline
\ \ \ \ \ \ \isakeyword{and}\ h{\isadigit{6}}{\isacharcolon}{\isachardoublequoteopen}ack\ t\ {\isacharequal}\ {\isacharbrackleft}connection{\isacharunderscore}ok{\isacharbrackright}{\isachardoublequoteclose}\isanewline
\ \ \ \ \ \ \isakeyword{and}\ h{\isadigit{7}}{\isacharcolon}{\isachardoublequoteopen}req\ {\isacharparenleft}Suc\ t{\isacharparenright}\ {\isacharequal}\ {\isacharbrackleft}send{\isacharbrackright}{\isachardoublequoteclose}\isanewline
\ \ \ \ \ \ \isakeyword{and}\ h{\isadigit{8}}{\isacharcolon}{\isachardoublequoteopen}{\isasymforall}k{\isasymle}Suc\ d{\isachardot}\ lose\ {\isacharparenleft}t\ {\isacharplus}\ k{\isacharparenright}\ {\isacharequal}\ {\isacharbrackleft}False{\isacharbrackright}{\isachardoublequoteclose}\ \isanewline
\ \ \isakeyword{shows}\ {\isachardoublequoteopen}ack\ {\isacharparenleft}Suc\ t{\isacharparenright}\ {\isacharequal}\ {\isacharbrackleft}sending{\isacharunderscore}data{\isacharbrackright}{\isachardoublequoteclose}\isanewline
\isadelimproof
\endisadelimproof
\isatagproof
\isacommand{proof}\isamarkupfalse%
\ {\isacharminus}\ \isanewline
\ \ \isacommand{from}\isamarkupfalse%
\ h{\isadigit{1}}\ \isacommand{obtain}\isamarkupfalse%
\ i{\isadigit{1}}\ i{\isadigit{2}}\ x\ y\isanewline
\ \ \ \ \isakeyword{where}\ a{\isadigit{1}}{\isacharcolon}{\isachardoublequoteopen}Sample\ req\ dt\ x\ stop\ lose\ ack\ i{\isadigit{1}}\ vc{\isachardoublequoteclose}\isanewline
\ \ \ \ \ \ \isakeyword{and}\ a{\isadigit{2}}{\isacharcolon}{\isachardoublequoteopen}Delay\ y\ i{\isadigit{1}}\ d\ x\ i{\isadigit{2}}{\isachardoublequoteclose}\isanewline
\ \ \ \ \ \ \isakeyword{and}\ a{\isadigit{3}}{\isacharcolon}{\isachardoublequoteopen}Loss\ lose\ a\ i{\isadigit{2}}\ y\ i{\isachardoublequoteclose}\isanewline
\ \ \ \ \isacommand{by}\isamarkupfalse%
\ {\isacharparenleft}simp\ only{\isacharcolon}\ Gateway{\isacharunderscore}def{\isacharcomma}\ auto{\isacharparenright}\ \isanewline
\ \ \isacommand{from}\isamarkupfalse%
\ a{\isadigit{2}}\ \isakeyword{and}\ a{\isadigit{3}}\ \isakeyword{and}\ h{\isadigit{3}}\ \isacommand{have}\isamarkupfalse%
\ sg{\isadigit{1}}{\isacharcolon}{\isachardoublequoteopen}msg\ {\isacharparenleft}Suc\ {\isadigit{0}}{\isacharparenright}\ x{\isachardoublequoteclose}\ \ \ \isanewline
\ \ \ \ \isacommand{by}\isamarkupfalse%
\ {\isacharparenleft}simp\ add{\isacharcolon}\ Loss{\isacharunderscore}Delay{\isacharunderscore}msg{\isacharunderscore}a{\isacharparenright}\ \isanewline
\ \ \isacommand{from}\isamarkupfalse%
\ a{\isadigit{1}}\ \isakeyword{and}\ h{\isadigit{2}}\ \isakeyword{and}\ h{\isadigit{4}}\ \isakeyword{and}\ sg{\isadigit{1}}\ \isacommand{obtain}\isamarkupfalse%
\ st\ buffer\ \isakeyword{where}\ a{\isadigit{4}}{\isacharcolon}\isanewline
\ \ \ \ {\isachardoublequoteopen}tiTable{\isacharunderscore}SampleT\ req\ x\ stop\ lose\ \isanewline
\ \ \ \ \ \ \ \ {\isacharparenleft}fin{\isacharunderscore}inf{\isacharunderscore}append\ {\isacharbrackleft}init{\isacharunderscore}state{\isacharbrackright}\ st{\isacharparenright}\ {\isacharparenleft}fin{\isacharunderscore}inf{\isacharunderscore}append\ {\isacharbrackleft}{\isacharbrackleft}{\isacharbrackright}{\isacharbrackright}\ buffer{\isacharparenright}\ ack\isanewline
\ \ \ \ \ \ \ \ \ i{\isadigit{1}}\ vc\ st{\isachardoublequoteclose}\isanewline
\ \ \ \ \ \isacommand{by}\isamarkupfalse%
\ {\isacharparenleft}simp\ add{\isacharcolon}\ Sample{\isacharunderscore}def\ Sample{\isacharunderscore}L{\isacharunderscore}def{\isacharcomma}\ auto{\isacharparenright}\isanewline
\ \ \isacommand{from}\isamarkupfalse%
\ a{\isadigit{4}}\ \isakeyword{and}\ h{\isadigit{5}}\ \isakeyword{and}\ sg{\isadigit{1}}\ \isakeyword{and}\ h{\isadigit{4}}\ \isacommand{have}\isamarkupfalse%
\ sg{\isadigit{2}}{\isacharcolon}{\isachardoublequoteopen}st\ t\ {\isacharequal}\ \ hd\ {\isacharparenleft}ack\ t{\isacharparenright}{\isachardoublequoteclose}\isanewline
\ \ \ \ \isacommand{by}\isamarkupfalse%
\ {\isacharparenleft}simp\ add{\isacharcolon}\ tiTable{\isacharunderscore}ack{\isacharunderscore}st{\isacharunderscore}hd{\isacharparenright}\ \ \isanewline
\ \ \isacommand{from}\isamarkupfalse%
\ sg{\isadigit{2}}\ \isakeyword{and}\ h{\isadigit{6}}\ \isacommand{have}\isamarkupfalse%
\ sg{\isadigit{3}}{\isacharcolon}{\isachardoublequoteopen}{\isacharparenleft}fin{\isacharunderscore}inf{\isacharunderscore}append\ {\isacharbrackleft}init{\isacharunderscore}state{\isacharbrackright}\ st{\isacharparenright}\ {\isacharparenleft}Suc\ t{\isacharparenright}\ {\isacharequal}\ connection{\isacharunderscore}ok{\isachardoublequoteclose}\isanewline
\ \ \ \ \isacommand{by}\isamarkupfalse%
\ {\isacharparenleft}simp\ add{\isacharcolon}\ correct{\isacharunderscore}fin{\isacharunderscore}inf{\isacharunderscore}append{\isadigit{1}}{\isacharparenright}\isanewline
\ \ \isacommand{from}\isamarkupfalse%
\ h{\isadigit{8}}\ \isacommand{have}\isamarkupfalse%
\ sg{\isadigit{4}}{\isacharcolon}{\isachardoublequoteopen}lose\ {\isacharparenleft}Suc\ t{\isacharparenright}\ {\isacharequal}\ {\isacharbrackleft}False{\isacharbrackright}{\isachardoublequoteclose}\ \isacommand{by}\isamarkupfalse%
\ auto\isanewline
\ \ \isacommand{from}\isamarkupfalse%
\ a{\isadigit{4}}\ \isakeyword{and}\ sg{\isadigit{3}}\ \isakeyword{and}\ sg{\isadigit{4}}\ \isakeyword{and}\ h{\isadigit{7}}\ \isacommand{have}\isamarkupfalse%
\ sg{\isadigit{5}}{\isacharcolon}{\isachardoublequoteopen}st\ {\isacharparenleft}Suc\ t{\isacharparenright}\ {\isacharequal}\ sending{\isacharunderscore}data{\isachardoublequoteclose}\isanewline
\ \ \ \ \isacommand{by}\isamarkupfalse%
\ {\isacharparenleft}simp\ add{\isacharcolon}\ tiTable{\isacharunderscore}SampleT{\isacharunderscore}def{\isacharparenright}\ \isanewline
\ \ \isacommand{from}\isamarkupfalse%
\ a{\isadigit{4}}\ \isakeyword{and}\ h{\isadigit{2}}\ \isakeyword{and}\ sg{\isadigit{1}}\ \isakeyword{and}\ h{\isadigit{4}}\ \isakeyword{and}\ h{\isadigit{5}}\ \isacommand{have}\isamarkupfalse%
\ sg{\isadigit{6}}{\isacharcolon}{\isachardoublequoteopen}ack\ {\isacharparenleft}Suc\ t{\isacharparenright}\ {\isacharequal}\ {\isacharbrackleft}st\ {\isacharparenleft}Suc\ t{\isacharparenright}{\isacharbrackright}{\isachardoublequoteclose}\isanewline
\ \ \ \ \isacommand{by}\isamarkupfalse%
\ {\isacharparenleft}simp\ add{\isacharcolon}\ tiTable{\isacharunderscore}ack{\isacharunderscore}st{\isacharparenright}\isanewline
\ \ \isacommand{from}\isamarkupfalse%
\ sg{\isadigit{5}}\ \isakeyword{and}\ sg{\isadigit{6}}\ \isacommand{show}\isamarkupfalse%
\ {\isacharquery}thesis\ \isacommand{by}\isamarkupfalse%
\ simp\ \ \isanewline
\isacommand{qed}\isamarkupfalse%
\endisatagproof
{\isafoldproof}%
\isadelimproof
\isanewline
\endisadelimproof
\isanewline
\isacommand{lemma}\isamarkupfalse%
\ Gateway{\isacharunderscore}L{\isadigit{4}}{\isacharcolon}\isanewline
\ \ \isakeyword{assumes}\ h{\isadigit{1}}{\isacharcolon}{\isachardoublequoteopen}Gateway\ req\ dt\ a\ stop\ lose\ d\ ack\ i\ vc{\isachardoublequoteclose}\isanewline
\ \ \ \ \ \ \isakeyword{and}\ h{\isadigit{2}}{\isacharcolon}{\isachardoublequoteopen}msg\ {\isacharparenleft}Suc\ {\isadigit{0}}{\isacharparenright}\ req{\isachardoublequoteclose}\isanewline
\ \ \ \ \ \ \isakeyword{and}\ h{\isadigit{3}}{\isacharcolon}{\isachardoublequoteopen}msg\ {\isacharparenleft}Suc\ {\isadigit{0}}{\isacharparenright}\ a{\isachardoublequoteclose}\isanewline
\ \ \ \ \ \ \isakeyword{and}\ h{\isadigit{4}}{\isacharcolon}{\isachardoublequoteopen}msg\ {\isacharparenleft}Suc\ {\isadigit{0}}{\isacharparenright}\ stop{\isachardoublequoteclose}\ \isanewline
\ \ \ \ \ \ \isakeyword{and}\ h{\isadigit{5}}{\isacharcolon}{\isachardoublequoteopen}ts\ lose{\isachardoublequoteclose}\isanewline
\ \ \ \ \ \ \isakeyword{and}\ h{\isadigit{6}}{\isacharcolon}{\isachardoublequoteopen}ack\ {\isacharparenleft}t\ {\isacharplus}\ d{\isacharparenright}\ {\isacharequal}\ {\isacharbrackleft}sending{\isacharunderscore}data{\isacharbrackright}{\isachardoublequoteclose}\isanewline
\ \ \ \ \ \ \isakeyword{and}\ h{\isadigit{7}}{\isacharcolon}{\isachardoublequoteopen}a\ {\isacharparenleft}Suc\ t{\isacharparenright}\ {\isacharequal}\ {\isacharbrackleft}sc{\isacharunderscore}ack{\isacharbrackright}{\isachardoublequoteclose}\isanewline
\ \ \ \ \ \ \isakeyword{and}\ h{\isadigit{8}}{\isacharcolon}{\isachardoublequoteopen}{\isasymforall}k{\isasymle}Suc\ d{\isachardot}\ lose\ {\isacharparenleft}t\ {\isacharplus}\ k{\isacharparenright}\ {\isacharequal}\ {\isacharbrackleft}False{\isacharbrackright}{\isachardoublequoteclose}\ \isanewline
\ \ \isakeyword{shows}\ {\isachardoublequoteopen}vc\ {\isacharparenleft}Suc\ {\isacharparenleft}t\ {\isacharplus}\ d{\isacharparenright}{\isacharparenright}\ {\isacharequal}\ {\isacharbrackleft}vc{\isacharunderscore}com{\isacharbrackright}{\isachardoublequoteclose}\isanewline
\isadelimproof
\endisadelimproof
\isatagproof
\isacommand{proof}\isamarkupfalse%
\ {\isacharminus}\ \isanewline
\ \ \isacommand{from}\isamarkupfalse%
\ h{\isadigit{1}}\ \isacommand{obtain}\isamarkupfalse%
\ i{\isadigit{1}}\ i{\isadigit{2}}\ x\ y\isanewline
\ \ \ \ \isakeyword{where}\ a{\isadigit{1}}{\isacharcolon}{\isachardoublequoteopen}Sample\ req\ dt\ x\ stop\ lose\ ack\ i{\isadigit{1}}\ vc{\isachardoublequoteclose}\isanewline
\ \ \ \ \ \ \isakeyword{and}\ a{\isadigit{2}}{\isacharcolon}{\isachardoublequoteopen}Delay\ y\ i{\isadigit{1}}\ d\ x\ i{\isadigit{2}}{\isachardoublequoteclose}\isanewline
\ \ \ \ \ \ \isakeyword{and}\ a{\isadigit{3}}{\isacharcolon}{\isachardoublequoteopen}Loss\ lose\ a\ i{\isadigit{2}}\ y\ i{\isachardoublequoteclose}\isanewline
\ \ \ \ \isacommand{by}\isamarkupfalse%
\ {\isacharparenleft}simp\ only{\isacharcolon}\ Gateway{\isacharunderscore}def{\isacharcomma}\ auto{\isacharparenright}\ \isanewline
\ \ \isacommand{from}\isamarkupfalse%
\ a{\isadigit{2}}\ \isakeyword{and}\ a{\isadigit{3}}\ \isakeyword{and}\ h{\isadigit{3}}\ \isacommand{have}\isamarkupfalse%
\ sg{\isadigit{1}}{\isacharcolon}{\isachardoublequoteopen}msg\ {\isacharparenleft}Suc\ {\isadigit{0}}{\isacharparenright}\ x{\isachardoublequoteclose}\ \ \ \isanewline
\ \ \ \ \isacommand{by}\isamarkupfalse%
\ {\isacharparenleft}simp\ add{\isacharcolon}\ Loss{\isacharunderscore}Delay{\isacharunderscore}msg{\isacharunderscore}a{\isacharparenright}\ \isanewline
\ \ \isacommand{from}\isamarkupfalse%
\ a{\isadigit{1}}\ \isakeyword{and}\ h{\isadigit{2}}\ \isakeyword{and}\ h{\isadigit{4}}\ \isakeyword{and}\ sg{\isadigit{1}}\ \isacommand{obtain}\isamarkupfalse%
\ st\ buffer\ \isakeyword{where}\ a{\isadigit{4}}{\isacharcolon}\isanewline
\ \ \ \ {\isachardoublequoteopen}tiTable{\isacharunderscore}SampleT\ req\ x\ stop\ lose\ \isanewline
\ \ \ \ \ \ \ \ {\isacharparenleft}fin{\isacharunderscore}inf{\isacharunderscore}append\ {\isacharbrackleft}init{\isacharunderscore}state{\isacharbrackright}\ st{\isacharparenright}\ {\isacharparenleft}fin{\isacharunderscore}inf{\isacharunderscore}append\ {\isacharbrackleft}{\isacharbrackleft}{\isacharbrackright}{\isacharbrackright}\ buffer{\isacharparenright}\ ack\isanewline
\ \ \ \ \ \ \ \ \ i{\isadigit{1}}\ vc\ st{\isachardoublequoteclose}\isanewline
\ \ \ \ \ \isacommand{by}\isamarkupfalse%
\ {\isacharparenleft}simp\ add{\isacharcolon}\ Sample{\isacharunderscore}def\ Sample{\isacharunderscore}L{\isacharunderscore}def{\isacharcomma}\ auto{\isacharparenright}\isanewline
\ \ \isacommand{from}\isamarkupfalse%
\ a{\isadigit{4}}\ \isakeyword{and}\ h{\isadigit{5}}\ \isakeyword{and}\ sg{\isadigit{1}}\ \isakeyword{and}\ h{\isadigit{4}}\ \isacommand{have}\isamarkupfalse%
\ sg{\isadigit{2}}{\isacharcolon}{\isachardoublequoteopen}st\ {\isacharparenleft}t{\isacharplus}d{\isacharparenright}\ {\isacharequal}\ \ hd\ {\isacharparenleft}ack\ {\isacharparenleft}t{\isacharplus}d{\isacharparenright}{\isacharparenright}{\isachardoublequoteclose}\isanewline
\ \ \ \ \isacommand{by}\isamarkupfalse%
\ {\isacharparenleft}simp\ add{\isacharcolon}\ tiTable{\isacharunderscore}ack{\isacharunderscore}st{\isacharunderscore}hd{\isacharparenright}\ \ \isanewline
\ \ \isacommand{from}\isamarkupfalse%
\ sg{\isadigit{2}}\ \isakeyword{and}\ h{\isadigit{6}}\ \isacommand{have}\isamarkupfalse%
\ sg{\isadigit{3}}{\isacharcolon}{\isachardoublequoteopen}{\isacharparenleft}fin{\isacharunderscore}inf{\isacharunderscore}append\ {\isacharbrackleft}init{\isacharunderscore}state{\isacharbrackright}\ st{\isacharparenright}\ {\isacharparenleft}Suc\ {\isacharparenleft}t{\isacharplus}d{\isacharparenright}{\isacharparenright}\ {\isacharequal}\ sending{\isacharunderscore}data{\isachardoublequoteclose}\isanewline
\ \ \ \ \isacommand{by}\isamarkupfalse%
\ {\isacharparenleft}simp\ add{\isacharcolon}\ correct{\isacharunderscore}fin{\isacharunderscore}inf{\isacharunderscore}append{\isadigit{1}}{\isacharparenright}\isanewline
\ \ \isacommand{from}\isamarkupfalse%
\ a{\isadigit{3}}\ \isakeyword{and}\ h{\isadigit{8}}\ \isacommand{have}\isamarkupfalse%
\ sg{\isadigit{4}}{\isacharcolon}{\isachardoublequoteopen}y\ {\isacharparenleft}Suc\ t{\isacharparenright}\ \ {\isacharequal}\ \ a\ {\isacharparenleft}Suc\ t{\isacharparenright}{\isachardoublequoteclose}\isanewline
\ \ \ \ \isacommand{by}\isamarkupfalse%
\ {\isacharparenleft}simp\ add{\isacharcolon}\ Loss{\isacharunderscore}def{\isacharcomma}\ auto{\isacharparenright}\isanewline
\ \ \isacommand{from}\isamarkupfalse%
\ a{\isadigit{2}}\ \isakeyword{and}\ sg{\isadigit{1}}\ \isacommand{have}\isamarkupfalse%
\ sg{\isadigit{5}}{\isacharcolon}{\isachardoublequoteopen}x\ {\isacharparenleft}{\isacharparenleft}Suc\ t{\isacharparenright}\ {\isacharplus}\ d{\isacharparenright}\ \ {\isacharequal}\ \ y\ {\isacharparenleft}Suc\ t{\isacharparenright}{\isachardoublequoteclose}\isanewline
\ \ \ \ \isacommand{by}\isamarkupfalse%
\ {\isacharparenleft}simp\ add{\isacharcolon}\ Delay{\isacharunderscore}def{\isacharparenright}\ \isanewline
\ \ \isacommand{from}\isamarkupfalse%
\ sg{\isadigit{5}}\ \isakeyword{and}\ sg{\isadigit{4}}\ \isakeyword{and}\ h{\isadigit{7}}\ \isacommand{have}\isamarkupfalse%
\ sg{\isadigit{6}}{\isacharcolon}{\isachardoublequoteopen}\ x\ {\isacharparenleft}Suc\ {\isacharparenleft}t\ {\isacharplus}\ d{\isacharparenright}{\isacharparenright}\ {\isacharequal}\ {\isacharbrackleft}sc{\isacharunderscore}ack{\isacharbrackright}{\isachardoublequoteclose}\ \isacommand{by}\isamarkupfalse%
\ simp\isanewline
\ \ \isacommand{from}\isamarkupfalse%
\ h{\isadigit{8}}\ \isacommand{have}\isamarkupfalse%
\ sg{\isadigit{7}}{\isacharcolon}{\isachardoublequoteopen}lose\ {\isacharparenleft}Suc\ {\isacharparenleft}t\ {\isacharplus}\ d{\isacharparenright}{\isacharparenright}\ {\isacharequal}\ {\isacharbrackleft}False{\isacharbrackright}{\isachardoublequoteclose}\ \isacommand{by}\isamarkupfalse%
\ auto\isanewline
\ \ \isacommand{from}\isamarkupfalse%
\ sg{\isadigit{6}}\ \isakeyword{and}\ a{\isadigit{4}}\ \isakeyword{and}\ h{\isadigit{2}}\ \isakeyword{and}\ sg{\isadigit{1}}\ \isakeyword{and}\ h{\isadigit{4}}\ \isakeyword{and}\ h{\isadigit{5}}\ \isakeyword{and}\ sg{\isadigit{7}}\ \isakeyword{and}\ sg{\isadigit{3}}\ \isacommand{show}\isamarkupfalse%
\ {\isacharquery}thesis\isanewline
\ \ \ \ \isacommand{by}\isamarkupfalse%
\ {\isacharparenleft}simp\ add{\isacharcolon}\ tiTable{\isacharunderscore}SampleT{\isacharunderscore}def{\isacharparenright}\ \isanewline
\isacommand{qed}\isamarkupfalse%
\endisatagproof
{\isafoldproof}%
\isadelimproof
\isanewline
\endisadelimproof
\isanewline
\isacommand{lemma}\isamarkupfalse%
\ Gateway{\isacharunderscore}L{\isadigit{5}}{\isacharcolon}\isanewline
\ \ \isakeyword{assumes}\ h{\isadigit{1}}{\isacharcolon}{\isachardoublequoteopen}Gateway\ req\ dt\ a\ stop\ lose\ d\ ack\ i\ vc{\isachardoublequoteclose}\isanewline
\ \ \ \ \ \ \isakeyword{and}\ h{\isadigit{2}}{\isacharcolon}{\isachardoublequoteopen}msg\ {\isacharparenleft}Suc\ {\isadigit{0}}{\isacharparenright}\ req{\isachardoublequoteclose}\isanewline
\ \ \ \ \ \ \isakeyword{and}\ h{\isadigit{3}}{\isacharcolon}{\isachardoublequoteopen}msg\ {\isacharparenleft}Suc\ {\isadigit{0}}{\isacharparenright}\ a{\isachardoublequoteclose}\isanewline
\ \ \ \ \ \ \isakeyword{and}\ h{\isadigit{4}}{\isacharcolon}{\isachardoublequoteopen}msg\ {\isacharparenleft}Suc\ {\isadigit{0}}{\isacharparenright}\ stop{\isachardoublequoteclose}\ \isanewline
\ \ \ \ \ \ \isakeyword{and}\ h{\isadigit{5}}{\isacharcolon}{\isachardoublequoteopen}ts\ lose{\isachardoublequoteclose}\isanewline
\ \ \ \ \ \ \isakeyword{and}\ h{\isadigit{6}}{\isacharcolon}{\isachardoublequoteopen}ack\ {\isacharparenleft}t\ {\isacharplus}\ d{\isacharparenright}\ {\isacharequal}\ {\isacharbrackleft}sending{\isacharunderscore}data{\isacharbrackright}{\isachardoublequoteclose}\isanewline
\ \ \ \ \ \ \isakeyword{and}\ h{\isadigit{7}}{\isacharcolon}{\isachardoublequoteopen}{\isasymforall}\ j\ {\isasymle}\ Suc\ d{\isachardot}\ a\ {\isacharparenleft}t{\isacharplus}j{\isacharparenright}\ {\isacharequal}\ {\isacharbrackleft}{\isacharbrackright}{\isachardoublequoteclose}\isanewline
\ \ \ \ \ \ \isakeyword{and}\ h{\isadigit{8}}{\isacharcolon}{\isachardoublequoteopen}{\isasymforall}k{\isasymle}\ {\isacharparenleft}d\ {\isacharplus}\ d{\isacharparenright}{\isachardot}\ lose\ {\isacharparenleft}t\ {\isacharplus}\ k{\isacharparenright}\ {\isacharequal}\ {\isacharbrackleft}False{\isacharbrackright}{\isachardoublequoteclose}\ \isanewline
\ \ \isakeyword{shows}\ {\isachardoublequoteopen}j\ {\isasymle}\ d\ {\isasymlongrightarrow}\ ack\ {\isacharparenleft}t{\isacharplus}d{\isacharplus}j{\isacharparenright}\ {\isacharequal}\ {\isacharbrackleft}sending{\isacharunderscore}data{\isacharbrackright}{\isachardoublequoteclose}\isanewline
\isadelimproof
\endisadelimproof
\isatagproof
\isacommand{proof}\isamarkupfalse%
\ {\isacharminus}\ \isanewline
\ \ \isacommand{from}\isamarkupfalse%
\ h{\isadigit{1}}\ \isacommand{obtain}\isamarkupfalse%
\ i{\isadigit{1}}\ i{\isadigit{2}}\ x\ y\isanewline
\ \ \ \ \isakeyword{where}\ a{\isadigit{1}}{\isacharcolon}{\isachardoublequoteopen}Sample\ req\ dt\ x\ stop\ lose\ ack\ i{\isadigit{1}}\ vc{\isachardoublequoteclose}\isanewline
\ \ \ \ \ \ \isakeyword{and}\ a{\isadigit{2}}{\isacharcolon}{\isachardoublequoteopen}Delay\ y\ i{\isadigit{1}}\ d\ x\ i{\isadigit{2}}{\isachardoublequoteclose}\isanewline
\ \ \ \ \ \ \isakeyword{and}\ a{\isadigit{3}}{\isacharcolon}{\isachardoublequoteopen}Loss\ lose\ a\ i{\isadigit{2}}\ y\ i{\isachardoublequoteclose}\isanewline
\ \ \ \ \isacommand{by}\isamarkupfalse%
\ {\isacharparenleft}simp\ only{\isacharcolon}\ Gateway{\isacharunderscore}def{\isacharcomma}\ auto{\isacharparenright}\ \isanewline
\ \ \isacommand{from}\isamarkupfalse%
\ a{\isadigit{2}}\ \isakeyword{and}\ a{\isadigit{3}}\ \isakeyword{and}\ h{\isadigit{3}}\ \isacommand{have}\isamarkupfalse%
\ sg{\isadigit{1}}{\isacharcolon}{\isachardoublequoteopen}msg\ {\isacharparenleft}Suc\ {\isadigit{0}}{\isacharparenright}\ x{\isachardoublequoteclose}\ \ \ \isanewline
\ \ \ \ \isacommand{by}\isamarkupfalse%
\ {\isacharparenleft}simp\ add{\isacharcolon}\ Loss{\isacharunderscore}Delay{\isacharunderscore}msg{\isacharunderscore}a{\isacharparenright}\ \isanewline
\ \ \isacommand{from}\isamarkupfalse%
\ a{\isadigit{1}}\ \isakeyword{and}\ h{\isadigit{2}}\ \isakeyword{and}\ h{\isadigit{4}}\ \isakeyword{and}\ sg{\isadigit{1}}\ \isacommand{obtain}\isamarkupfalse%
\ st\ buffer\ \isakeyword{where}\ a{\isadigit{4}}{\isacharcolon}\isanewline
\ \ \ \ {\isachardoublequoteopen}tiTable{\isacharunderscore}SampleT\ req\ x\ stop\ lose\ \isanewline
\ \ \ \ \ \ \ \ {\isacharparenleft}fin{\isacharunderscore}inf{\isacharunderscore}append\ {\isacharbrackleft}init{\isacharunderscore}state{\isacharbrackright}\ st{\isacharparenright}\ {\isacharparenleft}fin{\isacharunderscore}inf{\isacharunderscore}append\ {\isacharbrackleft}{\isacharbrackleft}{\isacharbrackright}{\isacharbrackright}\ buffer{\isacharparenright}\ ack\isanewline
\ \ \ \ \ \ \ \ \ i{\isadigit{1}}\ vc\ st{\isachardoublequoteclose}\isanewline
\ \ \ \ \ \isacommand{by}\isamarkupfalse%
\ {\isacharparenleft}simp\ add{\isacharcolon}\ Sample{\isacharunderscore}def\ Sample{\isacharunderscore}L{\isacharunderscore}def{\isacharcomma}\ auto{\isacharparenright}\isanewline
\ \ \isacommand{from}\isamarkupfalse%
\ assms\ \isakeyword{and}\ a{\isadigit{2}}\ \isakeyword{and}\ a{\isadigit{3}}\ \isakeyword{and}\ sg{\isadigit{1}}\ \isakeyword{and}\ a{\isadigit{4}}\ \isacommand{show}\isamarkupfalse%
\ {\isacharquery}thesis\isanewline
\ \ \isacommand{proof}\isamarkupfalse%
\ {\isacharparenleft}induct\ j{\isacharparenright}\isanewline
\ \ \ \ \isacommand{case}\isamarkupfalse%
\ {\isadigit{0}}\isanewline
\ \ \ \ \isacommand{from}\isamarkupfalse%
\ {\isadigit{0}}\ \isacommand{show}\isamarkupfalse%
\ {\isacharquery}case\ \isacommand{by}\isamarkupfalse%
\ simp\isanewline
\ \ \isacommand{next}\isamarkupfalse%
\isanewline
\ \ \ \ \isacommand{case}\isamarkupfalse%
\ {\isacharparenleft}Suc\ j{\isacharparenright}\isanewline
\ \ \ \ \isacommand{from}\isamarkupfalse%
\ Suc\ \isacommand{show}\isamarkupfalse%
\ {\isacharquery}case\isanewline
\ \ \ \ \isacommand{proof}\isamarkupfalse%
\ {\isacharparenleft}cases\ {\isachardoublequoteopen}Suc\ j\ {\isasymle}\ d{\isachardoublequoteclose}{\isacharparenright}\isanewline
\ \ \ \ \ \ \isacommand{assume}\isamarkupfalse%
\ {\isachardoublequoteopen}{\isasymnot}\ Suc\ j\ {\isasymle}\ d{\isachardoublequoteclose}\ \isacommand{from}\isamarkupfalse%
\ this\ \isacommand{show}\isamarkupfalse%
\ {\isacharquery}thesis\ \isacommand{by}\isamarkupfalse%
\ simp\isanewline
\ \ \ \ \isacommand{next}\isamarkupfalse%
\isanewline
\ \ \ \ \ \ \isacommand{assume}\isamarkupfalse%
\ a{\isadigit{0}}{\isacharcolon}{\isachardoublequoteopen}Suc\ j\ {\isasymle}\ d{\isachardoublequoteclose}\isanewline
\ \ \ \ \ \ \isacommand{from}\isamarkupfalse%
\ a{\isadigit{0}}\ \ \isacommand{have}\isamarkupfalse%
\ sg{\isadigit{2}}{\isacharcolon}{\isachardoublequoteopen}d\ {\isacharplus}\ Suc\ j\ {\isasymle}\ d\ {\isacharplus}\ d{\isachardoublequoteclose}\ \ \ \isacommand{by}\isamarkupfalse%
\ arith\isanewline
\ \ \ \ \ \ \isacommand{from}\isamarkupfalse%
\ sg{\isadigit{2}}\ \isacommand{have}\isamarkupfalse%
\ sg{\isadigit{3}}{\isacharcolon}{\isachardoublequoteopen}Suc\ {\isacharparenleft}d\ {\isacharplus}\ j{\isacharparenright}\ {\isasymle}\ d\ {\isacharplus}\ d{\isachardoublequoteclose}\ \isacommand{by}\isamarkupfalse%
\ arith\isanewline
\ \ \ \ \ \ \isacommand{from}\isamarkupfalse%
\ a{\isadigit{4}}\ \isakeyword{and}\ h{\isadigit{2}}\ \isakeyword{and}\ sg{\isadigit{1}}\ \isakeyword{and}\ h{\isadigit{4}}\ \isakeyword{and}\ h{\isadigit{5}}\ \isacommand{have}\isamarkupfalse%
\ sg{\isadigit{4}}{\isacharcolon}\ \isanewline
\ \ \ \ \ \ \ {\isachardoublequoteopen}st\ {\isacharparenleft}t{\isacharplus}d{\isacharplus}j{\isacharparenright}\ {\isacharequal}\ \ hd\ {\isacharparenleft}ack\ {\isacharparenleft}t{\isacharplus}d{\isacharplus}j{\isacharparenright}{\isacharparenright}{\isachardoublequoteclose}\isanewline
\ \ \ \ \ \ \ \ \isacommand{by}\isamarkupfalse%
\ {\isacharparenleft}simp\ add{\isacharcolon}\ tiTable{\isacharunderscore}ack{\isacharunderscore}st{\isacharunderscore}hd{\isacharparenright}\ \ \isanewline
\ \ \ \ \ \ \isacommand{from}\isamarkupfalse%
\ Suc\ \isakeyword{and}\ a{\isadigit{0}}\ \isakeyword{and}\ sg{\isadigit{4}}\ \isacommand{have}\isamarkupfalse%
\ sg{\isadigit{5}}{\isacharcolon}\isanewline
\ \ \ \ \ \ \ {\isachardoublequoteopen}{\isacharparenleft}fin{\isacharunderscore}inf{\isacharunderscore}append\ {\isacharbrackleft}init{\isacharunderscore}state{\isacharbrackright}\ st{\isacharparenright}\ {\isacharparenleft}Suc\ {\isacharparenleft}t{\isacharplus}d{\isacharplus}j{\isacharparenright}{\isacharparenright}\ {\isacharequal}\ sending{\isacharunderscore}data{\isachardoublequoteclose}\isanewline
\ \ \ \ \ \ \ \ \isacommand{by}\isamarkupfalse%
\ {\isacharparenleft}simp\ add{\isacharcolon}\ correct{\isacharunderscore}fin{\isacharunderscore}inf{\isacharunderscore}append{\isadigit{1}}{\isacharparenright}\ \isanewline
\ \ \ \ \ \ \isacommand{from}\isamarkupfalse%
\ h{\isadigit{7}}\ \isakeyword{and}\ a{\isadigit{0}}\ \ \isacommand{have}\isamarkupfalse%
\ sg{\isadigit{6}}{\isacharcolon}{\isachardoublequoteopen}{\isasymforall}j{\isasymle}\ d{\isachardot}\ a\ {\isacharparenleft}t\ {\isacharplus}\ Suc\ j{\isacharparenright}\ {\isacharequal}\ {\isacharbrackleft}{\isacharbrackright}{\isachardoublequoteclose}\ \isacommand{by}\isamarkupfalse%
\ auto\ \isanewline
\ \ \ \ \ \ \isacommand{from}\isamarkupfalse%
\ sg{\isadigit{6}}\ \isakeyword{and}\ a{\isadigit{3}}\ \isakeyword{and}\ a{\isadigit{0}}\ \isakeyword{and}\ h{\isadigit{5}}\ \isacommand{have}\isamarkupfalse%
\ sg{\isadigit{7}}{\isacharcolon}{\isachardoublequoteopen}y\ {\isacharparenleft}t\ {\isacharplus}\ {\isacharparenleft}Suc\ j{\isacharparenright}{\isacharparenright}\ {\isacharequal}\ {\isacharbrackleft}{\isacharbrackright}{\isachardoublequoteclose}\ \isanewline
\ \ \ \ \ \ \ \ \isacommand{by}\isamarkupfalse%
\ {\isacharparenleft}rule\ Loss{\isacharunderscore}L{\isadigit{5}}Suc{\isacharparenright}\isanewline
\ \ \ \ \ \ \isacommand{from}\isamarkupfalse%
\ sg{\isadigit{7}}\ \isakeyword{and}\ a{\isadigit{2}}\ \isacommand{have}\isamarkupfalse%
\ sg{\isadigit{8}}a{\isacharcolon}{\isachardoublequoteopen}x\ {\isacharparenleft}t\ {\isacharplus}\ d\ {\isacharplus}\ {\isacharparenleft}Suc\ j{\isacharparenright}{\isacharparenright}\ {\isacharequal}\ {\isacharbrackleft}{\isacharbrackright}{\isachardoublequoteclose}\isanewline
\ \ \ \ \ \ \ \ \isacommand{by}\isamarkupfalse%
\ {\isacharparenleft}simp\ add{\isacharcolon}\ Delay{\isacharunderscore}def{\isacharparenright}\isanewline
\ \ \ \ \ \ \isacommand{from}\isamarkupfalse%
\ sg{\isadigit{8}}a\ \ \isacommand{have}\isamarkupfalse%
\ sg{\isadigit{8}}{\isacharcolon}{\isachardoublequoteopen}x\ {\isacharparenleft}Suc\ {\isacharparenleft}t\ {\isacharplus}\ d\ {\isacharplus}\ j{\isacharparenright}{\isacharparenright}\ {\isacharequal}\ {\isacharbrackleft}{\isacharbrackright}{\isachardoublequoteclose}\ \isacommand{by}\isamarkupfalse%
\ simp\isanewline
\ \ \ \ \ \ \isacommand{have}\isamarkupfalse%
\ sg{\isadigit{9}}{\isacharcolon}{\isachardoublequoteopen}Suc\ {\isacharparenleft}t\ {\isacharplus}\ d\ {\isacharplus}\ j{\isacharparenright}\ {\isacharequal}\ Suc\ {\isacharparenleft}t\ {\isacharplus}\ {\isacharparenleft}d\ {\isacharplus}\ j{\isacharparenright}{\isacharparenright}{\isachardoublequoteclose}\ \isacommand{by}\isamarkupfalse%
\ arith\ \ \isanewline
\ \ \ \ \ \ \isacommand{from}\isamarkupfalse%
\ a{\isadigit{4}}\ \isacommand{have}\isamarkupfalse%
\ sg{\isadigit{1}}{\isadigit{0}}{\isacharcolon}\isanewline
\ \ \ \ \ \ \ \ {\isachardoublequoteopen}fin{\isacharunderscore}inf{\isacharunderscore}append\ {\isacharbrackleft}init{\isacharunderscore}state{\isacharbrackright}\ st\ {\isacharparenleft}Suc\ {\isacharparenleft}t\ {\isacharplus}\ d\ {\isacharplus}\ j{\isacharparenright}{\isacharparenright}\ {\isacharequal}\ sending{\isacharunderscore}data\ {\isasymand}\ \isanewline
\ \ \ \ \ \ \ \ \ x\ {\isacharparenleft}Suc\ {\isacharparenleft}t\ {\isacharplus}\ d\ {\isacharplus}\ j{\isacharparenright}{\isacharparenright}\ {\isacharequal}\ {\isacharbrackleft}{\isacharbrackright}\ {\isasymand}\ \isanewline
\ \ \ \ \ \ \ \ \ lose\ {\isacharparenleft}Suc\ {\isacharparenleft}t\ {\isacharplus}\ d\ {\isacharplus}\ j{\isacharparenright}{\isacharparenright}\ {\isacharequal}\ {\isacharbrackleft}False{\isacharbrackright}\ {\isasymlongrightarrow}\isanewline
\ \ \ \ \ \ \ \ \ ack\ {\isacharparenleft}Suc\ {\isacharparenleft}t\ {\isacharplus}\ d\ {\isacharplus}\ j{\isacharparenright}{\isacharparenright}\ {\isacharequal}\ {\isacharbrackleft}sending{\isacharunderscore}data{\isacharbrackright}{\isachardoublequoteclose}\isanewline
\ \ \ \ \ \ \ \ \isacommand{by}\isamarkupfalse%
\ {\isacharparenleft}simp\ add{\isacharcolon}\ tiTable{\isacharunderscore}SampleT{\isacharunderscore}def{\isacharparenright}\ \ \isanewline
\ \ \ \ \ \ \isacommand{from}\isamarkupfalse%
\ h{\isadigit{8}}\ \isakeyword{and}\ sg{\isadigit{3}}\ \isacommand{have}\isamarkupfalse%
\ sg{\isadigit{1}}{\isadigit{1}}{\isacharcolon}{\isachardoublequoteopen}lose\ {\isacharparenleft}t\ {\isacharplus}\ Suc\ {\isacharparenleft}d\ {\isacharplus}\ j{\isacharparenright}{\isacharparenright}\ {\isacharequal}\ {\isacharbrackleft}False{\isacharbrackright}{\isachardoublequoteclose}\ \isacommand{by}\isamarkupfalse%
\ blast\isanewline
\ \ \ \ \ \ \isacommand{have}\isamarkupfalse%
\ sg{\isadigit{1}}{\isadigit{2}}{\isacharcolon}{\isachardoublequoteopen}Suc\ {\isacharparenleft}t\ {\isacharplus}\ d\ {\isacharplus}\ j{\isacharparenright}\ {\isacharequal}\ t\ {\isacharplus}\ Suc\ {\isacharparenleft}d\ {\isacharplus}\ j{\isacharparenright}{\isachardoublequoteclose}\ \ \isacommand{by}\isamarkupfalse%
\ arith\isanewline
\ \ \ \ \ \ \isacommand{from}\isamarkupfalse%
\ sg{\isadigit{1}}{\isadigit{2}}\ \isakeyword{and}\ sg{\isadigit{1}}{\isadigit{1}}\ \isacommand{have}\isamarkupfalse%
\ sg{\isadigit{1}}{\isadigit{3}}{\isacharcolon}{\isachardoublequoteopen}lose\ {\isacharparenleft}Suc\ {\isacharparenleft}t\ {\isacharplus}\ d\ {\isacharplus}\ j{\isacharparenright}{\isacharparenright}\ {\isacharequal}\ {\isacharbrackleft}False{\isacharbrackright}{\isachardoublequoteclose}\ \isanewline
\ \ \ \ \ \ \ \ \isacommand{by}\isamarkupfalse%
\ \ {\isacharparenleft}simp\ {\isacharparenleft}no{\isacharunderscore}asm{\isacharunderscore}simp{\isacharparenright}{\isacharcomma}\ simp{\isacharparenright}\ \isanewline
\ \ \ \ \ \isacommand{from}\isamarkupfalse%
\ sg{\isadigit{1}}{\isadigit{0}}\ \isakeyword{and}\ sg{\isadigit{5}}\ \isakeyword{and}\ sg{\isadigit{8}}a\ \isakeyword{and}\ sg{\isadigit{1}}{\isadigit{3}}\ \isacommand{show}\isamarkupfalse%
\ {\isacharquery}thesis\ \isacommand{by}\isamarkupfalse%
\ simp\isanewline
\ \ \ \ \isacommand{qed}\isamarkupfalse%
\isanewline
\ \ \isacommand{qed}\isamarkupfalse%
\isanewline
\isacommand{qed}\isamarkupfalse%
\endisatagproof
{\isafoldproof}%
\isadelimproof
\isanewline
\endisadelimproof
\ \isanewline 
\isacommand{lemma}\isamarkupfalse%
\ Gateway{\isacharunderscore}L{\isadigit{6}}{\isacharunderscore}induction{\isacharcolon}\isanewline
\ \isakeyword{assumes}\ h{\isadigit{1}}{\isacharcolon}{\isachardoublequoteopen}msg\ {\isacharparenleft}Suc\ {\isadigit{0}}{\isacharparenright}\ req{\isachardoublequoteclose}\isanewline
\ \ \ \ \ \isakeyword{and}\ h{\isadigit{2}}{\isacharcolon}{\isachardoublequoteopen}msg\ {\isacharparenleft}Suc\ {\isadigit{0}}{\isacharparenright}\ x{\isachardoublequoteclose}\ \isanewline
\ \ \ \ \ \isakeyword{and}\ h{\isadigit{3}}{\isacharcolon}{\isachardoublequoteopen}msg\ {\isacharparenleft}Suc\ {\isadigit{0}}{\isacharparenright}\ stop{\isachardoublequoteclose}\isanewline
\ \ \ \ \ \isakeyword{and}\ h{\isadigit{4}}{\isacharcolon}{\isachardoublequoteopen}ts\ lose{\isachardoublequoteclose}\isanewline
\ \ \ \ \ \isakeyword{and}\ h{\isadigit{5}}{\isacharcolon}{\isachardoublequoteopen}{\isasymforall}j{\isasymle}\ k{\isachardot}\ lose\ {\isacharparenleft}t\ {\isacharplus}\ j{\isacharparenright}\ {\isacharequal}\ {\isacharbrackleft}False{\isacharbrackright}{\isachardoublequoteclose}\ \isanewline
\ \ \ \ \ \isakeyword{and}\ h{\isadigit{6}}{\isacharcolon}{\isachardoublequoteopen}{\isasymforall}m\ {\isasymle}\ k{\isachardot}\ req\ {\isacharparenleft}t\ {\isacharplus}\ m{\isacharparenright}\ {\isasymnoteq}\ {\isacharbrackleft}send{\isacharbrackright}{\isachardoublequoteclose}\isanewline
\ \ \ \ \ \isakeyword{and}\ h{\isadigit{7}}{\isacharcolon}{\isachardoublequoteopen}ack\ t\ {\isacharequal}\ {\isacharbrackleft}connection{\isacharunderscore}ok{\isacharbrackright}{\isachardoublequoteclose}\isanewline
\ \ \ \ \ \isakeyword{and}\ h{\isadigit{8}}{\isacharcolon}{\isachardoublequoteopen}Sample\ req\ dt\ x{\isadigit{1}}\ stop\ lose\ ack\ i{\isadigit{1}}\ vc{\isachardoublequoteclose}\ \isanewline
\ \ \ \ \ \isakeyword{and}\ h{\isadigit{9}}{\isacharcolon}{\isachardoublequoteopen}Delay\ x{\isadigit{2}}\ i{\isadigit{1}}\ d\ x{\isadigit{1}}\ i{\isadigit{2}}{\isachardoublequoteclose}\isanewline
\ \ \ \ \ \isakeyword{and}\ h{\isadigit{1}}{\isadigit{0}}{\isacharcolon}{\isachardoublequoteopen}Loss\ lose\ x\ i{\isadigit{2}}\ x{\isadigit{2}}\ i{\isachardoublequoteclose}\isanewline
\ \ \ \ \ \isakeyword{and}\ h{\isadigit{1}}{\isadigit{1}}{\isacharcolon}{\isachardoublequoteopen}m\ {\isasymle}\ k{\isachardoublequoteclose}\isanewline
\ \isakeyword{shows}\ {\isachardoublequoteopen}ack\ {\isacharparenleft}t\ {\isacharplus}\ m{\isacharparenright}\ {\isacharequal}\ {\isacharbrackleft}connection{\isacharunderscore}ok{\isacharbrackright}{\isachardoublequoteclose}\isanewline
\isadelimproof
\endisadelimproof
\isatagproof
\isacommand{using}\isamarkupfalse%
\ assms\isanewline
\isacommand{proof}\isamarkupfalse%
\ {\isacharparenleft}induct\ m{\isacharparenright}\isanewline
\ \ \isacommand{case}\isamarkupfalse%
\ {\isadigit{0}}\
 \isacommand{from}\isamarkupfalse%
\ this\ \isacommand{show}\isamarkupfalse%
\ {\isacharquery}case\ \ \isacommand{by}\isamarkupfalse%
\ simp\isanewline
\isacommand{next}\isamarkupfalse%
\isanewline
\ \ \isacommand{case}\isamarkupfalse%
\ {\isacharparenleft}Suc\ m{\isacharparenright}\isanewline
\ \ \isacommand{from}\isamarkupfalse%
\ Suc\ \isacommand{have}\isamarkupfalse%
\ sg{\isadigit{1}}{\isacharcolon}{\isachardoublequoteopen}msg\ {\isacharparenleft}Suc\ {\isadigit{0}}{\isacharparenright}\ x{\isadigit{1}}{\isachardoublequoteclose}\ \isacommand{by}\isamarkupfalse%
\ {\isacharparenleft}simp\ add{\isacharcolon}\ \ Loss{\isacharunderscore}Delay{\isacharunderscore}msg{\isacharunderscore}a{\isacharparenright}\isanewline
\ \ \isacommand{from}\isamarkupfalse%
\ Suc\ \isakeyword{and}\ sg{\isadigit{1}}\ \isacommand{obtain}\isamarkupfalse%
\ st\ buffer\ \isakeyword{where}\isanewline
\ \ \ \ a{\isadigit{1}}{\isacharcolon}{\isachardoublequoteopen}tiTable{\isacharunderscore}SampleT\ req\ x{\isadigit{1}}\ stop\ lose\ {\isacharparenleft}fin{\isacharunderscore}inf{\isacharunderscore}append\ {\isacharbrackleft}init{\isacharunderscore}state{\isacharbrackright}\ st{\isacharparenright}\ \isanewline
\ \ \ \ \ \ \ \ \ {\isacharparenleft}fin{\isacharunderscore}inf{\isacharunderscore}append\ {\isacharbrackleft}{\isacharbrackleft}{\isacharbrackright}{\isacharbrackright}\ buffer{\isacharparenright}\ ack\ i{\isadigit{1}}\ vc\ st{\isachardoublequoteclose}\ \ \isakeyword{and}\ \isanewline
\ \ \ \ a{\isadigit{2}}{\isacharcolon}{\isachardoublequoteopen}{\isasymforall}\ t{\isachardot}\ buffer\ t\ {\isacharequal}\ {\isacharparenleft}if\ dt\ t\ {\isacharequal}\ {\isacharbrackleft}{\isacharbrackright}\ then\ fin{\isacharunderscore}inf{\isacharunderscore}append\ {\isacharbrackleft}{\isacharbrackleft}{\isacharbrackright}{\isacharbrackright}\ buffer\ t\ else\ dt\ t{\isacharparenright}{\isachardoublequoteclose}\isanewline
\ \ \ \ \isacommand{by}\isamarkupfalse%
\ {\isacharparenleft}simp\ add{\isacharcolon}\ Sample{\isacharunderscore}def\ Sample{\isacharunderscore}L{\isacharunderscore}def{\isacharcomma}\ auto{\isacharparenright}\isanewline
\ \ \isacommand{from}\isamarkupfalse%
\ a{\isadigit{1}}\ \isakeyword{and}\ sg{\isadigit{1}}\ \isakeyword{and}\ h{\isadigit{3}}\ \isakeyword{and}\ h{\isadigit{4}}\ \isacommand{have}\isamarkupfalse%
\ sg{\isadigit{2}}{\isacharcolon}{\isachardoublequoteopen}st\ {\isacharparenleft}t\ {\isacharplus}\ m{\isacharparenright}\ {\isacharequal}\ \ hd\ {\isacharparenleft}ack\ {\isacharparenleft}t\ {\isacharplus}\ m{\isacharparenright}{\isacharparenright}{\isachardoublequoteclose}\ \ \isanewline
\ \ \ \ \isacommand{by}\isamarkupfalse%
\ {\isacharparenleft}simp\ add{\isacharcolon}\ tiTable{\isacharunderscore}ack{\isacharunderscore}st{\isacharunderscore}hd{\isacharparenright}\isanewline
\ \ \isacommand{from}\isamarkupfalse%
\ Suc\ \isacommand{have}\isamarkupfalse%
\ sg{\isadigit{3}}{\isacharcolon}{\isachardoublequoteopen}ack\ {\isacharparenleft}t\ {\isacharplus}\ m{\isacharparenright}\ {\isacharequal}\ {\isacharbrackleft}connection{\isacharunderscore}ok{\isacharbrackright}{\isachardoublequoteclose}\ \isacommand{by}\isamarkupfalse%
\ simp\isanewline
\ \ \isacommand{from}\isamarkupfalse%
\ a{\isadigit{1}}\ \isakeyword{and}\ sg{\isadigit{2}}\ \isakeyword{and}\ sg{\isadigit{3}}\ \isacommand{have}\isamarkupfalse%
\ sg{\isadigit{4}}{\isacharcolon}\isanewline
\ \ {\isachardoublequoteopen}{\isacharparenleft}fin{\isacharunderscore}inf{\isacharunderscore}append\ {\isacharbrackleft}init{\isacharunderscore}state{\isacharbrackright}\ st{\isacharparenright}\ {\isacharparenleft}Suc\ {\isacharparenleft}t\ {\isacharplus}\ m{\isacharparenright}{\isacharparenright}\ {\isacharequal}\ connection{\isacharunderscore}ok{\isachardoublequoteclose}\isanewline
\ \ \ \ \isacommand{by}\isamarkupfalse%
\ {\isacharparenleft}simp\ add{\isacharcolon}\ fin{\isacharunderscore}inf{\isacharunderscore}append{\isacharunderscore}def{\isacharparenright}\isanewline
\ \ \isacommand{from}\isamarkupfalse%
\ Suc\ \isacommand{have}\isamarkupfalse%
\ sg{\isadigit{5}}{\isacharcolon}{\isachardoublequoteopen}Suc\ m\ {\isasymle}\ k{\isachardoublequoteclose}\ \isacommand{by}\isamarkupfalse%
\ simp\isanewline
\ \ \isacommand{from}\isamarkupfalse%
\ sg{\isadigit{5}}\ \isakeyword{and}\ h{\isadigit{5}}\ \isacommand{have}\isamarkupfalse%
\ sg{\isadigit{6}}{\isacharcolon}{\isachardoublequoteopen}lose\ {\isacharparenleft}Suc\ {\isacharparenleft}t\ {\isacharplus}\ m{\isacharparenright}{\isacharparenright}\ {\isacharequal}\ {\isacharbrackleft}False{\isacharbrackright}{\isachardoublequoteclose}\ \isacommand{by}\isamarkupfalse%
\ auto\isanewline
\ \ \isacommand{from}\isamarkupfalse%
\ h{\isadigit{6}}\ \isakeyword{and}\ sg{\isadigit{5}}\ \isacommand{have}\isamarkupfalse%
\ sg{\isadigit{7}}{\isacharcolon}{\isachardoublequoteopen}req\ {\isacharparenleft}Suc\ {\isacharparenleft}t\ {\isacharplus}\ m{\isacharparenright}{\isacharparenright}\ {\isasymnoteq}\ {\isacharbrackleft}send{\isacharbrackright}{\isachardoublequoteclose}\ \isacommand{by}\isamarkupfalse%
\ auto\isanewline
\ \ \isacommand{from}\isamarkupfalse%
\ a{\isadigit{1}}\ \isakeyword{and}\ sg{\isadigit{3}}\ \isakeyword{and}\ sg{\isadigit{4}}\ \isakeyword{and}\ sg{\isadigit{5}}\ \isakeyword{and}\ sg{\isadigit{6}}\ \isakeyword{and}\ sg{\isadigit{7}}\ \isacommand{show}\isamarkupfalse%
\ {\isacharquery}case\isanewline
\ \ \ \ \isacommand{by}\isamarkupfalse%
\ {\isacharparenleft}simp\ add{\isacharcolon}\ tiTable{\isacharunderscore}SampleT{\isacharunderscore}def{\isacharparenright}\isanewline
\isacommand{qed}\isamarkupfalse%
\endisatagproof
{\isafoldproof}%
\isadelimproof
\isanewline
\endisadelimproof
\isanewline 
\isacommand{lemma}\isamarkupfalse%
\ Gateway{\isacharunderscore}L{\isadigit{6}}{\isacharcolon}\isanewline
\ \isakeyword{assumes}\ h{\isadigit{1}}{\isacharcolon}{\isachardoublequoteopen}Gateway\ req\ dt\ a\ stop\ lose\ d\ ack\ i\ vc{\isachardoublequoteclose}\isanewline
\ \ \ \ \ \isakeyword{and}\ h{\isadigit{2}}{\isacharcolon}{\isachardoublequoteopen}{\isasymforall}m{\isasymle}k{\isachardot}\ req\ {\isacharparenleft}t\ {\isacharplus}\ m{\isacharparenright}\ {\isasymnoteq}\ {\isacharbrackleft}send{\isacharbrackright}{\isachardoublequoteclose}\isanewline
\ \ \ \ \ \isakeyword{and}\ h{\isadigit{3}}{\isacharcolon}{\isachardoublequoteopen}{\isasymforall}j{\isasymle}k{\isachardot}\ lose\ {\isacharparenleft}t\ {\isacharplus}\ j{\isacharparenright}\ {\isacharequal}\ {\isacharbrackleft}False{\isacharbrackright}{\isachardoublequoteclose}\isanewline
\ \ \ \ \ \isakeyword{and}\ h{\isadigit{4}}{\isacharcolon}{\isachardoublequoteopen}ack\ t\ {\isacharequal}\ {\isacharbrackleft}connection{\isacharunderscore}ok{\isacharbrackright}{\isachardoublequoteclose}\isanewline
\ \ \ \ \ \isakeyword{and}\ h{\isadigit{5}}{\isacharcolon}{\isachardoublequoteopen}msg\ {\isacharparenleft}Suc\ {\isadigit{0}}{\isacharparenright}\ req{\isachardoublequoteclose}\isanewline
\ \ \ \ \ \isakeyword{and}\ h{\isadigit{6}}{\isacharcolon}{\isachardoublequoteopen}msg\ {\isacharparenleft}Suc\ {\isadigit{0}}{\isacharparenright}\ stop{\isachardoublequoteclose}\isanewline
\ \ \ \ \ \isakeyword{and}\ h{\isadigit{7}}{\isacharcolon}{\isachardoublequoteopen}msg\ {\isacharparenleft}Suc\ {\isadigit{0}}{\isacharparenright}\ a{\isachardoublequoteclose}\isanewline
\ \ \ \ \ \isakeyword{and}\ h{\isadigit{8}}{\isacharcolon}{\isachardoublequoteopen}ts\ lose{\isachardoublequoteclose}\isanewline
\ \isakeyword{shows}\ {\isachardoublequoteopen}{\isasymforall}m{\isasymle}k{\isachardot}\ ack\ {\isacharparenleft}t\ {\isacharplus}\ m{\isacharparenright}\ {\isacharequal}\ {\isacharbrackleft}connection{\isacharunderscore}ok{\isacharbrackright}{\isachardoublequoteclose}\isanewline
\isadelimproof
\endisadelimproof
\isatagproof
\isacommand{using}\isamarkupfalse%
\ assms\ \isanewline
\isacommand{by}\isamarkupfalse%
\ {\isacharparenleft}simp\ add{\isacharcolon}\ Gateway{\isacharunderscore}def{\isacharcomma}\ clarify{\isacharcomma}\ simp\ add{\isacharcolon}\ Gateway{\isacharunderscore}L{\isadigit{6}}{\isacharunderscore}induction{\isacharparenright}%
\endisatagproof
{\isafoldproof}%
\isadelimproof
\ \isanewline
\endisadelimproof
\isanewline 
\isacommand{lemma}\isamarkupfalse%
\ Gateway{\isacharunderscore}L{\isadigit{6}}a{\isacharcolon}\isanewline
\ \isakeyword{assumes}\ h{\isadigit{1}}{\isacharcolon}{\isachardoublequoteopen}Gateway\ req\ dt\ a\ stop\ lose\ d\ ack\ i\ vc{\isachardoublequoteclose}\isanewline
\ \ \ \ \ \isakeyword{and}\ h{\isadigit{2}}{\isacharcolon}{\isachardoublequoteopen}{\isasymforall}m{\isasymle}k{\isachardot}\ req\ {\isacharparenleft}t\ {\isacharplus}\ {\isadigit{2}}\ {\isacharplus}\ m{\isacharparenright}\ {\isasymnoteq}\ {\isacharbrackleft}send{\isacharbrackright}{\isachardoublequoteclose}\isanewline
\ \ \ \ \ \isakeyword{and}\ h{\isadigit{3}}{\isacharcolon}{\isachardoublequoteopen}{\isasymforall}j{\isasymle}k{\isachardot}\ lose\ {\isacharparenleft}t\ {\isacharplus}\ {\isadigit{2}}\ {\isacharplus}\ j{\isacharparenright}\ {\isacharequal}\ {\isacharbrackleft}False{\isacharbrackright}{\isachardoublequoteclose}\isanewline
\ \ \ \ \ \isakeyword{and}\ h{\isadigit{4}}{\isacharcolon}{\isachardoublequoteopen}ack\ {\isacharparenleft}t\ {\isacharplus}\ {\isadigit{2}}{\isacharparenright}\ {\isacharequal}\ {\isacharbrackleft}connection{\isacharunderscore}ok{\isacharbrackright}{\isachardoublequoteclose}\isanewline
\ \ \ \ \ \isakeyword{and}\ h{\isadigit{5}}{\isacharcolon}{\isachardoublequoteopen}msg\ {\isacharparenleft}Suc\ {\isadigit{0}}{\isacharparenright}\ req{\isachardoublequoteclose}\isanewline
\ \ \ \ \ \isakeyword{and}\ h{\isadigit{6}}{\isacharcolon}{\isachardoublequoteopen}msg\ {\isacharparenleft}Suc\ {\isadigit{0}}{\isacharparenright}\ stop{\isachardoublequoteclose}\isanewline
\ \ \ \ \ \isakeyword{and}\ h{\isadigit{7}}{\isacharcolon}{\isachardoublequoteopen}msg\ {\isacharparenleft}Suc\ {\isadigit{0}}{\isacharparenright}\ a{\isachardoublequoteclose}\isanewline
\ \ \ \ \ \isakeyword{and}\ h{\isadigit{8}}{\isacharcolon}{\isachardoublequoteopen}ts\ lose{\isachardoublequoteclose}\isanewline
\ \isakeyword{shows}\ {\isachardoublequoteopen}{\isasymforall}m{\isasymle}k{\isachardot}\ ack\ {\isacharparenleft}t\ {\isacharplus}\ {\isadigit{2}}\ {\isacharplus}\ m{\isacharparenright}\ {\isacharequal}\ {\isacharbrackleft}connection{\isacharunderscore}ok{\isacharbrackright}{\isachardoublequoteclose}\isanewline
\isadelimproof
\endisadelimproof
\isatagproof
\isacommand{using}\isamarkupfalse%
\ assms\ \isacommand{by}\isamarkupfalse%
\ {\isacharparenleft}rule\ Gateway{\isacharunderscore}L{\isadigit{6}}{\isacharparenright}%
\endisatagproof
{\isafoldproof}%
\isadelimproof
\isanewline
\endisadelimproof
\isanewline 
\isacommand{lemma}\isamarkupfalse%
\ aux{\isacharunderscore}k{\isadigit{3}}req{\isacharcolon}\isanewline
\ \isakeyword{assumes}\ h{\isadigit{1}}{\isacharcolon}{\isachardoublequoteopen}{\isasymforall}m{\isacharless}k\ {\isacharplus}\ {\isadigit{3}}{\isachardot}\ req\ {\isacharparenleft}t\ {\isacharplus}\ m{\isacharparenright}\ {\isasymnoteq}\ {\isacharbrackleft}send{\isacharbrackright}{\isachardoublequoteclose}\
\  \isakeyword{and}\ h{\isadigit{2}}{\isacharcolon}{\isachardoublequoteopen}m\ {\isasymle}\ k{\isachardoublequoteclose}\isanewline
\ \isakeyword{shows}\ {\isachardoublequoteopen}req\ {\isacharparenleft}Suc\ {\isacharparenleft}Suc\ {\isacharparenleft}t\ {\isacharplus}\ m{\isacharparenright}{\isacharparenright}{\isacharparenright}\ {\isasymnoteq}\ {\isacharbrackleft}send{\isacharbrackright}{\isachardoublequoteclose}\isanewline
\isadelimproof
\endisadelimproof
\isatagproof
\isacommand{proof}\isamarkupfalse%
\ {\isacharminus}\ \isanewline
\ \ \isacommand{from}\isamarkupfalse%
\ h{\isadigit{2}}\ \isacommand{have}\isamarkupfalse%
\ {\isachardoublequoteopen}m\ {\isacharplus}\ {\isadigit{2}}\ {\isacharless}\ k\ {\isacharplus}\ {\isadigit{3}}{\isachardoublequoteclose}\ \isacommand{by}\isamarkupfalse%
\ arith\isanewline
\ \ \isacommand{from}\isamarkupfalse%
\ h{\isadigit{1}}\ \isakeyword{and}\ this\ \isacommand{have}\isamarkupfalse%
\ {\isachardoublequoteopen}req\ {\isacharparenleft}t\ {\isacharplus}\ {\isacharparenleft}m\ {\isacharplus}\ {\isadigit{2}}{\isacharparenright}{\isacharparenright}\ {\isasymnoteq}\ {\isacharbrackleft}send{\isacharbrackright}{\isachardoublequoteclose}\ \isacommand{by}\isamarkupfalse%
\ blast\isanewline
\ \ \isacommand{from}\isamarkupfalse%
\ this\ \isacommand{show}\isamarkupfalse%
\ {\isacharquery}thesis\ \isacommand{by}\isamarkupfalse%
\ simp\isanewline
\isacommand{qed}\isamarkupfalse%
\endisatagproof
{\isafoldproof}%
\isadelimproof
\isanewline
\endisadelimproof
\isanewline
\isacommand{lemma}\isamarkupfalse%
\ aux{\isadigit{3}}lose{\isacharcolon}\isanewline
\ \ \isakeyword{assumes}\ h{\isadigit{1}}{\isacharcolon}{\isachardoublequoteopen}{\isasymforall}j{\isasymle}k\ {\isacharplus}\ d\ {\isacharplus}\ {\isadigit{3}}{\isachardot}\ lose\ {\isacharparenleft}t\ {\isacharplus}\ j{\isacharparenright}\ {\isacharequal}\ {\isacharbrackleft}False{\isacharbrackright}{\isachardoublequoteclose}\isanewline
\ \ \ \ \ \ \isakeyword{and}\ h{\isadigit{2}}{\isacharcolon}{\isachardoublequoteopen}j\ {\isasymle}\ k{\isachardoublequoteclose}\isanewline
\ \ \isakeyword{shows}\ {\isachardoublequoteopen}lose\ {\isacharparenleft}Suc\ {\isacharparenleft}Suc\ {\isacharparenleft}t\ {\isacharplus}\ j{\isacharparenright}{\isacharparenright}{\isacharparenright}\ {\isacharequal}\ {\isacharbrackleft}False{\isacharbrackright}{\isachardoublequoteclose}\isanewline
\isadelimproof
\endisadelimproof
\isatagproof
\isacommand{proof}\isamarkupfalse%
\ {\isacharminus}\ \isanewline
\ \ \isacommand{from}\isamarkupfalse%
\ h{\isadigit{2}}\ \isacommand{have}\isamarkupfalse%
\ {\isachardoublequoteopen}j\ {\isacharplus}\ {\isadigit{2}}\ {\isasymle}k\ {\isacharplus}\ d\ {\isacharplus}\ {\isadigit{3}}{\isachardoublequoteclose}\ \isacommand{by}\isamarkupfalse%
\ arith\isanewline
\ \ \isacommand{from}\isamarkupfalse%
\ h{\isadigit{1}}\ \isakeyword{and}\ this\ \isacommand{have}\isamarkupfalse%
\ {\isachardoublequoteopen}lose\ {\isacharparenleft}t\ {\isacharplus}\ {\isacharparenleft}j\ {\isacharplus}\ {\isadigit{2}}{\isacharparenright}{\isacharparenright}\ {\isacharequal}\ {\isacharbrackleft}False{\isacharbrackright}{\isachardoublequoteclose}\ \isacommand{by}\isamarkupfalse%
\ blast\isanewline
\ \ \isacommand{from}\isamarkupfalse%
\ this\ \isacommand{show}\isamarkupfalse%
\ {\isacharquery}thesis\ \isacommand{by}\isamarkupfalse%
\ simp\isanewline
\isacommand{qed}\isamarkupfalse%
\endisatagproof
{\isafoldproof}%
\isadelimproof
\isanewline
\endisadelimproof
\isanewline 
\isacommand{lemma}\isamarkupfalse%
\ Gateway{\isacharunderscore}L{\isadigit{7}}{\isacharcolon}\isanewline
\ \isakeyword{assumes}\ h{\isadigit{1}}{\isacharcolon}{\isachardoublequoteopen}Gateway\ req\ dt\ a\ stop\ lose\ d\ ack\ i\ vc{\isachardoublequoteclose}\isanewline
\ \ \ \ \ \isakeyword{and}\ h{\isadigit{2}}{\isacharcolon}{\isachardoublequoteopen}ts\ lose{\isachardoublequoteclose}\isanewline
\ \ \ \ \ \isakeyword{and}\ h{\isadigit{3}}{\isacharcolon}{\isachardoublequoteopen}msg\ {\isacharparenleft}Suc\ {\isadigit{0}}{\isacharparenright}\ a{\isachardoublequoteclose}\isanewline
\ \ \ \ \ \isakeyword{and}\ h{\isadigit{4}}{\isacharcolon}{\isachardoublequoteopen}msg\ {\isacharparenleft}Suc\ {\isadigit{0}}{\isacharparenright}\ stop{\isachardoublequoteclose}\ \isanewline
\ \ \ \ \ \isakeyword{and}\ h{\isadigit{5}}{\isacharcolon}{\isachardoublequoteopen}msg\ {\isacharparenleft}Suc\ {\isadigit{0}}{\isacharparenright}\ req{\isachardoublequoteclose}\isanewline
\ \ \ \ \ \isakeyword{and}\ h{\isadigit{6}}{\isacharcolon}{\isachardoublequoteopen}req\ {\isacharparenleft}Suc\ t{\isacharparenright}\ {\isacharequal}\ {\isacharbrackleft}init{\isacharbrackright}{\isachardoublequoteclose}\isanewline
\ \ \ \ \ \isakeyword{and}\ h{\isadigit{7}}{\isacharcolon}{\isachardoublequoteopen}{\isasymforall}m\ {\isacharless}\ {\isacharparenleft}k\ {\isacharplus}\ {\isadigit{3}}{\isacharparenright}{\isachardot}\ req\ {\isacharparenleft}t\ {\isacharplus}\ m{\isacharparenright}\ {\isasymnoteq}\ {\isacharbrackleft}send{\isacharbrackright}{\isachardoublequoteclose}\isanewline
\ \ \ \ \ \isakeyword{and}\ h{\isadigit{8}}{\isacharcolon}{\isachardoublequoteopen}req\ {\isacharparenleft}t\ {\isacharplus}\ {\isadigit{3}}\ {\isacharplus}\ k{\isacharparenright}\ {\isacharequal}\ {\isacharbrackleft}send{\isacharbrackright}{\isachardoublequoteclose}\isanewline
\ \ \ \ \ \isakeyword{and}\ h{\isadigit{9}}{\isacharcolon}{\isachardoublequoteopen}ack\ t\ {\isacharequal}\ {\isacharbrackleft}init{\isacharunderscore}state{\isacharbrackright}{\isachardoublequoteclose}\isanewline
\ \ \ \ \ \isakeyword{and}\ h{\isadigit{1}}{\isadigit{0}}{\isacharcolon}{\isachardoublequoteopen}{\isasymforall}j{\isasymle}k\ {\isacharplus}\ d\ {\isacharplus}\ {\isadigit{3}}{\isachardot}\ lose\ {\isacharparenleft}t\ {\isacharplus}\ j{\isacharparenright}\ {\isacharequal}\ {\isacharbrackleft}False{\isacharbrackright}{\isachardoublequoteclose}\isanewline
\ \ \ \ \ \isakeyword{and}\ h{\isadigit{1}}{\isadigit{1}}{\isacharcolon}{\isachardoublequoteopen}{\isasymforall}\ t{\isadigit{1}}\ {\isasymle}\ t{\isachardot}\ req\ t{\isadigit{1}}\ {\isacharequal}\ {\isacharbrackleft}{\isacharbrackright}{\isachardoublequoteclose}\isanewline
\ \ \isakeyword{shows}\ {\isachardoublequoteopen}{\isasymforall}\ t{\isadigit{2}}\ {\isacharless}\ {\isacharparenleft}t\ {\isacharplus}\ {\isadigit{3}}\ {\isacharplus}\ k\ {\isacharplus}\ d{\isacharparenright}{\isachardot}\ i\ t{\isadigit{2}}\ {\isacharequal}\ {\isacharbrackleft}{\isacharbrackright}{\isachardoublequoteclose}\isanewline
\isadelimproof
\endisadelimproof
\isatagproof
\isacommand{proof}\isamarkupfalse%
\ {\isacharminus}\isanewline
\ \ \isacommand{have}\isamarkupfalse%
\ {\isachardoublequoteopen}Suc\ {\isadigit{0}}\ {\isasymle}\ k\ {\isacharplus}\ d\ {\isacharplus}\ {\isadigit{3}}{\isachardoublequoteclose}\ \isacommand{by}\isamarkupfalse%
\ arith\isanewline
\ \ \isacommand{from}\isamarkupfalse%
\ h{\isadigit{1}}{\isadigit{0}}\ \isakeyword{and}\ this\ \isacommand{have}\isamarkupfalse%
\ {\isachardoublequoteopen}lose\ {\isacharparenleft}t\ {\isacharplus}\ Suc\ {\isadigit{0}}{\isacharparenright}\ {\isacharequal}\ {\isacharbrackleft}False{\isacharbrackright}{\isachardoublequoteclose}\ \isacommand{by}\isamarkupfalse%
\ blast\isanewline
\ \ \isacommand{from}\isamarkupfalse%
\ this\ \isacommand{have}\isamarkupfalse%
\ sg{\isadigit{1}}{\isacharcolon}{\isachardoublequoteopen}lose\ {\isacharparenleft}Suc\ t{\isacharparenright}\ {\isacharequal}\ {\isacharbrackleft}False{\isacharbrackright}{\isachardoublequoteclose}\ \isacommand{by}\isamarkupfalse%
\ simp\isanewline
\ \ \isacommand{have}\isamarkupfalse%
\ {\isachardoublequoteopen}Suc\ {\isacharparenleft}Suc\ {\isadigit{0}}{\isacharparenright}{\isasymle}\ k\ {\isacharplus}\ d\ {\isacharplus}\ {\isadigit{3}}{\isachardoublequoteclose}\ \isacommand{by}\isamarkupfalse%
\ arith\isanewline
\ \ \isacommand{from}\isamarkupfalse%
\ h{\isadigit{1}}{\isadigit{0}}\ \isakeyword{and}\ this\ \isacommand{have}\isamarkupfalse%
\ {\isachardoublequoteopen}lose\ {\isacharparenleft}t\ {\isacharplus}\ Suc\ {\isacharparenleft}Suc\ {\isadigit{0}}{\isacharparenright}{\isacharparenright}\ {\isacharequal}\ {\isacharbrackleft}False{\isacharbrackright}{\isachardoublequoteclose}\ \isacommand{by}\isamarkupfalse%
\ blast\isanewline
\ \ \isacommand{from}\isamarkupfalse%
\ this\ \isacommand{have}\isamarkupfalse%
\ sg{\isadigit{2}}{\isacharcolon}{\isachardoublequoteopen}lose\ {\isacharparenleft}Suc\ {\isacharparenleft}Suc\ t{\isacharparenright}{\isacharparenright}\ {\isacharequal}\ {\isacharbrackleft}False{\isacharbrackright}{\isachardoublequoteclose}\ \isacommand{by}\isamarkupfalse%
\ simp\ \isanewline
\ \ \isacommand{from}\isamarkupfalse%
\ h{\isadigit{1}}\ \isakeyword{and}\ h{\isadigit{2}}\ \isakeyword{and}\ h{\isadigit{3}}\ \isakeyword{and}\ h{\isadigit{4}}\ \isakeyword{and}\ h{\isadigit{5}}\ \isakeyword{and}\ h{\isadigit{6}}\ \isakeyword{and}\ h{\isadigit{9}}\ \isakeyword{and}\ sg{\isadigit{1}}\ \isakeyword{and}\ sg{\isadigit{2}}\ \isacommand{have}\isamarkupfalse%
\ sg{\isadigit{3}}{\isacharcolon}\isanewline
\ \ \ {\isachardoublequoteopen}ack\ {\isacharparenleft}t\ {\isacharplus}\ {\isadigit{2}}{\isacharparenright}\ {\isacharequal}\ {\isacharbrackleft}connection{\isacharunderscore}ok{\isacharbrackright}{\isachardoublequoteclose}\isanewline
\ \ \ \ \isacommand{by}\isamarkupfalse%
\ {\isacharparenleft}simp\ add{\isacharcolon}\ Gateway{\isacharunderscore}L{\isadigit{1}}{\isacharparenright}\isanewline
\ \isacommand{from}\isamarkupfalse%
\ h{\isadigit{7}}\ \isakeyword{and}\ this\ \isacommand{have}\isamarkupfalse%
\ sg{\isadigit{4}}{\isacharcolon}{\isachardoublequoteopen}{\isasymforall}m{\isasymle}\ k{\isachardot}\ req\ {\isacharparenleft}{\isacharparenleft}t\ {\isacharplus}\ {\isadigit{2}}{\isacharparenright}\ {\isacharplus}\ m{\isacharparenright}\ {\isasymnoteq}\ {\isacharbrackleft}send{\isacharbrackright}{\isachardoublequoteclose}\ \isanewline
\ \ \ \ \isacommand{by}\isamarkupfalse%
\ {\isacharparenleft}auto{\isacharcomma}\ simp\ add{\isacharcolon}\ aux{\isacharunderscore}k{\isadigit{3}}req{\isacharparenright}\ \ \isanewline
\ \ \isacommand{from}\isamarkupfalse%
\ h{\isadigit{1}}{\isadigit{0}}\ \isacommand{have}\isamarkupfalse%
\ sg{\isadigit{5}}{\isacharcolon}{\isachardoublequoteopen}{\isasymforall}j{\isasymle}\ k{\isachardot}\ lose\ {\isacharparenleft}{\isacharparenleft}t\ {\isacharplus}\ {\isadigit{2}}{\isacharparenright}\ {\isacharplus}\ j{\isacharparenright}\ {\isacharequal}\ {\isacharbrackleft}False{\isacharbrackright}{\isachardoublequoteclose}\ \isanewline
\ \ \ \ \isacommand{by}\isamarkupfalse%
\ {\isacharparenleft}auto{\isacharcomma}\ simp\ add{\isacharcolon}\ aux{\isadigit{3}}lose{\isacharparenright}\ \isanewline
\ \ \isacommand{from}\isamarkupfalse%
\ h{\isadigit{1}}\ \isakeyword{and}\ sg{\isadigit{4}}\ \isakeyword{and}\ sg{\isadigit{5}}\ \isakeyword{and}\ sg{\isadigit{3}}\ \isakeyword{and}\ h{\isadigit{5}}\ \isakeyword{and}\ h{\isadigit{4}}\ \isakeyword{and}\ h{\isadigit{3}}\ \isakeyword{and}\ h{\isadigit{2}}\ \isacommand{have}\isamarkupfalse%
\ sg{\isadigit{6}}{\isacharcolon}\isanewline
\ \ \ {\isachardoublequoteopen}{\isasymforall}m\ {\isasymle}\ \ k{\isachardot}\ ack\ {\isacharparenleft}{\isacharparenleft}t\ {\isacharplus}\ {\isadigit{2}}{\isacharparenright}\ {\isacharplus}\ m{\isacharparenright}\ {\isacharequal}\ {\isacharbrackleft}connection{\isacharunderscore}ok{\isacharbrackright}{\isachardoublequoteclose}\isanewline
\ \ \ \ \isacommand{by}\isamarkupfalse%
\ {\isacharparenleft}rule\ Gateway{\isacharunderscore}L{\isadigit{6}}a{\isacharparenright}\isanewline
\ \ \isacommand{from}\isamarkupfalse%
\ sg{\isadigit{6}}\ \isacommand{have}\isamarkupfalse%
\ sg{\isadigit{7}}{\isacharcolon}{\isachardoublequoteopen}ack\ {\isacharparenleft}t\ {\isacharplus}\ {\isadigit{2}}\ {\isacharplus}\ k{\isacharparenright}\ {\isacharequal}\ {\isacharbrackleft}connection{\isacharunderscore}ok{\isacharbrackright}{\isachardoublequoteclose}\ \ \isacommand{by}\isamarkupfalse%
\ auto\isanewline
\ \ \isacommand{from}\isamarkupfalse%
\ h{\isadigit{1}}\ \isacommand{obtain}\isamarkupfalse%
\ i{\isadigit{1}}\ i{\isadigit{2}}\ x\ y\ \isakeyword{where}\isanewline
\ \ \ \ a{\isadigit{1}}{\isacharcolon}{\isachardoublequoteopen}Sample\ req\ dt\ x\ stop\ lose\ ack\ i{\isadigit{1}}\ vc{\isachardoublequoteclose}\ \isakeyword{and}\isanewline
\ \ \ \ a{\isadigit{2}}{\isacharcolon}{\isachardoublequoteopen}Delay\ y\ i{\isadigit{1}}\ d\ x\ i{\isadigit{2}}{\isachardoublequoteclose}\ \isakeyword{and}\ \isanewline
\ \ \ \ a{\isadigit{3}}{\isacharcolon}{\isachardoublequoteopen}Loss\ lose\ a\ i{\isadigit{2}}\ y\ i{\isachardoublequoteclose}\isanewline
\ \ \ \ \isacommand{by}\isamarkupfalse%
\ {\isacharparenleft}simp\ add{\isacharcolon}\ Gateway{\isacharunderscore}def{\isacharcomma}\ auto{\isacharparenright}\ \isanewline
\ \ \isacommand{from}\isamarkupfalse%
\ h{\isadigit{3}}\ \isakeyword{and}\ a{\isadigit{2}}\ \isakeyword{and}\ a{\isadigit{3}}\ \ \isacommand{have}\isamarkupfalse%
\ sg{\isadigit{8}}{\isacharcolon}{\isachardoublequoteopen}msg\ {\isacharparenleft}Suc\ {\isadigit{0}}{\isacharparenright}\ x{\isachardoublequoteclose}\isanewline
\ \ \ \ \isacommand{by}\isamarkupfalse%
\ {\isacharparenleft}simp\ add{\isacharcolon}\ Loss{\isacharunderscore}Delay{\isacharunderscore}msg{\isacharunderscore}a{\isacharparenright}\ \ \isanewline
\ \ \isacommand{from}\isamarkupfalse%
\ a{\isadigit{1}}\ \isakeyword{and}\ sg{\isadigit{8}}\ \isakeyword{and}\ h{\isadigit{4}}\ \isakeyword{and}\ h{\isadigit{5}}\ \isacommand{obtain}\isamarkupfalse%
\ st\ buffer\ \isakeyword{where}\isanewline
\ \ \ \ a{\isadigit{4}}{\isacharcolon}{\isachardoublequoteopen}tiTable{\isacharunderscore}SampleT\ req\ x\ stop\ lose\ {\isacharparenleft}fin{\isacharunderscore}inf{\isacharunderscore}append\ {\isacharbrackleft}init{\isacharunderscore}state{\isacharbrackright}\ st{\isacharparenright}\ \isanewline
\ \ \ \ \ \ \ \ \ {\isacharparenleft}fin{\isacharunderscore}inf{\isacharunderscore}append\ {\isacharbrackleft}{\isacharbrackleft}{\isacharbrackright}{\isacharbrackright}\ buffer{\isacharparenright}\ ack\ i{\isadigit{1}}\ vc\ st{\isachardoublequoteclose}\ \ \isakeyword{and}\ \isanewline
\ \ \ \ a{\isadigit{5}}{\isacharcolon}{\isachardoublequoteopen}{\isasymforall}\ t{\isachardot}\ buffer\ t\ {\isacharequal}\ {\isacharparenleft}if\ dt\ t\ {\isacharequal}\ {\isacharbrackleft}{\isacharbrackright}\ then\ fin{\isacharunderscore}inf{\isacharunderscore}append\ {\isacharbrackleft}{\isacharbrackleft}{\isacharbrackright}{\isacharbrackright}\ buffer\ t\ else\ dt\ t{\isacharparenright}{\isachardoublequoteclose}\isanewline
\ \ \ \ \isacommand{by}\isamarkupfalse%
\ {\isacharparenleft}simp\ add{\isacharcolon}\ Sample{\isacharunderscore}def\ Sample{\isacharunderscore}L{\isacharunderscore}def{\isacharcomma}\ auto{\isacharparenright}\isanewline
\ \ \isacommand{from}\isamarkupfalse%
\ a{\isadigit{4}}\ \isakeyword{and}\ h{\isadigit{2}}\ \isakeyword{and}\ sg{\isadigit{8}}\ \isakeyword{and}\ h{\isadigit{4}}\ \isakeyword{and}\ h{\isadigit{1}}{\isadigit{1}}\ \isakeyword{and}\ h{\isadigit{6}}\ \isakeyword{and}\ h{\isadigit{7}}\ \isakeyword{and}\ sg{\isadigit{6}}\ \isakeyword{and}\ h{\isadigit{1}}{\isadigit{0}}\ \isanewline
\ \ \isacommand{have}\isamarkupfalse%
\ sg{\isadigit{9}}{\isacharcolon}{\isachardoublequoteopen}{\isasymforall}\ t{\isadigit{1}}\ {\isacharless}\ {\isacharparenleft}t\ {\isacharplus}\ {\isadigit{3}}\ {\isacharplus}\ k{\isacharparenright}{\isachardot}\ i{\isadigit{1}}\ t{\isadigit{1}}\ {\isacharequal}\ {\isacharbrackleft}{\isacharbrackright}{\isachardoublequoteclose}\isanewline
\ \ \ \ \isacommand{by}\isamarkupfalse%
\ {\isacharparenleft}simp\ add{\isacharcolon}\ tiTable{\isacharunderscore}i{\isadigit{1}}{\isacharunderscore}{\isadigit{4}}{\isacharparenright}\isanewline
\ \ \isacommand{from}\isamarkupfalse%
\ sg{\isadigit{9}}\ \isakeyword{and}\ a{\isadigit{2}}\ \isacommand{have}\isamarkupfalse%
\ sg{\isadigit{1}}{\isadigit{0}}{\isacharcolon}{\isachardoublequoteopen}{\isasymforall}\ t{\isadigit{2}}\ {\isacharless}\ {\isacharparenleft}t\ {\isacharplus}\ {\isadigit{3}}\ {\isacharplus}\ k\ {\isacharplus}\ d{\isacharparenright}{\isachardot}\ i{\isadigit{2}}\ t{\isadigit{2}}\ {\isacharequal}\ {\isacharbrackleft}{\isacharbrackright}{\isachardoublequoteclose}\isanewline
\ \ \ \ \isacommand{by}\isamarkupfalse%
\ {\isacharparenleft}rule\ Delay{\isacharunderscore}L{\isadigit{2}}{\isacharparenright}\ \isanewline
\ \ \isacommand{from}\isamarkupfalse%
\ sg{\isadigit{1}}{\isadigit{0}}\ \isakeyword{and}\ a{\isadigit{3}}\ \isakeyword{and}\ h{\isadigit{2}}\ \isacommand{show}\isamarkupfalse%
\ {\isacharquery}thesis\ \isacommand{by}\isamarkupfalse%
\ {\isacharparenleft}rule\ Loss{\isacharunderscore}L{\isadigit{2}}{\isacharparenright}\isanewline
\isacommand{qed}\isamarkupfalse%
\endisatagproof
{\isafoldproof}%
\isadelimproof
\ \isanewline
\endisadelimproof
\isanewline 
\isacommand{lemma}\isamarkupfalse%
\ Gateway{\isacharunderscore}L{\isadigit{8}}a{\isacharcolon}\isanewline
\ \ \isakeyword{assumes}\ h{\isadigit{1}}{\isacharcolon}{\isachardoublequoteopen}Gateway\ req\ dt\ a\ stop\ lose\ d\ ack\ i\ vc{\isachardoublequoteclose}\ \isanewline
\ \ \ \ \ \ \isakeyword{and}\ h{\isadigit{2}}{\isacharcolon}{\isachardoublequoteopen}msg\ {\isacharparenleft}Suc\ {\isadigit{0}}{\isacharparenright}\ req{\isachardoublequoteclose}\isanewline
\ \ \ \ \ \ \isakeyword{and}\ h{\isadigit{3}}{\isacharcolon}{\isachardoublequoteopen}msg\ {\isacharparenleft}Suc\ {\isadigit{0}}{\isacharparenright}\ stop{\isachardoublequoteclose}\isanewline
\ \ \ \ \ \ \isakeyword{and}\ h{\isadigit{4}}{\isacharcolon}{\isachardoublequoteopen}msg\ {\isacharparenleft}Suc\ {\isadigit{0}}{\isacharparenright}\ a{\isachardoublequoteclose}\isanewline
\ \ \ \ \ \ \isakeyword{and}\ h{\isadigit{5}}{\isacharcolon}{\isachardoublequoteopen}ts\ lose{\isachardoublequoteclose}\isanewline
\ \ \ \ \ \ \isakeyword{and}\ h{\isadigit{6}}{\isacharcolon}{\isachardoublequoteopen}{\isasymforall}j{\isasymle}{\isadigit{2}}\ {\isacharasterisk}\ d{\isachardot}\ lose\ {\isacharparenleft}t\ {\isacharplus}\ j{\isacharparenright}\ {\isacharequal}\ {\isacharbrackleft}False{\isacharbrackright}{\isachardoublequoteclose}\ \isanewline
\ \ \ \ \ \ \isakeyword{and}\ h{\isadigit{7}}{\isacharcolon}{\isachardoublequoteopen}ack\ t\ {\isacharequal}\ {\isacharbrackleft}sending{\isacharunderscore}data{\isacharbrackright}{\isachardoublequoteclose}\isanewline
\ \ \ \ \ \ \isakeyword{and}\ h{\isadigit{8}}{\isacharcolon}{\isachardoublequoteopen}{\isasymforall}t{\isadigit{3}}\ {\isasymle}\ t\ {\isacharplus}\ d{\isachardot}\ a\ t{\isadigit{3}}\ {\isacharequal}\ {\isacharbrackleft}{\isacharbrackright}{\isachardoublequoteclose}\isanewline
\ \ \ \ \ \ \isakeyword{and}\ h{\isadigit{9}}{\isacharcolon}{\isachardoublequoteopen}x\ {\isasymle}\ d\ {\isacharplus}\ d{\isachardoublequoteclose}\isanewline
\ \ \isakeyword{shows}\ {\isachardoublequoteopen}ack\ {\isacharparenleft}t\ {\isacharplus}\ x{\isacharparenright}\ {\isacharequal}\ {\isacharbrackleft}sending{\isacharunderscore}data{\isacharbrackright}{\isachardoublequoteclose}\isanewline
\isadelimproof
\endisadelimproof
\isatagproof
\isacommand{proof}\isamarkupfalse%
\ {\isacharminus}\isanewline
\ \ \isacommand{from}\isamarkupfalse%
\ h{\isadigit{1}}\ \isacommand{obtain}\isamarkupfalse%
\ i{\isadigit{1}}\ i{\isadigit{2}}\ x\ y\ \isakeyword{where}\isanewline
\ \ \ \ a{\isadigit{1}}{\isacharcolon}{\isachardoublequoteopen}Sample\ req\ dt\ x\ stop\ lose\ ack\ i{\isadigit{1}}\ vc{\isachardoublequoteclose}\ \isakeyword{and}\isanewline
\ \ \ \ a{\isadigit{2}}{\isacharcolon}{\isachardoublequoteopen}Delay\ y\ i{\isadigit{1}}\ d\ x\ i{\isadigit{2}}{\isachardoublequoteclose}\ \isakeyword{and}\ \isanewline
\ \ \ \ a{\isadigit{3}}{\isacharcolon}{\isachardoublequoteopen}Loss\ lose\ a\ i{\isadigit{2}}\ y\ i{\isachardoublequoteclose}\isanewline
\ \ \ \ \isacommand{by}\isamarkupfalse%
\ {\isacharparenleft}simp\ add{\isacharcolon}\ Gateway{\isacharunderscore}def{\isacharcomma}\ auto{\isacharparenright}\ \isanewline
\ \ \isacommand{from}\isamarkupfalse%
\ h{\isadigit{8}}\ \isakeyword{and}\ a{\isadigit{3}}\ \isakeyword{and}\ h{\isadigit{5}}\ \isacommand{have}\isamarkupfalse%
\ sg{\isadigit{1}}{\isacharcolon}{\isachardoublequoteopen}{\isasymforall}t{\isadigit{3}}\ {\isasymle}\ t\ {\isacharplus}\ d{\isachardot}\ y\ t{\isadigit{3}}\ {\isacharequal}\ {\isacharbrackleft}{\isacharbrackright}{\isachardoublequoteclose}\ \isacommand{by}\isamarkupfalse%
\ {\isacharparenleft}rule\ Loss{\isacharunderscore}L{\isadigit{6}}{\isacharparenright}\isanewline
\ \ \isacommand{from}\isamarkupfalse%
\ sg{\isadigit{1}}\ \isakeyword{and}\ a{\isadigit{2}}\ \isacommand{have}\isamarkupfalse%
\ sg{\isadigit{2}}{\isacharcolon}{\isachardoublequoteopen}{\isasymforall}t{\isadigit{4}}\ {\isasymle}\ t\ {\isacharplus}\ d\ {\isacharplus}\ d{\isachardot}\ x\ t{\isadigit{4}}\ {\isacharequal}\ {\isacharbrackleft}{\isacharbrackright}{\isachardoublequoteclose}\ \ \ \isacommand{by}\isamarkupfalse%
\ {\isacharparenleft}rule\ Delay{\isacharunderscore}L{\isadigit{4}}{\isacharparenright}\ \isanewline
\ \ \isacommand{from}\isamarkupfalse%
\ h{\isadigit{4}}\ \isakeyword{and}\ a{\isadigit{2}}\ \isakeyword{and}\ a{\isadigit{3}}\ \isacommand{have}\isamarkupfalse%
\ sg{\isadigit{3}}{\isacharcolon}{\isachardoublequoteopen}msg\ {\isacharparenleft}Suc\ {\isadigit{0}}{\isacharparenright}\ x{\isachardoublequoteclose}\ \isacommand{by}\isamarkupfalse%
\ {\isacharparenleft}simp\ add{\isacharcolon}\ \ Loss{\isacharunderscore}Delay{\isacharunderscore}msg{\isacharunderscore}a{\isacharparenright}\isanewline
\ \ \isacommand{from}\isamarkupfalse%
\ h{\isadigit{3}}\ \isakeyword{and}\ h{\isadigit{5}}\ \isakeyword{and}\ h{\isadigit{2}}\ \isakeyword{and}\ sg{\isadigit{3}}\ \isakeyword{and}\ h{\isadigit{6}}\ \isakeyword{and}\ h{\isadigit{7}}\ \isakeyword{and}\ a{\isadigit{1}}\ \isakeyword{and}\ h{\isadigit{9}}\ \isakeyword{and}\ sg{\isadigit{2}}\ \isacommand{show}\isamarkupfalse%
\ {\isacharquery}thesis\isanewline
\ \ \ \ \isacommand{by}\isamarkupfalse%
\ {\isacharparenleft}simp\ add{\isacharcolon}\ Sample{\isacharunderscore}sending{\isacharunderscore}data{\isacharparenright}\isanewline
\isacommand{qed}\isamarkupfalse%
\endisatagproof
{\isafoldproof}%
\isadelimproof
\isanewline
\endisadelimproof
\isanewline 
\isacommand{lemma}\isamarkupfalse%
\ Gateway{\isacharunderscore}L{\isadigit{8}}{\isacharcolon}\isanewline
\ \ \isakeyword{assumes}\ h{\isadigit{1}}{\isacharcolon}{\isachardoublequoteopen}Gateway\ req\ dt\ a\ stop\ lose\ d\ ack\ i\ vc{\isachardoublequoteclose}\ \isanewline
\ \ \ \ \ \ \isakeyword{and}\ h{\isadigit{2}}{\isacharcolon}{\isachardoublequoteopen}msg\ {\isacharparenleft}Suc\ {\isadigit{0}}{\isacharparenright}\ req{\isachardoublequoteclose}\isanewline
\ \ \ \ \ \ \isakeyword{and}\ h{\isadigit{3}}{\isacharcolon}{\isachardoublequoteopen}msg\ {\isacharparenleft}Suc\ {\isadigit{0}}{\isacharparenright}\ stop{\isachardoublequoteclose}\isanewline
\ \ \ \ \ \ \isakeyword{and}\ h{\isadigit{4}}{\isacharcolon}{\isachardoublequoteopen}msg\ {\isacharparenleft}Suc\ {\isadigit{0}}{\isacharparenright}\ a{\isachardoublequoteclose}\isanewline
\ \ \ \ \ \ \isakeyword{and}\ h{\isadigit{5}}{\isacharcolon}{\isachardoublequoteopen}ts\ lose{\isachardoublequoteclose}\isanewline
\ \ \ \ \ \ \isakeyword{and}\ h{\isadigit{6}}{\isacharcolon}{\isachardoublequoteopen}{\isasymforall}j{\isasymle}{\isadigit{2}}\ {\isacharasterisk}\ d{\isachardot}\ lose\ {\isacharparenleft}t\ {\isacharplus}\ j{\isacharparenright}\ {\isacharequal}\ {\isacharbrackleft}False{\isacharbrackright}{\isachardoublequoteclose}\ \isanewline
\ \ \ \ \ \ \isakeyword{and}\ h{\isadigit{7}}{\isacharcolon}{\isachardoublequoteopen}ack\ t\ {\isacharequal}\ {\isacharbrackleft}sending{\isacharunderscore}data{\isacharbrackright}{\isachardoublequoteclose}\isanewline
\ \ \ \ \ \ \isakeyword{and}\ h{\isadigit{8}}{\isacharcolon}{\isachardoublequoteopen}{\isasymforall}t{\isadigit{3}}\ {\isasymle}\ t\ {\isacharplus}\ d{\isachardot}\ a\ t{\isadigit{3}}\ {\isacharequal}\ {\isacharbrackleft}{\isacharbrackright}{\isachardoublequoteclose}\isanewline
\ \ \isakeyword{shows}\ {\isachardoublequoteopen}{\isasymforall}x\ {\isasymle}\ d\ {\isacharplus}\ d{\isachardot}\ ack\ {\isacharparenleft}t\ {\isacharplus}\ x{\isacharparenright}\ {\isacharequal}\ {\isacharbrackleft}sending{\isacharunderscore}data{\isacharbrackright}{\isachardoublequoteclose}\isanewline
\isadelimproof
\endisadelimproof
\isatagproof
\isacommand{using}\isamarkupfalse%
\ assms\ \isanewline
\ \ \isacommand{by}\isamarkupfalse%
\ {\isacharparenleft}simp\ add{\isacharcolon}\ Gateway{\isacharunderscore}L{\isadigit{8}}a{\isacharparenright}%
\endisatagproof
{\isafoldproof}%
\isadelimproof
\endisadelimproof
\isamarkupsubsection{Proof of the Refinement Relation for the Gateway Requirements%
}
\isamarkuptrue%
\isacommand{lemma}\isamarkupfalse%
\ Gateway{\isacharunderscore}L{\isadigit{0}}{\isacharcolon}\isanewline
\ \isakeyword{assumes}\ h{\isadigit{1}}{\isacharcolon}{\isachardoublequoteopen}Gateway\ req\ dt\ a\ stop\ lose\ d\ ack\ i\ vc{\isachardoublequoteclose}\isanewline
\ \isakeyword{shows}\ \ \ \ \ \ {\isachardoublequoteopen}GatewayReq\ req\ dt\ a\ stop\ lose\ d\ ack\ i\ vc{\isachardoublequoteclose}\isanewline
\isadelimproof
\endisadelimproof
\isatagproof
\isacommand{using}\isamarkupfalse%
\ assms\isanewline
\ \ \isacommand{by}\isamarkupfalse%
\ {\isacharparenleft}simp\ add{\isacharcolon}\ GatewayReq{\isacharunderscore}def\ Gateway{\isacharunderscore}L{\isadigit{1}}\ Gateway{\isacharunderscore}L{\isadigit{2}}\ Gateway{\isacharunderscore}L{\isadigit{3}}\ Gateway{\isacharunderscore}L{\isadigit{4}}{\isacharparenright}%
\endisatagproof
{\isafoldproof}%
\isadelimproof
\endisadelimproof
\isamarkupsubsection{Lemmas about Gateway Requirements%
}
\isamarkuptrue%
\isacommand{lemma}\isamarkupfalse%
\ GatewayReq{\isacharunderscore}L{\isadigit{1}}{\isacharcolon}\isanewline
\ \ \isakeyword{assumes}\ h{\isadigit{1}}{\isacharcolon}{\isachardoublequoteopen}msg\ {\isacharparenleft}Suc\ {\isadigit{0}}{\isacharparenright}\ req{\isachardoublequoteclose}\isanewline
\ \ \ \ \ \ \isakeyword{and}\ h{\isadigit{2}}{\isacharcolon}{\isachardoublequoteopen}msg\ {\isacharparenleft}Suc\ {\isadigit{0}}{\isacharparenright}\ stop{\isachardoublequoteclose}\isanewline
\ \ \ \ \ \ \isakeyword{and}\ h{\isadigit{3}}{\isacharcolon}{\isachardoublequoteopen}msg\ {\isacharparenleft}Suc\ {\isadigit{0}}{\isacharparenright}\ a{\isachardoublequoteclose}\isanewline
\ \ \ \ \ \ \isakeyword{and}\ h{\isadigit{4}}{\isacharcolon}{\isachardoublequoteopen}ts\ lose{\isachardoublequoteclose}\isanewline
\ \ \ \ \ \ \isakeyword{and}\ h{\isadigit{6}}{\isacharcolon}{\isachardoublequoteopen}req\ {\isacharparenleft}t\ {\isacharplus}\ {\isadigit{3}}\ {\isacharplus}\ k{\isacharparenright}\ {\isacharequal}\ {\isacharbrackleft}send{\isacharbrackright}{\isachardoublequoteclose}\ \isanewline
\ \ \ \ \ \ \isakeyword{and}\ h{\isadigit{7}}{\isacharcolon}{\isachardoublequoteopen}{\isasymforall}j{\isasymle}{\isadigit{2}}\ {\isacharasterisk}\ d\ {\isacharplus}\ {\isacharparenleft}{\isadigit{4}}\ {\isacharplus}\ k{\isacharparenright}{\isachardot}\ lose\ {\isacharparenleft}t\ {\isacharplus}\ j{\isacharparenright}\ {\isacharequal}\ {\isacharbrackleft}False{\isacharbrackright}{\isachardoublequoteclose}\isanewline
\ \ \ \ \ \ \isakeyword{and}\ h{\isadigit{9}}{\isacharcolon}{\isachardoublequoteopen}{\isasymforall}m{\isasymle}\ k{\isachardot}\ ack\ {\isacharparenleft}t\ {\isacharplus}\ {\isadigit{2}}\ {\isacharplus}\ m{\isacharparenright}\ {\isacharequal}\ {\isacharbrackleft}connection{\isacharunderscore}ok{\isacharbrackright}{\isachardoublequoteclose}\isanewline
\ \ \ \ \ \ \isakeyword{and}\ h{\isadigit{1}}{\isadigit{0}}{\isacharcolon}{\isachardoublequoteopen}GatewayReq\ req\ dt\ a\ stop\ lose\ d\ ack\ i\ vc{\isachardoublequoteclose}\isanewline
\ \isakeyword{shows}\ {\isachardoublequoteopen}ack\ {\isacharparenleft}t\ {\isacharplus}\ {\isadigit{3}}\ {\isacharplus}\ k{\isacharparenright}\ {\isacharequal}\ {\isacharbrackleft}sending{\isacharunderscore}data{\isacharbrackright}{\isachardoublequoteclose}\isanewline
\isadelimproof
\endisadelimproof
\isatagproof
\isacommand{proof}\isamarkupfalse%
\ {\isacharminus}\ \isanewline
\ \ \isacommand{from}\isamarkupfalse%
\ h{\isadigit{9}}\ \isacommand{have}\isamarkupfalse%
\ sg{\isadigit{1}}{\isacharcolon}{\isachardoublequoteopen}ack\ {\isacharparenleft}Suc\ {\isacharparenleft}Suc\ {\isacharparenleft}t\ {\isacharplus}\ k{\isacharparenright}{\isacharparenright}{\isacharparenright}\ {\isacharequal}\ {\isacharbrackleft}connection{\isacharunderscore}ok{\isacharbrackright}{\isachardoublequoteclose}\ \isacommand{by}\isamarkupfalse%
\ auto\ \isanewline
\ \ \isacommand{from}\isamarkupfalse%
\ h{\isadigit{7}}\ \isacommand{have}\isamarkupfalse%
\ sg{\isadigit{2}}{\isacharcolon}\ \isanewline
\ \ \ {\isachardoublequoteopen}{\isasymforall}ka{\isasymle}Suc\ d{\isachardot}\ lose\ {\isacharparenleft}Suc\ {\isacharparenleft}Suc\ {\isacharparenleft}t\ {\isacharplus}\ k\ {\isacharplus}\ ka{\isacharparenright}{\isacharparenright}{\isacharparenright}\ {\isacharequal}\ {\isacharbrackleft}False{\isacharbrackright}{\isachardoublequoteclose}\isanewline
\ \ \ \ \isacommand{by}\isamarkupfalse%
\ {\isacharparenleft}simp\ add{\isacharcolon}\ aux{\isacharunderscore}lemma{\isacharunderscore}lose{\isacharunderscore}{\isadigit{1}}{\isacharparenright}\ \isanewline
\ \ \isacommand{from}\isamarkupfalse%
\ h{\isadigit{1}}\ \isakeyword{and}\ h{\isadigit{2}}\ \isakeyword{and}\ h{\isadigit{3}}\ \isakeyword{and}\ h{\isadigit{4}}\ \isakeyword{and}\ h{\isadigit{6}}\ \isakeyword{and}\ h{\isadigit{1}}{\isadigit{0}}\ \isakeyword{and}\ sg{\isadigit{1}}\ \isakeyword{and}\ sg{\isadigit{2}}\ \isacommand{have}\isamarkupfalse%
\ sg{\isadigit{3}}{\isacharcolon}\isanewline
\ \ \ {\isachardoublequoteopen}ack\ {\isacharparenleft}t\ {\isacharplus}\ {\isadigit{2}}\ {\isacharplus}\ k{\isacharparenright}\ {\isacharequal}\ {\isacharbrackleft}connection{\isacharunderscore}ok{\isacharbrackright}\ {\isasymand}\ \isanewline
\ \ \ \ req\ {\isacharparenleft}Suc\ {\isacharparenleft}t\ {\isacharplus}\ {\isadigit{2}}\ {\isacharplus}\ k{\isacharparenright}{\isacharparenright}\ {\isacharequal}\ {\isacharbrackleft}send{\isacharbrackright}\ {\isasymand}\ {\isacharparenleft}{\isasymforall}k{\isasymle}Suc\ d{\isachardot}\ lose\ {\isacharparenleft}t\ {\isacharplus}\ k{\isacharparenright}\ {\isacharequal}\ {\isacharbrackleft}False{\isacharbrackright}{\isacharparenright}\ {\isasymlongrightarrow}\isanewline
\ \ \ \ ack\ {\isacharparenleft}Suc\ {\isacharparenleft}t\ {\isacharplus}\ {\isadigit{2}}\ {\isacharplus}\ k{\isacharparenright}{\isacharparenright}\ {\isacharequal}\ {\isacharbrackleft}sending{\isacharunderscore}data{\isacharbrackright}{\isachardoublequoteclose}\isanewline
\ \ \ \ \isacommand{by}\isamarkupfalse%
\ {\isacharparenleft}simp\ add{\isacharcolon}\ GatewayReq{\isacharunderscore}def{\isacharparenright}\ \ \isanewline
\ \ \isacommand{have}\isamarkupfalse%
\ sg{\isadigit{4}}{\isacharcolon}{\isachardoublequoteopen}t\ {\isacharplus}\ {\isadigit{3}}\ {\isacharplus}\ k\ {\isacharequal}\ Suc\ {\isacharparenleft}Suc\ {\isacharparenleft}Suc\ {\isacharparenleft}t\ {\isacharplus}\ k{\isacharparenright}{\isacharparenright}{\isacharparenright}{\isachardoublequoteclose}\ \isacommand{by}\isamarkupfalse%
\ arith\isanewline
\ \ \isacommand{from}\isamarkupfalse%
\ sg{\isadigit{3}}\ \isakeyword{and}\ sg{\isadigit{1}}\ \isakeyword{and}\ h{\isadigit{6}}\ \isakeyword{and}\ h{\isadigit{7}}\ \isakeyword{and}\ sg{\isadigit{4}}\ \isacommand{show}\isamarkupfalse%
\ {\isacharquery}thesis\ \isanewline
\ \ \ \ \isacommand{by}\isamarkupfalse%
\ {\isacharparenleft}simp\ add{\isacharcolon}\ eval{\isacharunderscore}nat{\isacharunderscore}numeral{\isacharparenright}\ \ \isanewline
\isacommand{qed}\isamarkupfalse%
\endisatagproof
{\isafoldproof}%
\isadelimproof
\isanewline
\endisadelimproof
\isanewline
\isacommand{lemma}\isamarkupfalse%
\ GatewayReq{\isacharunderscore}L{\isadigit{2}}{\isacharcolon}\isanewline
\ \isakeyword{assumes}\ \ h{\isadigit{1}}{\isacharcolon}{\isachardoublequoteopen}msg\ {\isacharparenleft}Suc\ {\isadigit{0}}{\isacharparenright}\ req{\isachardoublequoteclose}\isanewline
\ \ \ \ \ \ \isakeyword{and}\ h{\isadigit{2}}{\isacharcolon}{\isachardoublequoteopen}msg\ {\isacharparenleft}Suc\ {\isadigit{0}}{\isacharparenright}\ stop{\isachardoublequoteclose}\isanewline
\ \ \ \ \ \ \isakeyword{and}\ h{\isadigit{3}}{\isacharcolon}{\isachardoublequoteopen}msg\ {\isacharparenleft}Suc\ {\isadigit{0}}{\isacharparenright}\ a{\isachardoublequoteclose}\isanewline
\ \ \ \ \ \ \isakeyword{and}\ h{\isadigit{4}}{\isacharcolon}{\isachardoublequoteopen}ts\ lose{\isachardoublequoteclose}\isanewline
\ \ \ \ \ \ \isakeyword{and}\ h{\isadigit{5}}{\isacharcolon}{\isachardoublequoteopen}GatewayReq\ req\ dt\ a\ stop\ lose\ d\ ack\ i\ vc{\isachardoublequoteclose}\isanewline
\ \ \ \ \ \ \isakeyword{and}\ h{\isadigit{6}}{\isacharcolon}{\isachardoublequoteopen}req\ {\isacharparenleft}t\ {\isacharplus}\ {\isadigit{3}}\ {\isacharplus}\ k{\isacharparenright}\ {\isacharequal}\ {\isacharbrackleft}send{\isacharbrackright}{\isachardoublequoteclose}\isanewline
\ \ \ \ \ \ \isakeyword{and}\ h{\isadigit{7}}{\isacharcolon}{\isachardoublequoteopen}inf{\isacharunderscore}last{\isacharunderscore}ti\ dt\ t\ {\isasymnoteq}\ {\isacharbrackleft}{\isacharbrackright}{\isachardoublequoteclose}\isanewline
\ \ \ \ \ \ \isakeyword{and}\ h{\isadigit{8}}{\isacharcolon}{\isachardoublequoteopen}{\isasymforall}j{\isasymle}{\isadigit{2}}\ {\isacharasterisk}\ d\ {\isacharplus}\ {\isacharparenleft}{\isadigit{4}}\ {\isacharplus}\ k{\isacharparenright}{\isachardot}\ lose\ {\isacharparenleft}t\ {\isacharplus}\ j{\isacharparenright}\ {\isacharequal}\ {\isacharbrackleft}False{\isacharbrackright}{\isachardoublequoteclose}\isanewline
\ \ \ \ \ \ \isakeyword{and}\ h{\isadigit{9}}{\isacharcolon}{\isachardoublequoteopen}{\isasymforall}m{\isasymle}k{\isachardot}\ ack\ {\isacharparenleft}t\ {\isacharplus}\ {\isadigit{2}}\ {\isacharplus}\ m{\isacharparenright}\ {\isacharequal}\ {\isacharbrackleft}connection{\isacharunderscore}ok{\isacharbrackright}{\isachardoublequoteclose}\isanewline
\ \ \isakeyword{shows}\ {\isachardoublequoteopen}i\ {\isacharparenleft}t\ {\isacharplus}\ {\isadigit{3}}\ {\isacharplus}\ k\ {\isacharplus}\ d{\isacharparenright}\ {\isasymnoteq}\ {\isacharbrackleft}{\isacharbrackright}{\isachardoublequoteclose}\isanewline
\isadelimproof
\endisadelimproof
\isatagproof
\isacommand{proof}\isamarkupfalse%
\ {\isacharminus}\isanewline
\ \ \isacommand{from}\isamarkupfalse%
\ h{\isadigit{8}}\ \isacommand{have}\isamarkupfalse%
\ sg{\isadigit{1}}{\isacharcolon}{\isachardoublequoteopen}{\isacharparenleft}{\isasymforall}\ {\isacharparenleft}x{\isacharcolon}{\isacharcolon}nat{\isacharparenright}{\isachardot}\ x\ {\isasymle}\ {\isacharparenleft}d{\isacharplus}{\isadigit{1}}{\isacharparenright}\ {\isasymlongrightarrow}\ lose\ {\isacharparenleft}t{\isacharplus}x{\isacharparenright}\ {\isacharequal}\ {\isacharbrackleft}False{\isacharbrackright}{\isacharparenright}{\isachardoublequoteclose}\isanewline
\ \ \ \ \isacommand{by}\isamarkupfalse%
\ {\isacharparenleft}simp\ add{\isacharcolon}\ aux{\isacharunderscore}lemma{\isacharunderscore}lose{\isacharunderscore}{\isadigit{2}}{\isacharparenright}\ \isanewline
\ \ \isacommand{from}\isamarkupfalse%
\ h{\isadigit{8}}\ \isacommand{have}\isamarkupfalse%
\ sg{\isadigit{2}}{\isacharcolon}{\isachardoublequoteopen}{\isasymforall}ka{\isasymle}Suc\ d{\isachardot}\ lose\ {\isacharparenleft}Suc\ {\isacharparenleft}Suc\ {\isacharparenleft}t\ {\isacharplus}\ k\ {\isacharplus}\ ka{\isacharparenright}{\isacharparenright}{\isacharparenright}\ {\isacharequal}\ {\isacharbrackleft}False{\isacharbrackright}{\isachardoublequoteclose}\isanewline
\ \ \ \ \isacommand{by}\isamarkupfalse%
\ {\isacharparenleft}simp\ add{\isacharcolon}\ aux{\isacharunderscore}lemma{\isacharunderscore}lose{\isacharunderscore}{\isadigit{1}}{\isacharparenright}\isanewline
\ \ \isacommand{from}\isamarkupfalse%
\ h{\isadigit{9}}\ \isacommand{have}\isamarkupfalse%
\ sg{\isadigit{3}}{\isacharcolon}{\isachardoublequoteopen}ack\ {\isacharparenleft}t\ {\isacharplus}\ {\isadigit{2}}\ {\isacharplus}\ k{\isacharparenright}\ {\isacharequal}\ {\isacharbrackleft}connection{\isacharunderscore}ok{\isacharbrackright}{\isachardoublequoteclose}\ \isacommand{by}\isamarkupfalse%
\ simp\isanewline
\ \ \isacommand{from}\isamarkupfalse%
\ h{\isadigit{1}}\ \isakeyword{and}\ h{\isadigit{2}}\ \isakeyword{and}\ h{\isadigit{3}}\ \isakeyword{and}\ h{\isadigit{4}}\ \isakeyword{and}\ h{\isadigit{5}}\ \isakeyword{and}\ h{\isadigit{6}}\ \isakeyword{and}\ sg{\isadigit{2}}\ \isakeyword{and}\ sg{\isadigit{3}}\ \isacommand{have}\isamarkupfalse%
\ sg{\isadigit{4}}{\isacharcolon}\isanewline
\ \ \ {\isachardoublequoteopen}ack\ {\isacharparenleft}t\ {\isacharplus}\ {\isadigit{2}}\ {\isacharplus}\ k{\isacharparenright}\ {\isacharequal}\ {\isacharbrackleft}connection{\isacharunderscore}ok{\isacharbrackright}\ {\isasymand}\ \isanewline
\ \ \ \ req\ {\isacharparenleft}Suc\ {\isacharparenleft}t\ {\isacharplus}\ {\isadigit{2}}\ {\isacharplus}\ k{\isacharparenright}{\isacharparenright}\ {\isacharequal}\ {\isacharbrackleft}send{\isacharbrackright}\ {\isasymand}\ {\isacharparenleft}{\isasymforall}k{\isasymle}Suc\ d{\isachardot}\ lose\ {\isacharparenleft}t\ {\isacharplus}\ k{\isacharparenright}\ {\isacharequal}\ {\isacharbrackleft}False{\isacharbrackright}{\isacharparenright}\ {\isasymlongrightarrow}\isanewline
\ \ \ \ i\ {\isacharparenleft}Suc\ {\isacharparenleft}t\ {\isacharplus}\ {\isadigit{2}}\ {\isacharplus}\ k\ {\isacharplus}\ d{\isacharparenright}{\isacharparenright}\ {\isacharequal}\ inf{\isacharunderscore}last{\isacharunderscore}ti\ dt\ {\isacharparenleft}t\ {\isacharplus}\ {\isadigit{2}}\ {\isacharplus}\ k{\isacharparenright}{\isachardoublequoteclose}\isanewline
\ \ \ \ \isacommand{by}\isamarkupfalse%
\ {\isacharparenleft}simp\ add{\isacharcolon}\ GatewayReq{\isacharunderscore}def{\isacharcomma}\ auto{\isacharparenright}\ \ \isanewline
\ \ \isacommand{from}\isamarkupfalse%
\ h{\isadigit{7}}\ \isacommand{have}\isamarkupfalse%
\ sg{\isadigit{5}}{\isacharcolon}{\isachardoublequoteopen}inf{\isacharunderscore}last{\isacharunderscore}ti\ dt\ {\isacharparenleft}t\ {\isacharplus}\ {\isadigit{2}}\ {\isacharplus}\ k{\isacharparenright}\ {\isasymnoteq}\ {\isacharbrackleft}{\isacharbrackright}{\isachardoublequoteclose}\isanewline
\ \ \ \ \isacommand{by}\isamarkupfalse%
\ {\isacharparenleft}simp\ add{\isacharcolon}\ inf{\isacharunderscore}last{\isacharunderscore}ti{\isacharunderscore}nonempty{\isacharunderscore}k{\isacharparenright}\isanewline
\ \ \isacommand{have}\isamarkupfalse%
\ sg{\isadigit{6}}{\isacharcolon}{\isachardoublequoteopen}t\ {\isacharplus}\ {\isadigit{3}}\ {\isacharplus}\ k\ {\isacharequal}\ Suc\ {\isacharparenleft}Suc\ {\isacharparenleft}Suc\ {\isacharparenleft}t\ {\isacharplus}\ k{\isacharparenright}{\isacharparenright}{\isacharparenright}{\isachardoublequoteclose}\ \isacommand{by}\isamarkupfalse%
\ arith\isanewline
\ \ \isacommand{have}\isamarkupfalse%
\ sg{\isadigit{7}}{\isacharcolon}{\isachardoublequoteopen}t\ {\isacharplus}\ {\isadigit{2}}\ {\isacharplus}\ k\ {\isacharequal}\ Suc\ {\isacharparenleft}Suc\ {\isacharparenleft}t\ {\isacharplus}\ k{\isacharparenright}{\isacharparenright}{\isachardoublequoteclose}\ \isacommand{by}\isamarkupfalse%
\ arith\isanewline
\ \ \isacommand{from}\isamarkupfalse%
\ sg{\isadigit{1}}\ \isakeyword{and}\ sg{\isadigit{2}}\ \isakeyword{and}\ sg{\isadigit{3}}\ \isakeyword{and}\ sg{\isadigit{4}}\ \isakeyword{and}\ sg{\isadigit{5}}\ \isakeyword{and}\ sg{\isadigit{6}}\ \isakeyword{and}\ sg{\isadigit{7}}\ \isakeyword{and}\ h{\isadigit{6}}\ \isacommand{show}\isamarkupfalse%
\ {\isacharquery}thesis\isanewline
\ \ \ \ \isacommand{by}\isamarkupfalse%
\ {\isacharparenleft}simp\ add{\isacharcolon}\ eval{\isacharunderscore}nat{\isacharunderscore}numeral{\isacharparenright}\isanewline
\isacommand{qed}\isamarkupfalse%
\endisatagproof
{\isafoldproof}%
\isadelimproof
\endisadelimproof
\isamarkupsubsection{Properties of the Gateway System%
}
\isamarkuptrue%
\isacommand{lemma}\isamarkupfalse%
\ GatewaySystem{\isacharunderscore}L{\isadigit{1}}aux{\isacharcolon}\isanewline
\ \ \isakeyword{assumes}\ h{\isadigit{1}}{\isacharcolon}{\isachardoublequoteopen}msg\ {\isacharparenleft}Suc\ {\isadigit{0}}{\isacharparenright}\ req{\isachardoublequoteclose}\isanewline
\ \ \ \ \ \ \isakeyword{and}\ h{\isadigit{2}}{\isacharcolon}{\isachardoublequoteopen}msg\ {\isacharparenleft}Suc\ {\isadigit{0}}{\isacharparenright}\ stop{\isachardoublequoteclose}\isanewline
\ \ \ \ \ \ \isakeyword{and}\ h{\isadigit{3}}{\isacharcolon}{\isachardoublequoteopen}msg\ {\isacharparenleft}Suc\ {\isadigit{0}}{\isacharparenright}\ a{\isachardoublequoteclose}\isanewline
\ \ \ \ \ \ \isakeyword{and}\ h{\isadigit{4}}{\isacharcolon}{\isachardoublequoteopen}ts\ lose{\isachardoublequoteclose}\isanewline
\ \ \ \ \ \ \isakeyword{and}\ h{\isadigit{5}}{\isacharcolon}{\isachardoublequoteopen}msg\ {\isacharparenleft}Suc\ {\isadigit{0}}{\isacharparenright}\ req\ {\isasymand}\ msg\ {\isacharparenleft}Suc\ {\isadigit{0}}{\isacharparenright}\ a\ {\isasymand}\ msg\ {\isacharparenleft}Suc\ {\isadigit{0}}{\isacharparenright}\ stop\ {\isasymand}\ ts\ lose\ {\isasymlongrightarrow}\ \isanewline
\ \ \ \ \ \ \ \ {\isacharparenleft}{\isasymforall}t{\isachardot}\ {\isacharparenleft}ack\ t\ {\isacharequal}\ {\isacharbrackleft}init{\isacharunderscore}state{\isacharbrackright}\ {\isasymand}\isanewline
\ \ \ \ \ \ \ \ \ \ req\ {\isacharparenleft}Suc\ t{\isacharparenright}\ {\isacharequal}\ {\isacharbrackleft}init{\isacharbrackright}\ {\isasymand}\ lose\ {\isacharparenleft}Suc\ t{\isacharparenright}\ {\isacharequal}\ {\isacharbrackleft}False{\isacharbrackright}\ {\isasymand}\ \isanewline
\ \ \ \ \ \ \ \ \ \ lose\ {\isacharparenleft}Suc\ {\isacharparenleft}Suc\ t{\isacharparenright}{\isacharparenright}\ {\isacharequal}\ {\isacharbrackleft}False{\isacharbrackright}\ {\isasymlongrightarrow}\isanewline
\ \ \ \ \ \ \ \ \ \ ack\ {\isacharparenleft}Suc\ {\isacharparenleft}Suc\ t{\isacharparenright}{\isacharparenright}\ {\isacharequal}\ {\isacharbrackleft}connection{\isacharunderscore}ok{\isacharbrackright}{\isacharparenright}\ {\isasymand}\isanewline
\ \ \ \ \ \ \ \ \ {\isacharparenleft}ack\ t\ {\isacharequal}\ {\isacharbrackleft}connection{\isacharunderscore}ok{\isacharbrackright}\ {\isasymand}\ req\ {\isacharparenleft}Suc\ t{\isacharparenright}\ {\isacharequal}\ {\isacharbrackleft}send{\isacharbrackright}\ {\isasymand}\ \isanewline
\ \ \ \ \ \ \ \ \ {\isacharparenleft}{\isasymforall}k{\isasymle}Suc\ d{\isachardot}\ lose\ {\isacharparenleft}t\ {\isacharplus}\ k{\isacharparenright}\ {\isacharequal}\ {\isacharbrackleft}False{\isacharbrackright}{\isacharparenright}\ {\isasymlongrightarrow}\isanewline
\ \ \ \ \ \ \ \ \ \ i\ {\isacharparenleft}Suc\ {\isacharparenleft}t\ {\isacharplus}\ d{\isacharparenright}{\isacharparenright}\ {\isacharequal}\ inf{\isacharunderscore}last{\isacharunderscore}ti\ dt\ t\ {\isasymand}\ ack\ {\isacharparenleft}Suc\ t{\isacharparenright}\ {\isacharequal}\ {\isacharbrackleft}sending{\isacharunderscore}data{\isacharbrackright}{\isacharparenright}\ {\isasymand}\isanewline
\ \ \ \ \ \ \ \ \ {\isacharparenleft}ack\ {\isacharparenleft}t\ {\isacharplus}\ d{\isacharparenright}\ {\isacharequal}\ {\isacharbrackleft}sending{\isacharunderscore}data{\isacharbrackright}\ {\isasymand}\ a\ {\isacharparenleft}Suc\ t{\isacharparenright}\ {\isacharequal}\ {\isacharbrackleft}sc{\isacharunderscore}ack{\isacharbrackright}\ {\isasymand}\ \isanewline
\ \ \ \ \ \ \ \ \ \ {\isacharparenleft}{\isasymforall}k{\isasymle}Suc\ d{\isachardot}\ lose\ {\isacharparenleft}t\ {\isacharplus}\ k{\isacharparenright}\ {\isacharequal}\ {\isacharbrackleft}False{\isacharbrackright}{\isacharparenright}\ {\isasymlongrightarrow}\isanewline
\ \ \ \ \ \ \ \ \ \ vc\ {\isacharparenleft}Suc\ {\isacharparenleft}t\ {\isacharplus}\ d{\isacharparenright}{\isacharparenright}\ {\isacharequal}\ {\isacharbrackleft}vc{\isacharunderscore}com{\isacharbrackright}{\isacharparenright}{\isacharparenright}{\isachardoublequoteclose}\isanewline
\ \ \isakeyword{shows}\ \ {\isachardoublequoteopen}ack\ {\isacharparenleft}t\ {\isacharplus}\ {\isadigit{3}}\ {\isacharplus}\ k\ {\isacharplus}\ d\ {\isacharplus}\ d{\isacharparenright}\ {\isacharequal}\ {\isacharbrackleft}sending{\isacharunderscore}data{\isacharbrackright}\ {\isasymand}\isanewline
\ \ \ \ \ \ \ \ \ \ a\ {\isacharparenleft}Suc\ {\isacharparenleft}t\ {\isacharplus}\ {\isadigit{3}}\ {\isacharplus}\ k\ {\isacharplus}\ d{\isacharparenright}{\isacharparenright}\ {\isacharequal}\ {\isacharbrackleft}sc{\isacharunderscore}ack{\isacharbrackright}\ {\isasymand}\ \isanewline
\ \ \ \ \ \ \ \ \ {\isacharparenleft}{\isasymforall}ka{\isasymle}Suc\ d{\isachardot}\ lose\ {\isacharparenleft}t\ {\isacharplus}\ {\isadigit{3}}\ {\isacharplus}\ k\ {\isacharplus}\ d\ {\isacharplus}\ ka{\isacharparenright}\ {\isacharequal}\ {\isacharbrackleft}False{\isacharbrackright}{\isacharparenright}\ {\isasymlongrightarrow}\isanewline
\ \ \ \ \ \ \ \ \ vc\ {\isacharparenleft}Suc\ {\isacharparenleft}t\ {\isacharplus}\ {\isadigit{3}}\ {\isacharplus}\ k\ {\isacharplus}\ d\ {\isacharplus}\ d{\isacharparenright}{\isacharparenright}\ {\isacharequal}\ {\isacharbrackleft}vc{\isacharunderscore}com{\isacharbrackright}{\isachardoublequoteclose}\isanewline
\isadelimproof
\endisadelimproof
\isatagproof
\isacommand{using}\isamarkupfalse%
\ assms\ \ \isacommand{by}\isamarkupfalse%
\ blast%
\endisatagproof
{\isafoldproof}%
\isadelimproof
\isanewline
\endisadelimproof
\isanewline
\isacommand{lemma}\isamarkupfalse%
\ GatewaySystem{\isacharunderscore}L{\isadigit{3}}aux{\isacharcolon}\isanewline
\ \ \isakeyword{assumes}\ h{\isadigit{1}}{\isacharcolon}{\isachardoublequoteopen}msg\ {\isacharparenleft}Suc\ {\isadigit{0}}{\isacharparenright}\ req{\isachardoublequoteclose}\isanewline
\ \ \ \ \ \ \isakeyword{and}\ h{\isadigit{2}}{\isacharcolon}{\isachardoublequoteopen}msg\ {\isacharparenleft}Suc\ {\isadigit{0}}{\isacharparenright}\ stop{\isachardoublequoteclose}\isanewline
\ \ \ \ \ \ \isakeyword{and}\ h{\isadigit{3}}{\isacharcolon}{\isachardoublequoteopen}msg\ {\isacharparenleft}Suc\ {\isadigit{0}}{\isacharparenright}\ a{\isachardoublequoteclose}\isanewline
\ \ \ \ \ \ \isakeyword{and}\ h{\isadigit{4}}{\isacharcolon}{\isachardoublequoteopen}ts\ lose{\isachardoublequoteclose}\isanewline
\ \ \ \ \ \ \isakeyword{and}\ h{\isadigit{5}}{\isacharcolon}{\isachardoublequoteopen}msg\ {\isacharparenleft}Suc\ {\isadigit{0}}{\isacharparenright}\ req\ {\isasymand}\ msg\ {\isacharparenleft}Suc\ {\isadigit{0}}{\isacharparenright}\ a\ {\isasymand}\ msg\ {\isacharparenleft}Suc\ {\isadigit{0}}{\isacharparenright}\ stop\ {\isasymand}\ ts\ lose\ {\isasymlongrightarrow}\ \isanewline
\ \ \ \ \ \ \ \ {\isacharparenleft}{\isasymforall}t{\isachardot}\ {\isacharparenleft}ack\ t\ {\isacharequal}\ {\isacharbrackleft}init{\isacharunderscore}state{\isacharbrackright}\ {\isasymand}\isanewline
\ \ \ \ \ \ \ \ \ \ req\ {\isacharparenleft}Suc\ t{\isacharparenright}\ {\isacharequal}\ {\isacharbrackleft}init{\isacharbrackright}\ {\isasymand}\ lose\ {\isacharparenleft}Suc\ t{\isacharparenright}\ {\isacharequal}\ {\isacharbrackleft}False{\isacharbrackright}\ {\isasymand}\ \isanewline
\ \ \ \ \ \ \ \ \ \ lose\ {\isacharparenleft}Suc\ {\isacharparenleft}Suc\ t{\isacharparenright}{\isacharparenright}\ {\isacharequal}\ {\isacharbrackleft}False{\isacharbrackright}\ {\isasymlongrightarrow}\isanewline
\ \ \ \ \ \ \ \ \ \ ack\ {\isacharparenleft}Suc\ {\isacharparenleft}Suc\ t{\isacharparenright}{\isacharparenright}\ {\isacharequal}\ {\isacharbrackleft}connection{\isacharunderscore}ok{\isacharbrackright}{\isacharparenright}\ {\isasymand}\isanewline
\ \ \ \ \ \ \ \ \ {\isacharparenleft}ack\ t\ {\isacharequal}\ {\isacharbrackleft}connection{\isacharunderscore}ok{\isacharbrackright}\ {\isasymand}\ req\ {\isacharparenleft}Suc\ t{\isacharparenright}\ {\isacharequal}\ {\isacharbrackleft}send{\isacharbrackright}\ {\isasymand}\ \isanewline
\ \ \ \ \ \ \ \ \ {\isacharparenleft}{\isasymforall}k{\isasymle}Suc\ d{\isachardot}\ lose\ {\isacharparenleft}t\ {\isacharplus}\ k{\isacharparenright}\ {\isacharequal}\ {\isacharbrackleft}False{\isacharbrackright}{\isacharparenright}\ {\isasymlongrightarrow}\isanewline
\ \ \ \ \ \ \ \ \ \ i\ {\isacharparenleft}Suc\ {\isacharparenleft}t\ {\isacharplus}\ d{\isacharparenright}{\isacharparenright}\ {\isacharequal}\ inf{\isacharunderscore}last{\isacharunderscore}ti\ dt\ t\ {\isasymand}\ ack\ {\isacharparenleft}Suc\ t{\isacharparenright}\ {\isacharequal}\ {\isacharbrackleft}sending{\isacharunderscore}data{\isacharbrackright}{\isacharparenright}\ {\isasymand}\isanewline
\ \ \ \ \ \ \ \ \ {\isacharparenleft}ack\ {\isacharparenleft}t\ {\isacharplus}\ d{\isacharparenright}\ {\isacharequal}\ {\isacharbrackleft}sending{\isacharunderscore}data{\isacharbrackright}\ {\isasymand}\ a\ {\isacharparenleft}Suc\ t{\isacharparenright}\ {\isacharequal}\ {\isacharbrackleft}sc{\isacharunderscore}ack{\isacharbrackright}\ {\isasymand}\ \isanewline
\ \ \ \ \ \ \ \ \ \ {\isacharparenleft}{\isasymforall}k{\isasymle}Suc\ d{\isachardot}\ lose\ {\isacharparenleft}t\ {\isacharplus}\ k{\isacharparenright}\ {\isacharequal}\ {\isacharbrackleft}False{\isacharbrackright}{\isacharparenright}\ {\isasymlongrightarrow}\isanewline
\ \ \ \ \ \ \ \ \ \ vc\ {\isacharparenleft}Suc\ {\isacharparenleft}t\ {\isacharplus}\ d{\isacharparenright}{\isacharparenright}\ {\isacharequal}\ {\isacharbrackleft}vc{\isacharunderscore}com{\isacharbrackright}{\isacharparenright}{\isacharparenright}{\isachardoublequoteclose}\isanewline
\ \ \isakeyword{shows}\ {\isachardoublequoteopen}ack\ {\isacharparenleft}t\ {\isacharplus}\ {\isadigit{2}}\ {\isacharplus}\ k{\isacharparenright}\ {\isacharequal}\ {\isacharbrackleft}connection{\isacharunderscore}ok{\isacharbrackright}\ {\isasymand}\isanewline
\ \ \ \ \ \ \ \ \ req\ {\isacharparenleft}Suc\ {\isacharparenleft}t\ {\isacharplus}\ {\isadigit{2}}\ {\isacharplus}\ k{\isacharparenright}{\isacharparenright}\ {\isacharequal}\ {\isacharbrackleft}send{\isacharbrackright}\ {\isasymand}\ \isanewline
\ \ \ \ \ \ \ \ \ {\isacharparenleft}{\isasymforall}j{\isasymle}Suc\ d{\isachardot}\ lose\ {\isacharparenleft}t\ {\isacharplus}\ {\isadigit{2}}\ {\isacharplus}\ k\ {\isacharplus}\ j{\isacharparenright}\ {\isacharequal}\ {\isacharbrackleft}False{\isacharbrackright}{\isacharparenright}\ {\isasymlongrightarrow}\isanewline
\ \ \ \ \ \ \ \ \ i\ {\isacharparenleft}Suc\ {\isacharparenleft}t\ {\isacharplus}\ {\isadigit{2}}\ {\isacharplus}\ k\ {\isacharplus}\ d{\isacharparenright}{\isacharparenright}\ {\isacharequal}\ inf{\isacharunderscore}last{\isacharunderscore}ti\ dt\ {\isacharparenleft}t\ {\isacharplus}\ {\isadigit{2}}\ {\isacharplus}\ k{\isacharparenright}{\isachardoublequoteclose}\isanewline
\isadelimproof
\endisadelimproof
\isatagproof
\isacommand{using}\isamarkupfalse%
\ assms\ \ \isacommand{by}\isamarkupfalse%
\ blast%
\endisatagproof
{\isafoldproof}%
\isadelimproof
\isanewline
\endisadelimproof
\isanewline
\isanewline
\isacommand{lemma}\isamarkupfalse%
\ GatewaySystem{\isacharunderscore}L{\isadigit{1}}{\isacharcolon}\isanewline
\ \isakeyword{assumes}\ \ h{\isadigit{2}}{\isacharcolon}{\isachardoublequoteopen}ServiceCenter\ i\ a{\isachardoublequoteclose}\ \isanewline
\ \ \ \ \ \isakeyword{and}\ h{\isadigit{3}}{\isacharcolon}{\isachardoublequoteopen}GatewayReq\ req\ dt\ a\ stop\ lose\ d\ ack\ i\ vc{\isachardoublequoteclose}\ \isanewline
\ \ \ \ \ \isakeyword{and}\ h{\isadigit{4}}{\isacharcolon}{\isachardoublequoteopen}msg\ {\isacharparenleft}Suc\ {\isadigit{0}}{\isacharparenright}\ req{\isachardoublequoteclose}\isanewline
\ \ \ \ \ \isakeyword{and}\ h{\isadigit{5}}{\isacharcolon}{\isachardoublequoteopen}msg\ {\isacharparenleft}Suc\ {\isadigit{0}}{\isacharparenright}\ stop{\isachardoublequoteclose}\isanewline
\ \ \ \ \ \isakeyword{and}\ h{\isadigit{6}}{\isacharcolon}{\isachardoublequoteopen}msg\ {\isacharparenleft}Suc\ {\isadigit{0}}{\isacharparenright}\ a{\isachardoublequoteclose}\isanewline
\ \ \ \ \ \isakeyword{and}\ h{\isadigit{7}}{\isacharcolon}{\isachardoublequoteopen}ts\ lose{\isachardoublequoteclose}\isanewline
\ \ \ \ \ \isakeyword{and}\ h{\isadigit{9}}{\isacharcolon}{\isachardoublequoteopen}{\isasymforall}j{\isasymle}{\isadigit{2}}\ {\isacharasterisk}\ d\ {\isacharplus}\ {\isacharparenleft}{\isadigit{4}}\ {\isacharplus}\ k{\isacharparenright}{\isachardot}\ lose\ {\isacharparenleft}t\ {\isacharplus}\ j{\isacharparenright}\ {\isacharequal}\ {\isacharbrackleft}False{\isacharbrackright}{\isachardoublequoteclose}\ \isanewline
\ \ \ \ \ \isakeyword{and}\ h{\isadigit{1}}{\isadigit{1}}{\isacharcolon}{\isachardoublequoteopen}i\ {\isacharparenleft}t\ {\isacharplus}\ {\isadigit{3}}\ {\isacharplus}\ k\ {\isacharplus}\ d{\isacharparenright}\ {\isasymnoteq}\ {\isacharbrackleft}{\isacharbrackright}{\isachardoublequoteclose}\isanewline
\ \ \ \ \ \isakeyword{and}\ h{\isadigit{1}}{\isadigit{4}}{\isacharcolon}{\isachardoublequoteopen}{\isasymforall}x\ {\isasymle}\ d\ {\isacharplus}\ d{\isachardot}\ ack\ {\isacharparenleft}t\ {\isacharplus}\ {\isadigit{3}}\ {\isacharplus}\ k\ {\isacharplus}\ x{\isacharparenright}\ {\isacharequal}\ {\isacharbrackleft}sending{\isacharunderscore}data{\isacharbrackright}{\isachardoublequoteclose}\isanewline
\ \isakeyword{shows}\ {\isachardoublequoteopen}vc\ {\isacharparenleft}{\isadigit{2}}\ {\isacharasterisk}\ d\ {\isacharplus}\ {\isacharparenleft}t\ {\isacharplus}\ {\isacharparenleft}{\isadigit{4}}\ {\isacharplus}\ k{\isacharparenright}{\isacharparenright}{\isacharparenright}\ {\isacharequal}\ {\isacharbrackleft}vc{\isacharunderscore}com{\isacharbrackright}{\isachardoublequoteclose}\isanewline
\isadelimproof
\endisadelimproof
\isatagproof
\isacommand{proof}\isamarkupfalse%
\ {\isacharminus}\ \isanewline
\ \ \isacommand{from}\isamarkupfalse%
\ h{\isadigit{2}}\ \isacommand{have}\isamarkupfalse%
\ {\isachardoublequoteopen}{\isasymforall}t{\isachardot}\ a\ {\isacharparenleft}Suc\ t{\isacharparenright}\ {\isacharequal}\ {\isacharparenleft}if\ i\ t\ {\isacharequal}\ {\isacharbrackleft}{\isacharbrackright}\ then\ {\isacharbrackleft}{\isacharbrackright}\ else\ {\isacharbrackleft}sc{\isacharunderscore}ack{\isacharbrackright}{\isacharparenright}{\isachardoublequoteclose}\isanewline
\ \ \ \ \isacommand{by}\isamarkupfalse%
\ {\isacharparenleft}simp\ add{\isacharcolon}ServiceCenter{\isacharunderscore}def{\isacharparenright}\ \isanewline
\ \ \isacommand{from}\isamarkupfalse%
\ this\ \isacommand{have}\isamarkupfalse%
\ sg{\isadigit{1}}{\isacharcolon}\isanewline
\ \ \ \ {\isachardoublequoteopen}a\ {\isacharparenleft}Suc\ {\isacharparenleft}t\ {\isacharplus}\ {\isadigit{3}}\ {\isacharplus}\ k\ {\isacharplus}\ d{\isacharparenright}{\isacharparenright}\ {\isacharequal}\ {\isacharparenleft}if\ i\ {\isacharparenleft}t\ {\isacharplus}\ {\isadigit{3}}\ {\isacharplus}\ k\ {\isacharplus}\ d{\isacharparenright}\ {\isacharequal}\ {\isacharbrackleft}{\isacharbrackright}\ then\ {\isacharbrackleft}{\isacharbrackright}\ else\ {\isacharbrackleft}sc{\isacharunderscore}ack{\isacharbrackright}{\isacharparenright}{\isachardoublequoteclose}\ \isanewline
\ \ \ \ \ \isacommand{by}\isamarkupfalse%
\ blast\isanewline
\ \ \isacommand{from}\isamarkupfalse%
\ sg{\isadigit{1}}\ \isakeyword{and}\ h{\isadigit{1}}{\isadigit{1}}\ \isacommand{have}\isamarkupfalse%
\ sg{\isadigit{2}}{\isacharcolon}{\isachardoublequoteopen}a\ {\isacharparenleft}Suc\ {\isacharparenleft}t\ {\isacharplus}\ {\isadigit{3}}\ {\isacharplus}\ k\ {\isacharplus}\ d{\isacharparenright}{\isacharparenright}\ {\isacharequal}\ {\isacharbrackleft}sc{\isacharunderscore}ack{\isacharbrackright}{\isachardoublequoteclose}\ \isacommand{by}\isamarkupfalse%
\ auto\isanewline
\ \ \isacommand{from}\isamarkupfalse%
\ h{\isadigit{1}}{\isadigit{4}}\ \isacommand{have}\isamarkupfalse%
\ sg{\isadigit{3}}{\isacharcolon}{\isachardoublequoteopen}ack\ {\isacharparenleft}t\ {\isacharplus}\ {\isadigit{3}}\ {\isacharplus}\ k\ {\isacharplus}\ {\isadigit{2}}{\isacharasterisk}d{\isacharparenright}\ {\isacharequal}\ {\isacharbrackleft}sending{\isacharunderscore}data{\isacharbrackright}{\isachardoublequoteclose}\ \isacommand{by}\isamarkupfalse%
\ simp\isanewline
\ \ \isacommand{from}\isamarkupfalse%
\ h{\isadigit{4}}\ \isakeyword{and}\ h{\isadigit{5}}\ \isakeyword{and}\ h{\isadigit{6}}\ \isakeyword{and}\ h{\isadigit{7}}\ \isakeyword{and}\ h{\isadigit{3}}\ \isacommand{have}\isamarkupfalse%
\ sg{\isadigit{4}}{\isacharcolon}\isanewline
\ \ \ \ {\isachardoublequoteopen}ack\ {\isacharparenleft}t\ {\isacharplus}\ {\isadigit{3}}\ {\isacharplus}\ k\ {\isacharplus}\ d\ {\isacharplus}\ d{\isacharparenright}\ {\isacharequal}\ {\isacharbrackleft}sending{\isacharunderscore}data{\isacharbrackright}\ {\isasymand}\ a\ {\isacharparenleft}Suc\ {\isacharparenleft}t\ {\isacharplus}\ {\isadigit{3}}\ {\isacharplus}\ k\ {\isacharplus}\ d{\isacharparenright}{\isacharparenright}\ {\isacharequal}\ {\isacharbrackleft}sc{\isacharunderscore}ack{\isacharbrackright}\ {\isasymand}\ \isanewline
\ \ \ \ \ {\isacharparenleft}{\isasymforall}ka{\isasymle}Suc\ d{\isachardot}\ lose\ {\isacharparenleft}t\ {\isacharplus}\ {\isadigit{3}}\ {\isacharplus}\ k\ {\isacharplus}\ d\ {\isacharplus}\ ka{\isacharparenright}\ {\isacharequal}\ {\isacharbrackleft}False{\isacharbrackright}{\isacharparenright}\ {\isasymlongrightarrow}\isanewline
\ \ \ \ \ vc\ {\isacharparenleft}Suc\ {\isacharparenleft}t\ {\isacharplus}\ {\isadigit{3}}\ {\isacharplus}\ k\ {\isacharplus}\ d\ {\isacharplus}\ d{\isacharparenright}{\isacharparenright}\ {\isacharequal}\ {\isacharbrackleft}vc{\isacharunderscore}com{\isacharbrackright}{\isachardoublequoteclose}\ \isanewline
\ \ \ \ \isacommand{apply}\isamarkupfalse%
\ {\isacharparenleft}simp\ only{\isacharcolon}\ GatewayReq{\isacharunderscore}def{\isacharparenright}\isanewline
\ \ \ \ \isacommand{by}\isamarkupfalse%
\ {\isacharparenleft}rule\ GatewaySystem{\isacharunderscore}L{\isadigit{1}}aux{\isacharcomma}\ auto{\isacharparenright}\ \isanewline
\ \ \isacommand{from}\isamarkupfalse%
\ h{\isadigit{9}}\ \isacommand{have}\isamarkupfalse%
\ sg{\isadigit{5}}{\isacharcolon}{\isachardoublequoteopen}{\isasymforall}ka{\isasymle}Suc\ d{\isachardot}\ lose\ {\isacharparenleft}d\ {\isacharplus}\ {\isacharparenleft}t\ {\isacharplus}\ {\isacharparenleft}{\isadigit{3}}\ {\isacharplus}\ k{\isacharparenright}{\isacharparenright}\ {\isacharplus}\ ka{\isacharparenright}\ {\isacharequal}\ {\isacharbrackleft}False{\isacharbrackright}{\isachardoublequoteclose}\isanewline
\ \ \ \ \isacommand{by}\isamarkupfalse%
\ {\isacharparenleft}simp\ add{\isacharcolon}\ aux{\isacharunderscore}lemma{\isacharunderscore}lose{\isacharunderscore}{\isadigit{3}}{\isacharparenright}\isanewline
\ \ \isacommand{have}\isamarkupfalse%
\ sg{\isadigit{5}}a{\isacharcolon}{\isachardoublequoteopen}d\ {\isacharplus}\ {\isacharparenleft}t\ {\isacharplus}\ {\isacharparenleft}{\isadigit{3}}\ {\isacharplus}\ k{\isacharparenright}{\isacharparenright}\ {\isacharequal}\ t\ {\isacharplus}\ {\isadigit{3}}\ {\isacharplus}\ k\ {\isacharplus}\ d{\isachardoublequoteclose}\ \isacommand{by}\isamarkupfalse%
\ arith\isanewline
\ \ \isacommand{from}\isamarkupfalse%
\ sg{\isadigit{5}}\ \isakeyword{and}\ sg{\isadigit{5}}a\ \isacommand{have}\isamarkupfalse%
\ sg{\isadigit{5}}b{\isacharcolon}{\isachardoublequoteopen}{\isasymforall}ka{\isasymle}Suc\ d{\isachardot}\ lose\ {\isacharparenleft}t\ {\isacharplus}\ {\isadigit{3}}\ {\isacharplus}\ k\ {\isacharplus}\ d\ {\isacharplus}\ ka{\isacharparenright}\ {\isacharequal}\ {\isacharbrackleft}False{\isacharbrackright}{\isachardoublequoteclose}\  
 \ \isacommand{by}\isamarkupfalse%
\ auto\ \isanewline
\ \ \isacommand{have}\isamarkupfalse%
\ sg{\isadigit{6}}{\isacharcolon}{\isachardoublequoteopen}{\isacharparenleft}t\ {\isacharplus}\ {\isadigit{3}}\ {\isacharplus}\ k\ {\isacharplus}\ {\isadigit{2}}\ {\isacharasterisk}\ d{\isacharparenright}\ \ {\isacharequal}\ {\isacharparenleft}{\isadigit{2}}\ {\isacharasterisk}\ d\ {\isacharplus}\ {\isacharparenleft}t\ {\isacharplus}\ {\isacharparenleft}{\isadigit{3}}\ {\isacharplus}\ k{\isacharparenright}{\isacharparenright}{\isacharparenright}{\isachardoublequoteclose}\ \isacommand{by}\isamarkupfalse%
\ arith\isanewline
\ \ \isacommand{have}\isamarkupfalse%
\ sg{\isadigit{7}}{\isacharcolon}{\isachardoublequoteopen}Suc\ {\isacharparenleft}Suc\ {\isacharparenleft}Suc\ {\isacharparenleft}t\ {\isacharplus}\ k\ {\isacharplus}\ {\isacharparenleft}d\ {\isacharplus}\ d{\isacharparenright}{\isacharparenright}{\isacharparenright}{\isacharparenright}\ {\isacharequal}\ Suc\ {\isacharparenleft}Suc\ {\isacharparenleft}Suc\ {\isacharparenleft}t\ {\isacharplus}\ k\ {\isacharplus}\ d\ {\isacharplus}\ d{\isacharparenright}{\isacharparenright}{\isacharparenright}{\isachardoublequoteclose}\    \ \isacommand{by}\isamarkupfalse%
\ arith\isanewline
\ \ \isacommand{have}\isamarkupfalse%
\ sg{\isadigit{8}}{\isacharcolon}{\isachardoublequoteopen}Suc\ {\isacharparenleft}Suc\ {\isacharparenleft}Suc\ {\isacharparenleft}Suc\ {\isacharparenleft}t\ {\isacharplus}\ k\ {\isacharplus}\ d\ {\isacharplus}\ d{\isacharparenright}{\isacharparenright}{\isacharparenright}{\isacharparenright}\ {\isacharequal}\ \isanewline
\ \ \ \ \ \ \ \ \ \ \ \ Suc\ {\isacharparenleft}Suc\ {\isacharparenleft}Suc\ {\isacharparenleft}Suc\ {\isacharparenleft}d\ {\isacharplus}\ d\ {\isacharplus}\ {\isacharparenleft}t\ {\isacharplus}\ k{\isacharparenright}{\isacharparenright}{\isacharparenright}{\isacharparenright}{\isacharparenright}{\isachardoublequoteclose}\ \isacommand{by}\isamarkupfalse%
\ arith\isanewline
\ \ \isacommand{from}\isamarkupfalse%
\ sg{\isadigit{4}}\ \isakeyword{and}\ sg{\isadigit{3}}\ \isakeyword{and}\ sg{\isadigit{2}}\ \isakeyword{and}\ sg{\isadigit{5}}b\ \isakeyword{and}\ sg{\isadigit{6}}\ \isakeyword{and}\ sg{\isadigit{7}}\ \isakeyword{and}\ sg{\isadigit{8}}\ \isacommand{show}\isamarkupfalse%
\ {\isacharquery}thesis\isanewline
\ \ \ \ \isacommand{by}\isamarkupfalse%
\ {\isacharparenleft}simp\ add{\isacharcolon}\ eval{\isacharunderscore}nat{\isacharunderscore}numeral{\isacharparenright}\isanewline
\isacommand{qed}\isamarkupfalse%
\endisatagproof
{\isafoldproof}%
\isadelimproof
\isanewline
\endisadelimproof
\isanewline
\isanewline
\isacommand{lemma}\isamarkupfalse%
\ aux{\isadigit{4}}lose{\isadigit{1}}{\isacharcolon}\isanewline
\ \ \isakeyword{assumes}\ h{\isadigit{1}}{\isacharcolon}{\isachardoublequoteopen}{\isasymforall}j{\isasymle}{\isadigit{2}}\ {\isacharasterisk}\ d\ {\isacharplus}\ {\isacharparenleft}{\isadigit{4}}\ {\isacharplus}\ k{\isacharparenright}{\isachardot}\ lose\ {\isacharparenleft}t\ {\isacharplus}\ j{\isacharparenright}\ {\isacharequal}\ {\isacharbrackleft}False{\isacharbrackright}{\isachardoublequoteclose}\isanewline
\ \ \ \ \ \ \isakeyword{and}\ h{\isadigit{2}}{\isacharcolon}{\isachardoublequoteopen}j\ {\isasymle}\ k{\isachardoublequoteclose}\isanewline
\ \ \isakeyword{shows}\ {\isachardoublequoteopen}lose\ {\isacharparenleft}t\ {\isacharplus}\ {\isacharparenleft}{\isadigit{2}}{\isacharcolon}{\isacharcolon}nat{\isacharparenright}\ {\isacharplus}\ j{\isacharparenright}\ {\isacharequal}\ {\isacharbrackleft}False{\isacharbrackright}{\isachardoublequoteclose}\isanewline
\isadelimproof
\endisadelimproof
\isatagproof
\isacommand{proof}\isamarkupfalse%
\ {\isacharminus}\ \isanewline
\ \ \isacommand{from}\isamarkupfalse%
\ h{\isadigit{2}}\ \isacommand{have}\isamarkupfalse%
\ {\isachardoublequoteopen}{\isacharparenleft}{\isadigit{2}}{\isacharcolon}{\isacharcolon}nat{\isacharparenright}\ {\isacharplus}\ j\ {\isasymle}\ {\isacharparenleft}{\isadigit{2}}{\isacharcolon}{\isacharcolon}nat{\isacharparenright}\ {\isacharasterisk}\ d\ {\isacharplus}\ {\isacharparenleft}{\isadigit{4}}\ {\isacharplus}\ k{\isacharparenright}{\isachardoublequoteclose}\ \isacommand{by}\isamarkupfalse%
\ arith\isanewline
\ \ \isacommand{from}\isamarkupfalse%
\ h{\isadigit{1}}\ \isakeyword{and}\ this\ \isacommand{have}\isamarkupfalse%
\ {\isachardoublequoteopen}lose\ {\isacharparenleft}t\ {\isacharplus}\ {\isacharparenleft}{\isadigit{2}}\ {\isacharplus}\ j{\isacharparenright}{\isacharparenright}\ {\isacharequal}\ {\isacharbrackleft}False{\isacharbrackright}{\isachardoublequoteclose}\ \isacommand{by}\isamarkupfalse%
\ blast\isanewline
\ \ \isacommand{from}\isamarkupfalse%
\ this\ \isacommand{show}\isamarkupfalse%
\ {\isacharquery}thesis\ \isacommand{by}\isamarkupfalse%
\ simp\isanewline
\isacommand{qed}\isamarkupfalse%
\endisatagproof
{\isafoldproof}%
\isadelimproof
\isanewline
\endisadelimproof
\isanewline
\isacommand{lemma}\isamarkupfalse%
\ aux{\isadigit{4}}lose{\isadigit{2}}{\isacharcolon}\isanewline
\ \ \isakeyword{assumes}\ h{\isadigit{1}}{\isacharcolon}{\isachardoublequoteopen}{\isasymforall}j{\isasymle}{\isadigit{2}}\ {\isacharasterisk}\ d\ {\isacharplus}\ {\isacharparenleft}{\isadigit{4}}\ {\isacharplus}\ k{\isacharparenright}{\isachardot}\ lose\ {\isacharparenleft}t\ {\isacharplus}\ j{\isacharparenright}\ {\isacharequal}\ {\isacharbrackleft}False{\isacharbrackright}{\isachardoublequoteclose}\isanewline
\ \ \ \ \ \ \isakeyword{and}\ h{\isadigit{2}}{\isacharcolon}{\isachardoublequoteopen}{\isadigit{3}}\ {\isacharplus}\ k\ {\isacharplus}\ d\ {\isasymle}\ {\isadigit{2}}\ {\isacharasterisk}\ d\ {\isacharplus}\ {\isacharparenleft}{\isadigit{4}}\ {\isacharplus}\ k{\isacharparenright}{\isachardoublequoteclose}\isanewline
\ \ \isakeyword{shows}\ {\isachardoublequoteopen}lose\ {\isacharparenleft}t\ {\isacharplus}\ {\isacharparenleft}{\isadigit{3}}{\isacharcolon}{\isacharcolon}nat{\isacharparenright}\ {\isacharplus}\ k\ {\isacharplus}\ d{\isacharparenright}\ {\isacharequal}\ {\isacharbrackleft}False{\isacharbrackright}{\isachardoublequoteclose}\isanewline
\isadelimproof
\endisadelimproof
\isatagproof
\isacommand{proof}\isamarkupfalse%
\ {\isacharminus}\isanewline
\ \ \isacommand{from}\isamarkupfalse%
\ assms\ \isacommand{have}\isamarkupfalse%
\ {\isachardoublequoteopen}lose\ {\isacharparenleft}t\ {\isacharplus}\ {\isacharparenleft}{\isacharparenleft}{\isadigit{3}}{\isacharcolon}{\isacharcolon}nat{\isacharparenright}\ {\isacharplus}\ k\ {\isacharplus}\ d{\isacharparenright}{\isacharparenright}\ {\isacharequal}\ {\isacharbrackleft}False{\isacharbrackright}{\isachardoublequoteclose}\ \isacommand{by}\isamarkupfalse%
\ blast\isanewline
\ \ \isacommand{from}\isamarkupfalse%
\ this\ \isacommand{show}\isamarkupfalse%
\ {\isacharquery}thesis\ \isacommand{by}\isamarkupfalse%
\ {\isacharparenleft}simp\ add{\isacharcolon}\ arith{\isacharunderscore}sum{\isadigit{1}}{\isacharparenright}\isanewline
\isacommand{qed}\isamarkupfalse%
\endisatagproof
{\isafoldproof}%
\isadelimproof
\isanewline
\endisadelimproof
\isanewline
\isacommand{lemma}\isamarkupfalse%
\ aux{\isadigit{4}}req{\isacharcolon}\isanewline
\ \ \isakeyword{assumes}\ h{\isadigit{1}}{\isacharcolon}{\isachardoublequoteopen}{\isasymforall}\ {\isacharparenleft}m{\isacharcolon}{\isacharcolon}nat{\isacharparenright}\ {\isasymle}\ k\ {\isacharplus}\ {\isadigit{2}}{\isachardot}\ req\ {\isacharparenleft}t\ {\isacharplus}\ m{\isacharparenright}\ {\isasymnoteq}\ {\isacharbrackleft}send{\isacharbrackright}{\isachardoublequoteclose}\isanewline
\ \ \ \ \ \ \isakeyword{and}\ h{\isadigit{2}}{\isacharcolon}{\isachardoublequoteopen}m\ {\isasymle}\ k{\isachardoublequoteclose}\isanewline
\ \ \ \ \ \ \isakeyword{and}\ h{\isadigit{3}}{\isacharcolon}{\isachardoublequoteopen}req\ {\isacharparenleft}t\ {\isacharplus}\ {\isadigit{2}}\ {\isacharplus}\ m{\isacharparenright}\ {\isacharequal}\ {\isacharbrackleft}send{\isacharbrackright}{\isachardoublequoteclose}\isanewline
\ \ \isakeyword{shows}\ {\isachardoublequoteopen}False{\isachardoublequoteclose}\isanewline
\isadelimproof
\endisadelimproof
\isatagproof
\isacommand{proof}\isamarkupfalse%
\ {\isacharminus}\ \isanewline
\ \ \isacommand{from}\isamarkupfalse%
\ h{\isadigit{2}}\ \isacommand{have}\isamarkupfalse%
\ {\isachardoublequoteopen}{\isacharparenleft}{\isadigit{2}}{\isacharcolon}{\isacharcolon}nat{\isacharparenright}\ {\isacharplus}\ m\ {\isasymle}\ k\ {\isacharplus}\ {\isacharparenleft}{\isadigit{2}}{\isacharcolon}{\isacharcolon}nat{\isacharparenright}{\isachardoublequoteclose}\ \isacommand{by}\isamarkupfalse%
\ arith\isanewline
\ \ \isacommand{from}\isamarkupfalse%
\ h{\isadigit{1}}\ \isakeyword{and}\ this\ \isacommand{have}\isamarkupfalse%
\ {\isachardoublequoteopen}req\ {\isacharparenleft}t\ {\isacharplus}\ {\isacharparenleft}{\isadigit{2}}\ {\isacharplus}\ m{\isacharparenright}{\isacharparenright}\ {\isasymnoteq}\ {\isacharbrackleft}send{\isacharbrackright}{\isachardoublequoteclose}\ \isacommand{by}\isamarkupfalse%
\ blast\isanewline
\ \ \isacommand{from}\isamarkupfalse%
\ this\ \isakeyword{and}\ h{\isadigit{3}}\ \isacommand{show}\isamarkupfalse%
\ {\isacharquery}thesis\ \isacommand{by}\isamarkupfalse%
\ simp\isanewline
\isacommand{qed}\isamarkupfalse%
\endisatagproof
{\isafoldproof}%
\isadelimproof
\isanewline
\endisadelimproof
\ \isanewline
\isanewline
\isacommand{lemma}\isamarkupfalse%
\ GatewaySystem{\isacharunderscore}L{\isadigit{2}}{\isacharcolon}\isanewline
\ \isakeyword{assumes}\ h{\isadigit{1}}{\isacharcolon}{\isachardoublequoteopen}Gateway\ req\ dt\ a\ stop\ lose\ d\ ack\ i\ vc{\isachardoublequoteclose}\ \isanewline
\ \ \ \ \ \isakeyword{and}\ h{\isadigit{2}}{\isacharcolon}{\isachardoublequoteopen}ServiceCenter\ i\ a{\isachardoublequoteclose}\ \isanewline
\ \ \ \ \ \isakeyword{and}\ h{\isadigit{3}}{\isacharcolon}{\isachardoublequoteopen}GatewayReq\ req\ dt\ a\ stop\ lose\ d\ ack\ i\ vc{\isachardoublequoteclose}\ \isanewline
\ \ \ \ \ \isakeyword{and}\ h{\isadigit{4}}{\isacharcolon}{\isachardoublequoteopen}msg\ {\isacharparenleft}Suc\ {\isadigit{0}}{\isacharparenright}\ req{\isachardoublequoteclose}\isanewline
\ \ \ \ \ \isakeyword{and}\ h{\isadigit{5}}{\isacharcolon}{\isachardoublequoteopen}msg\ {\isacharparenleft}Suc\ {\isadigit{0}}{\isacharparenright}\ stop{\isachardoublequoteclose}\isanewline
\ \ \ \ \ \isakeyword{and}\ h{\isadigit{6}}{\isacharcolon}{\isachardoublequoteopen}msg\ {\isacharparenleft}Suc\ {\isadigit{0}}{\isacharparenright}\ a{\isachardoublequoteclose}\isanewline
\ \ \ \ \ \isakeyword{and}\ h{\isadigit{7}}{\isacharcolon}{\isachardoublequoteopen}ts\ lose{\isachardoublequoteclose}\isanewline
\ \ \ \ \ \isakeyword{and}\ h{\isadigit{8}}{\isacharcolon}{\isachardoublequoteopen}ack\ t\ {\isacharequal}\ {\isacharbrackleft}init{\isacharunderscore}state{\isacharbrackright}{\isachardoublequoteclose}\isanewline
\ \ \ \ \ \isakeyword{and}\ h{\isadigit{9}}{\isacharcolon}{\isachardoublequoteopen}req\ {\isacharparenleft}Suc\ t{\isacharparenright}\ {\isacharequal}\ {\isacharbrackleft}init{\isacharbrackright}{\isachardoublequoteclose}\ \isanewline
\ \ \ \ \ \isakeyword{and}\ h{\isadigit{1}}{\isadigit{0}}{\isacharcolon}{\isachardoublequoteopen}{\isasymforall}t{\isadigit{1}}{\isasymle}t{\isachardot}\ req\ t{\isadigit{1}}\ {\isacharequal}\ {\isacharbrackleft}{\isacharbrackright}{\isachardoublequoteclose}\isanewline
\ \ \ \ \ \isakeyword{and}\ h{\isadigit{1}}{\isadigit{1}}{\isacharcolon}{\isachardoublequoteopen}{\isasymforall}m\ {\isasymle}\ k\ {\isacharplus}\ {\isadigit{2}}{\isachardot}\ req\ {\isacharparenleft}t\ {\isacharplus}\ m{\isacharparenright}\ {\isasymnoteq}\ {\isacharbrackleft}send{\isacharbrackright}{\isachardoublequoteclose}\isanewline
\ \ \ \ \ \isakeyword{and}\ h{\isadigit{1}}{\isadigit{2}}{\isacharcolon}{\isachardoublequoteopen}req\ {\isacharparenleft}t\ {\isacharplus}\ {\isadigit{3}}\ {\isacharplus}\ k{\isacharparenright}\ {\isacharequal}\ {\isacharbrackleft}send{\isacharbrackright}{\isachardoublequoteclose}\ \isanewline
\ \ \ \ \ \isakeyword{and}\ h{\isadigit{1}}{\isadigit{3}}{\isacharcolon}{\isachardoublequoteopen}inf{\isacharunderscore}last{\isacharunderscore}ti\ dt\ t\ {\isasymnoteq}\ {\isacharbrackleft}{\isacharbrackright}{\isachardoublequoteclose}\isanewline
\ \ \ \ \ \isakeyword{and}\ h{\isadigit{1}}{\isadigit{4}}{\isacharcolon}{\isachardoublequoteopen}{\isasymforall}j{\isasymle}{\isadigit{2}}\ {\isacharasterisk}\ d\ {\isacharplus}\ {\isacharparenleft}{\isadigit{4}}\ {\isacharplus}\ k{\isacharparenright}{\isachardot}\ lose\ {\isacharparenleft}t\ {\isacharplus}\ j{\isacharparenright}\ {\isacharequal}\ {\isacharbrackleft}False{\isacharbrackright}{\isachardoublequoteclose}\isanewline
\ \isakeyword{shows}\ {\isachardoublequoteopen}vc\ {\isacharparenleft}{\isadigit{2}}\ {\isacharasterisk}\ d\ {\isacharplus}\ {\isacharparenleft}t\ {\isacharplus}\ {\isacharparenleft}{\isadigit{4}}\ {\isacharplus}\ k{\isacharparenright}{\isacharparenright}{\isacharparenright}\ {\isacharequal}\ {\isacharbrackleft}vc{\isacharunderscore}com{\isacharbrackright}{\isachardoublequoteclose}\isanewline
\isadelimproof
\endisadelimproof
\isatagproof
\isacommand{proof}\isamarkupfalse%
\ {\isacharminus}\isanewline
\ \ \isacommand{have}\isamarkupfalse%
\ {\isachardoublequoteopen}Suc\ {\isadigit{0}}\ {\isasymle}\ {\isadigit{2}}\ {\isacharasterisk}\ d\ {\isacharplus}\ {\isacharparenleft}{\isadigit{4}}\ {\isacharplus}\ k{\isacharparenright}{\isachardoublequoteclose}\ \isacommand{by}\isamarkupfalse%
\ arith\isanewline
\ \ \isacommand{from}\isamarkupfalse%
\ h{\isadigit{1}}{\isadigit{4}}\ \isakeyword{and}\ this\ \isacommand{have}\isamarkupfalse%
\ {\isachardoublequoteopen}lose\ {\isacharparenleft}t\ {\isacharplus}\ Suc\ {\isadigit{0}}{\isacharparenright}\ {\isacharequal}\ {\isacharbrackleft}False{\isacharbrackright}{\isachardoublequoteclose}\ \isacommand{by}\isamarkupfalse%
\ blast\isanewline
\ \ \isacommand{from}\isamarkupfalse%
\ this\ \isacommand{have}\isamarkupfalse%
\ sg{\isadigit{1}}{\isacharcolon}{\isachardoublequoteopen}lose\ {\isacharparenleft}Suc\ t{\isacharparenright}\ {\isacharequal}\ {\isacharbrackleft}False{\isacharbrackright}{\isachardoublequoteclose}\ \isacommand{by}\isamarkupfalse%
\ simp\isanewline
\ \ \isacommand{have}\isamarkupfalse%
\ {\isachardoublequoteopen}Suc\ {\isacharparenleft}Suc\ {\isadigit{0}}{\isacharparenright}\ {\isasymle}\ {\isadigit{2}}\ {\isacharasterisk}\ d\ {\isacharplus}\ {\isacharparenleft}{\isadigit{4}}\ {\isacharplus}\ k{\isacharparenright}{\isachardoublequoteclose}\ \isacommand{by}\isamarkupfalse%
\ arith\isanewline
\ \ \isacommand{from}\isamarkupfalse%
\ h{\isadigit{1}}{\isadigit{4}}\ \isakeyword{and}\ this\ \isacommand{have}\isamarkupfalse%
\ {\isachardoublequoteopen}lose\ {\isacharparenleft}t\ {\isacharplus}\ Suc\ {\isacharparenleft}Suc\ {\isadigit{0}}{\isacharparenright}{\isacharparenright}\ {\isacharequal}\ {\isacharbrackleft}False{\isacharbrackright}{\isachardoublequoteclose}\ \isacommand{by}\isamarkupfalse%
\ blast\isanewline
\ \ \isacommand{from}\isamarkupfalse%
\ this\ \isacommand{have}\isamarkupfalse%
\ sg{\isadigit{2}}{\isacharcolon}{\isachardoublequoteopen}lose\ {\isacharparenleft}Suc\ {\isacharparenleft}Suc\ t{\isacharparenright}{\isacharparenright}\ {\isacharequal}\ {\isacharbrackleft}False{\isacharbrackright}{\isachardoublequoteclose}\ \isacommand{by}\isamarkupfalse%
\ simp\ \isanewline
\ \ \isacommand{from}\isamarkupfalse%
\ h{\isadigit{3}}\ \isakeyword{and}\ h{\isadigit{4}}\ \isakeyword{and}\ h{\isadigit{5}}\ \isakeyword{and}\ h{\isadigit{6}}\ \isakeyword{and}\ h{\isadigit{7}}\ \isakeyword{and}\ h{\isadigit{8}}\ \isakeyword{and}\ h{\isadigit{9}}\ \isakeyword{and}\ sg{\isadigit{1}}\ \isakeyword{and}\ sg{\isadigit{2}}\ \isacommand{have}\isamarkupfalse%
\ sg{\isadigit{3}}{\isacharcolon}\isanewline
\ \ \ {\isachardoublequoteopen}ack\ {\isacharparenleft}t\ {\isacharplus}\ {\isadigit{2}}{\isacharparenright}\ {\isacharequal}\ {\isacharbrackleft}connection{\isacharunderscore}ok{\isacharbrackright}{\isachardoublequoteclose}\ \isanewline
\ \ \ \ \isacommand{by}\isamarkupfalse%
\ {\isacharparenleft}simp\ add{\isacharcolon}\ GatewayReq{\isacharunderscore}def{\isacharparenright}\isanewline
\ \ \isacommand{from}\isamarkupfalse%
\ h{\isadigit{1}}{\isadigit{4}}\ \isacommand{have}\isamarkupfalse%
\ sg{\isadigit{4}}{\isacharcolon}{\isachardoublequoteopen}\ {\isasymforall}j{\isasymle}k{\isachardot}\ lose\ {\isacharparenleft}t\ {\isacharplus}\ {\isadigit{2}}\ {\isacharplus}\ j{\isacharparenright}\ {\isacharequal}\ {\isacharbrackleft}False{\isacharbrackright}{\isachardoublequoteclose}\ \isanewline
\ \ \ \ \isacommand{by}\isamarkupfalse%
\ {\isacharparenleft}clarify{\isacharcomma}\ rule\ aux{\isadigit{4}}lose{\isadigit{1}}{\isacharcomma}\ simp{\isacharparenright}\ \isanewline
\ \ \isacommand{from}\isamarkupfalse%
\ h{\isadigit{1}}{\isadigit{1}}\ \isacommand{have}\isamarkupfalse%
\ sg{\isadigit{5}}{\isacharcolon}{\isachardoublequoteopen}{\isasymforall}m\ {\isasymle}\ k{\isachardot}\ req\ {\isacharparenleft}t\ {\isacharplus}\ {\isadigit{2}}\ {\isacharplus}\ m{\isacharparenright}\ {\isasymnoteq}\ {\isacharbrackleft}send{\isacharbrackright}{\isachardoublequoteclose}\ \isanewline
\ \ \ \ \isacommand{by}\isamarkupfalse%
\ {\isacharparenleft}clarify{\isacharcomma}\ rule\ aux{\isadigit{4}}req{\isacharcomma}\ auto{\isacharparenright}\ \isanewline
\ \ \isanewline
\ \ \isacommand{from}\isamarkupfalse%
\ h{\isadigit{1}}\ \isakeyword{and}\ sg{\isadigit{5}}\ \isakeyword{and}\ sg{\isadigit{4}}\ \isakeyword{and}\ sg{\isadigit{3}}\ \isakeyword{and}\ h{\isadigit{4}}\ \isakeyword{and}\ h{\isadigit{5}}\ \isakeyword{and}\ h{\isadigit{6}}\ \isakeyword{and}\ h{\isadigit{7}}\ \isacommand{have}\isamarkupfalse%
\ sg{\isadigit{6}}{\isacharcolon}\isanewline
\ \ \ {\isachardoublequoteopen}{\isasymforall}m\ {\isasymle}\ k{\isachardot}\ ack\ {\isacharparenleft}t\ {\isacharplus}\ {\isadigit{2}}\ {\isacharplus}\ m{\isacharparenright}\ {\isacharequal}\ {\isacharbrackleft}connection{\isacharunderscore}ok{\isacharbrackright}{\isachardoublequoteclose}\isanewline
\ \ \ \ \isacommand{by}\isamarkupfalse%
\ {\isacharparenleft}rule\ Gateway{\isacharunderscore}L{\isadigit{6}}{\isacharparenright}\isanewline
\ \ \isanewline
\ \ \isacommand{from}\isamarkupfalse%
\ h{\isadigit{3}}\ \isakeyword{and}\ h{\isadigit{4}}\ \isakeyword{and}\ h{\isadigit{5}}\ \isakeyword{and}\ h{\isadigit{6}}\ \isakeyword{and}\ h{\isadigit{7}}\ \isakeyword{and}\ h{\isadigit{1}}{\isadigit{2}}\ \isakeyword{and}\ h{\isadigit{1}}{\isadigit{4}}\ \isakeyword{and}\ sg{\isadigit{6}}\ \isacommand{have}\isamarkupfalse%
\ sg{\isadigit{1}}{\isadigit{0}}{\isacharcolon}\isanewline
\ \ \ {\isachardoublequoteopen}ack\ {\isacharparenleft}t\ {\isacharplus}\ {\isadigit{3}}\ {\isacharplus}\ k{\isacharparenright}\ {\isacharequal}\ {\isacharbrackleft}sending{\isacharunderscore}data{\isacharbrackright}{\isachardoublequoteclose}\isanewline
\ \ \ \ \isacommand{by}\isamarkupfalse%
\ {\isacharparenleft}simp\ add{\isacharcolon}\ GatewayReq{\isacharunderscore}L{\isadigit{1}}{\isacharparenright}\isanewline
\ \ \isacommand{from}\isamarkupfalse%
\ h{\isadigit{3}}\ \isakeyword{and}\ h{\isadigit{4}}\ \isakeyword{and}\ h{\isadigit{5}}\ \isakeyword{and}\ h{\isadigit{6}}\ \isakeyword{and}\ h{\isadigit{7}}\ \isakeyword{and}\ h{\isadigit{1}}{\isadigit{2}}\ \isakeyword{and}\ h{\isadigit{1}}{\isadigit{3}}\ \isakeyword{and}\ h{\isadigit{1}}{\isadigit{4}}\ \isakeyword{and}\ sg{\isadigit{6}}\ \isacommand{have}\isamarkupfalse%
\ sg{\isadigit{1}}{\isadigit{1}}{\isacharcolon}\isanewline
\ \ \ {\isachardoublequoteopen}i\ {\isacharparenleft}t\ {\isacharplus}\ {\isadigit{3}}\ {\isacharplus}\ k\ {\isacharplus}\ d{\isacharparenright}\ {\isasymnoteq}\ {\isacharbrackleft}{\isacharbrackright}{\isachardoublequoteclose}\isanewline
\ \ \ \ \isacommand{by}\isamarkupfalse%
\ {\isacharparenleft}simp\ add{\isacharcolon}\ GatewayReq{\isacharunderscore}L{\isadigit{2}}{\isacharparenright}\isanewline
\ \ \isanewline
\ \ \isacommand{from}\isamarkupfalse%
\ h{\isadigit{1}}{\isadigit{1}}\ \isacommand{have}\isamarkupfalse%
\ sg{\isadigit{1}}{\isadigit{2}}{\isacharcolon}{\isachardoublequoteopen}{\isasymforall}m\ {\isacharless}\ k\ {\isacharplus}\ {\isadigit{3}}{\isachardot}\ req\ {\isacharparenleft}t\ {\isacharplus}\ m{\isacharparenright}\ {\isasymnoteq}\ {\isacharbrackleft}send{\isacharbrackright}{\isachardoublequoteclose}\ \isacommand{by}\isamarkupfalse%
\ auto\isanewline
\ \ \isacommand{from}\isamarkupfalse%
\ h{\isadigit{1}}{\isadigit{4}}\ \isacommand{have}\isamarkupfalse%
\ sg{\isadigit{1}}{\isadigit{3}}{\isacharcolon}{\isachardoublequoteopen}{\isasymforall}j{\isasymle}k\ {\isacharplus}\ d\ {\isacharplus}\ {\isadigit{3}}{\isachardot}\ lose\ {\isacharparenleft}t\ {\isacharplus}\ j{\isacharparenright}\ {\isacharequal}\ {\isacharbrackleft}False{\isacharbrackright}{\isachardoublequoteclose}\ \isacommand{by}\isamarkupfalse%
\ auto\isanewline
\ \ \isacommand{from}\isamarkupfalse%
\ h{\isadigit{1}}\ \isakeyword{and}\ h{\isadigit{7}}\ \isakeyword{and}\ h{\isadigit{6}}\ \isakeyword{and}\ h{\isadigit{5}}\ \isakeyword{and}\ h{\isadigit{4}}\ \isakeyword{and}\ h{\isadigit{9}}\ \isakeyword{and}\ sg{\isadigit{1}}{\isadigit{2}}\ \isakeyword{and}\ h{\isadigit{1}}{\isadigit{2}}\ \isakeyword{and}\ h{\isadigit{8}}\ \isakeyword{and}\ sg{\isadigit{1}}{\isadigit{3}}\ \isakeyword{and}\ h{\isadigit{1}}{\isadigit{0}}\isanewline
\ \ \ \ \isacommand{have}\isamarkupfalse%
\ sg{\isadigit{1}}{\isadigit{4}}{\isacharcolon}{\isachardoublequoteopen}{\isasymforall}\ t{\isadigit{2}}\ {\isacharless}\ {\isacharparenleft}t\ {\isacharplus}\ {\isadigit{3}}\ {\isacharplus}\ k\ {\isacharplus}\ d{\isacharparenright}{\isachardot}\ i\ t{\isadigit{2}}\ {\isacharequal}\ {\isacharbrackleft}{\isacharbrackright}{\isachardoublequoteclose}\isanewline
\ \ \ \ \isacommand{by}\isamarkupfalse%
\ {\isacharparenleft}simp\ add{\isacharcolon}\ Gateway{\isacharunderscore}L{\isadigit{7}}{\isacharparenright}\ \isanewline
\ \ \isacommand{from}\isamarkupfalse%
\ sg{\isadigit{1}}{\isadigit{4}}\ \isakeyword{and}\ h{\isadigit{2}}\ \isacommand{have}\isamarkupfalse%
\ sg{\isadigit{1}}{\isadigit{5}}{\isacharcolon}{\isachardoublequoteopen}{\isasymforall}\ t{\isadigit{3}}\ {\isasymle}\ {\isacharparenleft}t\ {\isacharplus}\ {\isadigit{3}}\ {\isacharplus}\ k\ {\isacharplus}\ d{\isacharparenright}{\isachardot}\ a\ t{\isadigit{3}}\ {\isacharequal}\ {\isacharbrackleft}{\isacharbrackright}{\isachardoublequoteclose}\isanewline
\ \ \ \ \isacommand{by}\isamarkupfalse%
\ {\isacharparenleft}simp\ add{\isacharcolon}\ ServiceCenter{\isacharunderscore}L{\isadigit{2}}{\isacharparenright}\isanewline
\ \ \isacommand{from}\isamarkupfalse%
\ h{\isadigit{1}}{\isadigit{4}}\ \isacommand{have}\isamarkupfalse%
\ sg{\isadigit{1}}{\isadigit{8}}{\isacharcolon}{\isachardoublequoteopen}{\isasymforall}j{\isasymle}{\isadigit{2}}\ {\isacharasterisk}\ d{\isachardot}\ lose\ {\isacharparenleft}{\isacharparenleft}t\ {\isacharplus}\ {\isadigit{3}}\ {\isacharplus}\ k{\isacharparenright}\ {\isacharplus}\ j{\isacharparenright}\ {\isacharequal}\ {\isacharbrackleft}False{\isacharbrackright}{\isachardoublequoteclose}\ \isanewline
\ \ \ \ \isacommand{by}\isamarkupfalse%
\ {\isacharparenleft}simp\ add{\isacharcolon}\ streamValue{\isadigit{4}}{\isadigit{3}}{\isacharparenright}\isanewline
\ \ \isacommand{from}\isamarkupfalse%
\ h{\isadigit{1}}{\isadigit{4}}\ \ \isacommand{have}\isamarkupfalse%
\ sg{\isadigit{1}}{\isadigit{6}}a{\isacharcolon}{\isachardoublequoteopen}{\isasymforall}j{\isasymle}{\isadigit{2}}\ {\isacharasterisk}\ d{\isachardot}\ lose\ {\isacharparenleft}t\ {\isacharplus}\ j\ {\isacharplus}\ {\isacharparenleft}{\isadigit{4}}\ {\isacharplus}\ k{\isacharparenright}{\isacharparenright}\ {\isacharequal}\ {\isacharbrackleft}False{\isacharbrackright}{\isachardoublequoteclose}\ \isanewline
\ \ \ \ \isacommand{by}\isamarkupfalse%
\ {\isacharparenleft}simp\ add{\isacharcolon}\ streamValue{\isadigit{2}}{\isacharparenright}\isanewline
\ \ \isacommand{have}\isamarkupfalse%
\ sg{\isadigit{1}}{\isadigit{6}}b{\isacharcolon}{\isachardoublequoteopen}Suc\ {\isacharparenleft}{\isadigit{3}}\ {\isacharplus}\ k{\isacharparenright}\ {\isacharequal}\ {\isacharparenleft}{\isadigit{4}}\ {\isacharplus}\ k{\isacharparenright}{\isachardoublequoteclose}\ \isacommand{by}\isamarkupfalse%
\ arith\isanewline
\ \ \isacommand{from}\isamarkupfalse%
\ sg{\isadigit{1}}{\isadigit{6}}a\ \isakeyword{and}\ sg{\isadigit{1}}{\isadigit{6}}b\ \isacommand{have}\isamarkupfalse%
\ sg{\isadigit{1}}{\isadigit{6}}{\isacharcolon}{\isachardoublequoteopen}{\isasymforall}j{\isasymle}{\isadigit{2}}\ {\isacharasterisk}\ d{\isachardot}\ lose\ {\isacharparenleft}t\ {\isacharplus}\ j\ {\isacharplus}\ Suc\ {\isacharparenleft}{\isadigit{3}}\ {\isacharplus}\ k{\isacharparenright}{\isacharparenright}\ {\isacharequal}\ {\isacharbrackleft}False{\isacharbrackright}{\isachardoublequoteclose}\ \isanewline
\ \ \ \ \isacommand{by}\isamarkupfalse%
\ {\isacharparenleft}simp\ {\isacharparenleft}no{\isacharunderscore}asm{\isacharunderscore}simp{\isacharparenright}{\isacharparenright}\ \ \isanewline
\ \ \isacommand{from}\isamarkupfalse%
\ h{\isadigit{1}}\ \isakeyword{and}\ h{\isadigit{4}}\ \isakeyword{and}\ h{\isadigit{5}}\ \isakeyword{and}\ h{\isadigit{6}}\ \isakeyword{and}\ h{\isadigit{7}}\ \isakeyword{and}\ sg{\isadigit{1}}{\isadigit{8}}\ \isakeyword{and}\ sg{\isadigit{1}}{\isadigit{0}}\ \isakeyword{and}\ sg{\isadigit{1}}{\isadigit{5}}\ \isacommand{have}\isamarkupfalse%
\ sg{\isadigit{1}}{\isadigit{9}}{\isacharcolon}\isanewline
\ \ \ \ {\isachardoublequoteopen}{\isasymforall}x\ {\isasymle}\ d\ {\isacharplus}\ d{\isachardot}\ ack\ {\isacharparenleft}t\ {\isacharplus}\ {\isadigit{3}}\ {\isacharplus}\ k\ {\isacharplus}\ x{\isacharparenright}\ {\isacharequal}\ {\isacharbrackleft}sending{\isacharunderscore}data{\isacharbrackright}{\isachardoublequoteclose}\isanewline
\ \ \ \ \ \isacommand{by}\isamarkupfalse%
\ {\isacharparenleft}simp\ add{\isacharcolon}\ Gateway{\isacharunderscore}L{\isadigit{8}}{\isacharparenright}\isanewline
\ \ \isacommand{from}\isamarkupfalse%
\ sg{\isadigit{1}}{\isadigit{9}}\ \isacommand{have}\isamarkupfalse%
\ sg{\isadigit{1}}{\isadigit{9}}a{\isacharcolon}{\isachardoublequoteopen}ack\ {\isacharparenleft}t\ {\isacharplus}\ {\isadigit{3}}\ {\isacharplus}\ k\ {\isacharplus}\ d\ {\isacharplus}\ d{\isacharparenright}\ {\isacharequal}\ {\isacharbrackleft}sending{\isacharunderscore}data{\isacharbrackright}{\isachardoublequoteclose}\ \isacommand{by}\isamarkupfalse%
\ auto\isanewline
\ \ \isacommand{from}\isamarkupfalse%
\ sg{\isadigit{1}}{\isadigit{6}}\ \isacommand{have}\isamarkupfalse%
\ sg{\isadigit{2}}{\isadigit{0}}a{\isacharcolon}{\isachardoublequoteopen}{\isasymforall}j{\isasymle}\ d{\isachardot}\ lose\ {\isacharparenleft}t\ {\isacharplus}\ {\isadigit{3}}\ {\isacharplus}\ k\ {\isacharplus}\ d\ {\isacharplus}\ {\isacharparenleft}Suc\ j{\isacharparenright}{\isacharparenright}\ {\isacharequal}\ {\isacharbrackleft}False{\isacharbrackright}{\isachardoublequoteclose}\ \isanewline
\ \ \ \ \isacommand{by}\isamarkupfalse%
\ {\isacharparenleft}rule\ streamValue{\isadigit{1}}{\isadigit{0}}{\isacharparenright}\ \isanewline
\ \ \isacommand{have}\isamarkupfalse%
\ sg{\isadigit{2}}{\isadigit{0}}b{\isacharcolon}{\isachardoublequoteopen}{\isadigit{3}}\ {\isacharplus}\ k\ {\isacharplus}\ d\ {\isasymle}\ {\isadigit{2}}\ {\isacharasterisk}\ d\ {\isacharplus}\ {\isacharparenleft}{\isadigit{4}}\ {\isacharplus}\ k{\isacharparenright}{\isachardoublequoteclose}\ \isacommand{by}\isamarkupfalse%
\ arith\isanewline
\ \ \isacommand{from}\isamarkupfalse%
\ h{\isadigit{1}}{\isadigit{4}}\ \isakeyword{and}\ sg{\isadigit{2}}{\isadigit{0}}b\ \isacommand{have}\isamarkupfalse%
\ sg{\isadigit{2}}{\isadigit{0}}c{\isacharcolon}{\isachardoublequoteopen}lose\ {\isacharparenleft}t\ {\isacharplus}\ {\isadigit{3}}\ {\isacharplus}\ k\ {\isacharplus}\ d{\isacharparenright}\ {\isacharequal}\ {\isacharbrackleft}False{\isacharbrackright}{\isachardoublequoteclose}\isanewline
\ \ \ \ \isacommand{by}\isamarkupfalse%
\ {\isacharparenleft}rule\ aux{\isadigit{4}}lose{\isadigit{2}}{\isacharparenright}\isanewline
\ \ \isacommand{from}\isamarkupfalse%
\ sg{\isadigit{2}}{\isadigit{0}}a\ \ \isakeyword{and}\ sg{\isadigit{2}}{\isadigit{0}}c\ \isacommand{have}\isamarkupfalse%
\ sg{\isadigit{2}}{\isadigit{0}}{\isacharcolon}{\isachardoublequoteopen}{\isasymforall}j{\isasymle}Suc\ d{\isachardot}\ lose\ {\isacharparenleft}t\ {\isacharplus}\ {\isadigit{3}}\ {\isacharplus}\ k\ {\isacharplus}\ d\ {\isacharplus}\ j{\isacharparenright}\ {\isacharequal}\ {\isacharbrackleft}False{\isacharbrackright}{\isachardoublequoteclose}\ \isanewline
\ \ \ \ \isacommand{by}\isamarkupfalse%
\ {\isacharparenleft}rule\ streamValue{\isadigit{8}}{\isacharparenright}\ \isanewline
\ \ \isacommand{from}\isamarkupfalse%
\ h{\isadigit{4}}\ \isakeyword{and}\ h{\isadigit{5}}\ \isakeyword{and}\ h{\isadigit{6}}\ \isakeyword{and}\ h{\isadigit{7}}\ \isakeyword{and}\ h{\isadigit{3}}\ \isacommand{have}\isamarkupfalse%
\ sg{\isadigit{2}}{\isadigit{1}}{\isacharcolon}\isanewline
\ \ \ \ \ {\isachardoublequoteopen}ack\ {\isacharparenleft}t\ {\isacharplus}\ {\isadigit{3}}\ {\isacharplus}\ k\ {\isacharplus}\ d\ {\isacharplus}\ d{\isacharparenright}\ {\isacharequal}\ {\isacharbrackleft}sending{\isacharunderscore}data{\isacharbrackright}\ {\isasymand}\ \isanewline
\ \ \ \ \ \ a\ {\isacharparenleft}Suc\ {\isacharparenleft}t\ {\isacharplus}\ {\isadigit{3}}\ {\isacharplus}\ k\ {\isacharplus}\ d{\isacharparenright}{\isacharparenright}\ {\isacharequal}\ {\isacharbrackleft}sc{\isacharunderscore}ack{\isacharbrackright}\ {\isasymand}\ \isanewline
\ \ \ \ \ \ {\isacharparenleft}{\isasymforall}j{\isasymle}Suc\ d{\isachardot}\ lose\ {\isacharparenleft}t\ {\isacharplus}\ {\isadigit{3}}\ {\isacharplus}\ k\ {\isacharplus}\ d\ {\isacharplus}\ j{\isacharparenright}\ {\isacharequal}\ {\isacharbrackleft}False{\isacharbrackright}{\isacharparenright}\ {\isasymlongrightarrow}\isanewline
\ \ \ \ \ \ vc\ {\isacharparenleft}Suc\ {\isacharparenleft}t\ {\isacharplus}\ {\isadigit{3}}\ {\isacharplus}\ k\ {\isacharplus}\ d\ {\isacharplus}\ d{\isacharparenright}{\isacharparenright}\ {\isacharequal}\ {\isacharbrackleft}vc{\isacharunderscore}com{\isacharbrackright}{\isachardoublequoteclose}\isanewline
\ \ \ \ \ \isacommand{apply}\isamarkupfalse%
\ {\isacharparenleft}simp\ only{\isacharcolon}\ GatewayReq{\isacharunderscore}def{\isacharparenright}\ \isanewline
\ \ \ \ \ \isacommand{by}\isamarkupfalse%
\ {\isacharparenleft}rule\ GatewaySystem{\isacharunderscore}L{\isadigit{1}}aux{\isacharcomma}\ auto{\isacharparenright}\ \ \isanewline
\ \ \ \isacommand{from}\isamarkupfalse%
\ h{\isadigit{2}}\ \isakeyword{and}\ sg{\isadigit{1}}{\isadigit{1}}\ \isacommand{have}\isamarkupfalse%
\ sg{\isadigit{2}}{\isadigit{2}}{\isacharcolon}{\isachardoublequoteopen}a\ {\isacharparenleft}Suc\ {\isacharparenleft}t\ {\isacharplus}\ {\isadigit{3}}\ {\isacharplus}\ k\ {\isacharplus}\ d{\isacharparenright}{\isacharparenright}\ {\isacharequal}\ {\isacharbrackleft}sc{\isacharunderscore}ack{\isacharbrackright}{\isachardoublequoteclose}\isanewline
\ \ \ \ \ \isacommand{by}\isamarkupfalse%
\ {\isacharparenleft}simp\ only{\isacharcolon}\ ServiceCenter{\isacharunderscore}def{\isacharcomma}\ auto{\isacharparenright}\isanewline
\ \ \ \isacommand{from}\isamarkupfalse%
\ sg{\isadigit{2}}{\isadigit{1}}\ \isakeyword{and}\ sg{\isadigit{1}}{\isadigit{9}}a\ \isakeyword{and}\ sg{\isadigit{2}}{\isadigit{2}}\ \isakeyword{and}\ sg{\isadigit{2}}{\isadigit{0}}\ \isacommand{have}\isamarkupfalse%
\ sg{\isadigit{2}}{\isadigit{3}}{\isacharcolon}\isanewline
\ \ \ \ \ {\isachardoublequoteopen}vc\ {\isacharparenleft}Suc\ {\isacharparenleft}t\ {\isacharplus}\ {\isadigit{3}}\ {\isacharplus}\ k\ {\isacharplus}\ d\ {\isacharplus}\ d{\isacharparenright}{\isacharparenright}\ {\isacharequal}\ {\isacharbrackleft}vc{\isacharunderscore}com{\isacharbrackright}{\isachardoublequoteclose}\ \isacommand{by}\isamarkupfalse%
\ simp\isanewline
\ \ \ \isacommand{have}\isamarkupfalse%
\ sg{\isadigit{2}}{\isadigit{4}}{\isacharcolon}{\isachardoublequoteopen}{\isadigit{2}}\ {\isacharasterisk}\ d\ {\isacharplus}\ {\isacharparenleft}t\ {\isacharplus}\ {\isacharparenleft}{\isadigit{4}}\ {\isacharplus}\ k{\isacharparenright}{\isacharparenright}\ {\isacharequal}\ {\isacharparenleft}Suc\ {\isacharparenleft}t\ {\isacharplus}\ {\isadigit{3}}\ {\isacharplus}\ k\ {\isacharplus}\ d\ {\isacharplus}\ d{\isacharparenright}{\isacharparenright}{\isachardoublequoteclose}\ \isacommand{by}\isamarkupfalse%
\ arith\isanewline
\ \ \ \isacommand{from}\isamarkupfalse%
\ sg{\isadigit{2}}{\isadigit{3}}\ \isakeyword{and}\ sg{\isadigit{2}}{\isadigit{4}}\ \isacommand{show}\isamarkupfalse%
\ {\isacharquery}thesis\ \isanewline
\ \ \ \ \ \isacommand{by}\isamarkupfalse%
\ {\isacharparenleft}simp\ {\isacharparenleft}no{\isacharunderscore}asm{\isacharunderscore}simp{\isacharparenright}{\isacharcomma}\ simp{\isacharparenright}\isanewline
\isacommand{qed}\isamarkupfalse%
\endisatagproof
{\isafoldproof}%
\isadelimproof
\isanewline
\endisadelimproof
\isanewline
\isanewline
\isacommand{lemma}\isamarkupfalse%
\ GatewaySystem{\isacharunderscore}L{\isadigit{3}}{\isacharcolon}\isanewline
\ \isakeyword{assumes}\ h{\isadigit{1}}{\isacharcolon}{\isachardoublequoteopen}Gateway\ req\ dt\ a\ stop\ lose\ d\ ack\ i\ vc{\isachardoublequoteclose}\isanewline
\ \ \ \ \ \isakeyword{and}\ h{\isadigit{2}}{\isacharcolon}{\isachardoublequoteopen}ServiceCenter\ i\ a{\isachardoublequoteclose}\isanewline
\ \ \ \ \ \isakeyword{and}\ h{\isadigit{3}}{\isacharcolon}{\isachardoublequoteopen}GatewayReq\ req\ dt\ a\ stop\ lose\ d\ ack\ i\ vc{\isachardoublequoteclose}\isanewline
\ \ \ \ \ \isakeyword{and}\ h{\isadigit{4}}{\isacharcolon}{\isachardoublequoteopen}msg\ {\isacharparenleft}Suc\ {\isadigit{0}}{\isacharparenright}\ req{\isachardoublequoteclose}\isanewline
\ \ \ \ \ \isakeyword{and}\ h{\isadigit{5}}{\isacharcolon}{\isachardoublequoteopen}msg\ {\isacharparenleft}Suc\ {\isadigit{0}}{\isacharparenright}\ stop{\isachardoublequoteclose}\isanewline
\ \ \ \ \ \isakeyword{and}\ h{\isadigit{6}}{\isacharcolon}{\isachardoublequoteopen}msg\ {\isacharparenleft}Suc\ {\isadigit{0}}{\isacharparenright}\ a{\isachardoublequoteclose}\isanewline
\ \ \ \ \ \isakeyword{and}\ h{\isadigit{7}}{\isacharcolon}{\isachardoublequoteopen}ts\ lose{\isachardoublequoteclose}\isanewline
\ \ \ \ \ \isakeyword{and}\ h{\isadigit{8}}{\isacharcolon}\ {\isachardoublequoteopen}dt\ {\isacharparenleft}Suc\ t{\isacharparenright}\ {\isasymnoteq}\ {\isacharbrackleft}{\isacharbrackright}\ {\isasymor}\ dt\ {\isacharparenleft}Suc\ {\isacharparenleft}Suc\ t{\isacharparenright}{\isacharparenright}\ {\isasymnoteq}\ {\isacharbrackleft}{\isacharbrackright}{\isachardoublequoteclose}\isanewline
\ \ \ \ \ \isakeyword{and}\ h{\isadigit{9}}{\isacharcolon}\ {\isachardoublequoteopen}ack\ t\ {\isacharequal}\ {\isacharbrackleft}init{\isacharunderscore}state{\isacharbrackright}{\isachardoublequoteclose}\isanewline
\ \ \ \ \ \isakeyword{and}\ h{\isadigit{1}}{\isadigit{0}}{\isacharcolon}{\isachardoublequoteopen}req\ {\isacharparenleft}Suc\ t{\isacharparenright}\ {\isacharequal}\ {\isacharbrackleft}init{\isacharbrackright}{\isachardoublequoteclose}\isanewline
\ \ \ \ \ \isakeyword{and}\ h{\isadigit{1}}{\isadigit{1}}{\isacharcolon}{\isachardoublequoteopen}{\isasymforall}t{\isadigit{1}}{\isasymle}t{\isachardot}\ req\ t{\isadigit{1}}\ {\isacharequal}\ {\isacharbrackleft}{\isacharbrackright}{\isachardoublequoteclose}\isanewline
\ \ \ \ \ \isakeyword{and}\ h{\isadigit{1}}{\isadigit{2}}{\isacharcolon}{\isachardoublequoteopen}{\isasymforall}m\ {\isasymle}\ k\ {\isacharplus}\ {\isadigit{2}}{\isachardot}\ req\ {\isacharparenleft}t\ {\isacharplus}\ m{\isacharparenright}\ {\isasymnoteq}\ {\isacharbrackleft}send{\isacharbrackright}{\isachardoublequoteclose}\ \isanewline
\ \ \ \ \ \isakeyword{and}\ h{\isadigit{1}}{\isadigit{3}}{\isacharcolon}{\isachardoublequoteopen}req\ {\isacharparenleft}t\ {\isacharplus}\ {\isadigit{3}}\ {\isacharplus}\ k{\isacharparenright}\ {\isacharequal}\ {\isacharbrackleft}send{\isacharbrackright}{\isachardoublequoteclose}\isanewline
\ \ \ \ \ \isakeyword{and}\ h{\isadigit{1}}{\isadigit{4}}{\isacharcolon}{\isachardoublequoteopen}{\isasymforall}j{\isasymle}{\isadigit{2}}\ {\isacharasterisk}\ d\ {\isacharplus}\ {\isacharparenleft}{\isadigit{4}}\ {\isacharplus}\ k{\isacharparenright}{\isachardot}\ lose\ {\isacharparenleft}t\ {\isacharplus}\ j{\isacharparenright}\ {\isacharequal}\ {\isacharbrackleft}False{\isacharbrackright}{\isachardoublequoteclose}\isanewline
\ \isakeyword{shows}\ {\isachardoublequoteopen}vc\ {\isacharparenleft}{\isadigit{2}}\ {\isacharasterisk}\ d\ {\isacharplus}\ {\isacharparenleft}t\ {\isacharplus}\ {\isacharparenleft}{\isadigit{4}}\ {\isacharplus}\ k{\isacharparenright}{\isacharparenright}{\isacharparenright}\ {\isacharequal}\ {\isacharbrackleft}vc{\isacharunderscore}com{\isacharbrackright}{\isachardoublequoteclose}\isanewline
\isadelimproof
\endisadelimproof
\isatagproof
\isacommand{proof}\isamarkupfalse%
\ {\isacharminus}\isanewline
\ \ \isacommand{have}\isamarkupfalse%
\ {\isachardoublequoteopen}Suc\ {\isadigit{0}}\ {\isasymle}\ {\isadigit{2}}\ {\isacharasterisk}\ d\ {\isacharplus}\ {\isacharparenleft}{\isadigit{4}}\ {\isacharplus}\ k{\isacharparenright}{\isachardoublequoteclose}\ \isacommand{by}\isamarkupfalse%
\ arith\isanewline
\ \ \isacommand{from}\isamarkupfalse%
\ h{\isadigit{1}}{\isadigit{4}}\ \isakeyword{and}\ this\ \isacommand{have}\isamarkupfalse%
\ {\isachardoublequoteopen}lose\ {\isacharparenleft}t\ {\isacharplus}\ Suc\ {\isadigit{0}}{\isacharparenright}\ {\isacharequal}\ {\isacharbrackleft}False{\isacharbrackright}{\isachardoublequoteclose}\ \isacommand{by}\isamarkupfalse%
\ blast\isanewline
\ \ \isacommand{from}\isamarkupfalse%
\ this\ \isacommand{have}\isamarkupfalse%
\ sg{\isadigit{1}}{\isacharcolon}{\isachardoublequoteopen}lose\ {\isacharparenleft}Suc\ t{\isacharparenright}\ {\isacharequal}\ {\isacharbrackleft}False{\isacharbrackright}{\isachardoublequoteclose}\ \isacommand{by}\isamarkupfalse%
\ simp\isanewline
\ \ \isacommand{have}\isamarkupfalse%
\ {\isachardoublequoteopen}Suc\ {\isacharparenleft}Suc\ {\isadigit{0}}{\isacharparenright}\ {\isasymle}\ {\isadigit{2}}\ {\isacharasterisk}\ d\ {\isacharplus}\ {\isacharparenleft}{\isadigit{4}}\ {\isacharplus}\ k{\isacharparenright}{\isachardoublequoteclose}\ \isacommand{by}\isamarkupfalse%
\ arith\isanewline
\ \ \isacommand{from}\isamarkupfalse%
\ h{\isadigit{1}}{\isadigit{4}}\ \isakeyword{and}\ this\ \isacommand{have}\isamarkupfalse%
\ {\isachardoublequoteopen}lose\ {\isacharparenleft}t\ {\isacharplus}\ Suc\ {\isacharparenleft}Suc\ {\isadigit{0}}{\isacharparenright}{\isacharparenright}\ {\isacharequal}\ {\isacharbrackleft}False{\isacharbrackright}{\isachardoublequoteclose}\ \isacommand{by}\isamarkupfalse%
\ blast\isanewline
\ \ \isacommand{from}\isamarkupfalse%
\ this\ \isacommand{have}\isamarkupfalse%
\ sg{\isadigit{2}}{\isacharcolon}{\isachardoublequoteopen}lose\ {\isacharparenleft}Suc\ {\isacharparenleft}Suc\ t{\isacharparenright}{\isacharparenright}\ {\isacharequal}\ {\isacharbrackleft}False{\isacharbrackright}{\isachardoublequoteclose}\ \isacommand{by}\isamarkupfalse%
\ simp\ \isanewline
\ \ \isacommand{from}\isamarkupfalse%
\ h{\isadigit{3}}\ \isakeyword{and}\ h{\isadigit{4}}\ \isakeyword{and}\ h{\isadigit{5}}\ \isakeyword{and}\ h{\isadigit{6}}\ \isakeyword{and}\ h{\isadigit{7}}\ \isakeyword{and}\ h{\isadigit{1}}{\isadigit{0}}\ \isakeyword{and}\ h{\isadigit{9}}\ \isakeyword{and}\ sg{\isadigit{1}}\ \isakeyword{and}\ sg{\isadigit{2}}\ \isacommand{have}\isamarkupfalse%
\ sg{\isadigit{3}}{\isacharcolon}\isanewline
\ \ \ {\isachardoublequoteopen}ack\ {\isacharparenleft}t\ {\isacharplus}\ {\isadigit{2}}{\isacharparenright}\ {\isacharequal}\ {\isacharbrackleft}connection{\isacharunderscore}ok{\isacharbrackright}{\isachardoublequoteclose}\ \isanewline
\ \ \ \ \isacommand{by}\isamarkupfalse%
\ {\isacharparenleft}simp\ add{\isacharcolon}\ GatewayReq{\isacharunderscore}def{\isacharparenright}\isanewline
\ \ \isacommand{from}\isamarkupfalse%
\ h{\isadigit{1}}{\isadigit{4}}\ \isacommand{have}\isamarkupfalse%
\ sg{\isadigit{4}}{\isacharcolon}{\isachardoublequoteopen}\ {\isasymforall}j{\isasymle}k{\isachardot}\ lose\ {\isacharparenleft}t\ {\isacharplus}\ {\isadigit{2}}\ {\isacharplus}\ j{\isacharparenright}\ {\isacharequal}\ {\isacharbrackleft}False{\isacharbrackright}{\isachardoublequoteclose}\ \isanewline
\ \ \ \ \isacommand{by}\isamarkupfalse%
\ {\isacharparenleft}clarify{\isacharcomma}\ rule\ aux{\isadigit{4}}lose{\isadigit{1}}{\isacharcomma}\ simp{\isacharparenright}\ \isanewline
\ \ \isacommand{from}\isamarkupfalse%
\ h{\isadigit{1}}{\isadigit{2}}\ \isacommand{have}\isamarkupfalse%
\ sg{\isadigit{5}}{\isacharcolon}{\isachardoublequoteopen}{\isasymforall}m\ {\isasymle}\ k{\isachardot}\ req\ {\isacharparenleft}t\ {\isacharplus}\ {\isadigit{2}}\ {\isacharplus}\ m{\isacharparenright}\ {\isasymnoteq}\ {\isacharbrackleft}send{\isacharbrackright}{\isachardoublequoteclose}\ \isanewline
\ \ \ \ \isacommand{by}\isamarkupfalse%
\ {\isacharparenleft}clarify{\isacharcomma}\ rule\ aux{\isadigit{4}}req{\isacharcomma}\ auto{\isacharparenright}\ \ \isanewline
\ \ \isanewline
\ \ \isacommand{from}\isamarkupfalse%
\ h{\isadigit{1}}\ \isakeyword{and}\ sg{\isadigit{5}}\ \isakeyword{and}\ sg{\isadigit{4}}\ \isakeyword{and}\ sg{\isadigit{3}}\ \isakeyword{and}\ h{\isadigit{4}}\ \isakeyword{and}\ h{\isadigit{5}}\ \isakeyword{and}\ h{\isadigit{6}}\ \isakeyword{and}\ h{\isadigit{7}}\ \isacommand{have}\isamarkupfalse%
\ sg{\isadigit{6}}{\isacharcolon}\isanewline
\ \ \ {\isachardoublequoteopen}{\isasymforall}m\ {\isasymle}\ k{\isachardot}\ ack\ {\isacharparenleft}t\ {\isacharplus}\ {\isadigit{2}}\ {\isacharplus}\ m{\isacharparenright}\ {\isacharequal}\ {\isacharbrackleft}connection{\isacharunderscore}ok{\isacharbrackright}{\isachardoublequoteclose}\isanewline
\ \ \ \ \isacommand{by}\isamarkupfalse%
\ {\isacharparenleft}rule\ Gateway{\isacharunderscore}L{\isadigit{6}}{\isacharparenright}\isanewline
\ \ \isacommand{from}\isamarkupfalse%
\ sg{\isadigit{6}}\ \isacommand{have}\isamarkupfalse%
\ sg{\isadigit{6}}a{\isacharcolon}{\isachardoublequoteopen}ack\ {\isacharparenleft}t\ {\isacharplus}\ {\isadigit{2}}\ {\isacharplus}\ k{\isacharparenright}\ {\isacharequal}\ {\isacharbrackleft}connection{\isacharunderscore}ok{\isacharbrackright}{\isachardoublequoteclose}\ \isacommand{by}\isamarkupfalse%
\ simp\isanewline
\ \ \isanewline
\ \ \isacommand{from}\isamarkupfalse%
\ h{\isadigit{3}}\ \isakeyword{and}\ h{\isadigit{4}}\ \isakeyword{and}\ h{\isadigit{5}}\ \isakeyword{and}\ h{\isadigit{6}}\ \isakeyword{and}\ h{\isadigit{7}}\ \isakeyword{and}\ h{\isadigit{1}}{\isadigit{3}}\ \isakeyword{and}\ h{\isadigit{1}}{\isadigit{4}}\ \isakeyword{and}\ sg{\isadigit{6}}\ \isacommand{have}\isamarkupfalse%
\ sg{\isadigit{1}}{\isadigit{0}}{\isacharcolon}\isanewline
\ \ \ {\isachardoublequoteopen}ack\ {\isacharparenleft}t\ {\isacharplus}\ {\isadigit{3}}\ {\isacharplus}\ k{\isacharparenright}\ {\isacharequal}\ {\isacharbrackleft}sending{\isacharunderscore}data{\isacharbrackright}{\isachardoublequoteclose}\isanewline
\ \ \ \ \isacommand{by}\isamarkupfalse%
\ {\isacharparenleft}simp\ add{\isacharcolon}\ GatewayReq{\isacharunderscore}L{\isadigit{1}}{\isacharparenright}\isanewline
\ \ \isacommand{from}\isamarkupfalse%
\ h{\isadigit{3}}\ \isakeyword{and}\ h{\isadigit{4}}\ \isakeyword{and}\ h{\isadigit{5}}\ \isakeyword{and}\ h{\isadigit{6}}\ \isakeyword{and}\ h{\isadigit{7}}\ \isacommand{have}\isamarkupfalse%
\ sg{\isadigit{1}}{\isadigit{1}}a{\isacharcolon}\isanewline
\ \ \ {\isachardoublequoteopen}ack\ {\isacharparenleft}t\ {\isacharplus}\ {\isadigit{2}}\ {\isacharplus}\ k{\isacharparenright}\ {\isacharequal}\ {\isacharbrackleft}connection{\isacharunderscore}ok{\isacharbrackright}\ {\isasymand}\isanewline
\ \ \ \ req\ {\isacharparenleft}Suc\ {\isacharparenleft}t\ {\isacharplus}\ {\isadigit{2}}\ {\isacharplus}\ k{\isacharparenright}{\isacharparenright}\ {\isacharequal}\ {\isacharbrackleft}send{\isacharbrackright}\ {\isasymand}\ \isanewline
\ \ \ \ {\isacharparenleft}{\isasymforall}j{\isasymle}Suc\ d{\isachardot}\ lose\ {\isacharparenleft}{\isacharparenleft}t\ {\isacharplus}\ {\isadigit{2}}\ {\isacharplus}\ k{\isacharparenright}\ {\isacharplus}\ j{\isacharparenright}\ {\isacharequal}\ {\isacharbrackleft}False{\isacharbrackright}{\isacharparenright}\ {\isasymlongrightarrow}\isanewline
\ \ \ \ i\ {\isacharparenleft}Suc\ {\isacharparenleft}t\ {\isacharplus}\ {\isacharparenleft}{\isadigit{2}}{\isacharcolon}{\isacharcolon}nat{\isacharparenright}\ {\isacharplus}\ k\ {\isacharplus}\ d{\isacharparenright}{\isacharparenright}\ {\isacharequal}\ inf{\isacharunderscore}last{\isacharunderscore}ti\ dt\ {\isacharparenleft}t\ {\isacharplus}\ {\isadigit{2}}\ {\isacharplus}\ k{\isacharparenright}{\isachardoublequoteclose}\isanewline
\ \ \ \ \isacommand{apply}\isamarkupfalse%
\ {\isacharparenleft}simp\ only{\isacharcolon}\ GatewayReq{\isacharunderscore}def{\isacharparenright}\isanewline
\ \ \ \ \isacommand{by}\isamarkupfalse%
\ {\isacharparenleft}rule\ GatewaySystem{\isacharunderscore}L{\isadigit{3}}aux{\isacharcomma}\ auto{\isacharparenright}\ \isanewline
\ \ \isacommand{have}\isamarkupfalse%
\ sg{\isadigit{1}}{\isadigit{2}}{\isacharcolon}{\isachardoublequoteopen}Suc\ {\isacharparenleft}t\ {\isacharplus}\ {\isadigit{2}}\ {\isacharplus}\ k{\isacharparenright}\ {\isacharequal}\ t\ {\isacharplus}\ {\isadigit{3}}\ {\isacharplus}\ k{\isachardoublequoteclose}\ \isacommand{by}\isamarkupfalse%
\ arith\isanewline
\ \ \isacommand{from}\isamarkupfalse%
\ h{\isadigit{1}}{\isadigit{3}}\ \isakeyword{and}\ sg{\isadigit{1}}{\isadigit{2}}\ \isacommand{have}\isamarkupfalse%
\ sg{\isadigit{1}}{\isadigit{2}}a{\isacharcolon}{\isachardoublequoteopen}req\ {\isacharparenleft}Suc\ {\isacharparenleft}t\ {\isacharplus}\ {\isadigit{2}}\ {\isacharplus}\ k{\isacharparenright}{\isacharparenright}\ {\isacharequal}\ {\isacharbrackleft}send{\isacharbrackright}{\isachardoublequoteclose}\ \isanewline
\ \ \ \ \isacommand{by}\isamarkupfalse%
\ {\isacharparenleft}simp\ add{\isacharcolon}\ eval{\isacharunderscore}nat{\isacharunderscore}numeral{\isacharparenright}\isanewline
\ \ \isacommand{from}\isamarkupfalse%
\ h{\isadigit{1}}{\isadigit{4}}\ \isacommand{have}\isamarkupfalse%
\ sg{\isadigit{1}}{\isadigit{3}}{\isacharcolon}{\isachardoublequoteopen}{\isasymforall}j{\isasymle}Suc\ d{\isachardot}\ lose\ {\isacharparenleft}{\isacharparenleft}t\ {\isacharplus}\ {\isadigit{2}}\ {\isacharplus}\ k{\isacharparenright}\ {\isacharplus}\ j{\isacharparenright}\ {\isacharequal}\ {\isacharbrackleft}False{\isacharbrackright}{\isachardoublequoteclose}\isanewline
\ \ \ \ \isacommand{by}\isamarkupfalse%
\ {\isacharparenleft}rule\ streamValue{\isadigit{1}}{\isadigit{2}}{\isacharparenright}\ \isanewline
\ \ \isacommand{from}\isamarkupfalse%
\ sg{\isadigit{1}}{\isadigit{1}}a\ \isakeyword{and}\ sg{\isadigit{6}}a\ \isakeyword{and}\ h{\isadigit{1}}{\isadigit{3}}\ \isakeyword{and}\ sg{\isadigit{1}}{\isadigit{2}}a\ \isakeyword{and}\ sg{\isadigit{1}}{\isadigit{3}}\ \isacommand{have}\isamarkupfalse%
\ sg{\isadigit{1}}{\isadigit{4}}{\isacharcolon}\isanewline
\ \ \ \ {\isachardoublequoteopen}i\ {\isacharparenleft}Suc\ {\isacharparenleft}t\ {\isacharplus}\ {\isacharparenleft}{\isadigit{2}}{\isacharcolon}{\isacharcolon}nat{\isacharparenright}\ {\isacharplus}\ k\ {\isacharplus}\ d{\isacharparenright}{\isacharparenright}\ {\isacharequal}\ inf{\isacharunderscore}last{\isacharunderscore}ti\ dt\ {\isacharparenleft}t\ {\isacharplus}\ {\isadigit{2}}\ {\isacharplus}\ k{\isacharparenright}{\isachardoublequoteclose}\ \isacommand{by}\isamarkupfalse%
\ simp\isanewline
\ \ \isacommand{from}\isamarkupfalse%
\ h{\isadigit{8}}\ \isacommand{have}\isamarkupfalse%
\ sg{\isadigit{1}}{\isadigit{5}}{\isacharcolon}{\isachardoublequoteopen}inf{\isacharunderscore}last{\isacharunderscore}ti\ dt\ {\isacharparenleft}t\ {\isacharplus}\ {\isadigit{2}}\ {\isacharplus}\ k{\isacharparenright}\ {\isasymnoteq}\ {\isacharbrackleft}{\isacharbrackright}{\isachardoublequoteclose}\ \isanewline
\ \ \ \ \isacommand{by}\isamarkupfalse%
\ {\isacharparenleft}rule\ inf{\isacharunderscore}last{\isacharunderscore}ti{\isacharunderscore}Suc{\isadigit{2}}{\isacharparenright}\ \isanewline
\ \ \isacommand{from}\isamarkupfalse%
\ sg{\isadigit{1}}{\isadigit{4}}\ \isakeyword{and}\ sg{\isadigit{1}}{\isadigit{5}}\ \isacommand{have}\isamarkupfalse%
\ sg{\isadigit{1}}{\isadigit{6}}{\isacharcolon}\ {\isachardoublequoteopen}i\ {\isacharparenleft}t\ {\isacharplus}\ {\isadigit{3}}\ {\isacharplus}\ k\ {\isacharplus}\ d{\isacharparenright}\ {\isasymnoteq}\ {\isacharbrackleft}{\isacharbrackright}{\isachardoublequoteclose}\isanewline
\ \ \ \ \isacommand{by}\isamarkupfalse%
\ {\isacharparenleft}simp\ add{\isacharcolon}\ arith{\isacharunderscore}sum{\isadigit{4}}{\isacharparenright}\isanewline
\ \ \isanewline
\ \ \isacommand{from}\isamarkupfalse%
\ h{\isadigit{1}}{\isadigit{4}}\ \isacommand{have}\isamarkupfalse%
\ sg{\isadigit{1}}{\isadigit{7}}{\isacharcolon}{\isachardoublequoteopen}{\isasymforall}j{\isasymle}k\ {\isacharplus}\ d\ {\isacharplus}\ {\isadigit{3}}{\isachardot}\ lose\ {\isacharparenleft}t\ {\isacharplus}\ j{\isacharparenright}\ {\isacharequal}\ {\isacharbrackleft}False{\isacharbrackright}{\isachardoublequoteclose}\ \isacommand{by}\isamarkupfalse%
\ auto\isanewline
\ \ \isacommand{from}\isamarkupfalse%
\ h{\isadigit{1}}{\isadigit{2}}\ \isacommand{have}\isamarkupfalse%
\ sg{\isadigit{1}}{\isadigit{8}}{\isacharcolon}{\isachardoublequoteopen}{\isasymforall}m\ {\isacharless}\ {\isacharparenleft}k\ {\isacharplus}\ {\isadigit{3}}{\isacharparenright}{\isachardot}\ req\ {\isacharparenleft}t\ {\isacharplus}\ m{\isacharparenright}\ {\isasymnoteq}\ {\isacharbrackleft}send{\isacharbrackright}{\isachardoublequoteclose}\ \ \ \isacommand{by}\isamarkupfalse%
\ auto\isanewline
\ \ \isacommand{from}\isamarkupfalse%
\ h{\isadigit{1}}\ \isakeyword{and}\ h{\isadigit{4}}\ \isakeyword{and}\ h{\isadigit{5}}\ \isakeyword{and}\ h{\isadigit{6}}\ \isakeyword{and}\ h{\isadigit{7}}\ \isakeyword{and}\ h{\isadigit{1}}{\isadigit{0}}\ \isakeyword{and}\ sg{\isadigit{1}}{\isadigit{8}}\ \isakeyword{and}\ h{\isadigit{1}}{\isadigit{3}}\ \isakeyword{and}\ h{\isadigit{9}}\ \isakeyword{and}\ sg{\isadigit{1}}{\isadigit{7}}\ \isakeyword{and}\ h{\isadigit{1}}{\isadigit{1}}\ \isanewline
\ \ \ \ \isacommand{have}\isamarkupfalse%
\ sg{\isadigit{1}}{\isadigit{9}}{\isacharcolon}{\isachardoublequoteopen}{\isasymforall}\ t{\isadigit{2}}\ {\isacharless}\ {\isacharparenleft}t\ {\isacharplus}\ {\isadigit{3}}\ {\isacharplus}\ k\ {\isacharplus}\ d{\isacharparenright}{\isachardot}\ i\ t{\isadigit{2}}\ {\isacharequal}\ {\isacharbrackleft}{\isacharbrackright}{\isachardoublequoteclose}\isanewline
\ \ \ \ \isacommand{by}\isamarkupfalse%
\ {\isacharparenleft}simp\ add{\isacharcolon}\ Gateway{\isacharunderscore}L{\isadigit{7}}{\isacharparenright}\ \isanewline
\ \ \isacommand{from}\isamarkupfalse%
\ h{\isadigit{2}}\ \isakeyword{and}\ sg{\isadigit{1}}{\isadigit{9}}\ \isacommand{have}\isamarkupfalse%
\ sg{\isadigit{2}}{\isadigit{0}}{\isacharcolon}{\isachardoublequoteopen}{\isasymforall}\ t{\isadigit{3}}\ {\isasymle}\ {\isacharparenleft}t\ {\isacharplus}\ {\isadigit{3}}\ {\isacharplus}\ k\ {\isacharplus}\ d{\isacharparenright}{\isachardot}\ a\ t{\isadigit{3}}\ {\isacharequal}\ {\isacharbrackleft}{\isacharbrackright}{\isachardoublequoteclose}\isanewline
\ \ \ \ \isacommand{by}\isamarkupfalse%
\ {\isacharparenleft}simp\ add{\isacharcolon}\ ServiceCenter{\isacharunderscore}L{\isadigit{2}}{\isacharparenright}\isanewline
\ \ \isacommand{from}\isamarkupfalse%
\ h{\isadigit{1}}{\isadigit{4}}\ \isacommand{have}\isamarkupfalse%
\ sg{\isadigit{2}}{\isadigit{1}}{\isacharcolon}{\isachardoublequoteopen}{\isasymforall}j{\isasymle}{\isadigit{2}}\ {\isacharasterisk}\ d{\isachardot}\ lose\ {\isacharparenleft}t\ {\isacharplus}\ {\isadigit{3}}\ {\isacharplus}\ k\ {\isacharplus}\ j{\isacharparenright}\ {\isacharequal}\ {\isacharbrackleft}False{\isacharbrackright}{\isachardoublequoteclose}\isanewline
\ \ \ \ \isacommand{by}\isamarkupfalse%
\ {\isacharparenleft}simp\ add{\isacharcolon}\ streamValue{\isadigit{4}}{\isadigit{3}}{\isacharparenright}\isanewline
\ \ \isacommand{from}\isamarkupfalse%
\ h{\isadigit{1}}\ \isakeyword{and}\ h{\isadigit{4}}\ \isakeyword{and}\ h{\isadigit{5}}\ \isakeyword{and}\ h{\isadigit{6}}\ \isakeyword{and}\ h{\isadigit{7}}\ \isakeyword{and}\ sg{\isadigit{2}}{\isadigit{1}}\ \isakeyword{and}\ sg{\isadigit{1}}{\isadigit{0}}\ \isakeyword{and}\ sg{\isadigit{2}}{\isadigit{0}}\ \isacommand{have}\isamarkupfalse%
\ sg{\isadigit{2}}{\isadigit{2}}{\isacharcolon}\isanewline
\ \ \ \ {\isachardoublequoteopen}{\isasymforall}x\ {\isasymle}\ d\ {\isacharplus}\ d{\isachardot}\ ack\ {\isacharparenleft}t\ {\isacharplus}\ {\isadigit{3}}\ {\isacharplus}\ k\ {\isacharplus}\ x{\isacharparenright}\ {\isacharequal}\ {\isacharbrackleft}sending{\isacharunderscore}data{\isacharbrackright}{\isachardoublequoteclose}\isanewline
\ \ \ \ \isacommand{by}\isamarkupfalse%
\ {\isacharparenleft}simp\ add{\isacharcolon}\ Gateway{\isacharunderscore}L{\isadigit{8}}{\isacharparenright}\isanewline
\ \ \isacommand{from}\isamarkupfalse%
\ h{\isadigit{2}}\ \isakeyword{and}\ h{\isadigit{3}}\ \isakeyword{and}\ h{\isadigit{4}}\ \isakeyword{and}\ h{\isadigit{5}}\ \isakeyword{and}\ h{\isadigit{6}}\ \isakeyword{and}\ h{\isadigit{7}}\ \isakeyword{and}\ h{\isadigit{1}}{\isadigit{4}}\ \isakeyword{and}\ sg{\isadigit{1}}{\isadigit{6}}\ \isakeyword{and}\ sg{\isadigit{2}}{\isadigit{2}}\ \isacommand{show}\isamarkupfalse%
\ {\isacharquery}thesis\isanewline
\ \ \ \ \isacommand{by}\isamarkupfalse%
\ {\isacharparenleft}simp\ add{\isacharcolon}\ GatewaySystem{\isacharunderscore}L{\isadigit{1}}{\isacharparenright}\isanewline
\isacommand{qed}\isamarkupfalse%
\endisatagproof
{\isafoldproof}%
\isadelimproof
\endisadelimproof
\isamarkupsubsection{Proof of the Refinement for the Gateway System%
}
\isamarkuptrue%
\isacommand{lemma}\isamarkupfalse%
\ GatewaySystem{\isacharunderscore}L{\isadigit{0}}{\isacharcolon}\isanewline
\ \isakeyword{assumes}\ h{\isadigit{1}}{\isacharcolon}{\isachardoublequoteopen}GatewaySystem\ req\ dt\ stop\ lose\ d\ ack\ vc{\isachardoublequoteclose}\isanewline
\ \isakeyword{shows}\ \ \ \ \ \ {\isachardoublequoteopen}GatewaySystemReq\ req\ dt\ stop\ lose\ d\ ack\ vc{\isachardoublequoteclose}\isanewline
\isadelimproof
\endisadelimproof
\isatagproof
\isacommand{proof}\isamarkupfalse%
\ {\isacharminus}\ \isanewline
\ \ \isacommand{from}\isamarkupfalse%
\ h{\isadigit{1}}\ \isacommand{obtain}\isamarkupfalse%
\ x\ i\ \isakeyword{where}\isanewline
\ \ \ \ a{\isadigit{1}}{\isacharcolon}{\isachardoublequoteopen}Gateway\ req\ dt\ x\ stop\ lose\ d\ ack\ i\ vc{\isachardoublequoteclose}\ \isakeyword{and}\ \isanewline
\ \ \ \ a{\isadigit{2}}{\isacharcolon}{\isachardoublequoteopen}ServiceCenter\ i\ x{\isachardoublequoteclose}\isanewline
\ \ \ \ \isacommand{by}\isamarkupfalse%
\ {\isacharparenleft}simp\ add{\isacharcolon}\ GatewaySystem{\isacharunderscore}def{\isacharcomma}\ auto{\isacharparenright}\isanewline
\ \ \isacommand{from}\isamarkupfalse%
\ a{\isadigit{1}}\ \isacommand{have}\isamarkupfalse%
\ sg{\isadigit{1}}{\isacharcolon}{\isachardoublequoteopen}GatewayReq\ req\ dt\ x\ stop\ lose\ d\ ack\ i\ vc{\isachardoublequoteclose}\isanewline
\ \ \ \ \isacommand{by}\isamarkupfalse%
\ {\isacharparenleft}simp\ add{\isacharcolon}\ Gateway{\isacharunderscore}L{\isadigit{0}}{\isacharparenright}\isanewline
\ \ \isacommand{from}\isamarkupfalse%
\ a{\isadigit{2}}\ \isacommand{have}\isamarkupfalse%
\ sg{\isadigit{2}}{\isacharcolon}{\isachardoublequoteopen}msg\ {\isacharparenleft}Suc\ {\isadigit{0}}{\isacharparenright}\ x{\isachardoublequoteclose}\isanewline
\ \ \ \ \isacommand{by}\isamarkupfalse%
\ {\isacharparenleft}simp\ add{\isacharcolon}\ ServiceCenter{\isacharunderscore}a{\isacharunderscore}msg{\isacharparenright}\isanewline
\ \ \isacommand{from}\isamarkupfalse%
\ h{\isadigit{1}}\ \isakeyword{and}\ a{\isadigit{1}}\ \isakeyword{and}\ a{\isadigit{2}}\ \isakeyword{and}\ sg{\isadigit{1}}\ \isakeyword{and}\ sg{\isadigit{2}}\ \isacommand{show}\isamarkupfalse%
\ {\isacharquery}thesis\isanewline
\ \ \ \ \isacommand{apply}\isamarkupfalse%
\ {\isacharparenleft}simp\ add{\isacharcolon}\ GatewaySystemReq{\isacharunderscore}def{\isacharcomma}\ auto{\isacharparenright}\isanewline
\ \ \ \ \isacommand{apply}\isamarkupfalse%
\ {\isacharparenleft}simp\ add{\isacharcolon}\ GatewaySystem{\isacharunderscore}L{\isadigit{3}}{\isacharparenright}\isanewline
\ \ \ \ \isacommand{apply}\isamarkupfalse%
\ {\isacharparenleft}simp\ add{\isacharcolon}\ GatewaySystem{\isacharunderscore}L{\isadigit{3}}{\isacharparenright}\isanewline
\ \ \ \ \isacommand{apply}\isamarkupfalse%
\ {\isacharparenleft}simp\ add{\isacharcolon}\ GatewaySystem{\isacharunderscore}L{\isadigit{3}}{\isacharparenright}\isanewline
\ \ \ \ \isacommand{by}\isamarkupfalse%
\ {\isacharparenleft}simp\ add{\isacharcolon}\ GatewaySystem{\isacharunderscore}L{\isadigit{2}}{\isacharparenright}\isanewline
\isacommand{qed}\isamarkupfalse%
\endisatagproof
{\isafoldproof}%
\isadelimproof
\isanewline
\endisadelimproof
\isanewline
\isadelimtheory
\isanewline
\endisadelimtheory
\isatagtheory
\isacommand{end}\isamarkupfalse%
\endisatagtheory
{\isafoldtheory}%
\isadelimtheory
\endisadelimtheory
\end{isabellebody}%

\newpage
\bibliographystyle{abbrv}

\begin{thebibliography}{1}
\bibitem{broy_refinement2}
M.~Broy.
\newblock Compositional refinement of interactive systems modelled by
  relations.
\newblock {\em COMPOS'97: Revised Lectures from the International Symposium on
  Compositionality: The Significant Difference}, pages 130--149, 1998.
  
\bibitem{focus}
M.~Broy and K.~St{\o}len.
\newblock {\em Specification and Development of Interactive Systems: Focus on
  Streams, Interfaces, and Refinement}.
\newblock Springer, 2001.

\bibitem{FlexRayConsortium}
{FlexRay Consortium}.
\newblock \url{http://www.flexray.com}.

\bibitem{FlexRayProt}
{FlexRay Consortium}.
\newblock {\em {FlexRay Communication System - Protocol Specification - Version
  2.0}}, 2004.

\bibitem{efts_book}
C.~K{\"u}hnel and M.~Spichkova.
\newblock {{Fault-Tolerant Communication for Distributed Embedded Systems}}.
\newblock In {\em Software Engineering and Fault Tolerance}, Series on Software
  Engineering and Knowledge Engineering, 2007.
  
\bibitem{kuhnel2006upcoming}
C.~K{\"u}hnel and M.~Spichkova.
\newblock {{Upcoming automotive standards for fault-tolerant communication: FlexRay and OSEKtime FTCom}}.
\newblock In {\em {EFTS  International Workshop on Engineering of Fault Tolerant Systems}}, 2006.

\bibitem{npw}
T.~Nipkow, L.~C. Paulson, and M.~Wenzel.
\newblock {\em {Isabelle/HOL -- A Proof Assistant for Higher-Order Logic}}.
\newblock LNCS. Springer, 2013.


  \bibitem{issec_cps2013}
M. Spichkova, H. Schmidt, and I. Peake.
\newblock {\em {From abstract modelling to remote cyberphysical integration/interoperability testing}}.
  \newblock In {\em{Improving Systems and Software Engineering Conference (iSSEC)}}, 2013.
  
\bibitem{Spichkova_Campetelli2012}
M. Spichkova and A. Campetelli.
\newblock {\em {Towards system development methodologies:
  From software to cyber-physical domain}}.
  \newblock In {\em{International
  Workshop on Formal Techniques for Safety-Critical Systems}}, 2012.
  

  \bibitem{spichkova2012towards}
M. Spichkova.
\newblock {Towards Focus on Time}. 
  \newblock In {\em{12th International Workshop on Automated Verification of Critical Systems (AVoCS'12)}}, 2012.

  \bibitem{ArchReqDecRef}
M. Spichkova.
\newblock { {Architecture: Requirements+ Decomposition+ Refinement}}.
  \newblock In {\em{Softwaretechnik-Trends 31 (4)}}, 2011.
 
 \bibitem{spichkova}
M.~Spichkova.
\newblock {\em {Specification and Seamless Verification of Embedded Real-Time
  Systems: FOCUS on Isabelle}}.
\newblock PhD thesis, 2007.

 \bibitem{spichkova2006flexray}
 M. Spichkova.
\newblock {\em {FlexRay: Verification of the FOCUS Specification in Isabelle/HOL. A Case Study}}.
\newblock {Technische Universit{\"a}t M{\"u}nchen, Tech. Rep.}, TUM-I0602, 2006.

  \bibitem{verisoft} 
\newblock Verisoft project. \url{http://www.verisoft.de}.

\bibitem{IsabelleManual}
M.~Wenzel.
\newblock {\em The Isabelle/Isar Reference Manual}.
\newblock TU M\"unchen, 2013.

\end{thebibliography}

\end{document}